\documentclass[twocolumn]{aastex63}

\shorttitle{A Census of Sub-kpc Scale Metallicity Gradients at High-$z$ from \hst}
\shortauthors{Wang et al. (2020)}

\hypersetup{linkcolor=dgreen,citecolor=lblue,filecolor=cyan,urlcolor=magenta}

\newcommand{\Ntot}{76\xspace}

\newcommand{\Nlowz}{37\xspace}
\newcommand{\Nmedz}{24\xspace}
\newcommand{\Nhiz}{15\xspace}

\newcommand{\NinvERsig}{7\xspace}
\newcommand{\NnegERsig}{15\xspace}

\newcommand{\NinvSANsig}{3\xspace}
\newcommand{\NnegSANsig}{7\xspace}

\newcommand{\Notherm}{35\xspace}
\newcommand{\Notherz}{43\xspace}
\newcommand{\Nall}{111\xspace}

\usepackage[english]{babel} 
\usepackage[utf8]{inputenc} 
\usepackage[T1]{fontenc}    
\usepackage{ae,aecompl}
\usefont{T1}{tnr}{m}{sl}
\usepackage{fancyhdr}       
\usepackage{pgf,pgfarrows,pgfnodes,pgfautomata,pgfheaps}
\usepackage{graphicx}       
\usepackage{natbib}         
\usepackage{url}            
\usepackage{grffile}        
\usepackage{mathtools}      
\usepackage{bbold}          

\usepackage{bbm}            

\usepackage{multirow}       
\usepackage{xspace}         
\usepackage{epstopdf}

\usepackage{amsmath,amssymb,amsxtra,amsfonts}   
\usepackage{txfonts}

\usepackage{pifont}
\usepackage{color}         
\definecolor{gold}{rgb}{1,0.80,0}
\definecolor{orange}{rgb}{1,0.5,0}
\definecolor{midgray}{gray}{0.3}
\definecolor{lblue}{rgb}{0,0.2,0.6}
\definecolor{dgreen}{rgb}{0.1,0.6,0.3}
\definecolor{purple}{rgb}{0.5019607843137255,0.0,0.5019607843137255}

\newcommand{\be}{\begin{equation}}
\newcommand{\ee}{\end{equation}}

\newcommand{\ba}{\begin{align}}
\newcommand{\ea}{\end{align}}

\newcommand{\defeq}{\vcentcolon=}

\newcommand{\Msun}{\ensuremath{M_\odot}\xspace}

\newcommand{\chisq}{\ensuremath{\chi^2}\xspace}

\newcommand{\Mstar}{\ensuremath{M_\ast}\xspace}

\newcommand{\Sstar}{\ensuremath{\Sigma_\ast}\xspace}
\newcommand{\oh}{\ensuremath{12+\log({\rm O/H})}\xspace}
\newcommand{\Av}{\ensuremath{A_{\rm V}}\xspace}
\newcommand{\Rv}{\ensuremath{R_{\rm V}}\xspace}

\newcommand{\SFR}{\ensuremath{{\rm SFR}}\xspace}

\newcommand{\kpc}{\ensuremath{\rm kpc}\xspace}

\newcommand{\K}{\ensuremath{\rm K}\xspace}

\newcommand{\Hunit}{\ensuremath{\rm km~s^{-1}~Mpc^{-1}}\xspace}
\newcommand{\Funit}{\ensuremath{\rm erg~s^{-1}~cm^{-2}}\xspace}

\newcommand{\SBunit}{\ensuremath{\rm erg~s^{-1}~cm^{-2}~arcsec^{-2}}\xspace}

\def\micron{\ensuremath{\mu\textrm{m}}\xspace}  

\newcommand{\Ha}{\textrm{H}\ensuremath{\alpha}\xspace}
\newcommand{\Hb}{\textrm{H}\ensuremath{\beta}\xspace}
\newcommand{\Hg}{\textrm{H}\ensuremath{\gamma}\xspace}
\newcommand{\HII}{\textrm{H}\textsc{ii}\xspace}

\newcommand{\OII}{[\textrm{O}~\textsc{ii}]\xspace}
\newcommand{\OIII}{[\textrm{O}~\textsc{iii}]\xspace}

\newcommand{\NII}{[\textrm{N}~\textsc{ii}]\xspace}
\newcommand{\SII}{[\textrm{S}~\textsc{ii}]\xspace}

\def\U{\ensuremath{U_{360}}\xspace}     
\def\B{\ensuremath{B_{435}}\xspace}
\def\V{\ensuremath{V_{606}}\xspace}
\def\I{\ensuremath{I_{814}}\xspace}
\def\Y{\ensuremath{Y_{105}}\xspace}
\def\J{\ensuremath{J_{125}}\xspace}
\def\JH{\ensuremath{JH_{140}}\xspace}
\def\H{\ensuremath{H_{160}}\xspace}

\newcommand{\cler}{Abell 2744\xspace}
\newcommand{\clsan}{Abell 370\xspace}
\newcommand{\clsi}{MACS0416.1-2403\xspace}
\newcommand{\clwu}{MACS0717.5+3745\xspace}
\newcommand{\clliu}{RXJ2248.7-4431\xspace}
\newcommand{\clqi}{RXJ1347.5-1145\xspace}
\newcommand{\clba}{MACS0744.9+3927\xspace}
\newcommand{\cljiu}{MACS2129.4-0741\xspace}
\newcommand{\clshi}{MACS1423.8+2404\xspace}

\newcommand{\sex}{\textsc{SExtractor}\xspace}
\newcommand{\emc}{\textsc{Emcee}\xspace}
\newcommand{\linmix}{\textsc{linmix}\xspace}
\newcommand{\adriz}{\textsc{AstroDrizzle}\xspace}

\newcommand{\fast}{\textsc{FAST}\xspace}

\newcommand{\SJ}{\textsc{Sharon \& Johnson}\xspace}
\newcommand{\grzl}{\textsc{Grizli}\xspace}

\newcommand{\jwst}{\textit{JWST}\xspace}

\newcommand{\hst}{\textit{HST}\xspace}

\newcommand{\glass}{\textit{GLASS}\xspace}

\newcommand{\clash}{\textit{CLASH}\xspace}

\newcommand{\hff}{\textit{HFF}\xspace}
\newcommand{\muse}{\textit{MUSE}\xspace}
\newcommand{\kmos}{\textit{KMOS}\xspace}
\newcommand{\keck}{\textit{Keck}\xspace}

\newcommand{\kd}{\textit{KMOS}$^{3\rm D}$\xspace}

\def\clash{\textit{CLASH}\xspace}

\newcommand{\vlt}{\textit{VLT}\xspace}
\newcommand{\osiris}{\textit{OSIRIS}\xspace}
\newcommand{\sinf}{\textit{SINFONI}\xspace}

\def\ie{i.e.\xspace}
\def\eg{e.g.\xspace}

\renewcommand\({\left(}
\renewcommand\){\right)}

\newcommand{\el}[1]{\ensuremath{\textrm{EL}_{#1}}}

\def\p{{\rm prior}}

\newcommand{\Om} {\ensuremath{\Omega_{\rm{m}}}\xspace}

\newcommand{\Ol} {\ensuremath{\Omega_{\Lambda}}\xspace}

\newcommand{\n}{\ensuremath{{\nu}\rm}\xspace}

\usepackage{etoolbox}
\makeatletter
\patchcmd{\NAT@citex}
  {\@citea\NAT@hyper@{\NAT@nmfmt{\NAT@nm}\NAT@date}}
  {\@citea\NAT@nmfmt{\NAT@nm}\NAT@hyper@{\NAT@date}}
  {}
  {}
\patchcmd{\NAT@citex}
  {\@citea\NAT@hyper@{%
     \NAT@nmfmt{\NAT@nm}%
     \hyper@natlinkbreak{\NAT@aysep\NAT@spacechar}{\@citeb\@extra@b@citeb}%
     \NAT@date}}
  {\@citea\NAT@nmfmt{\NAT@nm}%
   \NAT@aysep\NAT@spacechar%
   \NAT@hyper@{\NAT@date}}
  {}
  {}
\patchcmd{\NAT@citex}
  {\@citea\NAT@hyper@{%
     \NAT@nmfmt{\NAT@nm}%
     \hyper@natlinkbreak{\NAT@spacechar\NAT@@open\if*#1*\else#1\NAT@spacechar\fi}%
       {\@citeb\@extra@b@citeb}%
     \NAT@date}}
  {\@citea\NAT@nmfmt{\NAT@nm}%
   \NAT@spacechar\NAT@@open\if*#1*\else#1\NAT@spacechar\fi%
   \NAT@hyper@{\NAT@date}}
  {}
  {}
\makeatother

\begin{document}


\title{A Census of Sub-kiloparsec Resolution Metallicity Gradients in Star-forming Galaxies at Cosmic Noon 
from \hst Slitless Spectroscopy}

\correspondingauthor{Xin~Wang}
\email{wangxin@ipac.caltech.edu}

\author{Xin~Wang}
\affil{Infrared Processing and Analysis Center, Caltech, 1200 E. California Blvd., Pasadena, CA 91125, USA}
\affil{Department of Physics and Astronomy, University of California, Los Angeles, CA 90095-1547, USA}

\author{Tucker~A.~Jones}
\affil{University of California Davis, 1 Shields Avenue, Davis, CA 95616, USA}

\author{Tommaso~Treu}
\affil{Department of Physics and Astronomy, University of California, Los Angeles, CA 90095-1547, USA}

\author{Emanuele~Daddi}
\affil{Laboratoire AIM, CEA/DSM-CNRS-Universit\'e Paris Diderot, IRFU/Service d'Astrophysique, B\^at.  709, CEA 
Saclay, F-91191 Gif-sur-Yvette Cedex, France}

\author{Gabriel~B.~Brammer}
\affil{Cosmic Dawn Centre, University of Copenhagen, Blegdamsvej 17, 2100 Copenhagen, Denmark}

\author{Keren~Sharon}
\affil{Department of Astronomy, University of Michigan, 1085 S. University Avenue, Ann Arbor, MI 48109, USA}

\author{Takahiro~Morishita}
\affil{Space Telescope Science Institute, 3700 San Martin Drive, Baltimore, MD, 21218, USA}


\author{Louis~E.~Abramson}
\affil{The Observatories of the Carnegie Institution for Science, 813 Santa Barbara St., Pasadena, CA 91101, USA}

\author{James~W.~Colbert}
\affil{Infrared Processing and Analysis Center, Caltech, 1200 E. California Blvd., Pasadena, CA 91125, USA}

\author{Alaina~L.~Henry}
\affil{Space Telescope Science Institute, 3700 San Martin Drive, Baltimore, MD, 21218, USA}

\author{Philip~F.~Hopkins}
\affil{TAPIR, California Institute of Technology, Pasadena, CA 91125, USA}

\author{Matthew~A.~Malkan}
\affil{Department of Physics and Astronomy, University of California, Los Angeles, CA 90095-1547, USA}

\author{Kasper~B.~Schmidt}
\affil{Leibniz-Institut f\"ur Astrophysik Potsdam (AIP), An der Sternwarte 16, D-14482 Potsdam, Germany}

\author{Harry~I.~Teplitz}
\affil{Infrared Processing and Analysis Center, Caltech, 1200 E. California Blvd., Pasadena, CA 91125, USA}

\author{Benedetta~Vulcani}
\affil{INAF- Osservatorio Astronomico di Padova, Vicolo Osservatorio 5, I-35122 Padova, Italy 0000-0003-0980-1499}


\begin{abstract}
    We present hitherto the largest sample of gas-phase metallicity radial gradients measured at sub-kiloparsec 
    resolution in star-forming galaxies in the redshift range of $z\in[1.2, 2.3]$.  These measurements are 
    enabled by the synergy of slitless spectroscopy from the Hubble Space Telescope near-infrared channels and 
    the lensing magnification from foreground galaxy clusters.  Our sample consists of \Ntot galaxies with 
    stellar mass ranging from 10$^7$ to 10$^{10}$ \Msun, instantaneous star-formation rate in the range of [1, 
    100] \Msun/yr, and global metallicity [$\frac{1}{12}$, 2] solar.  At 2-$\sigma$ confidence level, 
    \NnegERsig/\Ntot galaxies in our sample show negative radial gradients, whereas \NinvERsig/\Ntot show 
    inverted gradients. Combining ours and all other metallicity gradients obtained at similar resolution 
    currently available in the literature, we measure a negative mass dependence of $\Delta\log({\rm O/H})/\Delta r~
    [\mathrm{dex~kpc^{-1}}] = \left(-0.020\pm0.007\right) + \left(-0.016\pm0.008\right) 
    \log(\Mstar/10^{9.4}\Msun)$ with the intrinsic scatter being $\sigma=0.060\pm0.006$ over four orders of 
    magnitude in stellar mass.
    Our result is consistent with strong feedback, not secular processes, being the primary governor of the 
    chemo-structural evolution of star-forming galaxies during the disk mass assembly at cosmic noon.
    We also find that the intrinsic scatter of metallicity gradients
    increases with decreasing stellar mass and increasing specific star-formation rate.
    This increase in the intrinsic scatter is likely caused by the combined effect of
    cold-mode gas accretion and merger-induced starbursts, with the latter more predominant in the dwarf mass 
    regime of $\Mstar\lesssim10^9\Msun$.
\end{abstract}


\section{Introduction}\label{sect:intro}

Metallicity is one of the most fundamental proxies of galaxy evolution at the peak of cosmic star formation and 
metal enrichment ($1\lesssim z \lesssim 3$), \ie, the cosmic noon epoch \citep{2014ARA&A..52..415M}.
The interstellar medium oxygen abundance relative to hydrogen --- metallicity\footnote{Throughout this paper, we 
use metallicity to stand for gas-phase oxygen abundance unless otherwise specified.} --- has been shown to 
correlate strongly with stellar mass (\Mstar), star-formation rate (\SFR) and gas fraction \citep[see the recent 
review by][and references therein]{Maiolino:2018vq}.
The cumulative history of the baryonic mass assembly, \eg, star formation, gas accretion, mergers, feedback and 
galactic winds, altogether governs the total amount of metals remaining in gas 
\citep{2008MNRAS.385.2181F,2012MNRAS.421...98D,Lilly:2013ko,Dekel:2014jm,Peng:2014hn}.
Moreover, these baryon cycling processes also tightly regulate the spatial distribution of metals in galaxies 
\citep{Ho:2015gq,SanchezMenguiano:2016gj,Belfiore:2019wu}.
Thus, a powerful way to learn about the baryon cycle is to use spatially resolved information.

The conventional way to obtain spatially resolved information is through integral field spectroscopy (IFS). IFS has dramatically 
expanded our vision of galaxies from spectroscopic measurements integrated through single slits/fibres to panoramic 2-dimensional (2D) 
views across their full surfaces, allowing for spatial variations of physical properties (including metallicity).
This facilitates several large ground-based surveys (e.g. CALIFA, MaNGA, SAMI) to constrain the radial profile of 
metallicity in hundreds of galaxies, successfully capturing the dynamic signatures of the baryon cycle \citep[see 
\eg,][]{2014A&A...563A..49S,Belfiore:2017bv,Poetrodjojo:2018fb}.
Meanwhile numerical simulations are now capable of making useful predictions for metallicity radial gradients and 
their evolution with redshift \citep[\eg][]{2017MNRAS.466.4780M,Tissera:2018vv}.
The main challenge for observations is that sub-kiloparsec (sub-kpc) spatial resolution is required for accurate 
results and meaningful comparison with theoretical predictions. While this spatial sampling is readily achieved 
for nearby galaxies ($z\lesssim0.3$), seeing-limited data are insufficient for galaxies at moderate to high 
redshift. Therefore we need an effective approach to achieve sub-kpc resolved spectroscopy for statistically 
representative samples of high-$z$ galaxies to compare meaningfully with cosmological zoom-in simulations.

The approach we take is space-based slitless spectroscopy.
Building upon our previous efforts \citep{2015AJ....149..107J,Wang:2016um,Wang:2019cf}, we exploit
grism spectroscopy from the Hubble Space Telescope (\hst).
\hst's diffraction limit in the near-infrared wavelengths is equivalent to a physical scale of $\sim$1 kpc at $z$$\sim$2.
Additional gain in resolution can be provided by gravitational lensing by foreground galaxies and/or galaxy 
clusters to fully satisfy the requirement for sufficiently resolving the chemical profiles of galaxies at that 
epoch. Lensing is thus essential for resolving the lowest-mass galaxies at high redshifts.
Recently, \citet{Curti:2020gx} derived metallicity maps and radial gradients in a sample of 28 lensed galaxies with 
stellar mass as low as $10^9$\Msun.
In this work, we measure radial gradients of metallicity in \Ntot star-forming galaxies at $1.2\lesssim 
z\lesssim2.3$ gravitationally lensed by foreground galaxy clusters, further extending to even lower stellar masses.
Our sample enables a detailed comparison between observed and simulated chemo-structural 
properties of galaxies, offering valuable insights into galaxy evolution.

This paper is organized as follows. In Section~\ref{sect:data}, we describe the data and galaxy sample analyzed 
in this work. The measurements of various physical quantities for our sample galaxies are presented in 
Section~\ref{sect:measure}.
Then two major pieces of our analysis results, \ie, the redshift evolution and mass dependence of sub-kpc 
resolution metallicity gradients,
are shown in Sections~\ref{sect:gradVSz} and \ref{sect:gradVSm}, respectively.
We finally conclude in Section~\ref{sect:conclu}.
Throughout this paper, the AB magnitude system and standard concordance cosmology (\Om=0.3, \Ol=0.7, 
$H_0=70\Hunit$) are used. Forbidden lines are indicated as follows: $\OIII\lambda5008\defeq\OIII$, 
$\OII\lambda\lambda3727,3730\defeq\OII$, $\NII~\lambda6585\defeq\NII$, $\SII~\lambda\lambda6718,6732\defeq\SII$,
if presented without wavelength values.

\section{Data and sample selection}\label{sect:data}

\begin{figure*}
    \centering
    \includegraphics[width=.33\textwidth, trim = 0cm 0cm 0cm 0cm, clip]{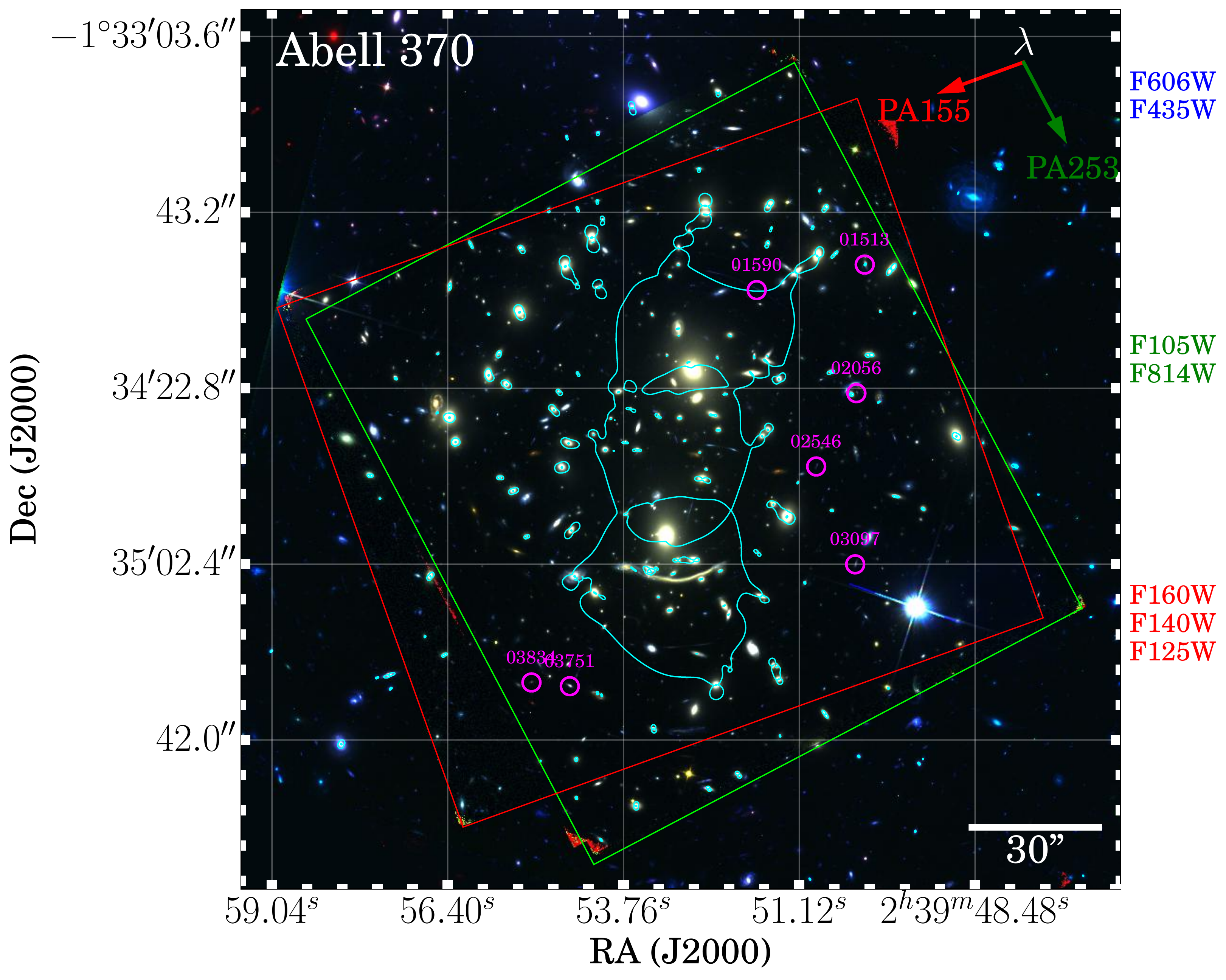}
    \includegraphics[width=.33\textwidth, trim = 0cm 0cm 0cm 0cm, clip]{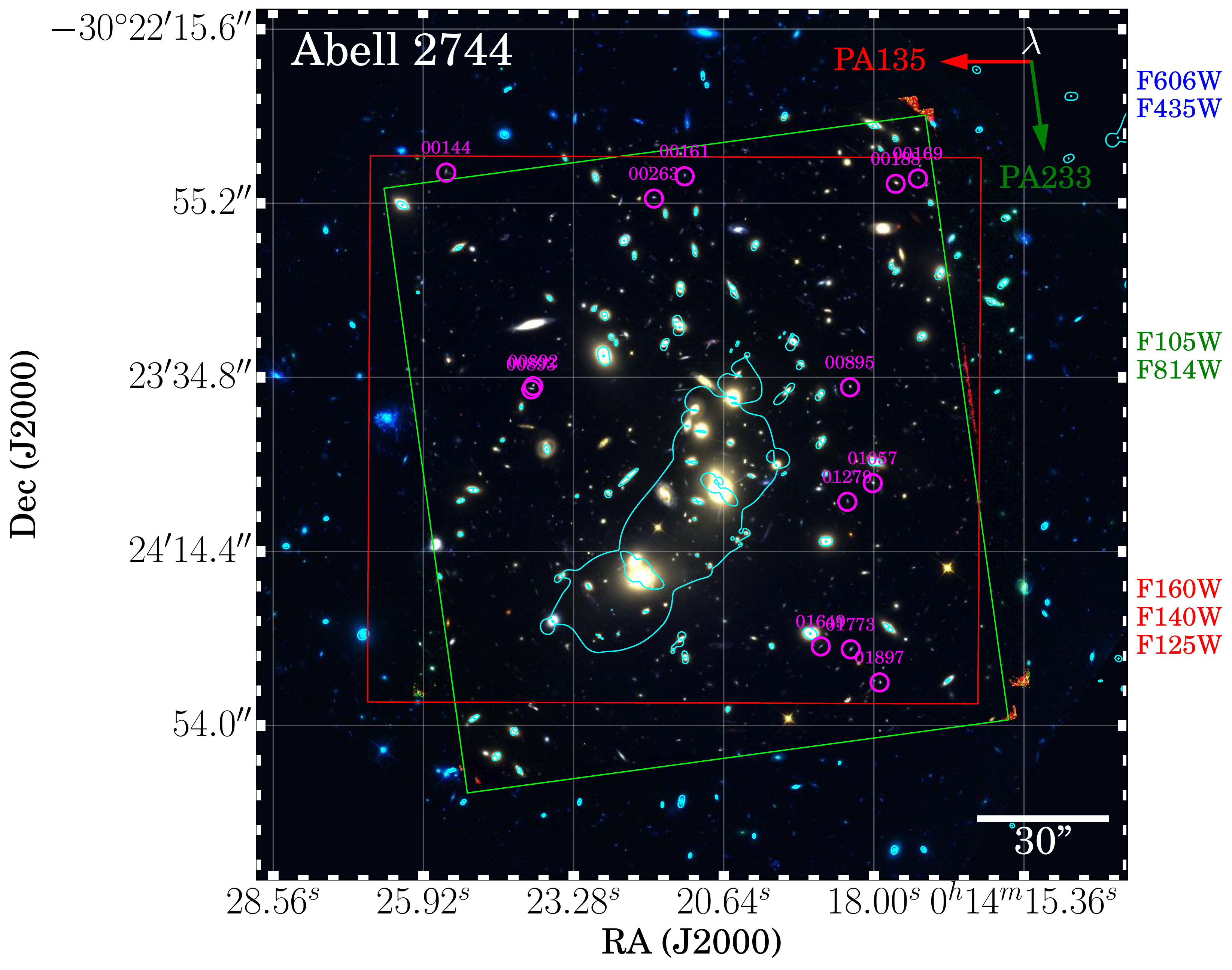}
    \includegraphics[width=.33\textwidth, trim = 0cm 0cm 0cm 0cm, clip]{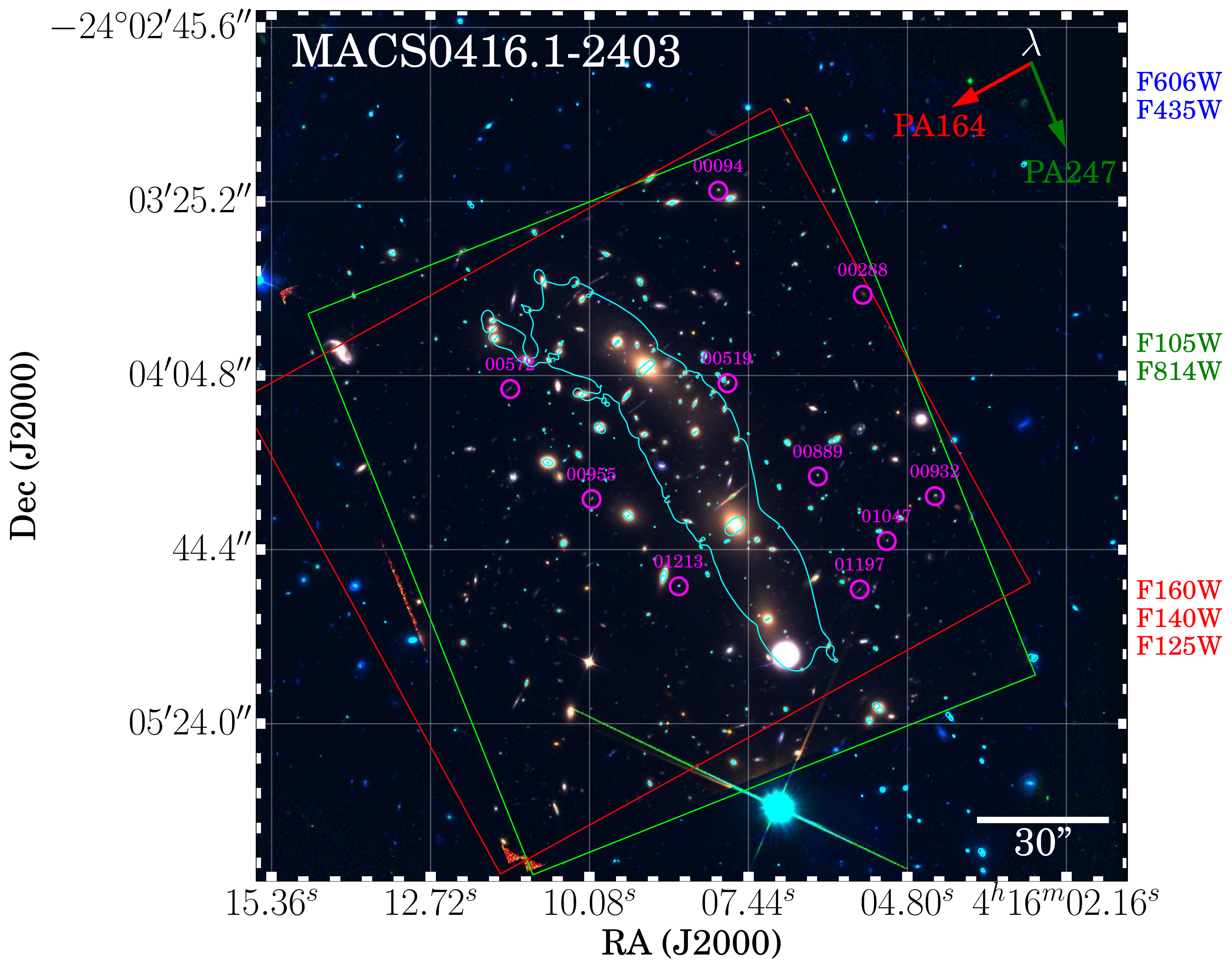}\\
    \includegraphics[width=.33\textwidth, trim = 0cm 0cm 0cm 0cm, clip]{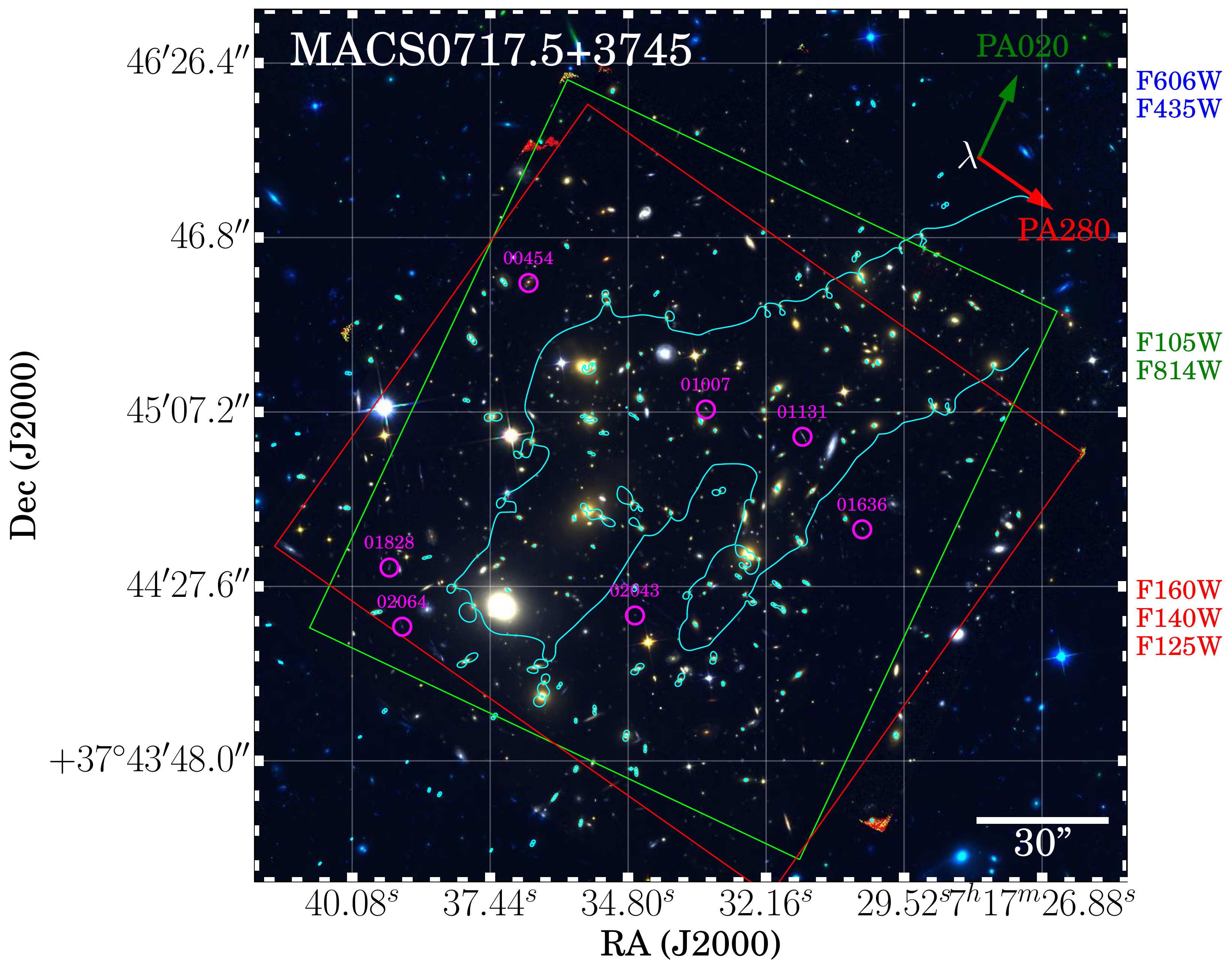}
    \includegraphics[width=.33\textwidth, trim = 0cm 0cm 0cm 0cm, clip]{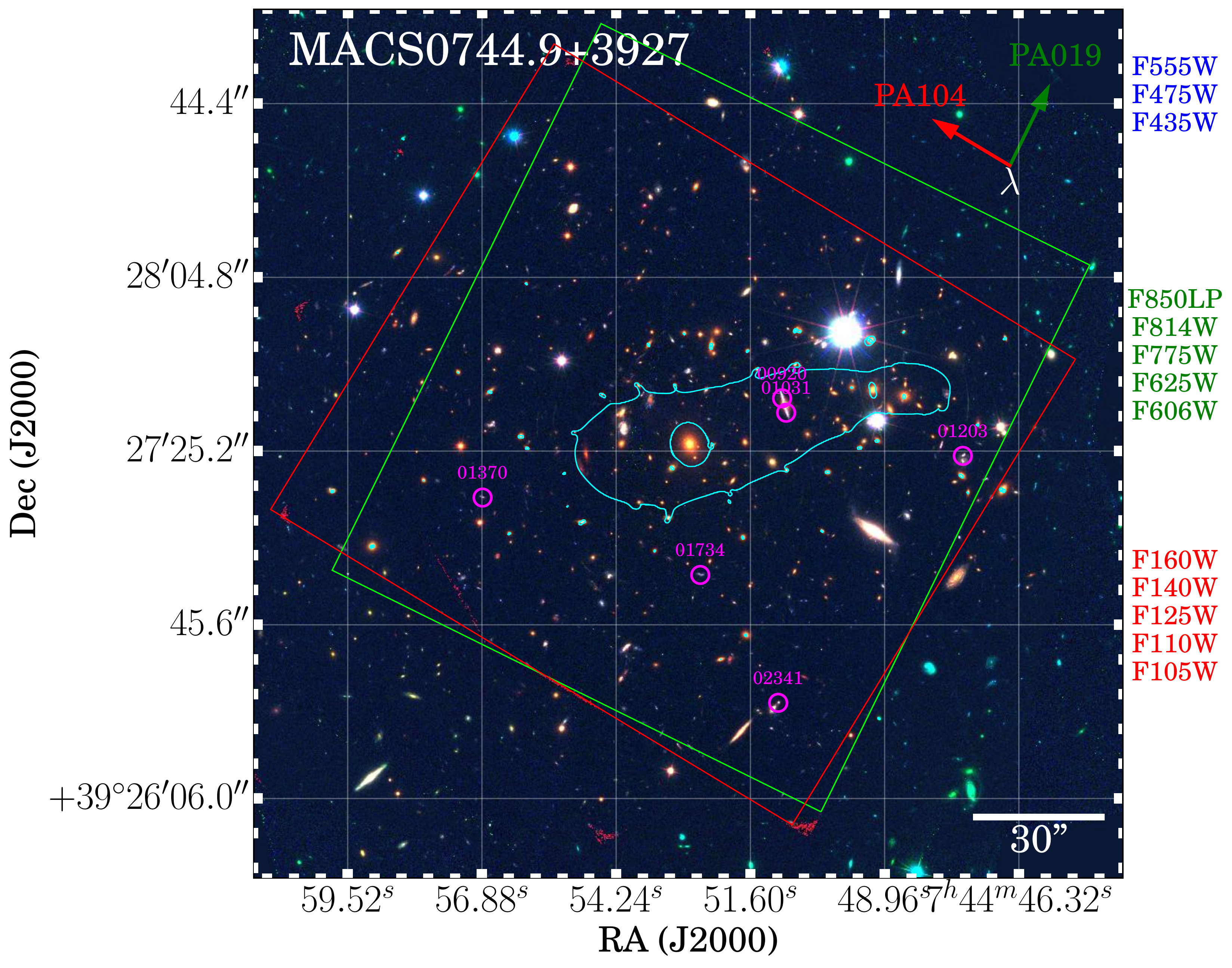}
    \includegraphics[width=.33\textwidth, trim = 0cm 0cm 0cm 0cm, clip]{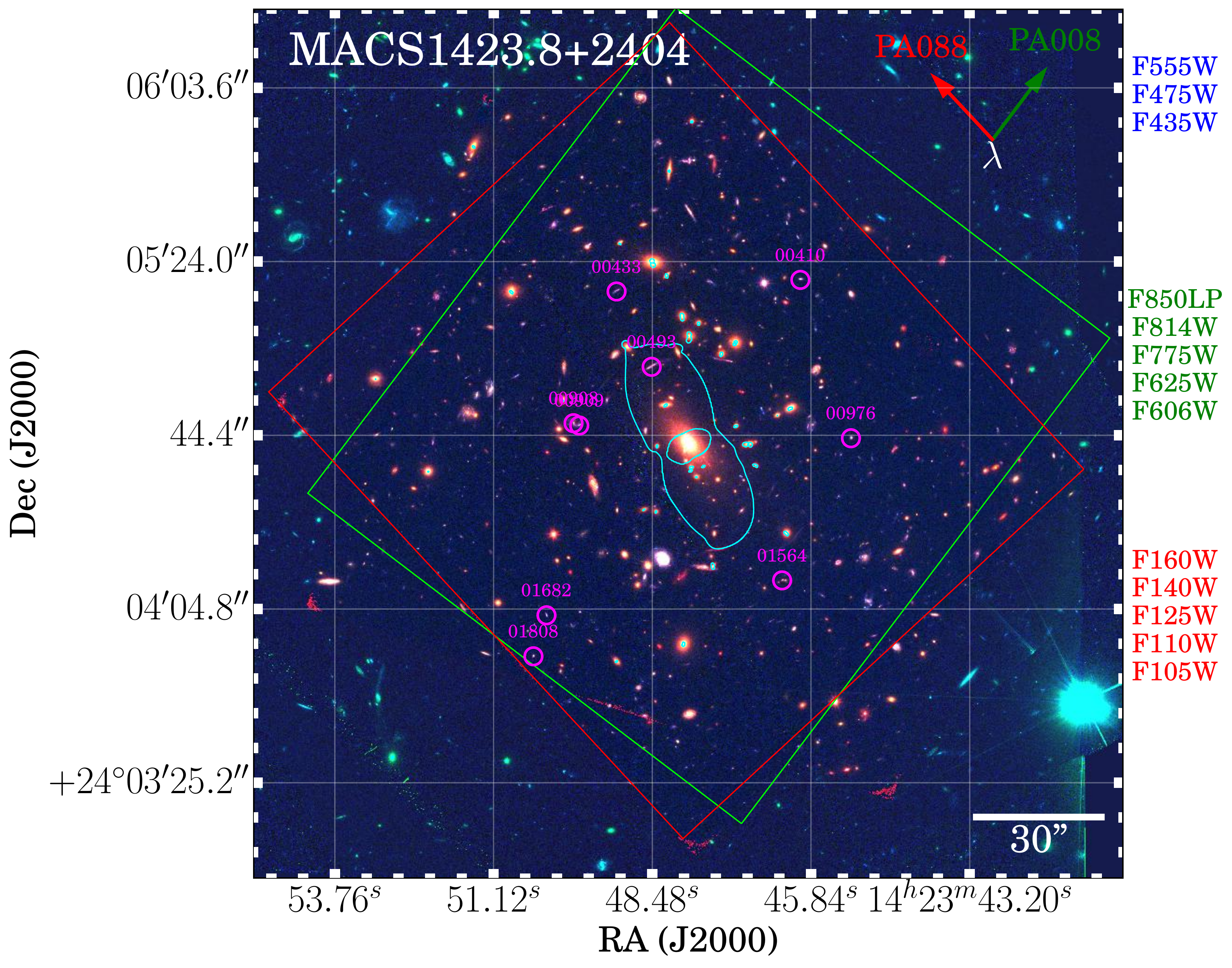}\\
    \includegraphics[width=.33\textwidth, trim = 0cm 0cm 0cm 0cm, clip]{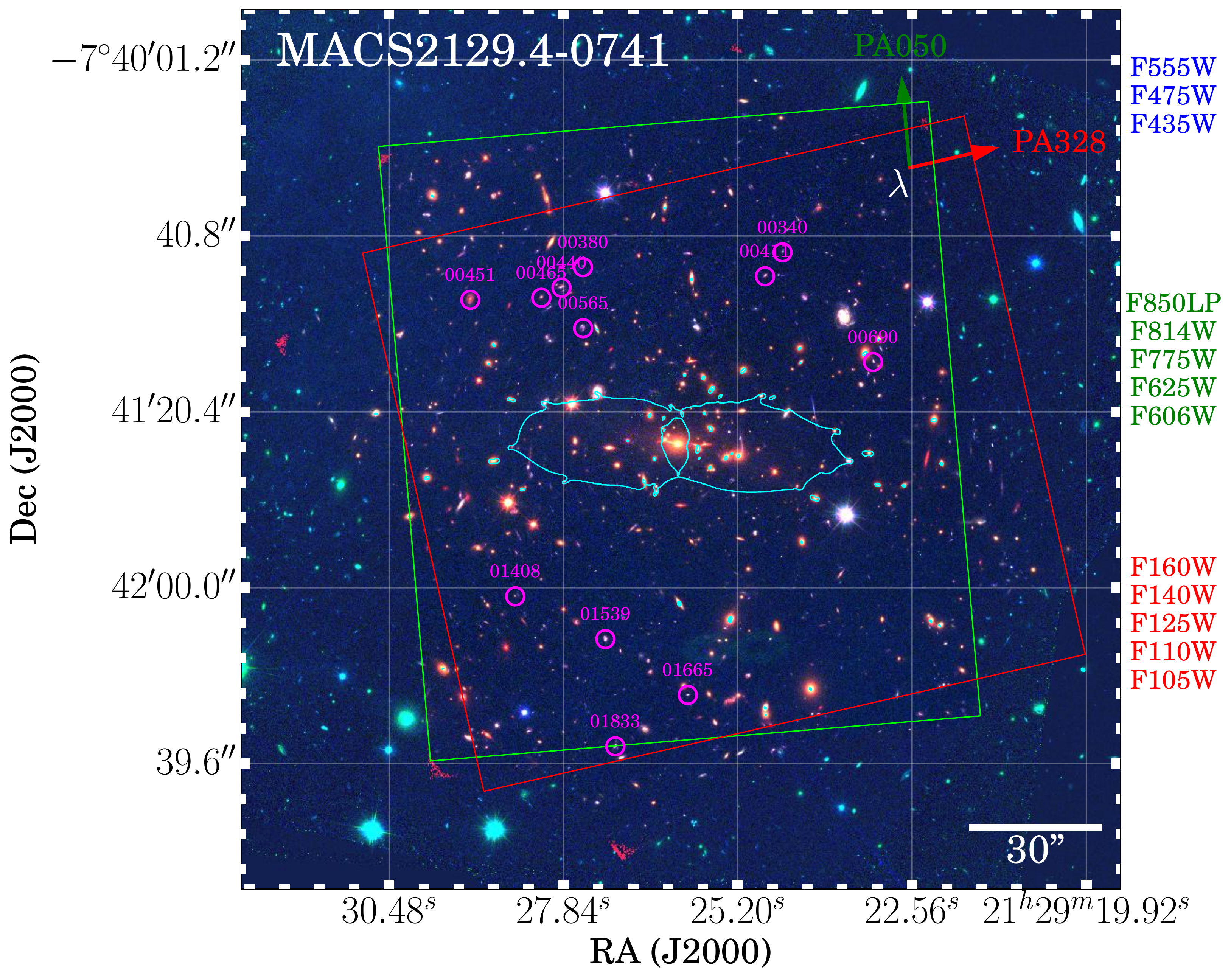}
    \includegraphics[width=.33\textwidth, trim = 0cm 0cm 0cm 0cm, clip]{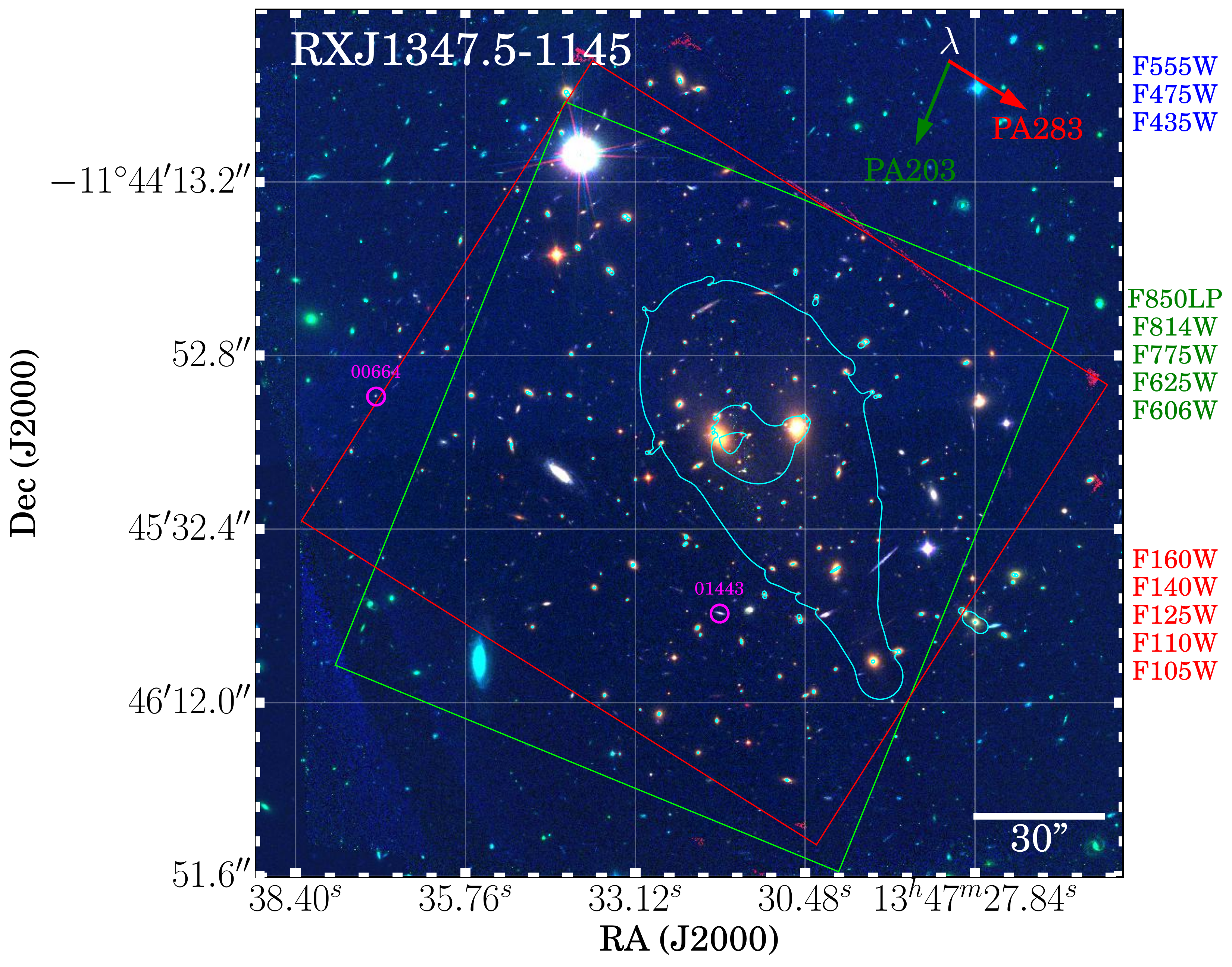}
    \includegraphics[width=.33\textwidth, trim = 0cm 0cm 0cm 0cm, clip]{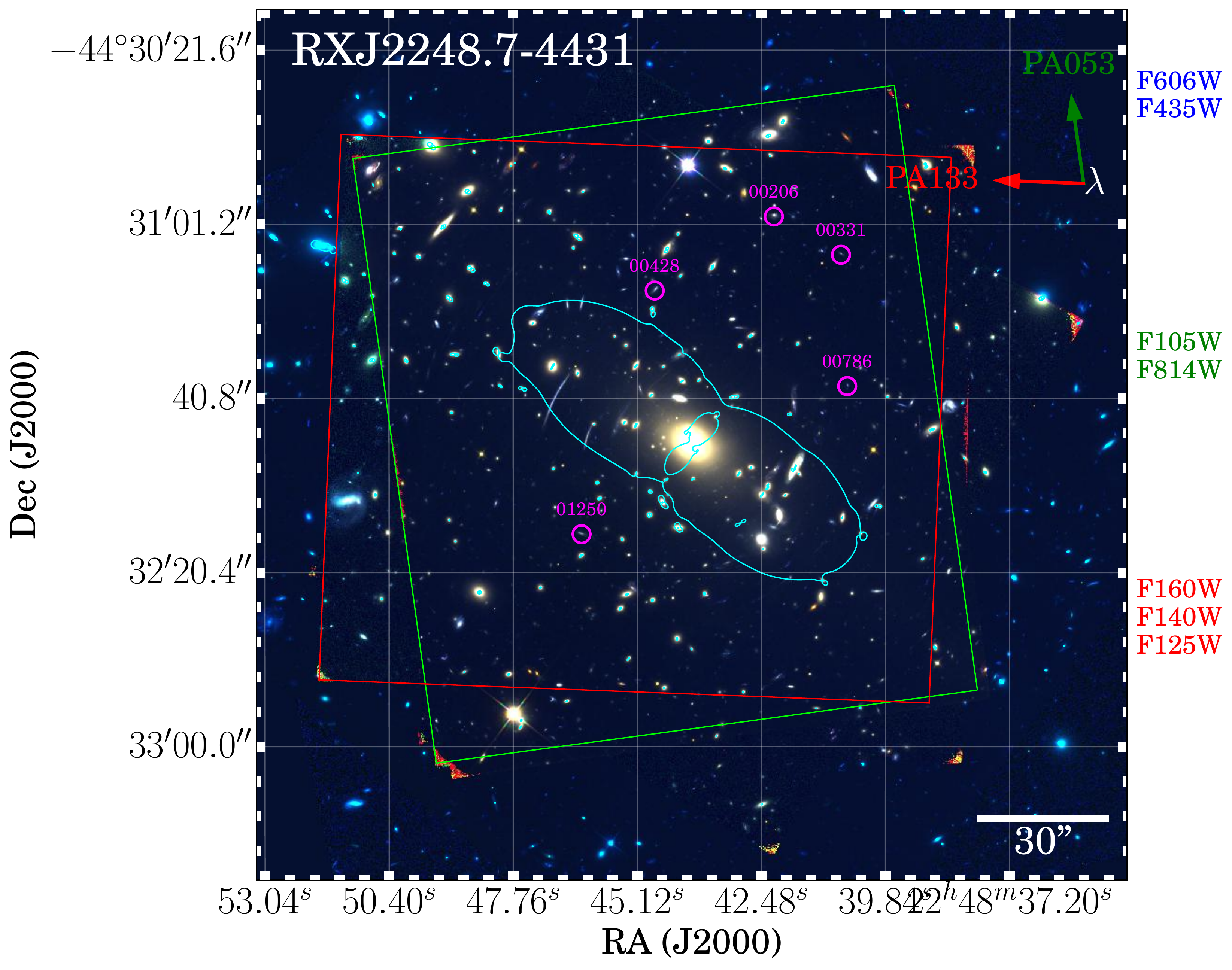}
    \caption{Color-composite images of the nine cluster center fields presented in this work (for the tenth 
    field, \ie, MACS1149, see Fig.~1 of \citet{Wang:2016um}).
    The blue, green and red channels are stacked images from the \hff/\clash mosaics taken at various filters, 
    shown on the right to each panel.
    The footprints of \hst WFC3 near-infrared grism pointings are denoted by the red and green squares, with the 
    corresponding wavelength dispersion directions marked by the arrows in the same color in the upper right 
    corner.
    The cyan contours represent the critical curves at sample median redshift ($z=1.63$) predicted by our default 
    macroscopic lens models (see Sect.~\ref{subsect:metal}).
    Our sources with sub-kpc resolution metallicity gradient measurements are marked by magenta circles.
    \label{fig:RGBfullFoV}}
\end{figure*}

The spectroscopic data analyzed in this work are acquired by the Grism Lens-Amplified Survey from 
Space\footnote{\url{https://archive.stsci.edu/prepds/glass/}} \citep[\glass; Proposal ID 13459;
P.I. Treu,][]{2014ApJ...782L..36S,2015ApJ...812..114T}.
It is a cycle-21 \hst large program allocated 140 orbits of Wide-Field Camera 3 (WFC3) near-infrared
slitless spectroscopy on the centers of 10 strong-lensing galaxy clusters. For each cluster center field, we have 
10 orbits of G102 (covering 0.8-1.15\micron) and 4 orbits of G141 (covering 1.1-1.7\micron) exposures, amounting 
to $\sim$22 kilo-seconds of G102 and $\sim$9 kilo-seconds of G141 in total, together with $\sim$7 kilo-seconds of 
F105W+F140W direct imaging for wavelength/flux calibration and astrometric alignment.
This exposure time is divided equally into two orients with almost orthogonal light dispersion directions, 
designed to disentangle contamination from neighbor objects.
As a result, we obtain two suites of G102+G141 spectra for each object, in an uninterrupted wavelength range of 
0.8-1.7 \micron with nearly uniform sensitivity, reaching 1-$\sigma$ surface brightness of $3\times10^{-16}$ 
\SBunit.
The 10 cluster fields are listed in Table~\ref{tab:obsdata} and shown in Fig.~\ref{fig:RGBfullFoV}.
Among these clusters, 6 have ultra-deep 7-filter imaging from the Hubble Frontier Fields (\hff) initiative 
\citep{Lotz:2016ca}. The other 4 have multi-band imaging from the Cluster Lensing And Supernova survey with 
Hubble (\clash) \citep{Postman:2012ca}.

We base our source selection on the redshift catalogs made public by the \glass collaboration.
From these catalogs, we select 327 galaxies with secure spectroscopic redshifts in the range of 
$z\in[1.2,2.3]$ with the redshift quality flag $\geq$3.
This redshift range is chosen to enable the grism coverage of the oxygen collisionally excited lines and the 
Balmer lines in rest-frame optical (\ie, \OIII, \Hb, \OII),
which are the most promising and frequently used metallicity diagnostics at extragalactic distances.
We also visually inspect the spatial extent and grism data quality of each source, to remove sources with 
compact morphology (\ie, with half-light radius $R_{50}<0\farcs25$ measured in \H-band imaging) and/or severe grism defects, not suitable for our analysis.
As a consequence, we compile a list of 93 objects with secure spectroscopic redshifts, relatively extended spatial profiles, 
and no severe defects nor lack of data in their grism spectra.
After further removing the sources with low signal-to-noise ratio (SNR) detections of emission lines (see 
Sect.~\ref{subsect:ELflux}), and those with ionization contamination from the active galactic nucleus (AGN; see Sect.~\ref{subsect:MEx}),
we obtain the final sample comprising a total of 76 star-forming galaxies at $z\in[1.2,2.3]$, on which we present the subsequent measurements.

\begin{deluxetable*}{llcccclcccccc}
    \tablecolumns{13}
    \tablewidth{0pt}
    \tablecaption{Summary of the \hst observations presented in this work}
\tablehead{
    \colhead{Cluster Field} &
    \colhead{Cluster Alias} &
    \colhead{Cluster Redshift} &
    \colhead{R.A.}  &
    \colhead{Decl.} &
    \colhead{Grism PA\tablenotemark{a}} & 
    \colhead{\hst imaging} & 
    \colhead{$N_{\rm source}$\tablenotemark{b}}\\
    & & &
    \colhead{(deg.)} &
    \colhead{(deg.)} &
    \colhead{(deg.)} &
    &
}
\startdata
Abell 370       & A370      &   0.375   &   02:39:52.9  &  -01:34:36.5  & 155, 253 & \hff &        7  \\
Abell 2744      & A2744     &   0.308   &   00:14:21.2  &  -30:23:50.1  & 135, 233 & \hff &        13 \\
MACS0416.1-2403 & MACS0416  &   0.420   &   04:16:08.9  &  -24:04:28.7  & 164, 247 & \hff/\clash & 9 \\
MACS0717.5+3745 & MACS0717  &   0.548   &   07:17:34.0  &  +37:44:49.0  & 020, 280 & \hff/\clash & 5 \\
MACS0744.9+3927 & MACS0744  &   0.686   &   07:44:52.8  &  +39:27:24.0  & 019, 104 & \clash &      6 \\
MACS1423.8+2404 & MACS1423  &   0.545   &   14:23:48.3  &  +24:04:47.0  & 008, 088 & \clash &      9 \\
MACS2129.4-0741 & MACS2129  &   0.570   &   21:29:26.0  &  -07:41:28.0  & 050, 328 & \clash &      10 \\
RXJ1347.5-1145  & RXJ1347   &   0.451   &   13:47:30.6  &  -11:45:10.0  & 203, 283 & \clash &      2 \\
RXJ2248.7-4431  & RXJ2248   &   0.348   &   22:48:44.4  &  -44:31:48.5  & 053, 133 & \hff/\clash & 5 \\
MACS1149.6+2223$\tablenotemark{c}$ & MACS1149  &   0.544   &   11:49:36.3  &  +22:23:58.1  & 032, 111, 119, 125 & \hff/\clash & 10
\enddata
    \tablecomments{Here we only list the primary pointings of the analyzed \hst slitless spectroscopy, covering the cluster centers
    with WFC3/NIR grisms.}
    \tablenotetext{a}{The position angles (PAs) are represented by the ``PA\_V3'' values reported in the
    corresponding raw image headers. The PA of the actual dispersion axis of slitless spectroscopy, in degrees
    east of north, is given by $\mathrm{PA_{disp}} \approx \mathrm{PA\_V3} - 45.2$.
    For each one of the \glass PAs (\ie excluding PAs 111 and 119 for MACS1149), 2 orbits of G141 and 5 orbits of
    G102 exposures have been taken, amounting to $\sim$4.5 and $\sim$11 kilo-seconds science exposure times for
    G141 and G102 respectively.}
    \tablenotetext{b}{The number of galaxies in which we secure sub-kpc resolution metallicity gradient
    measurements from \hst spectroscopy.}
    \tablenotetext{c}{The detailed analyses of gradient measurements have already been presented in our earlier
    paper \citep{Wang:2016um}. Here we update the SED fitting results associated with these galaxies.}
\label{tab:obsdata}
\end{deluxetable*}


\section{Methodology and measurements}\label{sect:measure}

\subsection{Emission line flux}\label{subsect:ELflux}

We adopt the Grism Redshift and Line Analysis software 
(\grzl\footnote{\url{https://github.com/gbrammer/grizli/}}; G. Brammer et al. in prep) to handle wide-field 
slitless spectroscopy data reduction.
\grzl is a state-of-the-art software that performs ``one-stop-shopping'' processing of paired direct and grism 
exposures acquired by space telescopes.
The entire procedure consists of five steps: 1) preprocessing of raw grism exposures\footnote{Specifically, step 
1) includes bad-pixel/persistence masking, bias correction, dark subtraction, cosmic ray flagging, 
relative/absolute astrometric alignment, flat fielding, master/variable sky background subtraction, geometric 
distortion correction, extraction of source catalogs and segmentation images at visit levels.}, 2) forward 
modeling full field-of-view (FoV) grism images, 3) redshift fitting via spectral template synthesis, 4) refining 
full FoV grism model, and 5) extracting 1D/2D spectra and emission line maps of individual targets.

\begin{figure*}
    \includegraphics[width=\textwidth]{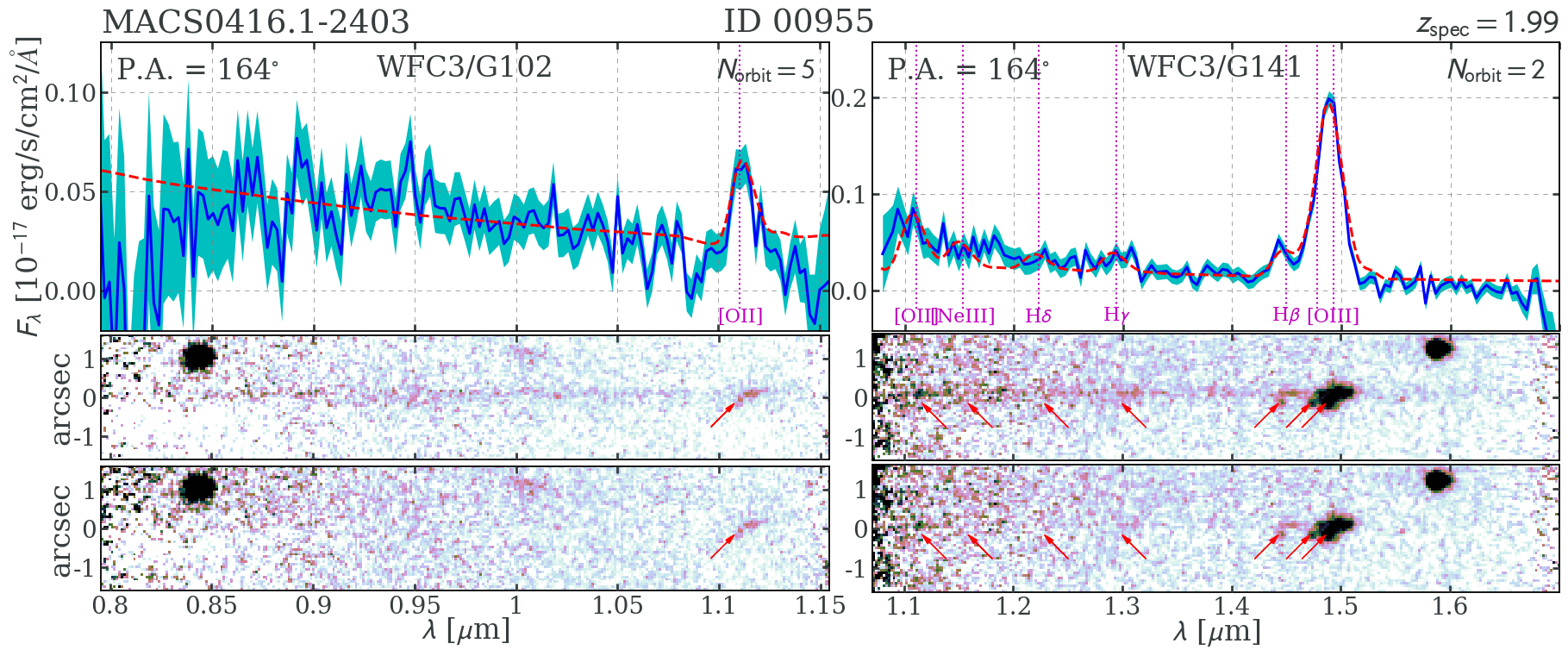}
    \includegraphics[width=\textwidth]{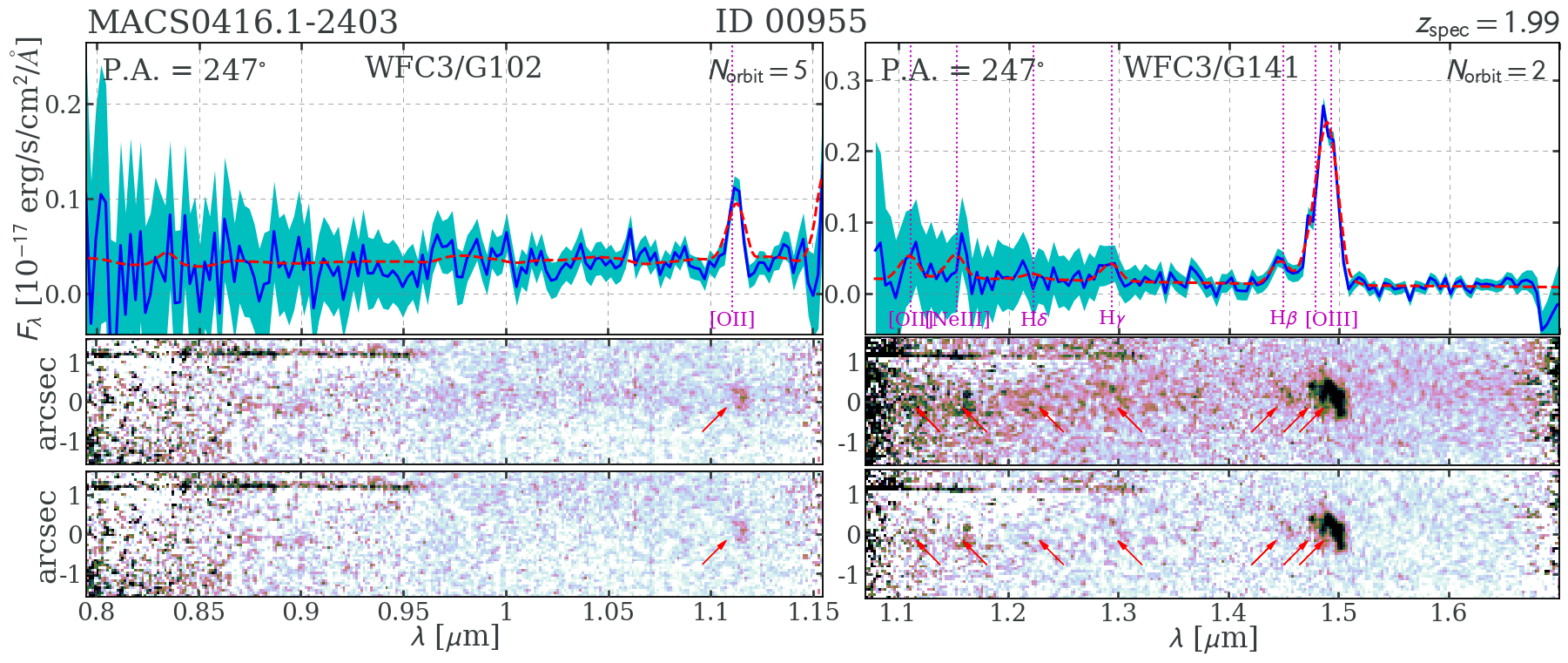}
    \caption{
    The \hst grism spectra for one exemplary object in our sample, MACS0416-ID00955 at $z\sim2$.
    The total on-target exposure time is equally split between two nearly orthogonal P.A.s (shown in the two sub-figures), 
    reaching 5 orbits of G102 and 2 orbits of G141 exposures per P.A..
    In each sub-figure, we show the optimally extracted 1D spectra and the full 2D spectra without and with source stellar 
    continuum subtraction, for both grism channels.
    The 1D observed $F_{\lambda}$ flux and its 1-$\sigma$ uncertainty are represented by the blue solid lines and the cyan shaded 
    bands, respectively.
    The wavelengths of the nebular line emission features are marked by magenta vertical dotted lines and red arrows, in 1D and 2D 
    spectra respectively.
    The red dashed curves show the 1D model spectra, combining both stellar continuum (given by spectral template synthesis) and 
    nebular line emission (modeled as Gaussian profiles), \emph{after} the source morphological broadening is already taken into account.
    We emphasize that the same best-fit spectral model is used for each individual source, yet the differences in continuum shape 
    and flux levels at the two P.A.s are originated from the varying source morphological kernels along the two light 
    dispersion directions.
    \label{fig:ID00955spec}}
\end{figure*}

In Step 3), we derive the best-fit redshift of our sources from spectral template fitting based on a library of spectral energy 
distributions (SED) of stellar populations with a range of characteristic ages \citep[see Appendix A in][for more 
details]{Wang:2019cf}.
We also fit the intrinsic nebular emission using 1D Gaussian functions centered at corresponding wavelengths and estimate the line 
fluxes.
The morphological broadening is taken into account with respect to the dispersion direction associated with each 
exposure.
Fig.~\ref{fig:ID00955spec} shows the typical 1D and 2D spectra of one of our target galaxies.
The majority (61/76) of our sample galaxies have \OIII detected with SNR$\gtrsim$10.
55, 35, and 15 within the entire sample have SNR$\gtrsim$5 detections of \OII, \Hb, and \Hg, respectively.
For galaxies at $z\leq1.6$, we also typically have access to their \Ha\footnote{31 out of the 37 sources in 
this redshift range have \Ha detected with SNR$\gtrsim$10.} and \SII, which help constrain metallicity and 
nebular extinction.
The best-fit redshifts and observed emission line fluxes for all our sources are presented in 
Table~\ref{tab:srcprop}.

\subsection{Emission line maps}\label{subsect:ELmap}

In addition to the measurements of integrated emission line fluxes, another key piece of information that we need to retrieve from grism 
spectroscopy is the spatial distribution of emission line surface brightnesses, \ie, the emission line maps.
The \hst WFC3 near-infrared grisms have limited spectral resolution: for point sources, $R$$\sim$210 and 130, for G102 and G141, respectively.
Yet this is actually an advantage in obtaining emission line maps.
Since the instrument full-width half-maximum (FWHM) is equivalently $\sim$700 km/s for G102, and $\sim$1200 km/s for G141, it is 
reasonable to assume that the source 1D spectral shapes and 2D emission line maps are not affected by any kinematic motions of gas ionized by
the star-forming regions, where outflows typically have speed <500 km/s \citep[see \eg][for a recent 
review]{Erb:2015fd}.
However, our sample galaxies are selected to be spatially extended, having their half-light radius 
$R_{50}\gtrsim0\farcs25$, measured from their continuum morphology in the \H-band imaging acquired by \hff or 
\clash.
Their spatial profiles along the light dispersion direction are convolved onto the wavelength axis, resulting in severe morphological broadening of the line-spread function FWHM \citep{vanDokkum:2011cq}.
This morphological broadening effect is already taken into account when estimating the best-fit grism redshift from the spectral 
template synthesis process described in Sect.~\ref{subsect:ELflux}.
It also poses a great challenge for obtaining spatial 2D maps of emission lines that have very close rest-frame wavelengths, 
in particular the line complex of \Hb+\OIII$\lambda\lambda$4960,5008 doublets.

\begin{figure*}
    \centering
    {\hspace{2.5em}\includegraphics[width=.9\textwidth,trim=0cm .3cm 0cm 0cm,clip=true]{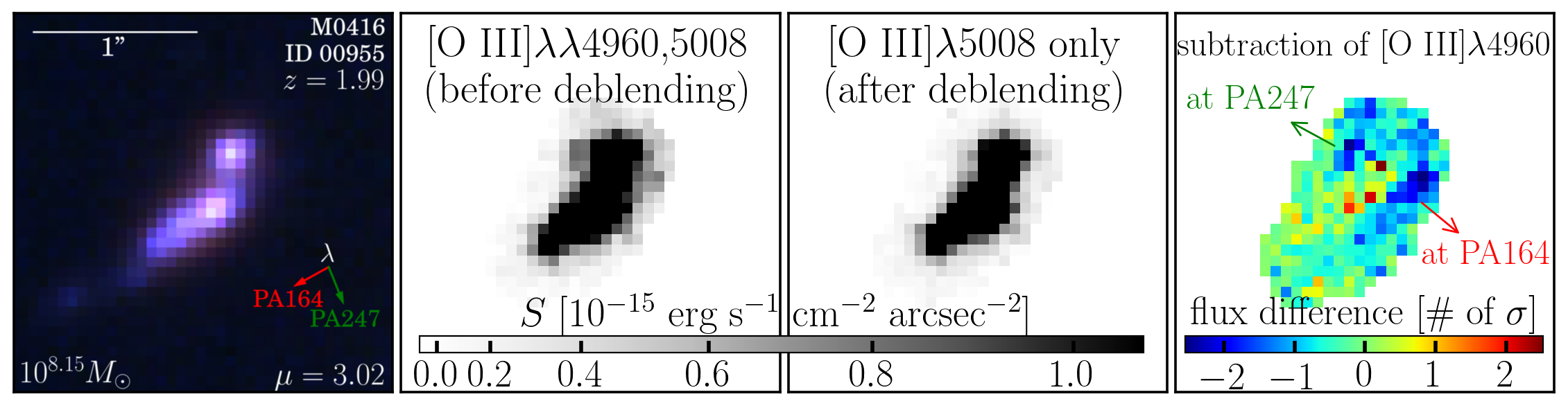}}\\
    \includegraphics[width=.95\textwidth,trim=0cm 0cm 0cm 0.1cm,clip=true]{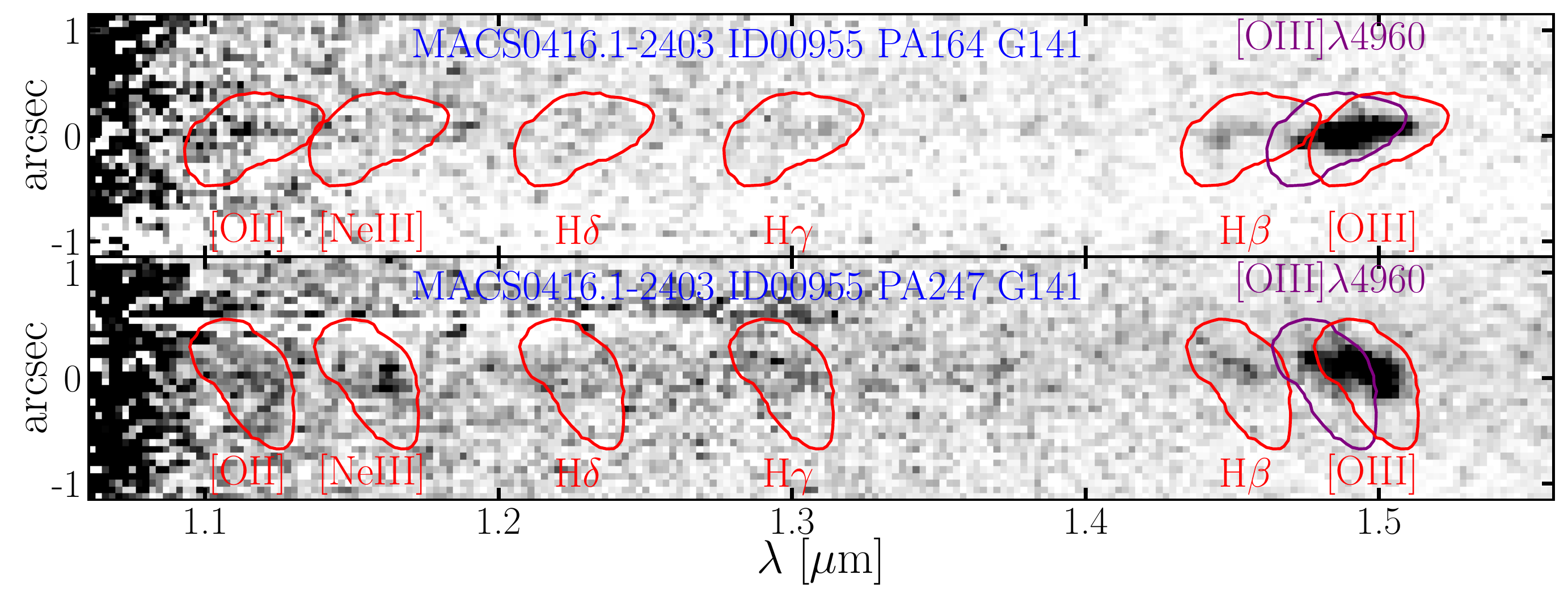}
    \caption{Our custom technique of obtaining pure \textrm{\OIII$\lambda$5008} maps combined from multiple orients of grism 
    exposures.
    {\bf Top}, from left to right: the color-composite image of object MACS0416-ID00955 (a $z$$\sim$2 dwarf galaxy with 
    $\Mstar$$\simeq$$10^8$\Msun), its \OIII map before deblending the \OIII doublets, its pure \OIII$\lambda$5008 map clean from 
    the partial contamination of \OIII$\lambda$4960 at two orients (PA164 and PA247), and the significance of difference between 
    these two \OIII maps.
    The significance is expressed as the flux differences divided by the corresponding flux uncertainties (\ie $\sigma$) of 
    \textrm{\OIII$\lambda$5008} in each spatial pixel.
    {\bf Middle and bottom}: 2D contamination and continuum subtracted G141 spectra of this dwarf galaxy at two orients (PA164 and 
    PA274) separately.
    Note that these 2D traces are basically cutouts from the continuum subtracted G141 spectra presented in 
    Fig.~\ref{fig:ID00955spec}.
    Due to the limited spectral resolutions of \hst grisms and extended source morphology, fluxes of \OIII$\lambda$4960 are 
    blended into \OIII$\lambda$5008 and \Hb in a spatially inhomogeneous fashion, specific to the light dispersion direction at 
    individual orients.
    \label{fig:deblend}}
\end{figure*}

We hence develop a custom technique to deblend the line complex as follows.
First, we measure the source broad-band isophotes that encompass over 90\% of the surface brightness in \JH and 
\Y-band, and overlay them on top of the source 2D G141 and G102 spectra respectively.
The 2D grism spectra are standard data products produced by our \grzl reduction with contamination and source 
continuum removed.
The positions of the overlaid isophotes on the 2D grism spectra mark the locations of the redshifted emission lines (see the 
middle and bottom rows of Fig.~\ref{fig:deblend}).
We rely on the pre-imaging (\ie \JH and \Y) paired with the grism (\ie G141 and G102) observations to measure the 
isophotes because they cover similar wavelength range, share comparable PSF properties, and are acquired at the 
same PA of the telescope.
In this step, the grism spectra taken at different PAs have to be processed separately, since the morphological 
broadening varies drastically amongst different PAs if the source has asymmetric radial profiles.
This broad-band isophote is used as an aperture for emission line map extraction.
Since the red (\ie more to the right on the wavelength axis in 2D spectra) portion of the aperture centered at 
the redshifted \OIII$\lambda$5008 is not contaminated by \OIII$\lambda$4960 and the flux ratio between the \OIII 
doublets is fixed ($f_{\OIII~5008}/f_{\OIII~4960}$ = 2.98:1, calculated by \citet{Storey:2000jd}), we can obtain 
the same red portion of the clean \OIII$\lambda$4960 2D map.
This red part of \OIII$\lambda$4960 map is contaminating slightly bluer part of the \OIII$\lambda$5008 map, and 
can be subtracted off, with flux errors properly propagated, therefore yielding cleaned \OIII$\lambda$5008 flux 
in those slightly bluer areas within the extraction aperture.
This procedure is then conducted iteratively, until the \OIII$\lambda$4960 fluxes in all spatial pixels within 
the aperture have been removed, and clean 2D maps of \OIII$\lambda$5008 and \Hb can be obtained, \emph{at 
individual PAs}.
Finally, we use \adriz \citep{Gonzaga:2012tj} to combine the clean \OIII$\lambda$5008 and \Hb maps extracted at 
multiple PAs.
The resultant 2D stamps are drizzled onto a 0\farcs06 grid, Nyquist sampling the FWHM of the WFC3 PSF, and 
astrometrically matched to the corresponding broad-band images.
Notably, our custom deblending technique does not rely on any models of the spatial profiles of \OIII 
emission\footnote{We note that in the most up-to-date version of \grzl, the subtraction of \OIII$\lambda$4960 is 
implemented. However \grzl's automatic subtraction is based on a spatial model of \OIII$\lambda$4960 emission,
which is different from our procedure presented here.}.
This is a critical procedure to account for the orient-specific contaminations of \OIII$\lambda$4960, which can 
be over 2-$\sigma$ in some spatial areas within the extraction aperture (see the upper right panel of 
Fig.~\ref{fig:deblend}).

\subsection{Stellar mass}\label{subsect:Mstar}

We perform SED fitting to the broad-band photometry of our galaxies from the \hst imaging data obtained by \hff or \clash.
The \fast software \citep{Kriek:2009eo} is used to infer stellar mass (\Mstar), star-formation rate (SFR$^{\rm 
S}$, see Sect.~\ref{subsect:SFR} for more details), and dust extinction of stellar continuum (A$_{\rm V}^{\rm 
S}$), based on the \citet{Bruzual:2003ck} (BC03) stellar population synthesis
models.
We assume the \citet{Chabrier:2003ki} initial mass function, constant star formation history, the \citet{Calzetti:2000iy} extinction law, and fixed
stellar metallicity being one-fifth solar.
Since the majority of our galaxies show strong nebular emission in their rest-frame optical, we need to subtract 
their \emph{nebular} emission from the corresponding broad-band fluxes to estimate more accurately the level of 
\emph{stellar} continuum.
We convolve the best-fit Gaussian profiles for each emission line at the source redshift with the \hst bandpass throughput, 
to derive the nebular flux, and then subtract it from the measured broad-band photometry.
In Table~\ref{tab:srcprop}, we show the observed \JH-band magnitude before this correction and the reduction factor, which is a ratio between the 
\JH-band flux after and before correcting for nebular emission.
We verify that this correction is essential for deriving reliable \Mstar estimates for galaxies on the low mass 
end ($\Mstar<5\times10^9\Msun$); without this correction \Mstar can be over-estimated by as much as 0.7 dex.
We present the results of our stellar continuum SED fitting in Table~\ref{tab:srcprop}.
Thanks to lensing magnification, our sample extends significantly into the low-mass regime at high $z$, highly 
complementary to the targets from ground-based surveys \citep[\eg, \kd,][]{2016ApJ...827...74W}.

\subsection{AGN contamination}\label{subsect:MEx}

\begin{figure}
    \centering
    \includegraphics[width=.495\textwidth]{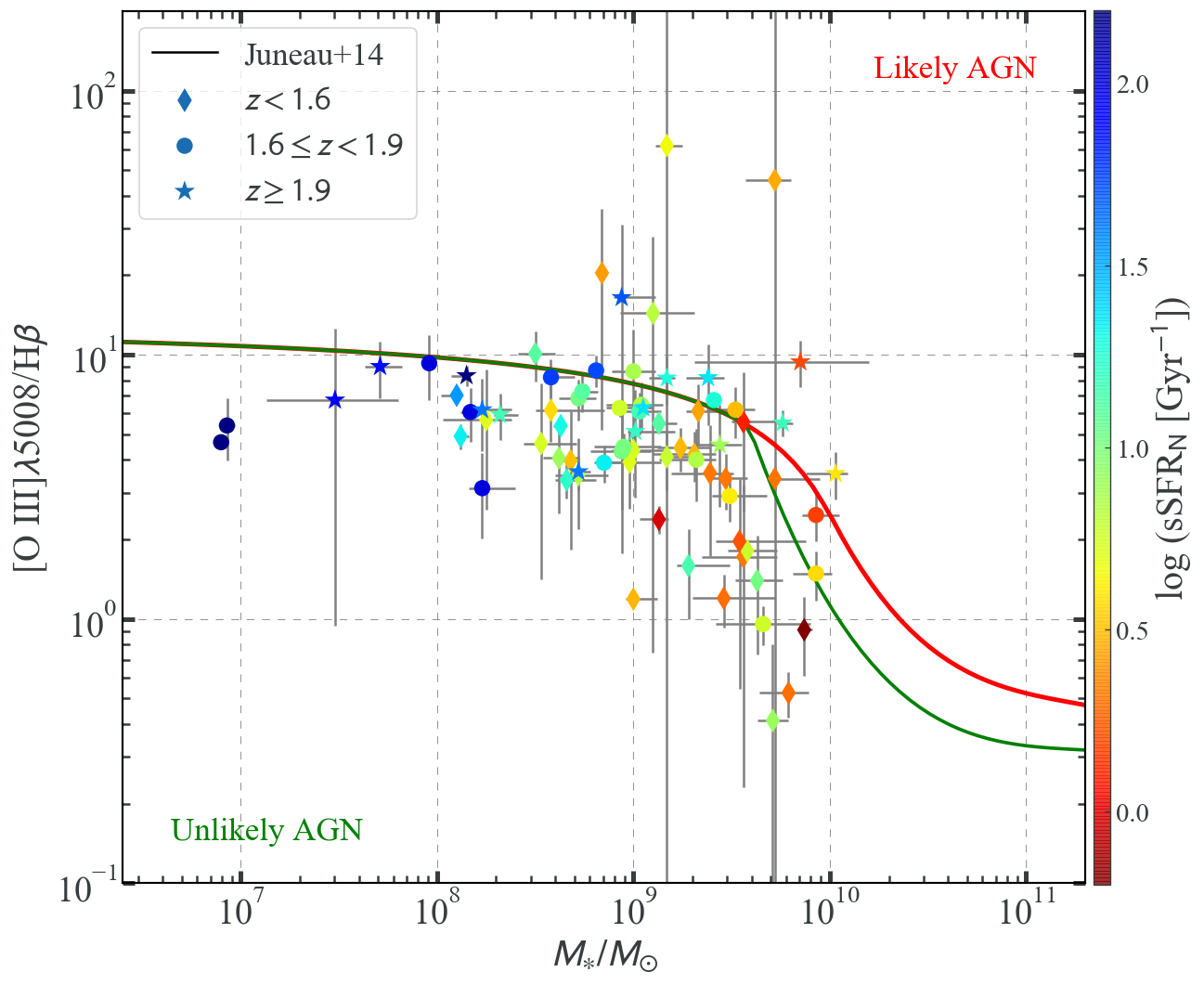}
    \caption{Mass-excitation diagram for our galaxies. The demarcation curves are from \citet{Juneau:2014ca} 
    based on the $z\sim0$ SDSS DR7 emission-line sample: AGNs are located mainly above the red curve, 
    star-forming galaxies are located below the green curve, and AGN/star-forming composites are in between.
    Our entire sample is separated into three redshift bins: $z<1.6$ (\Nlowz sources), $1.6\leq z<1.9$ (\Nmedz 
    sources), and $z\geq 1.9$ (\Nhiz sources), color-coded in sSFR.
    We show that the majority of our sources are located below the green curve, where the possibility of being 
    classified as AGNs is low (<10\%).
    \label{fig:MEx}}
\end{figure}

Before carrying out the metallicity inference on our sample, we check for contamination of active galactic nucleus (AGN) 
ionizations.
In Fig.~\ref{fig:MEx}, we rely on the mass-excitation diagram to exclude AGN candidates from our sample.
The demarcation scheme \citep{Juneau:2014ca} aims to separate AGN from star-forming galaxies, based on the SDSS 
DR7 emission-line galaxies at $z\sim0$.
This scheme has been shown to reproduce the bivariate distributions seen in a number of high-redshift galaxy 
samples out to $z\sim1.5$ \citep{Juneau:2014ca}.
We therefore discard sources in our sample that are 2-$\sigma$ away from the star-forming region in the diagram, 
given the measurement uncertainties on \Mstar and \OIII/\Hb.
To examine possible redshift-dependent trends in the future Sections, we subdivide our sample into three bins:
\Nlowz, \Nmedz, and \Nhiz galaxies at $z\in[1.2,1.6]$, $z\in[1.6,1.9]$, and $z\in[1.9,2.3]$, respectively, marked 
by different symbols in Fig.~\ref{fig:MEx}.

Moreover, \citet{Coil:2015dp} found that a +0.75 dex shift in \Mstar of the demarcation curves is necessary to 
match the loci of AGNs and star-forming galaxies in the MOSDEF surveys at $z\sim2.3$, to account for the redshift 
evolution of the mass-metallicity relation.
On part of the sample, we also obtained \Ha gas kinematics from the ground-based Keck OSIRIS observations 
\citep{Hirtenstein:2018tn}.
The integrated measurement of $f_{\NII}/f_{\Ha}$ is typically $\lesssim$0.1 at 3-$\sigma$ confidence level, 
indicative of star-forming regions with no significant AGN contamination.
We thus verify that there is no sign of significant AGN ionization in our sample.

\subsection{Star-formation rate}\label{subsect:SFR}

We have two methods for estimating star-formation rate (\SFR).
First of all, \SFR can be obtained from the stellar continuum SED fitting outlined in Sect.~\ref{subsect:Mstar}.
This method is sensitive to the underlying assumptions of star-formation history and stellar population synthesis 
models adopted in the fitting procedure.
Hereafter, we refer to these measurements as SFR$^{\rm S}$.

Secondly, \SFR can be derived from nebular emission after correcting for dust attenuation.
From our Bayesian inference method presented in Sect.~\ref{subsect:metal}, we obtain posterior probability 
distributions of the de-reddened \Hb flux, which can be converted to the intrinsic \Ha luminosity given source 
redshift.
As a consequence, \SFR (hereafter denoted as SFR$^{\rm N}$) can then be calculated following the widely used 
calibration \citep{Kennicutt:1998ki}, \ie,
\begin{align}\label{eq:sfr}
    \mathrm{SFR^N} = 4.6\times10^{-42}\frac{L(\Ha)}{\rm erg/s} ~~~~ (\Msun/\mathrm{yr}),
\end{align}
appropriate for the \citet{Chabrier:2003ki} initial mass function.
Unlike the measurements from SED fitting, this method provides a proxy of instantaneous
star-forming activities on the time scale of $\sim$10 Myrs.
This short time scale is relevant to probe the highly dynamic feedback processes which are effective in 
re-distributing metals \citep[see \eg,][]{GalaxiesonFIREFe:2014dn}.
Therefore, we quote the values of SFR$^{\rm N}$ as our fiducial \SFR measurements if not stated otherwise.

We note that for our low-$z$ sample (\Nlowz galaxies at $z\in[1.2,1.6]$), \Ha is covered by the WFC3/G141 grism. 
However, due to the low spectral resolution, it is heavily blended with \NII.
We hence rely on the empirical prescription of \citet{Faisst:2017wv} to subtract the contribution of \NII fluxes
from the measured \Ha flux, based on stellar mass and redshift of our galaxies (see Table~\ref{tab:srcprop} for 
the calculated \NII/\Ha flux ratios).
This ensures a more reliable estimate of SFR$^{\rm N}$, less impacted by dust than the \Hb-based measurements.

\begin{figure*}
    \centering
    \includegraphics[width=.495\textwidth]{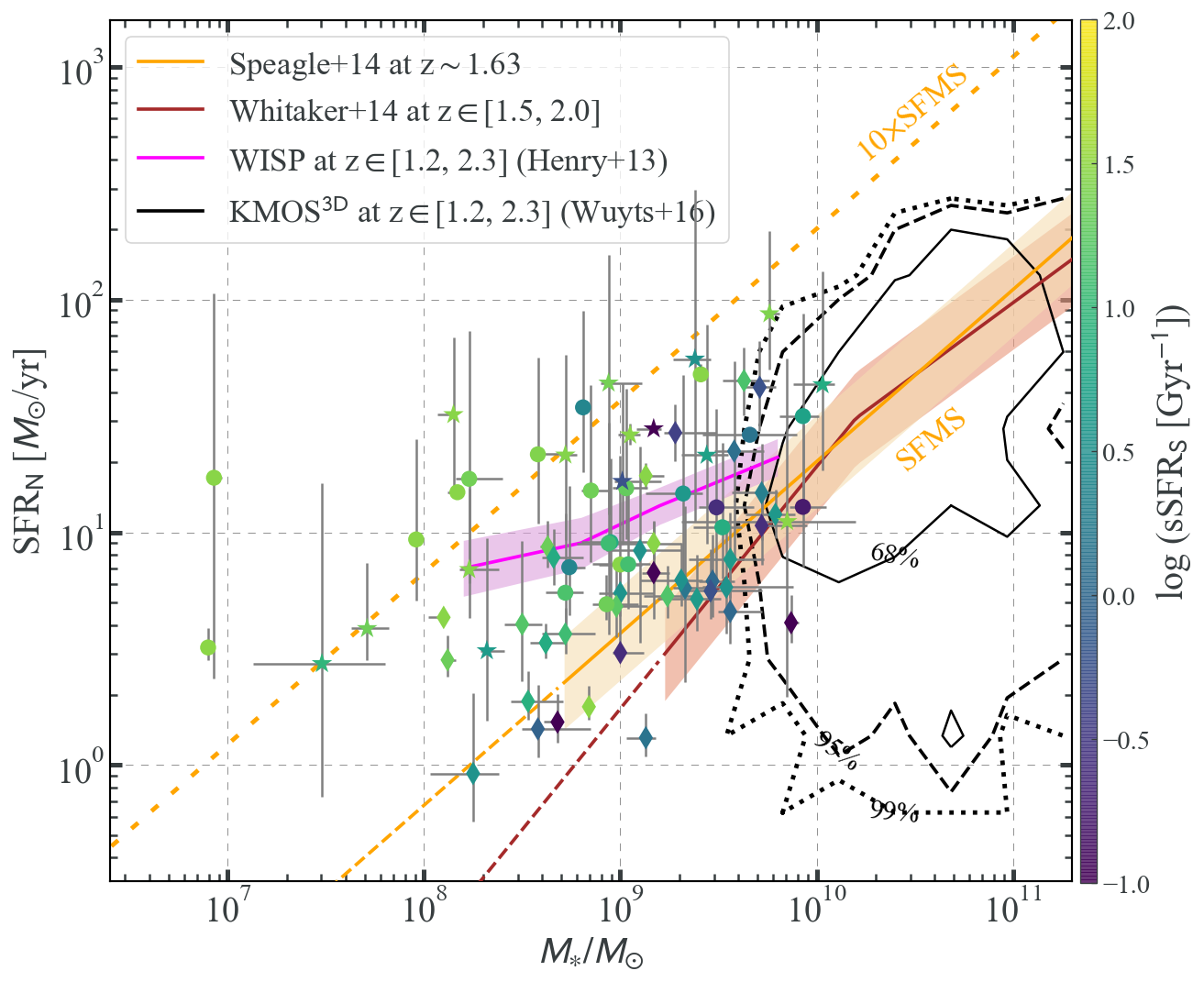}
    \includegraphics[width=.495\textwidth]{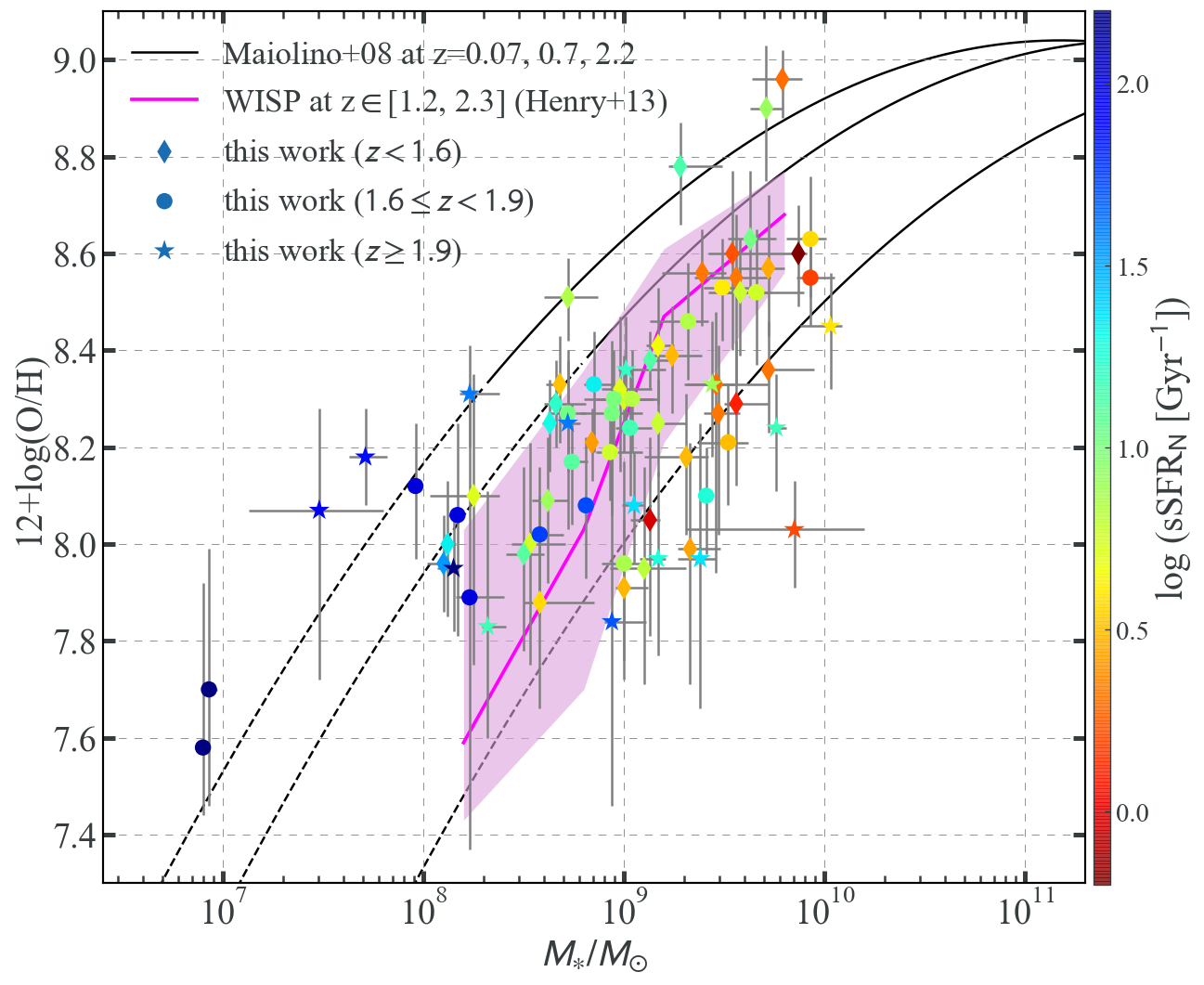}
    \caption{Global properties of our sample.
    \textbf{Left}: SFR as a function of \Mstar for galaxies at cosmic noon. Our galaxies are represented by the 
    symbols following the scheme in Fig.~\ref{fig:MEx} corresponding to different $z$ bins. However the color 
    coding reflects the specific SFR derived from stellar continuum SED fitting, after subtracting emission line fluxes (see 
    Sect.~\ref{subsect:Mstar}).
    The loci of our galaxies are consistent with that of the WISP survey \citep{Henry:2013gx}, if the SFR 
    inferred from dust-corrected nebular emission is adopted.
    We also show that in terms of mass coverage, our sample is highly complementary to the ground-based 
    mass-complete sample of \kd, which can only probe down to $\sim5\times10^{9}\Msun$ at $z\sim2$.
    \textbf{Right}: mass-metallicity relations for high-$z$ galaxies. The symbols of our sample now have the same 
    color-coding as in Fig.~\ref{fig:MEx}.
    Our galaxies follow similar trends of the MZRs from the WISP survey and \citet{2008A&A...488..463M}.
    In the low mass regime ($\Mstar\lesssim10^{8}\Msun$), our galaxies are more metal enriched than the simple extensions of those MZRs.
    These metal-enriched galaxies also have higher sSFR than the sample average.
    \label{fig:SFMS_MZR}}
\end{figure*}

On the left panel of Fig.~\ref{fig:SFMS_MZR}, we show the loci of our galaxies in the diagram of \SFR vs. \Mstar.
By selecting lensed galaxies via their nebular emission line flux, our sample reaches an instantaneous SFR limit of $\sim$1 
\Msun/yr at $z\sim2$.  In comparison to mass-complete samples \citep[from \eg, the \kd 
survey,][]{2016ApJ...827...74W} and galaxies from the star-forming main sequence 
\citep[SFMS,][]{Speagle:2014dd,Whitaker:2014ko}, we push the exploration of star-forming galaxies at the cosmic 
noon by 1-2 dex deeper into the low-mass regime.
We also show the loci of the spectral stacks from the WFC3 Infrared Spectroscopic Parallel (WISP) Survey 
\citep{Henry:2013gx}, very close to that of our galaxies given similar observing strategies.
We gain over 1 dex in \Mstar thanks to lensing magnification and the 14-orbit depth of the \glass data in each 
field.

\subsection{Metallicity and its radial gradient}\label{subsect:metal}

\begin{figure*}
    \centering
    \includegraphics[width=.495\textwidth]{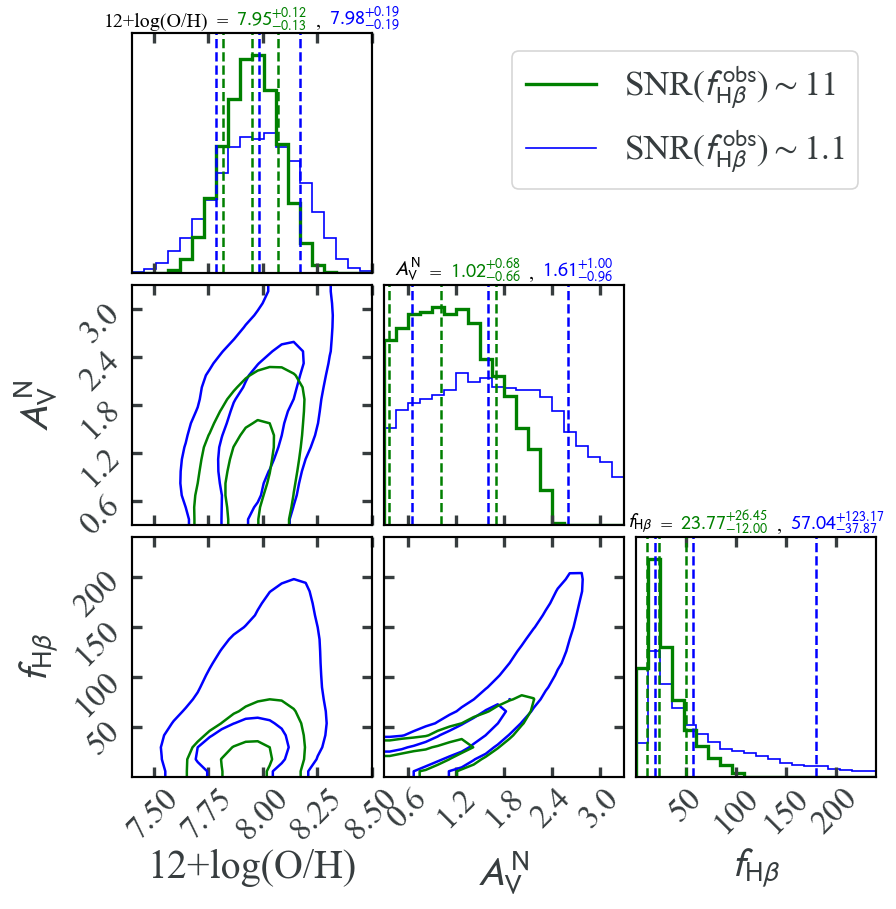}
    \includegraphics[width=.495\textwidth]{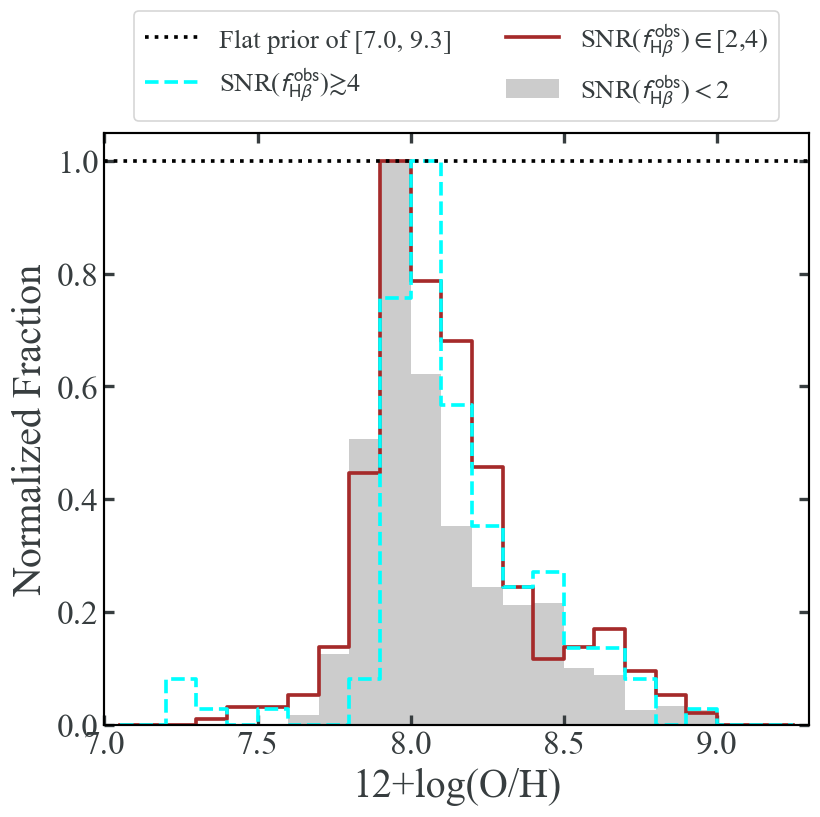}
    \caption{Rigorous constraint on metallicity from our forward-modeling Bayesian inference method.
    \textbf{Left}: marginalized 1D/2D constraints on metallicity (\oh), nebular dust extinction (A$_{\rm V}^{\rm N}$) and de-reddened 
    \Hb flux ($f_{\Hb}$) based on the integrated emission line fluxes of object MACS0416-ID00955, presented in Table~\ref{tab:srcprop} (also 
    see Figs.~\ref{fig:ID00955spec} and \ref{fig:deblend}).
    The parameter inference values shown on top of each column are medians with 1-$\sigma$ uncertainties drawn from the [16, 50, 
    84] percentiles, marked by the vertical dashed lines in the 1D histograms.
    The green color corresponds to the inference from the actual observed emission line fluxes where \Hb is detected at SNR$\sim$11 (\ie 
    $f_{\Hb}^{\rm obs}=6.88\pm0.60$).
    The blue-colored results are derived with the uncertainty of \Hb artificially increased by a factor of 10 (\ie SNR$\sim$1.1), 
    and other emission line flux measurements unchanged.
    The comparison between the green and blue colored results shows that although the constraints on A$_{\rm V}^{\rm N}$ and 
    $f_{\Hb}$ are severely worsened by the decrease in SNR of \Hb, the inference on \oh remains largely unchanged.
    This comparison thereby testifies that our forward-modeling Bayesian inference of metallicity does not require high SNR 
    detection of \Hb.
    \textbf{Right}: histograms of median metallicities measured using our forward-modeling method,
    in all individual Voronoi cells from our entire galaxy sample.
    The distribution of metallicity measurements is divided into three groups corresponding to three different 
    ranges of \Hb observed SNR in the corresponding Voronoi cells.
    The horizontal dotted line overlaid shows the flat prior of \oh$\in[7.0, 9.3]$ used in our Bayesian inference.
    We demonstrate that the returned metallicity estimates do not simply revert to 
    the prior and there is no systematic offset in our metallicity inference,
    even in the low SNR regime of \Hb.
    \label{fig:uncertHb}}
\end{figure*}

Following our previous work \citep{Wang:2016um,Wang:2019cf}, we adopt a \emph{forward-modeling} Bayesian method to infer simultaneously
metallicity (\oh), nebular dust extinction (A$_{\rm V}^{\rm N}$) and de-reddened \Hb flux ($f_{\Hb}$), based on 
observed emission line fluxes directly, as measured in Sect.~\ref{subsect:ELflux}.
We use flat priors for \oh and A$_{\rm V}^{\rm N}$, in the range of [7.0, 9.3] and [0, 4], respectively, which are 
appropriate for the \citet{2008A&A...488..463M} strong line calibrations adopted in our inference.
For $f_{\Hb}$, we use the Jeffrey's prior in the range of [0, 1000], in unit of $10^{-17}\Funit$.
The MCMC sampler \emc is used to explore the parameter space, with the likelihood function defined as
\begin{align}\label{eq:chi2}
    \mathrm{L}\propto\exp(-\chisq/2)=\exp\left(-\frac{1}{2}\cdot\sum_i \frac{\(f_{\el{i}} - R_i \cdot 
    f_{\Hb}\)^2}
        {\(\sigma_{\el{i}}\)^2 + \(f_{\Hb}\)^2\cdot\(\sigma_{R_i}\)^2}\right).
\end{align}
Here \el{i} represents each of the available emission lines, among the set of \OII, \Hg, \Hb, \OIII, \Ha, \SII.
$f_{\el{i}}$ and $\sigma_{\el{i}}$ denote the \el{i} flux and its uncertainty, de-reddened given a value of 
$\Av^{\rm N}$ drawn from the MCMC sampling.
The \citet{1989ApJ...345..245C} galactic extinction law with \Rv=3.1 is adopted to correct for dust reddening.
$R_i$ is the expected flux ratio between \el{i} and \Hb, with $\sigma_{R_i}$ being its intrinsic scatter.
The content of $R_i$ varies from strong-line diagnostic to Balmer decrement depending on the associated \el{i}. 
In practice, if $\el{i}$ is one of the Balmer lines, $R_i$ is given by $\Ha/\Hb=2.86$ and $\Hg/\Hb=0.47$, \ie, 
the Balmer decrement ratios assuming case B recombination under fiducial \HII region situations.
Instead, if $\el{i}$ is one of the oxygen collisionally excited lines, we take the strong-line flux ratios (\ie 
$f_{\OIII}/f_{\Hb}$ and $f_{\OII}/f_{\Hb}$) calibrated by \citet{2008A&A...488..463M} as $R_i$.
Last, if $\el{i}$ is \SII, we rely on our strong-line calibration of \SII/\Ha presented in \citet{Wang:2016um}.

This forward-modeling approach is superior to converting emission line flux ratios (\eg, $R_{23}$, $O_{32}$) to metallicity,
because it properly takes into account any weak nebular emission that falls short of the detection limit,
and avoids double counting information as it happens when combining multiple flux ratios that involve the same 
line.
All sources in our sample have SNR$\gtrsim$10 in at least one of the oxygen collisionally excited lines and/or \Ha (if source 
redshift is $z\lesssim1.6$), yet the \Hb detection is usually not as strong given its intrinsic faintness.
By not calculating observed emission line flux ratios but forward modeling observed line fluxes directly, 
we avoid compromising the high SNR detections of the bright \OIII and \OII lines by the faint \Hb lines.
As a result, our forward-modeling methodology improves our ability of accurate metallicity inference based on high SNR detections 
of strong nebular lines (\ie \OIII and \OII), and does not necessarily require high SNR detections of faint emission lines.
In the left panel of Fig.~\ref{fig:uncertHb}, we show the joint constraints on (\oh, A$_{\rm V}^{\rm N}$, 
$f_{\Hb}$), derived from the observed integrated emission line fluxes for an exemplary object 
MACS0416-ID00955, whose 1D/2D spectra are shown in Figs.~\ref{fig:ID00955spec} and \ref{fig:deblend}.  
Together we also simulate a scenario for this observation with much worse SNR detection of \Hb, \ie, 
artificially increasing the observed \Hb uncertainty by a factor of 10 while keeping other measurements 
unchanged.  It is found that the resultant constraint on \oh for this simulated scenario stays similar.  This 
test demonstrates that our metallicity inference method can still ensure reasonable constraints on \oh, even 
in cases where some emission lines such as \Hb are only marginally detected.
In the right panel of Fig.~\ref{fig:uncertHb}, we show the histograms of metallicity inferences (median values) 
given by our forward-modeling technique, in all individual Voronoi cells from our entire galaxy sample. We 
divide all these metallicity measurements in terms of the observed SNR of \Hb in the corresponding Voronoi 
cells.
Regardless of \Hb SNR, the three histograms all peak at \oh~$\sim$8.0, consistent with the integrated 
metallicity measurements from other work in similar ranges of redshift and mass \citep[see \eg 
Fig.~\ref{fig:SFMS_MZR} and][]{2008A&A...488..463M,Henry:2013gx}.
From this test, we show that there is no systematic offset of the distribution of inferred metallicities among 
the three groups divided by \Hb SNR, and our metallicity estimates do not simply revert to the prior used in 
our Bayesian inference.

Our forward-modeling Bayesian inference is first performed on the integrated emission line fluxes measured for each galaxy, to yield global 
metallicity.
On the right panel of Fig.~\ref{fig:SFMS_MZR}, we show the mass-metallicity relation (MZR) from our sample, 
color-coded by the specific \SFR ($\mathrm{sSFR}=\mathrm{SFR}/\Mstar$) obtained from the aforementioned analyses.
We also overlay the MZR from the WISP survey derived using the same strong-line calibrations 
\citep{Henry:2013gx}. It is encouraging to see that the two MZRs follow similar trends, due to similar source 
selection technique and observing strategy. Notably, our galaxies at the extreme low-mass end 
($\Mstar\lesssim10^9\Msun$) show both elevated metallicity and sSFR.
This is consistent with the hypothesis that these low-mass systems are in the phase of early mass assembly with 
efficient metal enrichment and minimum dilution from pristine gas infall.

In addition to the integrated emission line fluxes, from the procedures described in Sect.~\ref{subsect:ELmap}, we also 
obtain 2D spatial distributions of emission line surface brightnesses.
We utilize Voronoi tessellation as in \citet{Wang:2019cf} to divide spatial bins with nearly uniform SNRs of the 
strongest emission line available (usually \OIII).
Our spatially resolved analysis based on Voronoi tessellation is superior to averaging the signals in radial 
annuli, because of azimuthal variations (as large as 0.2 dex) in metallicity spatial distribution in nearby 
spiral galaxies \citep{2015ApJ...806...16B,Ho:2017he}.
Our Bayesian inference is then executed in each of the Voronoi bins for all sources, yielding their metallicity 
maps at sub-kpc resolution.

\begin{figure*}
    \centering
    \includegraphics[width=.16\textwidth]{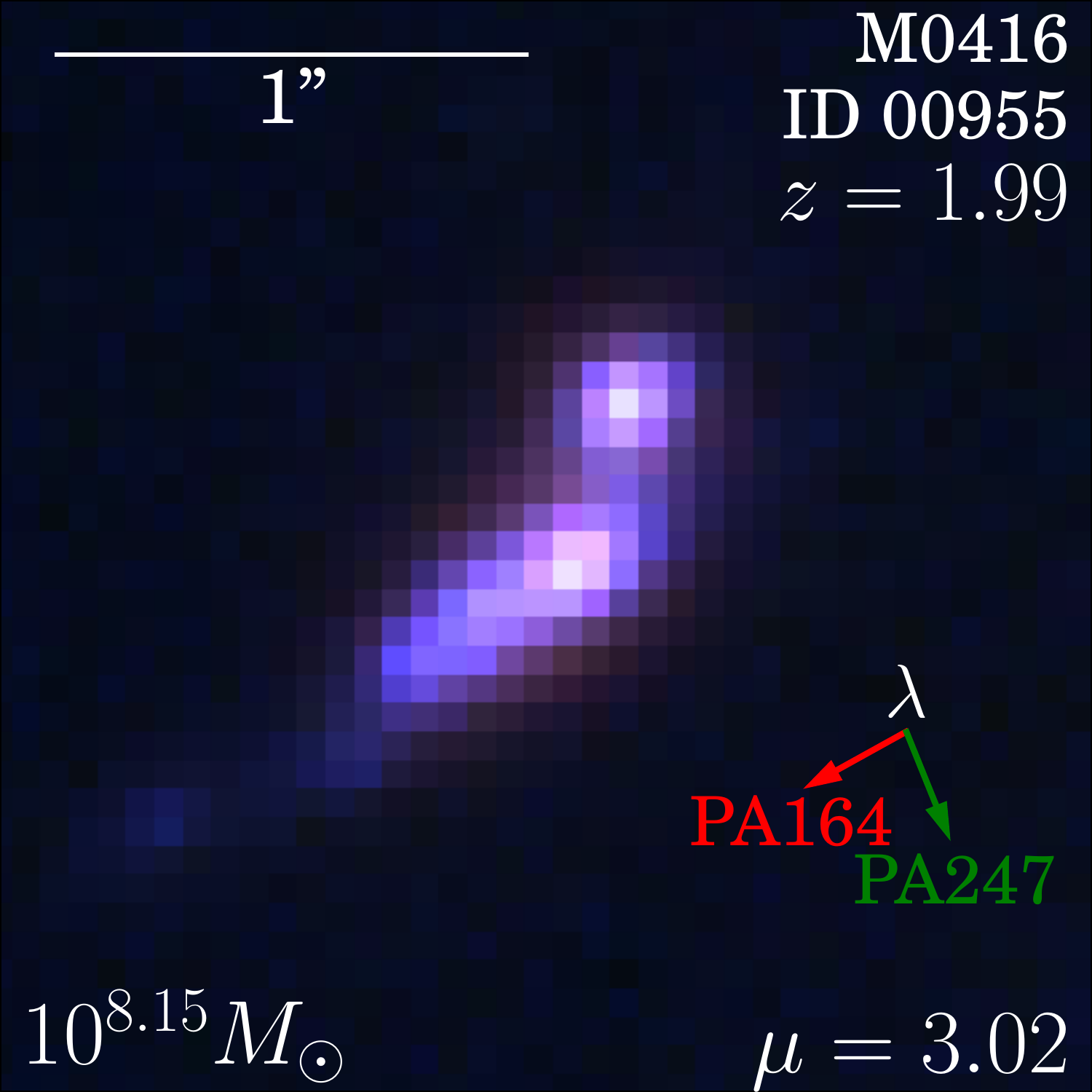}
    \includegraphics[width=.16\textwidth]{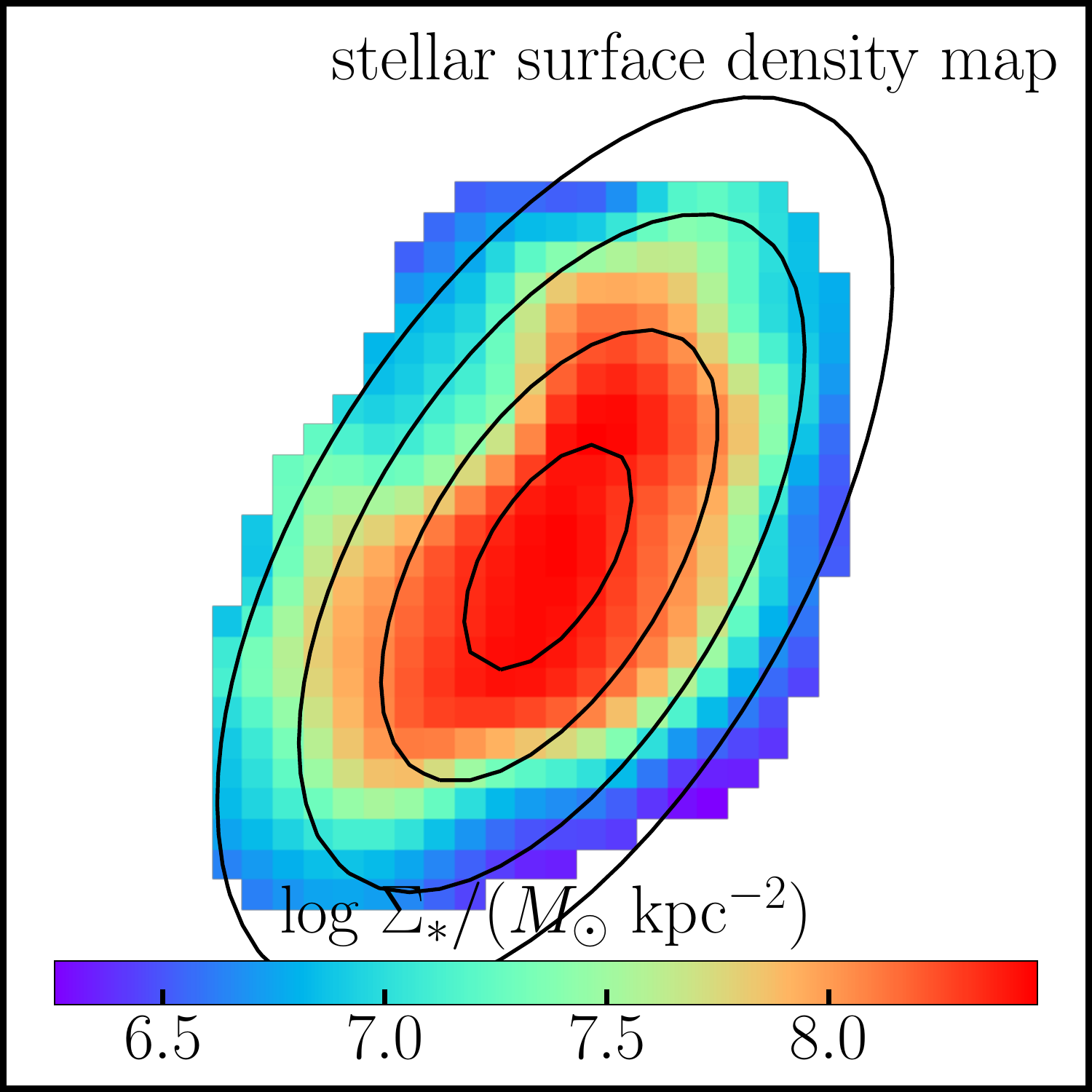}
    \includegraphics[width=.16\textwidth]{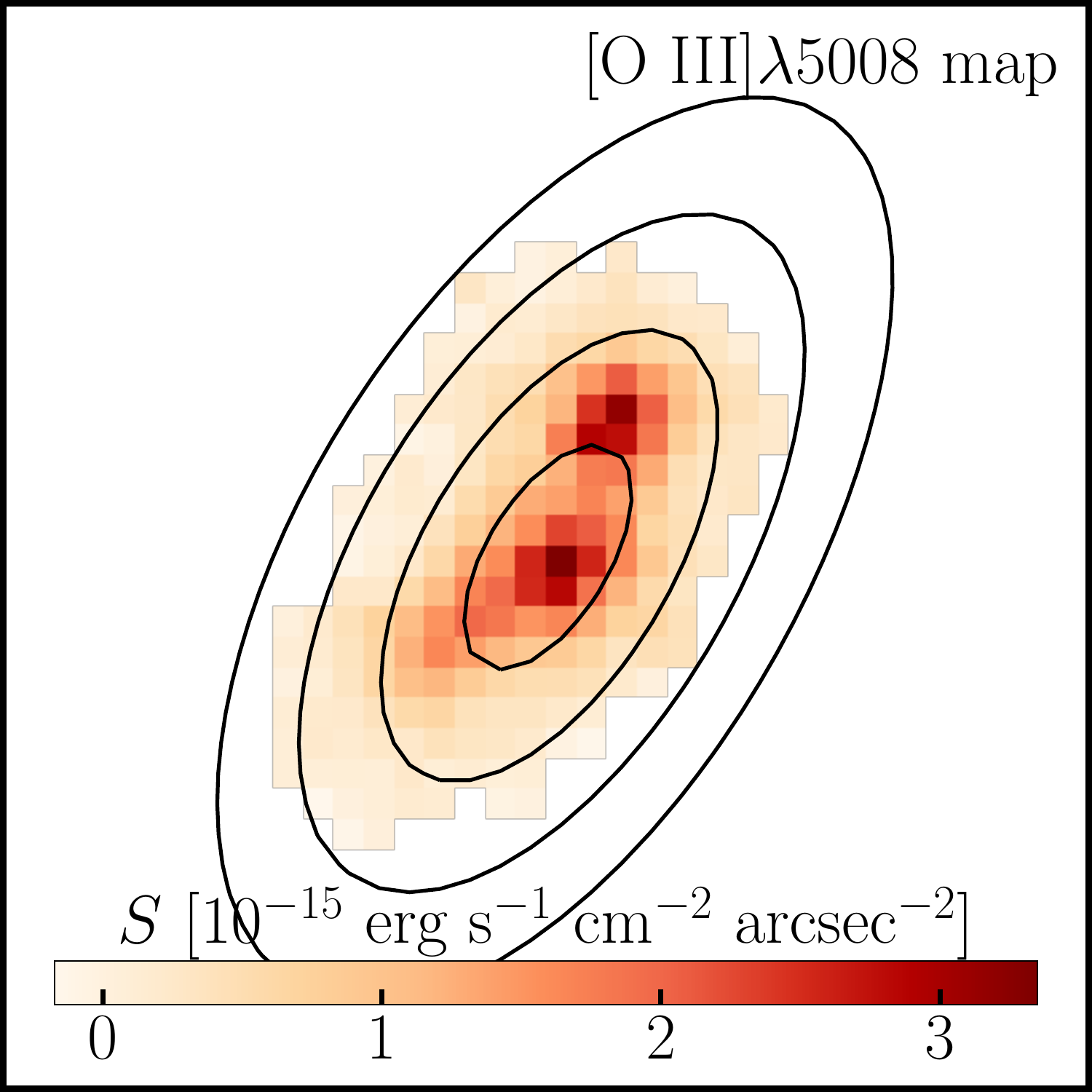}
    \includegraphics[width=.16\textwidth]{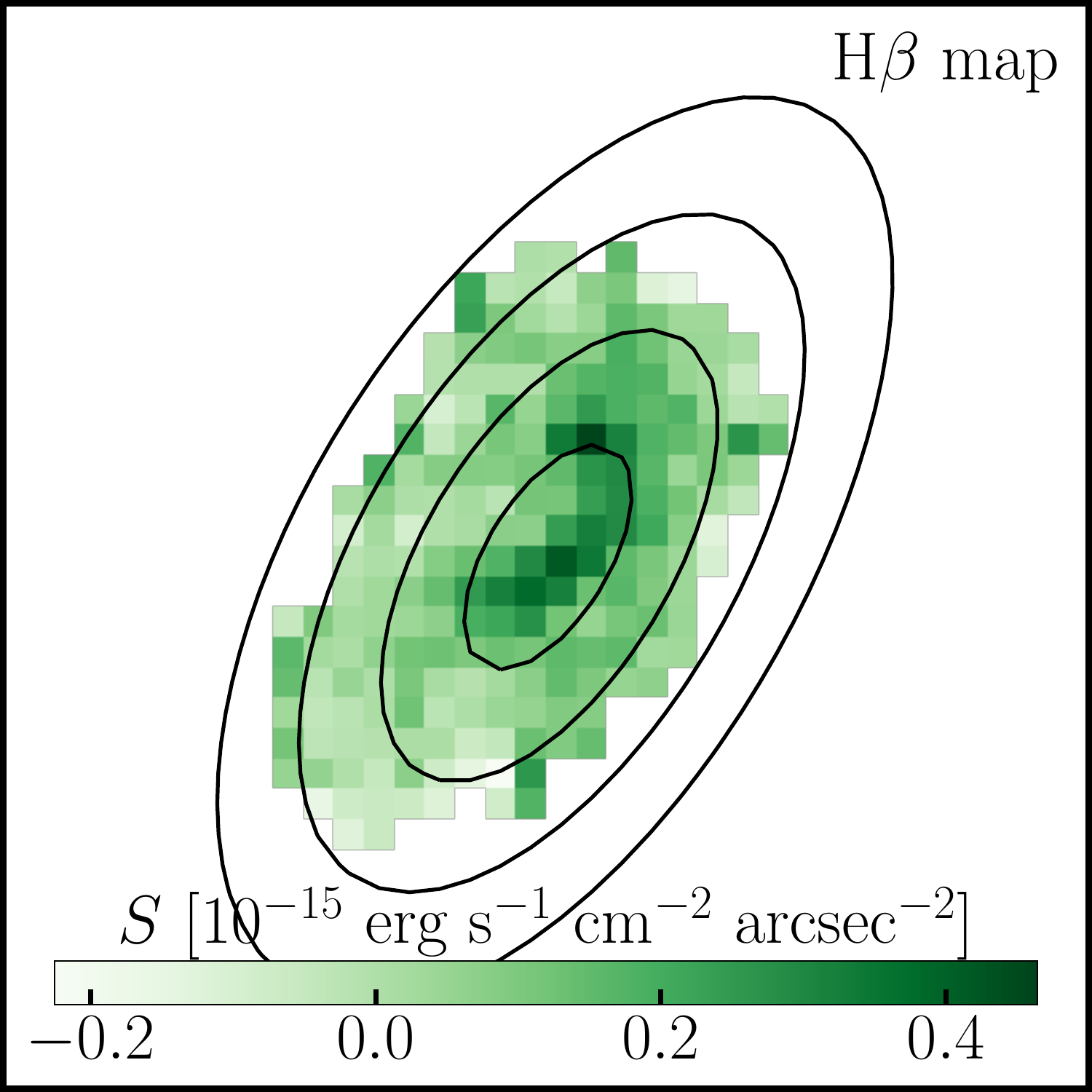}
    \includegraphics[width=.16\textwidth]{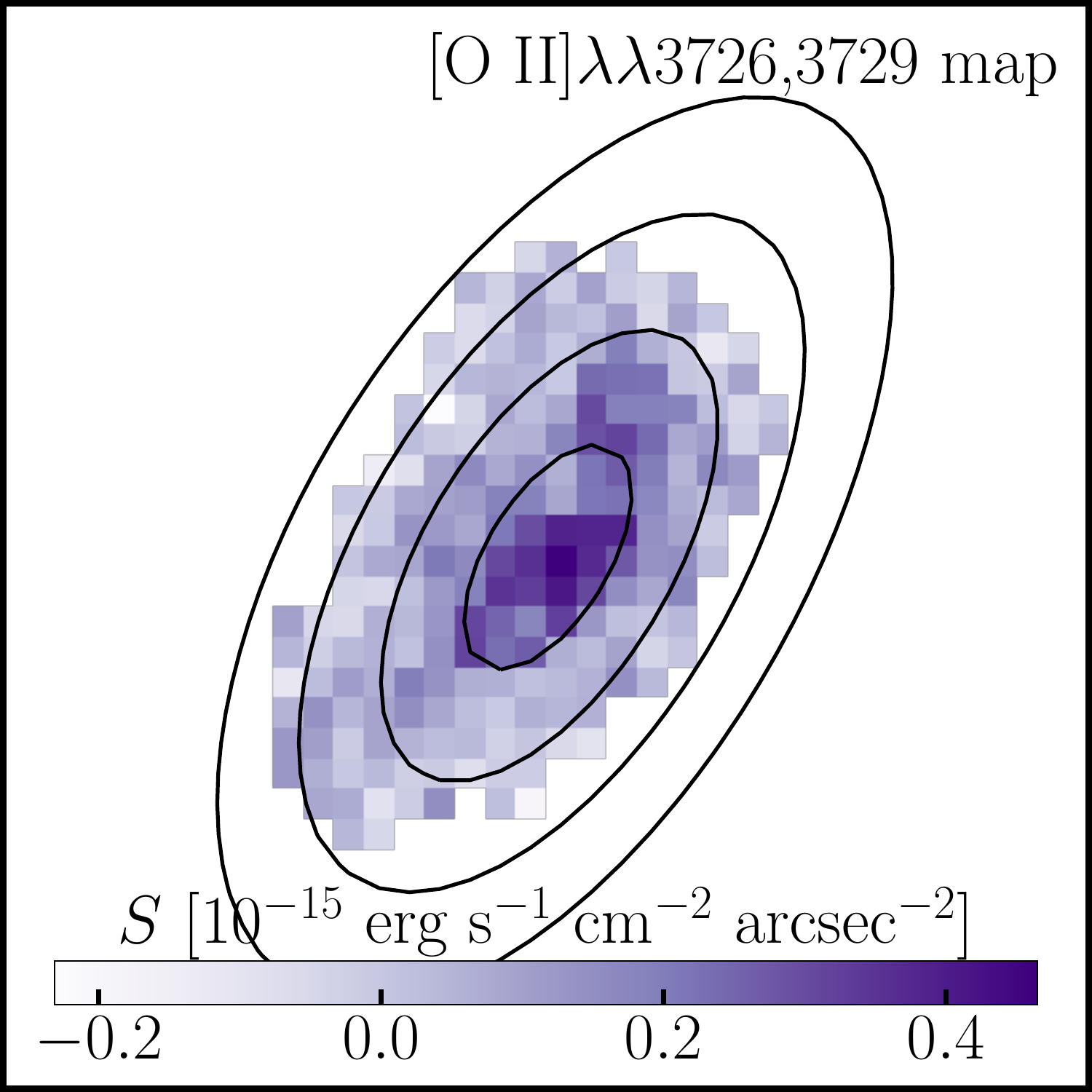}
    \includegraphics[width=.16\textwidth]{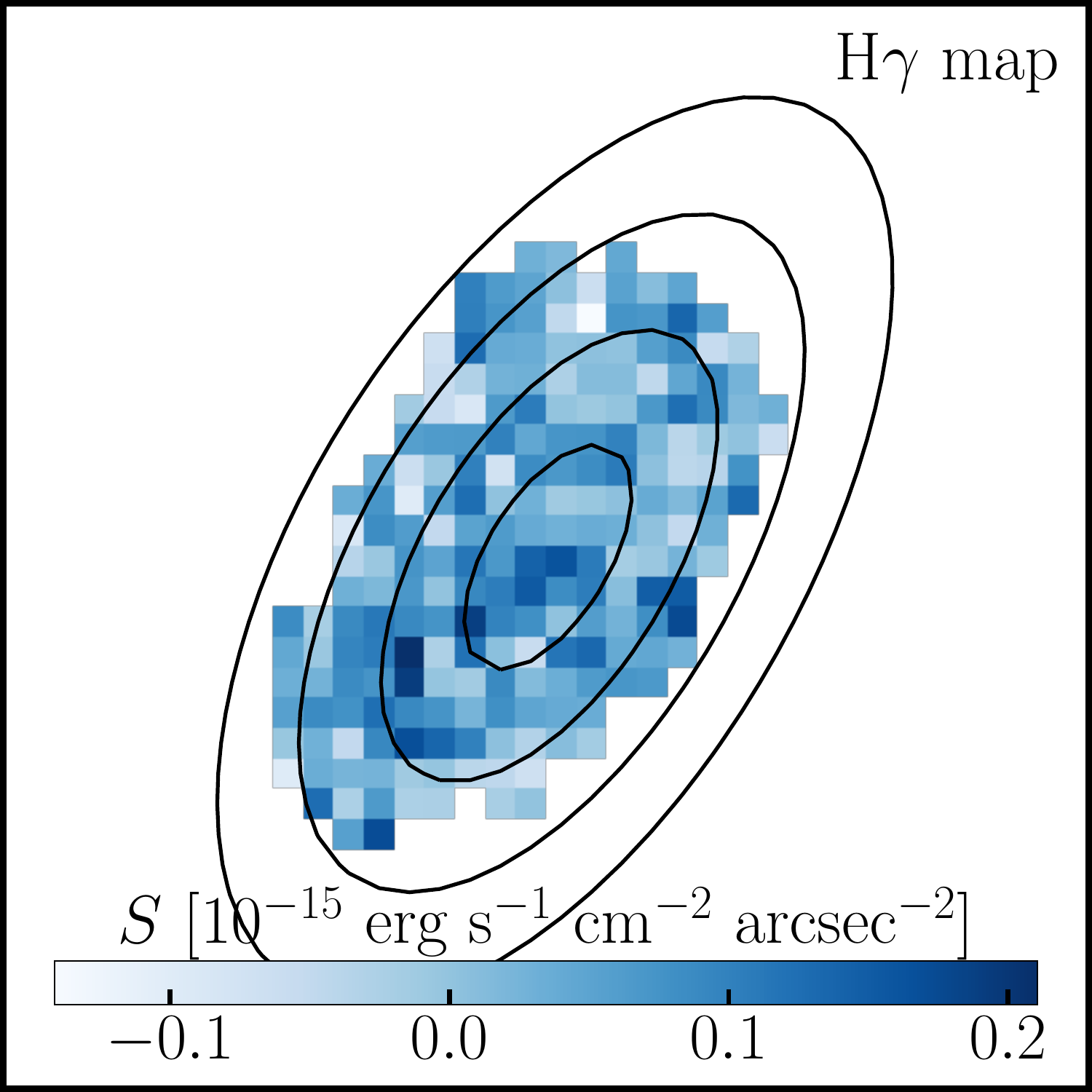}\\
    \includegraphics[width=\textwidth]{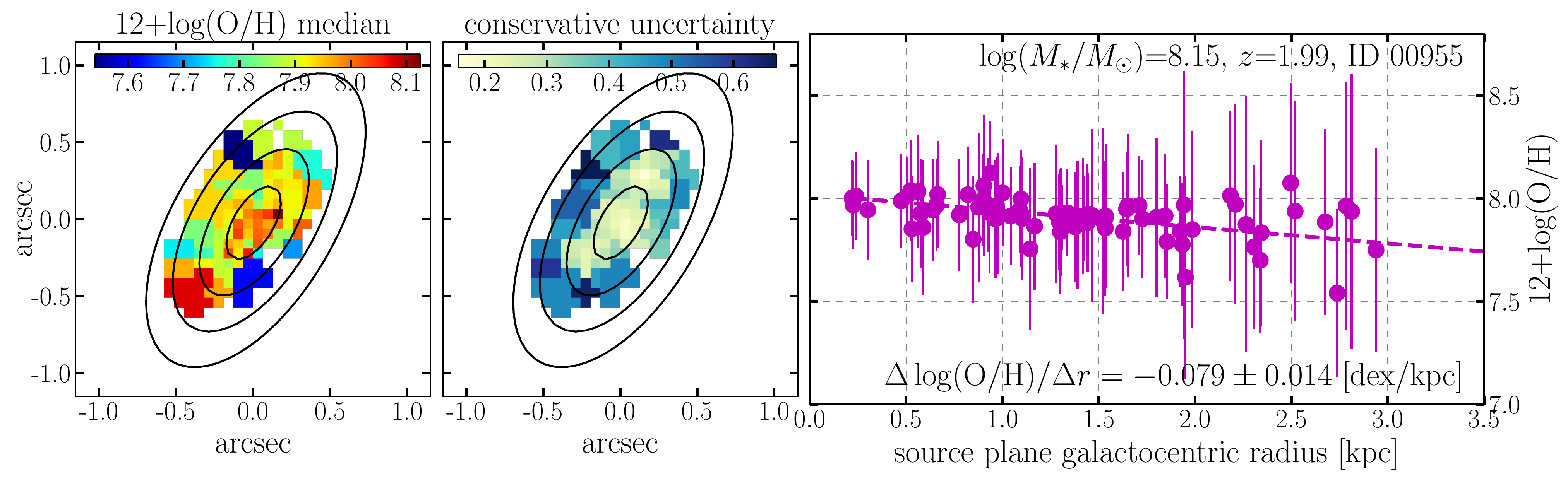}
    \caption{A $z$$\sim$2 star-forming dwarf galaxy ($\Mstar$$\simeq$$10^8$\Msun) with a negative metallicity radial gradient, similar 
    to that measured in our Milky Way \citep[\ie, $-0.07\pm0.01$,][]{Smartt:1997fm}. We show this as an example of the analysis 
    procedures applied to our entire sample.
    {\bf Top}, from left to right: color composite stamp (from the \hff imaging), stellar surface density (\Sstar) map (obtained from 
    pixel-by-pixel SED fitting to \hff photometry), and surface brightness maps of emission lines \OIII, \Hb, \OII, and \Hg.
    We use the technique demonstrated in Fig.~\ref{fig:deblend} to obtain pure \OIII$\lambda$5008 and \Hb maps for the source.
    The black contours mark the de-lensed de-projected galacto-centric radii with 1 kpc interval, given by our source plane 
    morphological reconstruction described in Sect.~\ref{subsect:metal}.
    {\bf Bottom}: metallicity map and radial gradient determination for this galaxy. The weighted Voronoi tessellation technique 
    \citep{Cappellari:2003eu,Diehl:2006cz} is adopted to divide the surface into spatial bins with a constant SNR of 5 on \OIII.
    In the right panel, the metallicity measurements in these Voronoi bins are plotted as magenta points.
    The dashed magenta line denotes the linear regression, with the corresponding slope shown at the bottom.
    The spatial extent and orientation remain unchanged throughout all the 2D maps in both rows, with north up and east to the left.
    \label{fig:combEL_metalgrad}}
\end{figure*}

To get the intrinsic deprojected galactocentric distance scale for each Voronoi bin, we conduct detailed 
reconstructions of the source-plane morphology of each galaxy in our sample.
We first obtain a 2D map of stellar surface density (\Sstar, \eg, as shown in Fig.~\ref{fig:combEL_metalgrad}) 
for each source through pixel-by-pixel SED fitting following the prescription described in 
Sect.~\ref{subsect:Mstar}.
Then the pixels in this map are ray-traced back to their source plane positions, according to the deflection 
fields given by the macroscopic cluster lens models.
For all the \hff clusters, we use the \SJ version 4corr models \citep{Johnson:2014cf}.
For the \clash-only clusters except RXJ1347, we use the Zitrin PIEMD+eNFW version 2 model.
For RXJ1347, we use our own model built following closely the approach in \citet{Johnson:2014cf}.
To this de-lensed 2D \Sstar map, we fit a 2D elliptical Gaussian function, to determine the galaxy's inclination, axis ratio, 
and major axis orientation, so that the source intrinsic morphology is recovered from lensing distortion.

Since we have measured both metallicity and source-plane de-projected galactocentric radius for each Voronoi bin, 
we can estimate a radial gradient slope via linear regression (see Appendix~\ref{sect:srcRecon} for the gradient measurements
based on metallicities derived in \emph{source-plane} Voronoi bins and related discussions about the effect of anisotropic lensing distortion).
Fig.~\ref{fig:combEL_metalgrad} demonstrates the entire process for measuring the metallicity radial gradient of 
a $z\sim2$ star-forming dwarf galaxy.
As a sanity check, we also measure its radial gradient using metallicity inferences derived in each individual spatial pixel and 
radial annulus.
We verified that the differences among the three methods are $\lesssim$0.03 dex/kpc, within the measurement uncertainties.

In the end, we secure a total of \Ntot galaxies in the redshift range of $1.2\lesssim z\lesssim2.3$ with sub-kpc 
resolution metallicity gradients (see Table~\ref{tab:obsdata} for the numbers of sources in individual cluster 
center fields).
This is hitherto the largest sample of such measurements in the distant Universe.
This sample enables robust measures of both average gradient slopes and scatter in the population.

\section{The cosmic evolution of metallicity gradients at high redshifts}\label{sect:gradVSz}

In this section, we collect published results on radial gradients of metallicity measured in the distant 
Universe.
We focus on the measurements that are derived with sub-kpc resolution, because insufficient spatial sampling is 
shown to cause spuriously flat gradient measurements \citep{2013ApJ...767..106Y}.
This poses a real challenge for ground-based observations, given the optimal seeing condition is $\sim$0\farcs6, 
equivalent to 5 \kpc at $z\sim2$.
There have been a number of attempts to overcome this beam smearing through correcting the distorted light wave 
front with the adaptive optics (AO) technique.
Using the \sinf instrument on the \vlt under the AO mode, \citet{2012MNRAS.426..935S} measured 7 gradients at 
$z\sim1.5$.
Following the same strategy, \citet{ForsterSchreiber:2018uq} expanded the sample by adding 21 new
measurements at $z\sim2$ from the SINS/zC-SINF survey\footnote{Note that \citet{ForsterSchreiber:2018uq} only 
published the radial gradients of \NII/\Ha measured in their sample galaxies. We convert those measurements into 
metallicity gradients following the widely adopted strong line calibration of \citet{2004MNRAS.348L..59P}.}.
Lensing can also help increase the spatial sampling rate.
\citet{2010ApJ...725L.176J,2013ApJ...765...48J} brought forward
this approach by securing 4 gradients at $z\sim2$ in galaxy-galaxy lensing 
systems using the AO-assisted \osiris instrument on the \keck telescope, with resolution further boosted 
$\gtrsim$3x by lensing magnification.
\citet{2015arXiv150901279L} carried out similar analyses and measured 11 new gradients at similar redshifts.
To recap, there exist a total of \Notherz metallicity gradient measurements with sub-kpc spatial resolution at 
cosmic noon before our work.

\begin{figure*}
    \centering
    \includegraphics[width=\textwidth]{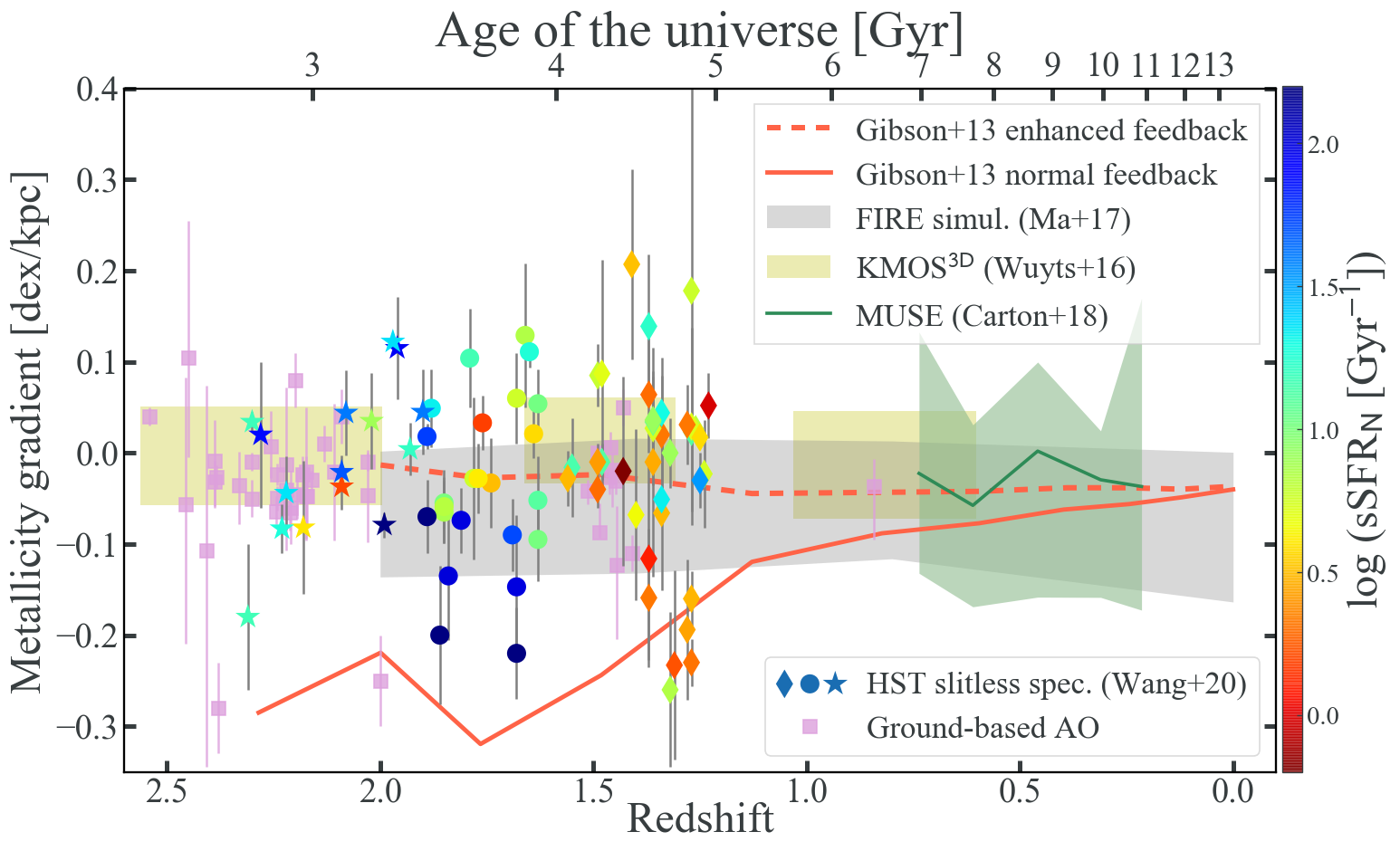}
    \caption{Overview of metallicity gradients in the distant Universe. Our measurements are represented by three symbols, corresponding to different 
    $z$ ranges as in Fig.~\ref{fig:MEx}, color-coded in sSFR.
    As a comparison, we also include individual measurements at similar resolution ($\lesssim$kpc scale) from 
    ground-based AO-assisted observations, marked by magenta squares 
    \citep{2012MNRAS.426..935S,2013ApJ...765...48J,2015arXiv150901279L,ForsterSchreiber:2018uq}.
    The 2-$\sigma$ spreads of measurements from \kd \citep{2016ApJ...827...74W} and \muse \citep{Carton:2018kv}, and the simulation results from FIRE 
    \citep{2017MNRAS.466.4780M}, are shown as shaded regions in green, yellow, and grey, respectively.
    The evolutionary tracks of two simulated disk galaxies (Milky Way analogs at $z\sim0$) with different 
    feedback strength but otherwise identical numerical setup are denoted by the two orange curves.
    \label{fig:gradVSz}}
\end{figure*}

In this work, we \emph{triple} the sample size by presenting \Ntot sub-kpc resolution metallicity radial 
gradients in star-forming galaxies at cosmic noon.
This is by far the largest homogeneous sample with sufficient spatial resolution, which enables a uniform 
analysis.
In Fig.~\ref{fig:gradVSz}, our results are highlighted by three sets of symbols --- corresponding to the three 
$z$ sub-groups --- color-coded in sSFR.
From a total of \Ntot galaxies in our sample with sub-kpc resolution gradient measurements, there are \NnegERsig and \NinvERsig 
sources showing negative and positive (\ie inverted) gradients greater than 2-$\sigma$ away from being flat, respectively.
At 3-$\sigma$ confidence level, the number of galaxies showing negative and inverted gradients are \NnegSANsig and \NinvSANsig, 
respectively.
Notably, two of the \NinvSANsig inverted gradients (A370-ID03751 and MACS0744-ID01203) have already been reported in detail in \citet{Wang:2019cf}.
All individual ground-based measurements at similar resolution ($\lesssim$kpc scale) are represented by magenta squares.
Recently \citet{Curti:2020gx} analyzed the \kmos observations in the field of RXJ2248, and measured metallicity 
gradients in 12 background galaxies lensed by RXJ2248, out of which three are in overlap with our sample (\ie ID00206, 
ID00428, ID01205).
We verified that the gradient results measured from both works are compatible at 1-$\sigma$ confidence level.
It is encouraging to see that the metallicity gradients derived using different methods and datasets are in good agreement.

Some theoretical trends are overlaid in Fig.~\ref{fig:gradVSz}.
In particular, two numerical simulations with different galactic feedback strengths but otherwise identical 
settings by \citet{Gibson:2013jw} are shown as the orange curves.
The comparison between these two trends demonstrates that enhanced feedback can be highly efficient in erasing 
metal inhomogeneity. Therefore resolved chemical properties, if measured accurately, can shed light on the 
strength of galactic feedback in the early phase of disk growth.

Fig.~\ref{fig:gradVSz} also shows the spread of the \kd gradient measurements by \citet{2016ApJ...827...74W}, 
which is highly clustered to flatness.
Without AO support nor lensing magnification gain on the spatial sampling rate, these gradients are usually obtained 
at a FWHM angular resolution of $\sim$0\farcs6, imposed by the natural seeing.
For a $z\sim1.5$ star-forming galaxy with intrinsically negative metallicity gradient
($\Delta\log({\rm O/H})/\Delta r= -0.16\pm0.02~[\mathrm{dex~kpc^{-1}}]$),
\citet{2013ApJ...767..106Y} show that from seeing-limited observations with a FWHM angular scale of $\sim$0\farcs5, 
its radial metallicity gradient is instead measured to be 
$\Delta\log({\rm O/H})/\Delta r= -0.01\pm0.03~[\mathrm{dex~kpc^{-1}}]$),
significantly biased towards flatness caused by beam smearing.
To mitigate the potential bias from beam smearing, \citet{Carton:2018kv} conducted a forward-modeling analysis to 
recover 65 gradients at $0.1\lesssim z\lesssim0.8$ from the seeing-limited \muse observations (marked in green in 
Fig.~\ref{fig:gradVSz}).

The 2-$\sigma$ interval of the FIRE simulations \citep{2017MNRAS.466.4780M} is shown as the grey-shaded region in 
Fig.~\ref{fig:gradVSz}.
We see that the scatter predicted by the FIRE simulations matches well that from low-$z$ observations \citep[at 
$z\lesssim1$, \eg, from][]{Carton:2018kv}, but it is smaller by a factor of 2 at higher redshifts, especially at 
$z\gtrsim1.3$.
This likely reflects that galaxies display more diverse chemo-structural properties at the peak epoch of cosmic 
structure formation and metal enrichment, when star formation is more episodic and vigorous \citep[see 
\eg.,][]{GalaxiesonFIREFe:2014dn}.

\section{The mass dependence of metallicity gradients at sub-kpc resolution: testing theories over 4 dex of 
\Mstar}\label{sect:gradVSm}

\begin{figure*}
    \centering
    \includegraphics[width=\textwidth]{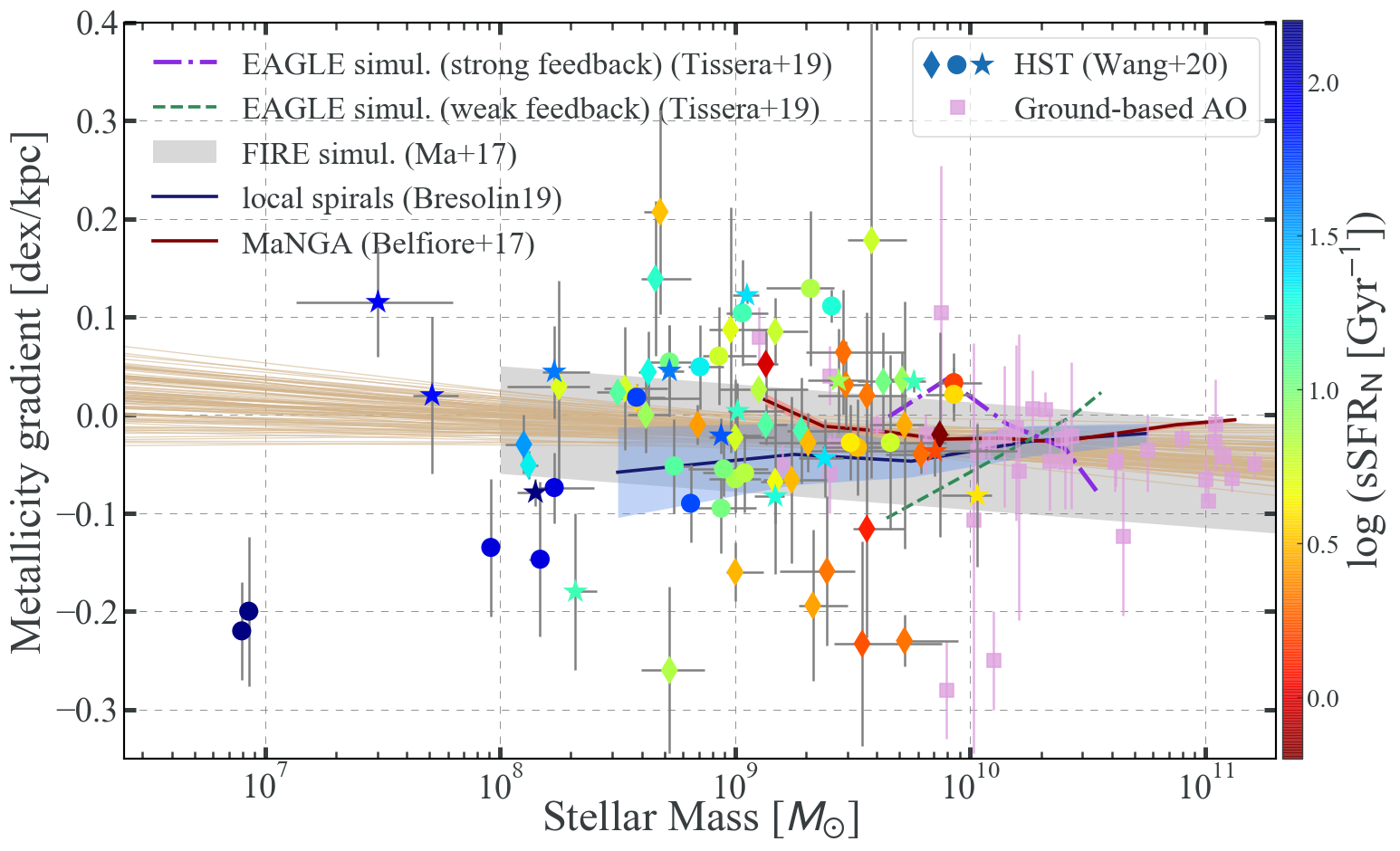}
    \caption{Metallicity gradient as a function of stellar mass for high-$z$ and local star-forming galaxies. As 
    in Fig.~\ref{fig:gradVSz}, our measurements are represented by three types of symbols regarding three $z$ 
    bins colored coded in sSFR, whereas high-$z$ ground-based measurements with similar resolution are denoted by 
    magenta squares.
    For comparison, we also show the median measurements with 1-$\sigma$ interval of local measurements 
    \citep{Belfiore:2017bv,Bresolin:2019wv}, the 2-$\sigma$ spread of the FIRE simulations 
    \citep{2017MNRAS.466.4780M}, and two mass dependencies derived from the EAGLE simulations assuming different 
    feedback settings \citep{Tissera:2018vv}.
    Combining all available high-$z$ gradients measured at sufficient spatial resolution ($\lesssim$kpc), we 
    obtain a weakly negative mass dependence over four orders of magnitude in \Mstar: $\Delta\log({\rm 
    O/H})/\Delta r~ [\mathrm{dex~kpc^{-1}}] = \left(-0.020\pm0.007\right) + \left(-0.014\pm0.008\right) 
    \log(\Mstar/10^{9.4}\Msun)$ with the intrinsic scatter being $\sigma=0.060\pm0.006$.
    The thin lines in tan mark 100 random draws from the linear regression.
    This observed mass dependence is in remarkable agreement with the predictions of the FIRE simulations.
    However, as shown in Table~\ref{tab:linreg}, we also observe an increase of the intrinsic scatter from 
    high-mass to low-mass systems, not captured by theoretical predictions.
    \label{fig:gradVSm}}
\end{figure*}

With the sample statistics greatly improved, we can quantify the mass dependence of reliably measured metallicity 
gradients at high redshifts, as a test of theoretical predictions.
The combined sample includes our \Ntot measurements at $z\in[1.2,2.3]$, and \Notherm\footnote{Only 3/11 gradients 
reported in \citet{2015arXiv150901279L} have \Mstar measured.} others' as given in Sect.~\ref{sect:gradVSz}.
Following the same color/marker styles as in Fig.~\ref{fig:gradVSz}, we plot these high-resolution gradient 
measurements as a function of their associated \Mstar in Fig.~\ref{fig:gradVSm}.
It is remarkable that now the observational data cover \emph{four orders of magnitude} in \Mstar.
Notably, over half of our gradient measurements reside in the dwarf mass regime 
($\Mstar\lesssim2\times10^9\Msun$), probing $\gtrsim$2 dex deeper into the low-mass end, compared with the 
ground-based AO results (magenta squares).

We perform linear regression on all these measurements of metallicity gradient and stellar mass, with errors on 
both quantities taken into account, using the following formula,
\begin{align}\label{eq:linreg}
    \Delta\log({\rm O/H})/\Delta r~[\mathrm{dex~kpc^{-1}}] = \alpha + \beta \log(\Mstar/M_{\rm med}) + 
    \mathrm{N}(0, \sigma^2).
\end{align}
Here $\alpha$ and $\beta$ are the intercept and the slope of the linear function, respectively.
$\mathrm{N}(0, \sigma^2)$ represents a normal distribution with $\sigma$ being the intrinsic scatter in units of 
$\mathrm{dex~kpc^{-1}}$. $M_{\rm med}$ is the median of the input stellar masses taken as normalization.
For the entire mass range (where $M_{\rm med}=10^{9.4}\Msun$), we obtain the following estimates:
$\alpha=-0.020\pm0.007$, $\beta=-0.016\pm0.008$, $\sigma=0.060\pm0.006$ (see the result of Case I in 
Table~\ref{tab:linreg}).
This shows a weak negative correction between metallicity gradient and stellar mass for these \Nall high-$z$ 
star-forming galaxies.

To understand this negative mass dependence, we show two theoretical predictions from the EAGLE simulations
in Fig.~\ref{fig:gradVSm}, corresponding to two suites of numerical simulations implementing different strengths 
of supernova feedback \citep{Tissera:2018vv}.
We see a drastic difference in the slope of the mass dependence of metallicity gradients predicted by different 
feedback settings in EAGLE, albeit the short \Mstar coverage.
This difference is largely caused by the bifurcations seen in the temporal evolutions of radial chemical profiles 
for individual galaxies, exemplified by the two simulation tracks shown in Fig.~\ref{fig:gradVSz}.
Under the assumption of weak feedback, galaxies evolve according to secular processes, and their radial gradients 
flatten over time \citep{Pilkington:2012ib}. Given mass assembly down-sizing, more massive galaxies are in a 
later phase of disk growth than less massive ones \citep{Brinchmann:2004hy}. Collectively, a positive mass 
dependence of radial gradients manifests.
However, when feedback is enhanced, feedback-driven gas flows can efficiently mix stellar nucleosynthesis yields 
and prevent any metal inhomogeneity from emerging \citep{2017MNRAS.466.4780M}.
This effect is more pronounced in lower mass galaxies living in smaller dark matter halos with shallower 
gravitational potentials.
As a result, a generally negative mass dependence (flat/inverted gradients at low-mass end and negative gradients 
at high-mass end) can be anticipated.

Fig.~\ref{fig:gradVSm} also shows the 2-$\sigma$ spread of the mass dependence from the FIRE simulations 
\citep{2017MNRAS.466.4780M}.
Given the relatively strong feedback scheme implemented in FIRE, we expect a negative mass dependence, which is 
indeed seen.
Remarkably, the predictions of the FIRE simulations match very well the linear regression fit based on the 
combined high-$z$ metallicity gradient sample.
Our result is thus in better agreement with enhanced feedback --- rather than secular processes --- playing a significant role in 
shaping the chemical enrichment and structural evolution during the disk mass assembly 
\citep{GalaxiesonFIREFe:2014dn,2014Natur.509..177V}.


To verify that the observed trends are robust, we subdivide the gradient measurements into three mass bins and 
perform a separate linear regression analysis to each bin. The results are given in Table~\ref{tab:linreg}.
We see that the intercept value becomes more negative as \Mstar increases, with the slopes all consistent with 
zero, confirming the negative mass dependence of metallicity gradients over the entire mass range.
More importantly, we observe an increase in the intrinsic scatter of metallicity gradients with \Mstar from 
high-mass to low-mass regimes, consistent with the findings in local spiral galaxies by \citet{Bresolin:2019wv}.
This increase in scatter can also be found if separating the galaxies based on their sSFR\footnote{Here we use 
the SED-derived SFR, \ie, SFR$^{\rm S}$ in Table~\ref{tab:srcprop}, for our galaxies to be self-consistent 
throughout the combined sample.}.
For galaxies in the combined sample with $\mathrm{sSFR}\gtrsim5~\mathrm{Gyr^{-1}}$, the scatter is constrained to 
be $\sigma=0.082_{-0.011}^{+0.012}$ (case IVa in Table~\ref{tab:linreg})
whereas for galaxies with $\mathrm{sSFR}\lesssim5~\mathrm{Gyr^{-1}}$, the scatter is instead 
$\sigma=0.046_{-0.006}^{+0.007}$ (case IVb in Table~\ref{tab:linreg}).

The increase of sSFR in low-mass systems can be ascribed to the accretion of low-metallicity gas from the cosmic 
filaments \citep[\ie, cold-mode gas accretion,][]{Dekel:2009fz}, or gravitational interaction events amplifying 
the star-formation efficiency \citep[\ie, merger-induced starbursts,][]{Stott:2013bc}. Both of them can bring 
about large dispersions in the radial chemical profiles.
To investigate which one of the two effects is more dominant in boosting the chemo-structural diversity in 
low-mass high-sSFR galaxies, we turn to the global MZR of our sample, presented in Section~\ref{subsect:metal} 
(see Fig.~\ref{fig:SFMS_MZR}).
We rely on the WISP measurements as the control sample, because of the similar source selection criteria, mass 
coverage, redshift range, and consistent techniques in estimating SFR (based on Balmer line fluxes) and 
metallicity (assuming the \citet{2008A&A...488..463M} calibrations).

We find that in the medium-mass bin ($\Mstar/\Msun\in[10^9,10^{10}]$), the galaxies in our sample with higher SFR 
than that of the WISP stacks, are more metal-poor by 0.15 dex than the WISP metallicities in the corresponding 
mass range.
This is supportive of the cold-mode accretion diluting the global metallicity of our galaxies, stimulating star 
formation and increasing the intrinsic scatter of metallicity gradients.
However, in the low-mass bin ($\Mstar/\Msun\in[10^8,10^9]$), our galaxies with higher SFR than WISP show 
significant metal-enrichment, \ie, higher by 0.27 dex than the corresponding WISP metallicities.
We hence argue that in the dwarf-mass regime of $\Mstar\lesssim10^9\Msun$, merger-driven starbursts play a more 
predominant role than the cold-mode accretion does to boost the chemo-structural diversity.
Our result is consistent with the sharp increase of the merger fraction --- from 10\% to over 50\% for galaxies 
at $\Mstar\sim10^{10}$ to $10^{8.5}$ at $z\sim1.5$ --- found by the HiZELS survey 
\citep{Stott:2013bc,2014MNRAS.443.2695S}.
For part of our dwarf galaxies on which we have mapped their gas kinematics using Keck OSIRIS, we also found that 
the velocity field becomes more turbulent (\ie with lower ratios of rotational speed versus velocity dispersion) 
for galaxies with higher sSFR \citep{Hirtenstein:2018tn}.
This kinematic evidence further reinforces the scenario that mergers boost the star-formation efficiency, random 
motions, and the chemo-structural diversity in dwarf galaxies at cosmic noon.

Lastly, the combined high-$z$ metallicity gradient sample reveals that inverted gradients are almost exclusively 
found in the low-mass range, \ie, $\Mstar\lesssim3\times10^9$ \citep[also see][]{Carton:2018kv}.
This feature is also seen in the local Universe: only the lowest \Mstar bin (at $\sim10^9\Msun$) from the MaNGA 
survey shows positive gradient slope \citep{Belfiore:2017bv}.
The reason for inverted gradients in isolated systems is still under debate, with possible causes ranging from 
centrally-directed cold-mode accretion \citep{Cresci:2010hr}, or metal-loaded outflows triggered by galactic 
winds \citep{Wang:2019cf}.
In any case, these processes should be more pronounced in low-mass systems, suggested by the occurrence rate of 
this inverted gradient phenomenon.

\begin{deluxetable*}{lcrccclcccccc}
    \tablecolumns{13}
    \tablewidth{0pt}
    \tablecaption{Linear regression results of the mass dependence of metallicity radial gradients at cosmic noon}
\tablehead{
    \colhead{Case} &
    \colhead{$\alpha$} &
    \colhead{$\beta$} &
    \colhead{$\sigma$}  &
    \colhead{$M_{\rm med} [\Msun]$}  &
    \colhead{$N_{\rm source}$}  &
    \colhead{Notes}
}
\startdata
    I       &   $-0.0203_{-0.0068}^{+0.0070}$    &   $-0.0156_{-0.0077}^{+0.0076}$    &   $0.0601_{-0.0057}^{+0.0065}$    &   $10^{9.4}$   &   111   &   all metallicity gradients measured at sub-kpc resolution \\
    II      &   $-0.0128_{-0.0097}^{+0.0099}$    &   $-0.0025_{-0.0181}^{+0.0181}$    &   $0.0677_{-0.0077}^{+0.0091}$    &   $10^{9.0}$   &    76   &   all metallicity gradients from \hst spectroscopy \\
    IIIa    &   $-0.0430_{-0.0085}^{+0.0078}$    &   $-0.0009_{-0.0179}^{+0.0186}$    &   $0.0342_{-0.0060}^{+0.0075}$    &   $10^{10.4}$  &    27   &   high-mass bin: $\Mstar/\Msun\gtrsim10^{10}$   \\
    IIIb    &   $-0.0082_{-0.0137}^{+0.0134}$    &   $-0.0520_{-0.0552}^{+0.0565}$    &   $0.0785_{-0.0104}^{+0.0122}$    &   $10^{9.5}$   &    47   &   medium-mass bin: $10^{9}\lesssim\Mstar/\Msun\lesssim10^{10}$    \\
    IIIc    &   $-0.0198_{-0.0139}^{+0.0140}$    &    $0.0270_{-0.0512}^{+0.0524}$    &   $0.0607_{-0.0125}^{+0.0154}$    &   $10^{8.6}$   &    31   &   low-mass bin: $10^{8}\lesssim\Mstar/\Msun\lesssim10^{9}$ \\
    IVa     &   $-0.0129_{-0.0132}^{+0.0128}$    &    $0.0207_{-0.0166}^{+0.0164}$    &   $0.0823_{-0.0108}^{+0.0125}$    &   $10^{8.9}$   &    50   &   high-sSFR bin: $\mathrm{sSFR}\gtrsim 5\mathrm{Gyr}^{-1}$ \\
    IVb     &   $-0.0344_{-0.0080}^{+0.0078}$    &   $-0.0074_{-0.0094}^{+0.0094}$    &   $0.0464_{-0.0062}^{+0.0074}$    &   $10^{9.7}$   &    61   &   low-sSFR bin: $\mathrm{sSFR}\lesssim 5\mathrm{Gyr}^{-1}$
\enddata
    \tablecomments{The linear regression is performed using the \linmix software\footnote{https://github.com/jmeyers314/linmix}
    taking into account the measurement uncertainties on both stellar mass (\Mstar) and metallicity gradient ($\Delta\log({\rm
    O/H})/\Delta r$), following the Bayeisan method proposed by \citet{Kelly:2007bv}. The following function form (Eq.~\ref{eq:linreg}) is adopted:
    $\Delta\log({\rm O/H})/\Delta r~[\mathrm{dex~kpc^{-1}}] = \alpha + \beta \log(\Mstar/M_{\rm med})+\mathrm{N}(0, \sigma^2)$.
    As given in the rightmost column, Cases I corresponds to the linear regression result based on all sub-kpc scale metallicity
    gradient measurements at the cosmic noon epoch, whereas Case II shows the result from our gradient measurements only.
    We divide the entire sample into three \Mstar bins and conduct linear regressions separately, with results represented by
    Cases IIIa,b,c.
    Cases IVa,b show the results if the entire sample is divided based on sSFR, instead of \Mstar.
    The number of sources ($N_{\rm source}$) involved in each case is shown in the second rightmost column.
    }
\label{tab:linreg}
\end{deluxetable*}


\section{Conclusion} \label{sect:conclu}

To summarize, we have presented an unprecedentedly large sample of sub-kpc resolution metallicity radial 
gradients in \Ntot gravitationally-lensed star-forming galaxies at $1.2\lesssim z\lesssim2.3$, using \hst near-infrared
slitless spectroscopy.
We performed state-of-the-art reduction of grism data, careful stellar continuum SED fitting after subtracting 
nebular emission from broad-band photometry, and Bayesian inferences of metallicity and \SFR based on emission line fluxes.
Our sample spans a \Mstar range of [10$^7$, 10$^{10}$] \Msun, an instantaneous \SFR range of [1, 100] \Msun/yr, 
and a global metallicity range of $7.6\lesssim\oh\lesssim9.0$, \ie, [$\frac{1}{12}$, 2] solar.
At 2-$\sigma$ confidence level, we secured \NnegERsig and \NinvERsig galaxies that show negative and inverted gradients, 
respectively.
Collecting all high resolution gradient measurements at high redshifts currently existing (where results 
presented in this work constitute 2/3 of all measurements), we measure a weak negative mass dependence over four 
orders of magnitude in \Mstar: $\Delta\log({\rm O/H})/\Delta r~ [\mathrm{dex~kpc^{-1}}] = 
\left(-0.020\pm0.007\right) + \left(-0.016\pm0.008\right) \log(\Mstar/10^{9.4}\Msun)$ with $\sigma=0.060\pm0.006$
being the intrinsic scatter.
This supports enhanced feedback as the main driver of the chemo-structural evolution of star-forming galaxies at 
cosmic noon.
Moreover, we also find that the intrinsic scatter of metallicity gradients increases with decreasing \Mstar and 
increasing sSFR.
Combined with the global metallicity measurements, our result is consistent with the hypothesis that the combined 
effect of cold-mode gas accretion and merger-induced starbursts strongly boosts the chemo-structural diversity of 
low-mass star-forming galaxies at cosmic noon, with mergers playing a much more predominant role in the 
dwarf-mass regime of $\Mstar\lesssim10^9\Msun$.
This work demonstrates that by accurately mapping the radial chemical profiles of star-forming galaxies at high 
redshifts, we can cast strong constraints on the role that feedback, gas flows and mergers play in the early 
phase of disk mass assembly.
The observed trends between metallicity and galaxy properties, while weak, are nonetheless very well measured 
over a wide dynamic range of mass. This census offers a stringent test for theoretical models and cosmological 
simulations, as the resulting trends are highly sensitive to baryon cycling processes at the peak of cosmic star 
formation ($1.2\lesssim z\lesssim2.3$).
Using the Near-Infrared Imager and Slitless Spectrograph (NIRISS) onboard the soon-to-be-launched James Webb 
Space Telescope (\jwst), the GLASS-\jwst ERS program (PI Treu, ID 1324) and the CAnadian NIRISS Unbiased Cluster 
Survey (CANUCS) GTO program (PI Willott) will conduct $K$-band slitless spectroscopy on several galaxy cluster 
center fields.
The data acquired by these programs will enable sub-kpc resolution measurements of metallicity gradients to 
$z\lesssim3.5$, and thus extend the test for theoretical predictions to even higher redshifts.

\acknowledgements
We thank the anonymous referee for careful reading and constructive comments that improve the quality of our paper.
This work is supported by NASA through HST grant HST-GO-13459.
XW acknowledges support by UCLA through a dissertation year fellowship.
XW is greatly indebted to his family, \ie, Dr. Xiaolei Meng, SX Wang and ST Wang, for their tremendous love, care and support during the COVID-19 pandemic, without which this work could not have been completed.

{\it Software:}
APLpy \citep{Robitaille:2012wl}, 
\adriz \citep{Gonzaga:2012tj}, 
Astropy \citep{astropy:2018ti}, 
\emc \citep{ForemanMackey:2013io}, 
\fast \citep{Kriek:2009eo}, 
Grizli (G. Brammer et al. in prep), 
\sex \citep{Bertin:1996hf}, 
VorBin \citep{Cappellari:2003eu}.

\vspace{0.1em}

\begin{appendix}

\section{Measuring metallicity radial gradients using Voronoi tessellation in the source plane}\label{sect:srcRecon}

As explained in Section~\ref{subsect:metal}, we use Voronoi tessellation to divide the spatial extent of our sample galaxies into
sub-regions, where we measure metallicities individually to estimate their radial gradients. This tessellation process is by default
performed in the image plane, since the noise properties of the observed emission line fluxes are well defined in the image plane.
Furthermore, the majority of our sample galaxies have magnifications less than 4 and the sample median value is $\mu=2.69$
(see the results presented in Table~\ref{tab:srcprop}), which indicates that the highly anisotropic lensing phenomenon is
relatively rare in our sample. 

\begin{figure*}
    \centering
    \includegraphics[width=.19\textwidth]{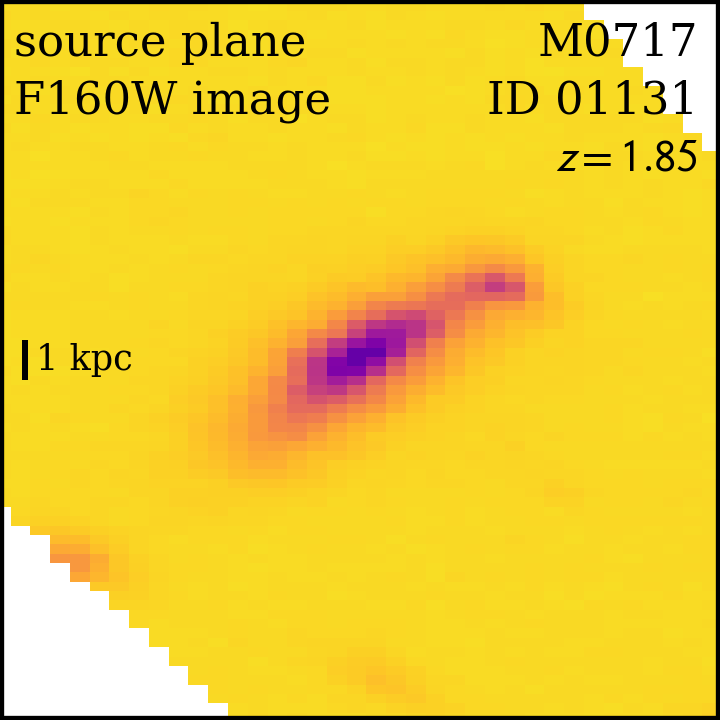}
    \includegraphics[width=.19\textwidth]{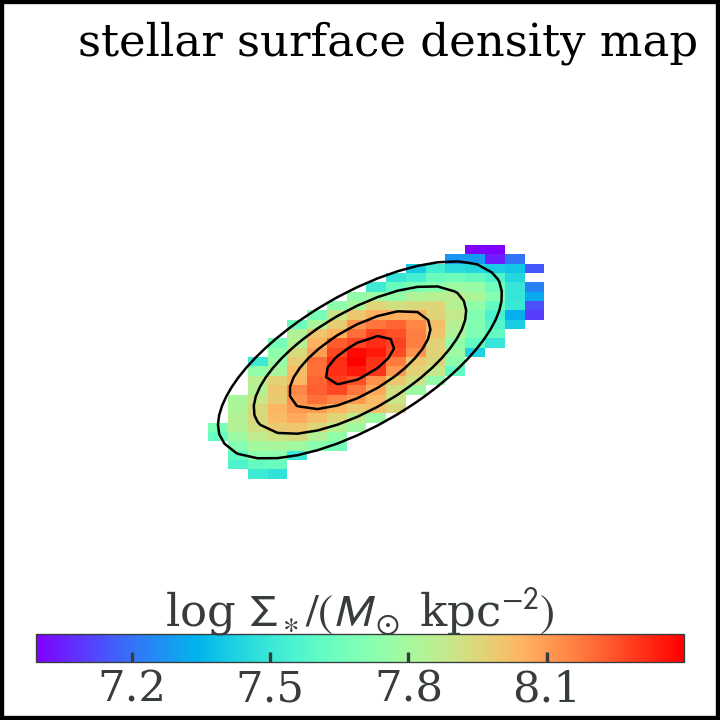}
    \includegraphics[width=.19\textwidth]{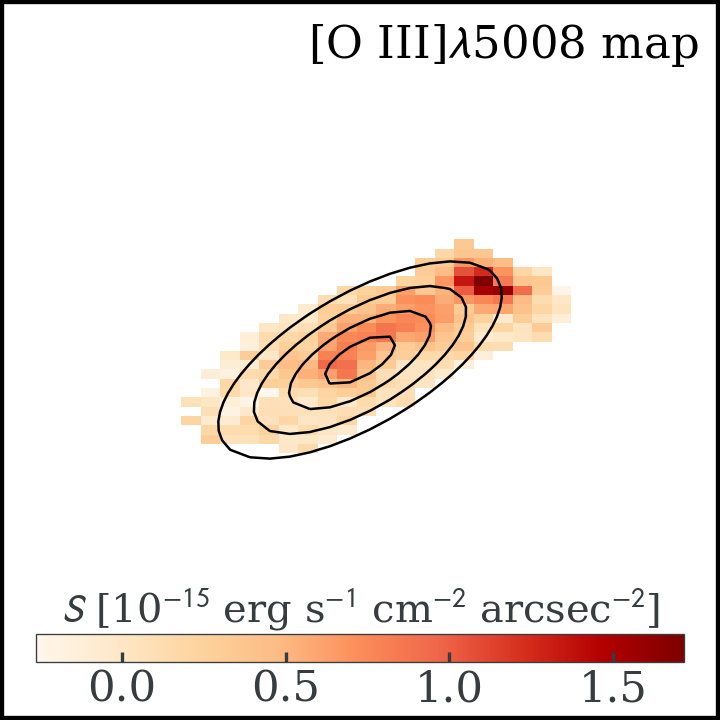}
    \includegraphics[width=.19\textwidth]{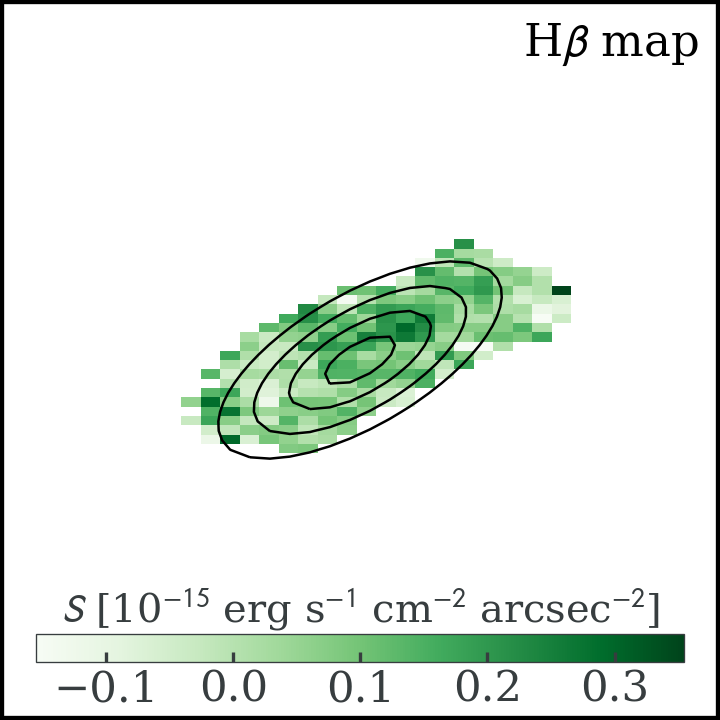}
    \includegraphics[width=.19\textwidth]{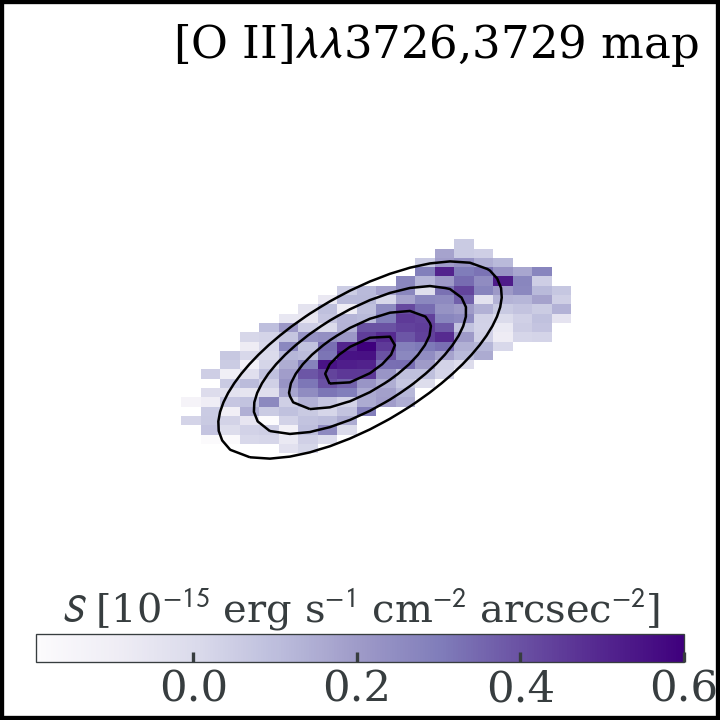}\\
    \includegraphics[width=\textwidth]{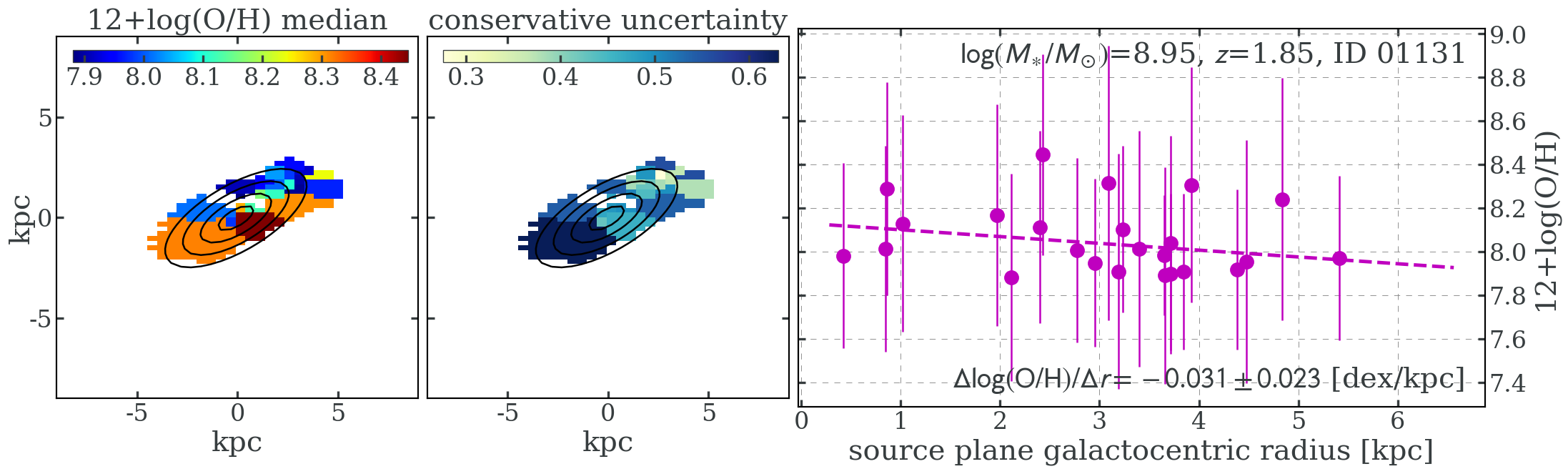}
    \caption{
    Source-plane metallicity radial gradient measurement of one highly anisotropically magnified galaxy (MACS0717-ID01131) in our sample.
    Its image-plane metallicity radial gradient measurement is presented in Figure~\ref{fig:clM0717_ID01131_figs}.
    {\bf Top}, from left to right: the source-plane reconstructed 2D maps of \H-band surface brightness, stellar surface density
    (\Sstar), and surface brightness of emission line \OIII, \Hb, and \OII.
    These 2D maps are arranged on the same spatial scale with a scale bar of 1 kpc shown in the leftmost panel.
    The black contours mark the de-projected galacto-centric radii with 1 kpc interval, with galaxy inclination taken into
    account. Note that the black radius contours in Figure~\ref{fig:clM0717_ID01131_figs} are obtained from re-lensing the contours shown
    in this figure to the image plane of this galaxy.
    {\bf Bottom}: metallicity map and radial gradient determination for this galaxy in its source plane.
    We again use Voronoi tessellation to divide its source-plane reconstructed spatial extent into bins with a constant
    SNR of 5 on \OIII, the same as used in the gradient measurement in the image plane.
    In the right panel, the metallicity measurements in these source-plane Voronoi bins are plotted as magenta points.
    The dashed magenta line denotes the linear regression, with the corresponding slope shown at the bottom.
    The metallicity gradient measured in the source plane is $\Delta\log({\rm O/H})/\Delta r= -0.031\pm0.023~
    [\mathrm{dex~kpc^{-1}}]$, in agreement with the gradient measured in the image plane, \ie, $\Delta\log({\rm O/H})/\Delta
    r=-0.055\pm0.028~[\mathrm{dex~kpc^{-1}}]$.
    \label{fig:metalgrad_srcpix}}
\end{figure*}

Nevertheless, there indeed exist some galaxies in our sample that are highly anisotropically magnified. It is thus important to
verify that the highly anisotropic lensing effect does not introduce significant systematic offset into their radial gradients measured
in the image plane. For that purpose, we measure their radial gradients in the source plane, using similar methods outlined in
Section~\ref{subsect:metal}.
Figure~\ref{fig:metalgrad_srcpix} shows such analysis of one exemplary source in our sample, \ie, MACS0717-ID01131 at $z=1.85$ with
$\mu=5.88$. The image-plane morphology, shown in Figure~\ref{fig:clM0717_ID01131_figs}, indicates that one of the two spatial
directions is preferentially magnified.
Here, unlike the procedures given in Figure~\ref{fig:clM0717_ID01131_figs}, we first transform the observed 2D emission line maps
of this galaxy into its
source plane, by ray-tracing each pixel to their source-plane positions according to the lensing deflection fields given by the adopted macroscopic lens model.
Then the weighted Voronoi tessellation technique \citep{Cappellari:2003eu,Diehl:2006cz} is again adopted to divide the source-plane surface
into spatial bins with a constant SNR of 5 on \OIII, the same as used in the gradient measurement in the image plane.
Our forward-modeling Bayesian method of metallicity inference is conducted in each individual source-plane Voronoi bin to yield
metallicity maps in the source plane.

For the galacto-centric distance scale of each individual source-plane Voronoi bin, we again rely on the 2D elliptical Gaussian
function fit to the source-plane stellar mass surface density map of this galaxy\footnote{Note that this fitting is always
performed for metallicity gradient measurements in the image plane, such that the black contours shown in all image-plane 2D maps
for our sample galaxies are obtained by re-lensing the corresponding best-fit source-plane de-projected galacto-centric radius contours.}
(see the second to the left panel in Figure~\ref{fig:metalgrad_srcpix}).
This fitting procedure yields the best-fit inclination, axis ratio, and major axis orientation of the galaxy, so that 
we not only take out the effect of lensing distortion, but also take into account the projection effect when determining the source intrinsic morphology.
At last, the radial gradient can be given by a linear regression to the metallicity estimates in all source-plane Voronoi bins.

For our exemplary differentially magnified source MACS0717-ID01131, the metallicity gradient measured in its source plane
is $\Delta\log({\rm O/H})/\Delta r= -0.031\pm0.023~[\mathrm{dex~kpc^{-1}}]$, in agreement at 1-$\sigma$ confidence level with the
gradient measured in the image plane, \ie, $\Delta\log({\rm O/H})/\Delta r= -0.055\pm0.028~[\mathrm{dex~kpc^{-1}}]$ (shown in
Figure~\ref{fig:clM0717_ID01131_figs} and given in Table~\ref{tab:srcprop}).
We verified that this difference ($\lesssim$0.03 $\mathrm{dex~kpc^{-1}}$, compatible within 1-$\sigma$) is typical for the few
highly anisotropically magnified galaxies in our sample.
In fact, the metallicity radial gradient measurement in this galaxy has been presented in our previous work
\citep{2015AJ....149..107J}. Using half of the grism data (2 orbits of G141 and 5 orbits of G102) available at that time,
\citet{2015AJ....149..107J} estimated the radial metallicity gradient of this galaxy to be $-0.03\pm0.03~\mathrm{dex~kpc^{-1}}$
from metallicities measured in individual spatial pixels, and $-0.05\pm0.05~\mathrm{dex~kpc^{-1}}$ from metallicities derived in
radial annuli.
We see that our updated results derived in both the image and source planes presented in this work are compatible with previous
measurements within measurement uncertainties.

\section{A summary of the data products and analysis results for the full metallicity gradient sample
presented in this paper (online material)}\label{sect:online}

\begin{figure*}
    \centering
    \includegraphics[width=\textwidth]{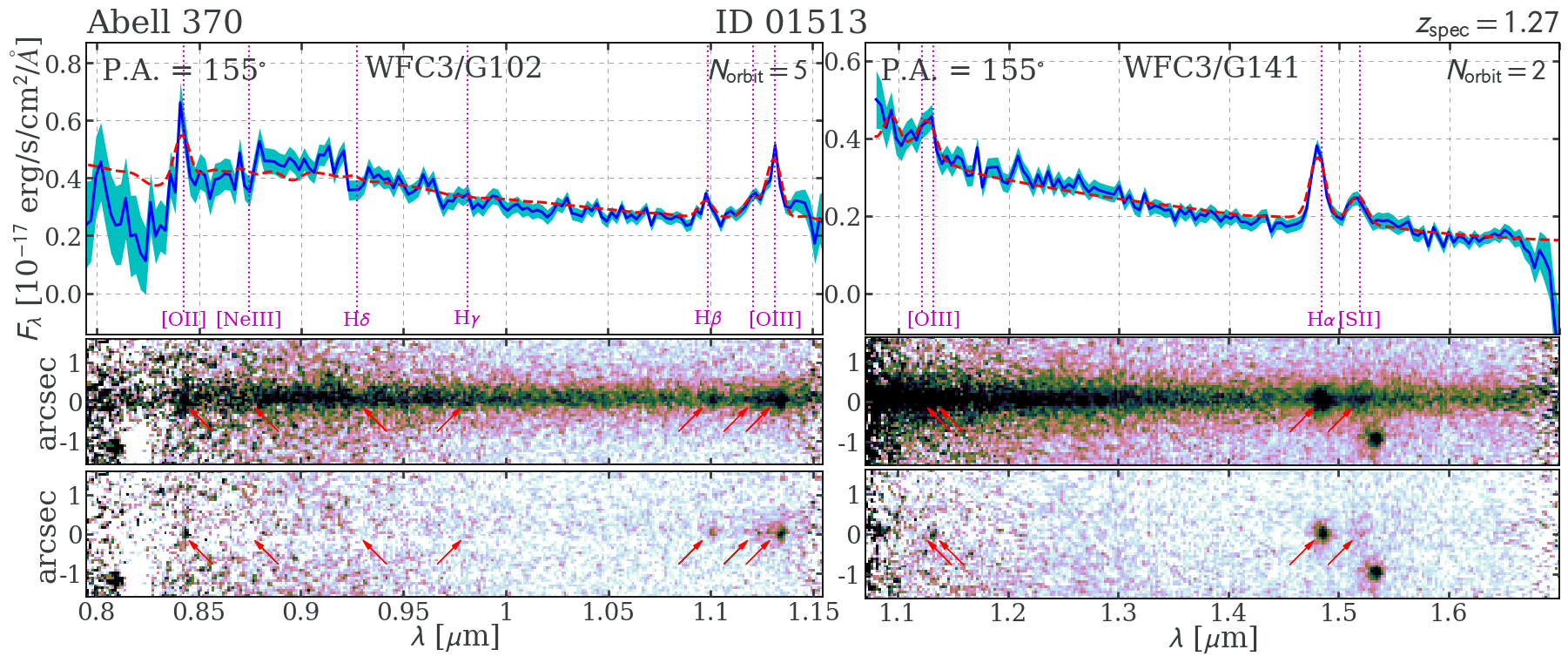}\\
    \includegraphics[width=\textwidth]{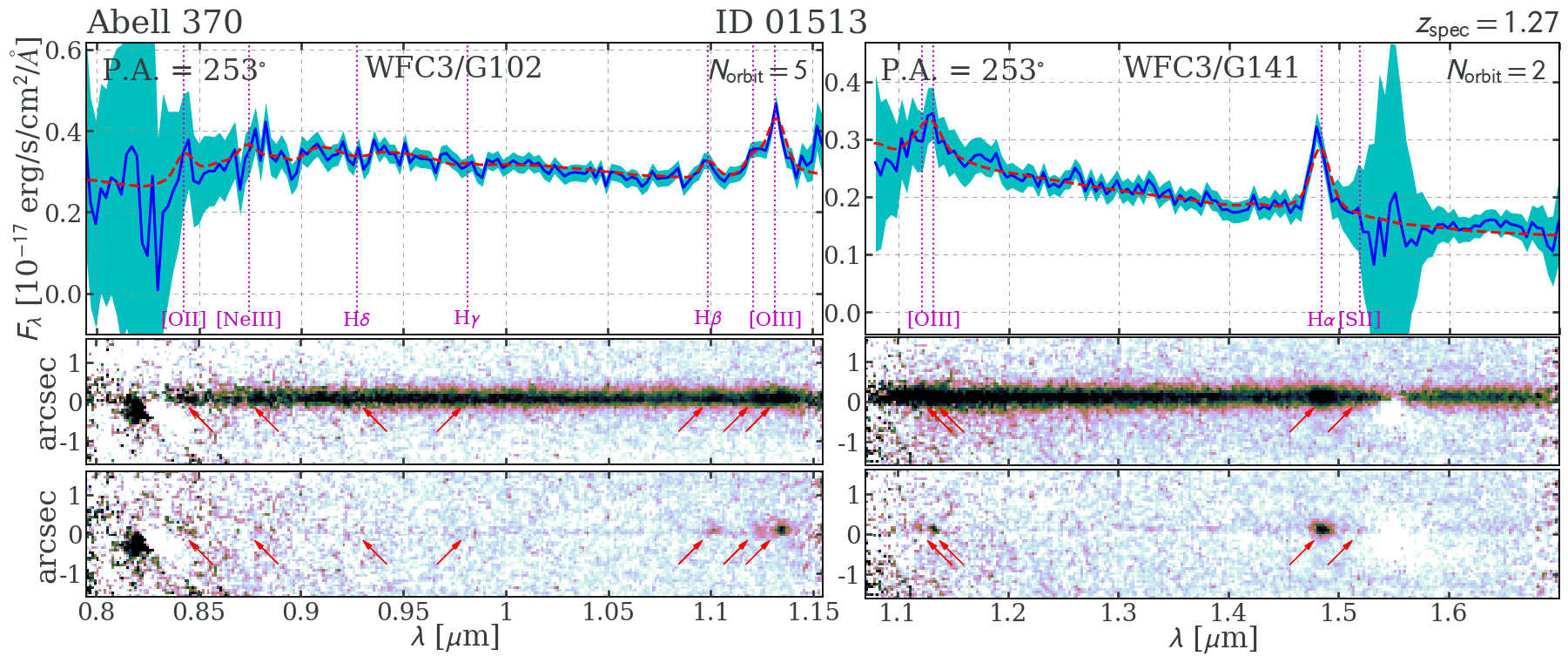}\\
    \includegraphics[width=.16\textwidth]{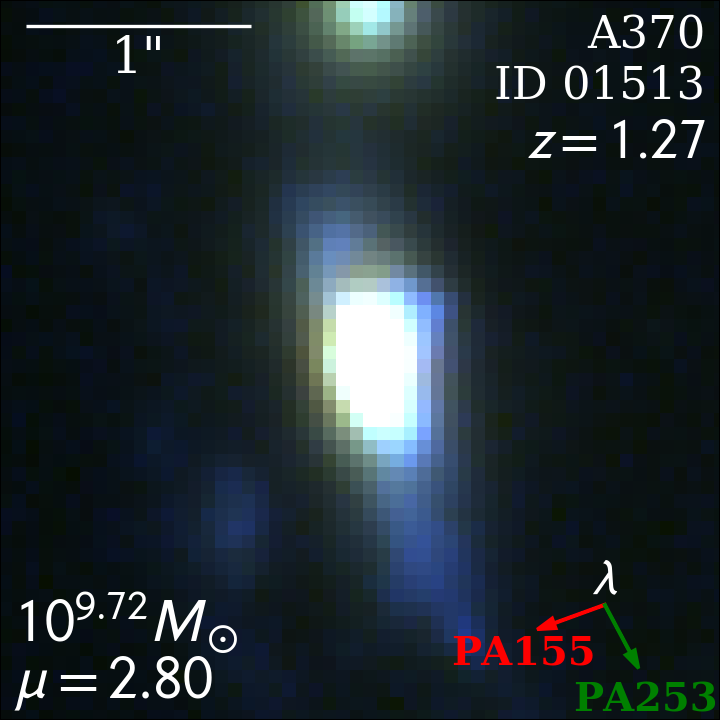}
    \includegraphics[width=.16\textwidth]{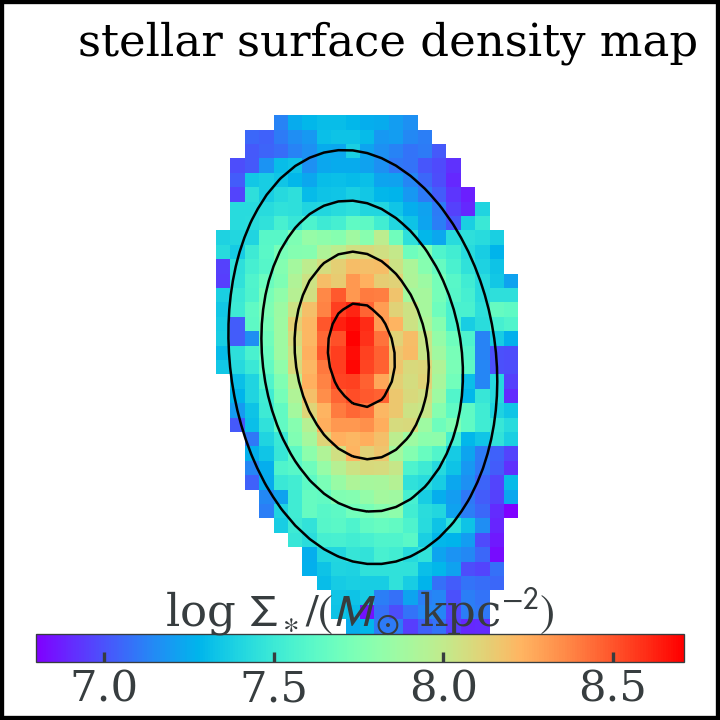}
    \includegraphics[width=.16\textwidth]{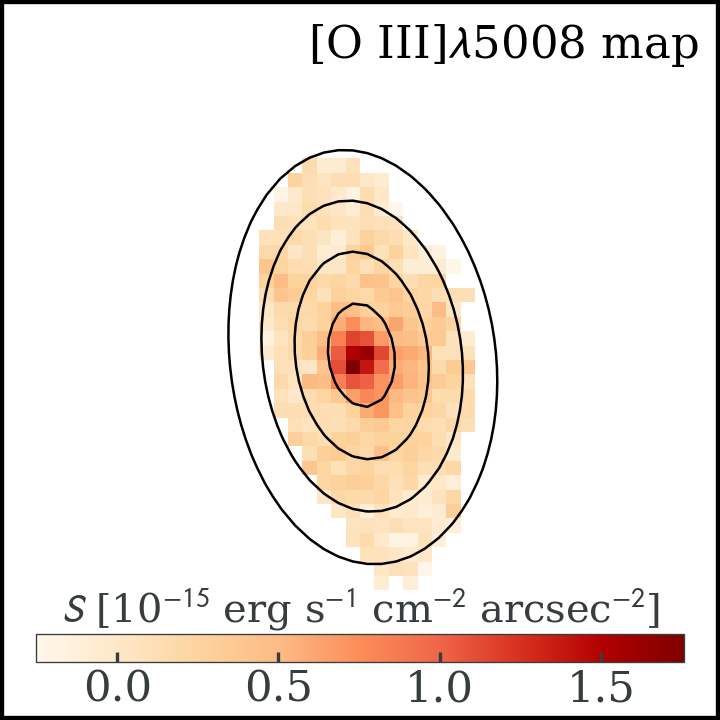}
    \includegraphics[width=.16\textwidth]{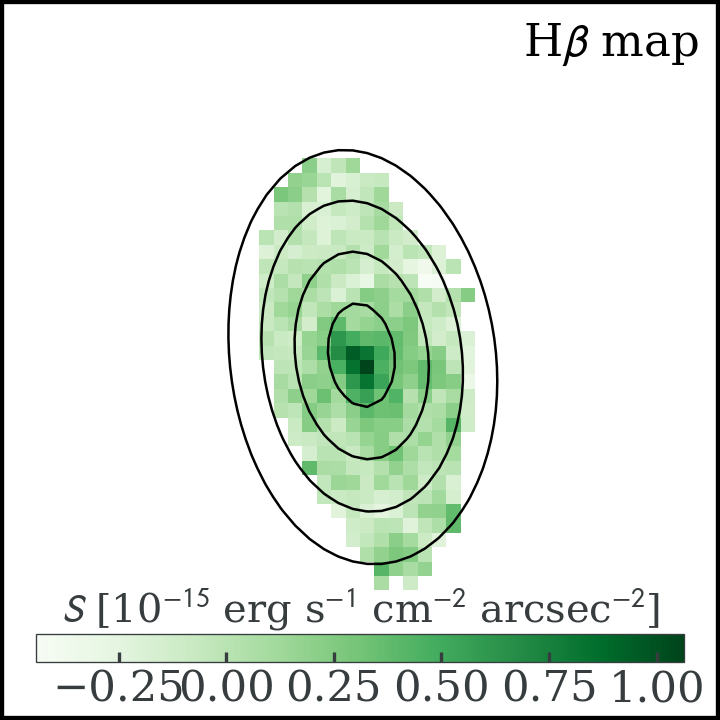}
    \includegraphics[width=.16\textwidth]{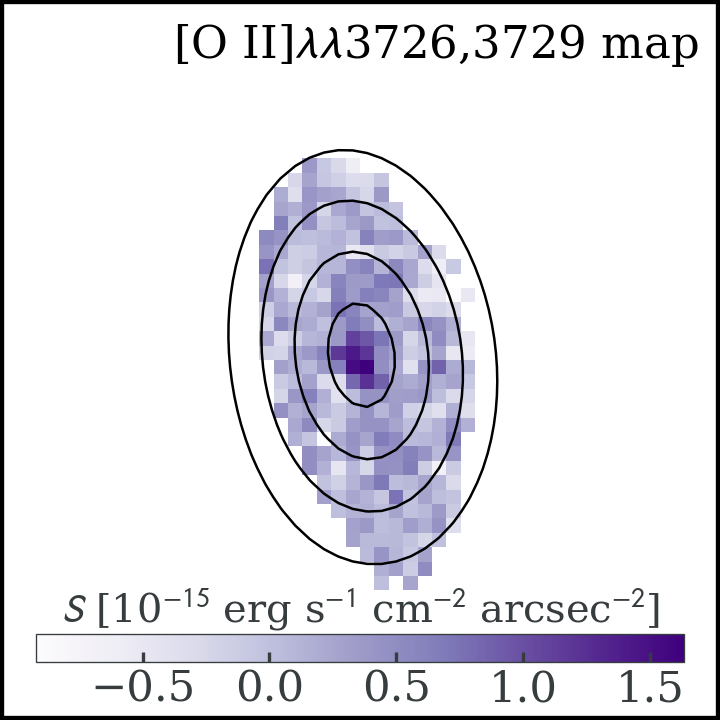}
    \includegraphics[width=.16\textwidth]{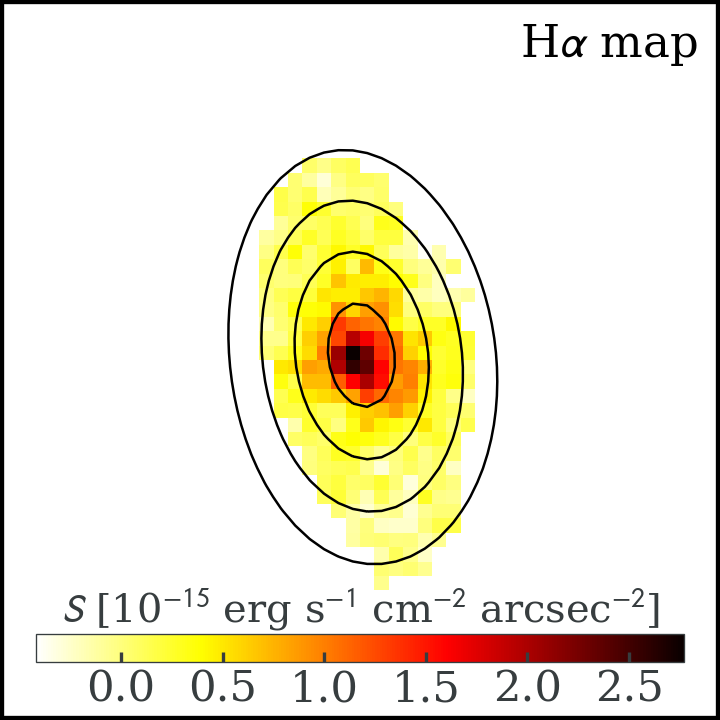}\\
    \includegraphics[width=\textwidth]{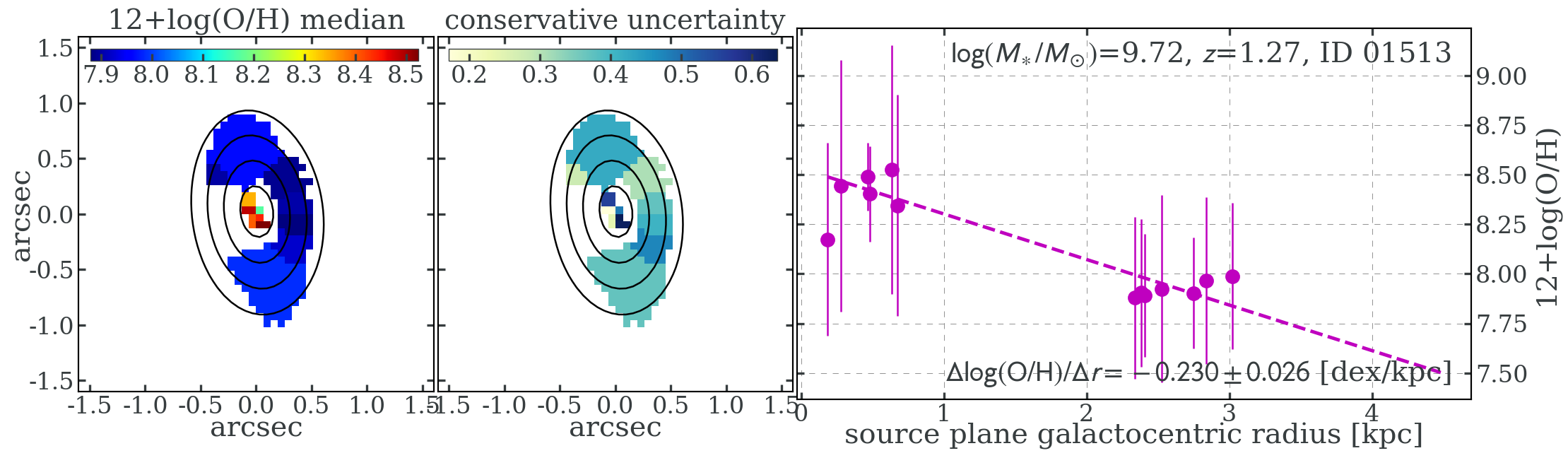}
    \caption{The source ID01513 in the field of \clsan is shown.}
    \label{fig:clA370_ID01513_figs}
\end{figure*}

\begin{figure*}
    \centering
    \includegraphics[width=\textwidth]{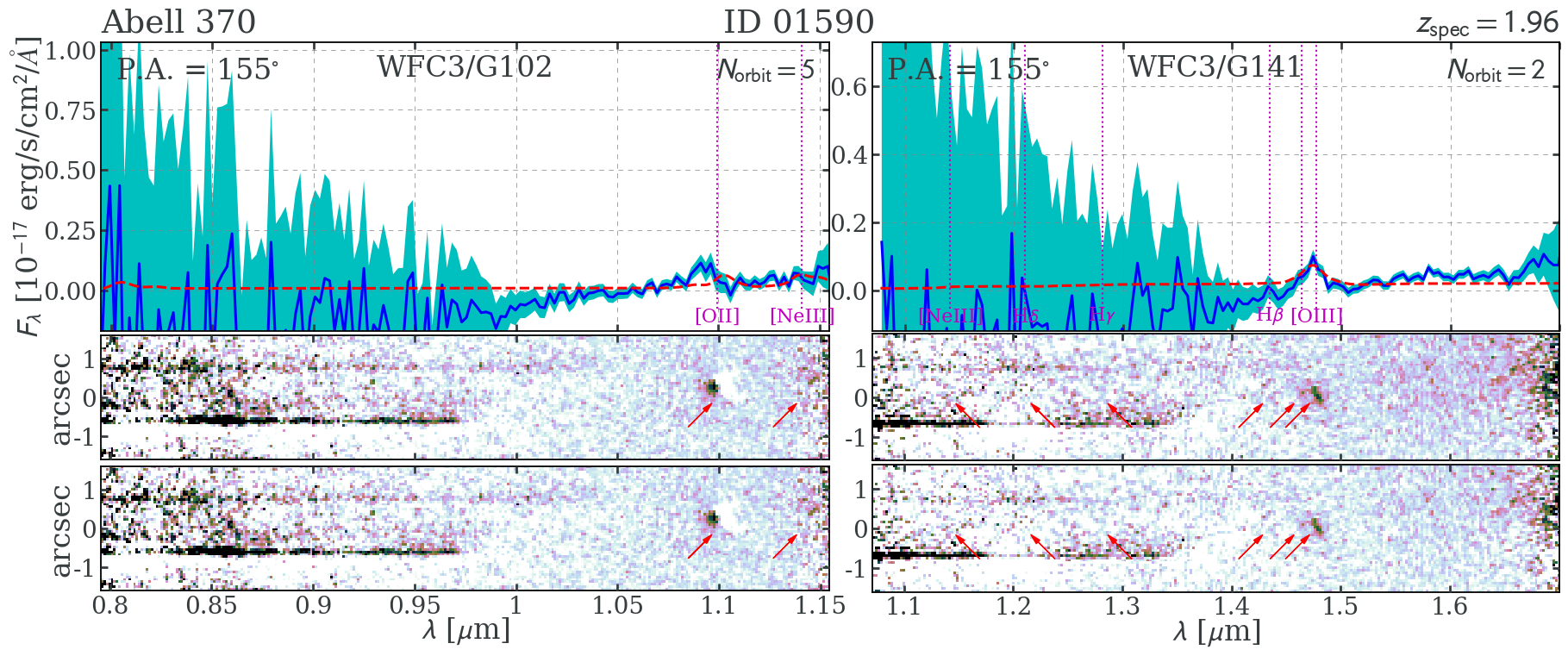}\\
    \includegraphics[width=\textwidth]{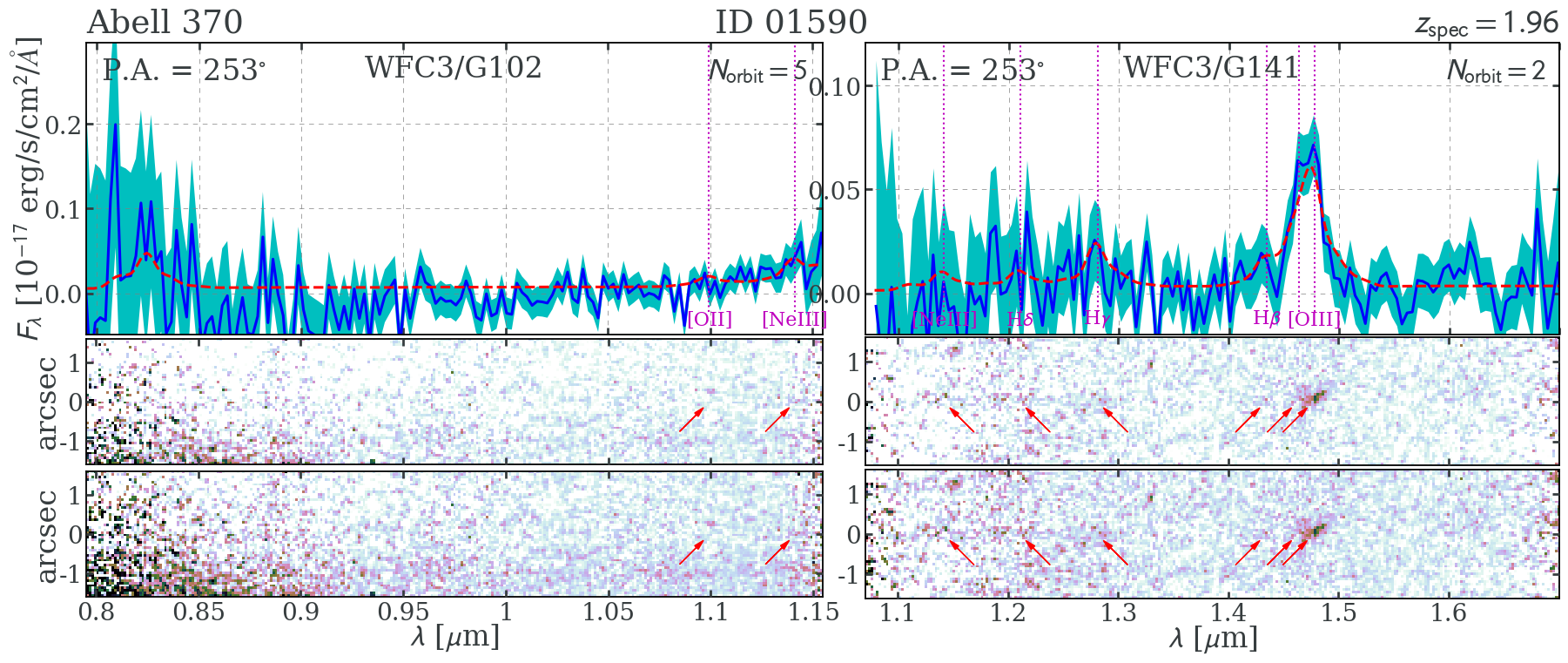}\\
    \includegraphics[width=.16\textwidth]{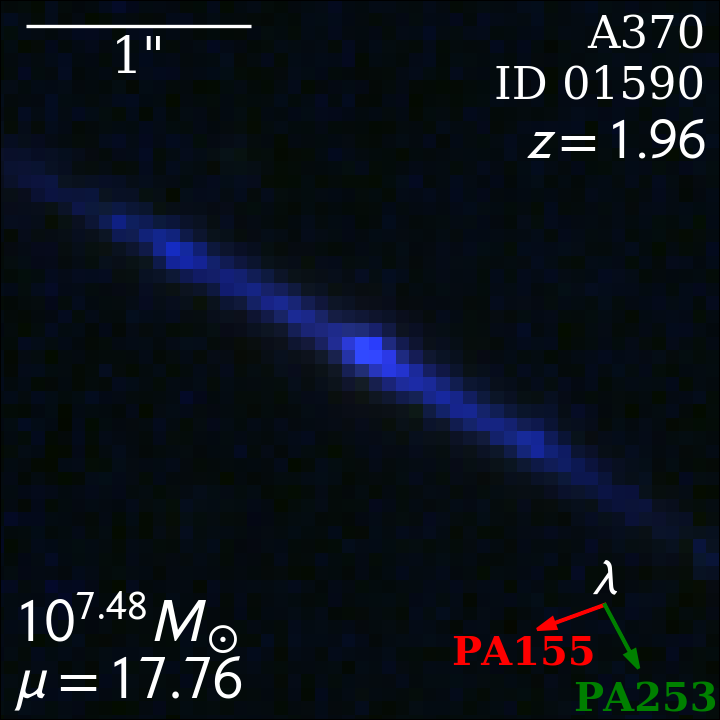}
    \includegraphics[width=.16\textwidth]{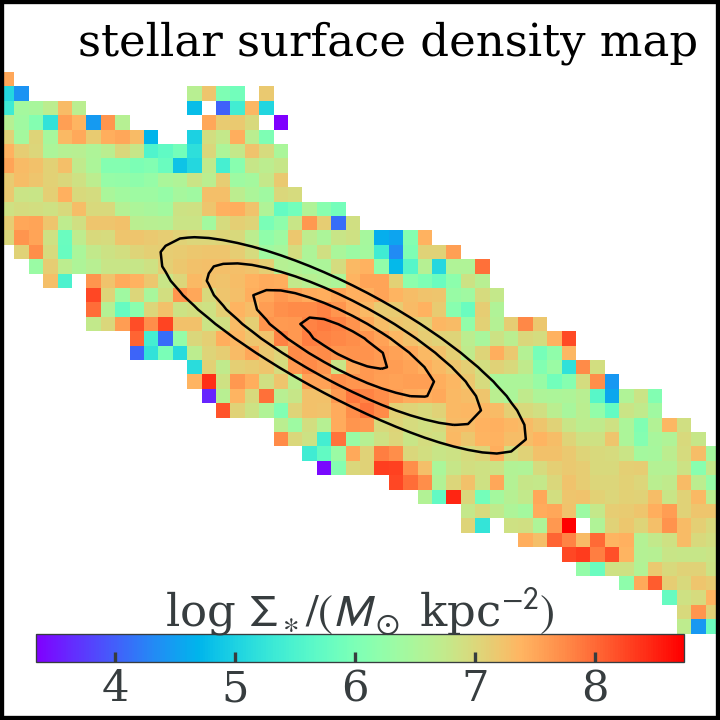}
    \includegraphics[width=.16\textwidth]{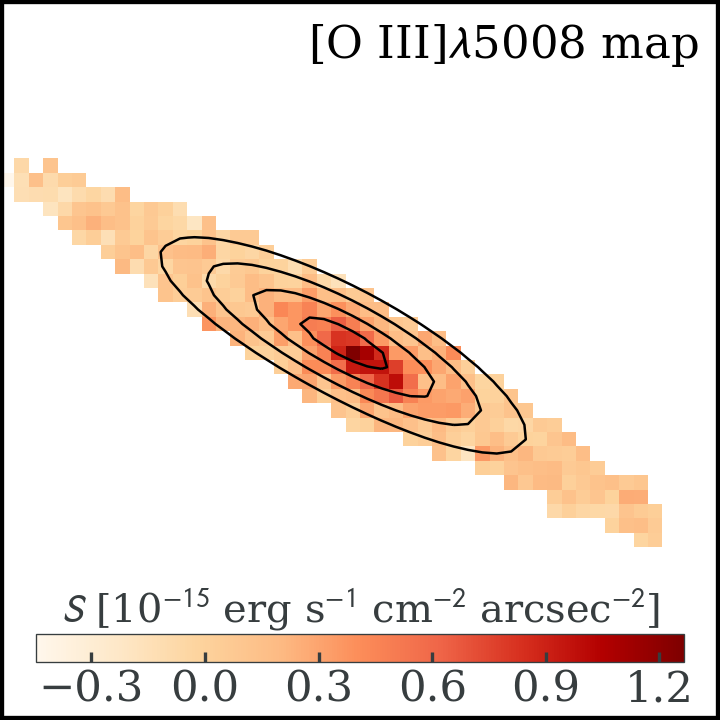}
    \includegraphics[width=.16\textwidth]{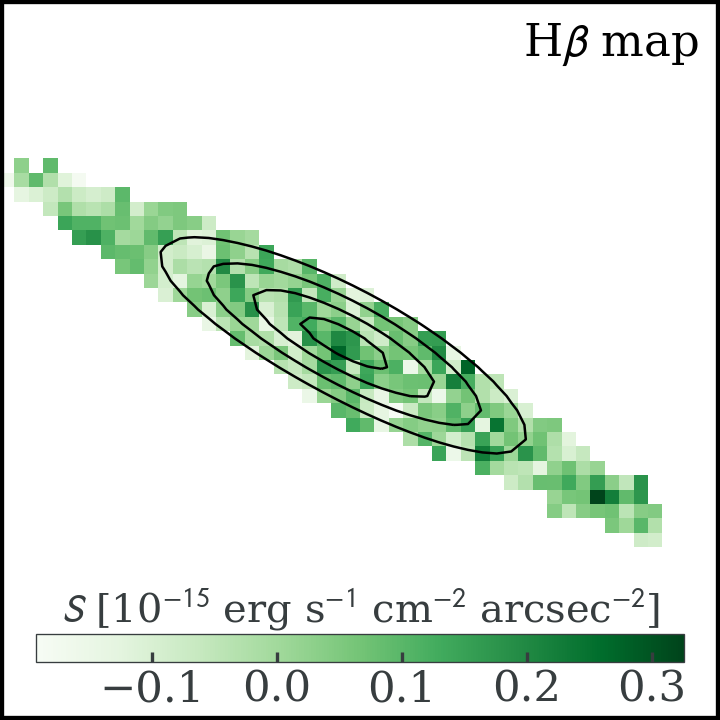}
    \includegraphics[width=.16\textwidth]{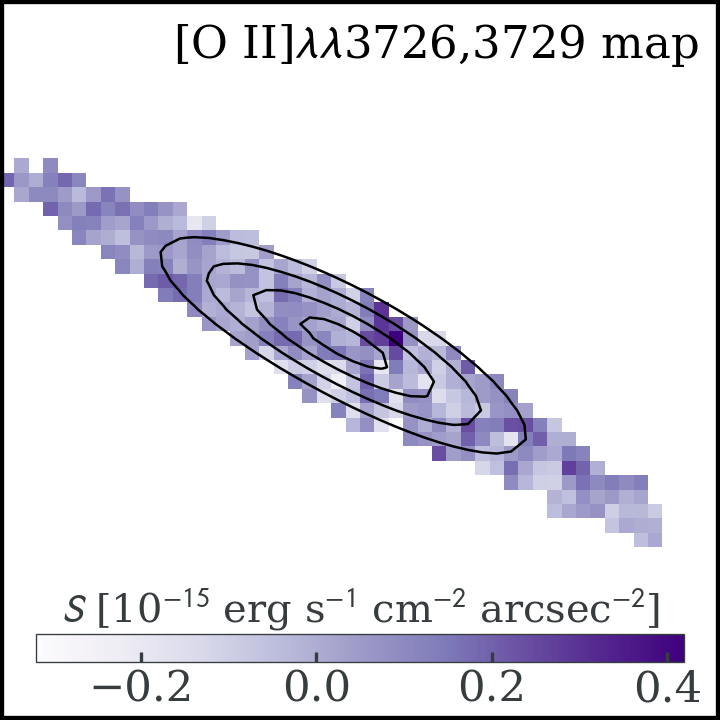}
    \includegraphics[width=.16\textwidth]{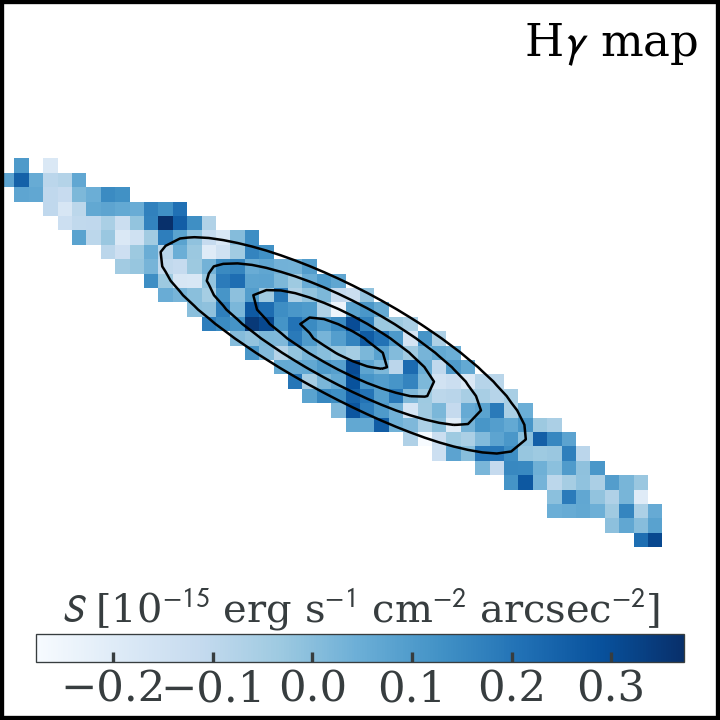}\\
    \includegraphics[width=\textwidth]{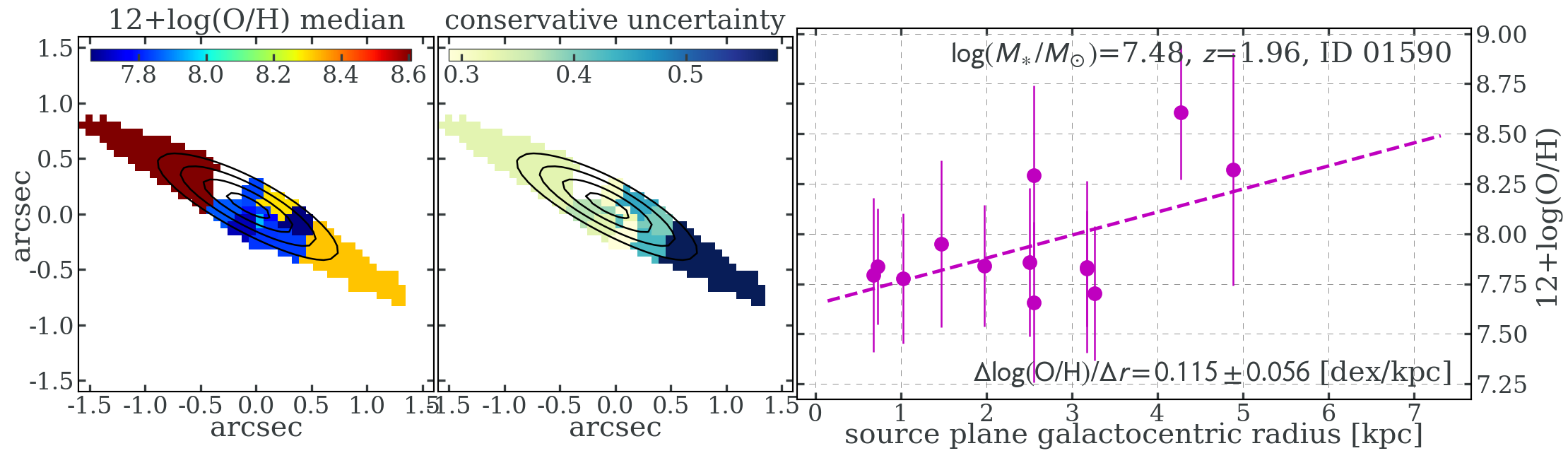}
    \caption{The source ID01590 in the field of \clsan is shown.}
    \label{fig:clA370_ID01590_figs}
\end{figure*}

\begin{figure*}
    \centering
    \includegraphics[width=\textwidth]{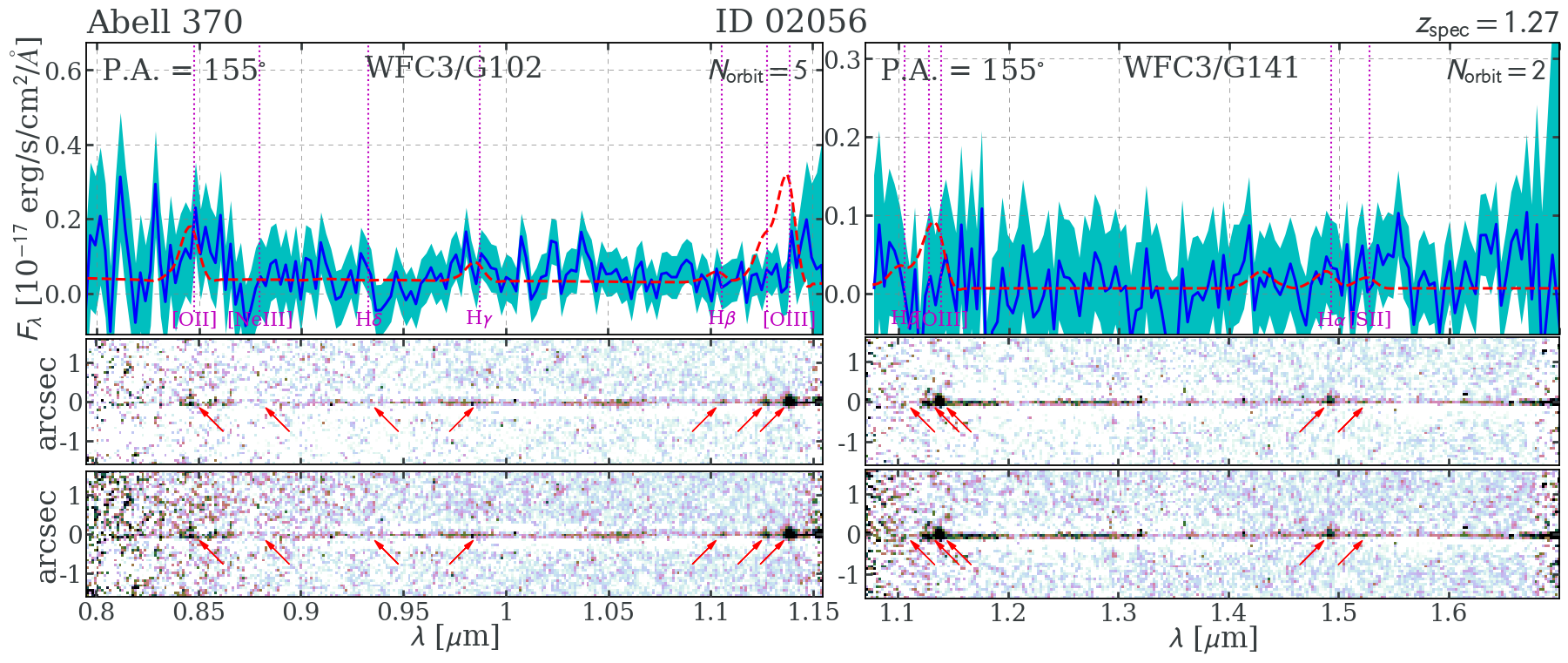}\\
    \includegraphics[width=\textwidth]{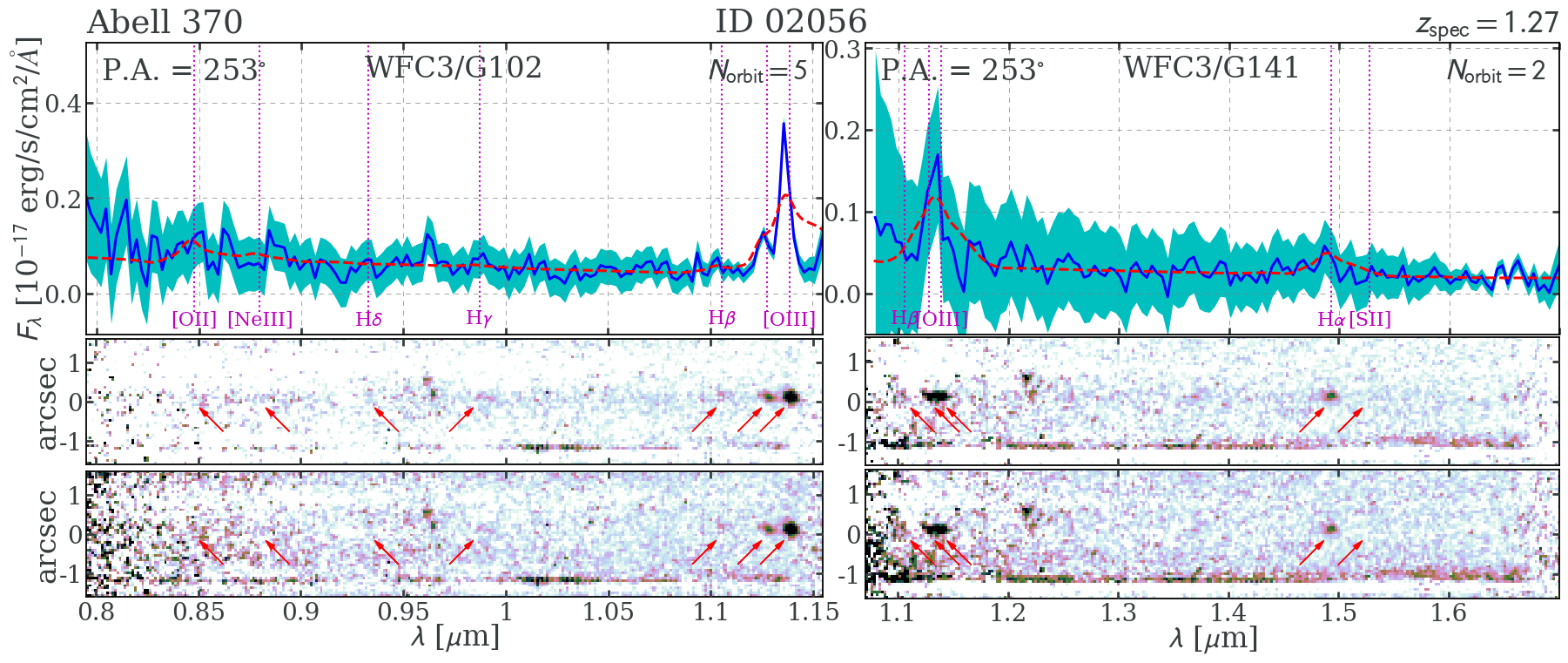}\\
    \includegraphics[width=.16\textwidth]{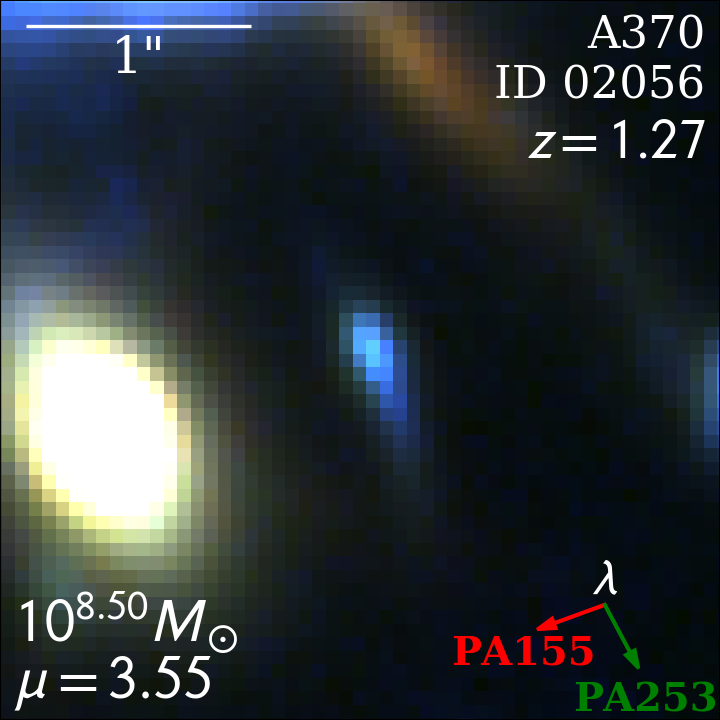}
    \includegraphics[width=.16\textwidth]{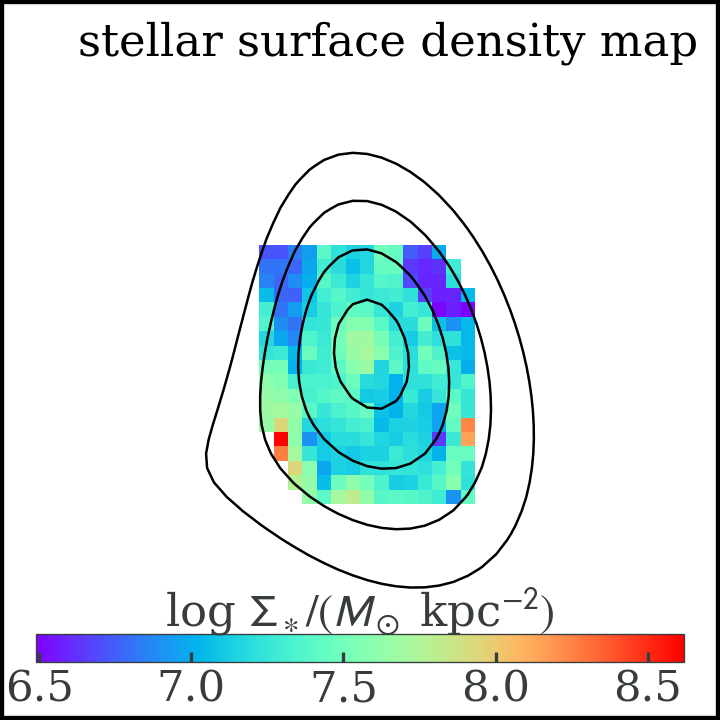}
    \includegraphics[width=.16\textwidth]{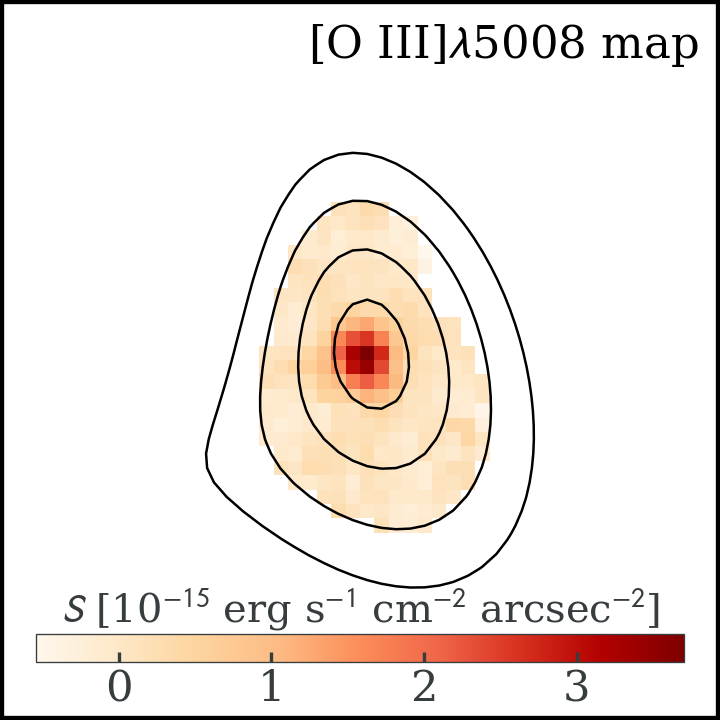}
    \includegraphics[width=.16\textwidth]{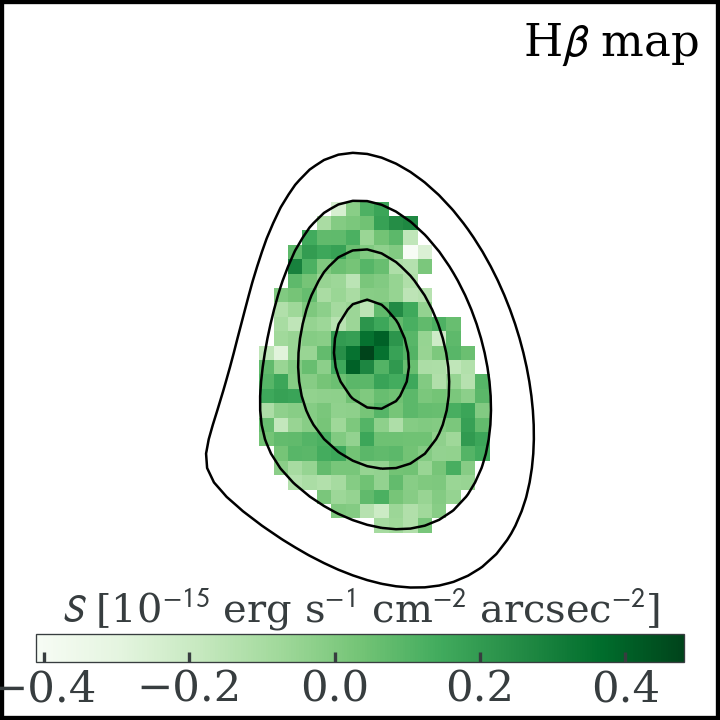}
    \includegraphics[width=.16\textwidth]{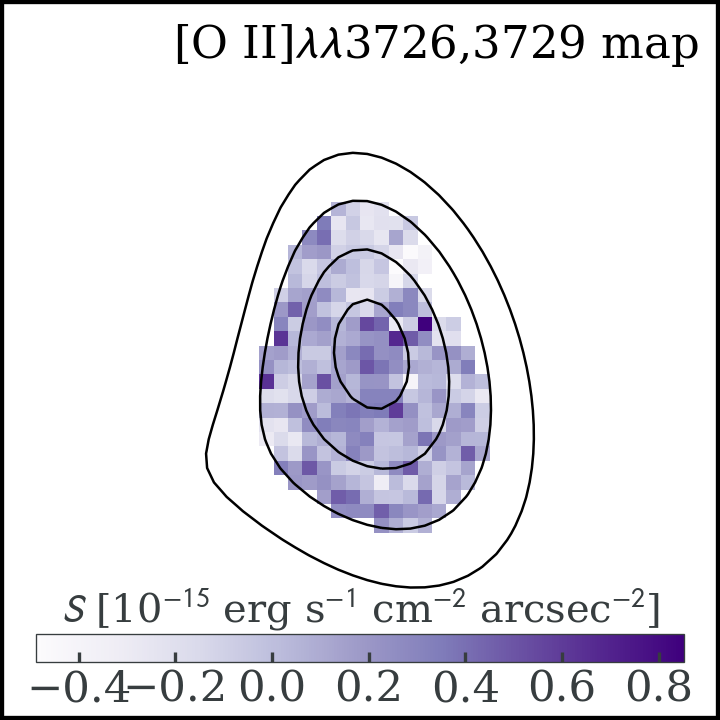}
    \includegraphics[width=.16\textwidth]{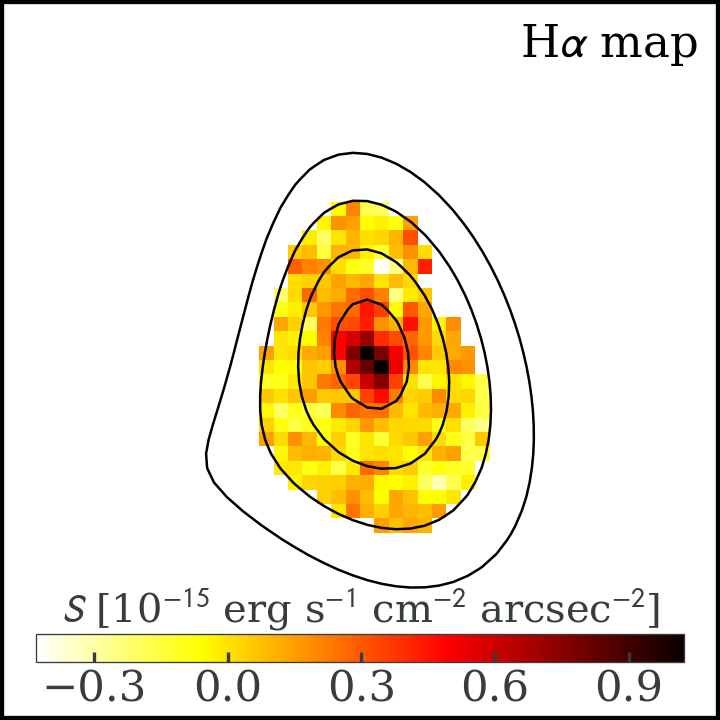}\\
    \includegraphics[width=\textwidth]{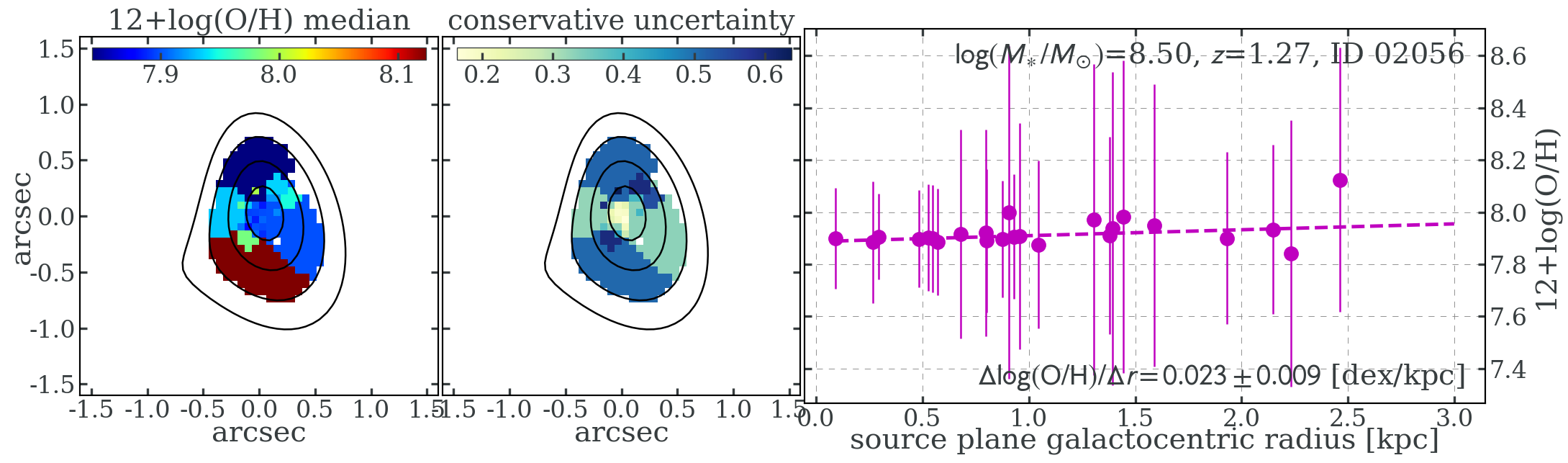}
    \caption{The source ID02056 in the field of \clsan is shown.}
    \label{fig:clA370_ID02056_figs}
\end{figure*}

\begin{figure*}
    \centering
    \includegraphics[width=\textwidth]{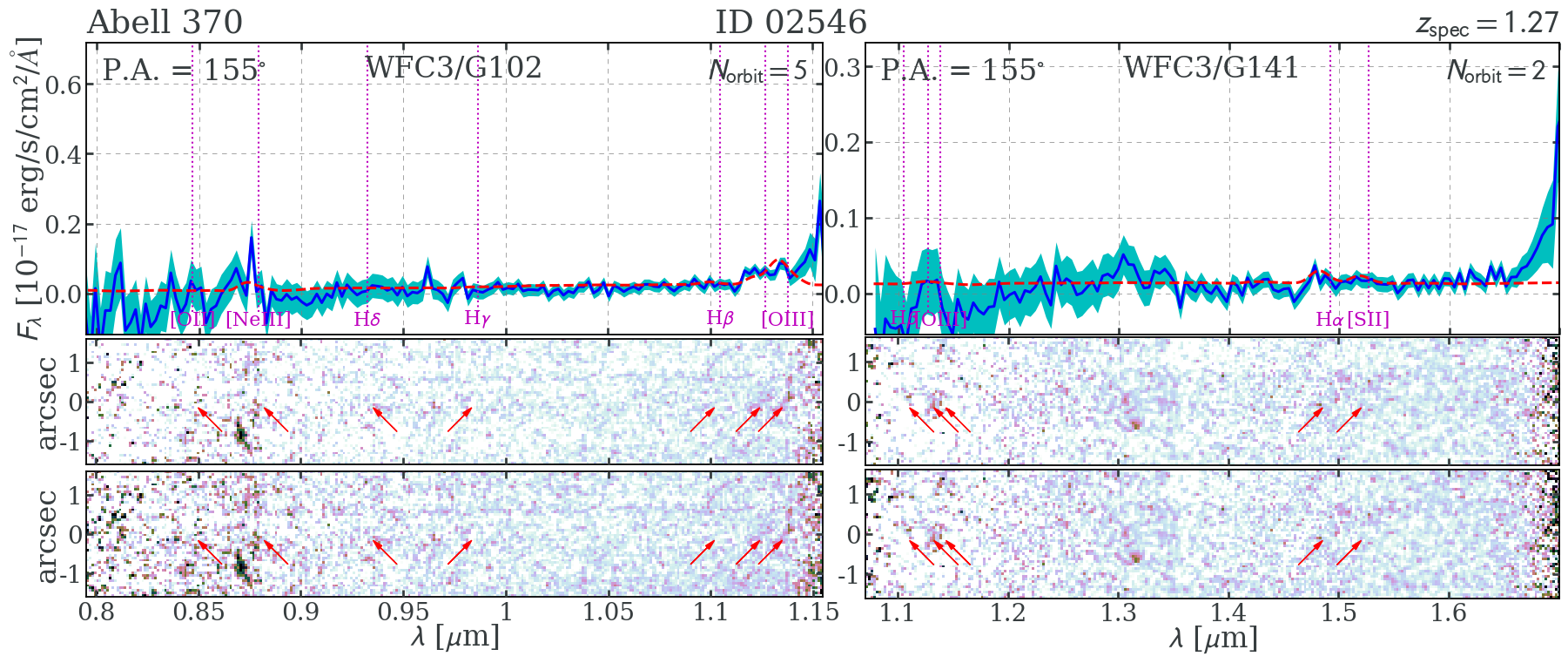}\\
    \includegraphics[width=\textwidth]{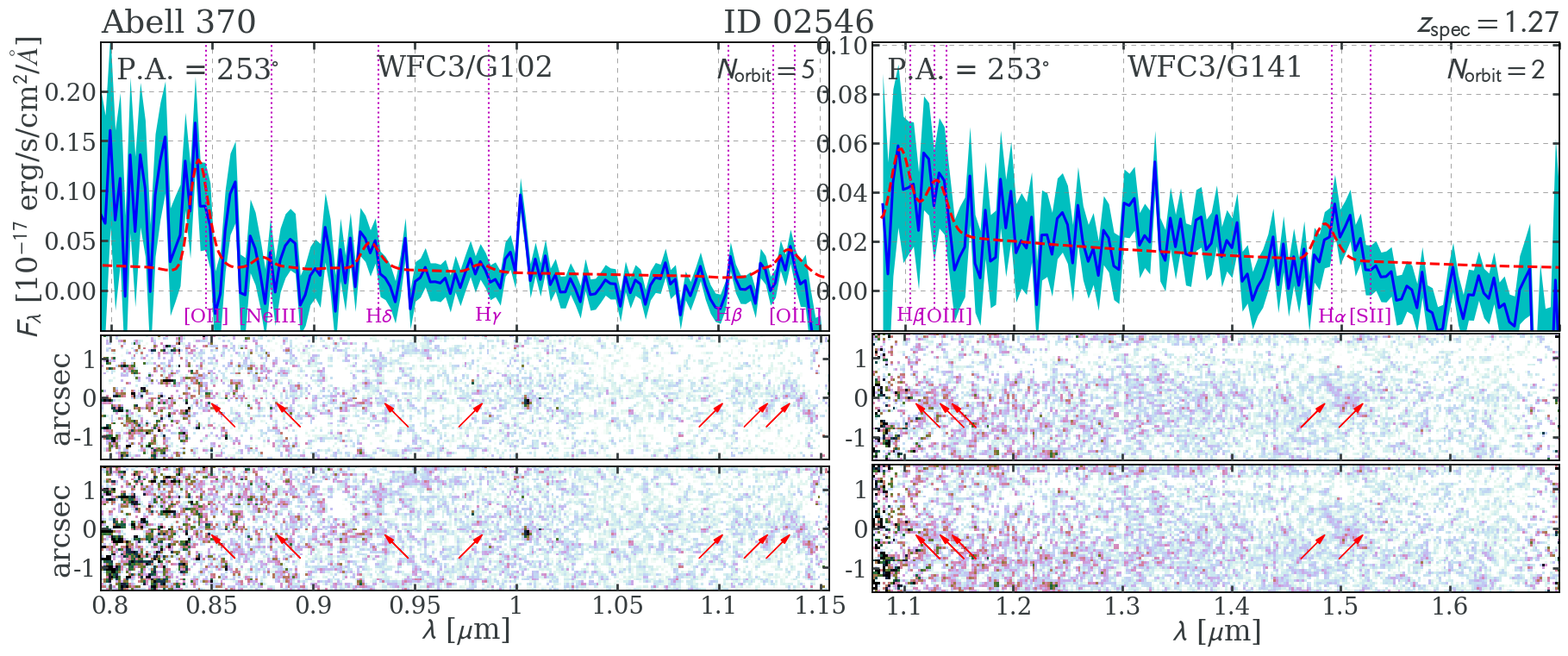}\\
    \includegraphics[width=.16\textwidth]{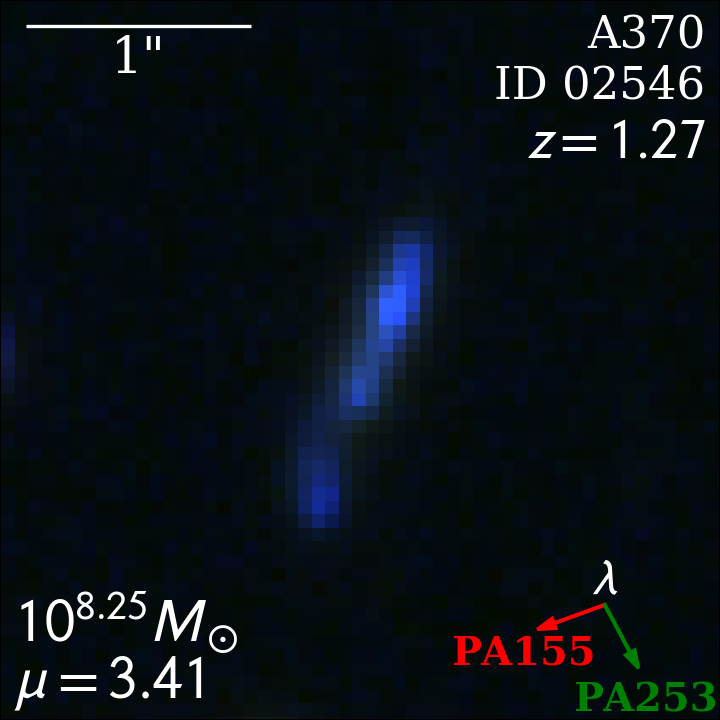}
    \includegraphics[width=.16\textwidth]{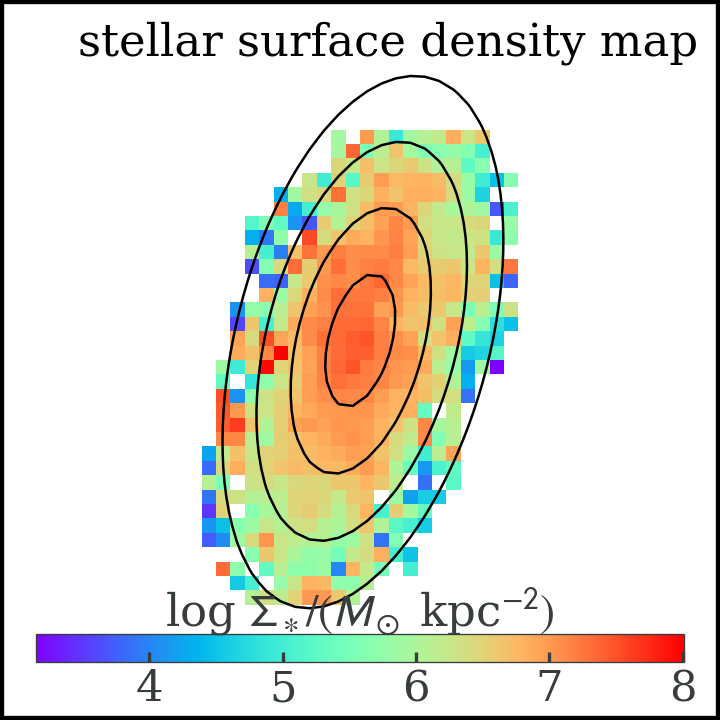}
    \includegraphics[width=.16\textwidth]{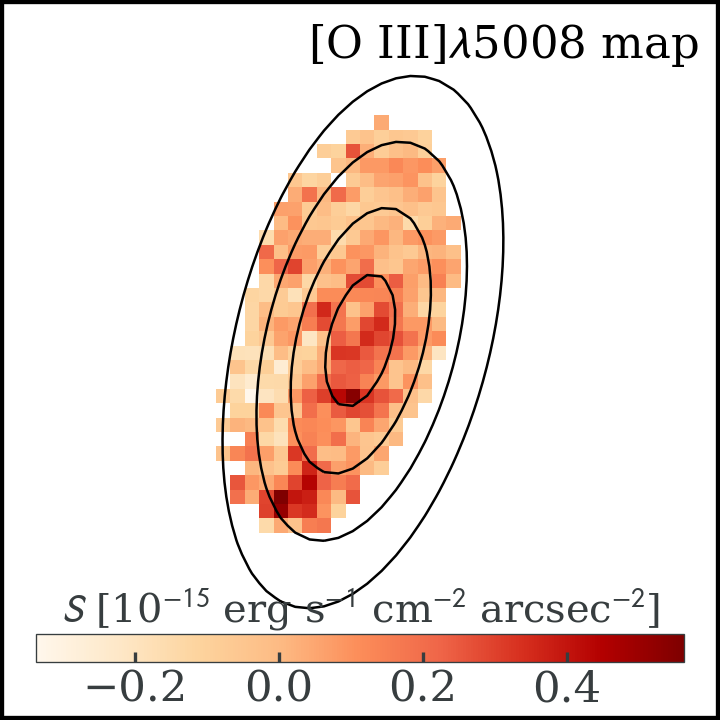}
    \includegraphics[width=.16\textwidth]{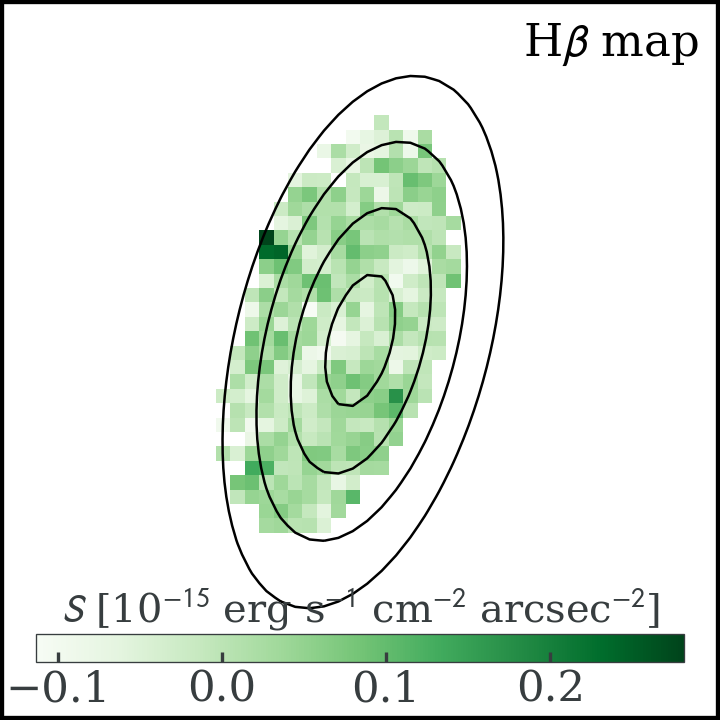}
    \includegraphics[width=.16\textwidth]{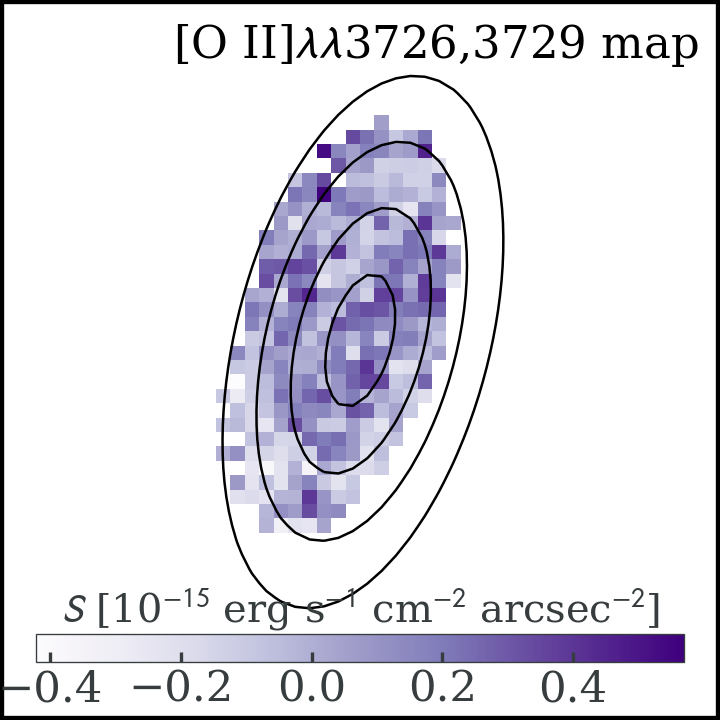}
    \includegraphics[width=.16\textwidth]{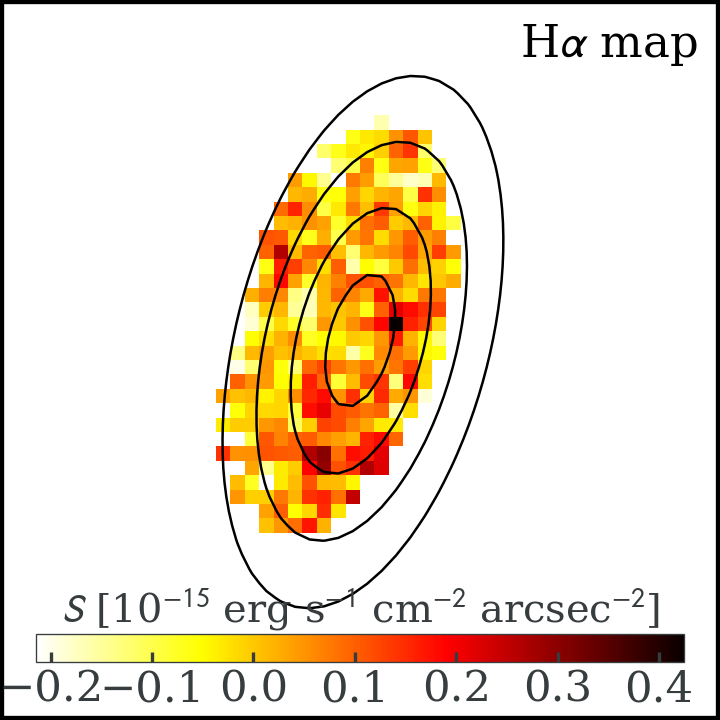}\\
    \includegraphics[width=\textwidth]{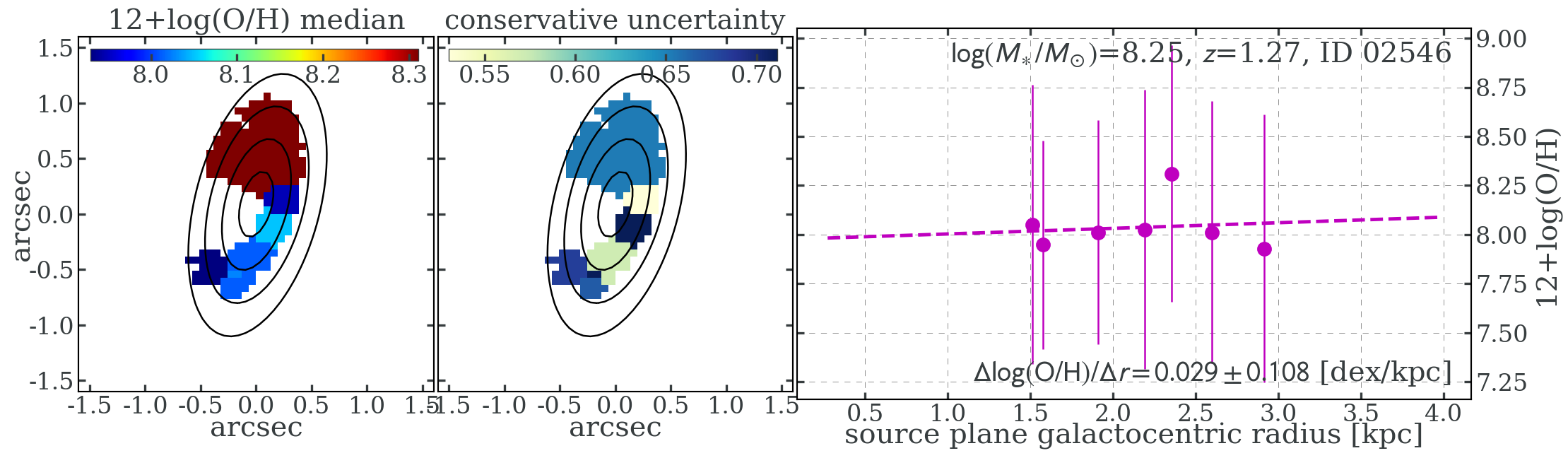}
    \caption{The source ID02546 in the field of \clsan is shown.}
    \label{fig:clA370_ID02546_figs}
\end{figure*}

\begin{figure*}
    \centering
    \includegraphics[width=\textwidth]{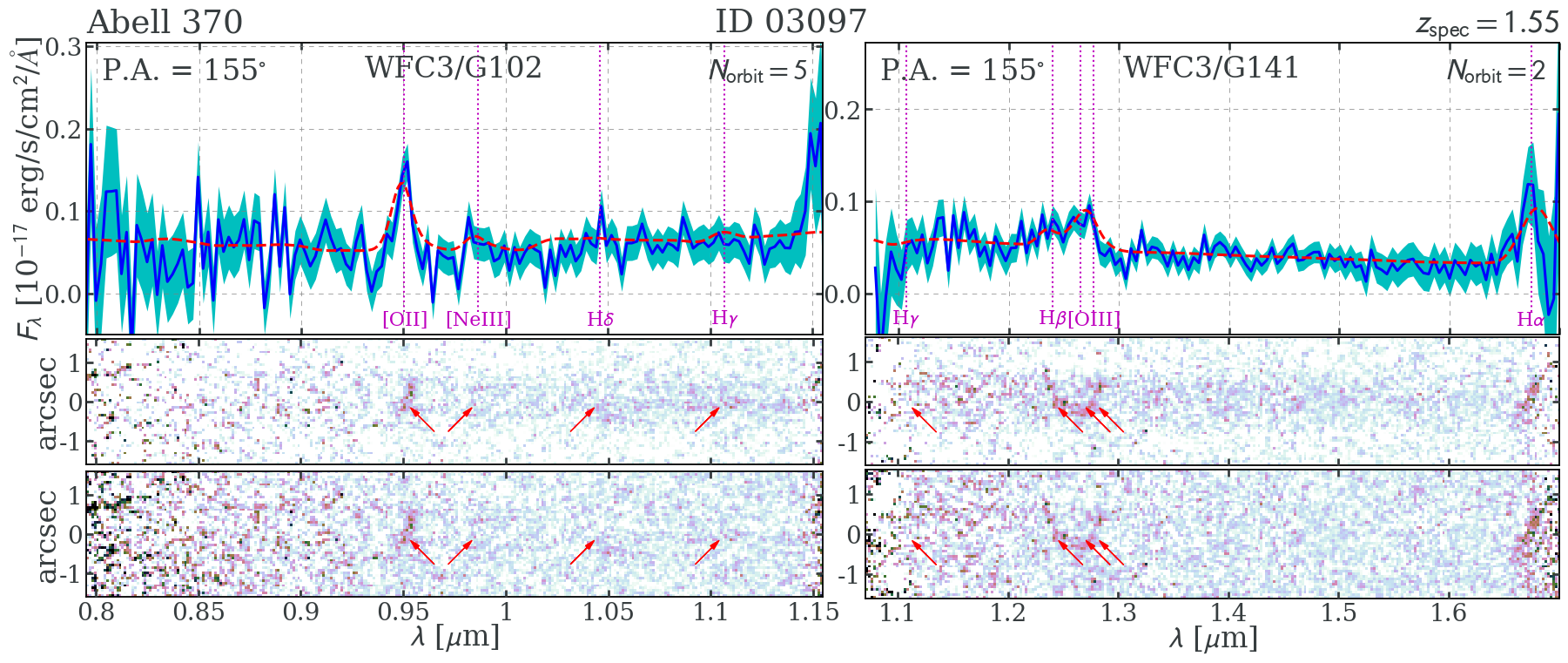}\\
    \includegraphics[width=\textwidth]{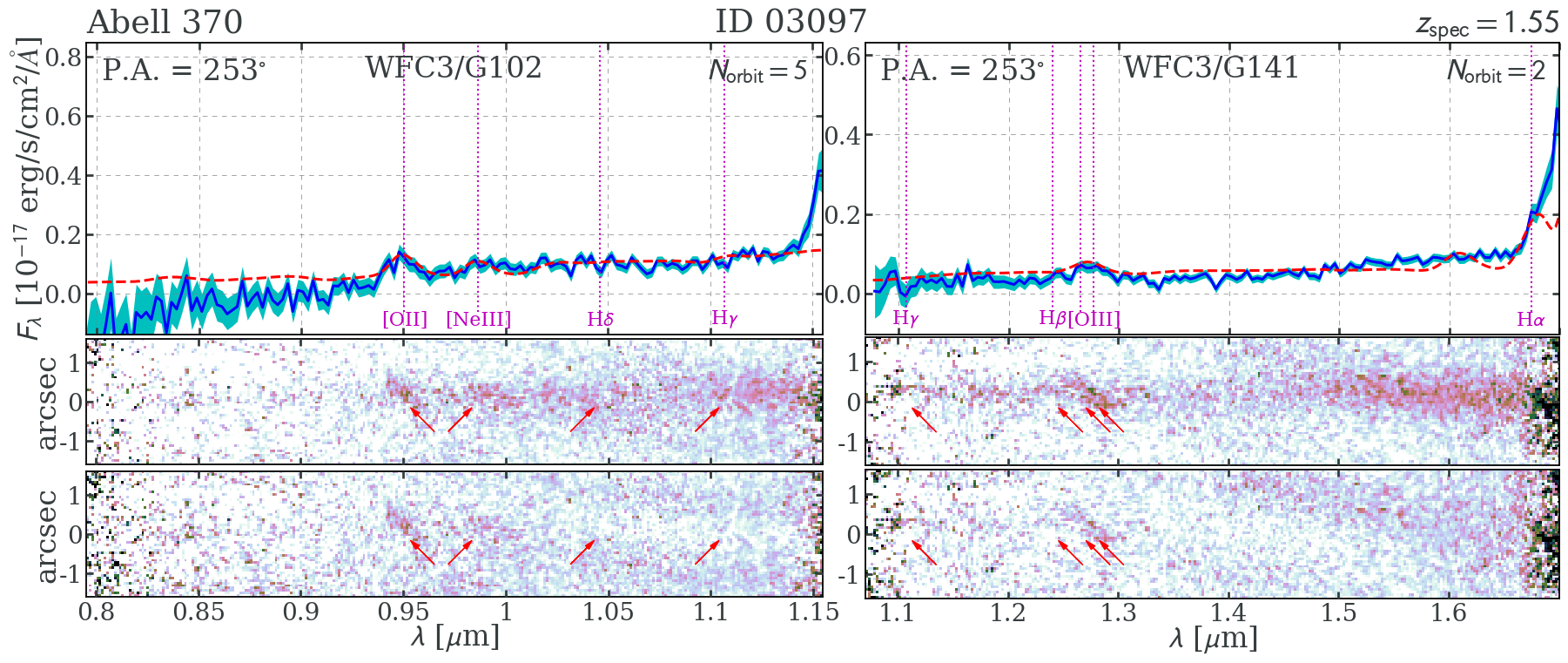}\\
    \includegraphics[width=.16\textwidth]{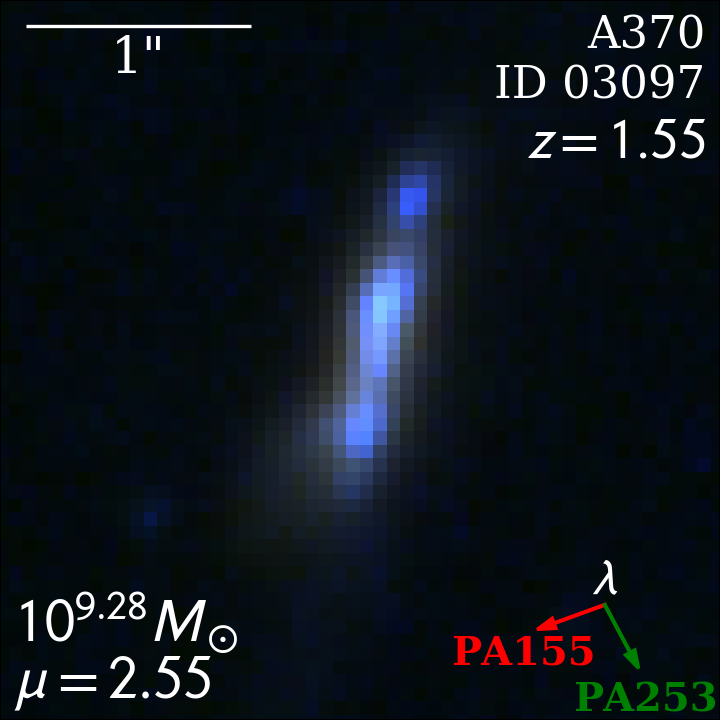}
    \includegraphics[width=.16\textwidth]{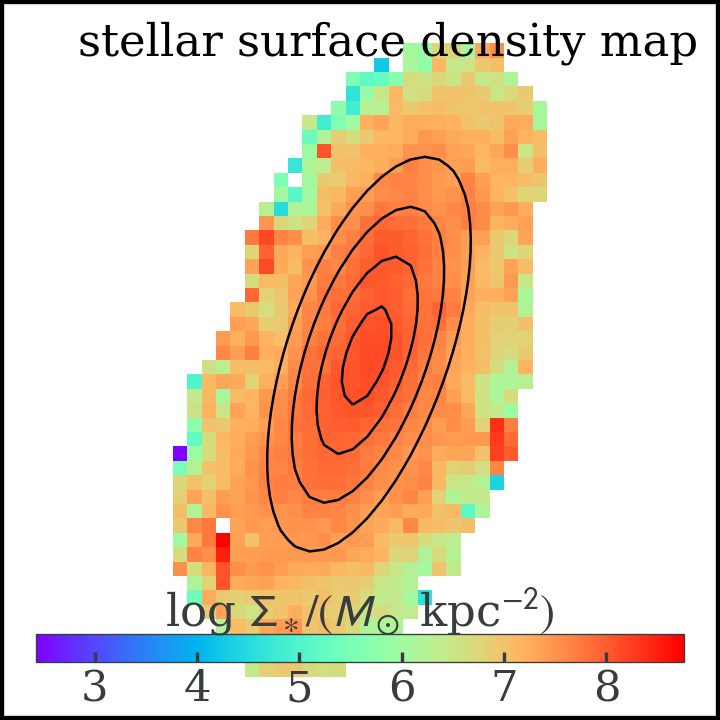}
    \includegraphics[width=.16\textwidth]{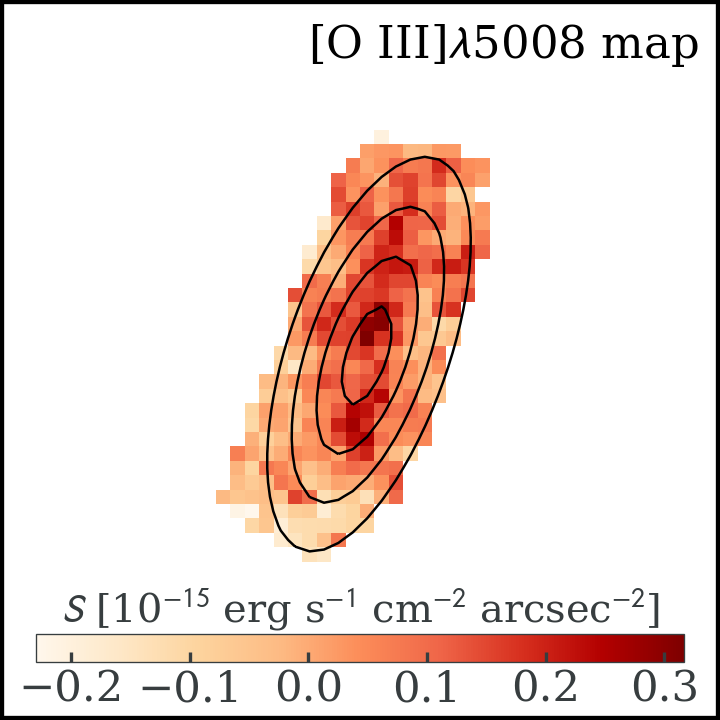}
    \includegraphics[width=.16\textwidth]{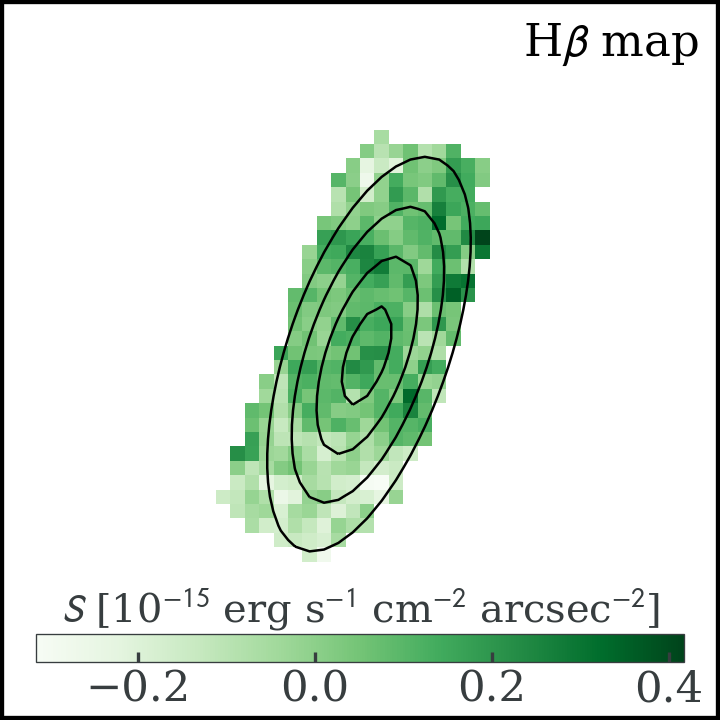}
    \includegraphics[width=.16\textwidth]{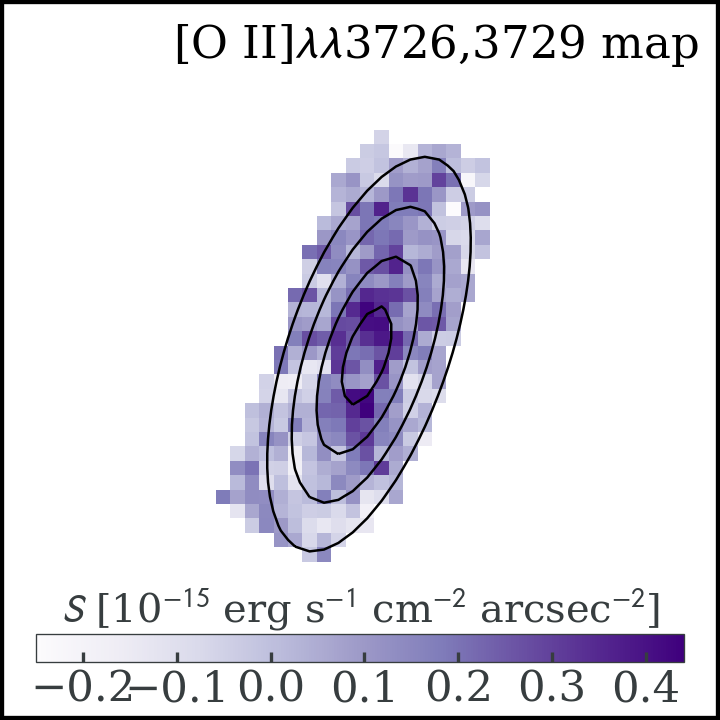}
    \includegraphics[width=.16\textwidth]{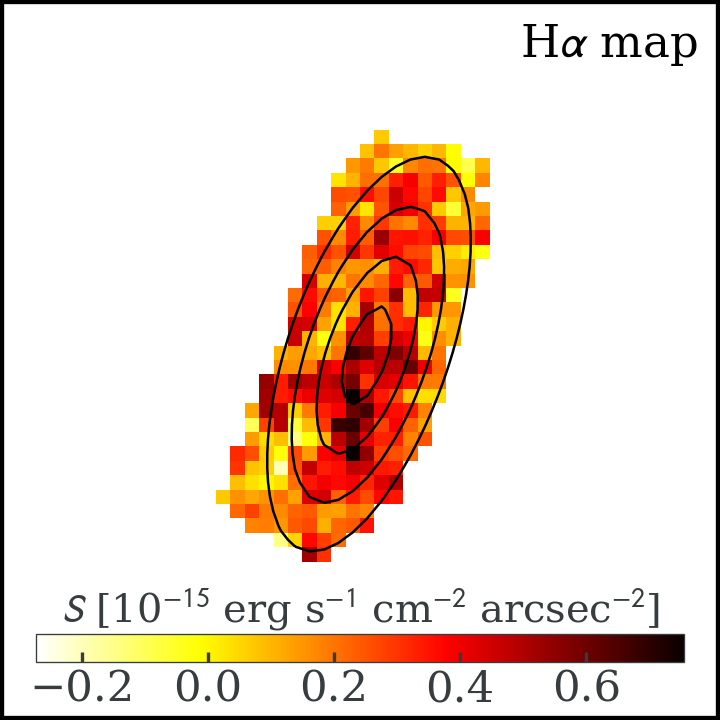}\\
    \includegraphics[width=\textwidth]{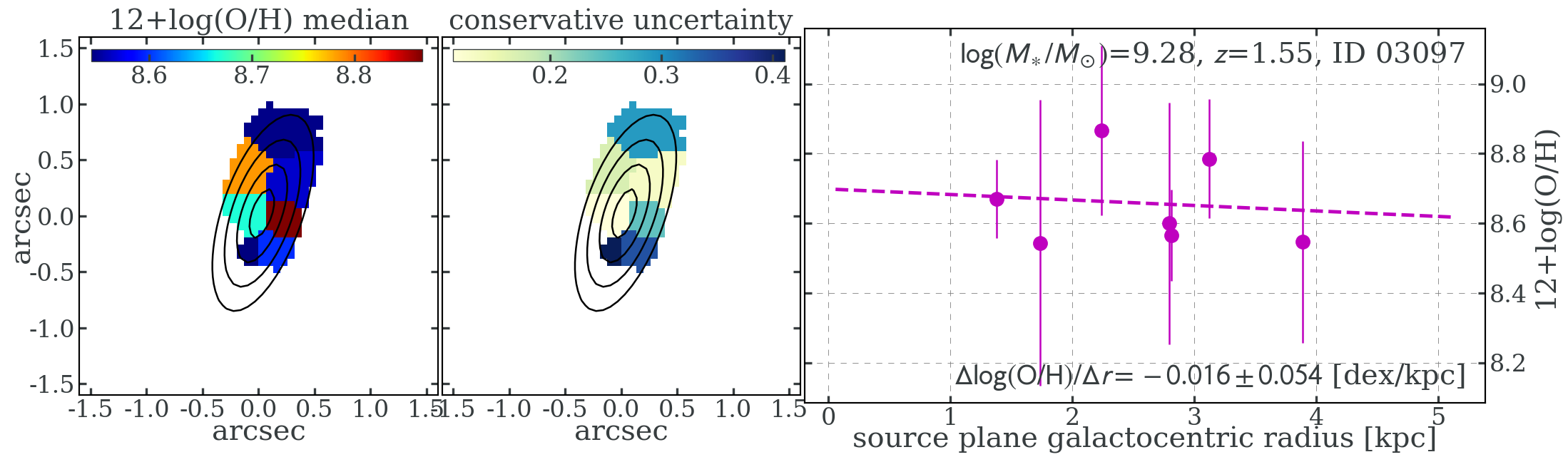}
    \caption{The source ID03097 in the field of \clsan is shown.}
    \label{fig:clA370_ID03097_figs}
\end{figure*}

\begin{figure*}
    \centering
    \includegraphics[width=\textwidth]{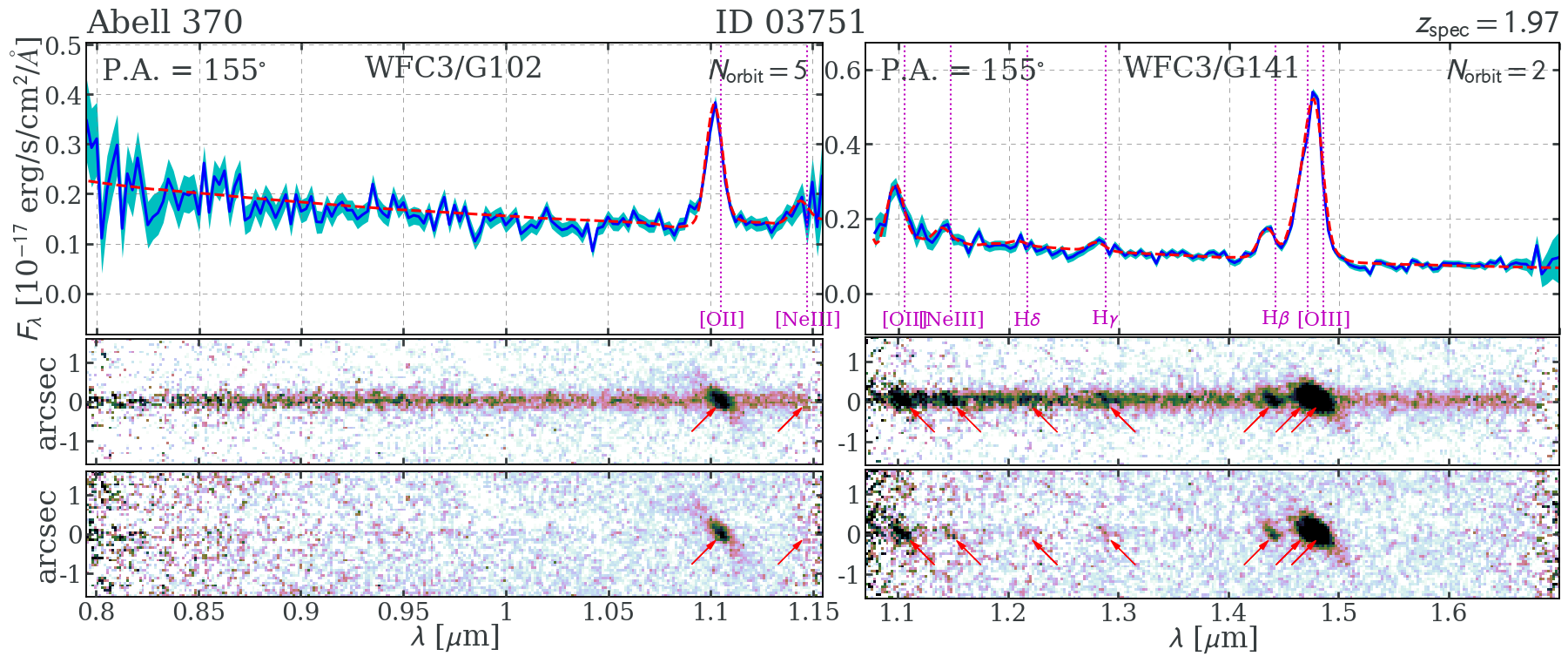}\\
    \includegraphics[width=\textwidth]{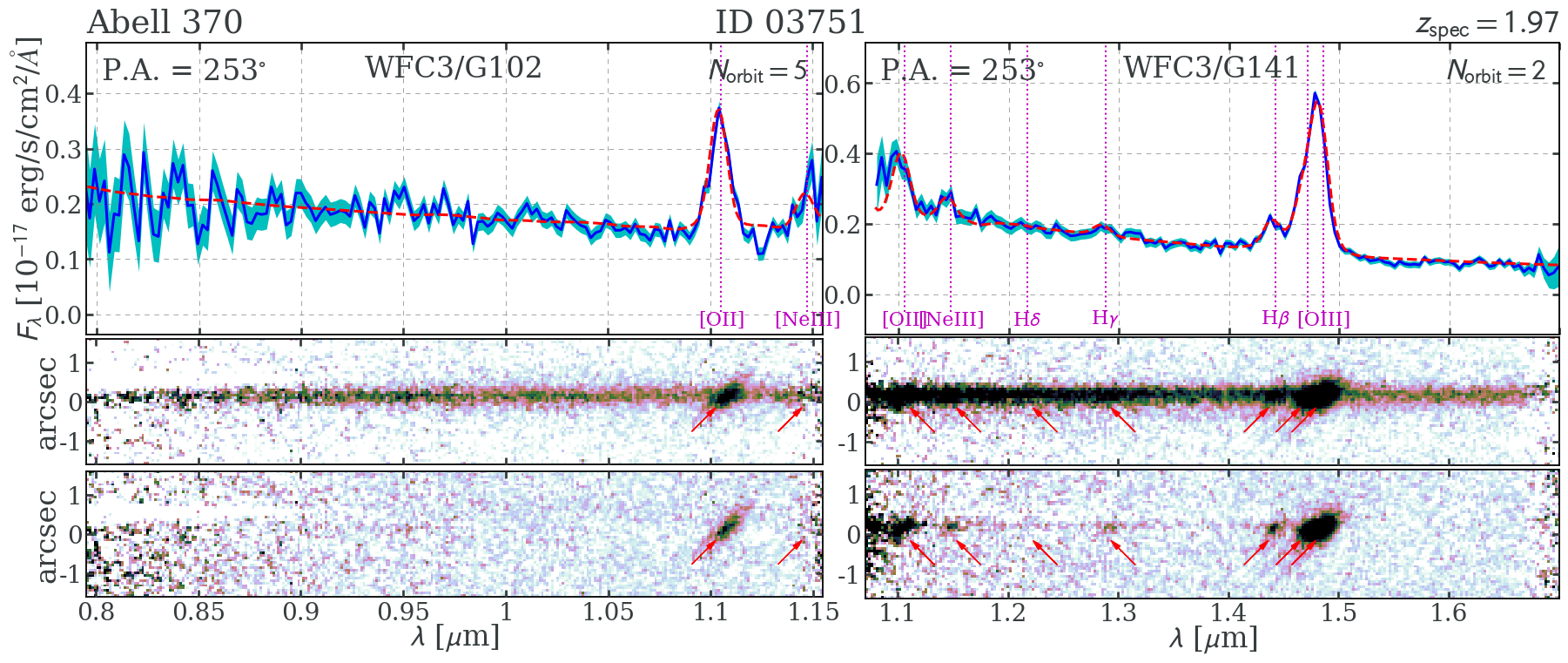}\\
    \includegraphics[width=.16\textwidth]{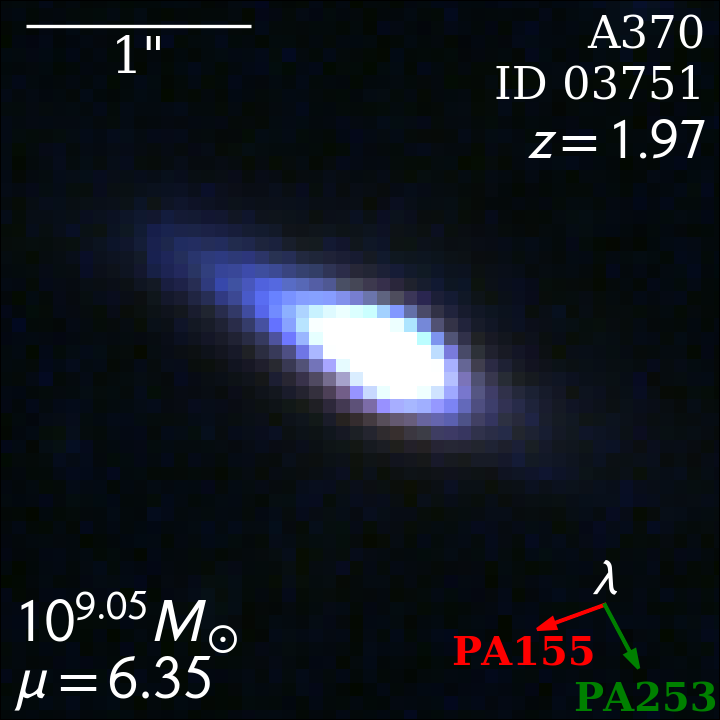}
    \includegraphics[width=.16\textwidth]{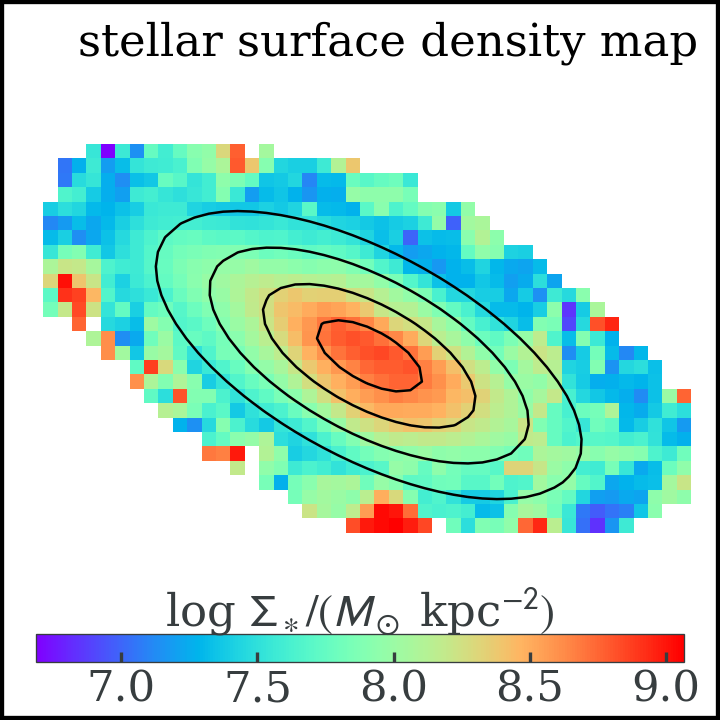}
    \includegraphics[width=.16\textwidth]{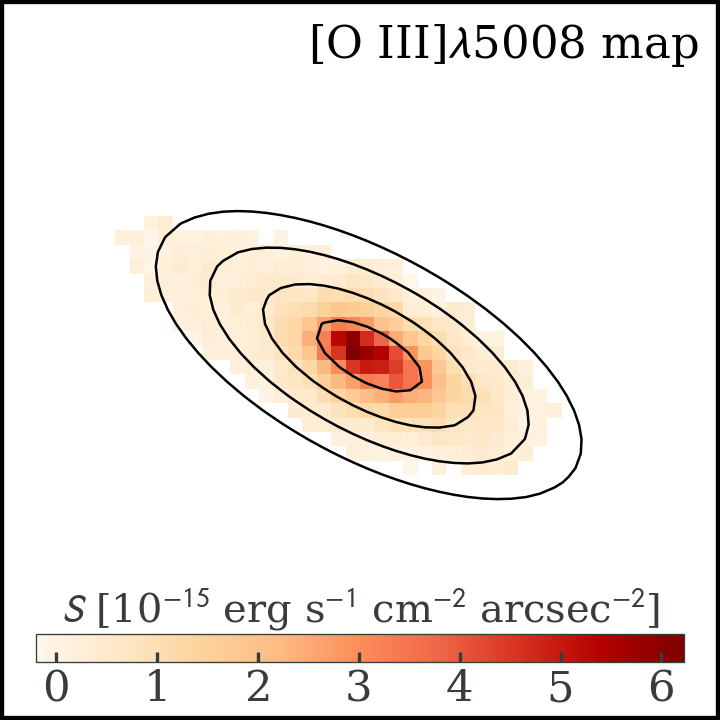}
    \includegraphics[width=.16\textwidth]{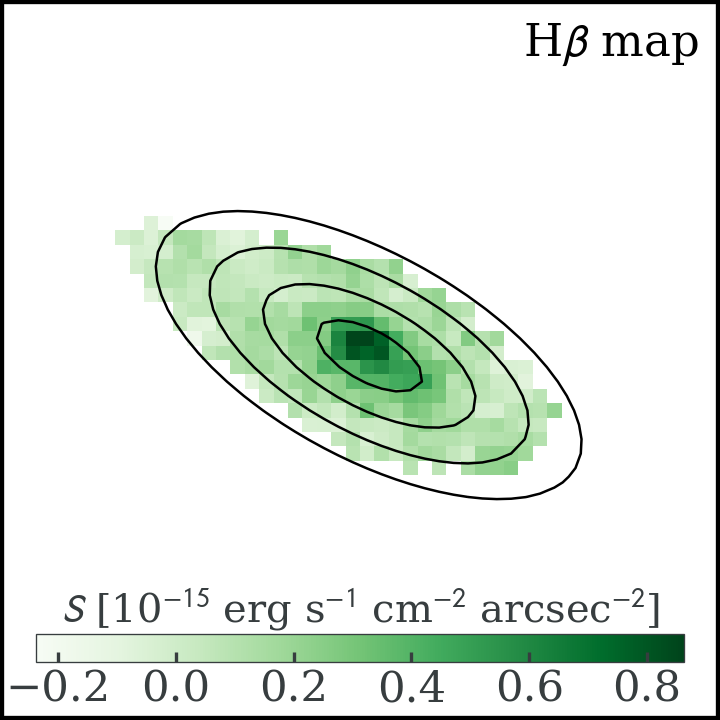}
    \includegraphics[width=.16\textwidth]{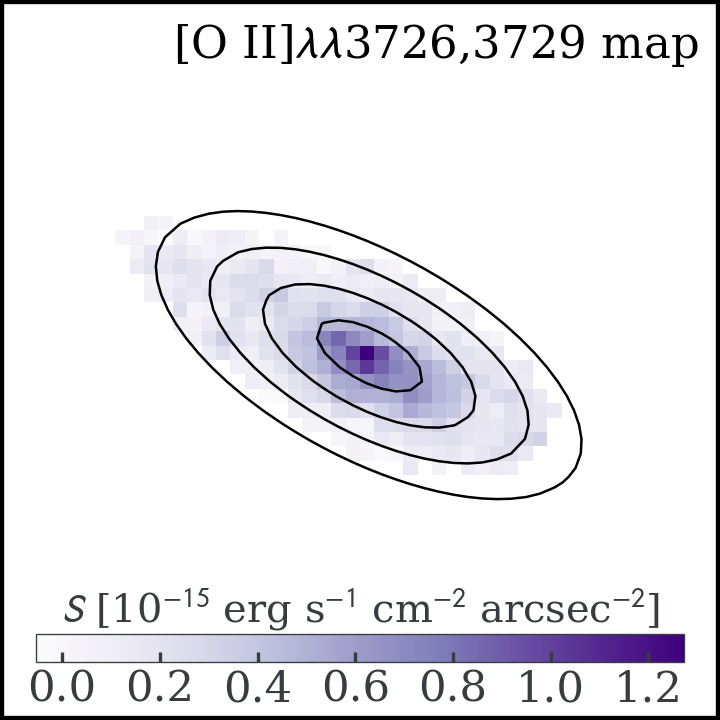}
    \includegraphics[width=.16\textwidth]{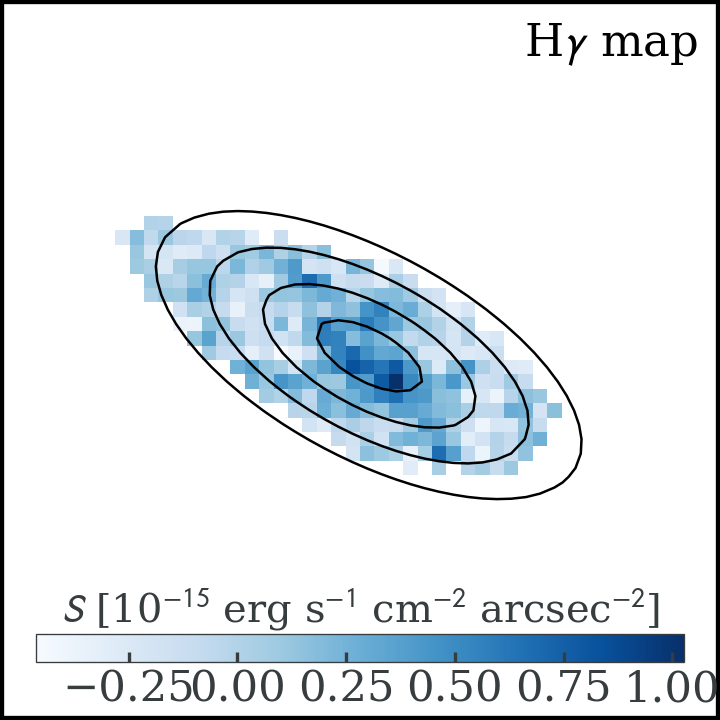}\\
    \includegraphics[width=\textwidth]{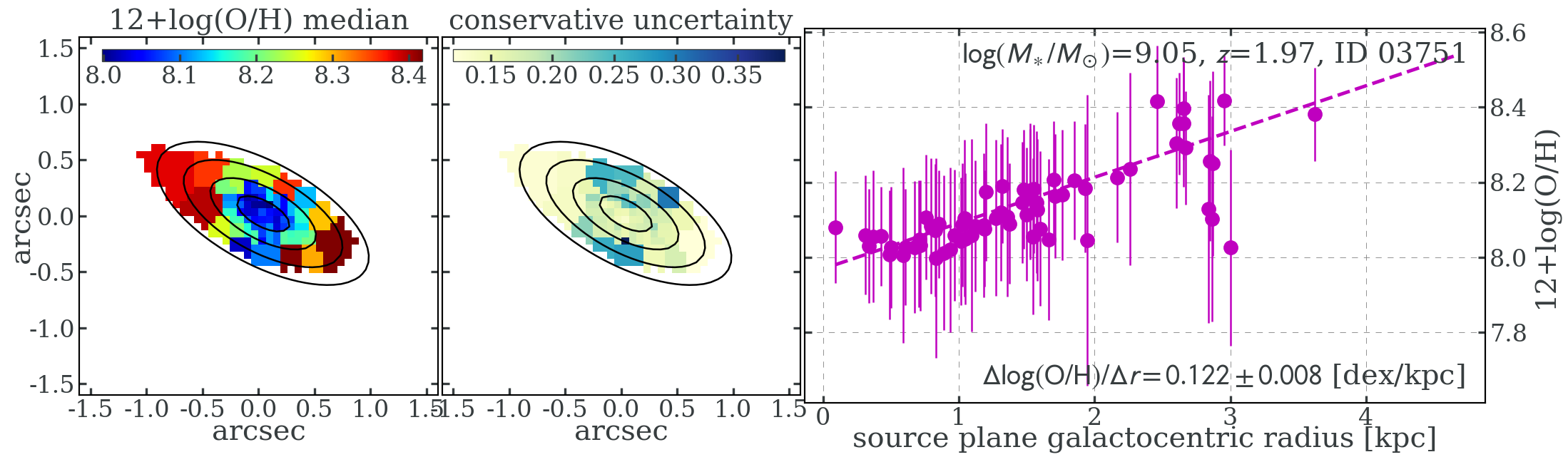}
    \caption{The source ID03751 in the field of \clsan is shown.}
    \label{fig:clA370_ID03751_figs}
\end{figure*}

\begin{figure*}
    \centering
    \includegraphics[width=\textwidth]{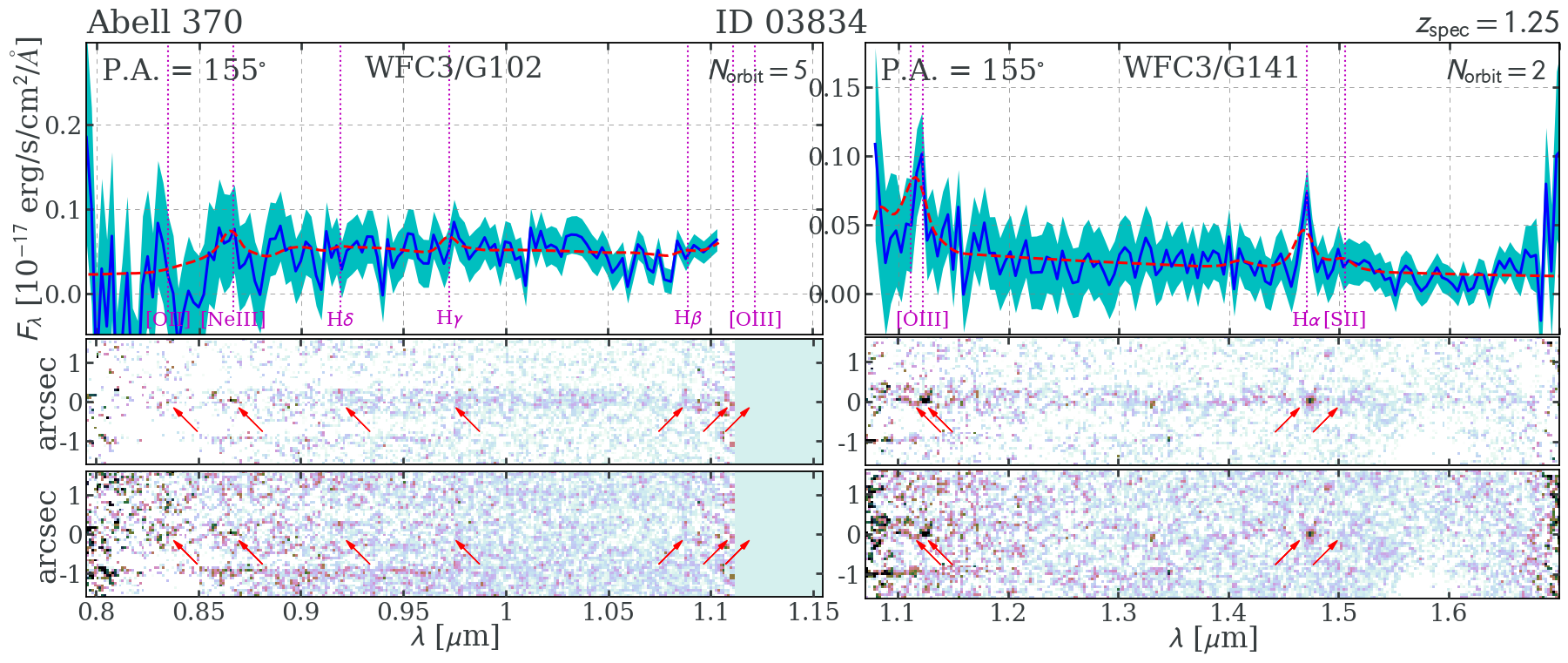}\\
    \includegraphics[width=\textwidth]{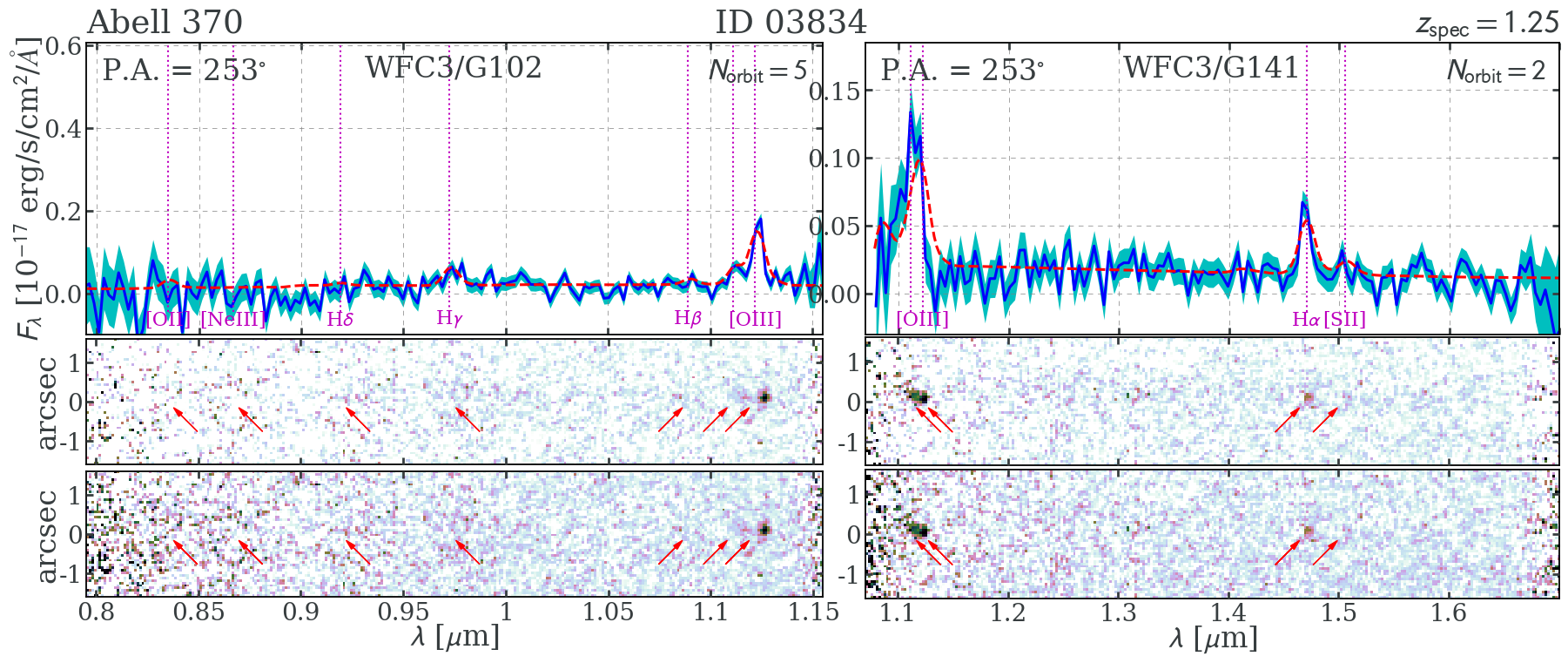}\\
    \includegraphics[width=.16\textwidth]{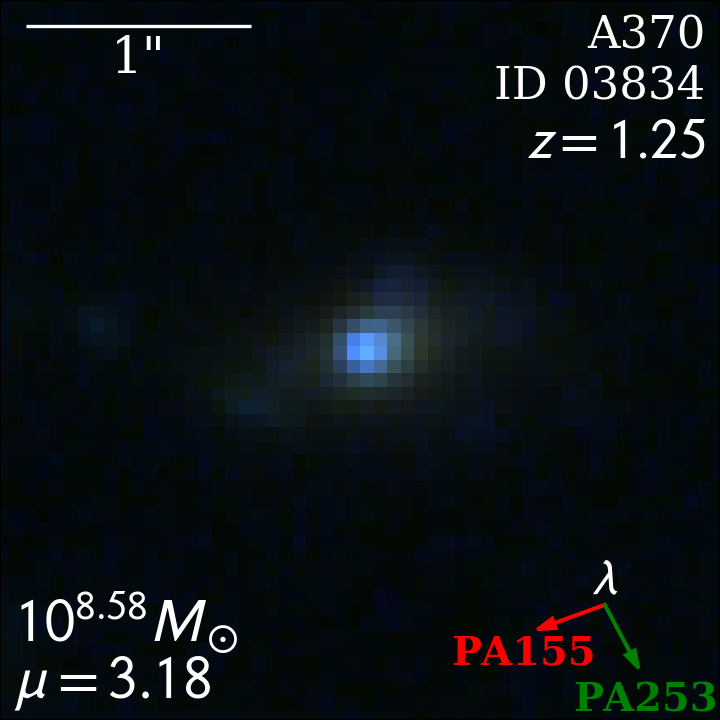}
    \includegraphics[width=.16\textwidth]{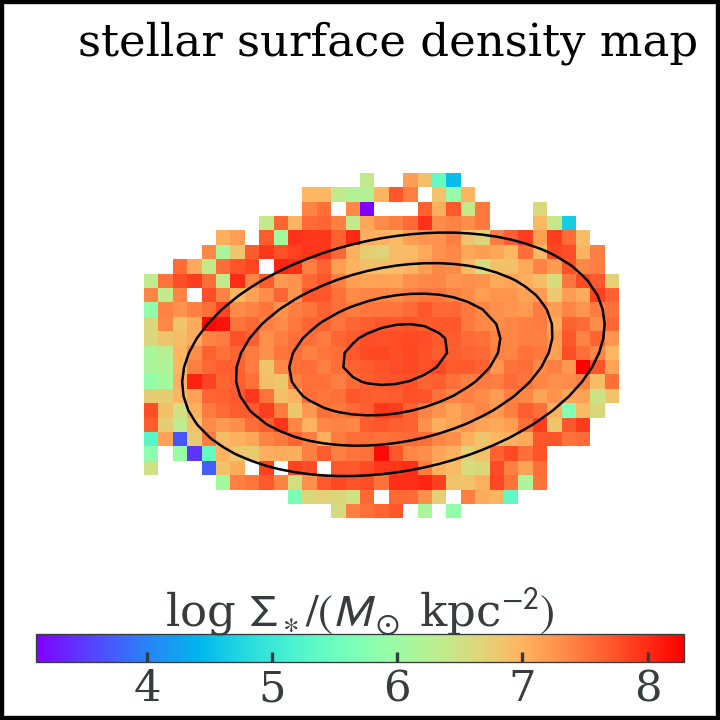}
    \includegraphics[width=.16\textwidth]{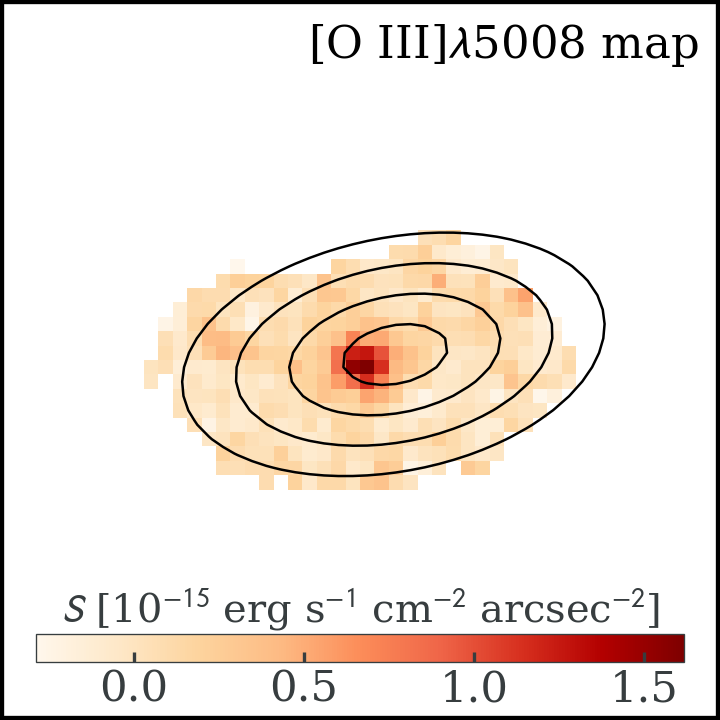}
    \includegraphics[width=.16\textwidth]{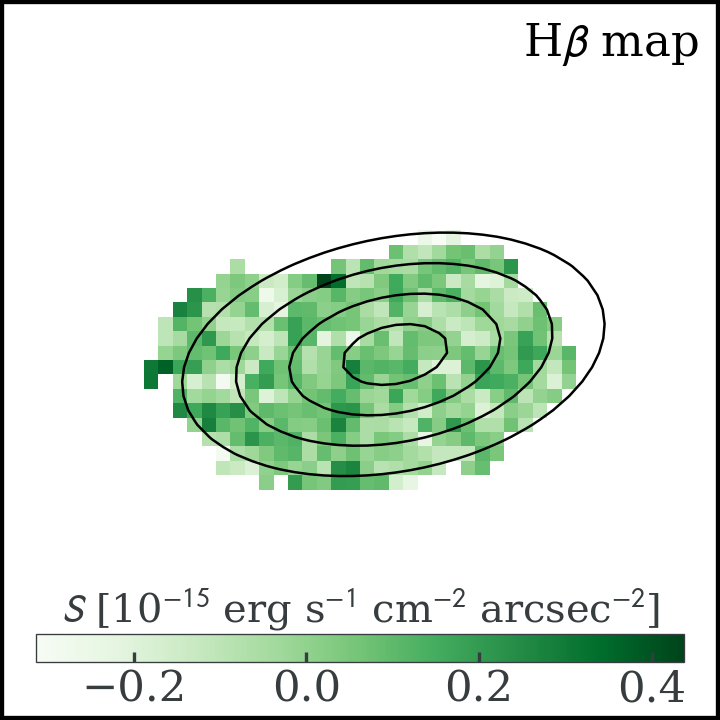}
    \includegraphics[width=.16\textwidth]{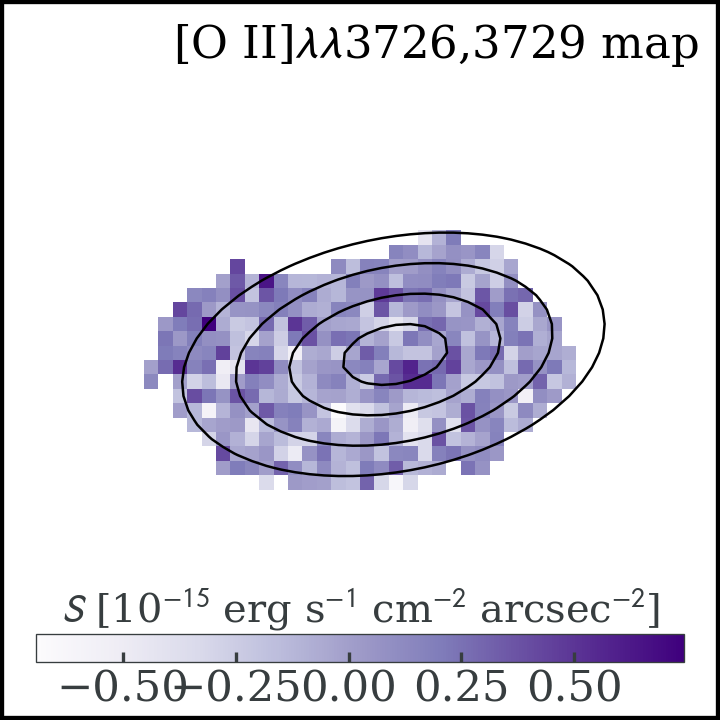}
    \includegraphics[width=.16\textwidth]{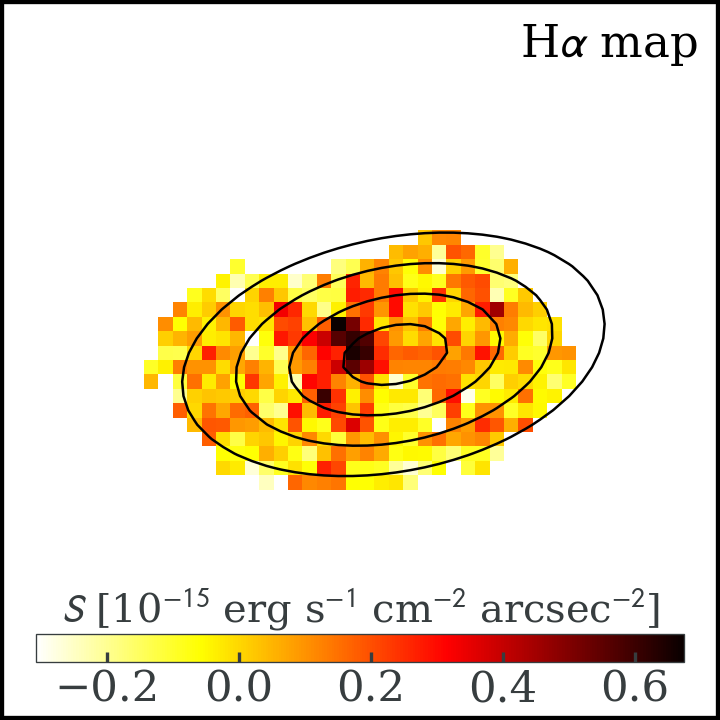}\\
    \includegraphics[width=\textwidth]{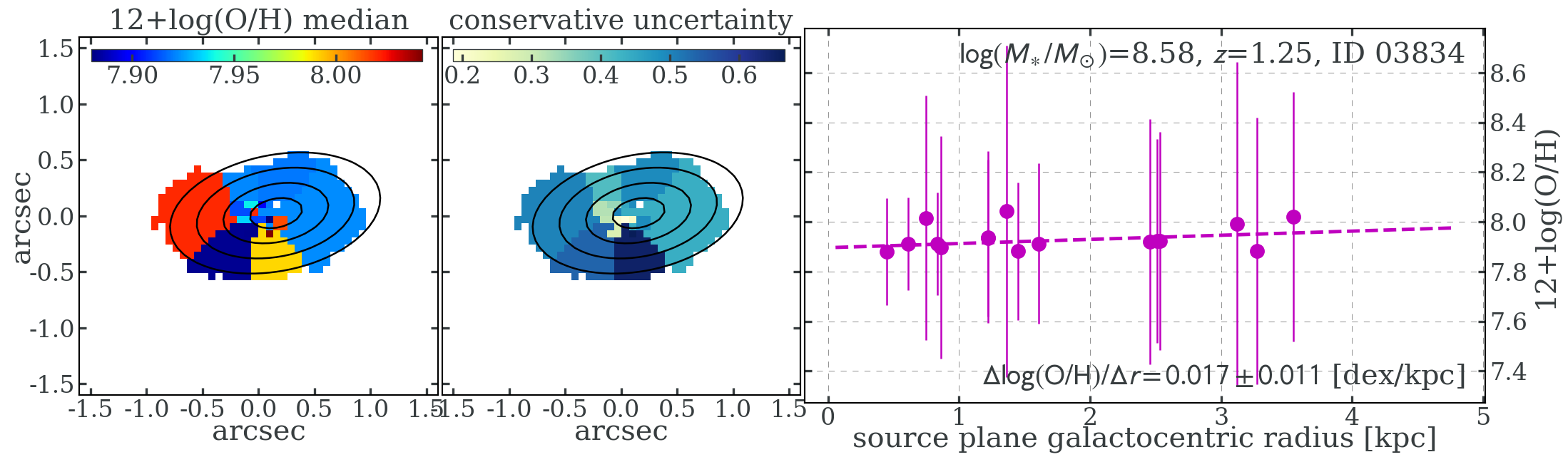}
    \caption{The source ID03834 in the field of \clsan is shown.}
    \label{fig:clA370_ID03834_figs}
\end{figure*}

\begin{figure*}
    \centering
    \includegraphics[width=\textwidth]{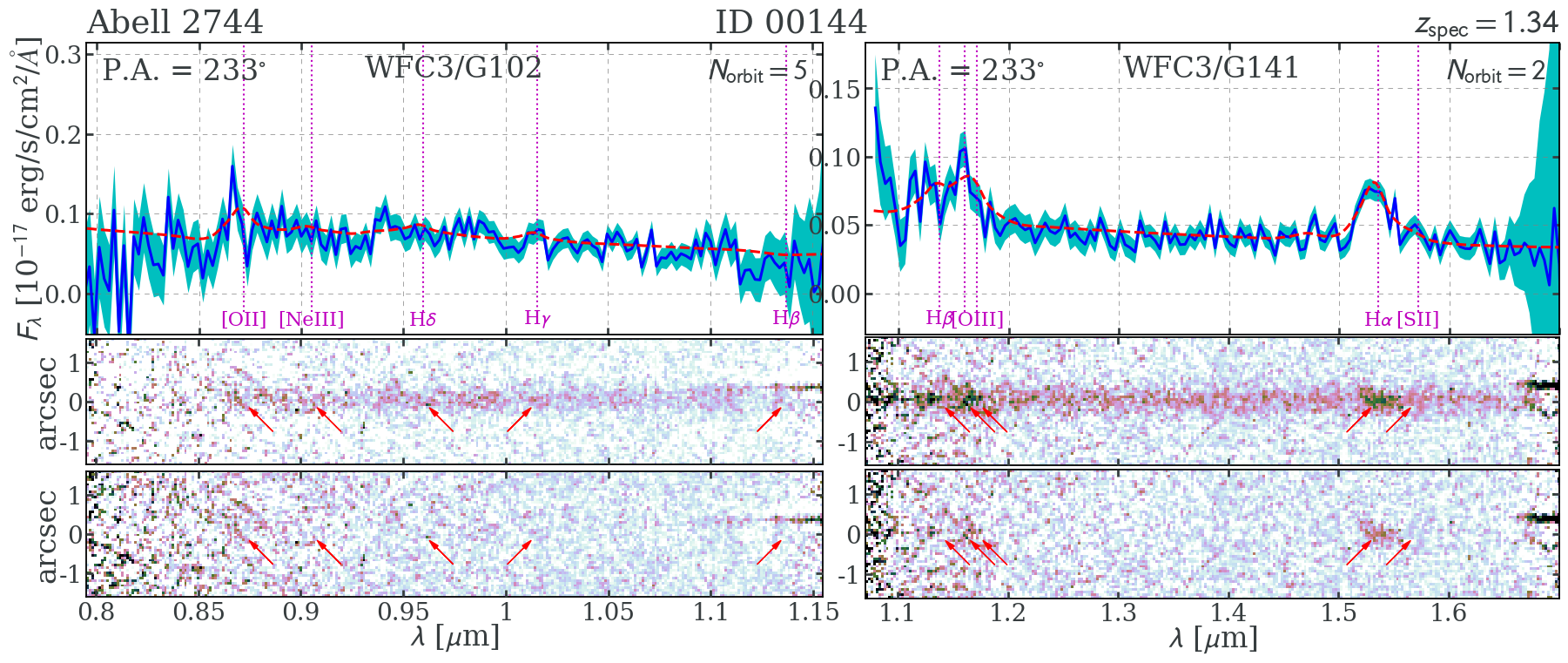}\\
    \includegraphics[width=.16\textwidth]{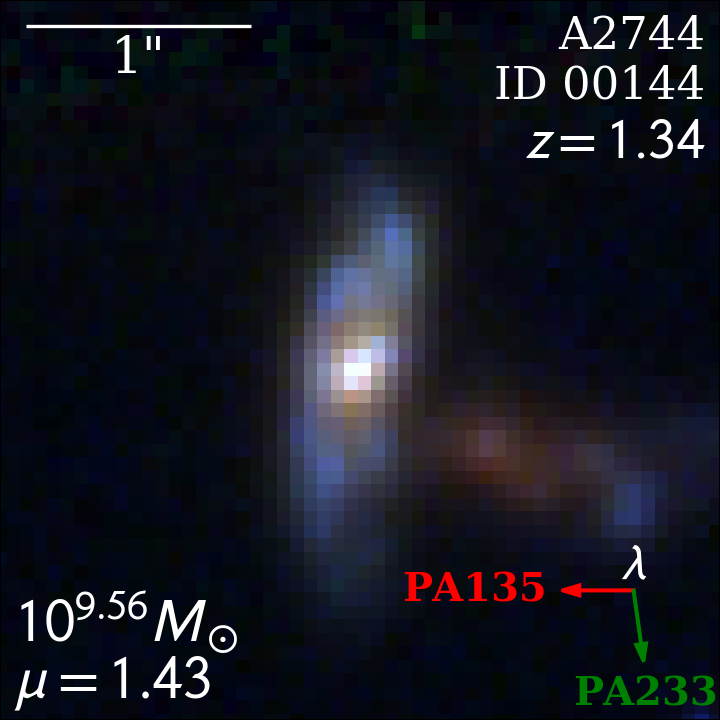}
    \includegraphics[width=.16\textwidth]{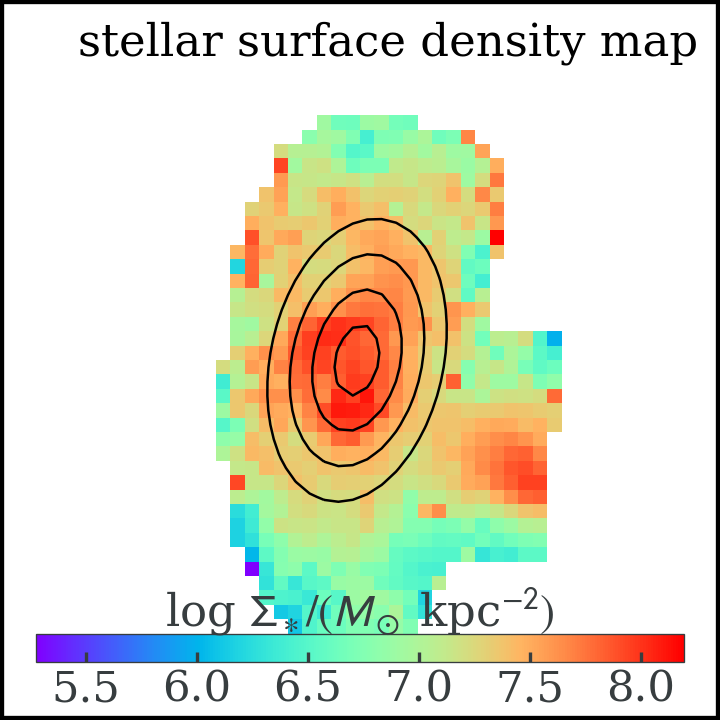}
    \includegraphics[width=.16\textwidth]{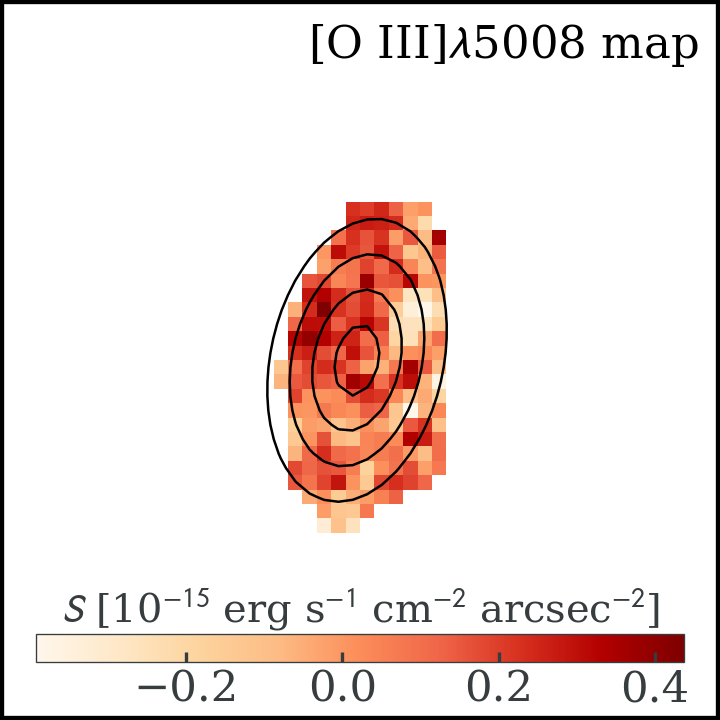}
    \includegraphics[width=.16\textwidth]{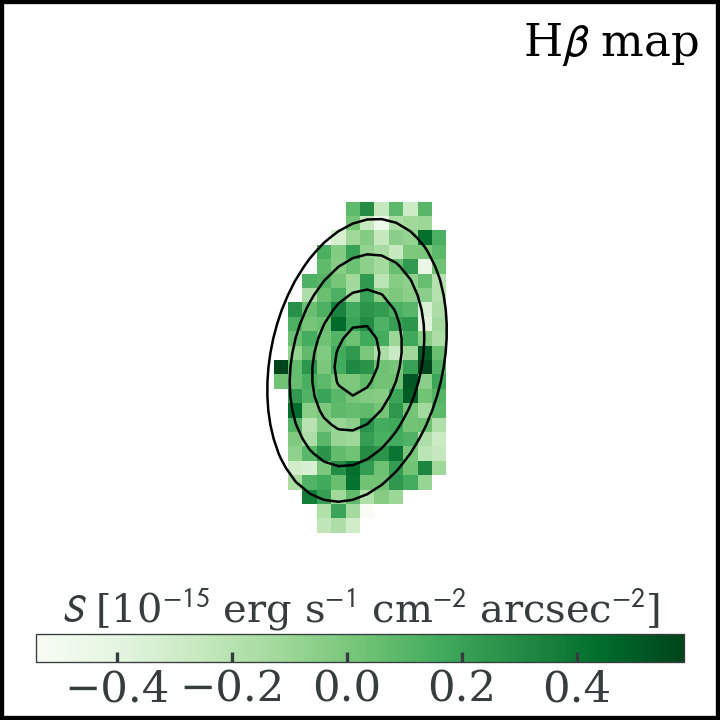}
    \includegraphics[width=.16\textwidth]{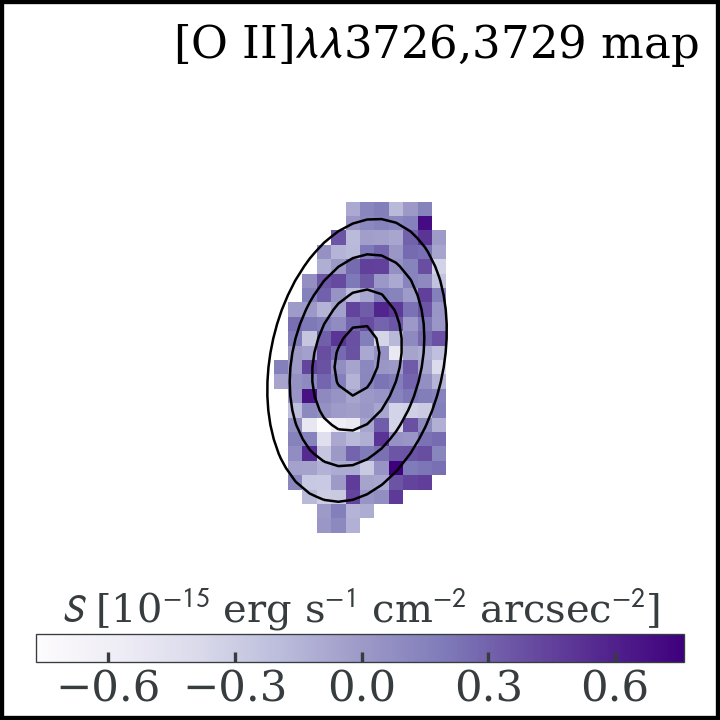}
    \includegraphics[width=.16\textwidth]{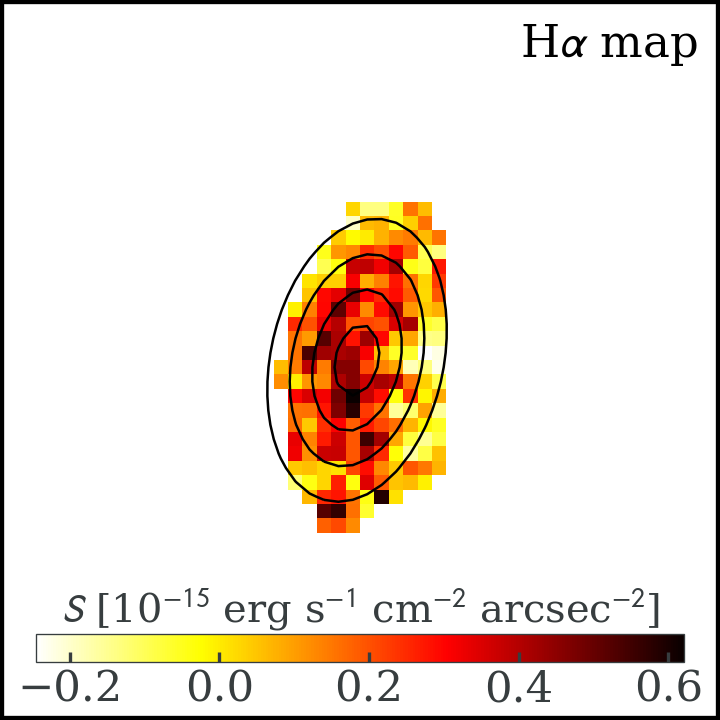}\\
    \includegraphics[width=\textwidth]{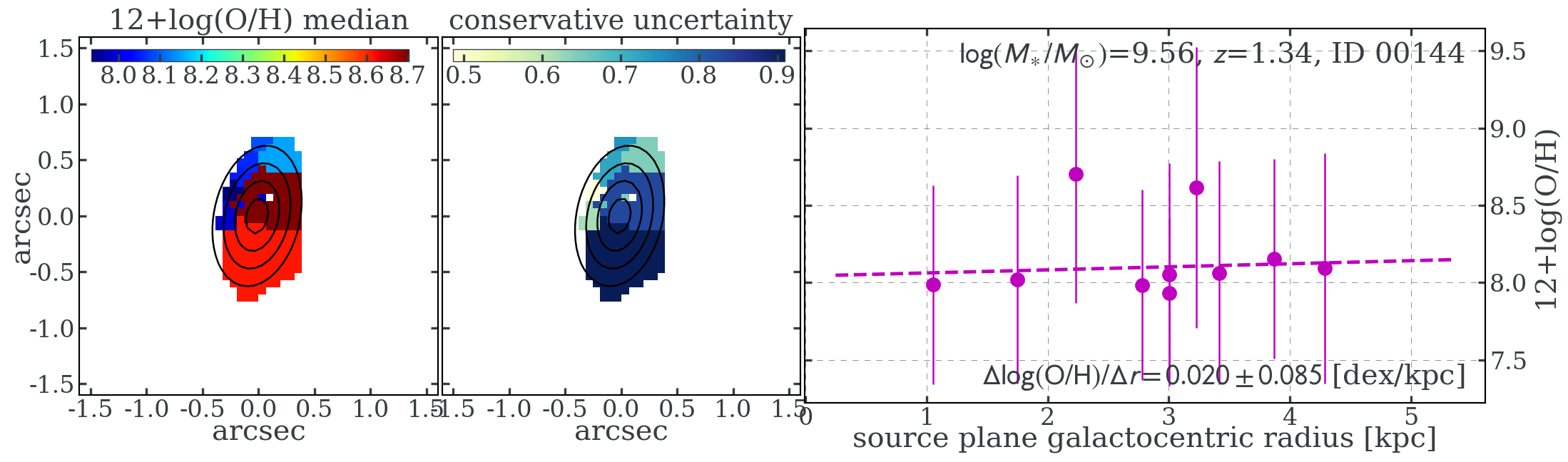}
    \caption{The source ID00144 in the field of \cler is shown.}
    \label{fig:clA2744_ID00144_figs}
\end{figure*}

\begin{figure*}
    \centering
    \includegraphics[width=\textwidth]{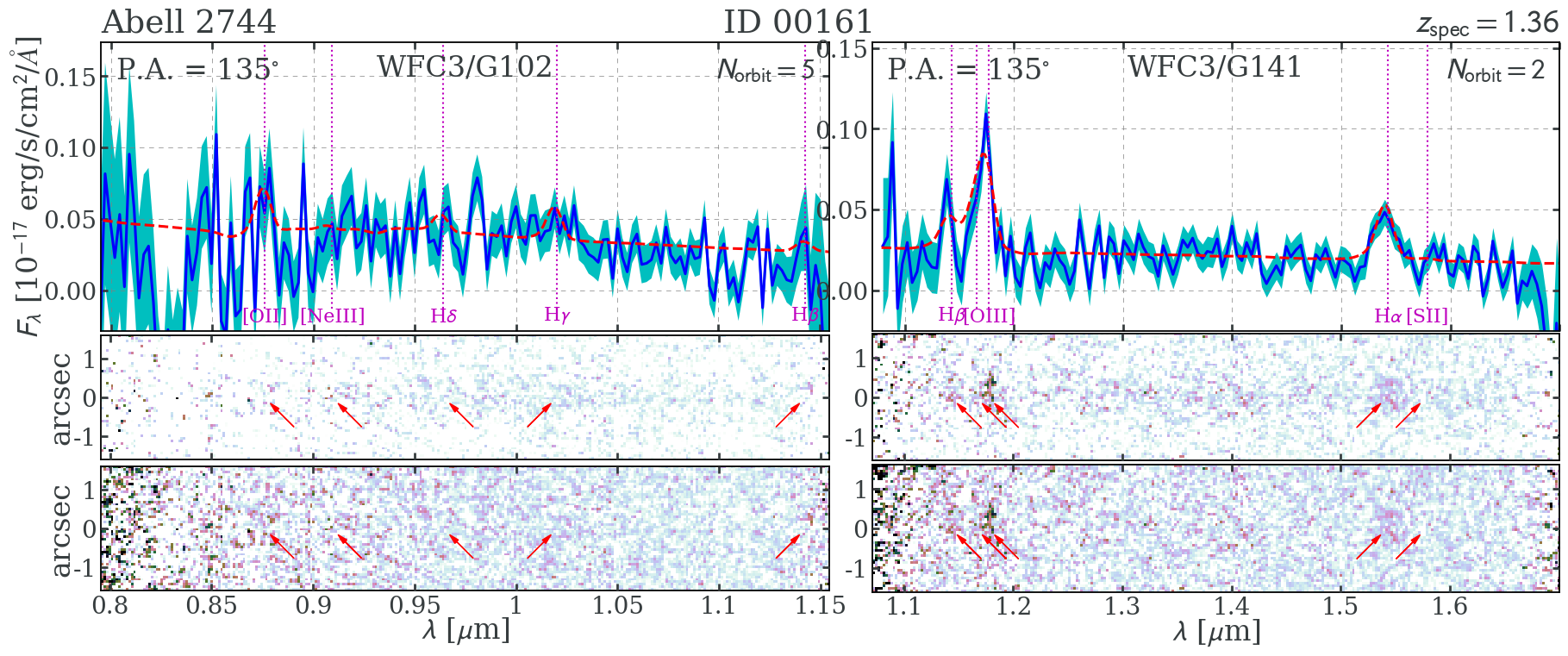}\\
    \includegraphics[width=\textwidth]{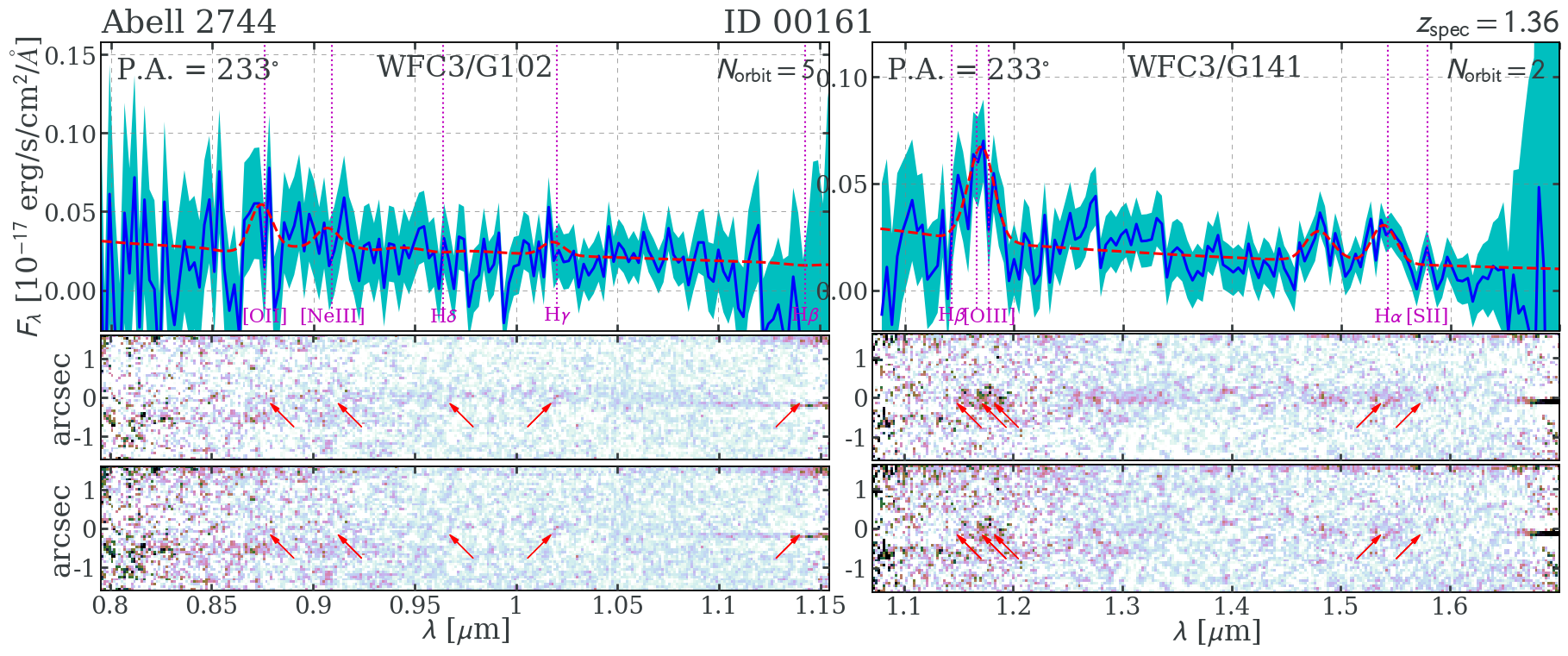}\\
    \includegraphics[width=.16\textwidth]{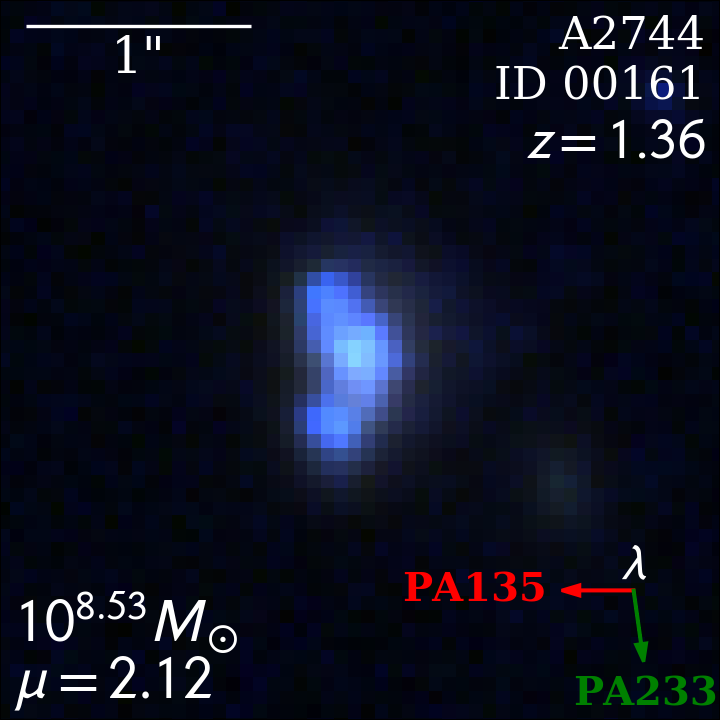}
    \includegraphics[width=.16\textwidth]{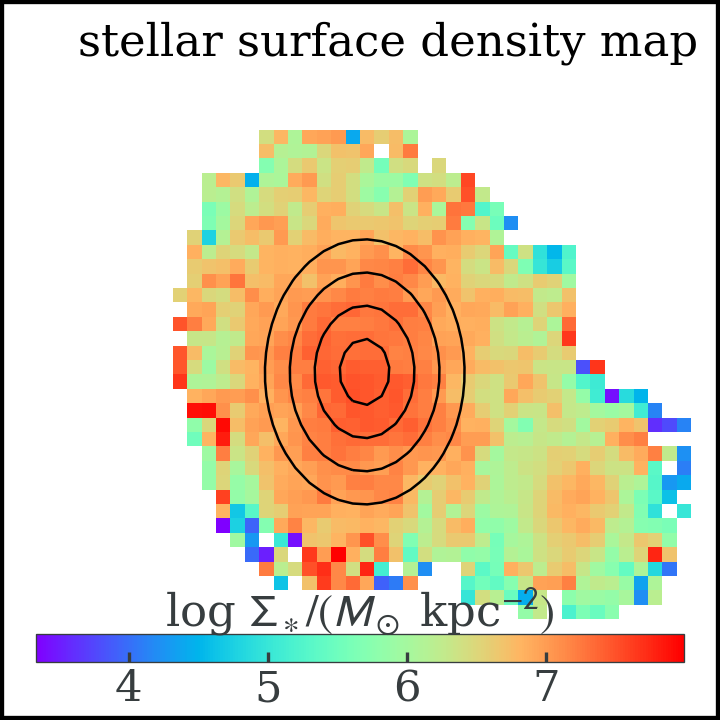}
    \includegraphics[width=.16\textwidth]{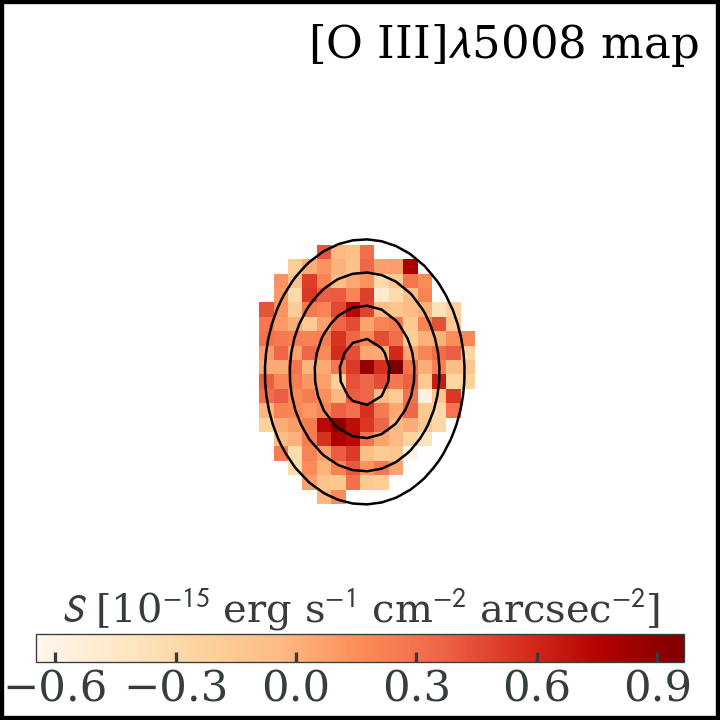}
    \includegraphics[width=.16\textwidth]{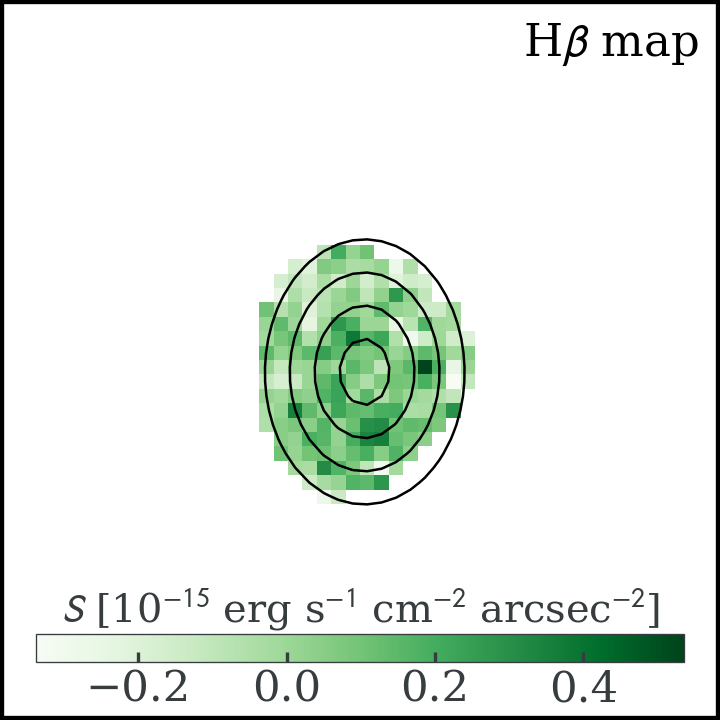}
    \includegraphics[width=.16\textwidth]{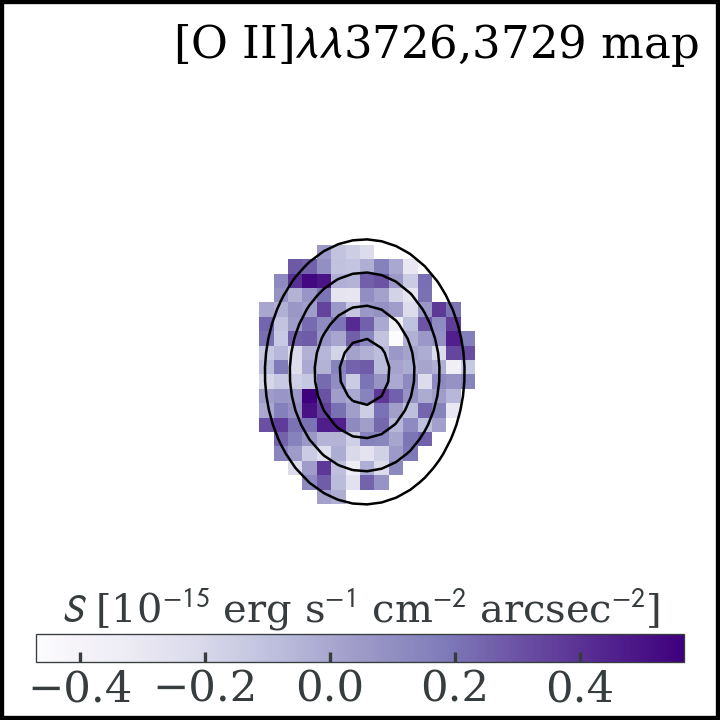}
    \includegraphics[width=.16\textwidth]{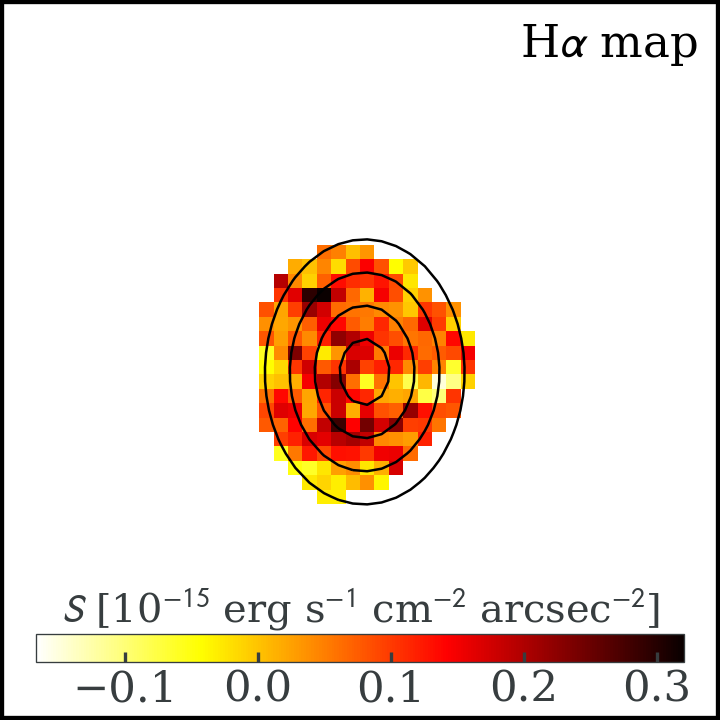}\\
    \includegraphics[width=\textwidth]{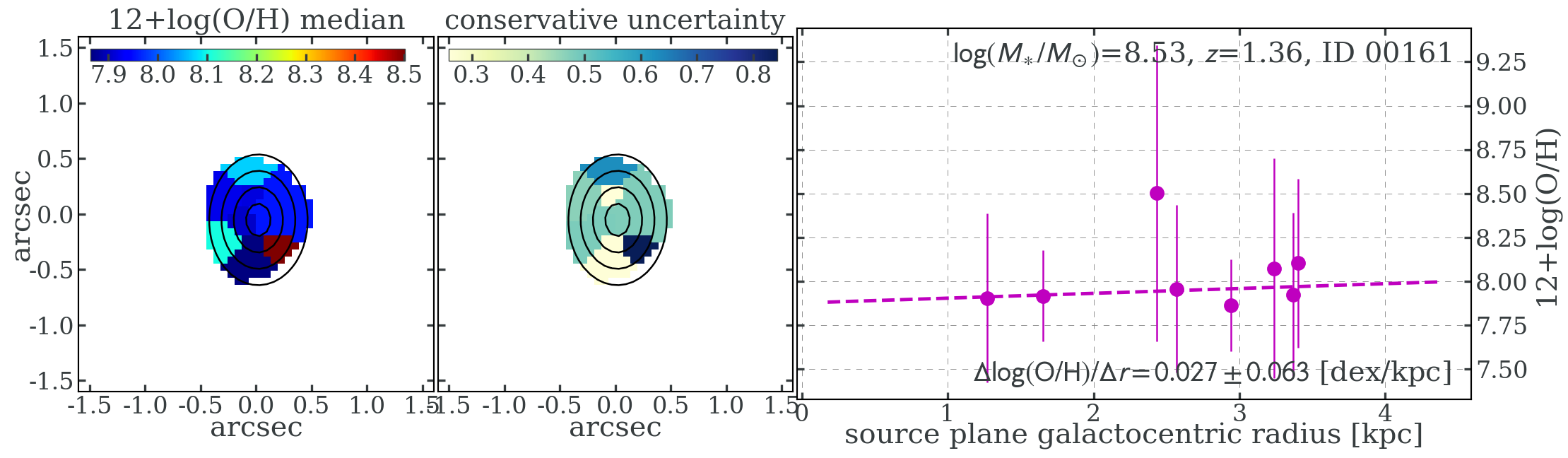}
    \caption{The source ID00161 in the field of \cler is shown.}
    \label{fig:clA2744_ID00161_figs}
\end{figure*}

\begin{figure*}
    \centering
    \includegraphics[width=\textwidth]{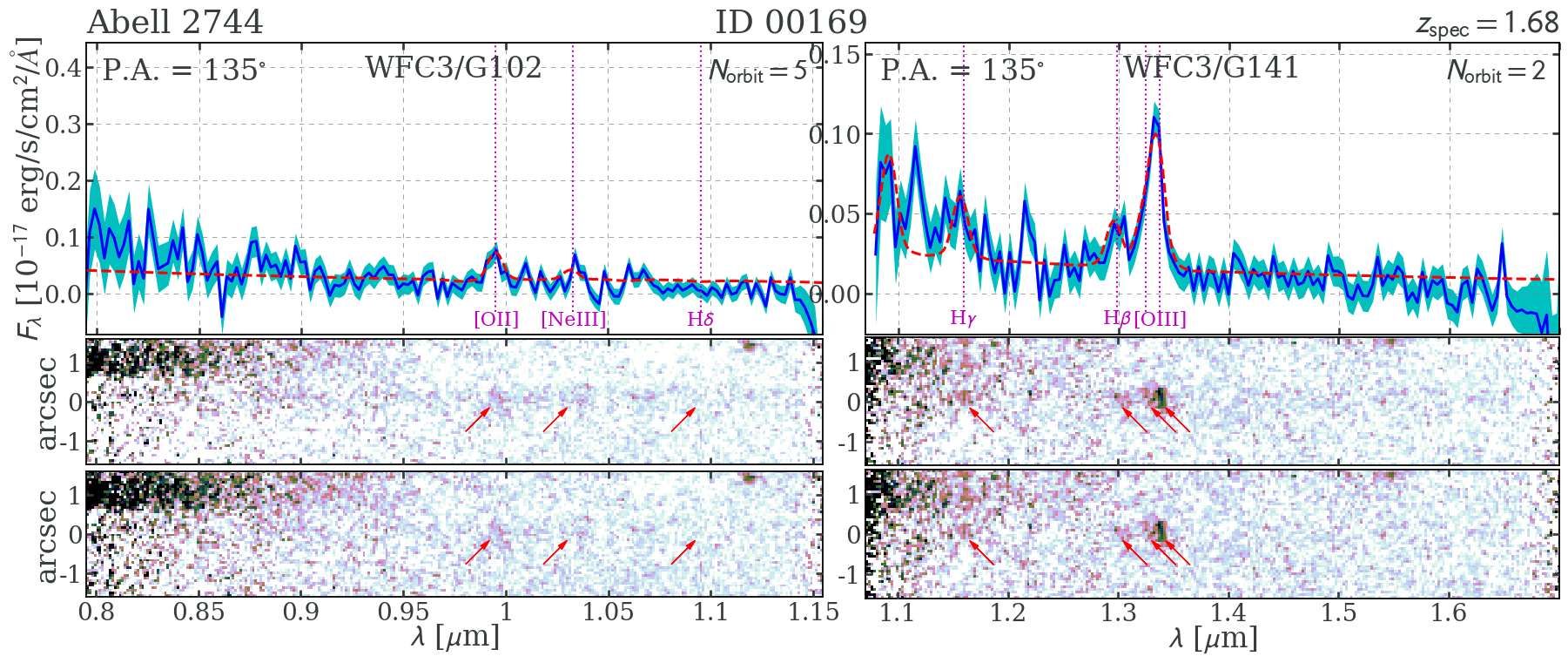}\\
    \includegraphics[width=\textwidth]{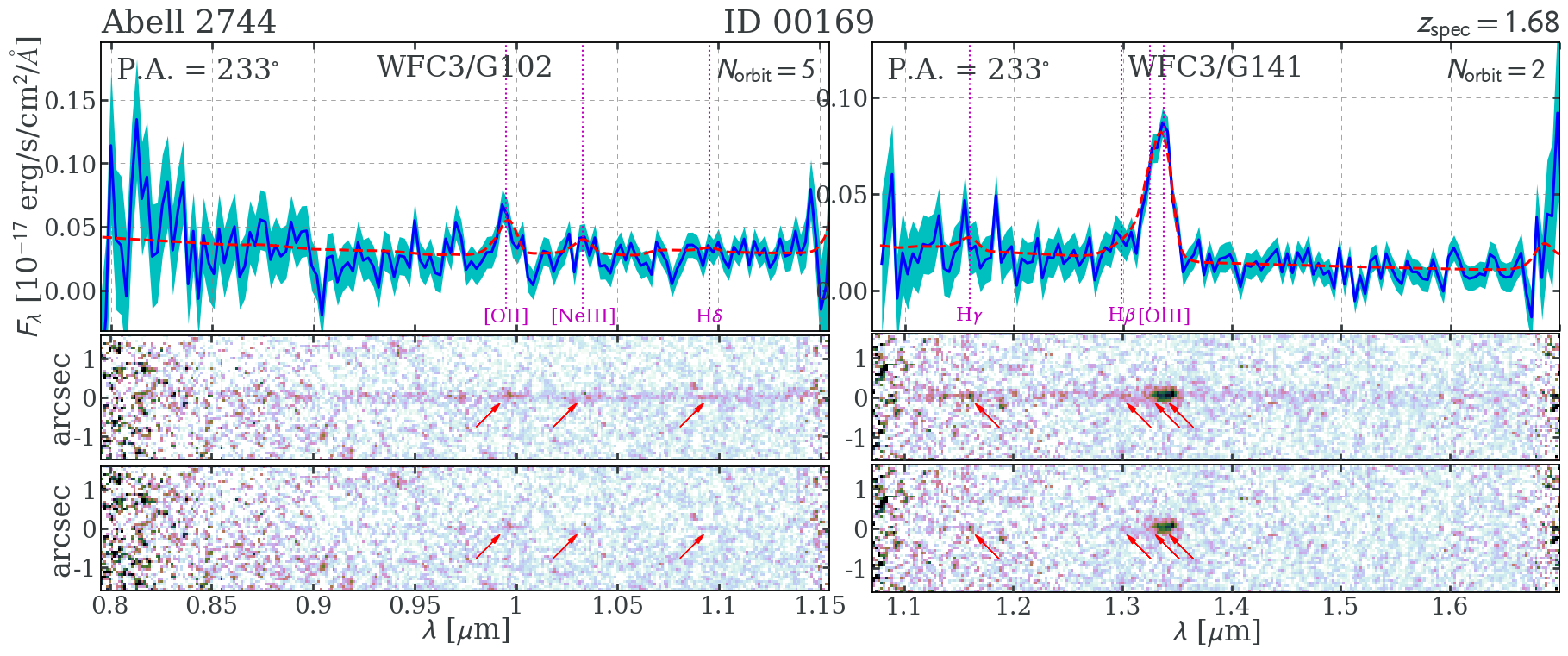}\\
    \includegraphics[width=.16\textwidth]{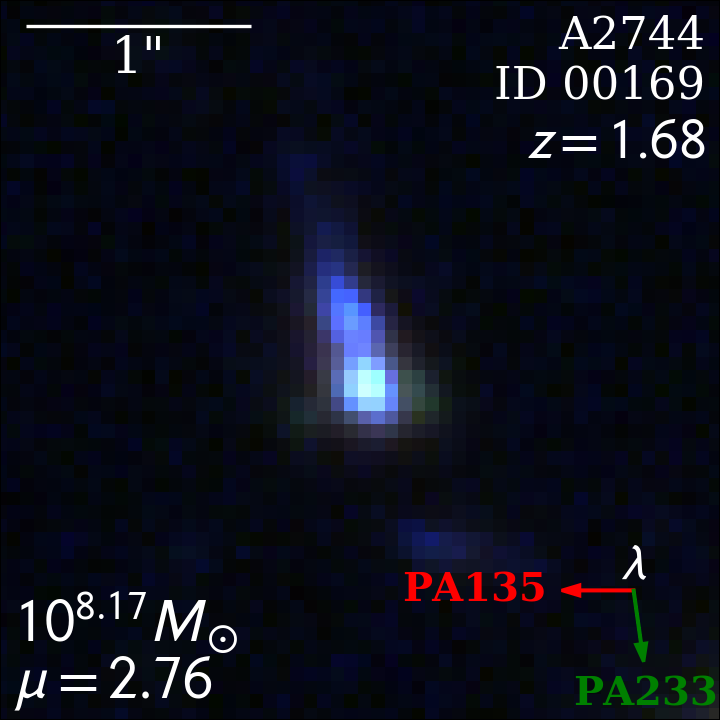}
    \includegraphics[width=.16\textwidth]{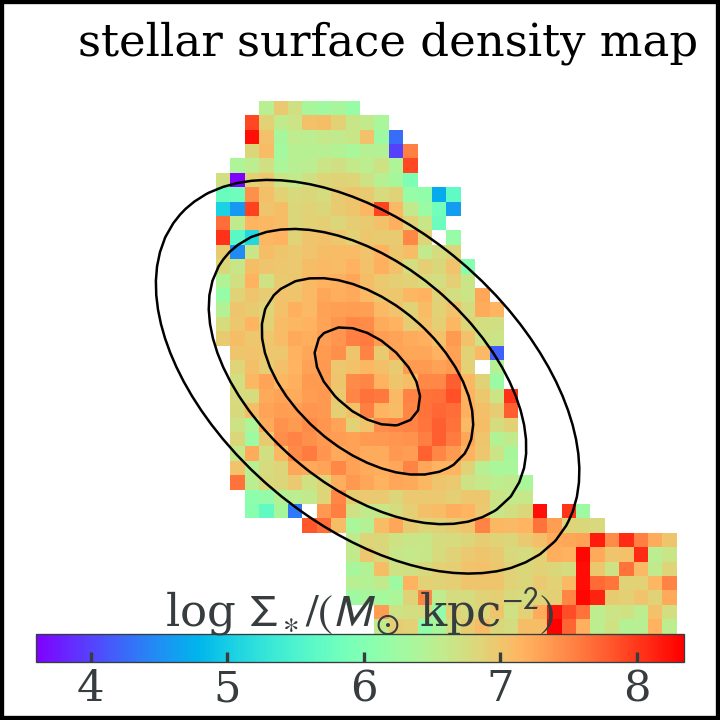}
    \includegraphics[width=.16\textwidth]{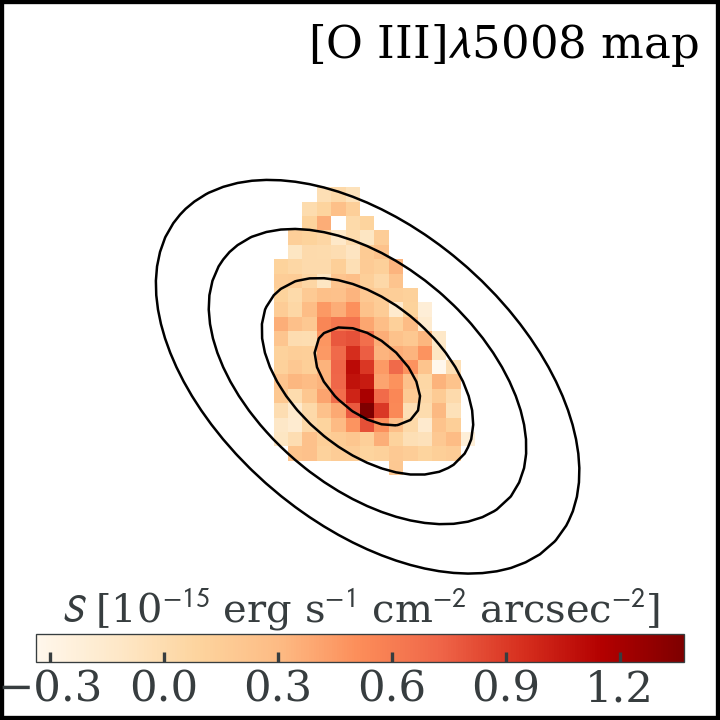}
    \includegraphics[width=.16\textwidth]{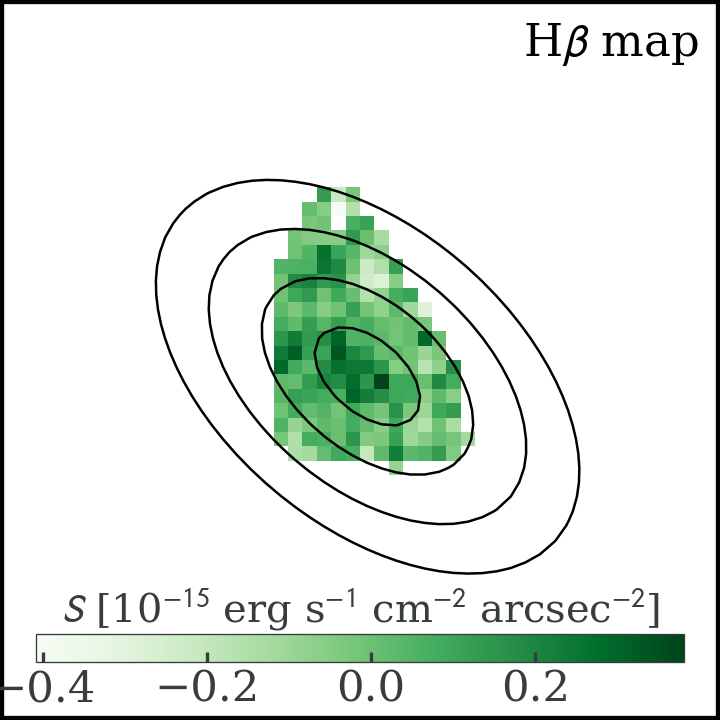}
    \includegraphics[width=.16\textwidth]{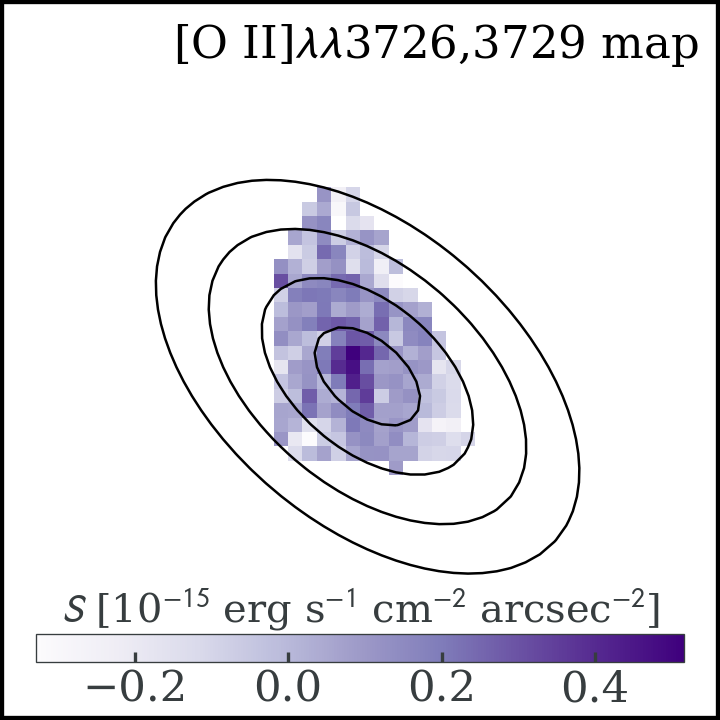}
    \includegraphics[width=.16\textwidth]{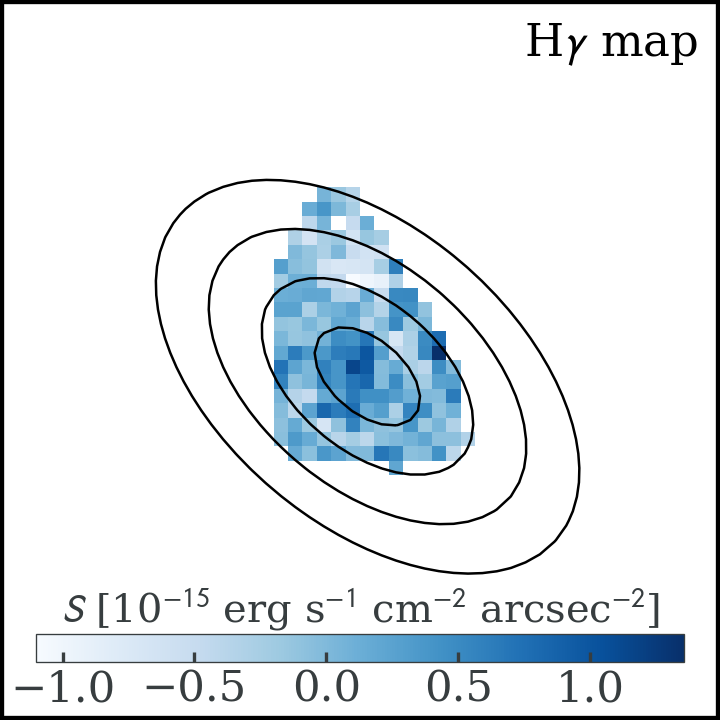}\\
    \includegraphics[width=\textwidth]{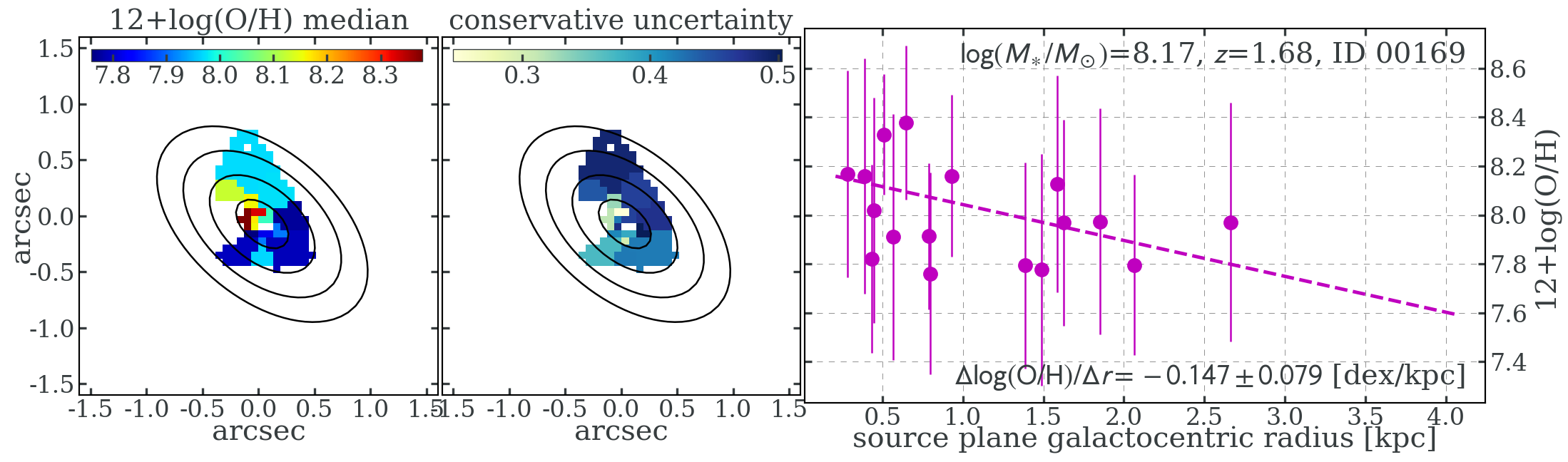}
    \caption{The source ID00169 in the field of \cler is shown.}
    \label{fig:clA2744_ID00169_figs}
\end{figure*}

\begin{figure*}
    \centering
    \includegraphics[width=\textwidth]{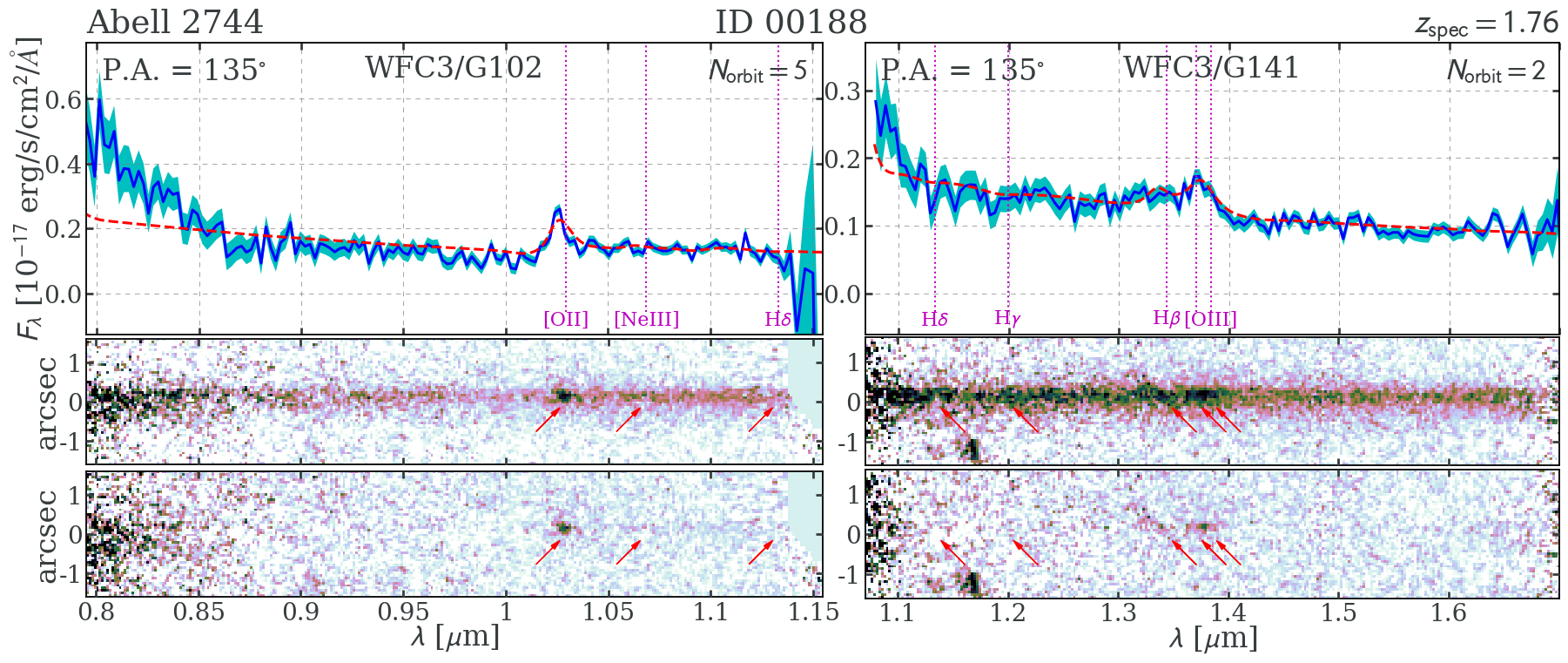}\\
    \includegraphics[width=\textwidth]{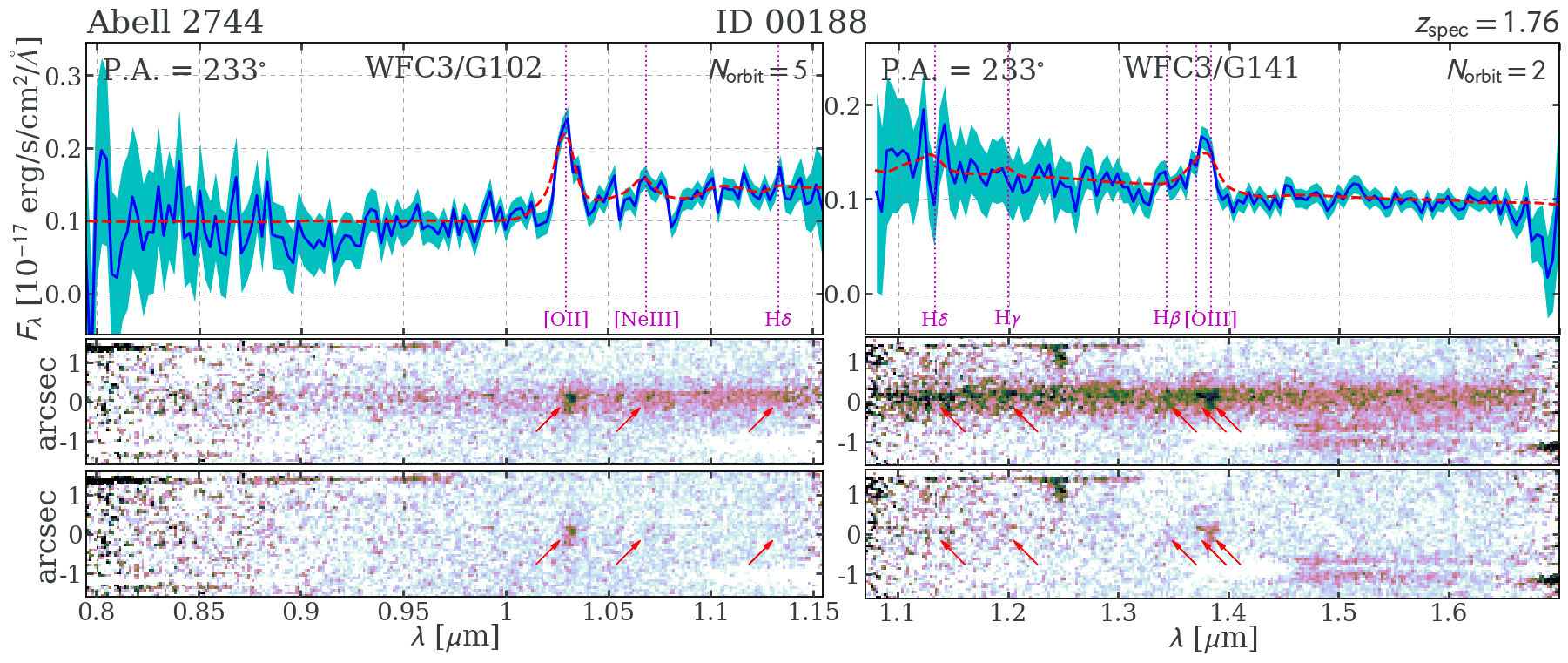}\\
    \includegraphics[width=.16\textwidth]{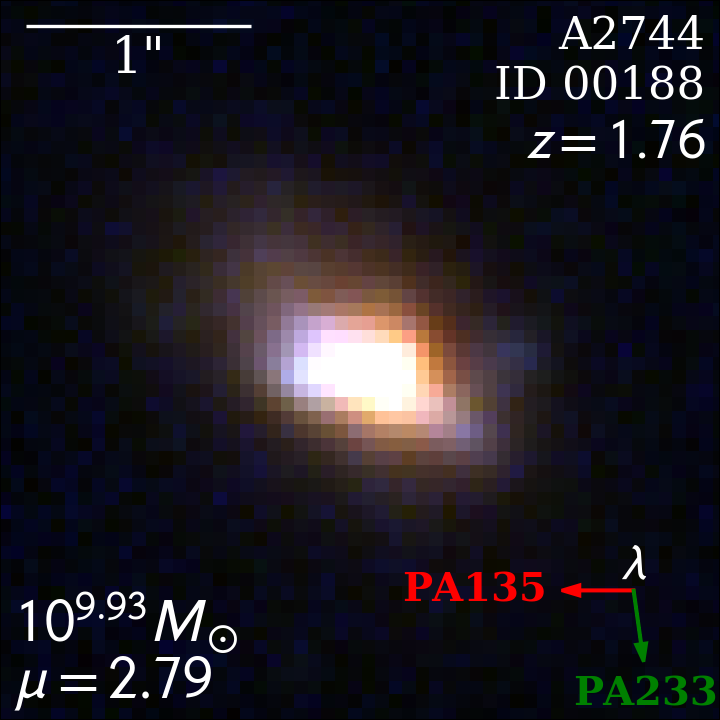}
    \includegraphics[width=.16\textwidth]{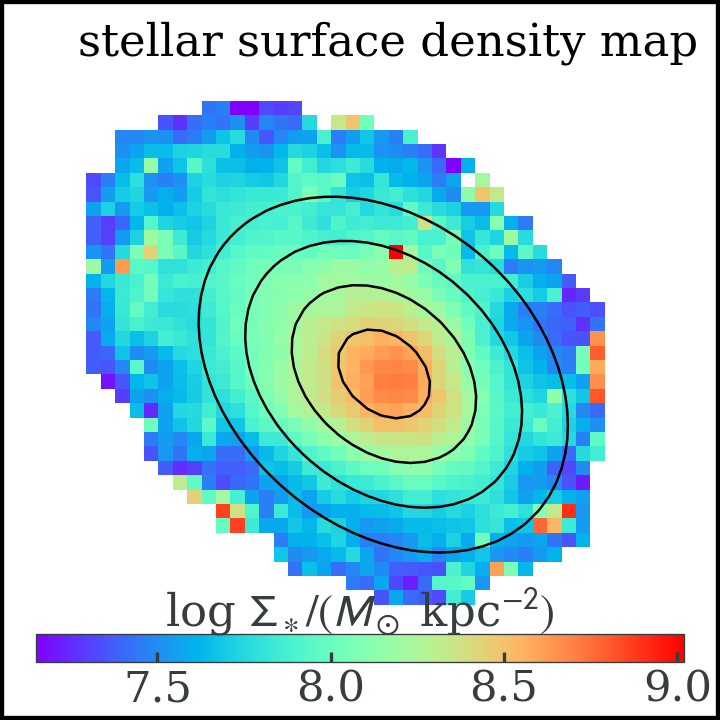}
    \includegraphics[width=.16\textwidth]{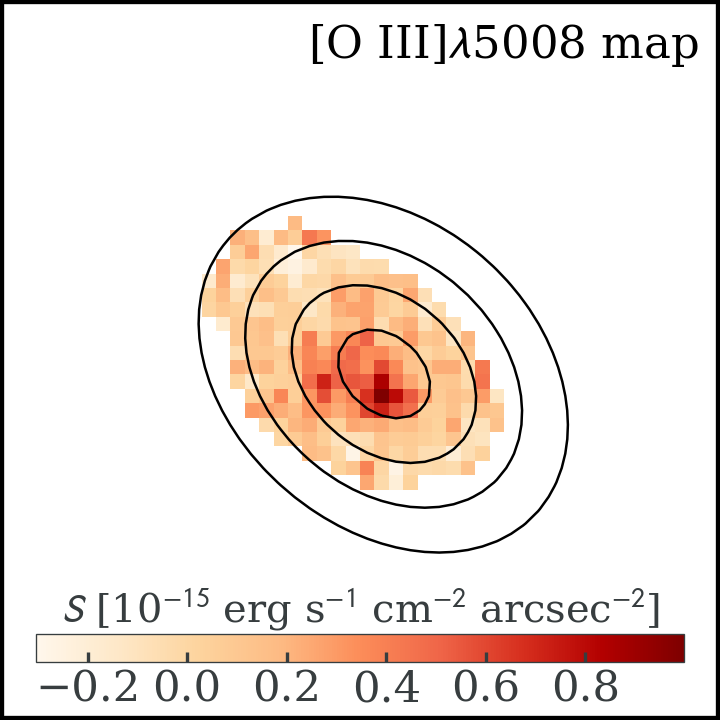}
    \includegraphics[width=.16\textwidth]{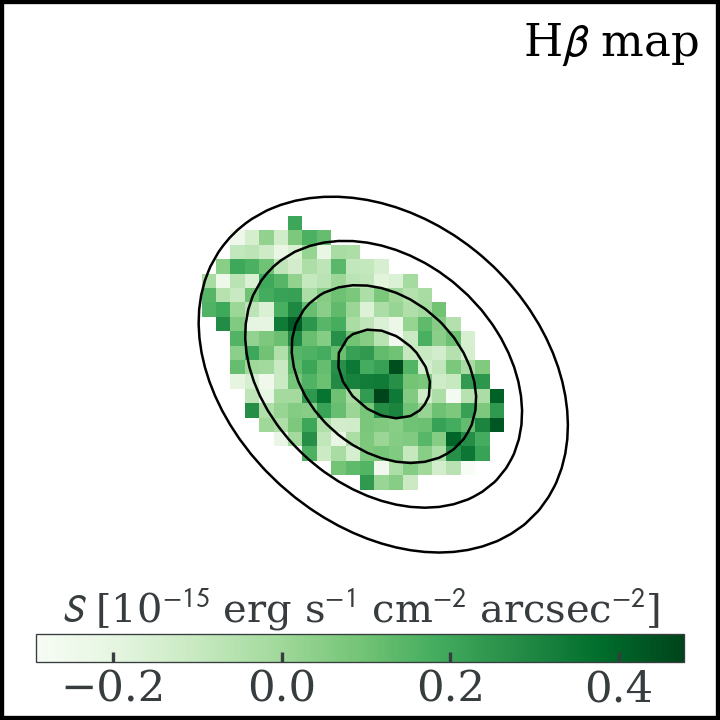}
    \includegraphics[width=.16\textwidth]{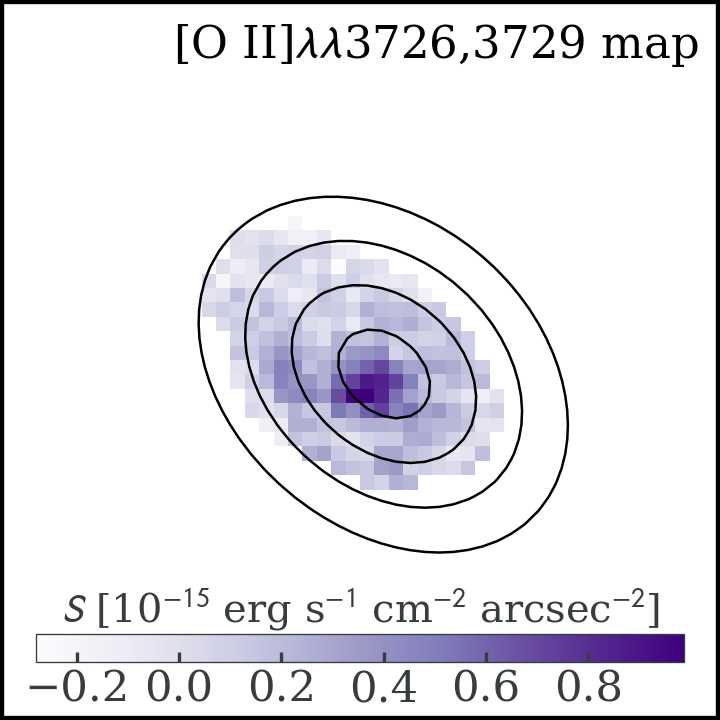}
    \includegraphics[width=.16\textwidth]{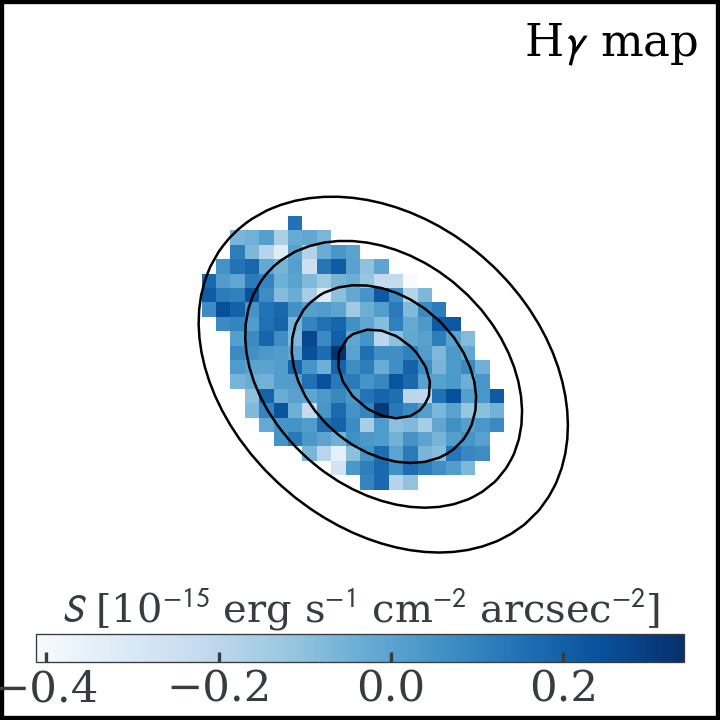}\\
    \includegraphics[width=\textwidth]{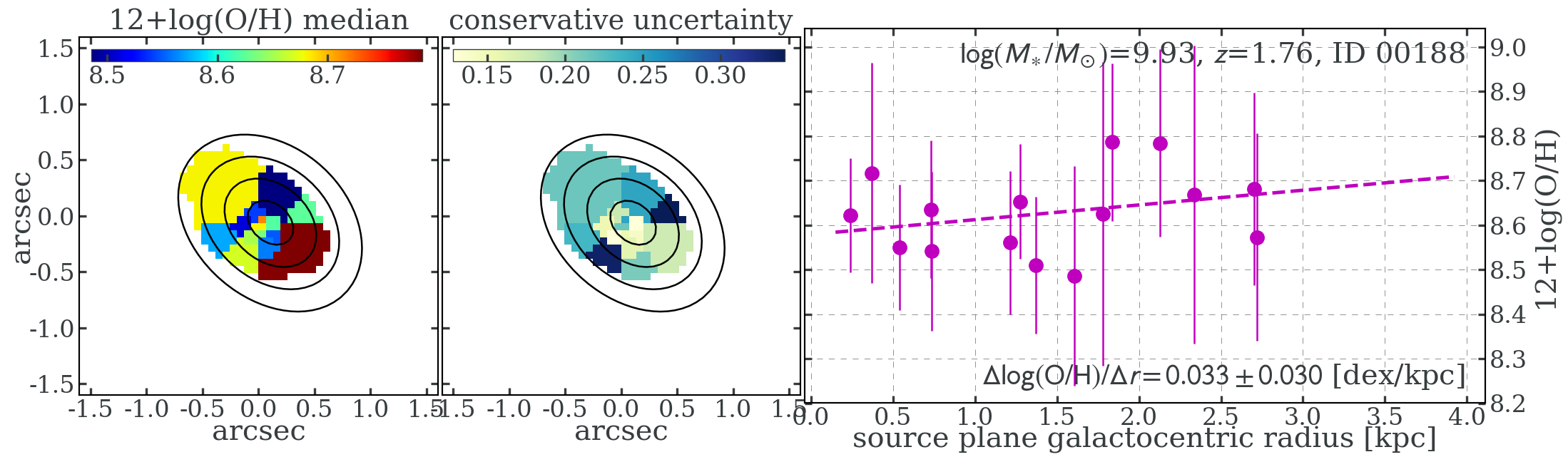}
    \caption{The source ID00188 in the field of \cler is shown.}
    \label{fig:clA2744_ID00188_figs}
\end{figure*}

\begin{figure*}
    \centering
    \includegraphics[width=\textwidth]{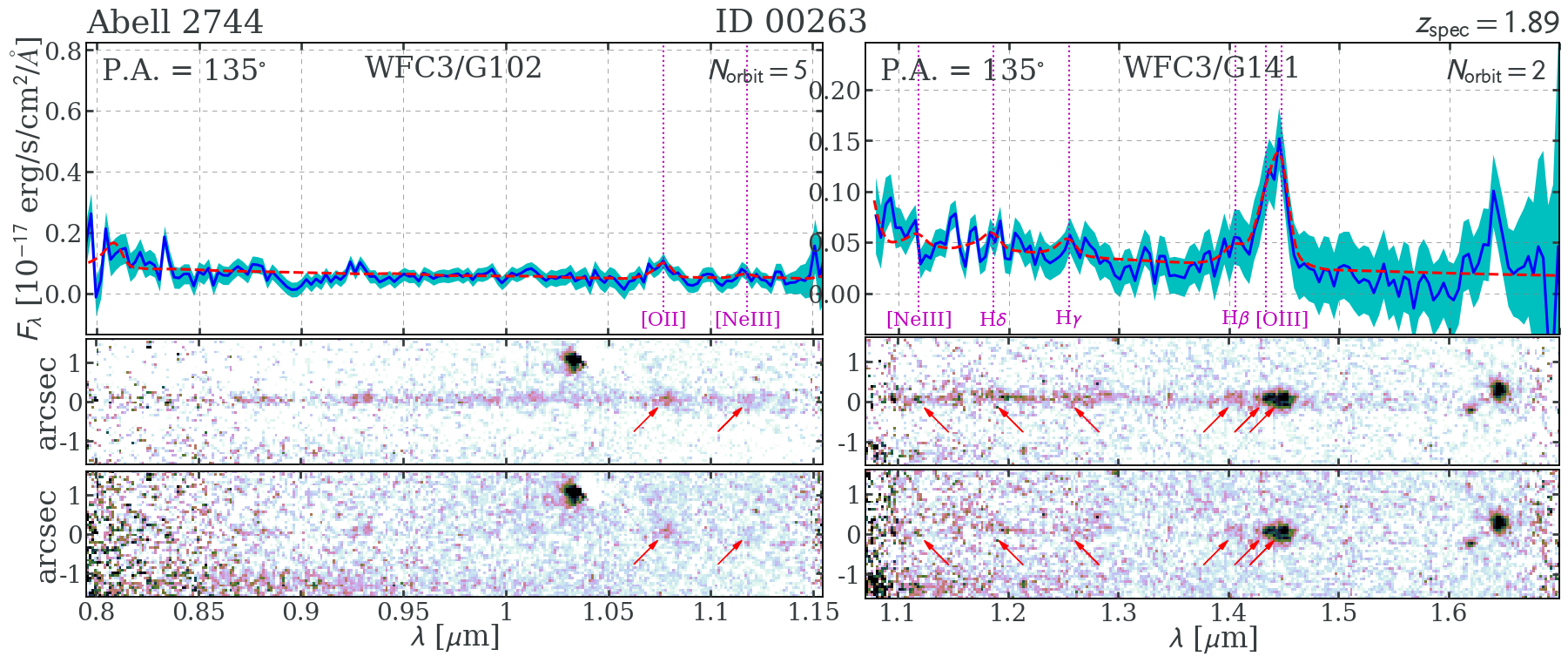}\\
    \includegraphics[width=\textwidth]{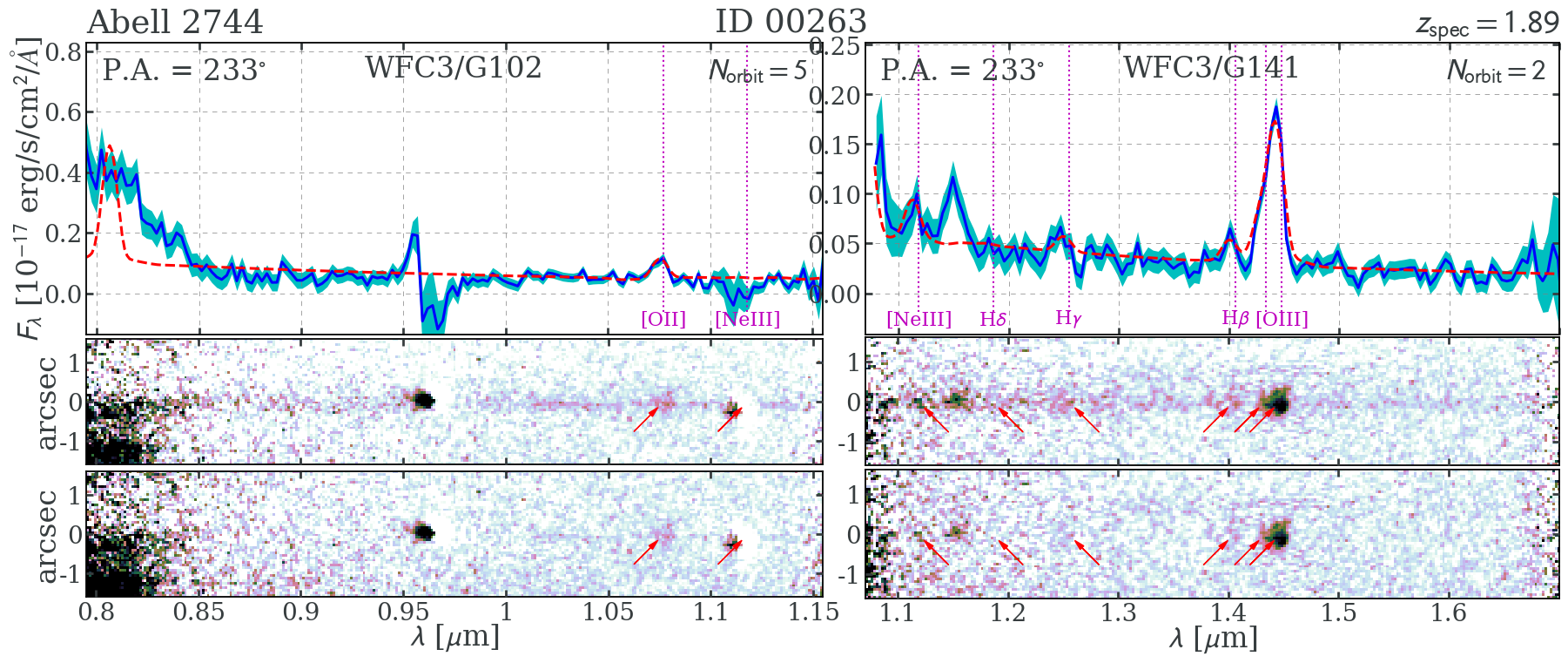}\\
    \includegraphics[width=.16\textwidth]{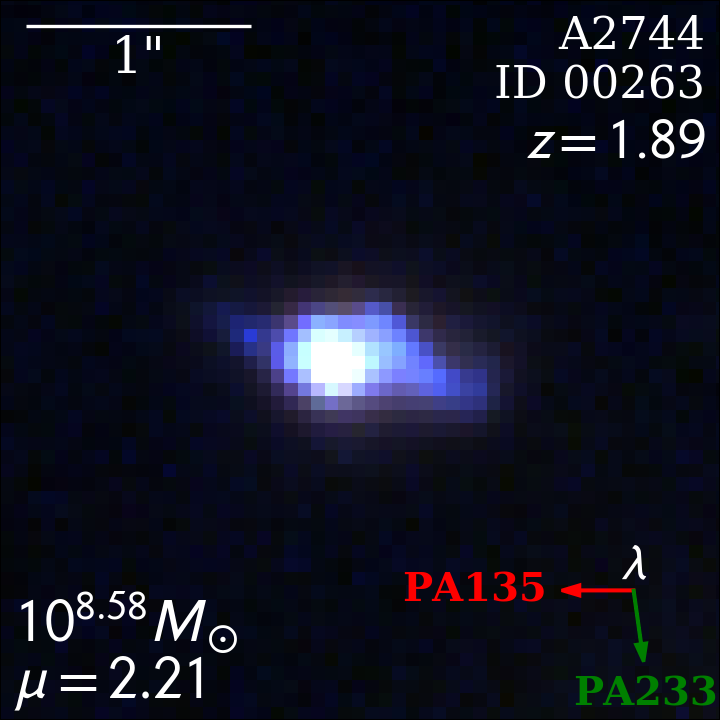}
    \includegraphics[width=.16\textwidth]{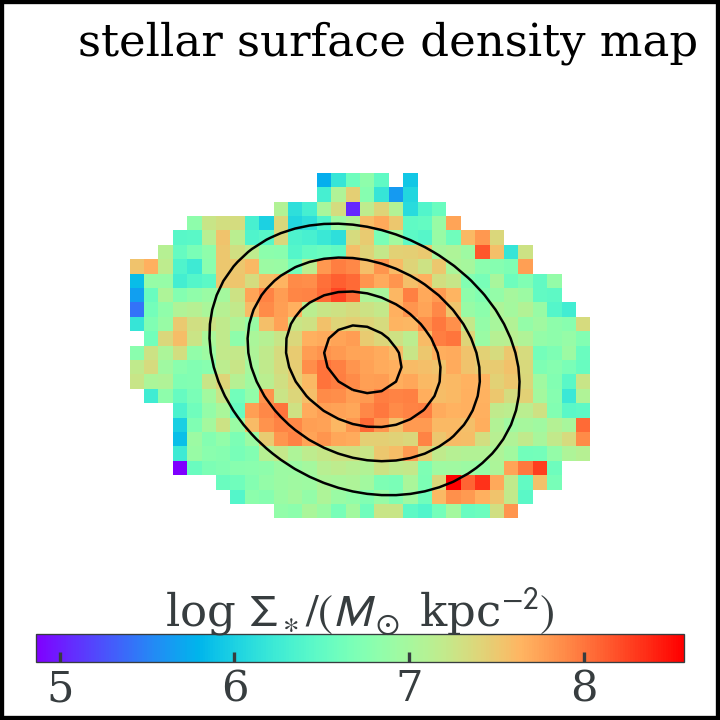}
    \includegraphics[width=.16\textwidth]{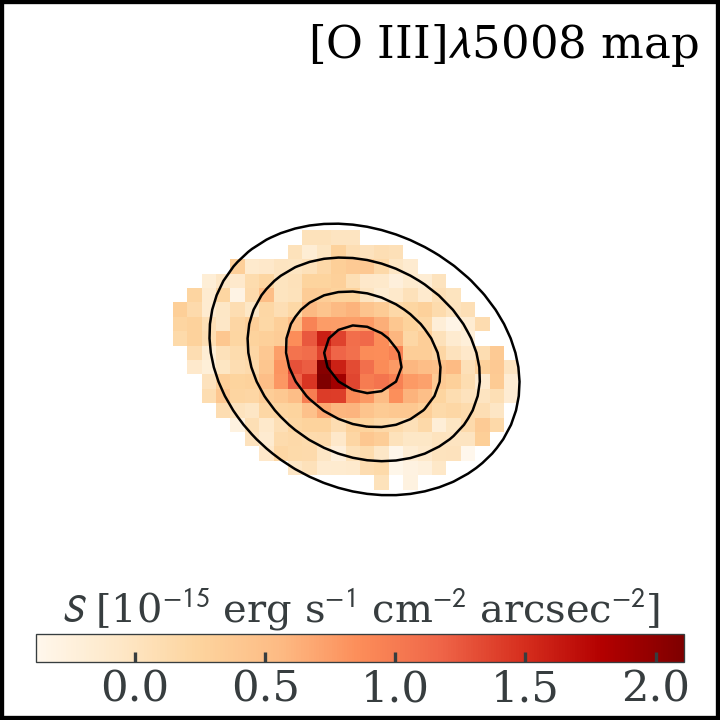}
    \includegraphics[width=.16\textwidth]{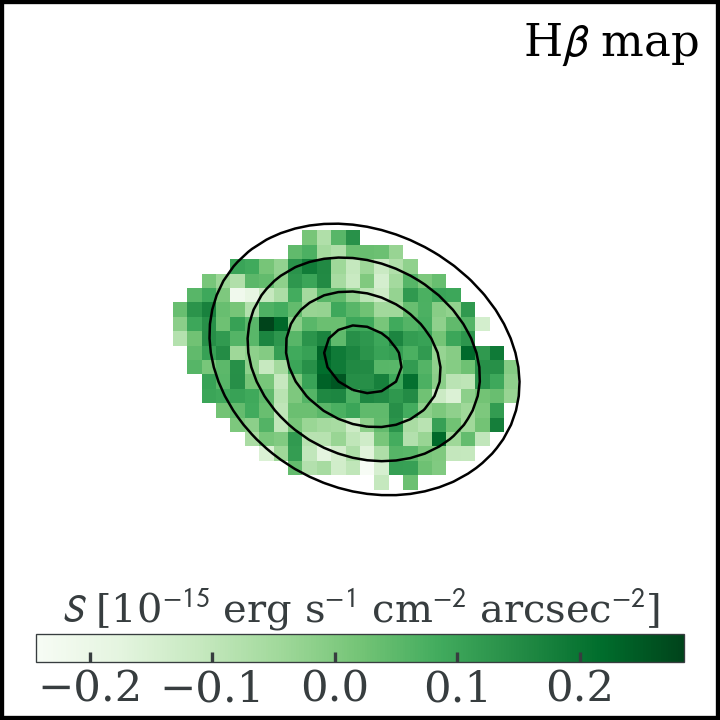}
    \includegraphics[width=.16\textwidth]{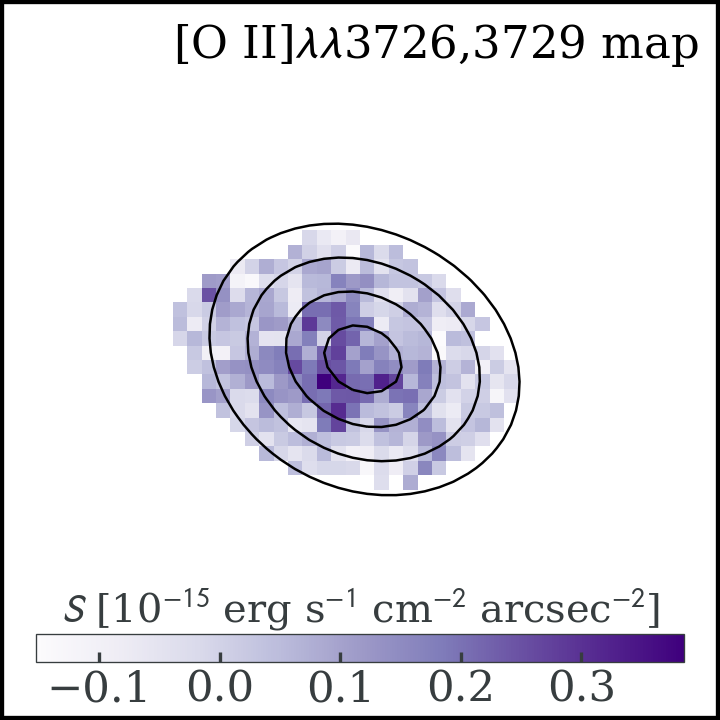}
    \includegraphics[width=.16\textwidth]{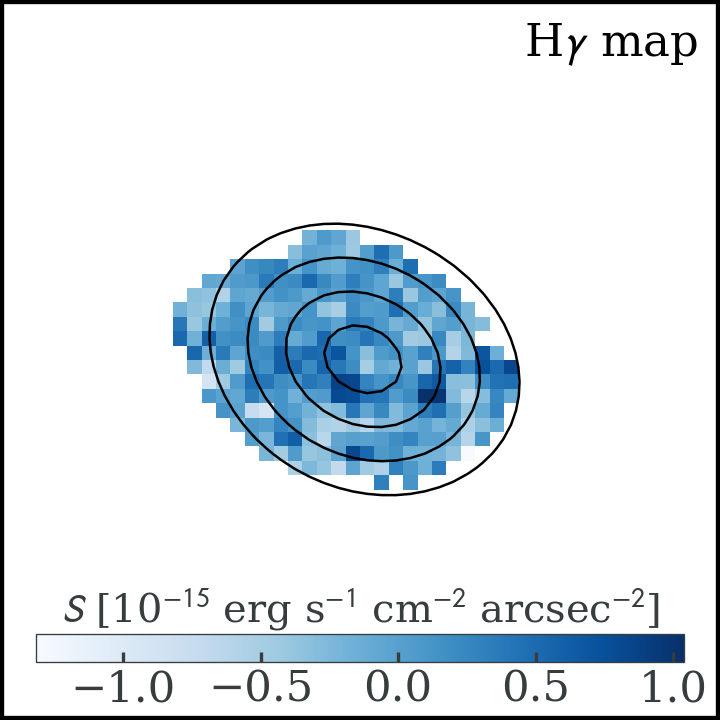}\\
    \includegraphics[width=\textwidth]{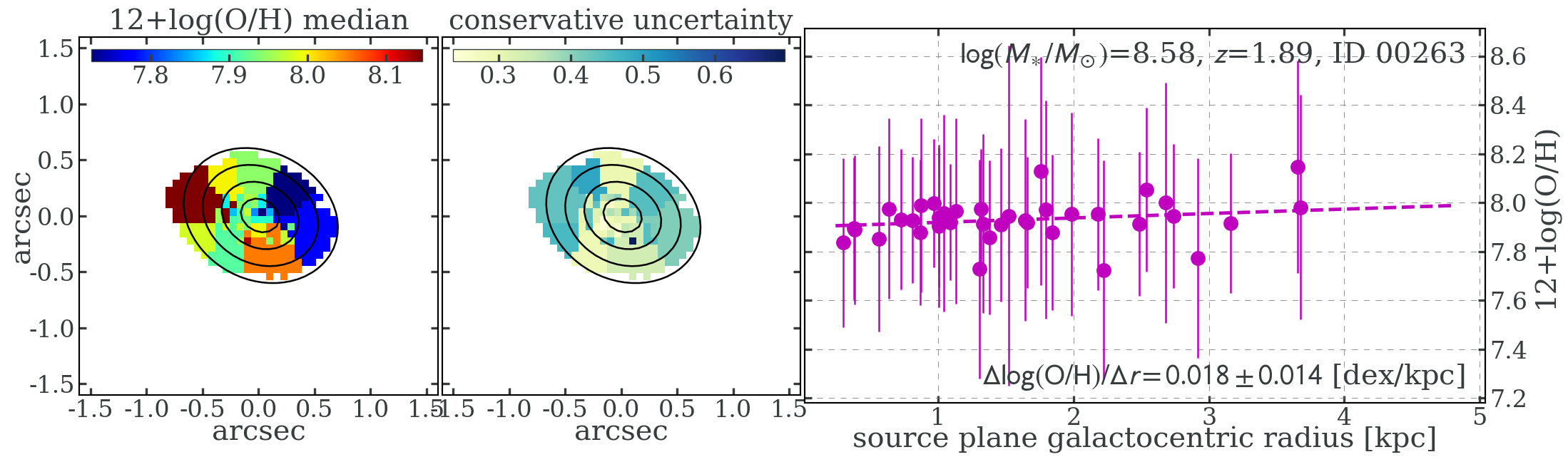}
    \caption{The source ID00263 in the field of \cler is shown.}
    \label{fig:clA2744_ID00263_figs}
\end{figure*}

\begin{figure*}
    \centering
    \includegraphics[width=\textwidth]{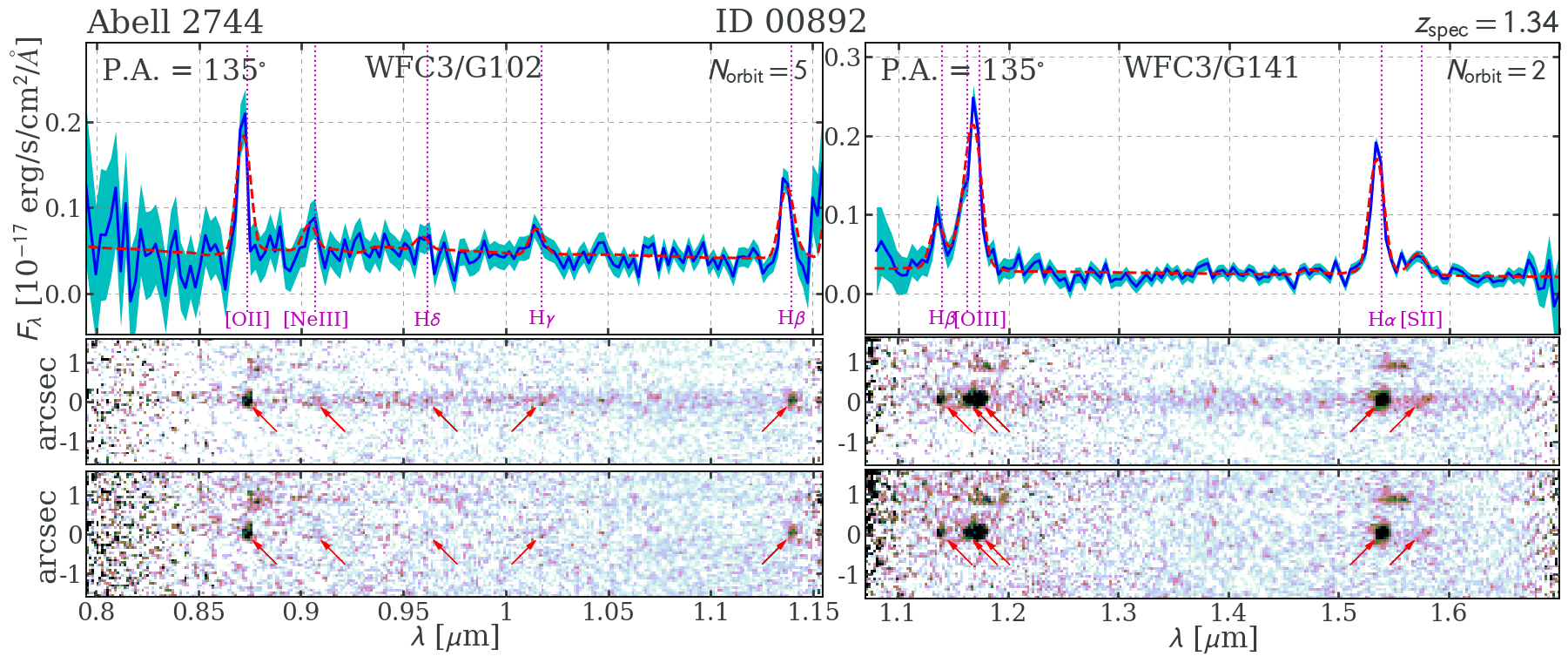}\\
    \includegraphics[width=\textwidth]{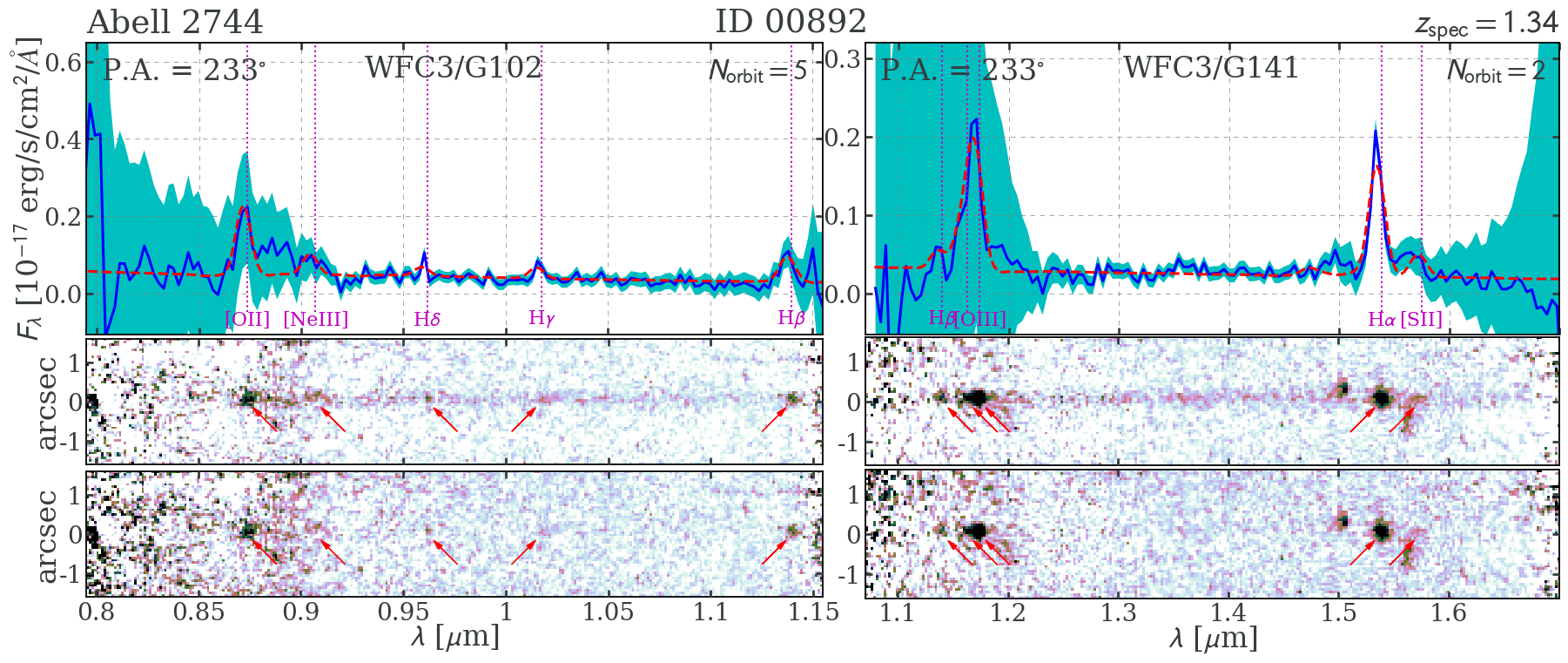}\\
    \includegraphics[width=.16\textwidth]{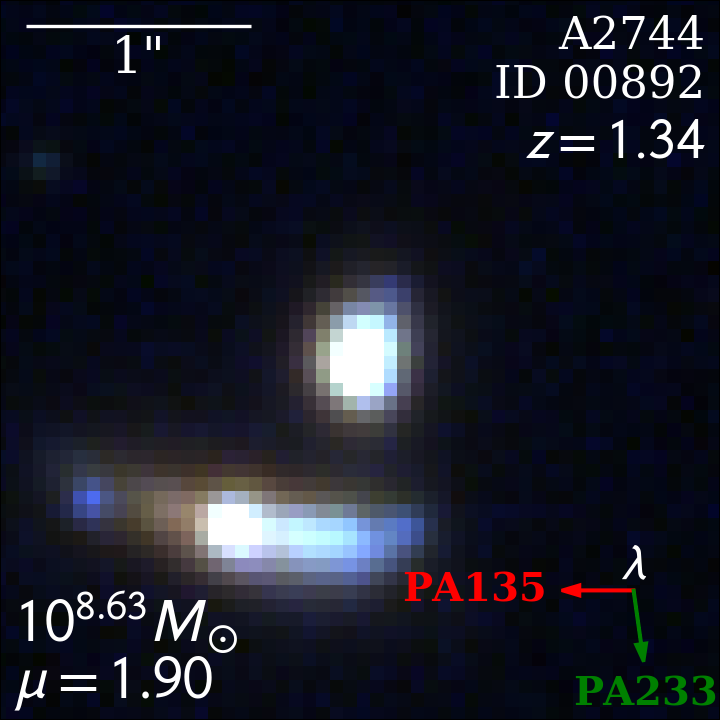}
    \includegraphics[width=.16\textwidth]{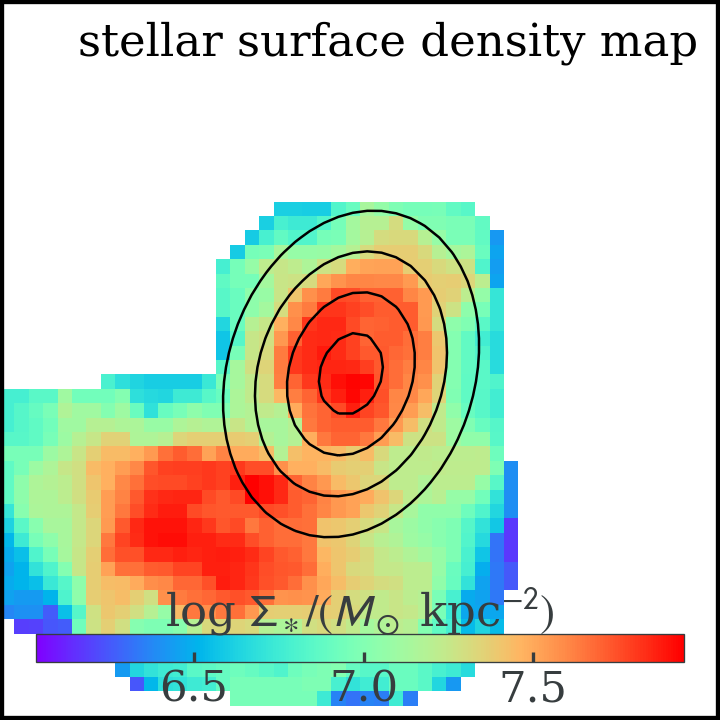}
    \includegraphics[width=.16\textwidth]{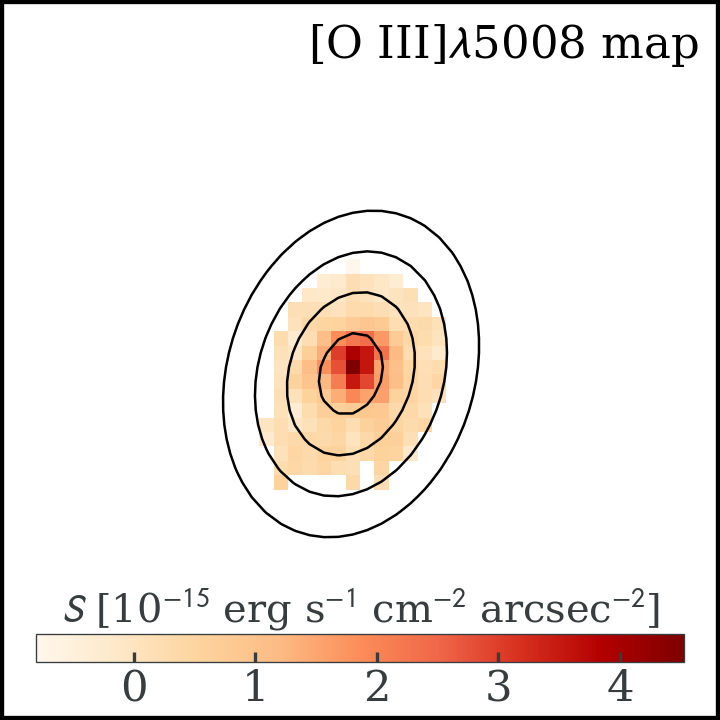}
    \includegraphics[width=.16\textwidth]{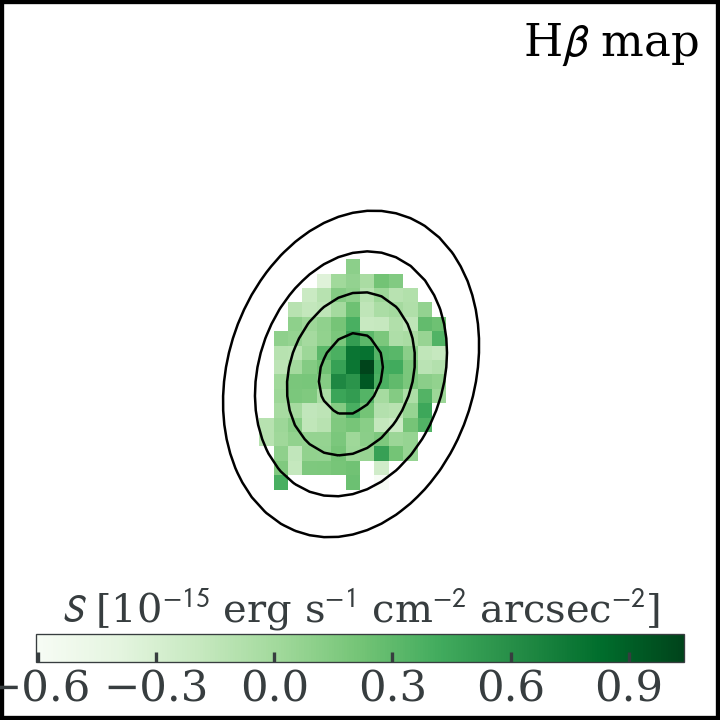}
    \includegraphics[width=.16\textwidth]{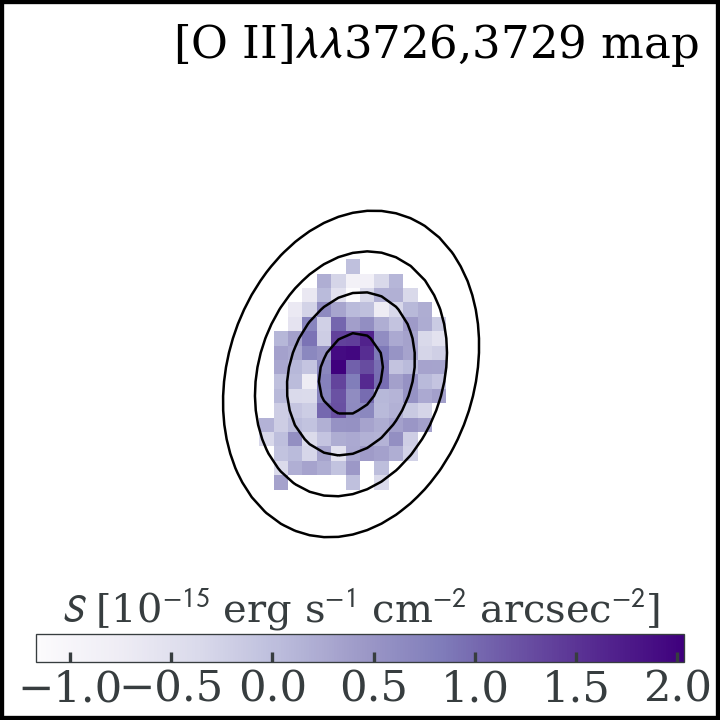}
    \includegraphics[width=.16\textwidth]{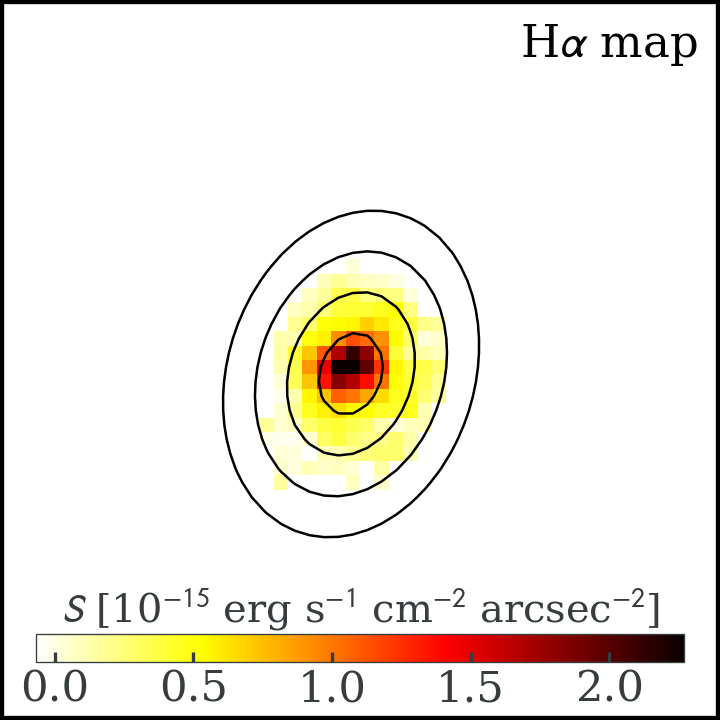}\\
    \includegraphics[width=\textwidth]{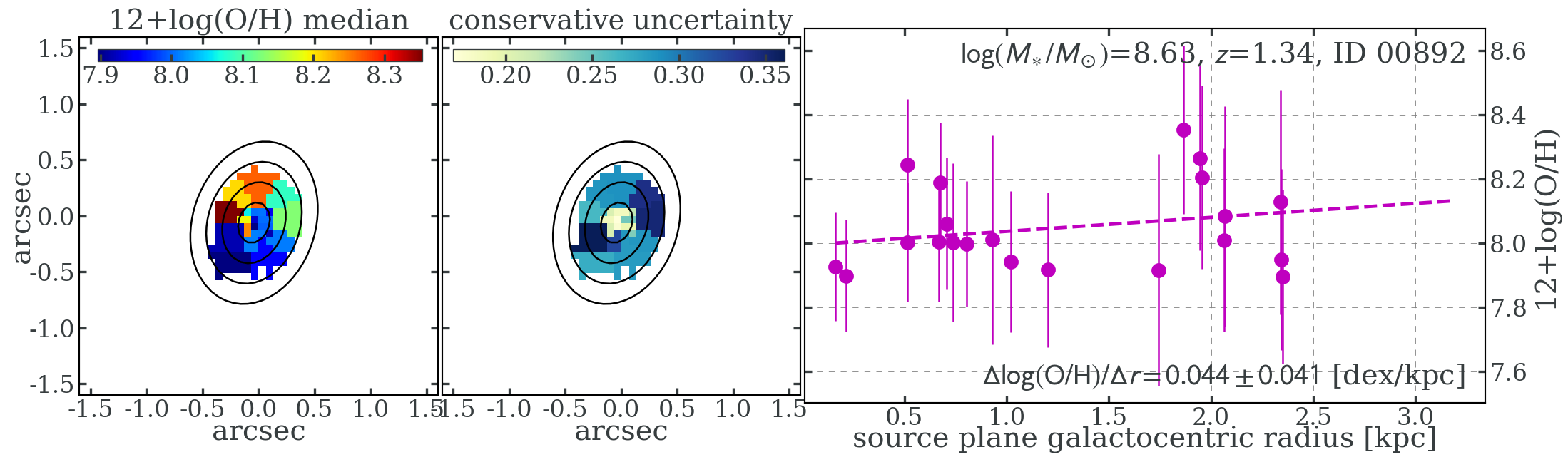}
    \caption{The source ID00892 in the field of \cler is shown.}
    \label{fig:clA2744_ID00892_figs}
\end{figure*}

\begin{figure*}
    \centering
    \includegraphics[width=\textwidth]{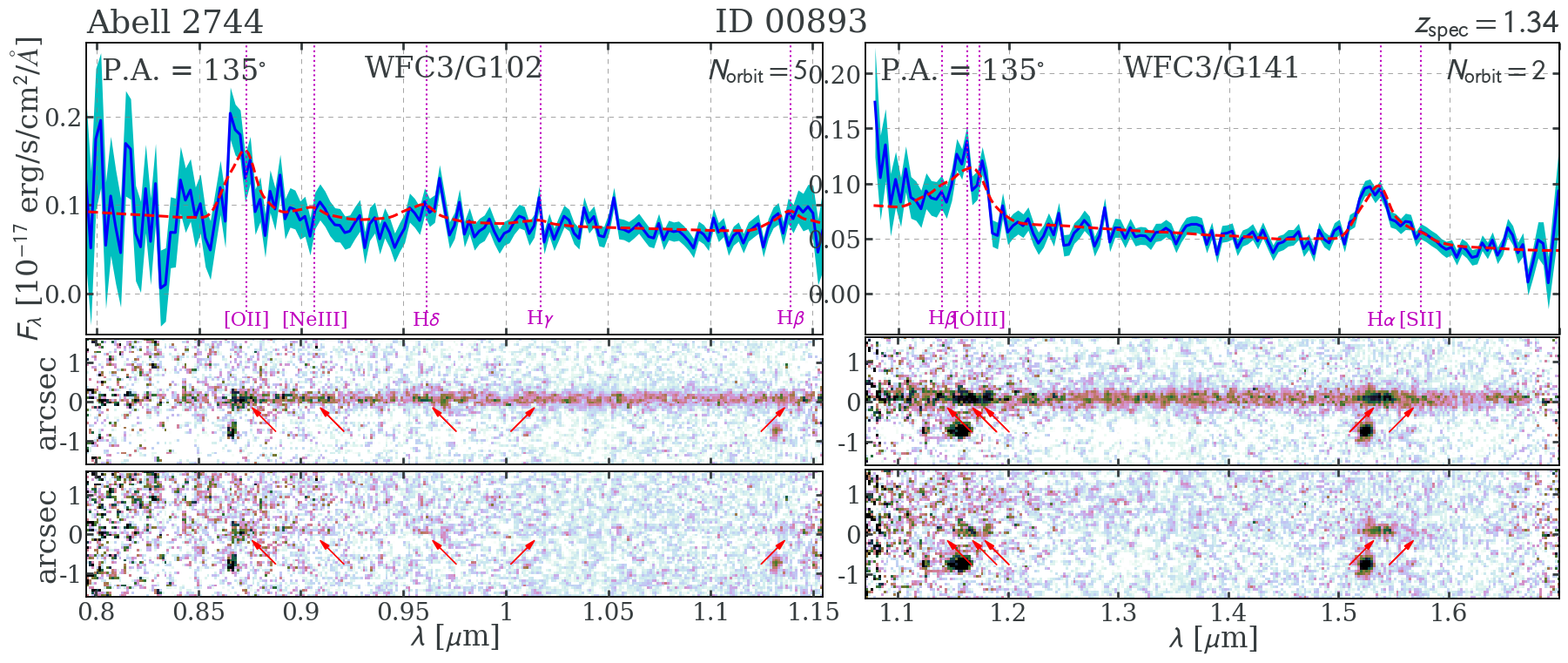}\\
    \includegraphics[width=\textwidth]{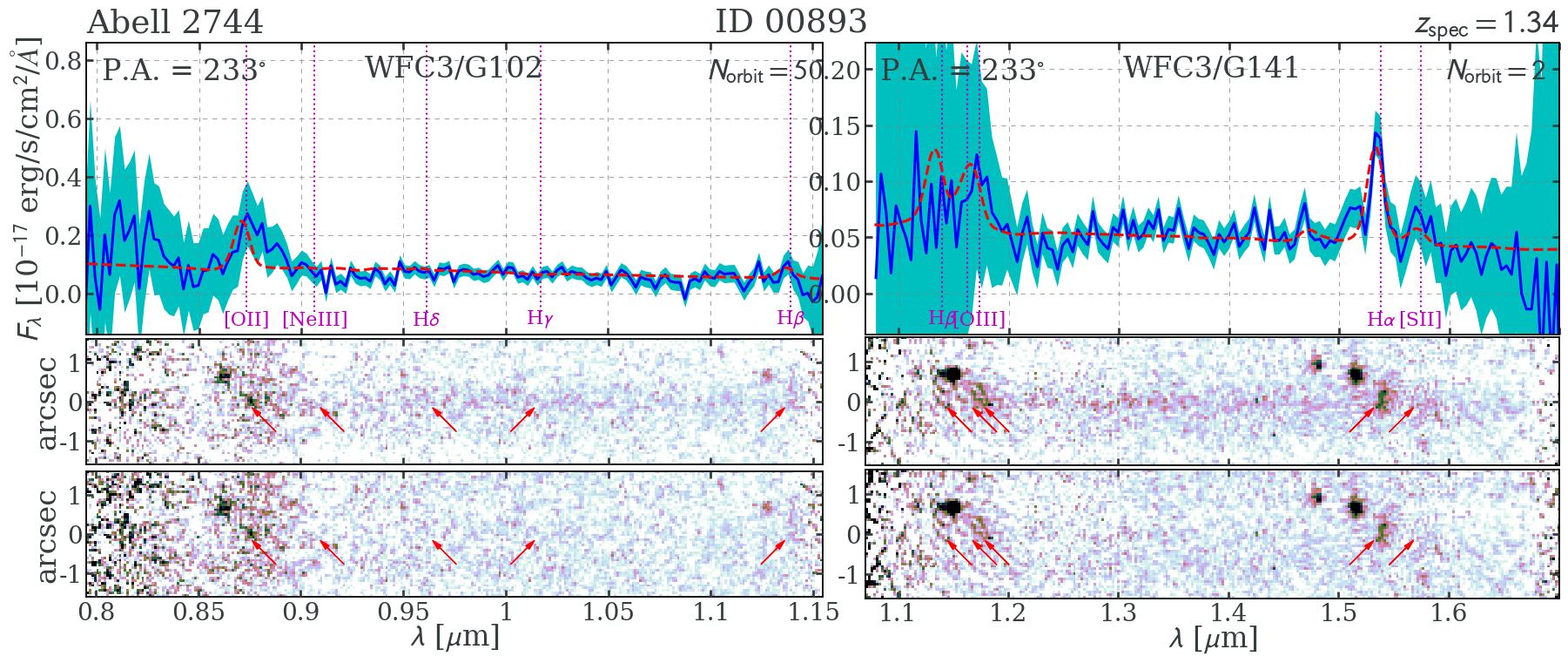}\\
    \includegraphics[width=.16\textwidth]{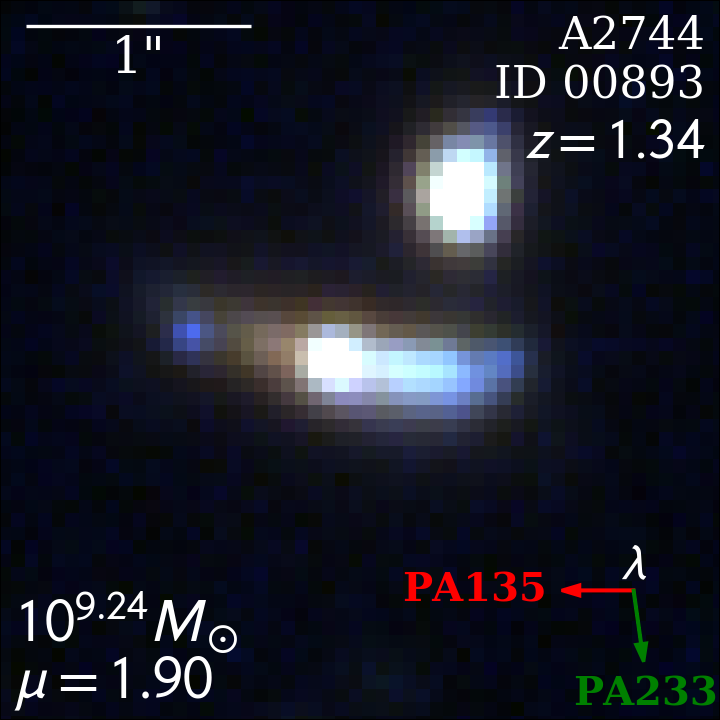}
    \includegraphics[width=.16\textwidth]{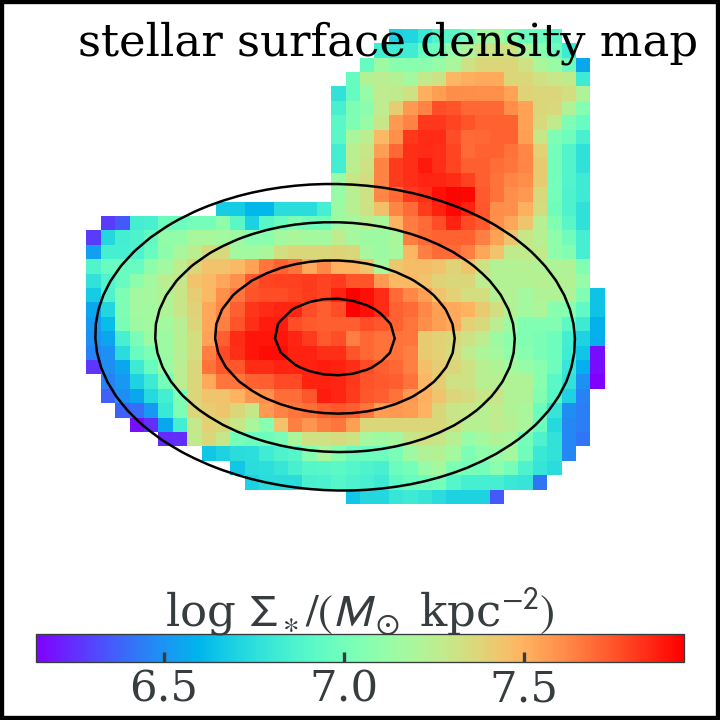}
    \includegraphics[width=.16\textwidth]{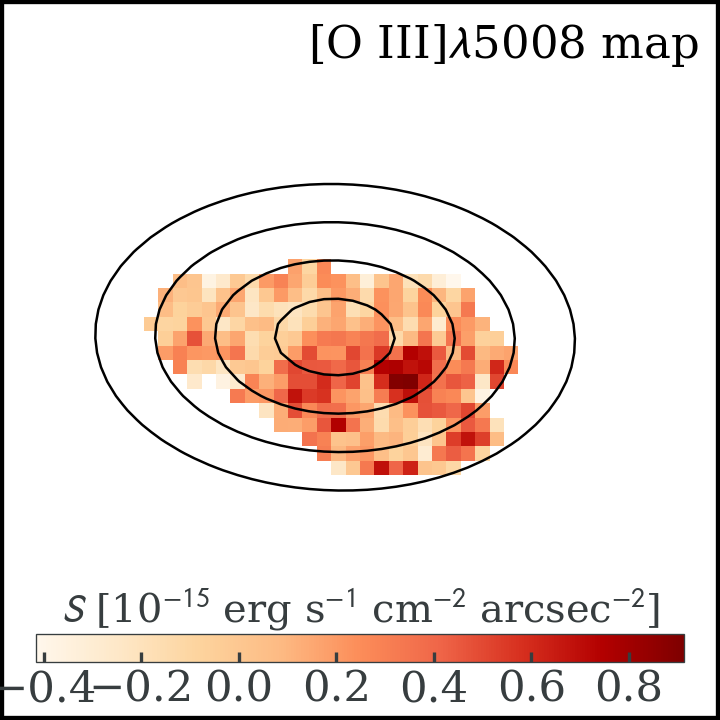}
    \includegraphics[width=.16\textwidth]{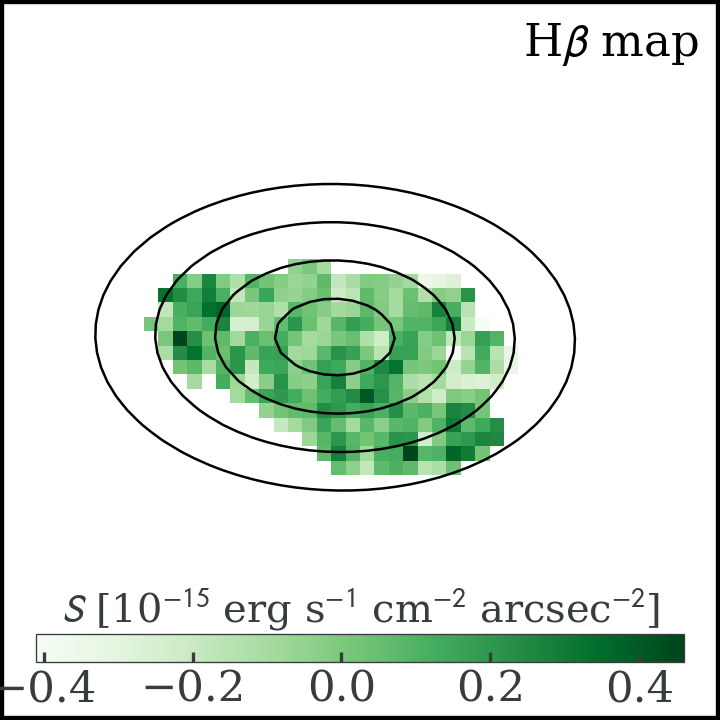}
    \includegraphics[width=.16\textwidth]{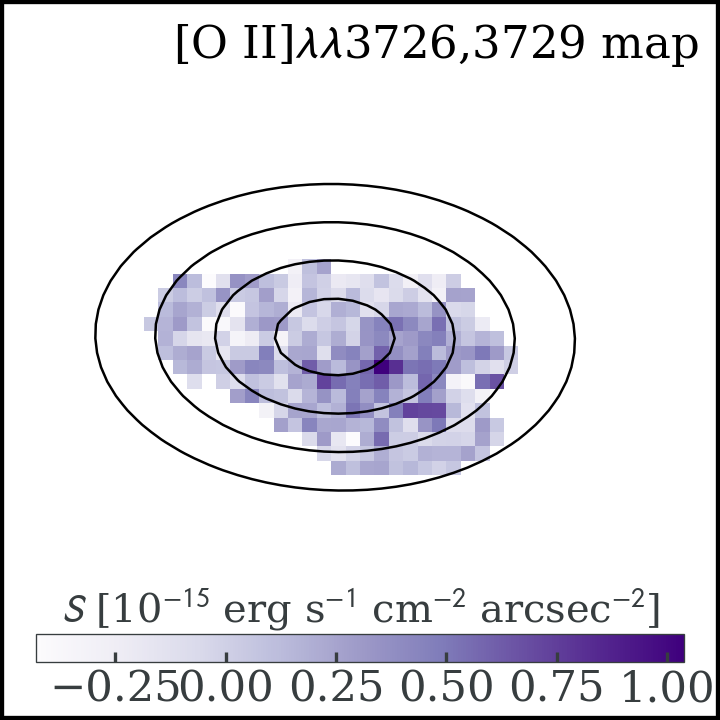}
    \includegraphics[width=.16\textwidth]{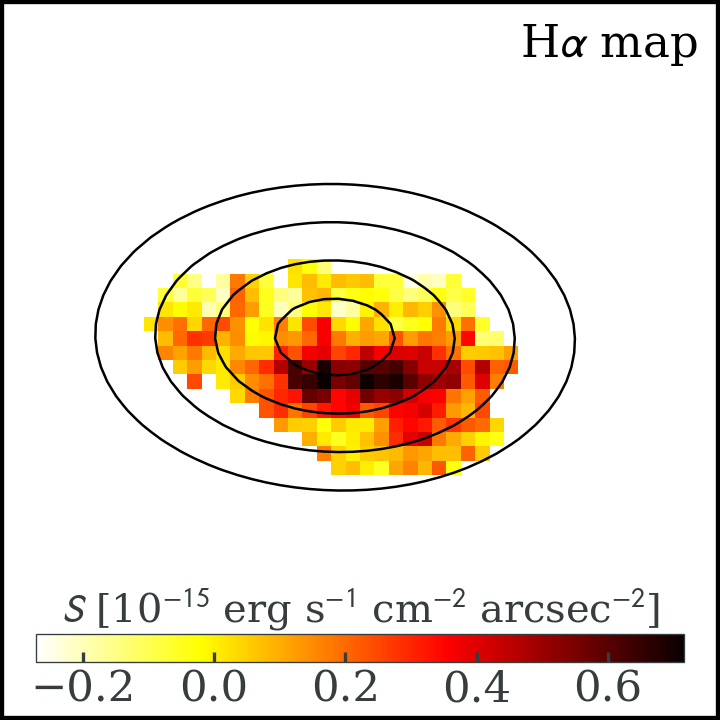}\\
    \includegraphics[width=\textwidth]{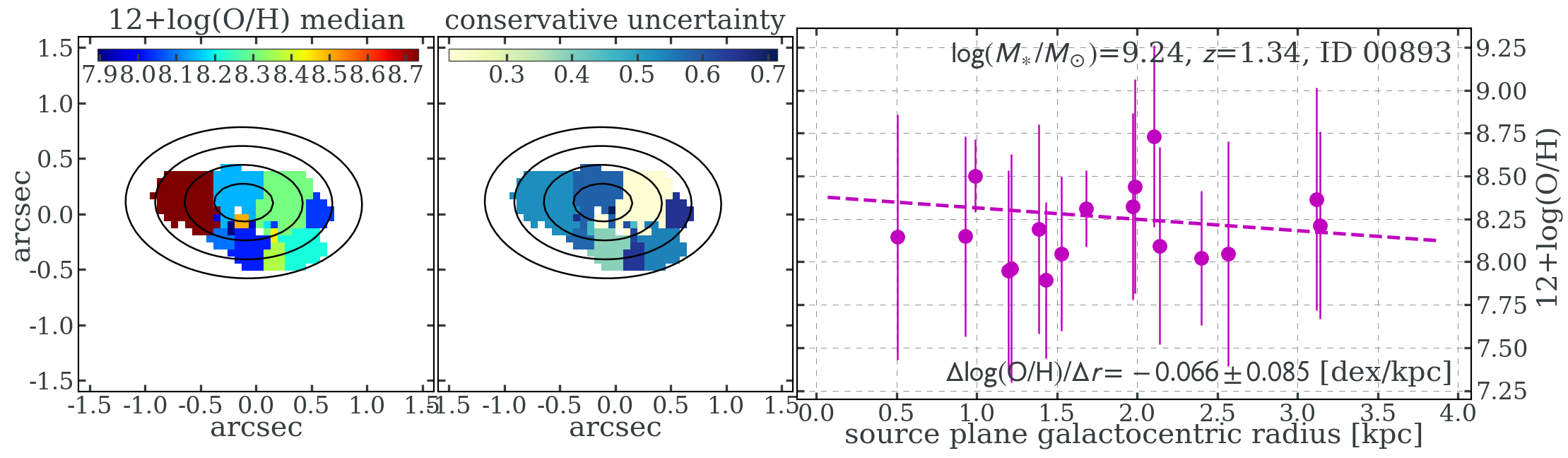}
    \caption{The source ID00893 in the field of \cler is shown.}
    \label{fig:clA2744_ID00893_figs}
\end{figure*}

\begin{figure*}
    \centering
    \includegraphics[width=\textwidth]{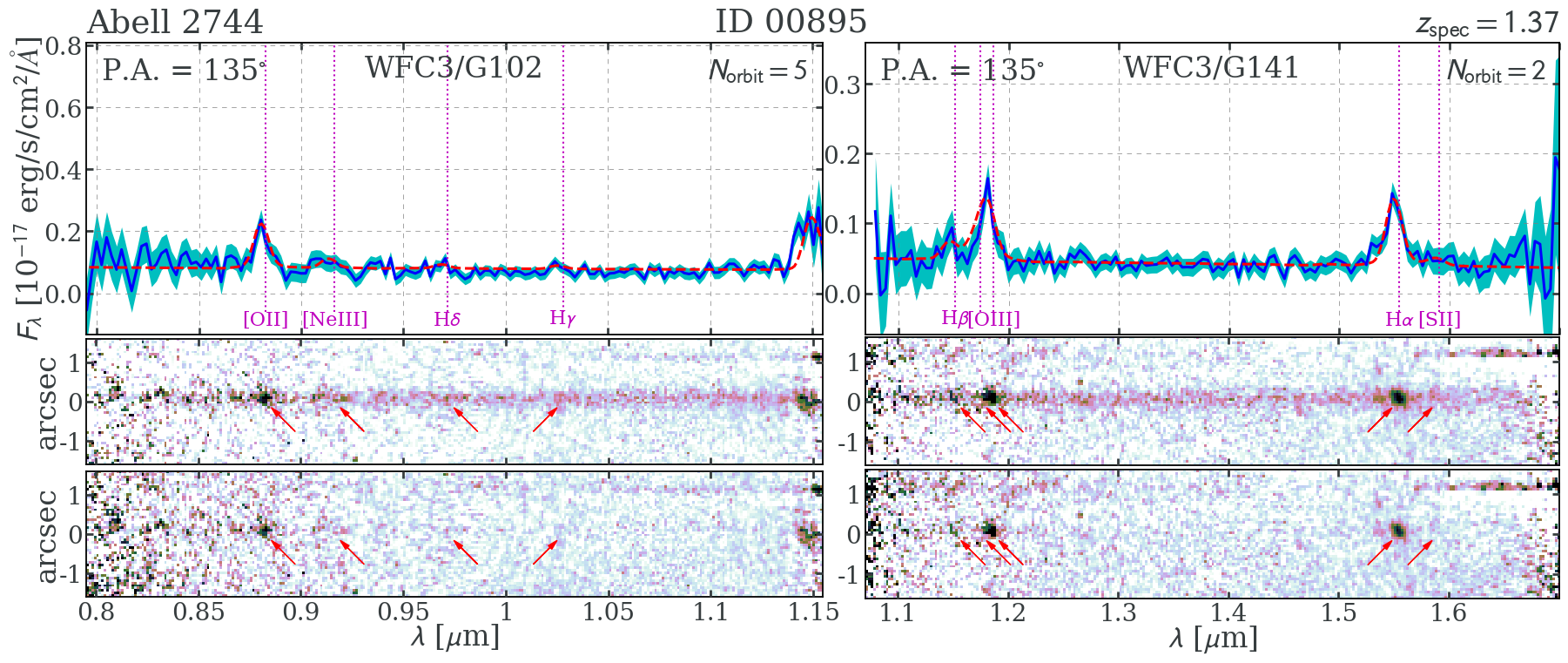}\\
    \includegraphics[width=\textwidth]{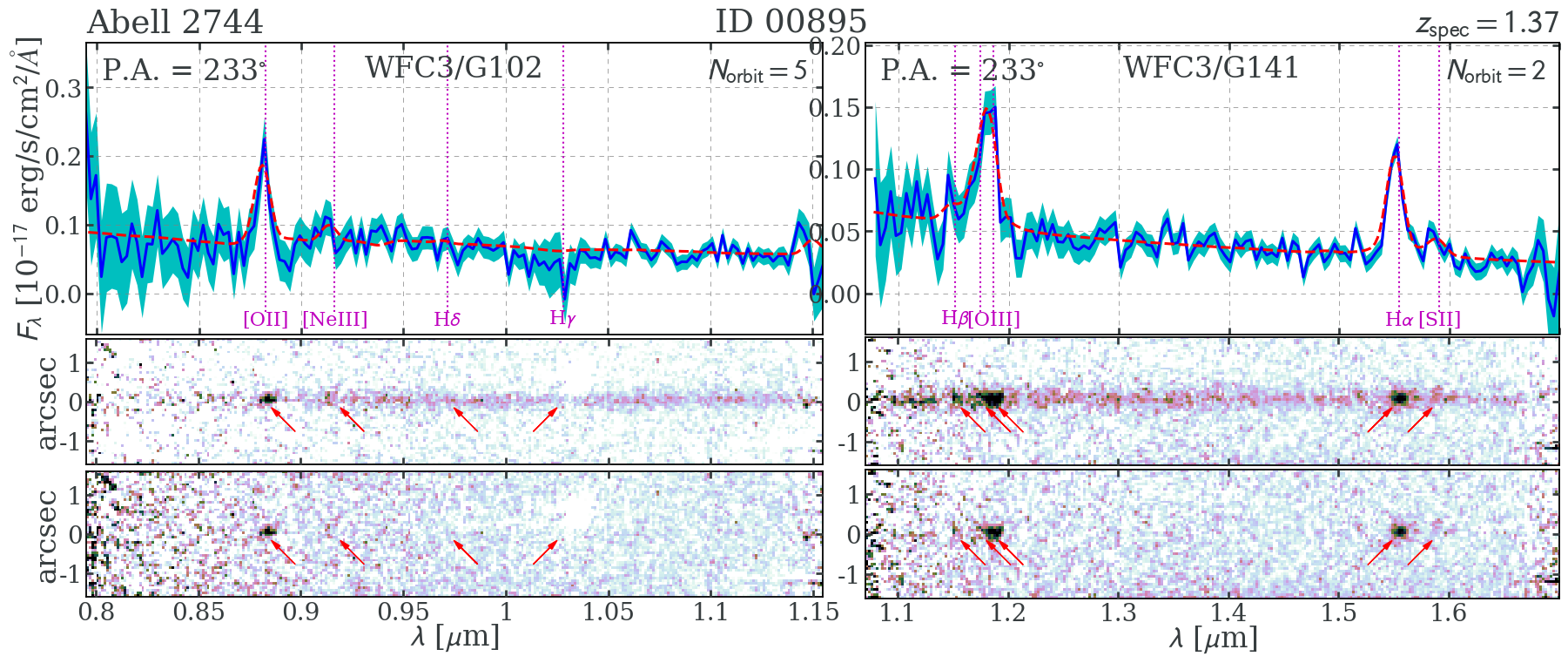}\\
    \includegraphics[width=.16\textwidth]{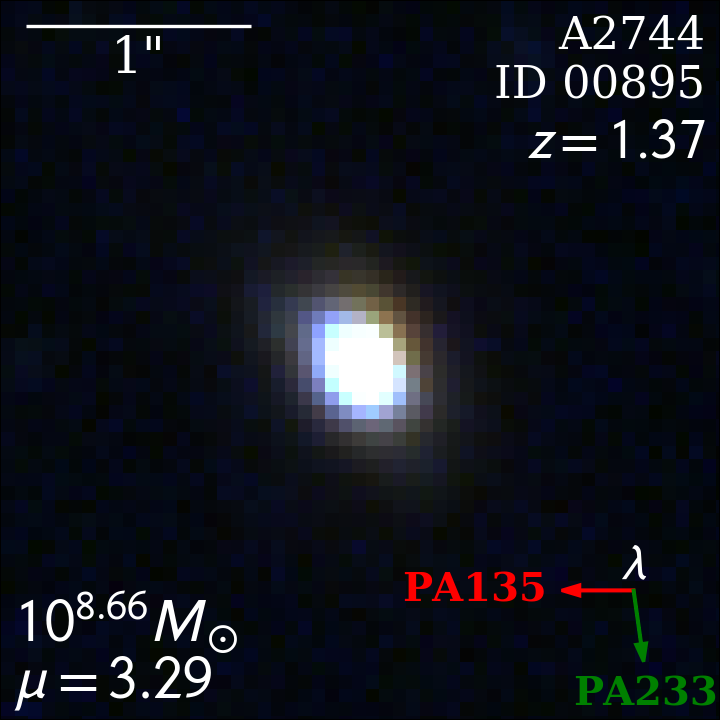}
    \includegraphics[width=.16\textwidth]{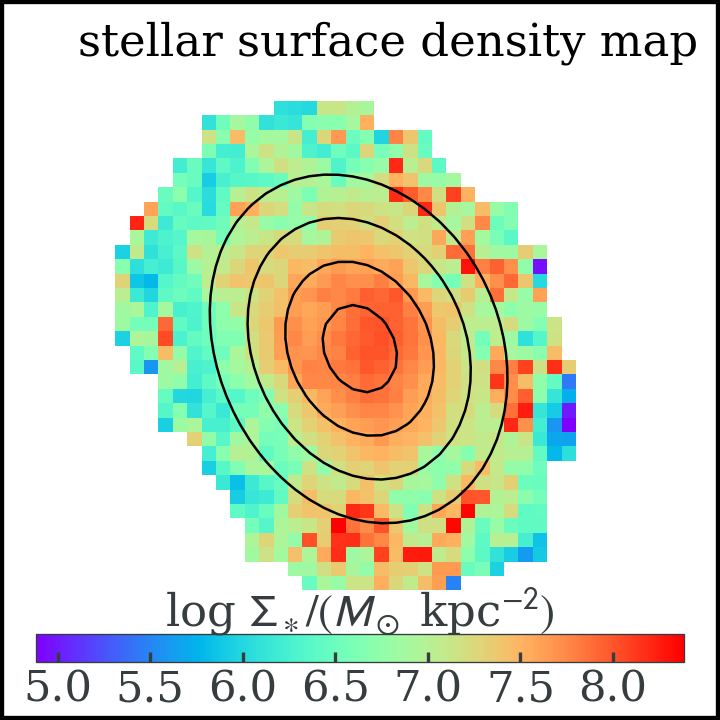}
    \includegraphics[width=.16\textwidth]{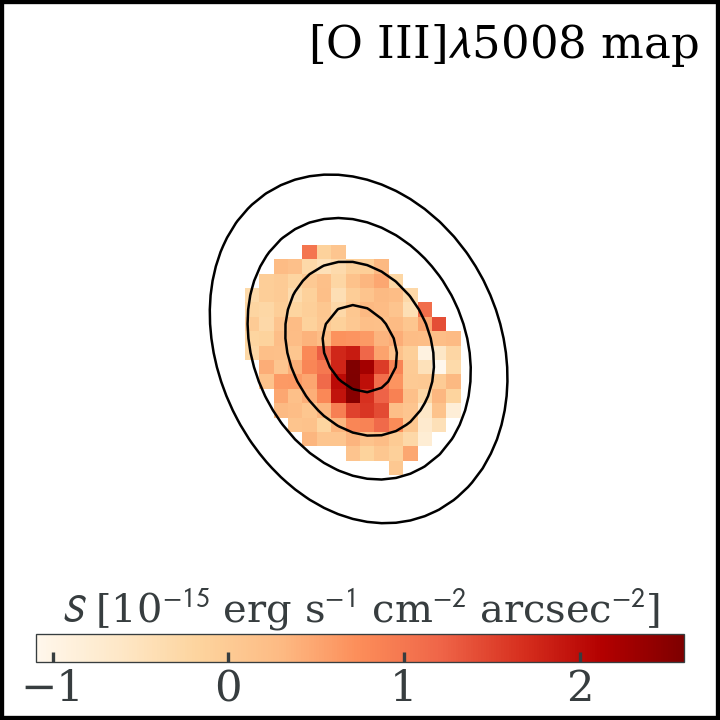}
    \includegraphics[width=.16\textwidth]{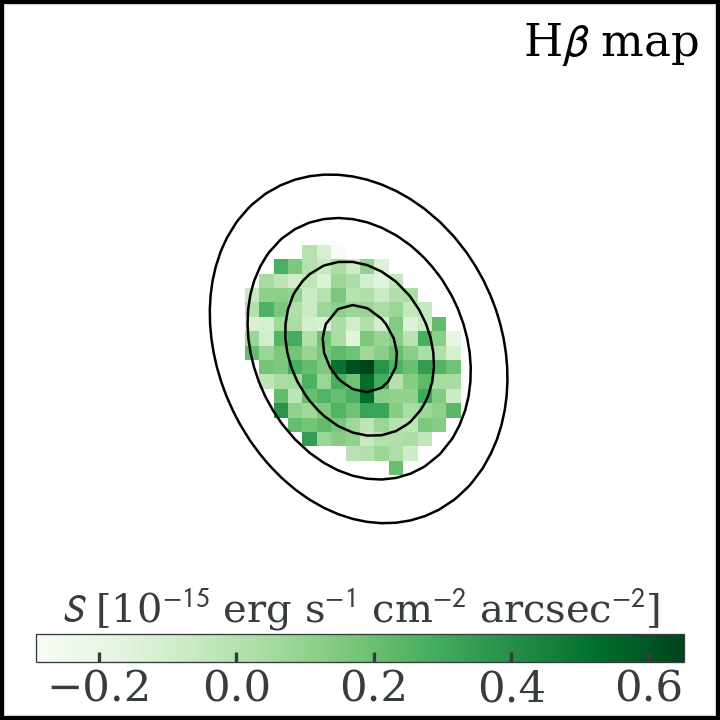}
    \includegraphics[width=.16\textwidth]{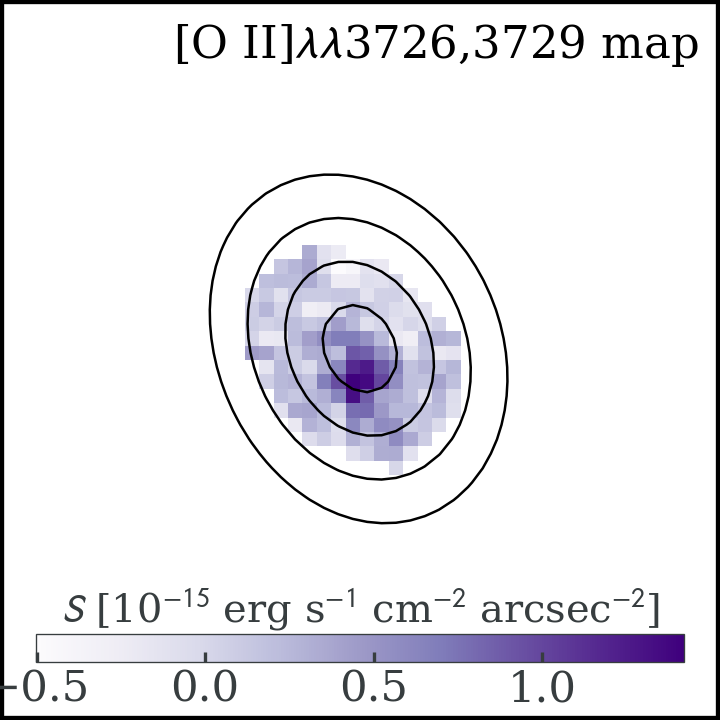}
    \includegraphics[width=.16\textwidth]{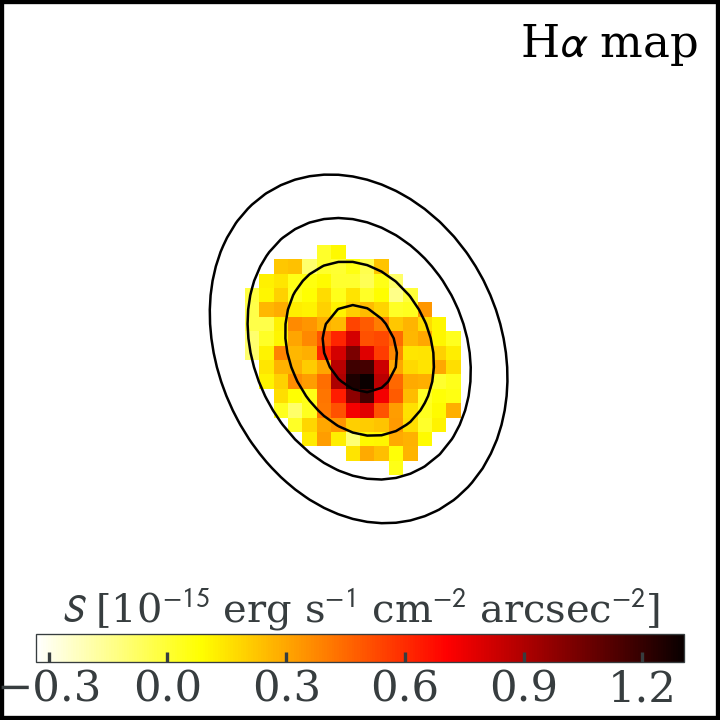}\\
    \includegraphics[width=\textwidth]{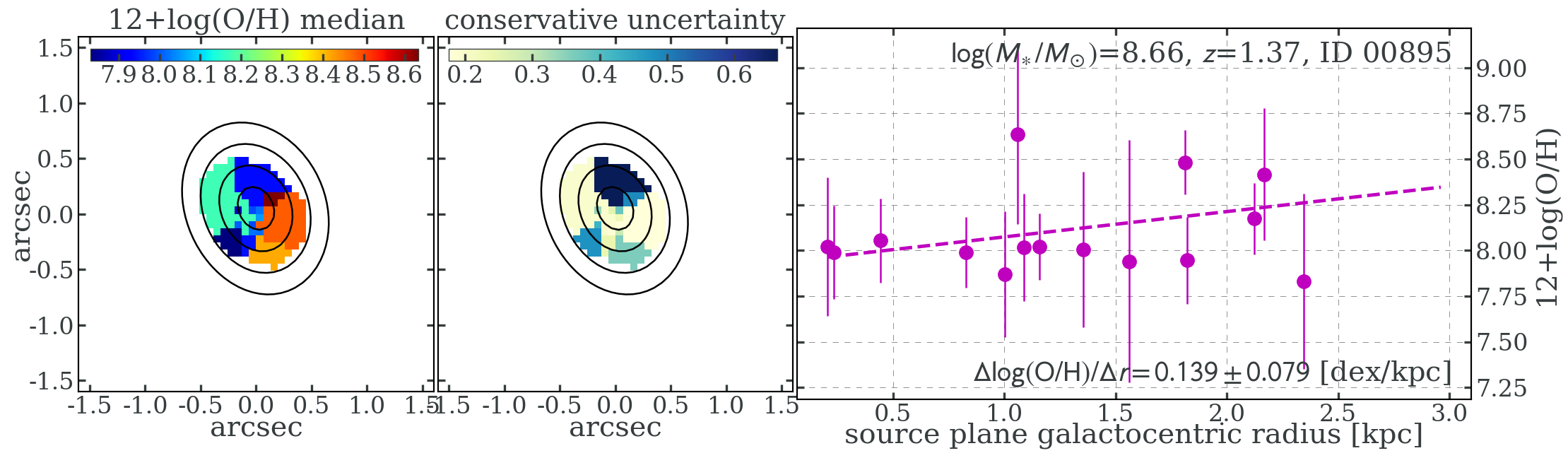}
    \caption{The source ID00895 in the field of \cler is shown.}
    \label{fig:clA2744_ID00895_figs}
\end{figure*}

\begin{figure*}
    \centering
    \includegraphics[width=\textwidth]{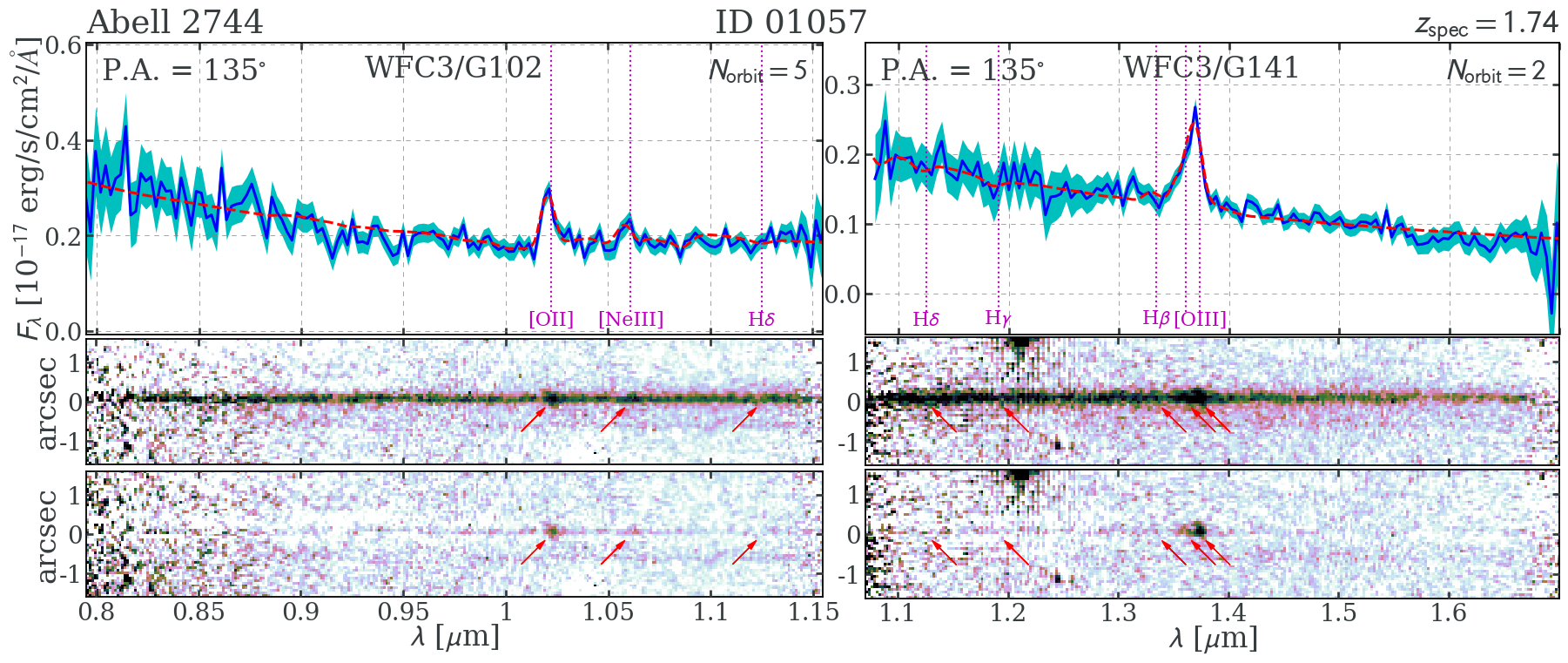}\\
    \includegraphics[width=\textwidth]{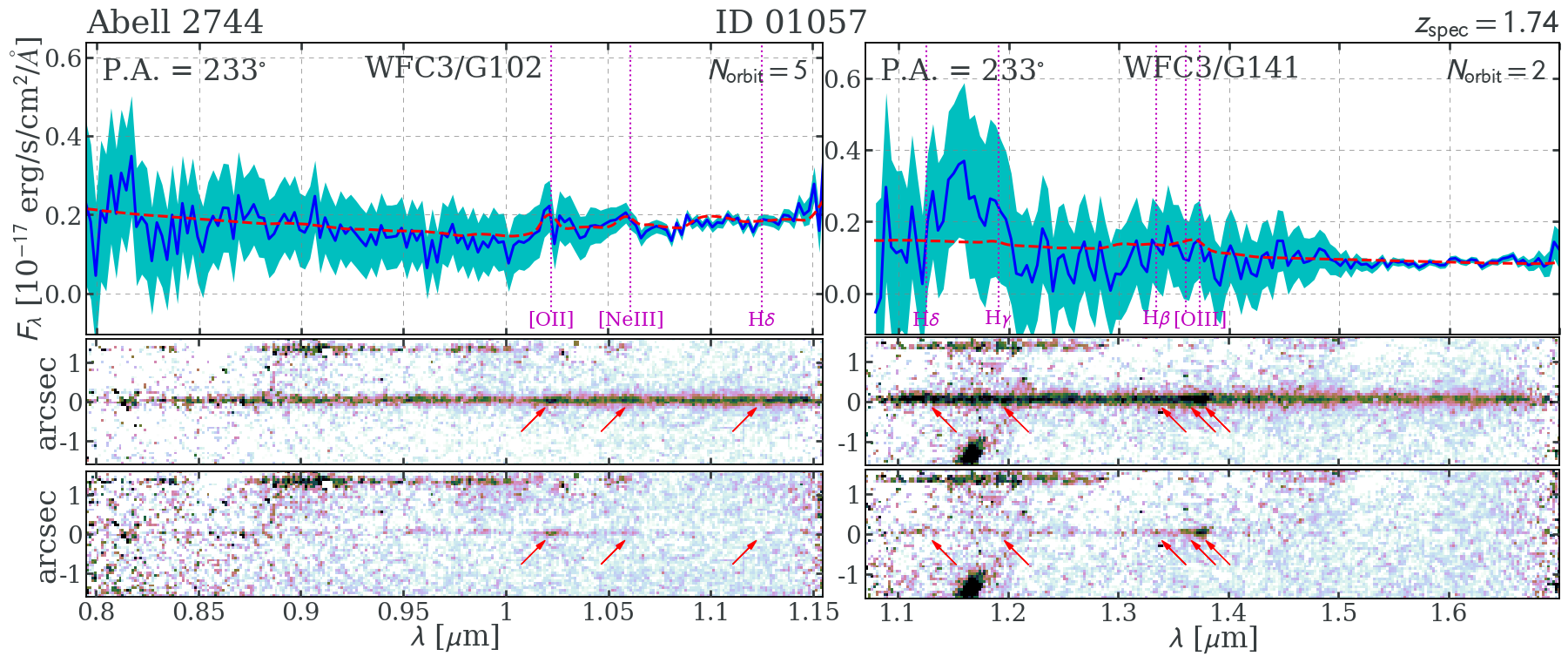}\\
    \includegraphics[width=.16\textwidth]{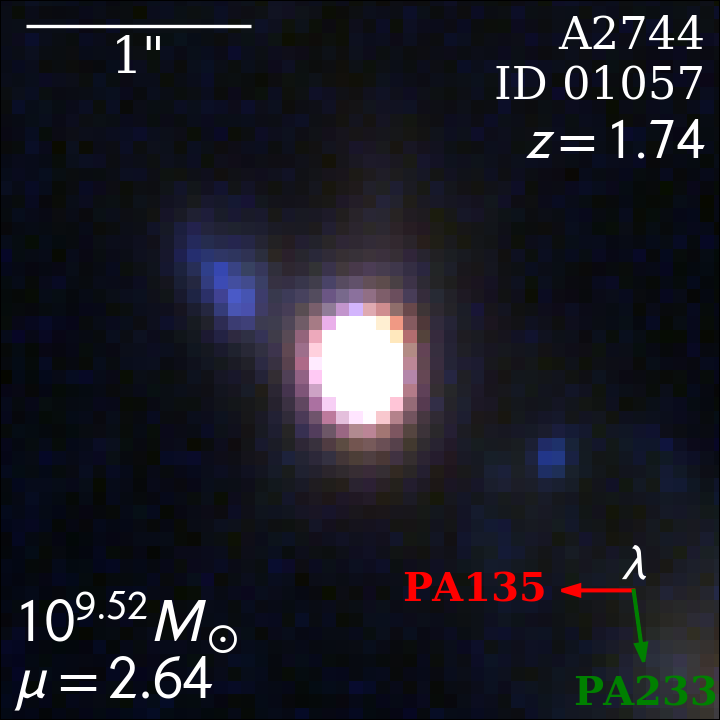}
    \includegraphics[width=.16\textwidth]{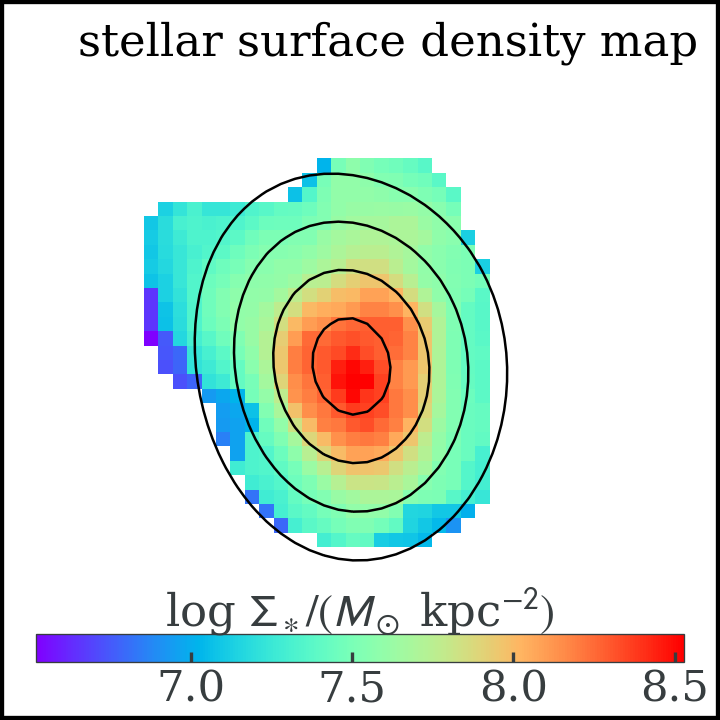}
    \includegraphics[width=.16\textwidth]{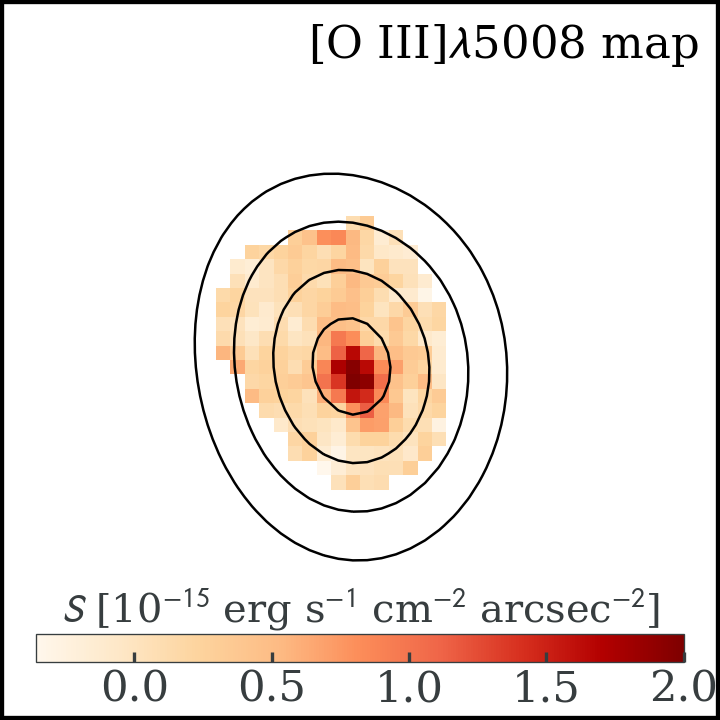}
    \includegraphics[width=.16\textwidth]{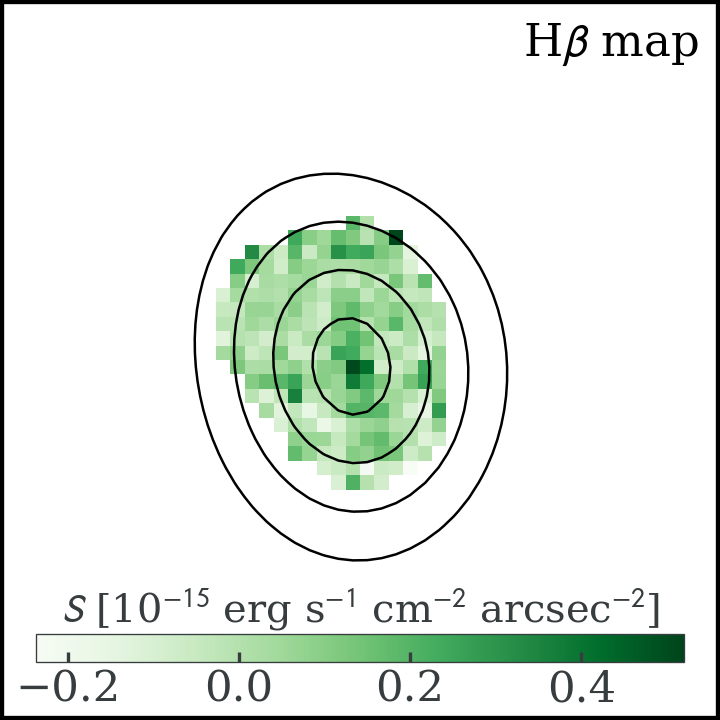}
    \includegraphics[width=.16\textwidth]{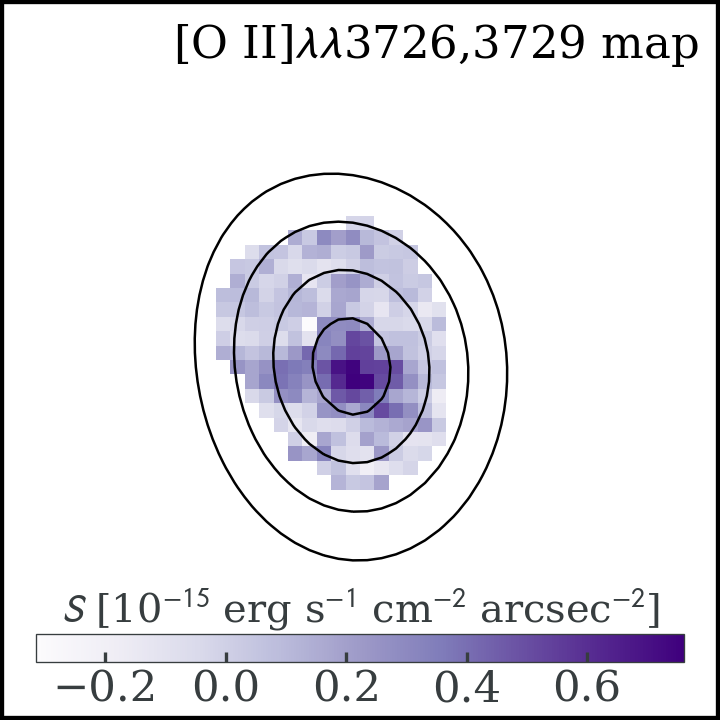}
    \includegraphics[width=.16\textwidth]{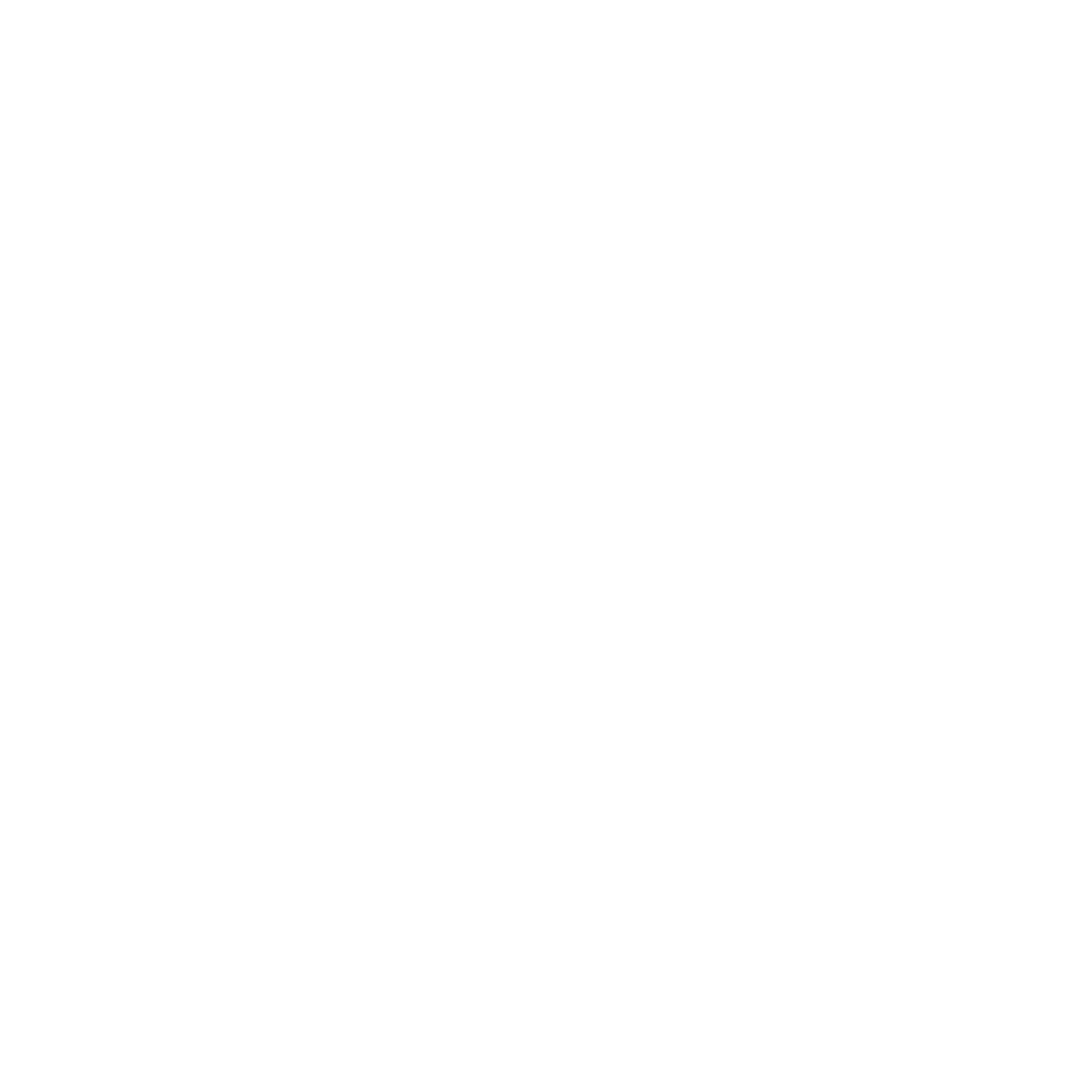}\\
    \includegraphics[width=\textwidth]{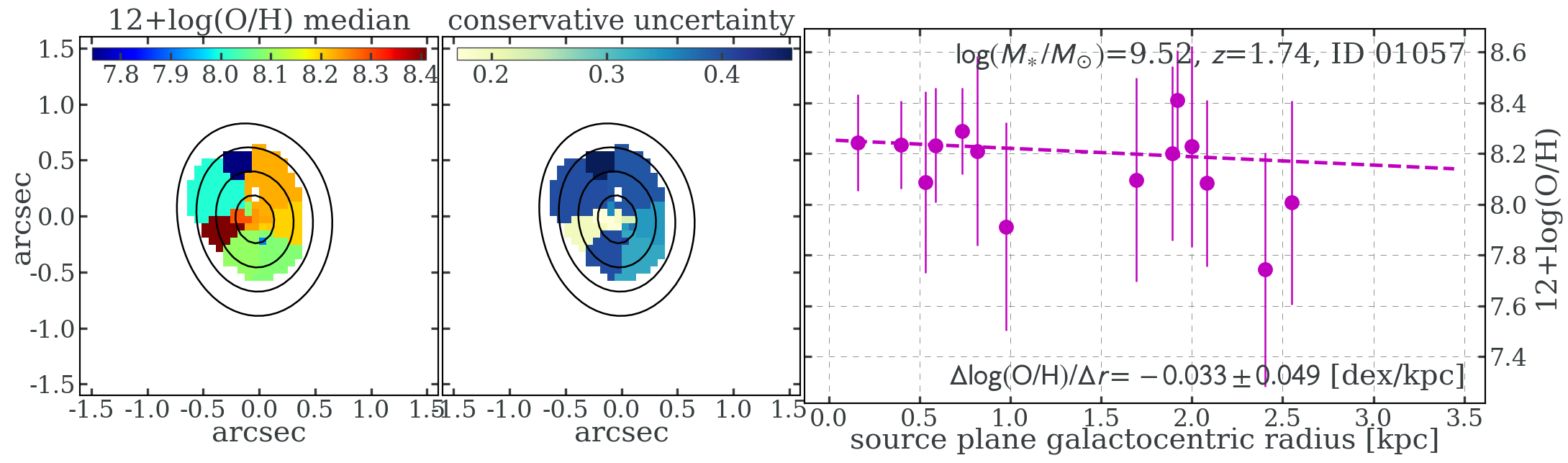}
    \caption{The source ID01057 in the field of \cler is shown.}
    \label{fig:clA2744_ID01057_figs}
\end{figure*}

\begin{figure*}
    \centering
    \includegraphics[width=\textwidth]{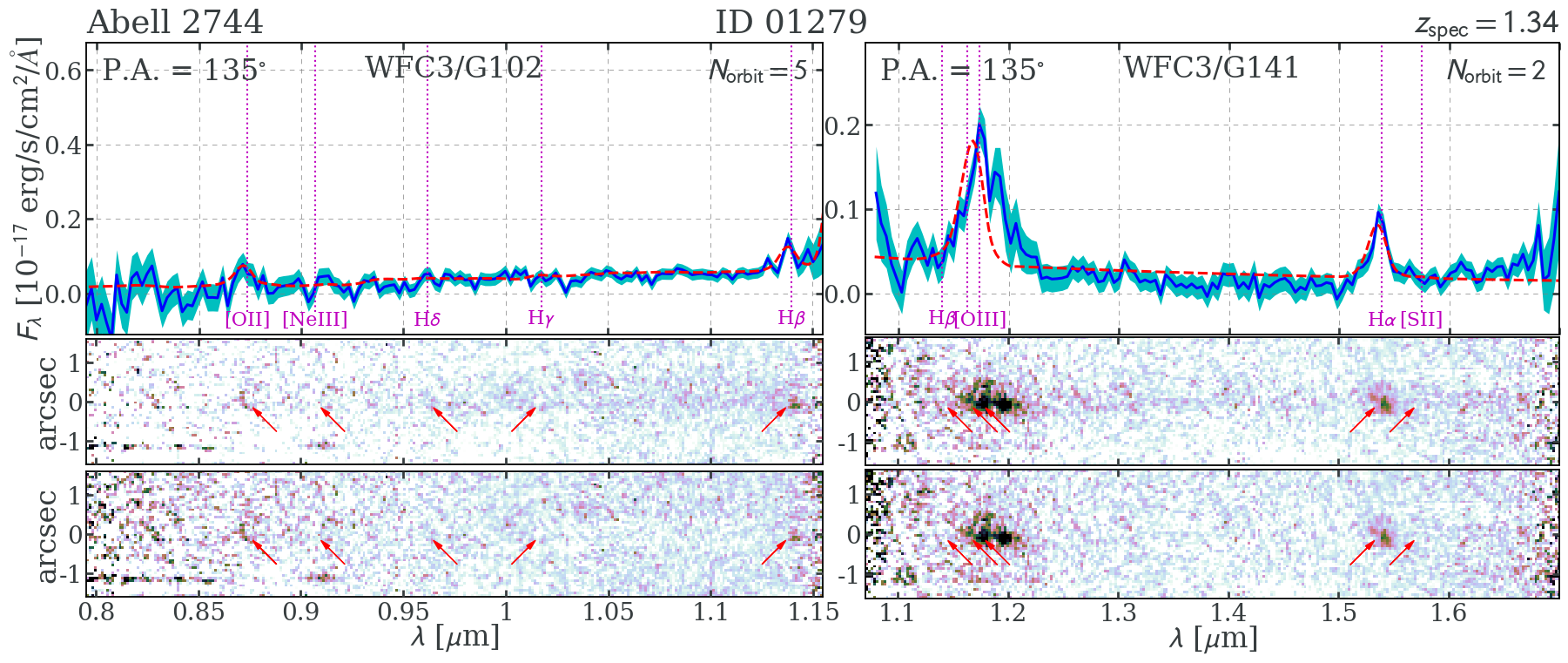}\\
    \includegraphics[width=\textwidth]{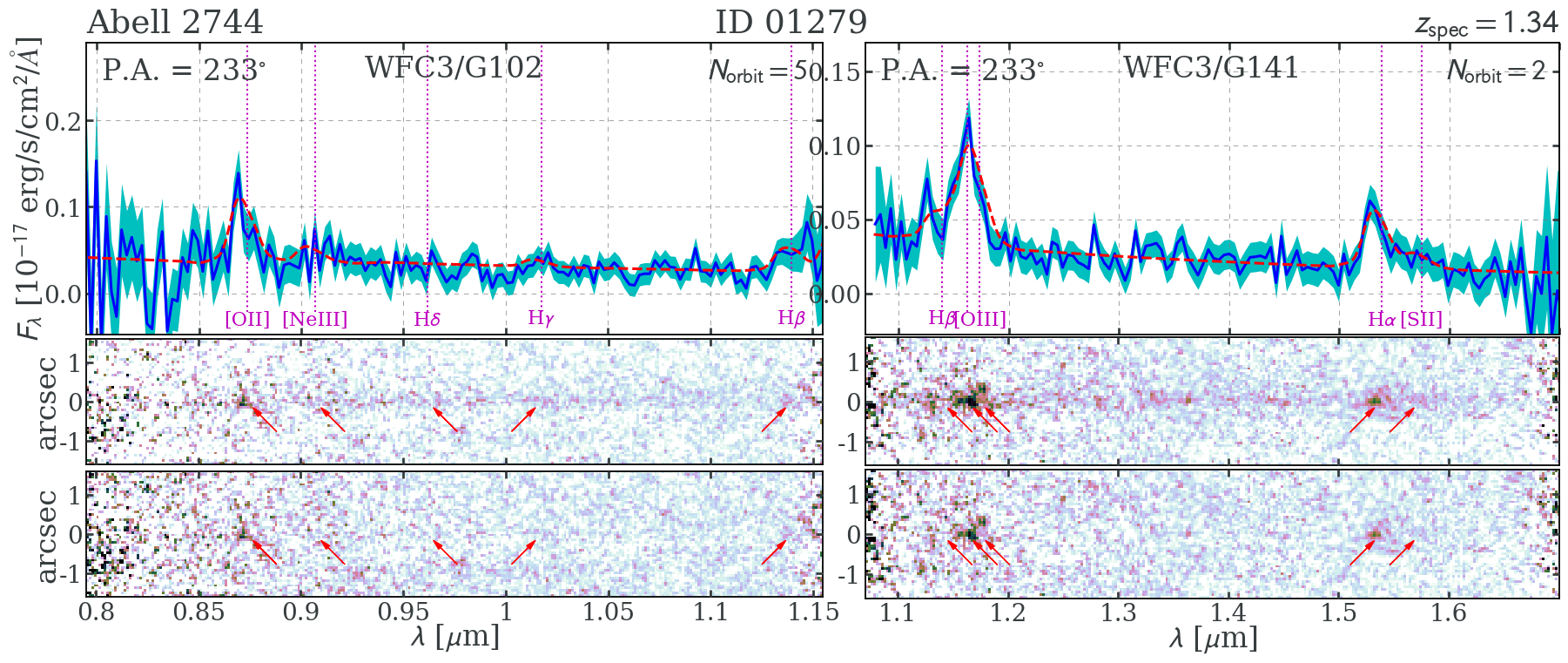}\\
    \includegraphics[width=.16\textwidth]{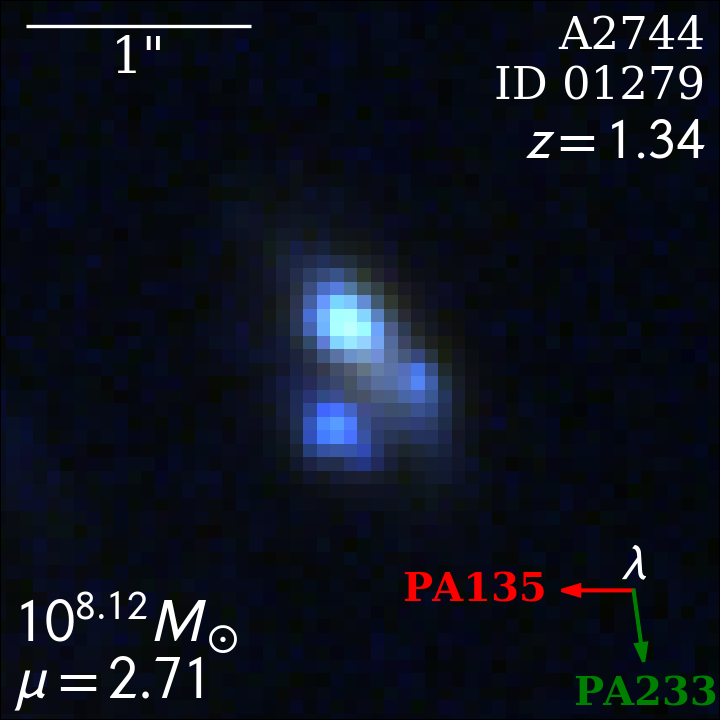}
    \includegraphics[width=.16\textwidth]{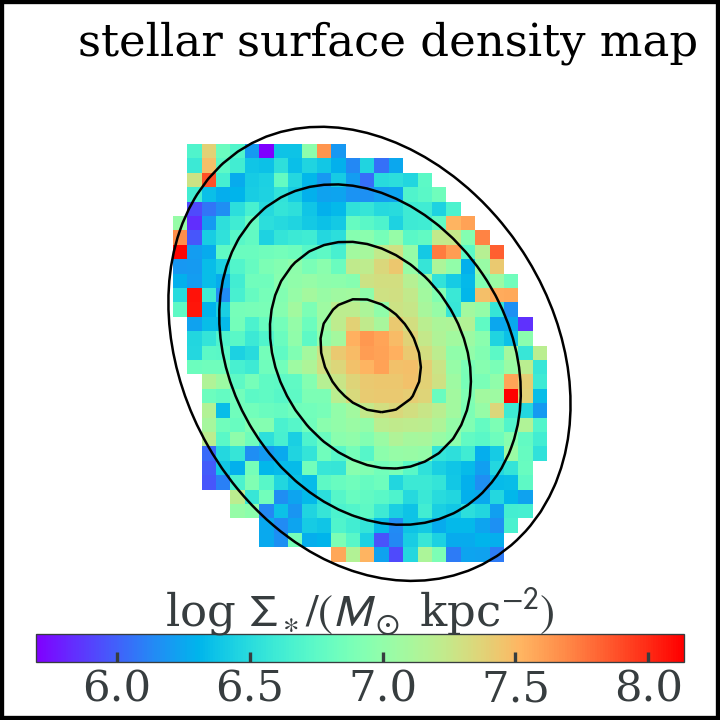}
    \includegraphics[width=.16\textwidth]{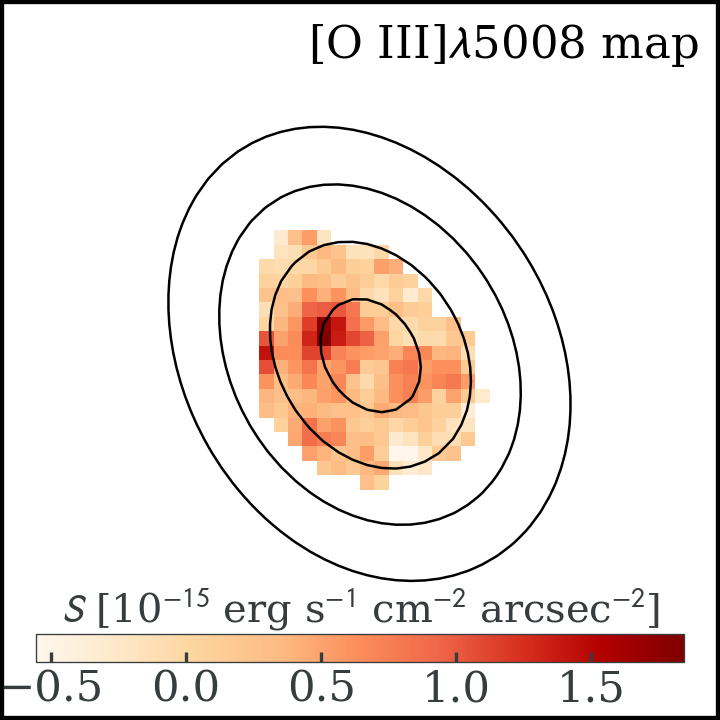}
    \includegraphics[width=.16\textwidth]{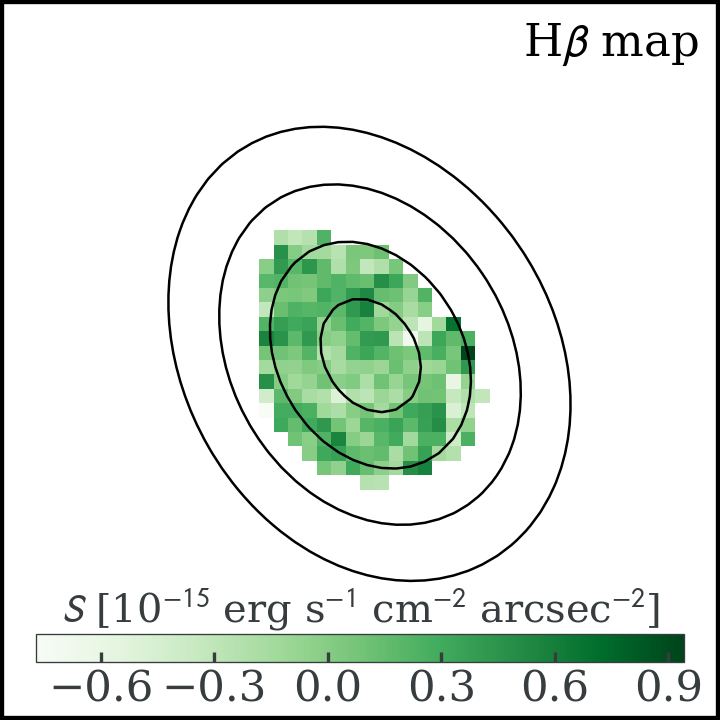}
    \includegraphics[width=.16\textwidth]{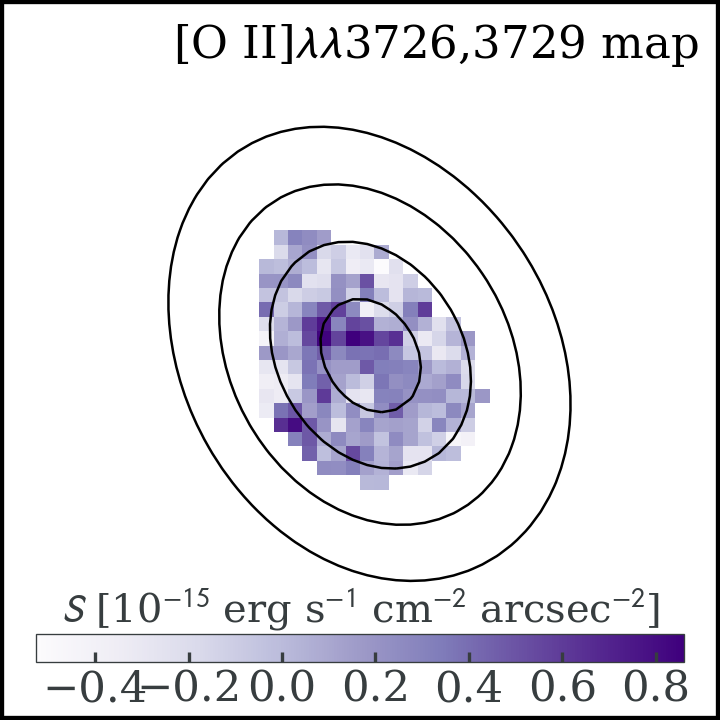}
    \includegraphics[width=.16\textwidth]{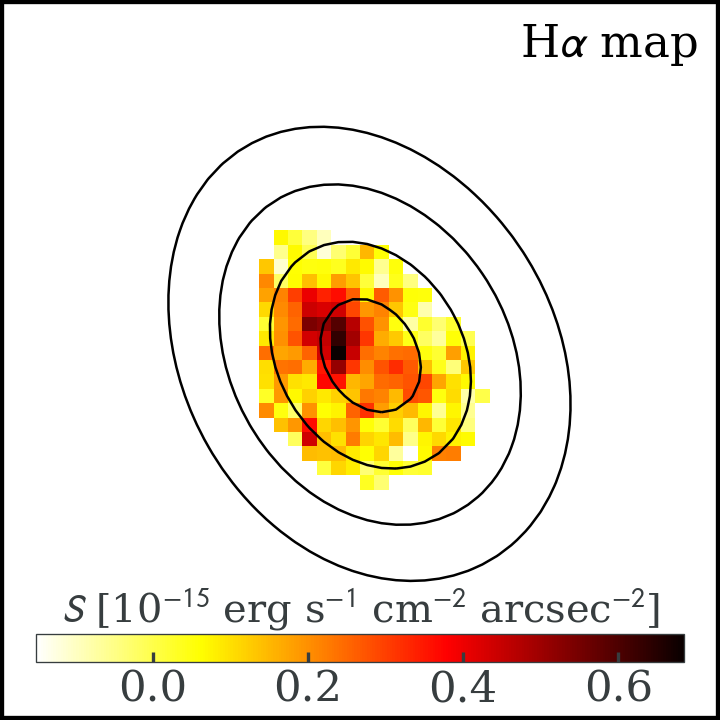}\\
    \includegraphics[width=\textwidth]{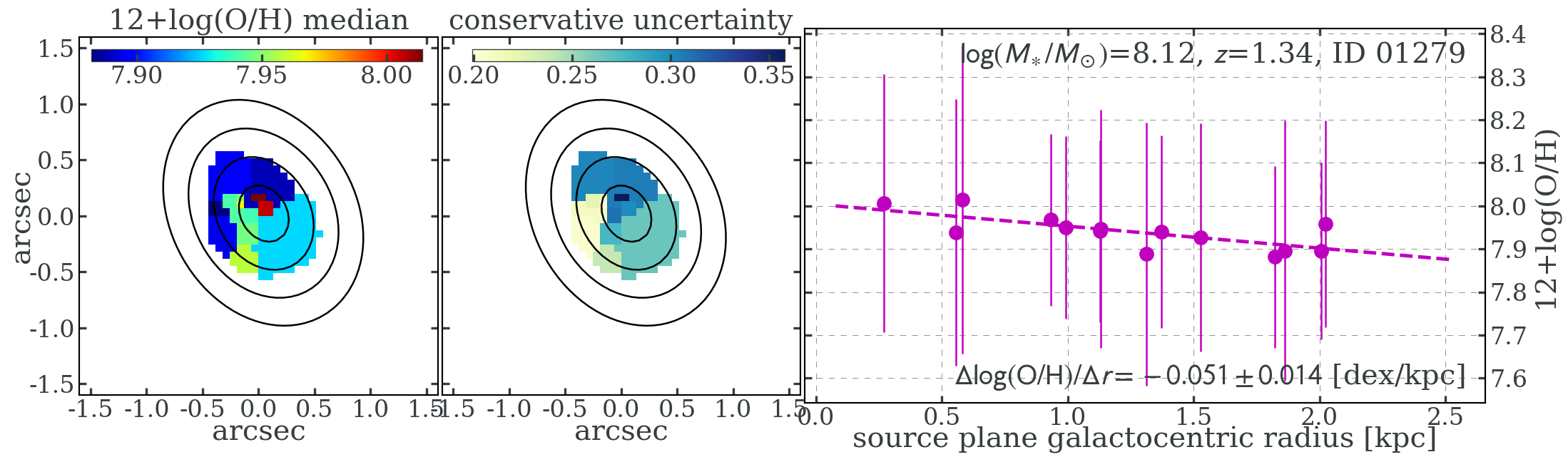}
    \caption{The source ID01279 in the field of \cler is shown.}
    \label{fig:clA2744_ID01279_figs}
\end{figure*}

\begin{figure*}
    \centering
    \includegraphics[width=\textwidth]{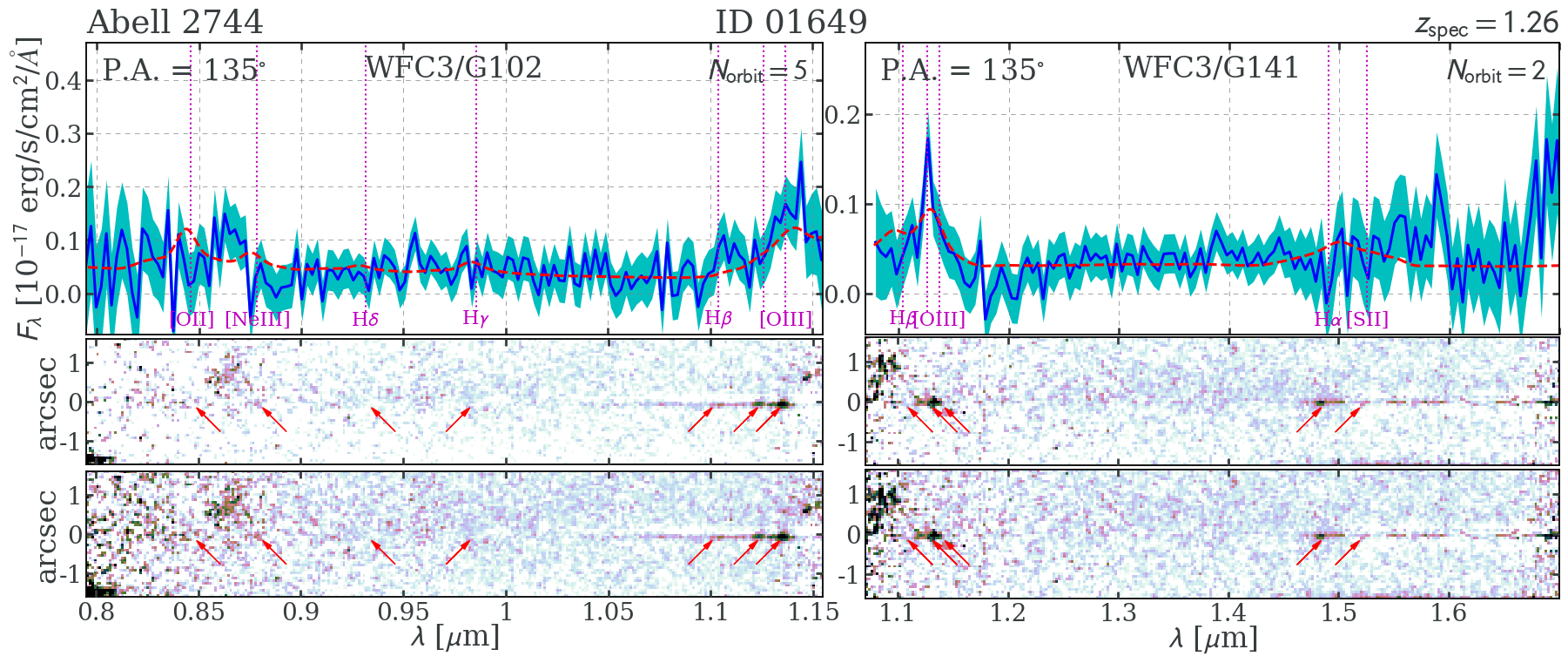}\\
    \includegraphics[width=\textwidth]{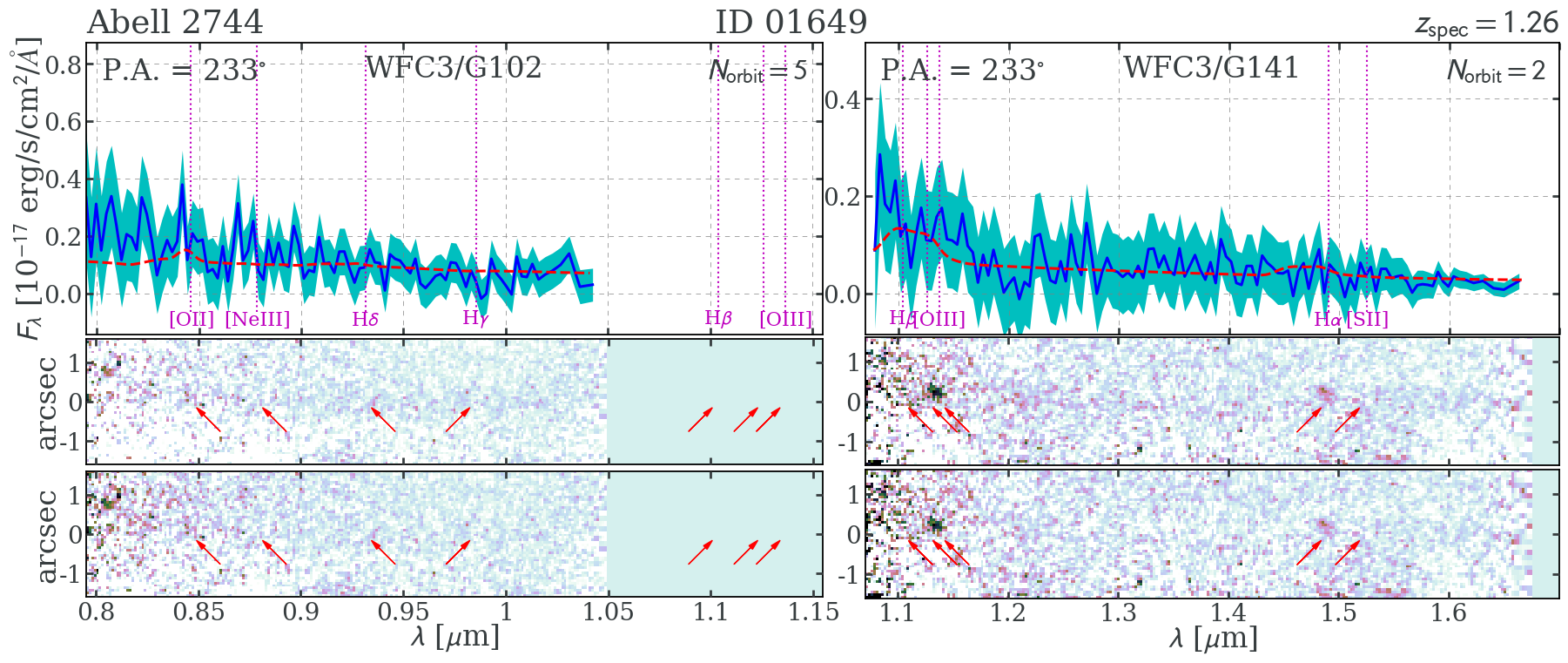}\\
    \includegraphics[width=.16\textwidth]{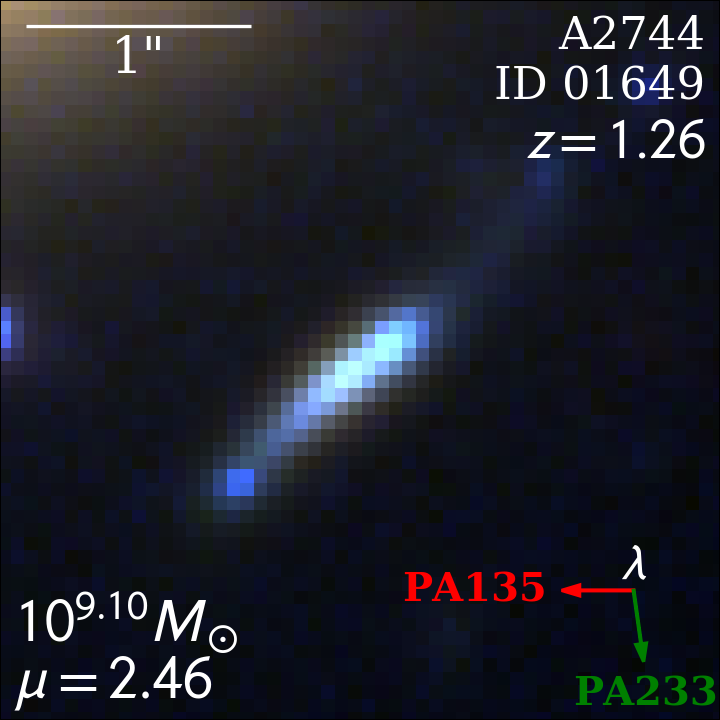}
    \includegraphics[width=.16\textwidth]{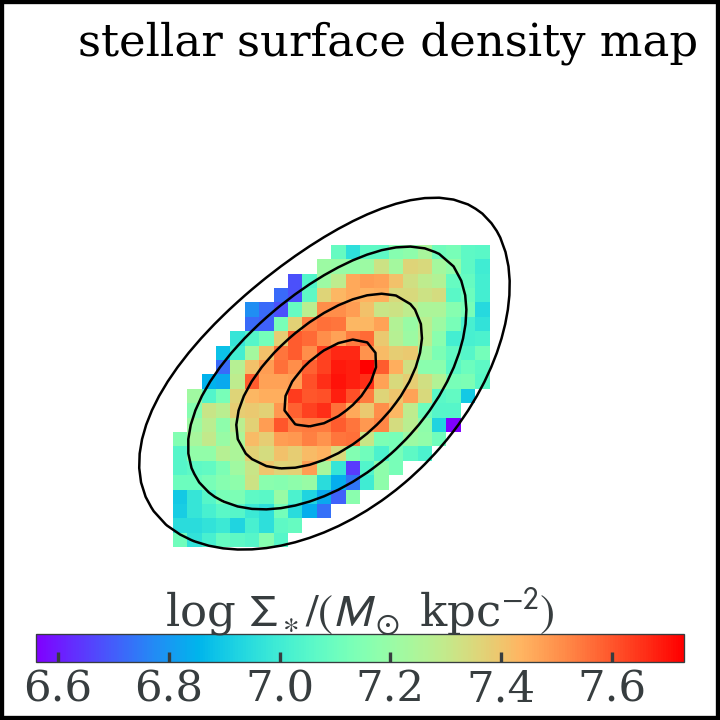}
    \includegraphics[width=.16\textwidth]{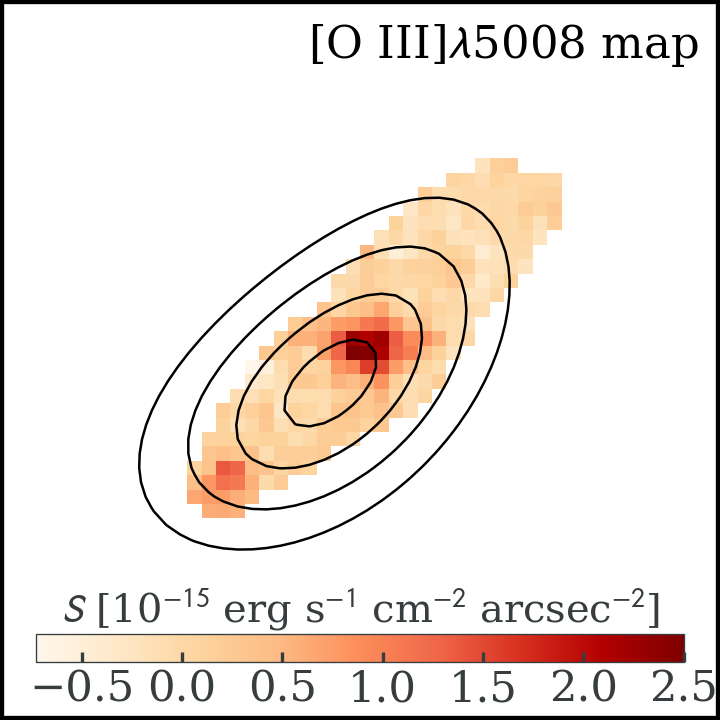}
    \includegraphics[width=.16\textwidth]{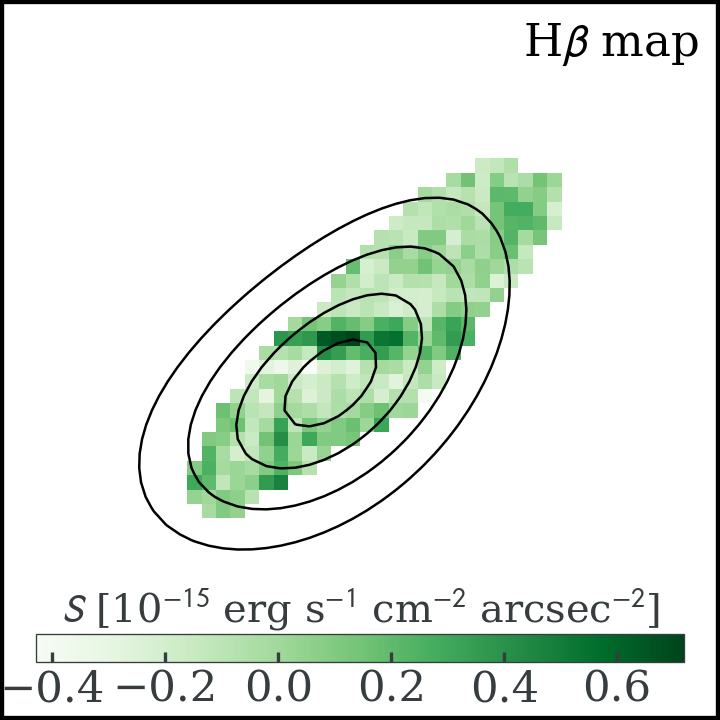}
    \includegraphics[width=.16\textwidth]{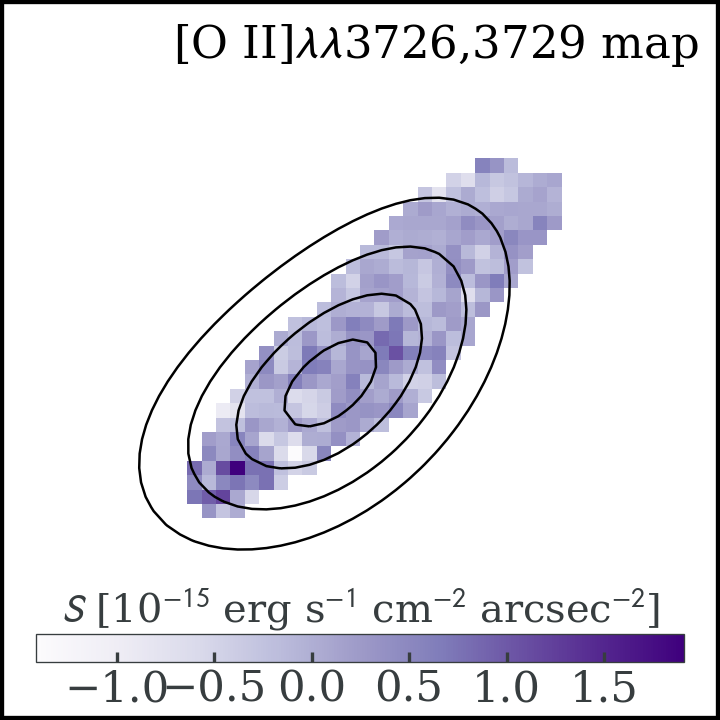}
    \includegraphics[width=.16\textwidth]{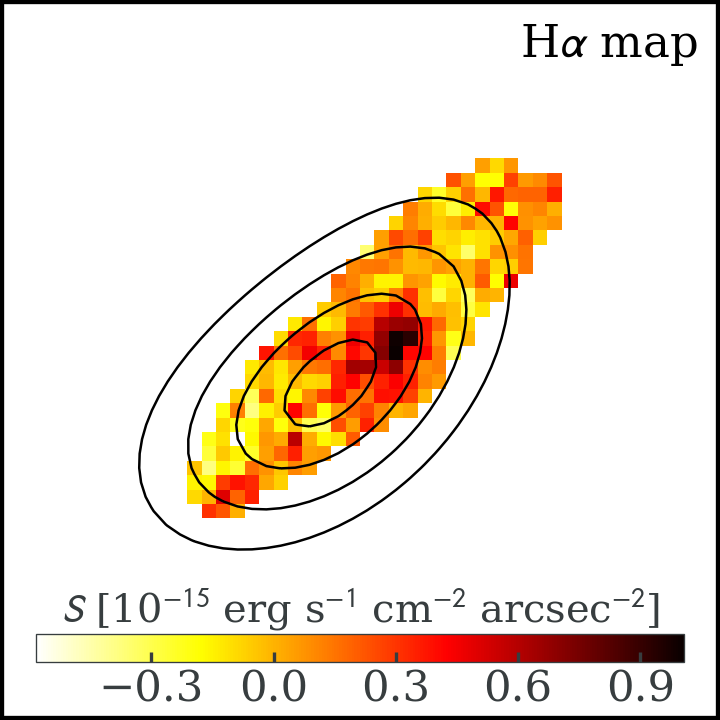}\\
    \includegraphics[width=\textwidth]{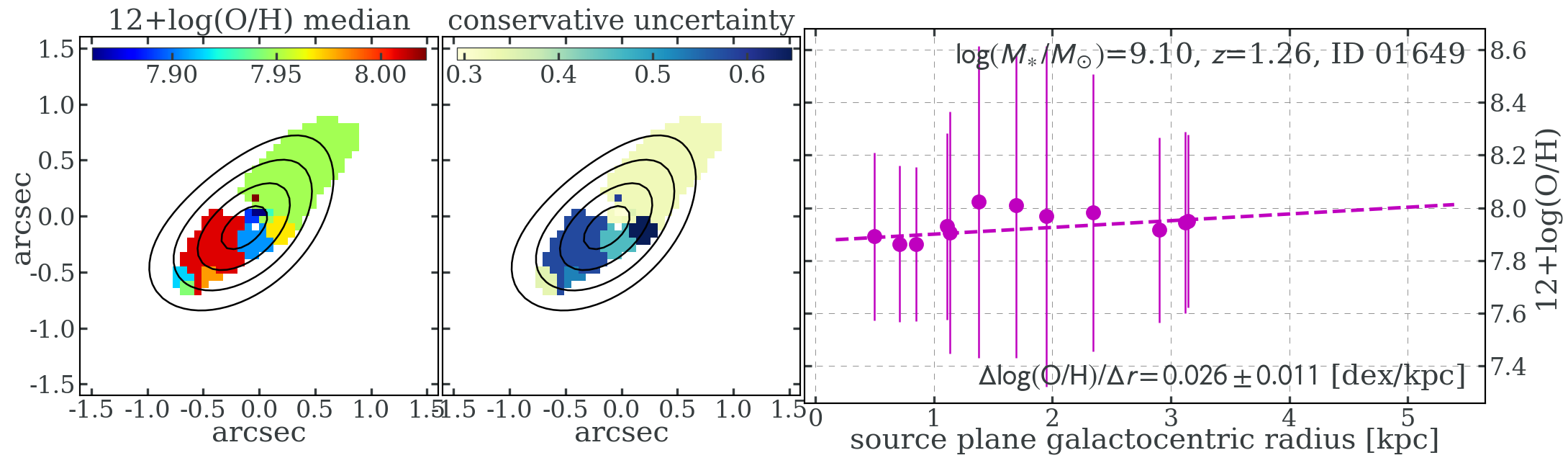}
    \caption{The source ID01649 in the field of \cler is shown.}
    \label{fig:clA2744_ID01649_figs}
\end{figure*}

\begin{figure*}
    \centering
    \includegraphics[width=\textwidth]{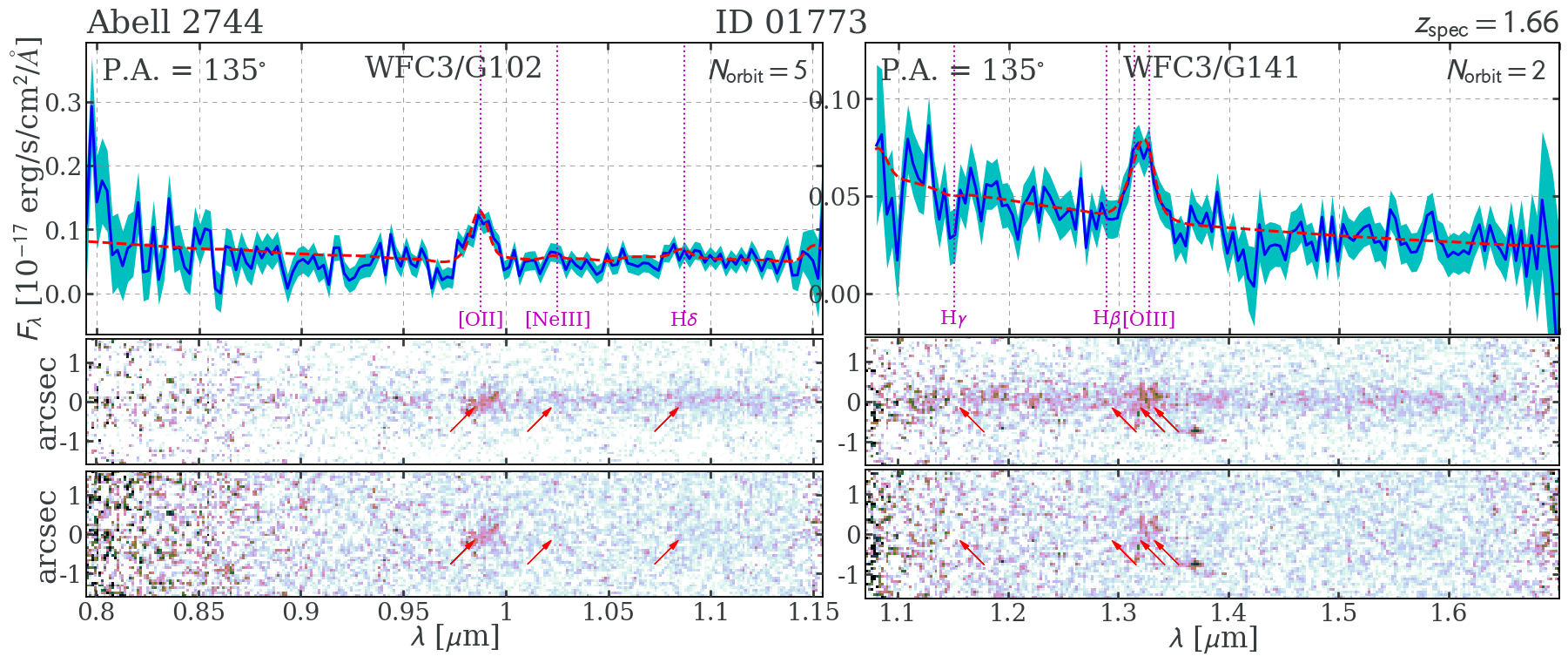}\\
    \includegraphics[width=\textwidth]{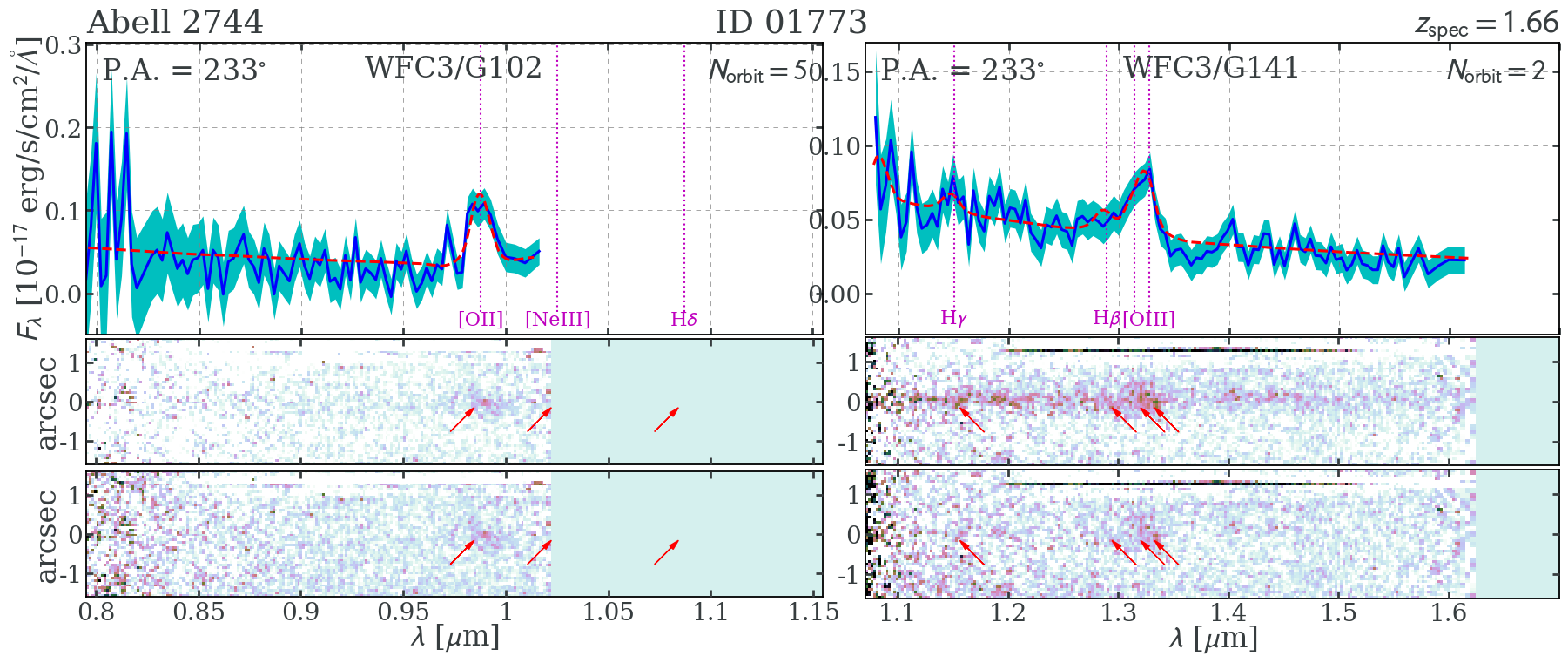}\\
    \includegraphics[width=.16\textwidth]{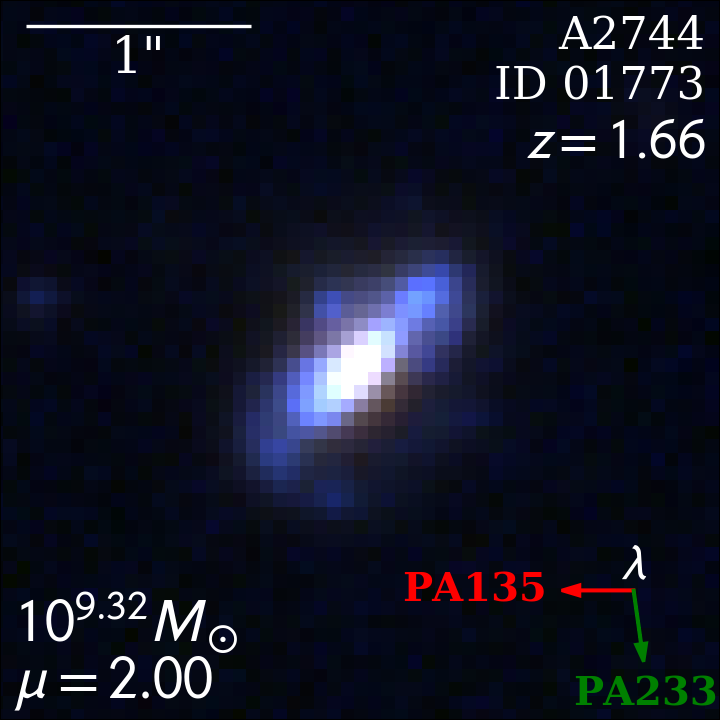}
    \includegraphics[width=.16\textwidth]{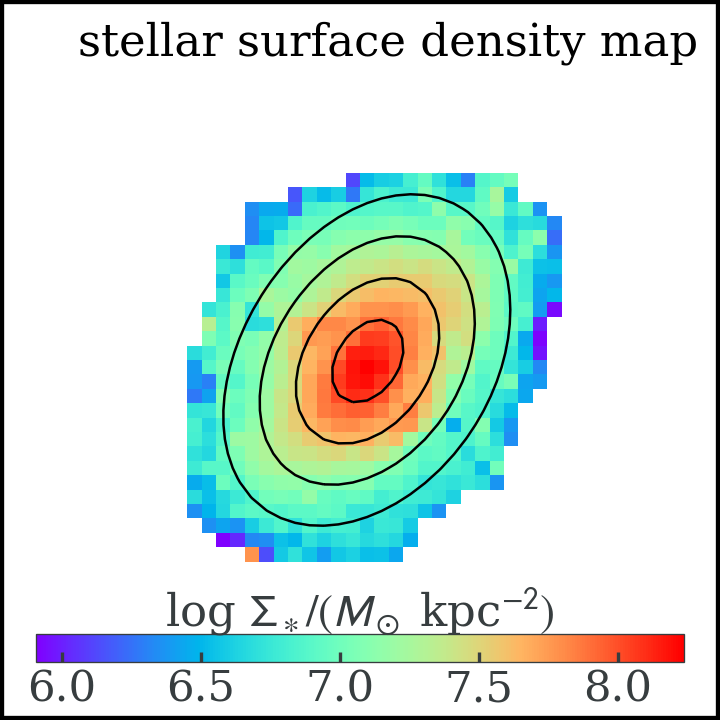}
    \includegraphics[width=.16\textwidth]{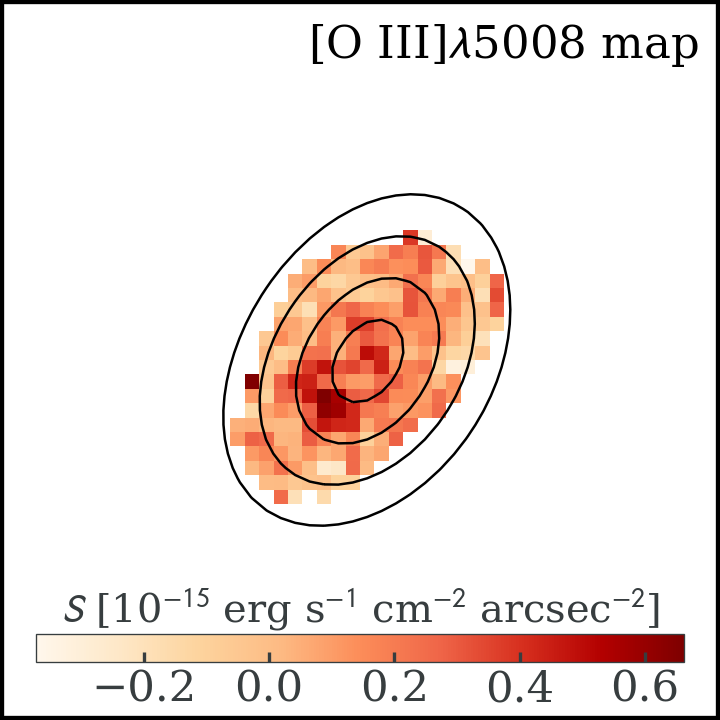}
    \includegraphics[width=.16\textwidth]{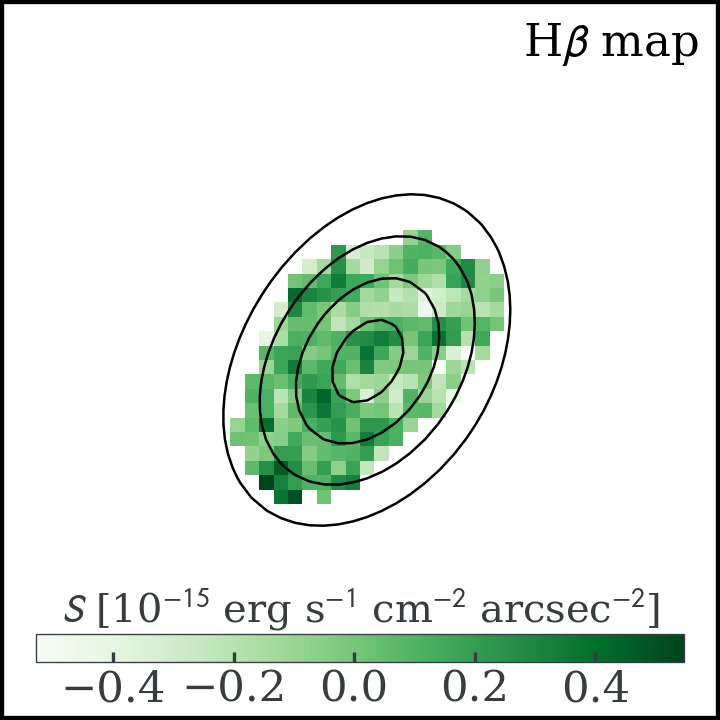}
    \includegraphics[width=.16\textwidth]{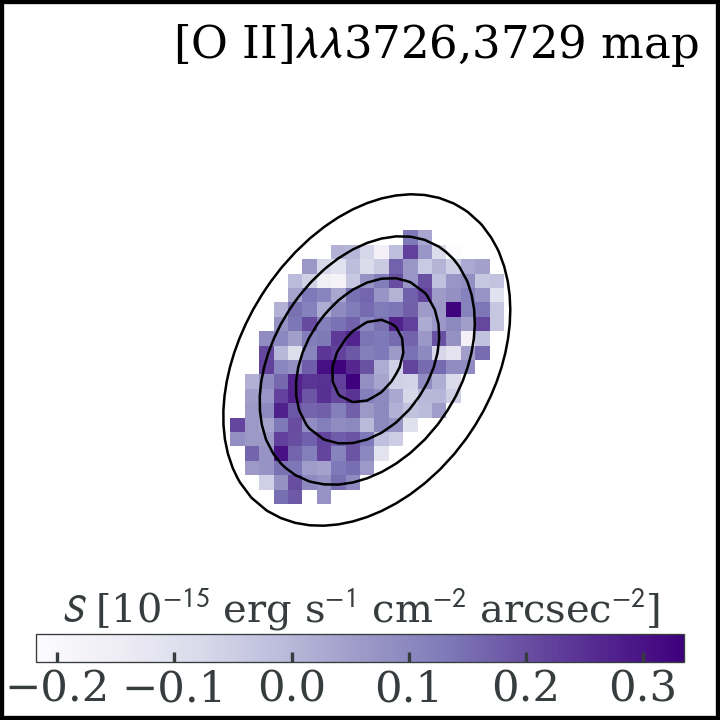}
    \includegraphics[width=.16\textwidth]{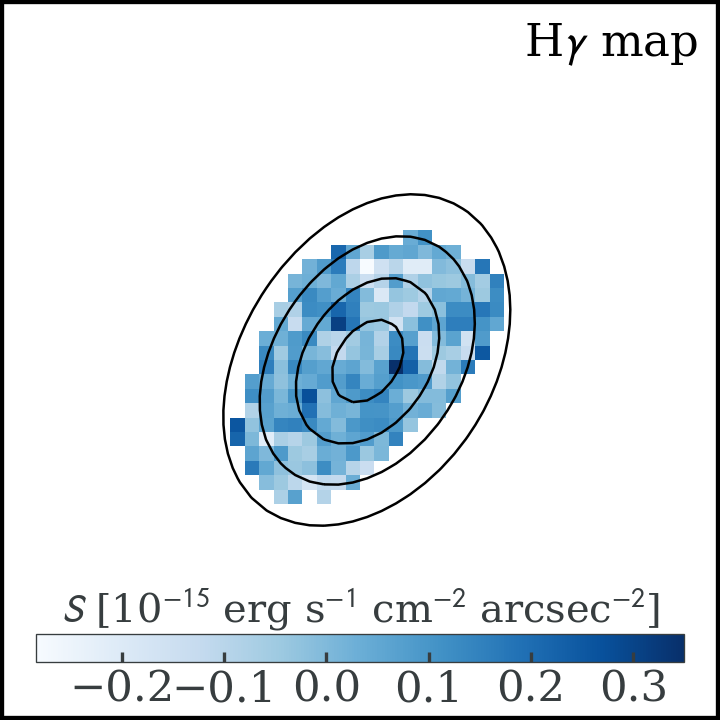}\\
    \includegraphics[width=\textwidth]{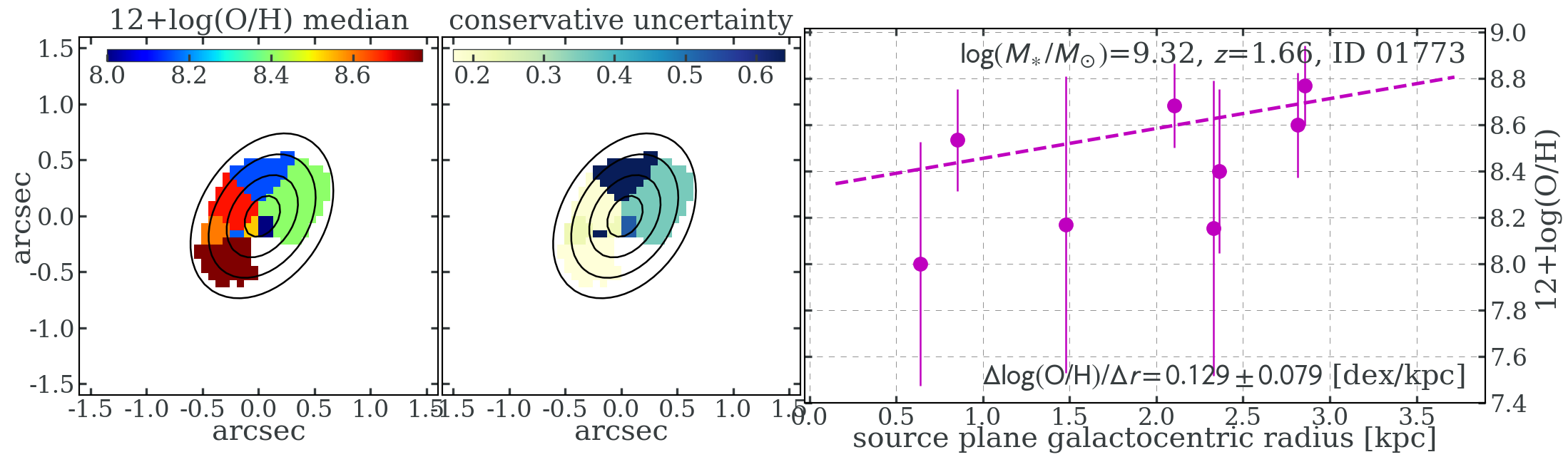}
    \caption{The source ID01773 in the field of \cler is shown.}
    \label{fig:clA2744_ID01773_figs}
\end{figure*}

\begin{figure*}
    \centering
    \includegraphics[width=\textwidth]{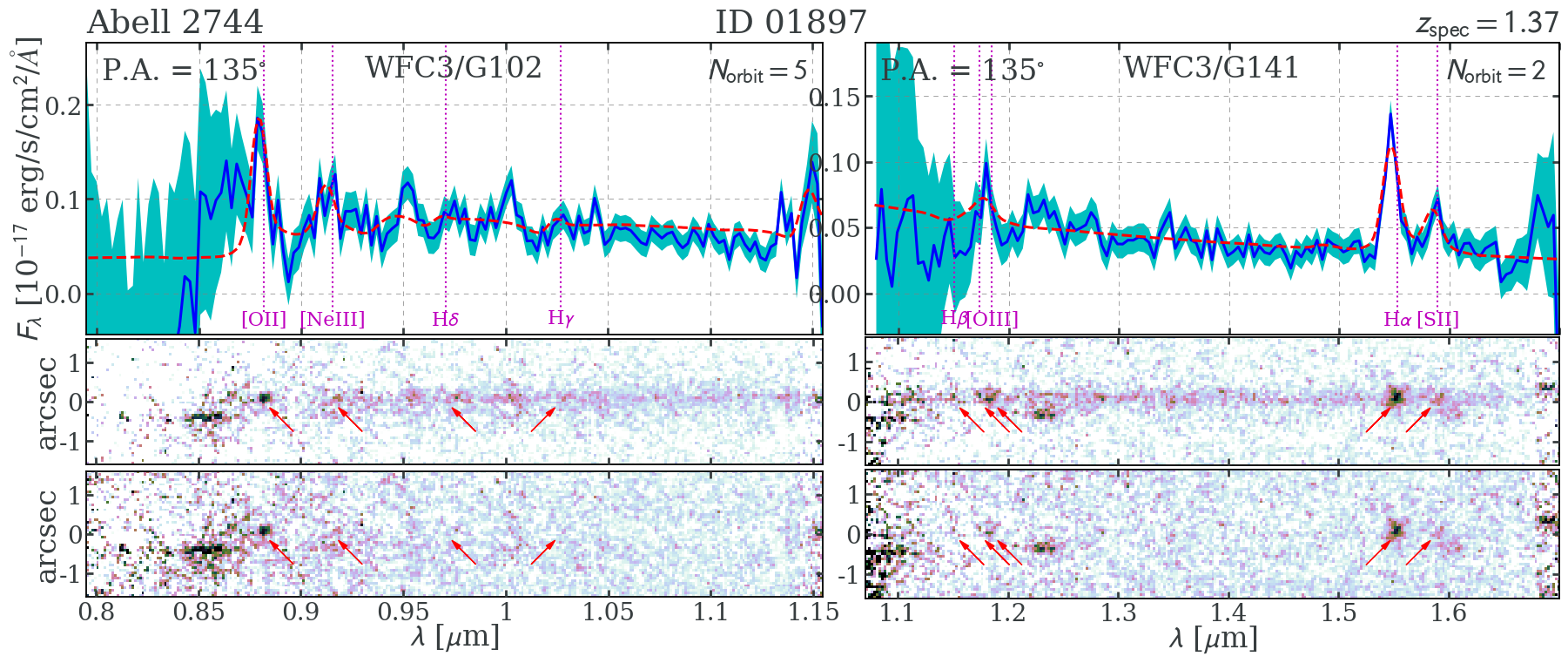}\\
    \includegraphics[width=.16\textwidth]{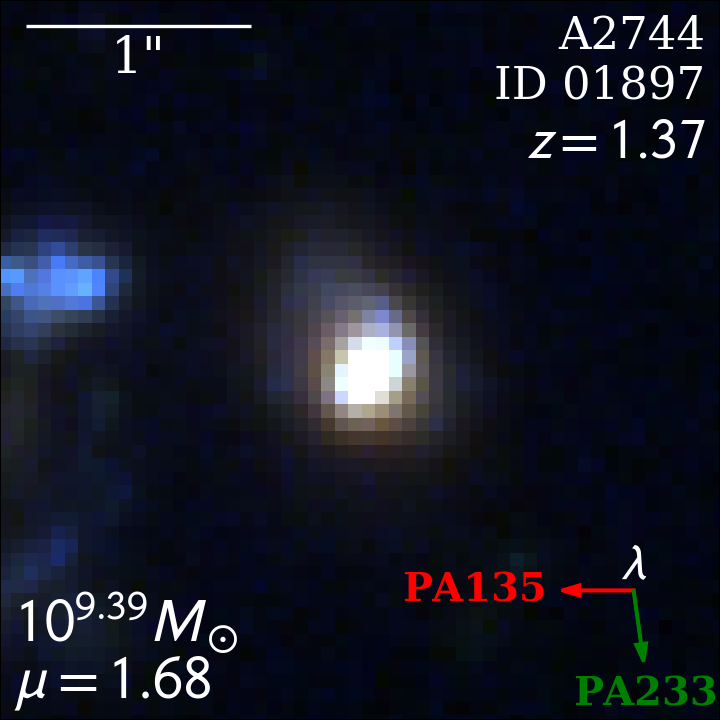}
    \includegraphics[width=.16\textwidth]{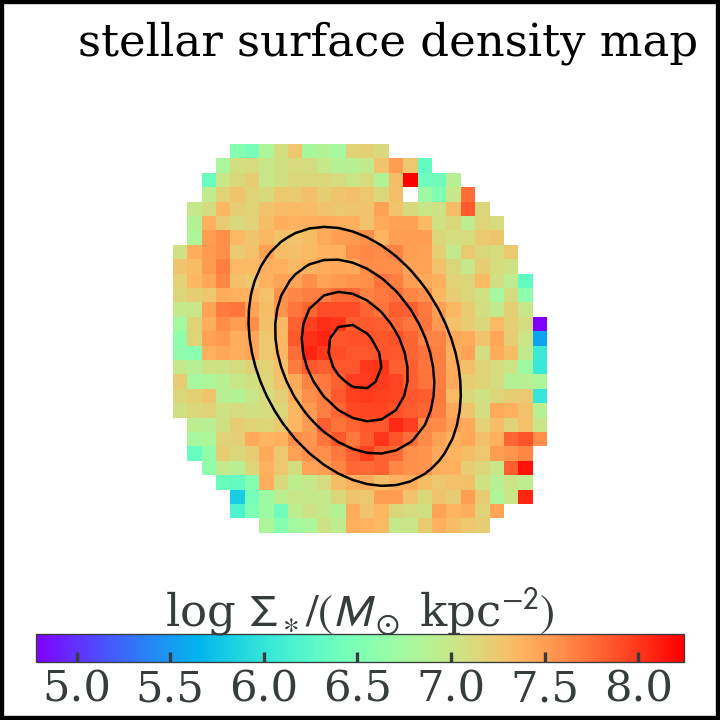}
    \includegraphics[width=.16\textwidth]{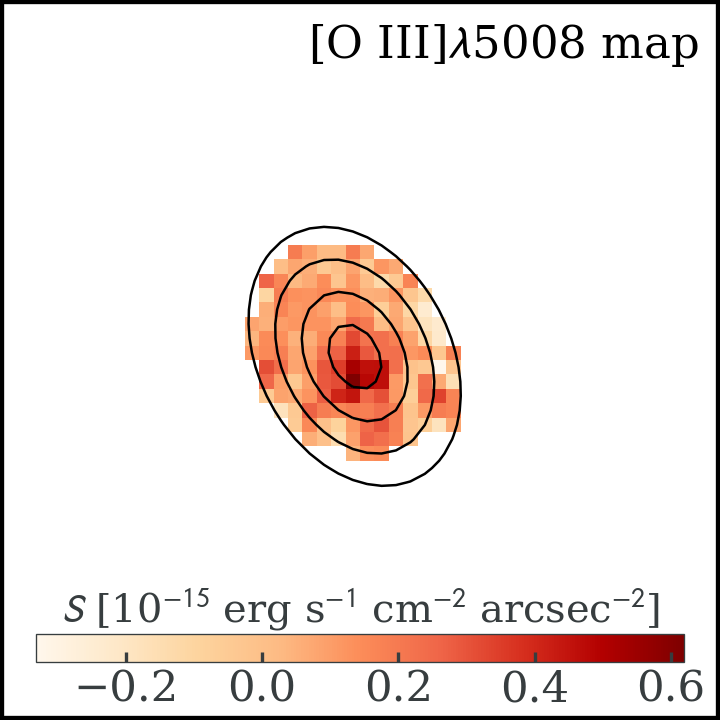}
    \includegraphics[width=.16\textwidth]{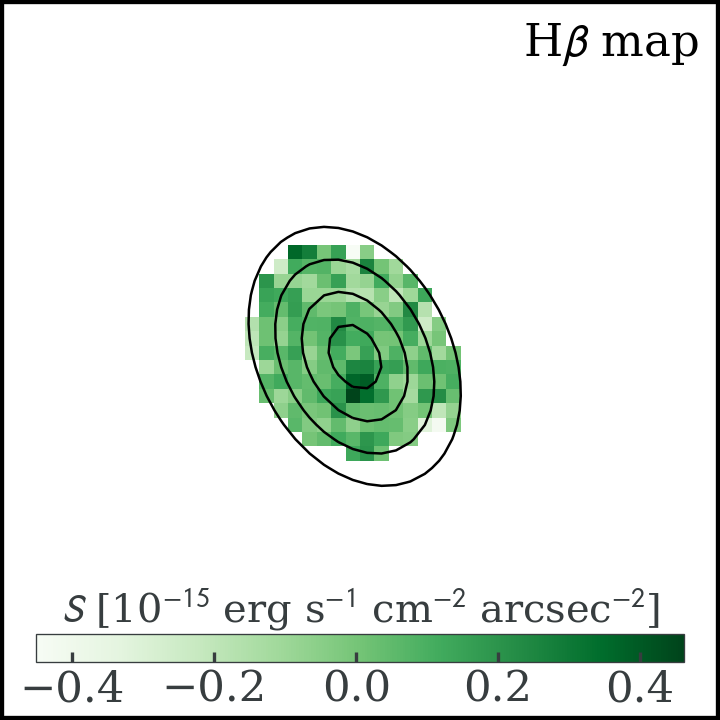}
    \includegraphics[width=.16\textwidth]{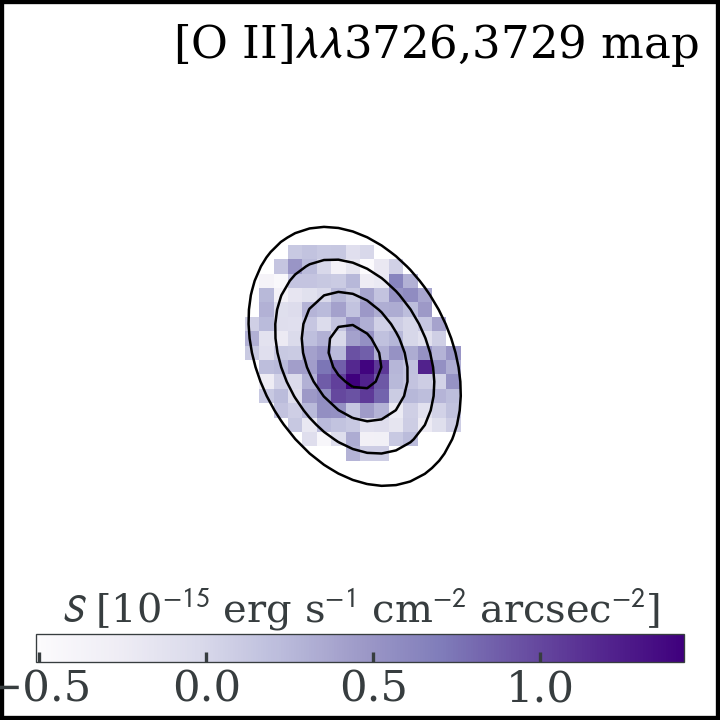}
    \includegraphics[width=.16\textwidth]{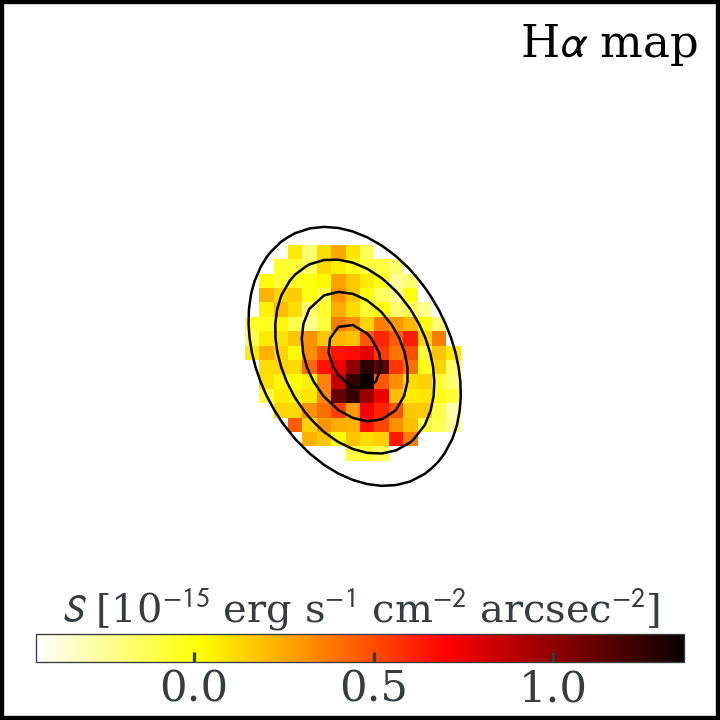}\\
    \includegraphics[width=\textwidth]{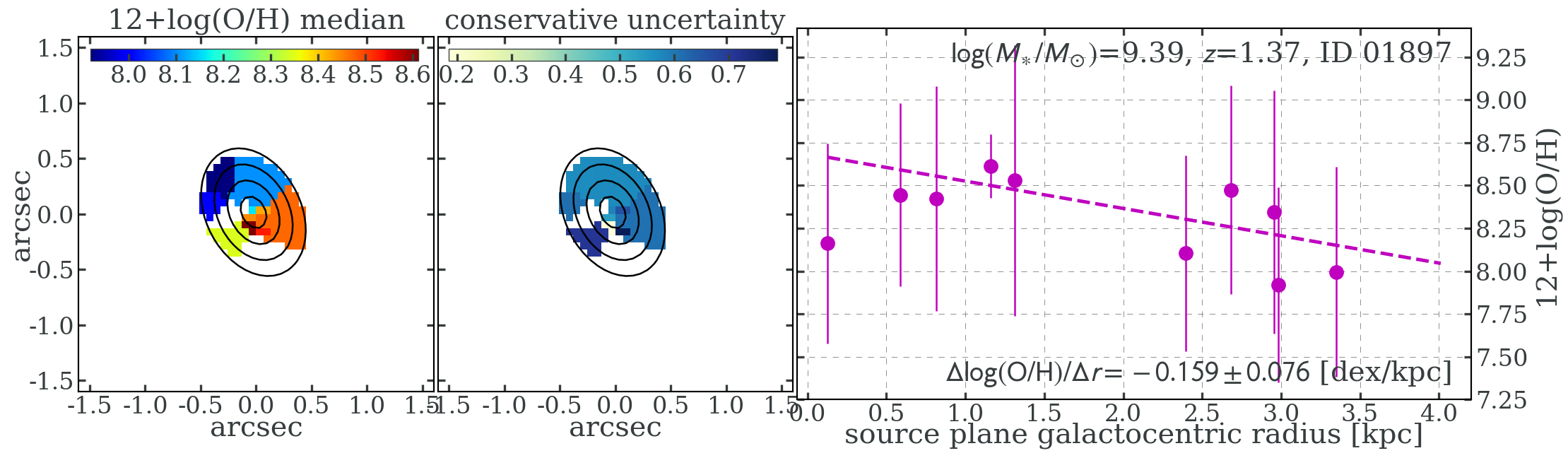}
    \caption{The source ID01897 in the field of \cler is shown.}
    \label{fig:clA2744_ID01897_figs}
\end{figure*}

\begin{figure*}
    \centering
    \includegraphics[width=\textwidth]{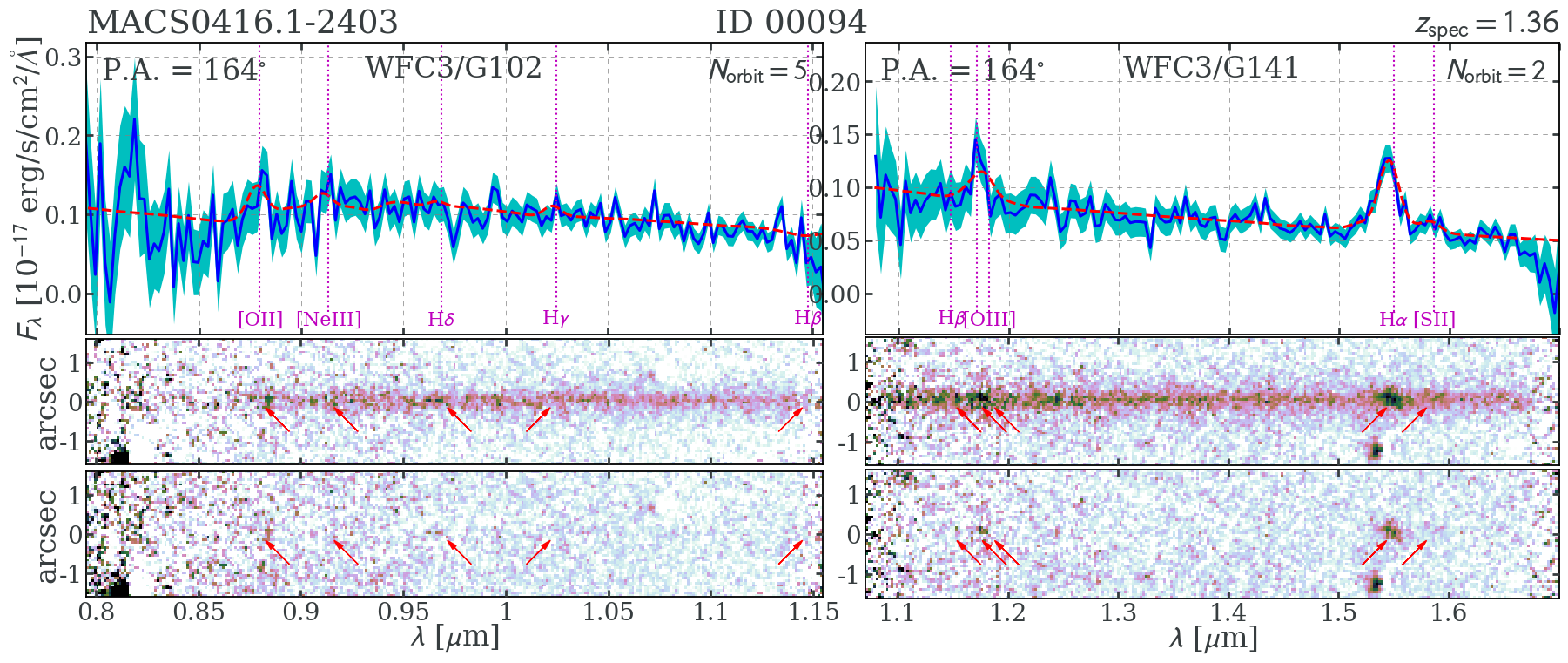}\\
    \includegraphics[width=\textwidth]{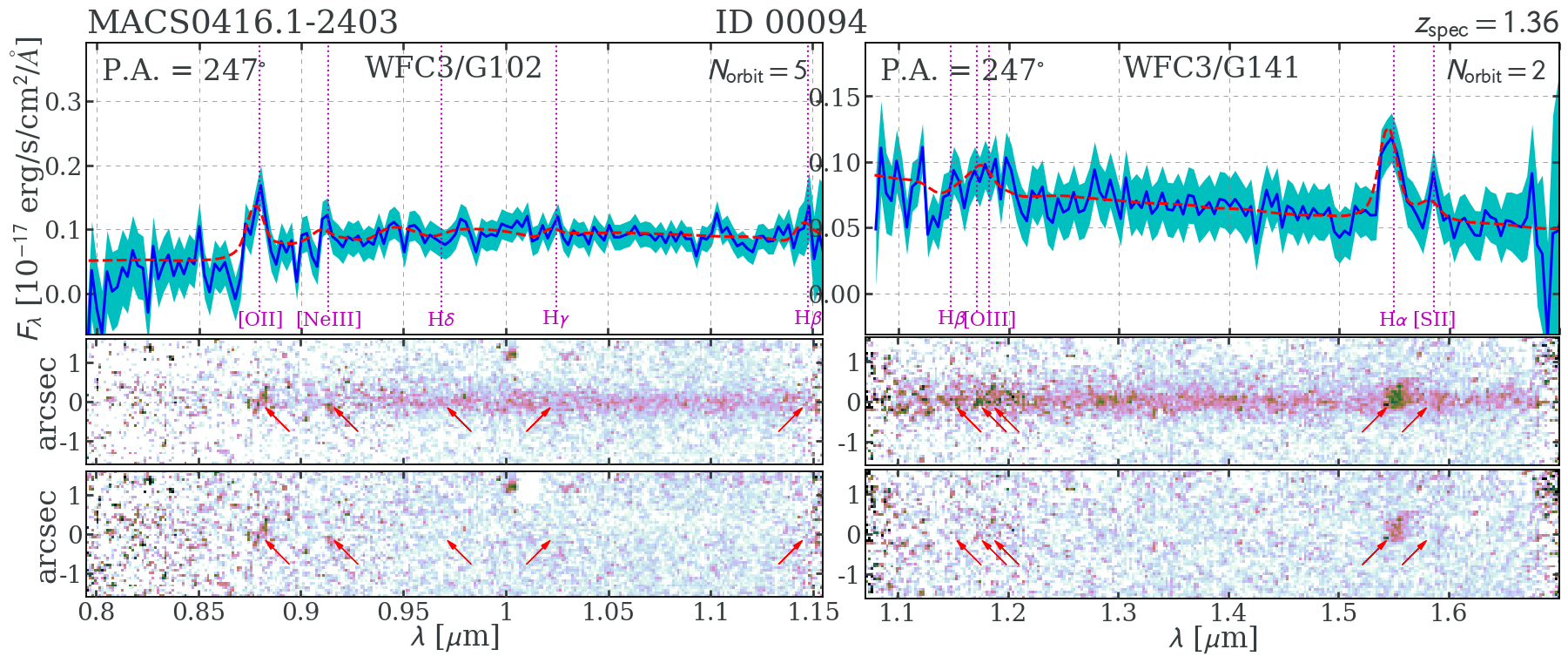}\\
    \includegraphics[width=.16\textwidth]{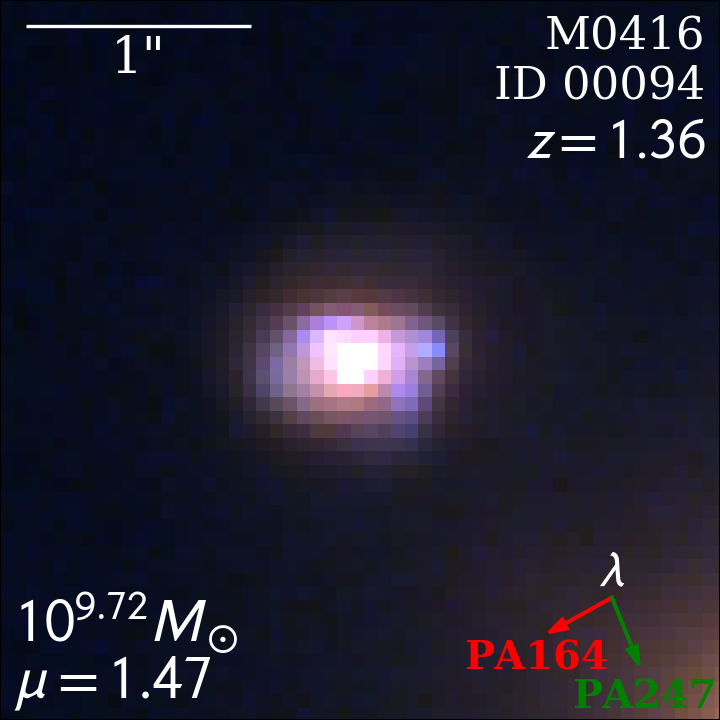}
    \includegraphics[width=.16\textwidth]{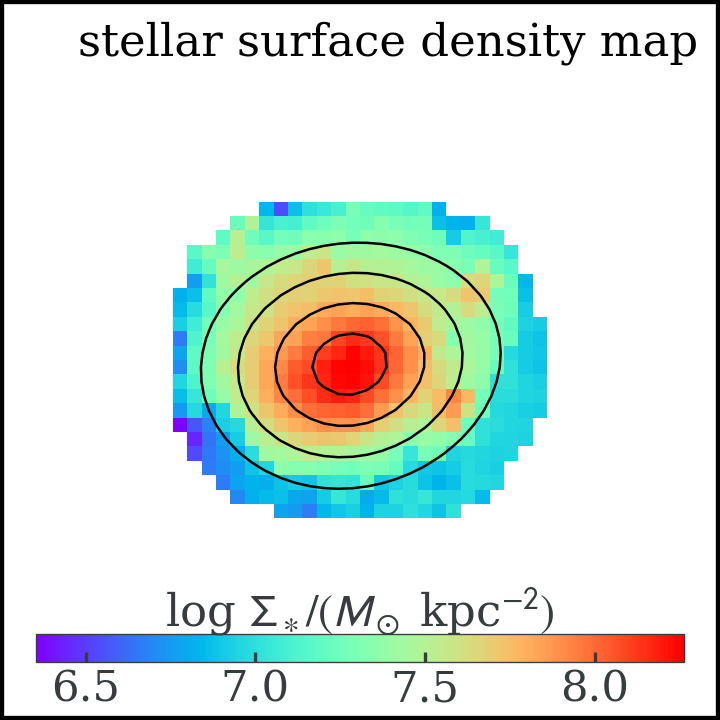}
    \includegraphics[width=.16\textwidth]{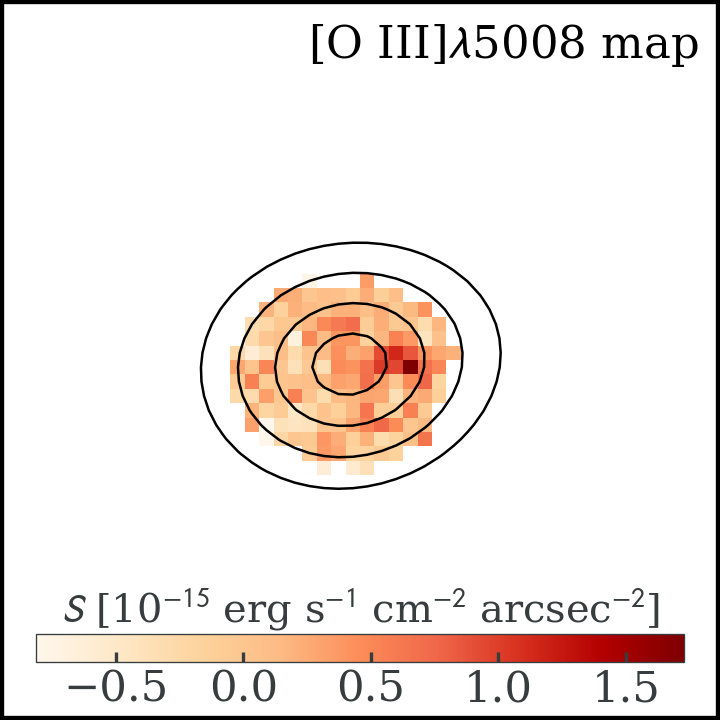}
    \includegraphics[width=.16\textwidth]{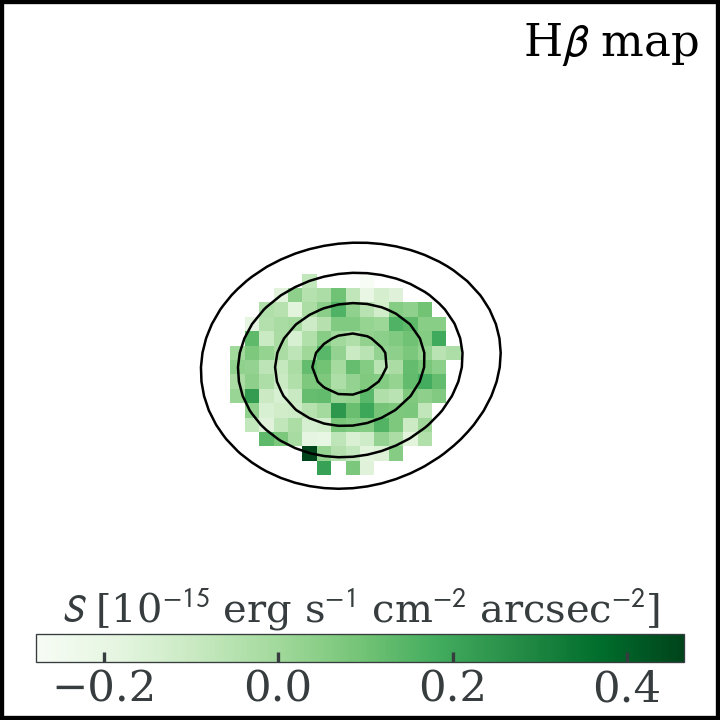}
    \includegraphics[width=.16\textwidth]{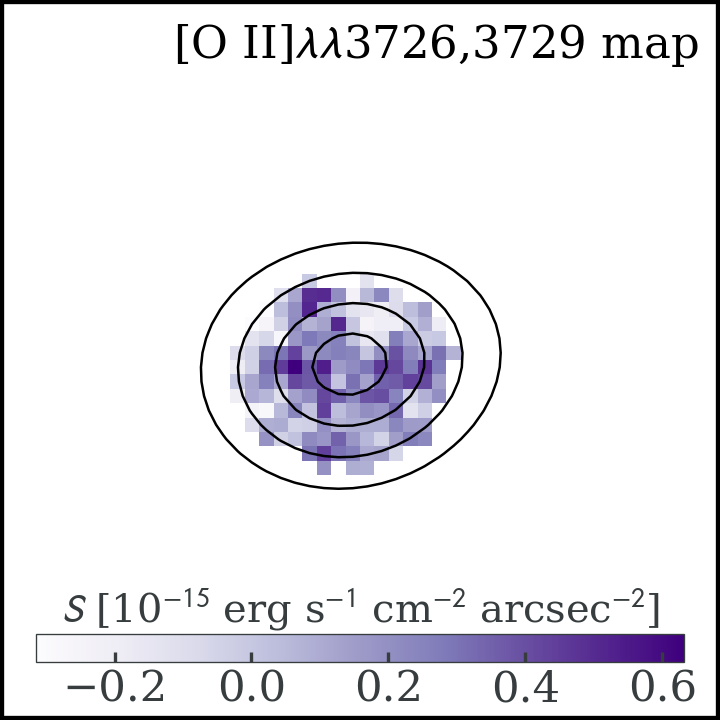}
    \includegraphics[width=.16\textwidth]{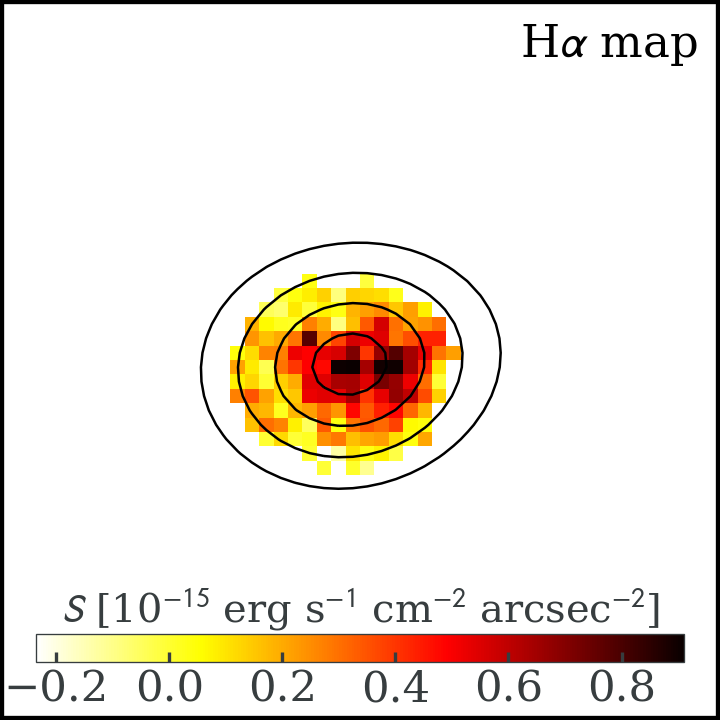}\\
    \includegraphics[width=\textwidth]{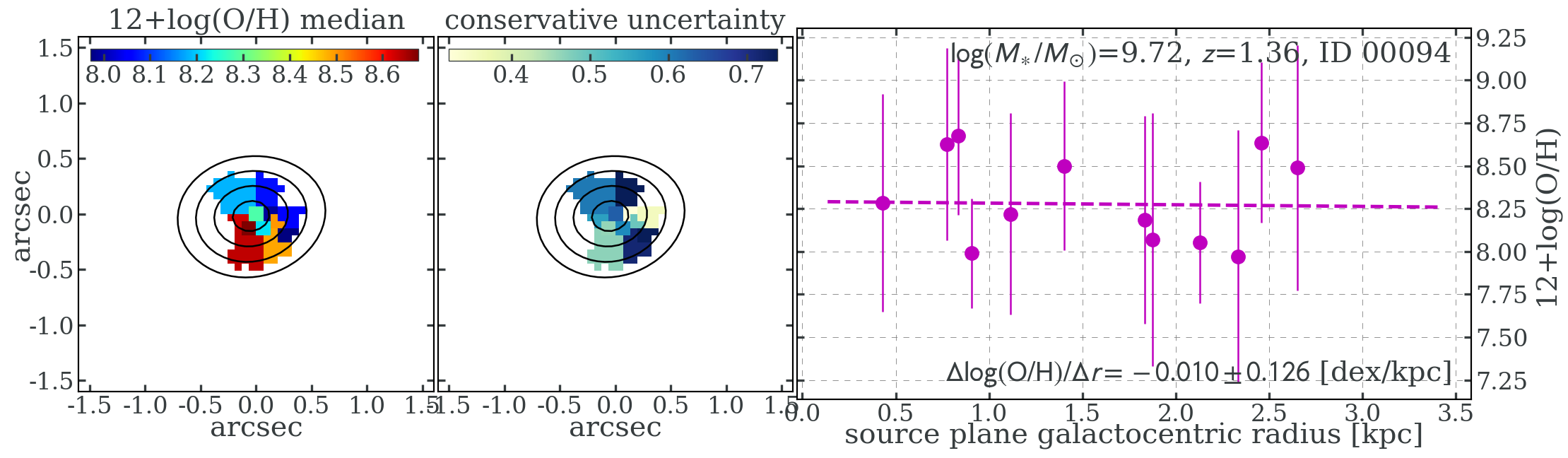}
    \caption{The source ID00094 in the field of \clsi is shown.}
    \label{fig:clM0416_ID00094_figs}
\end{figure*}

\begin{figure*}
    \centering
    \includegraphics[width=\textwidth]{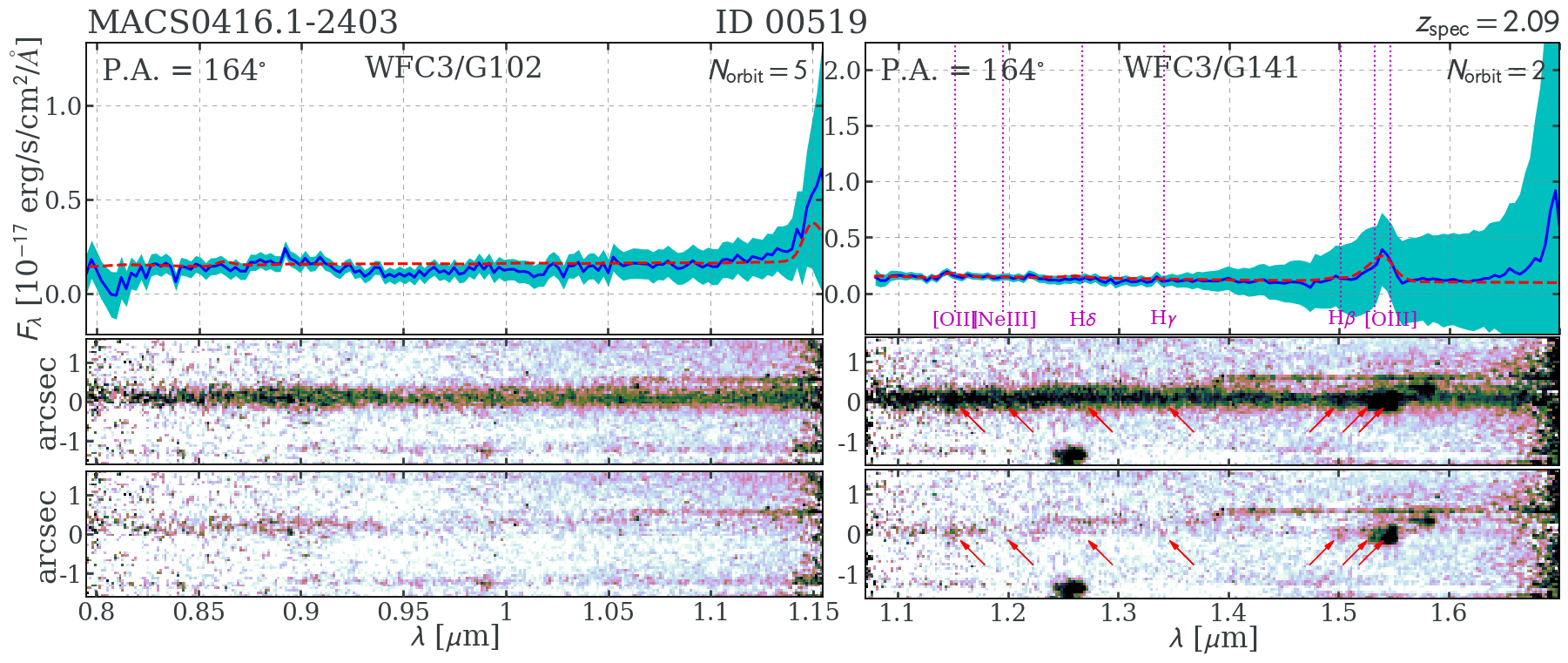}\\
    \includegraphics[width=\textwidth]{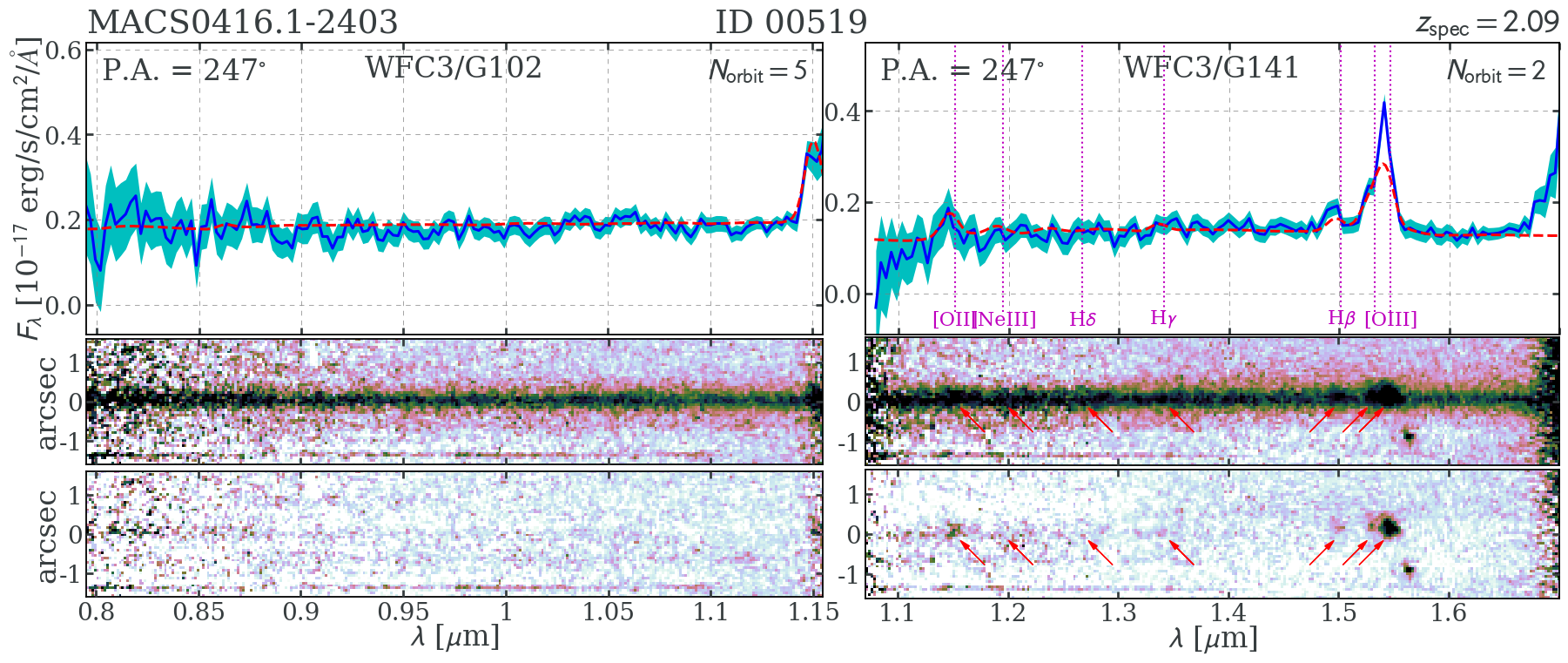}\\
    \includegraphics[width=.16\textwidth]{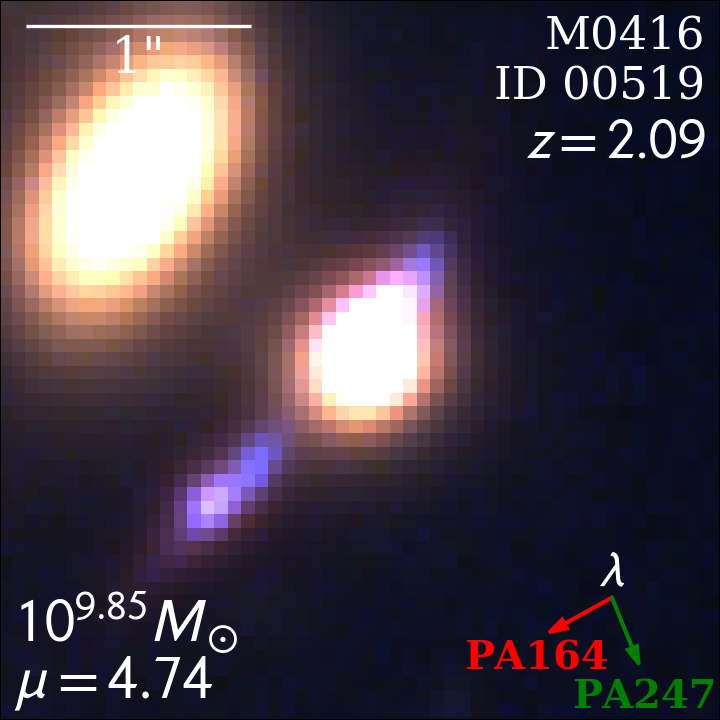}
    \includegraphics[width=.16\textwidth]{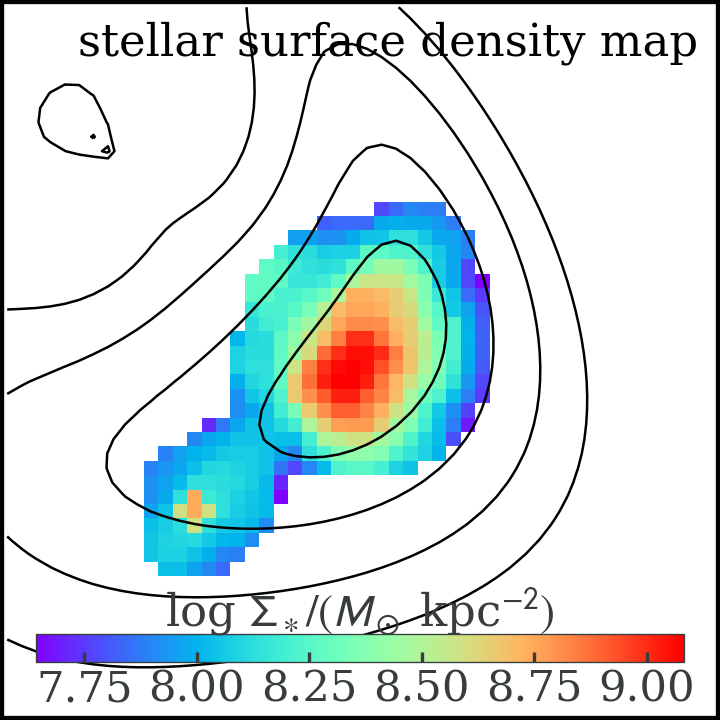}
    \includegraphics[width=.16\textwidth]{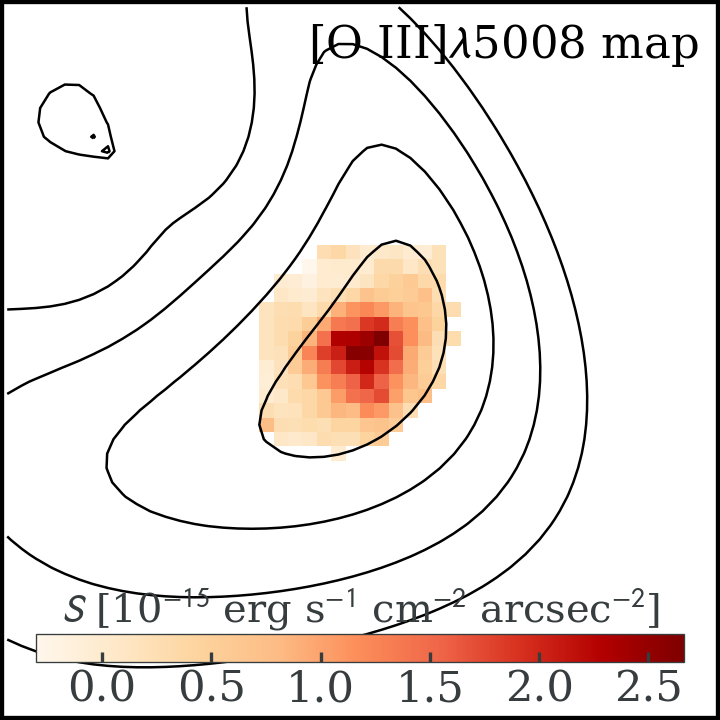}
    \includegraphics[width=.16\textwidth]{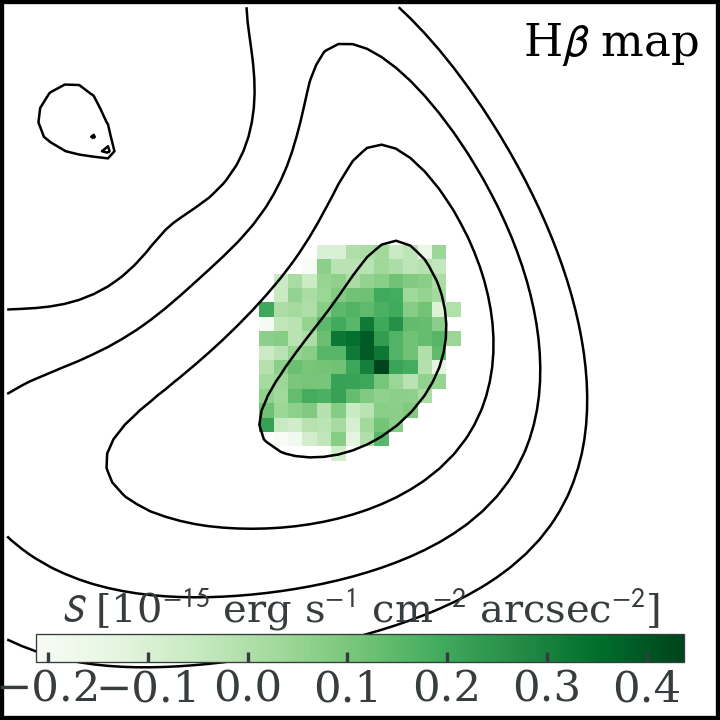}
    \includegraphics[width=.16\textwidth]{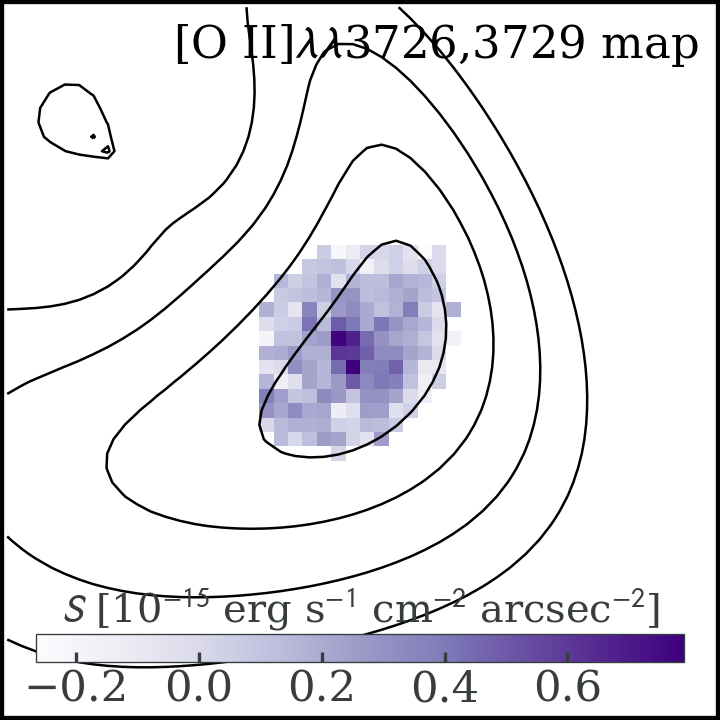}
    \includegraphics[width=.16\textwidth]{fig_ELmaps/baiban.png}\\
    \includegraphics[width=\textwidth]{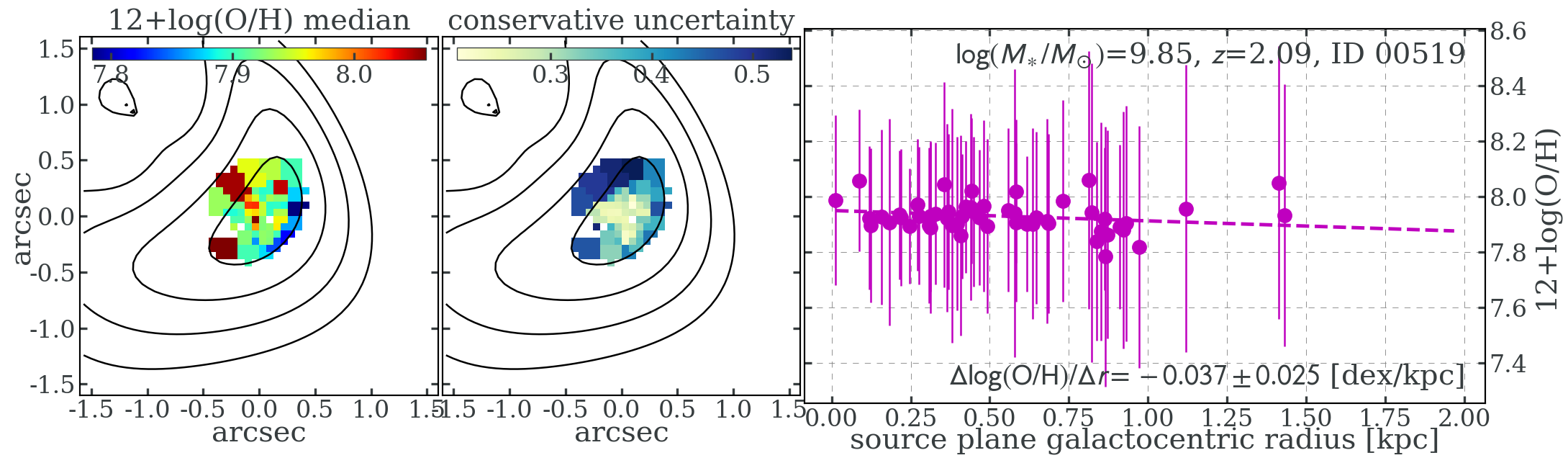}
    \caption{The source ID00519 in the field of \clsi is shown.}
    \label{fig:clM0416_ID00519_figs}
\end{figure*}

\begin{figure*}
    \centering
    \includegraphics[width=\textwidth]{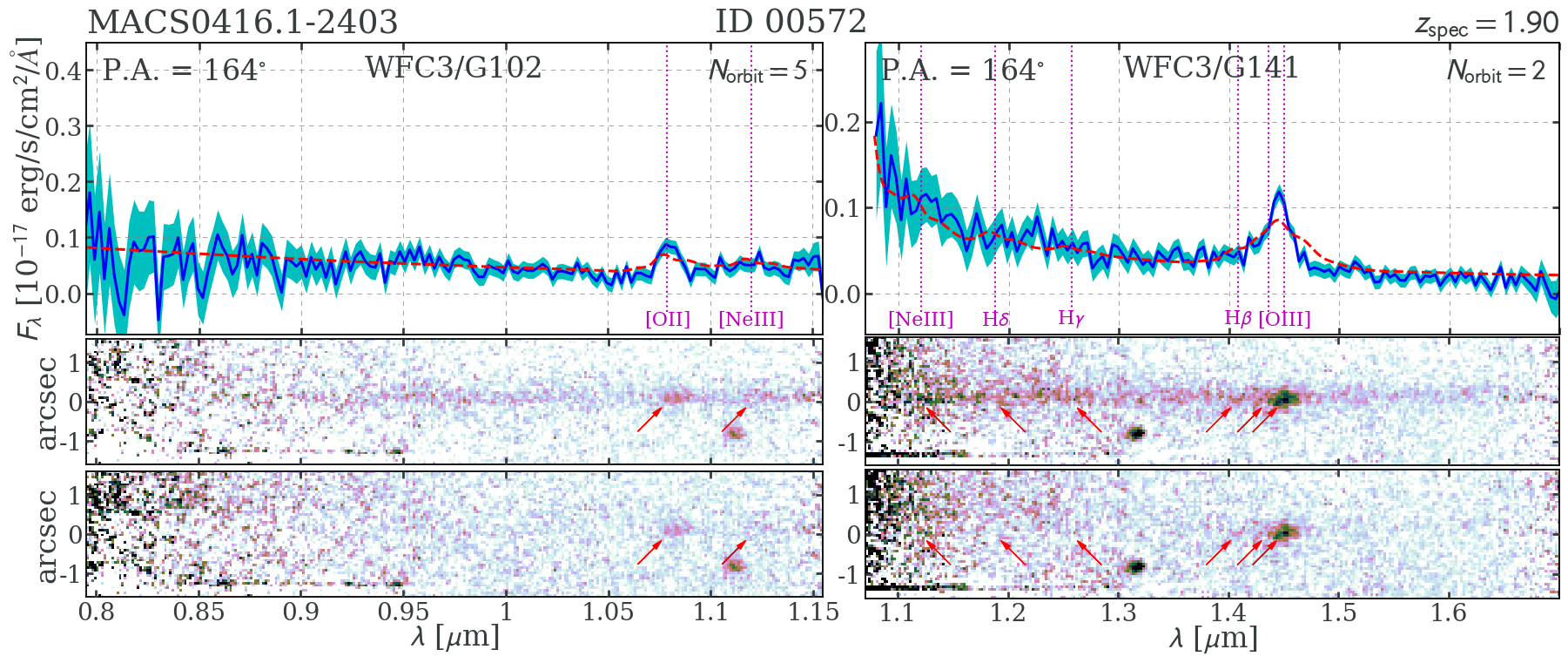}\\
    \includegraphics[width=\textwidth]{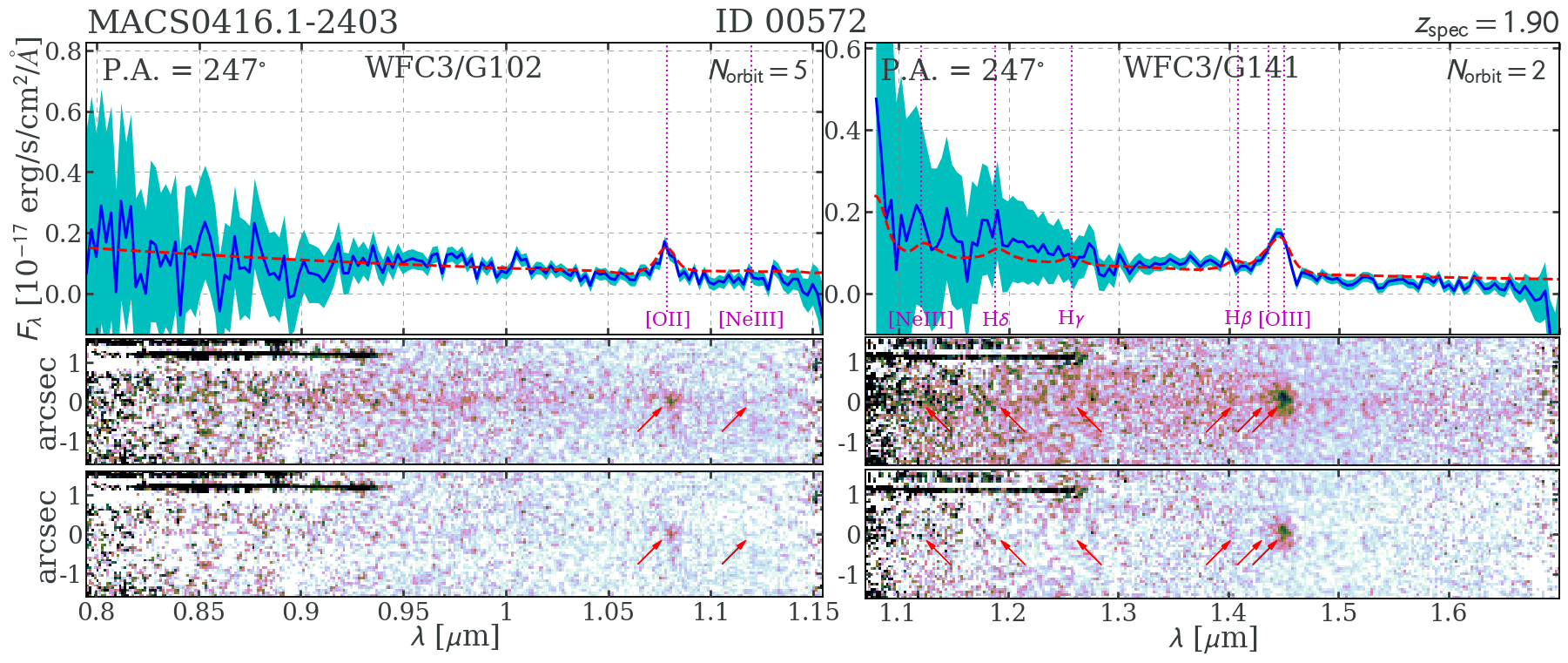}\\
    \includegraphics[width=.16\textwidth]{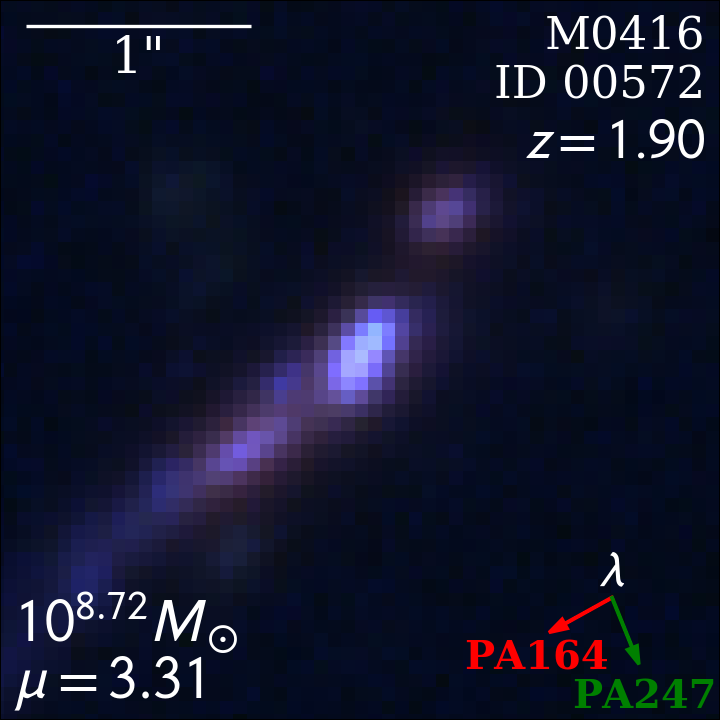}
    \includegraphics[width=.16\textwidth]{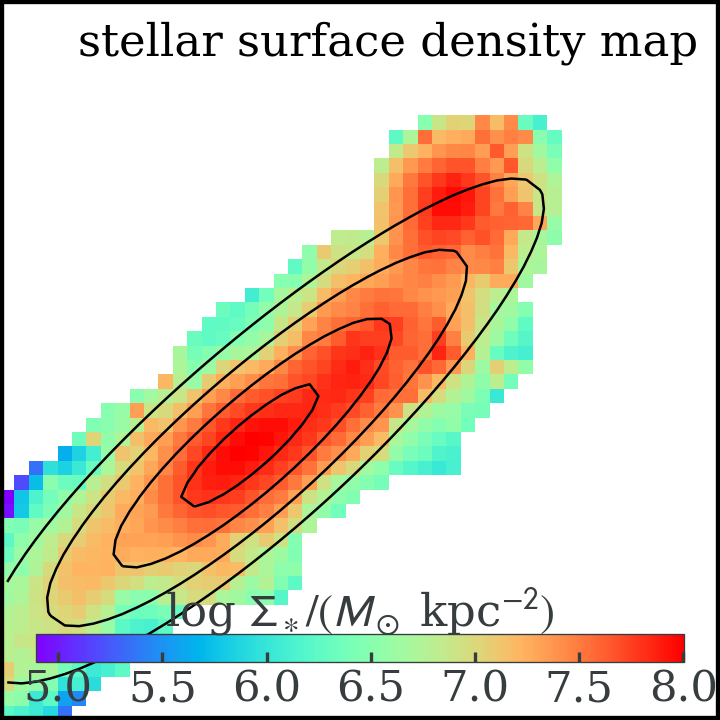}
    \includegraphics[width=.16\textwidth]{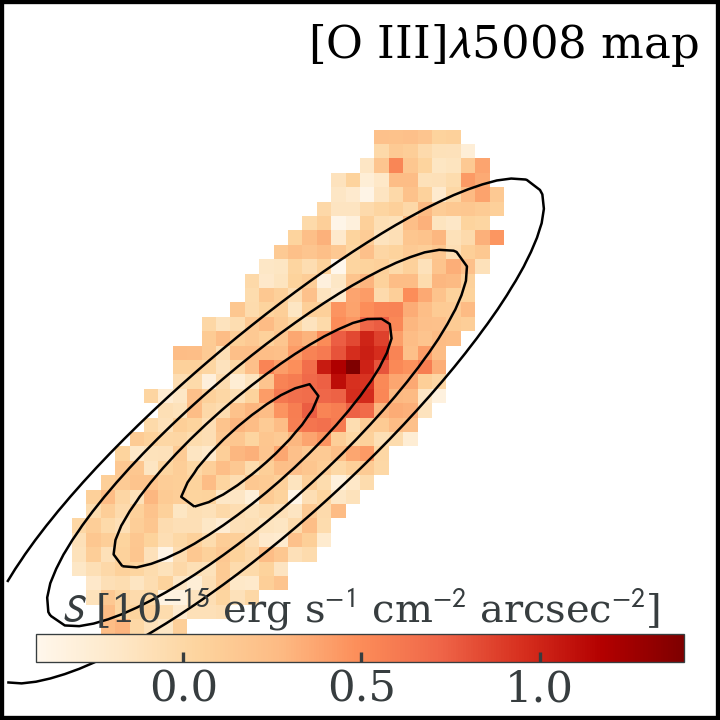}
    \includegraphics[width=.16\textwidth]{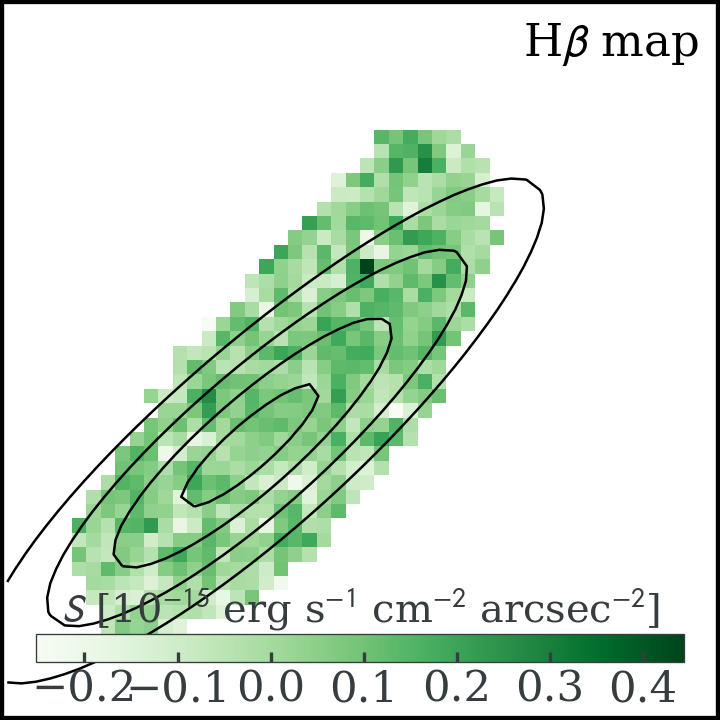}
    \includegraphics[width=.16\textwidth]{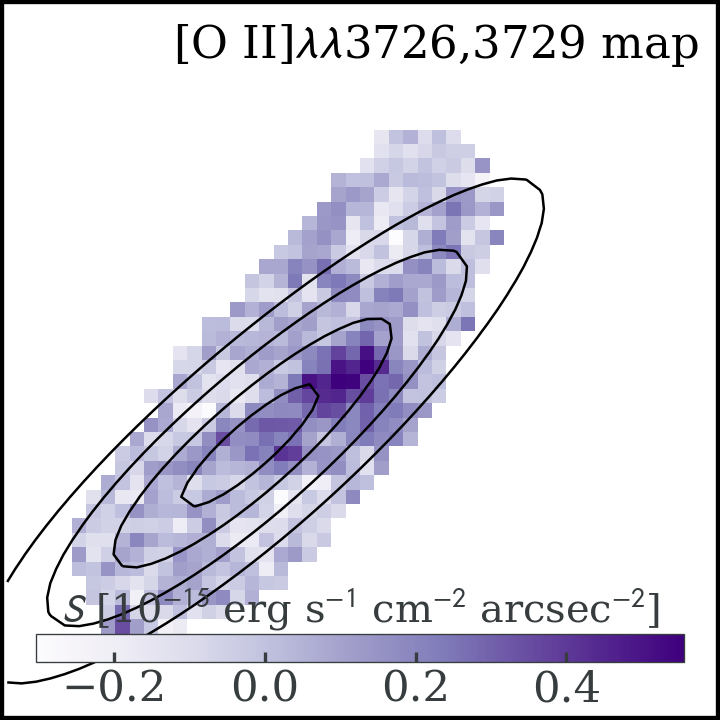}
    \includegraphics[width=.16\textwidth]{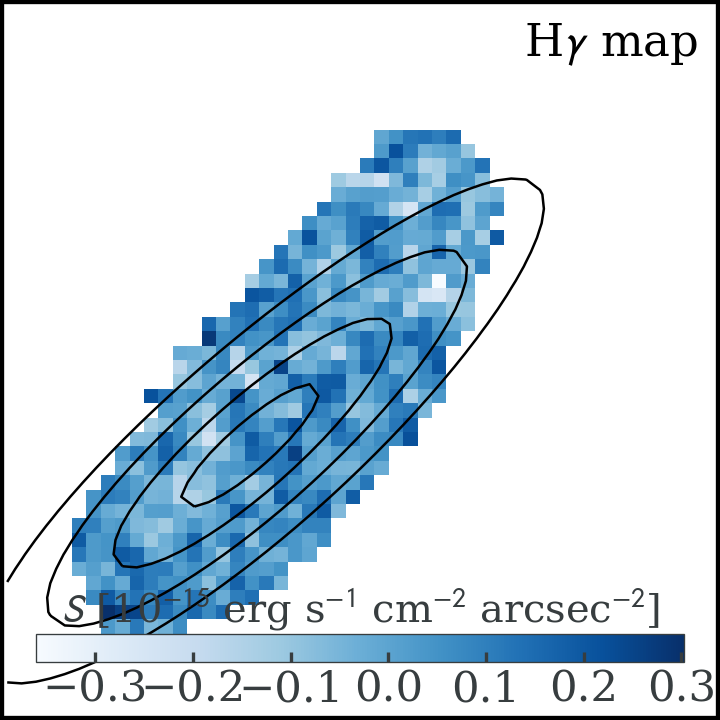}\\
    \includegraphics[width=\textwidth]{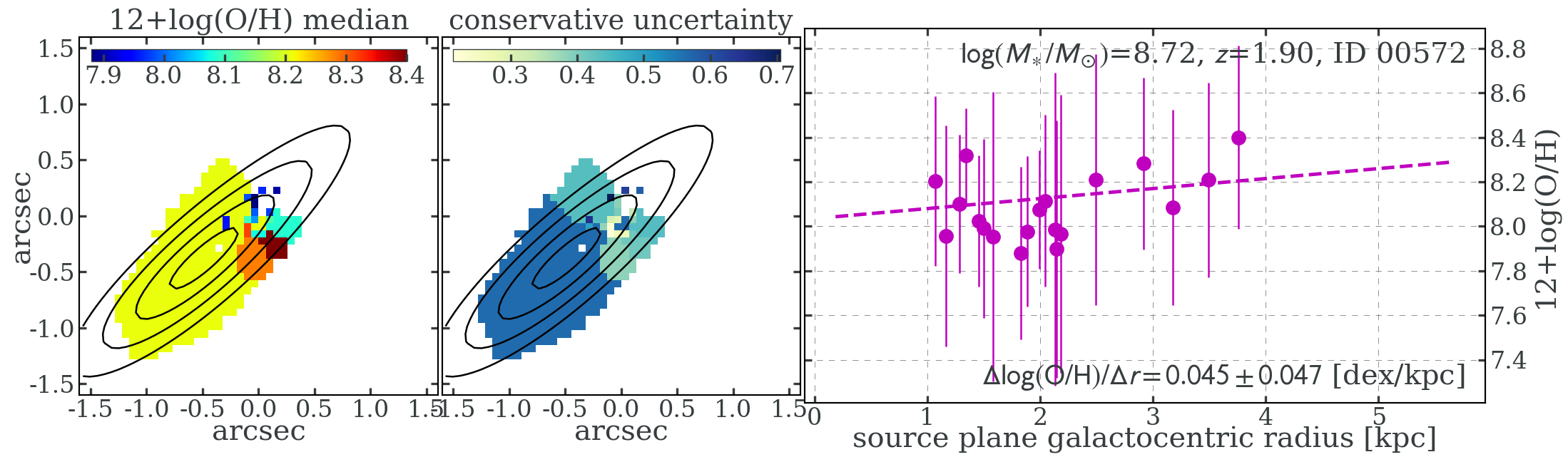}
    \caption{The source ID00572 in the field of \clsi is shown.}
    \label{fig:clM0416_ID00572_figs}
\end{figure*}

\begin{figure*}
    \centering
    \includegraphics[width=\textwidth]{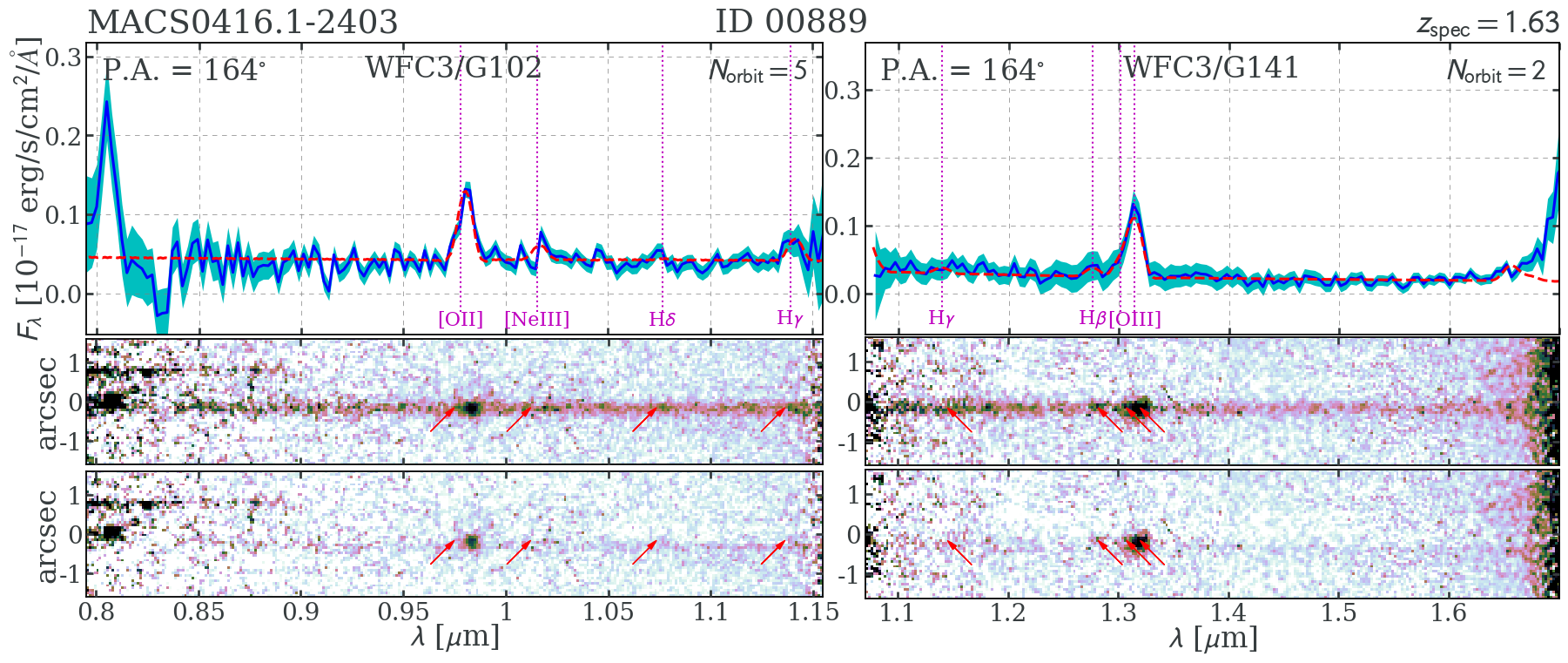}\\
    \includegraphics[width=\textwidth]{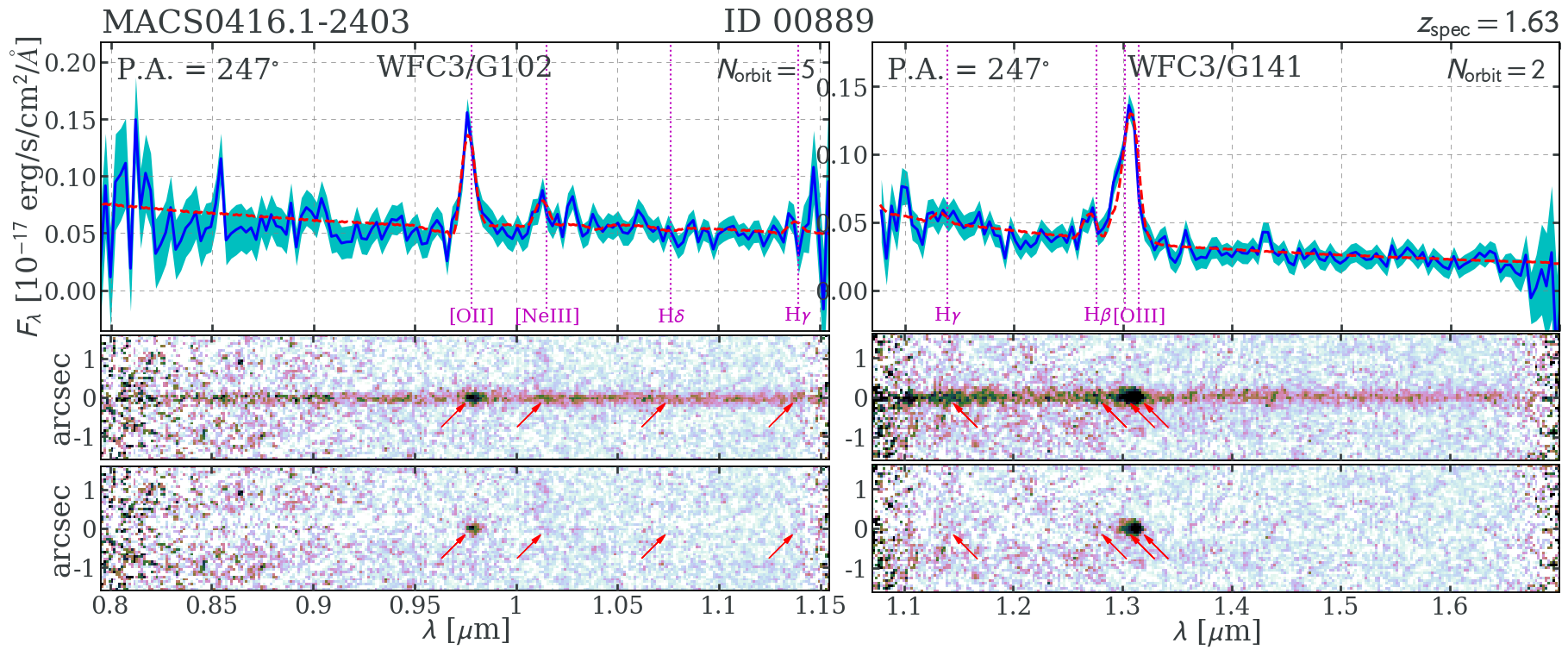}\\
    \includegraphics[width=.16\textwidth]{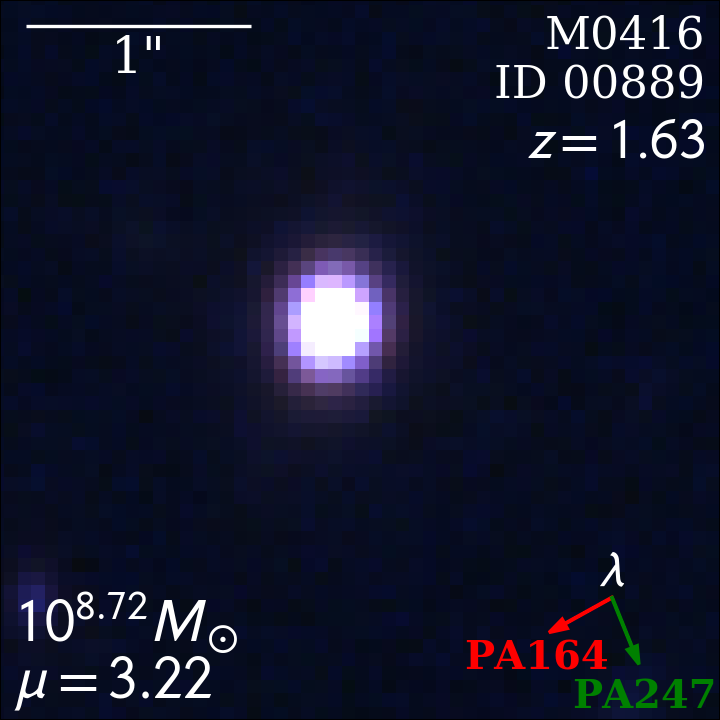}
    \includegraphics[width=.16\textwidth]{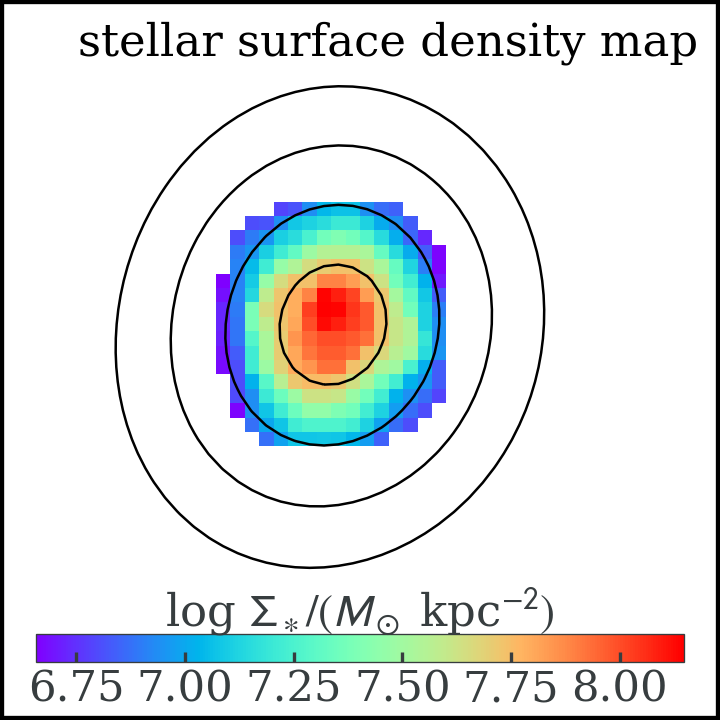}
    \includegraphics[width=.16\textwidth]{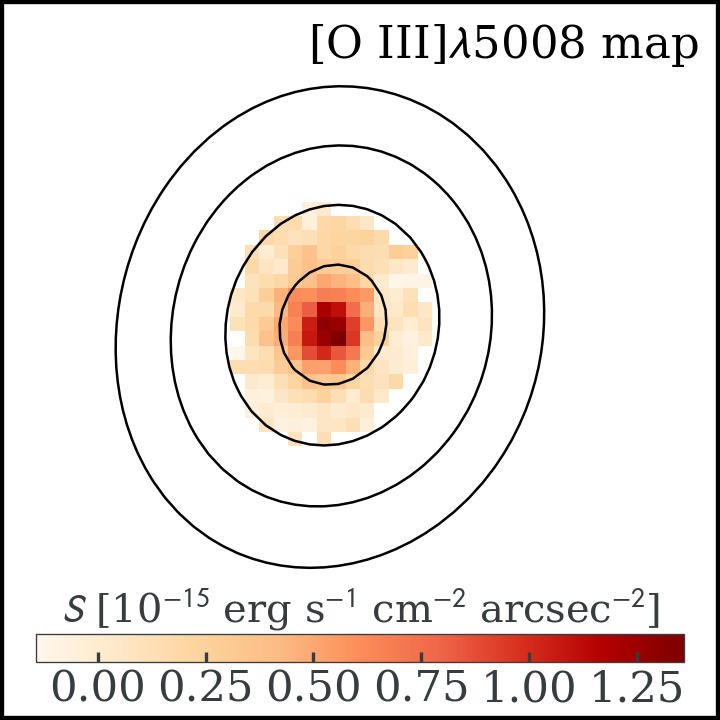}
    \includegraphics[width=.16\textwidth]{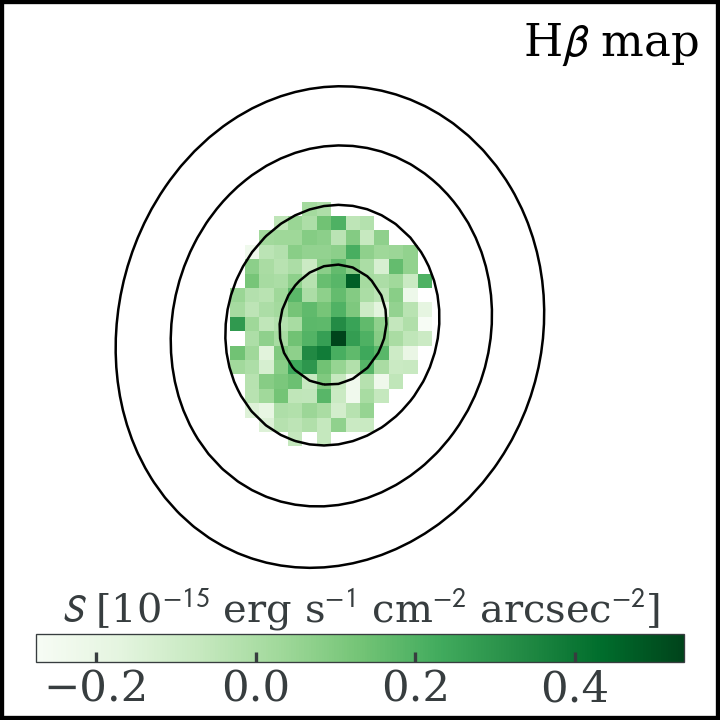}
    \includegraphics[width=.16\textwidth]{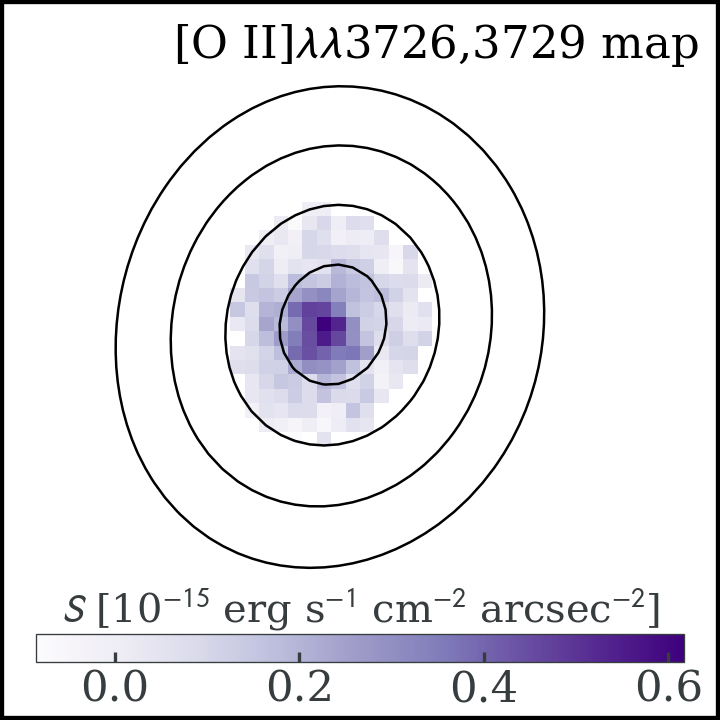}
    \includegraphics[width=.16\textwidth]{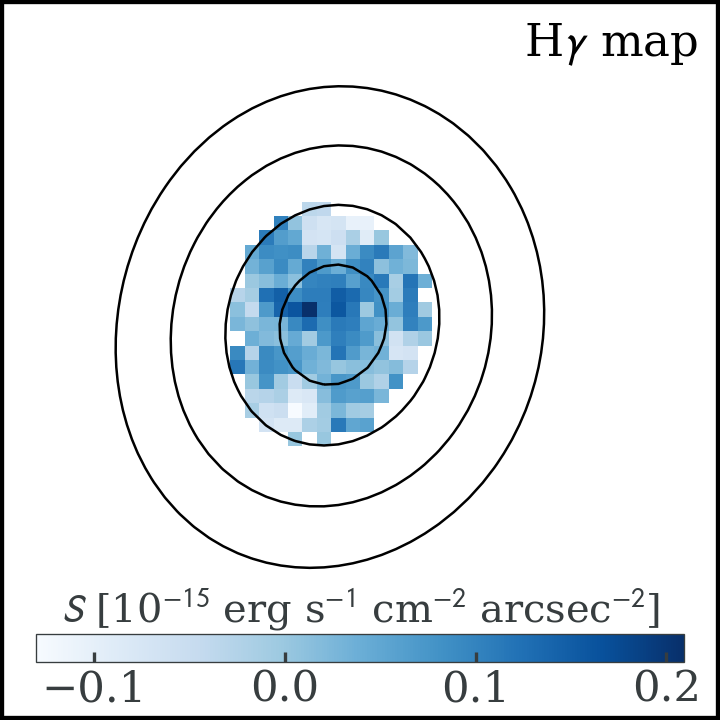}\\
    \includegraphics[width=\textwidth]{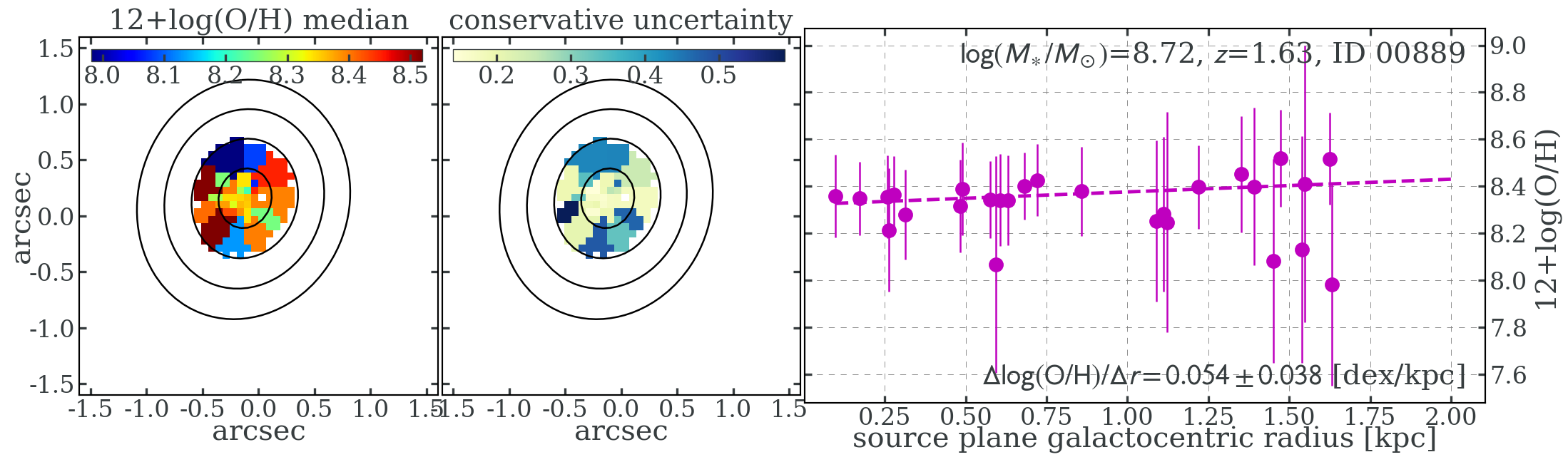}
    \caption{The source ID00889 in the field of \clsi is shown.}
    \label{fig:clM0416_ID00889_figs}
\end{figure*}

\begin{figure*}
    \centering
    \includegraphics[width=\textwidth]{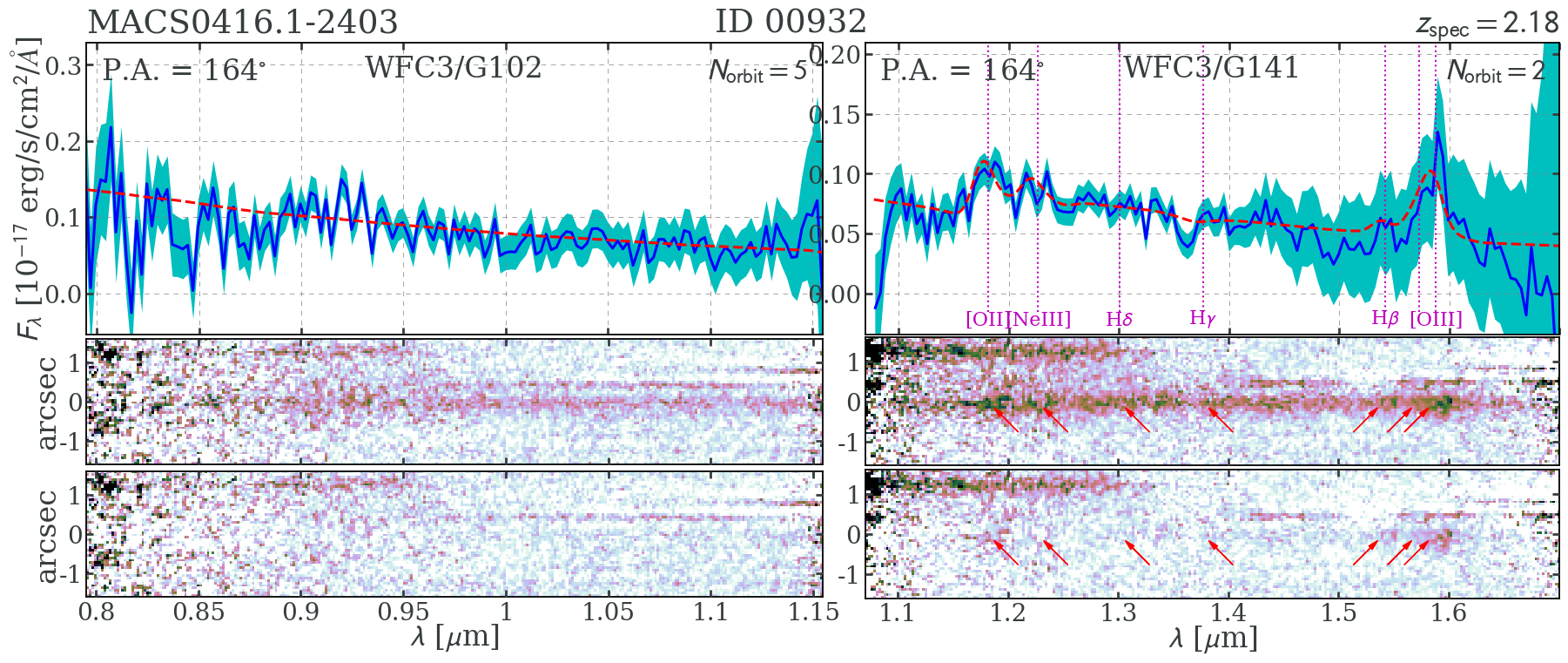}\\
    \includegraphics[width=\textwidth]{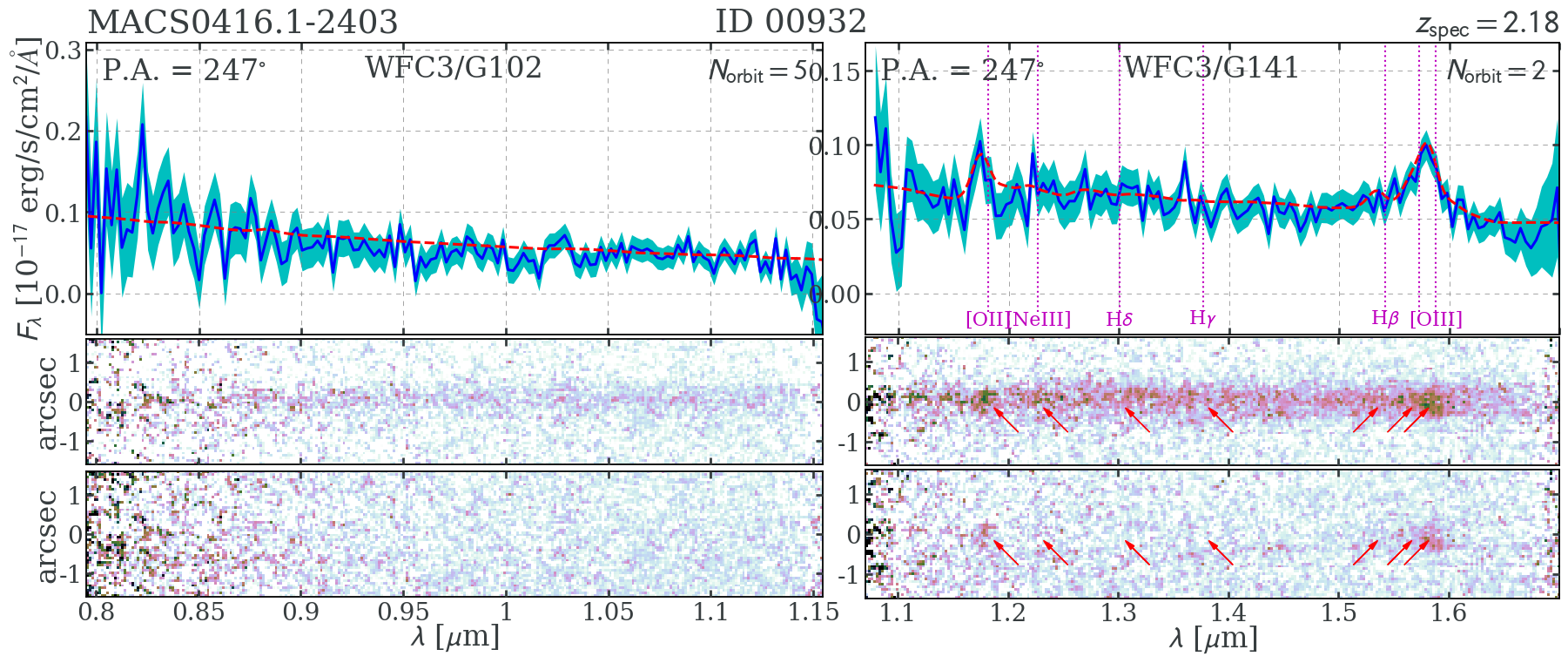}\\
    \includegraphics[width=.16\textwidth]{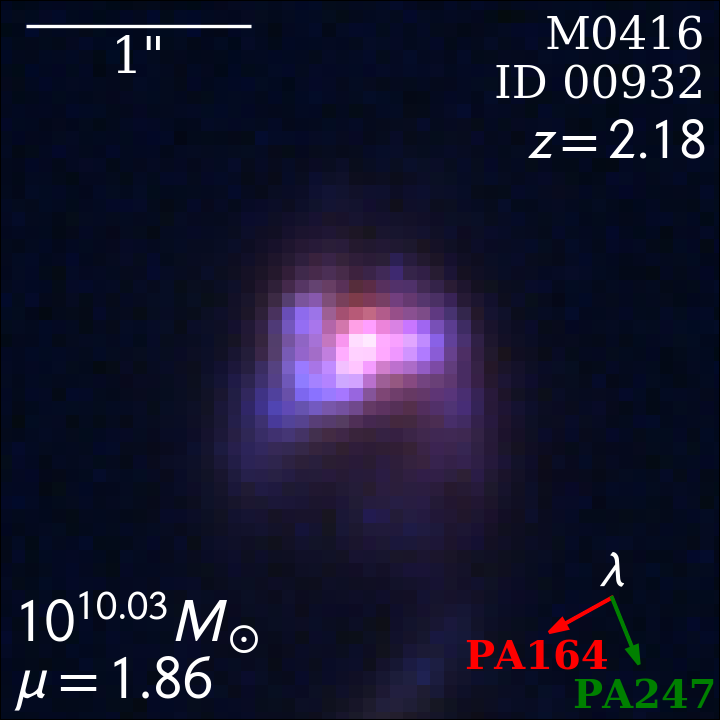}
    \includegraphics[width=.16\textwidth]{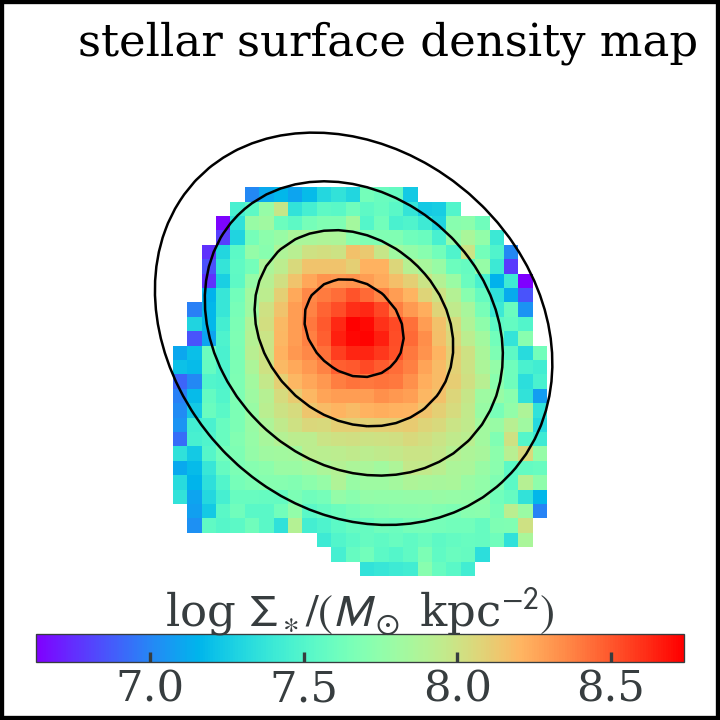}
    \includegraphics[width=.16\textwidth]{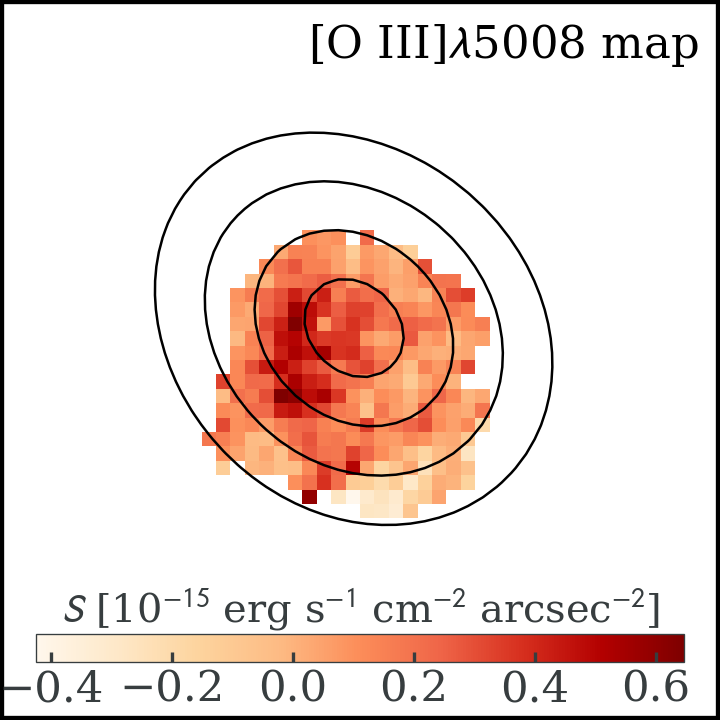}
    \includegraphics[width=.16\textwidth]{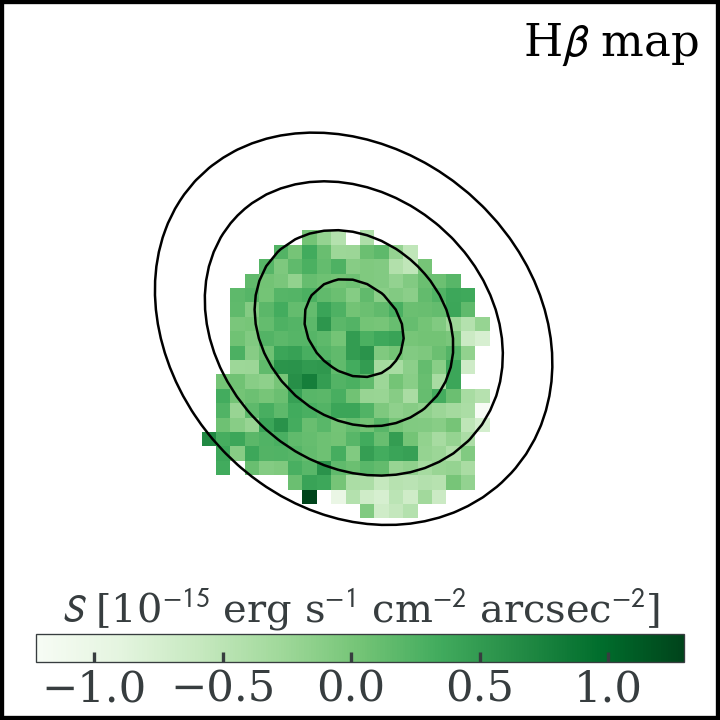}
    \includegraphics[width=.16\textwidth]{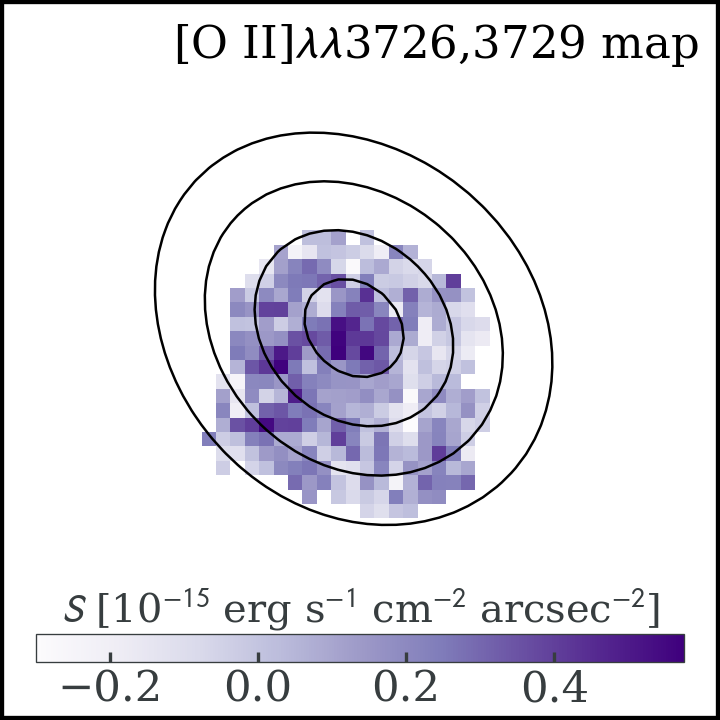}
    \includegraphics[width=.16\textwidth]{fig_ELmaps/baiban.png}\\
    \includegraphics[width=\textwidth]{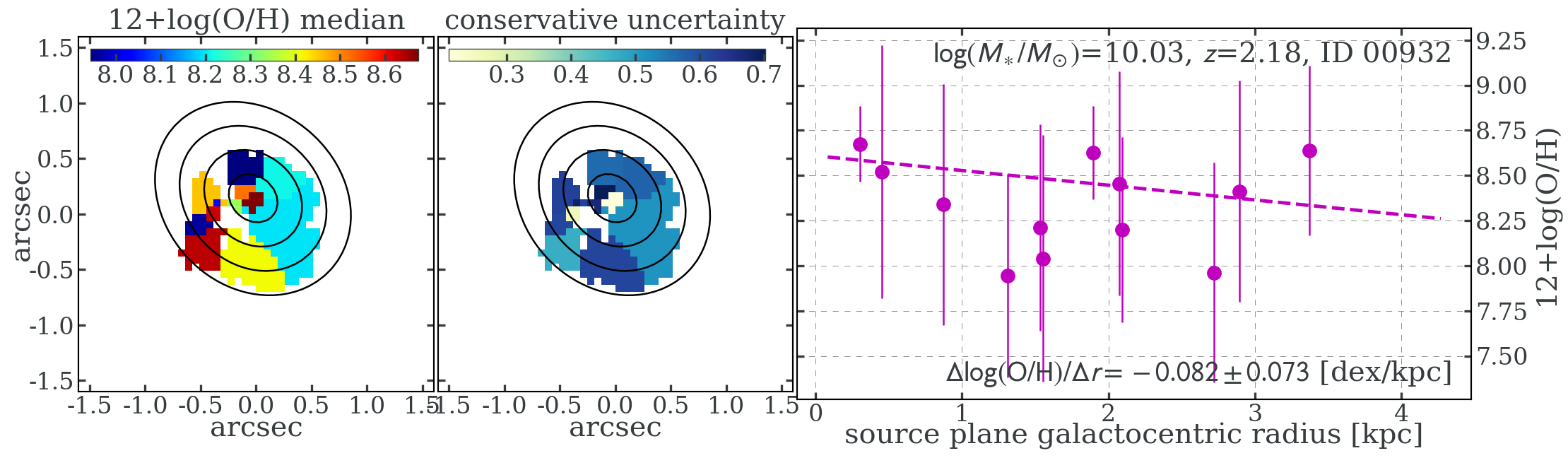}
    \caption{The source ID00932 in the field of \clsi is shown.}
    \label{fig:clM0416_ID00932_figs}
\end{figure*}

\begin{figure*}
    \centering
    \includegraphics[width=\textwidth]{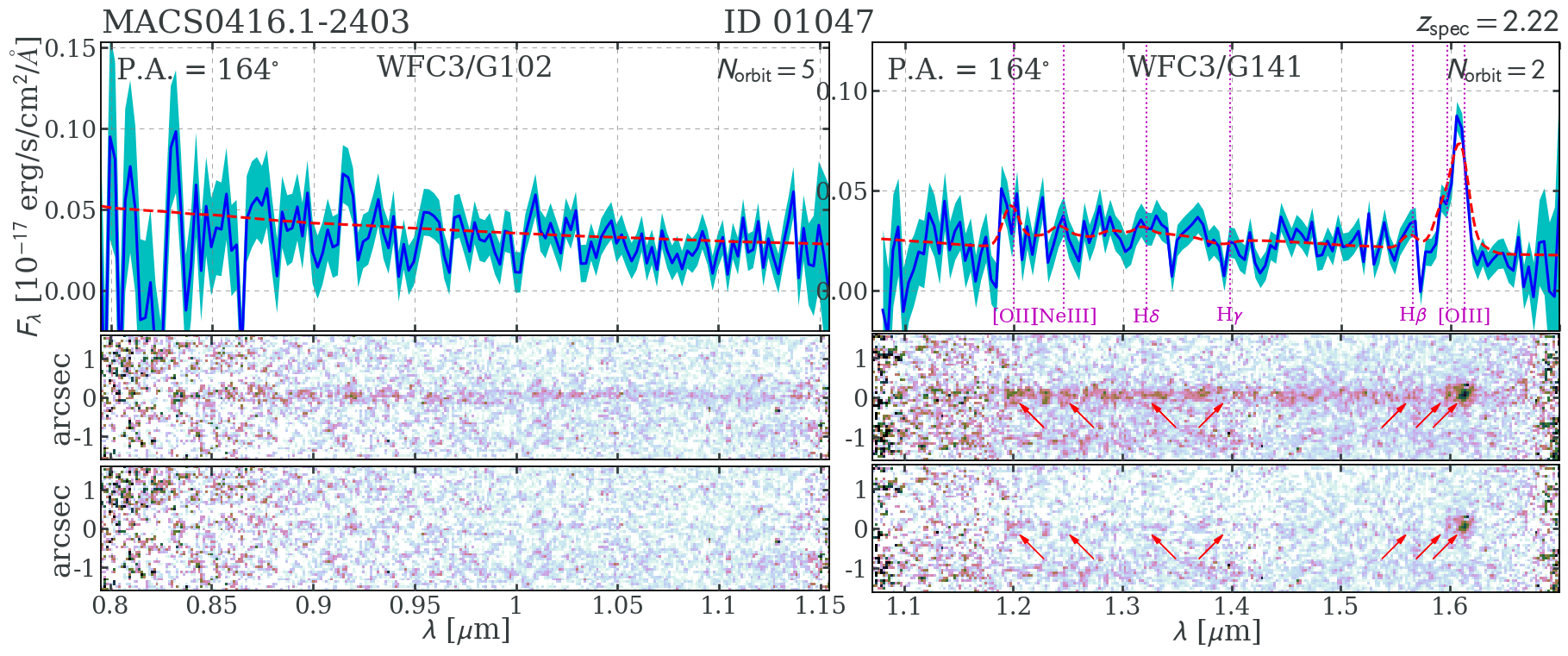}\\
    \includegraphics[width=\textwidth]{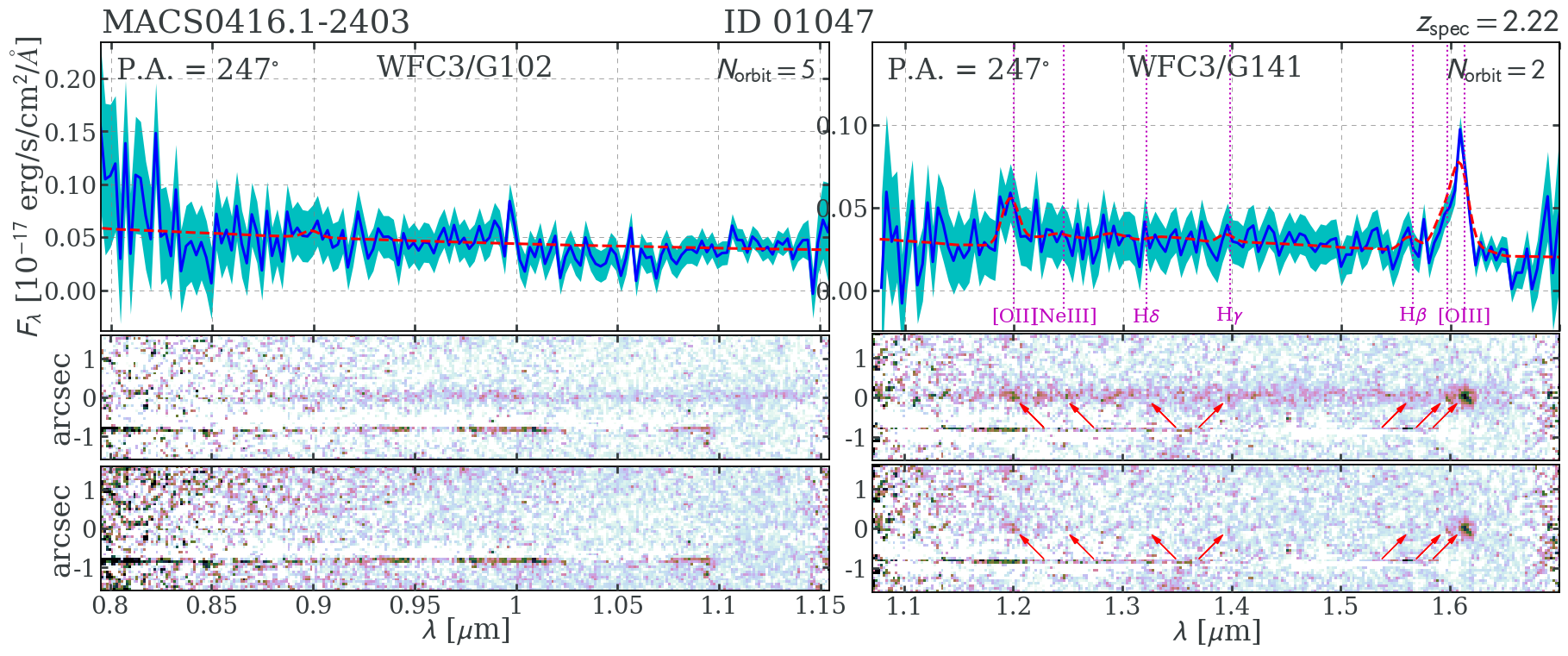}\\
    \includegraphics[width=.16\textwidth]{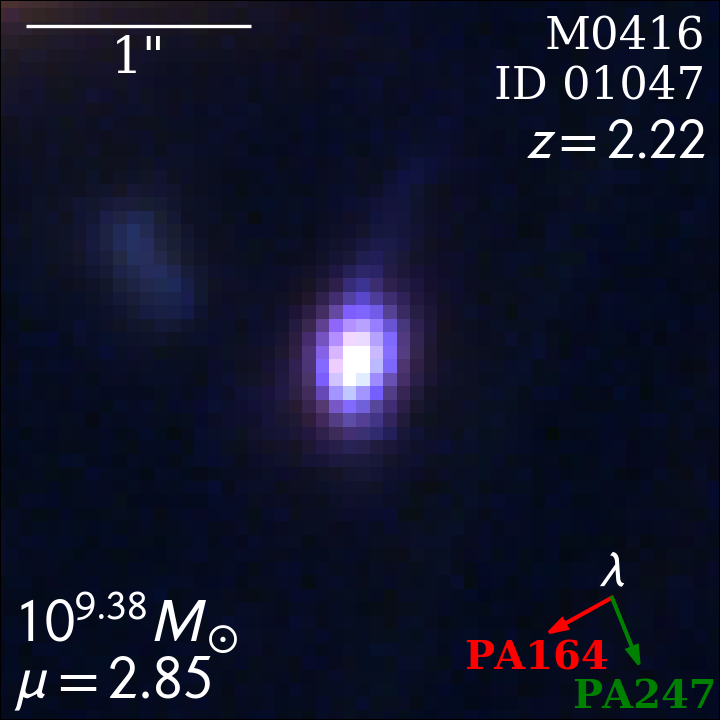}
    \includegraphics[width=.16\textwidth]{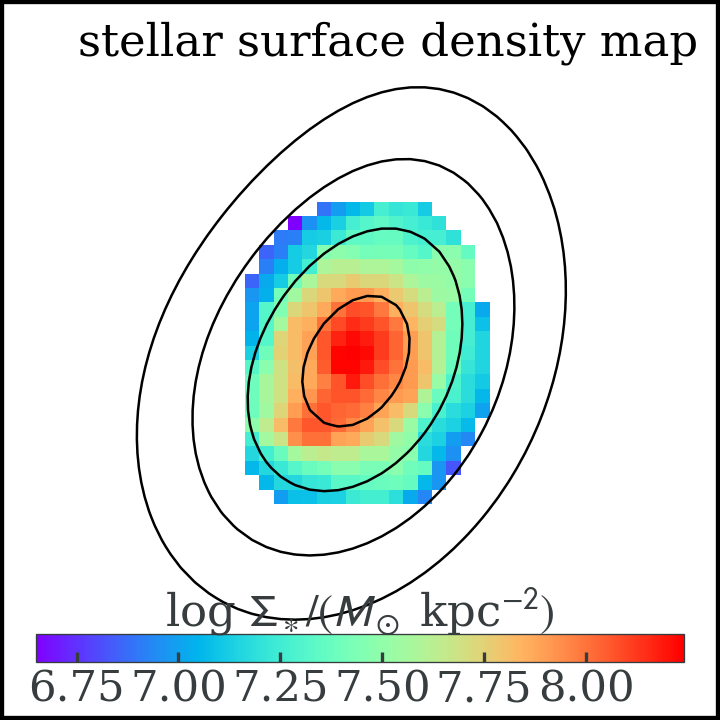}
    \includegraphics[width=.16\textwidth]{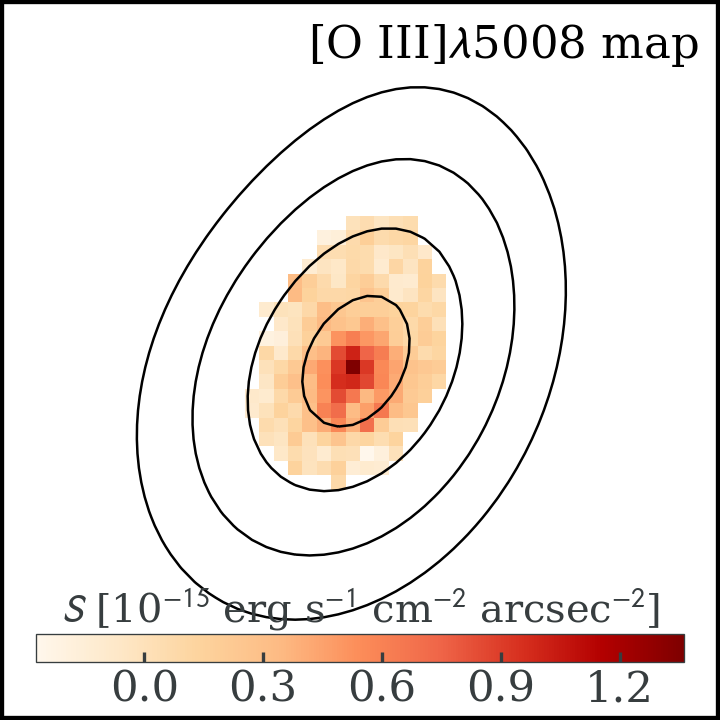}
    \includegraphics[width=.16\textwidth]{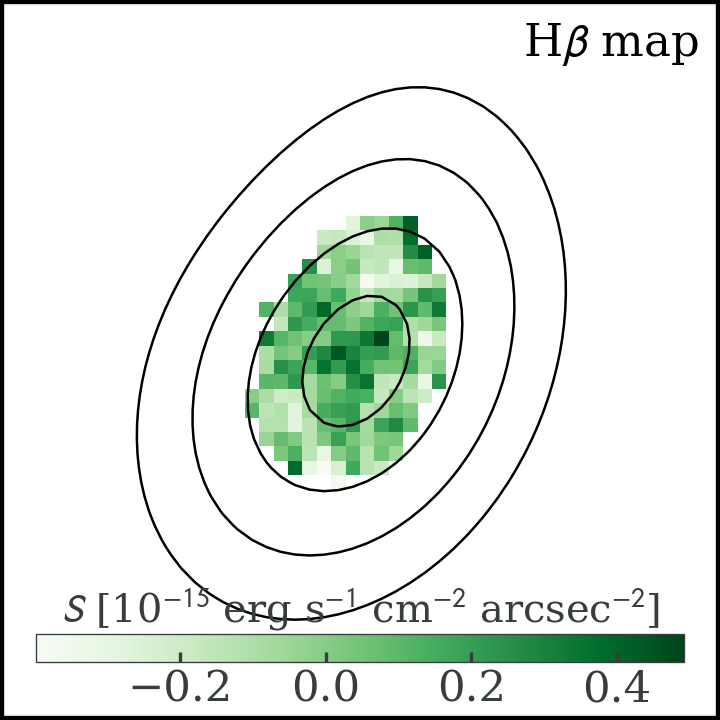}
    \includegraphics[width=.16\textwidth]{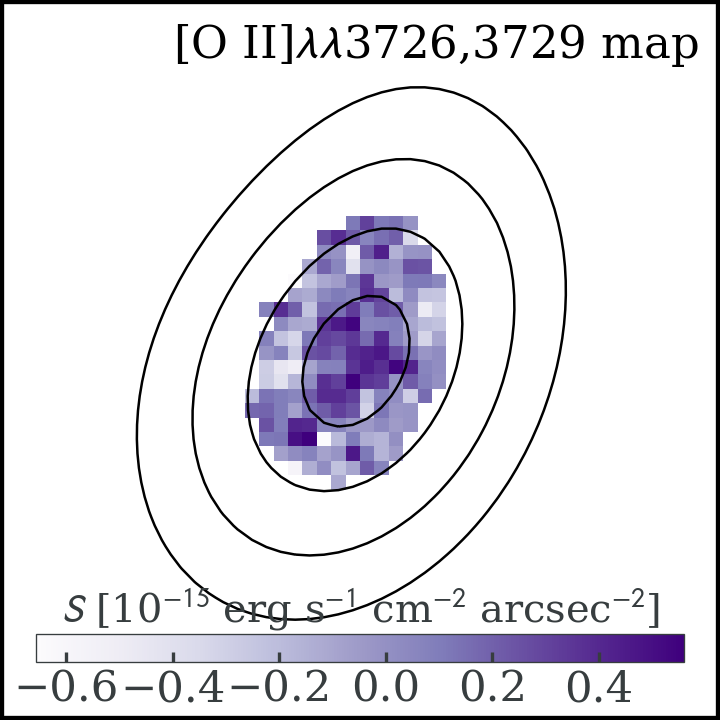}
    \includegraphics[width=.16\textwidth]{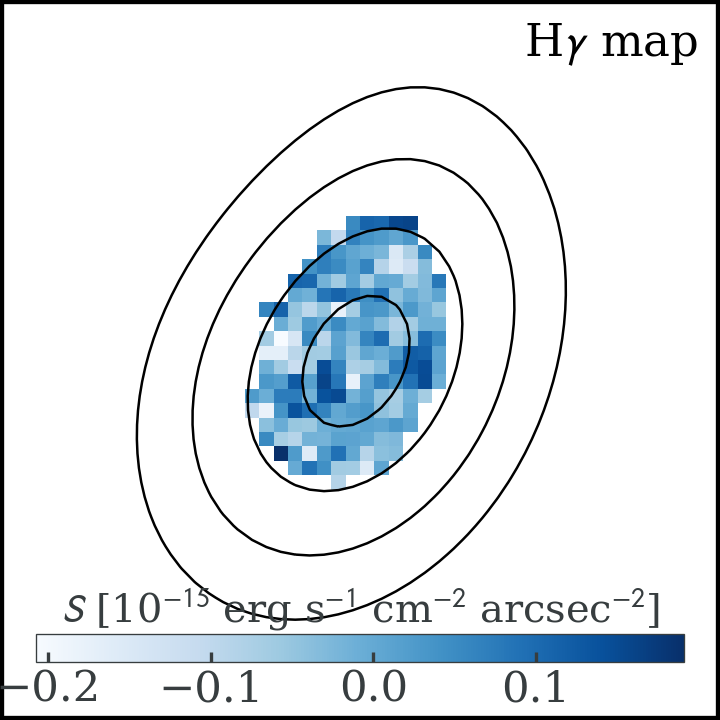}\\
    \includegraphics[width=\textwidth]{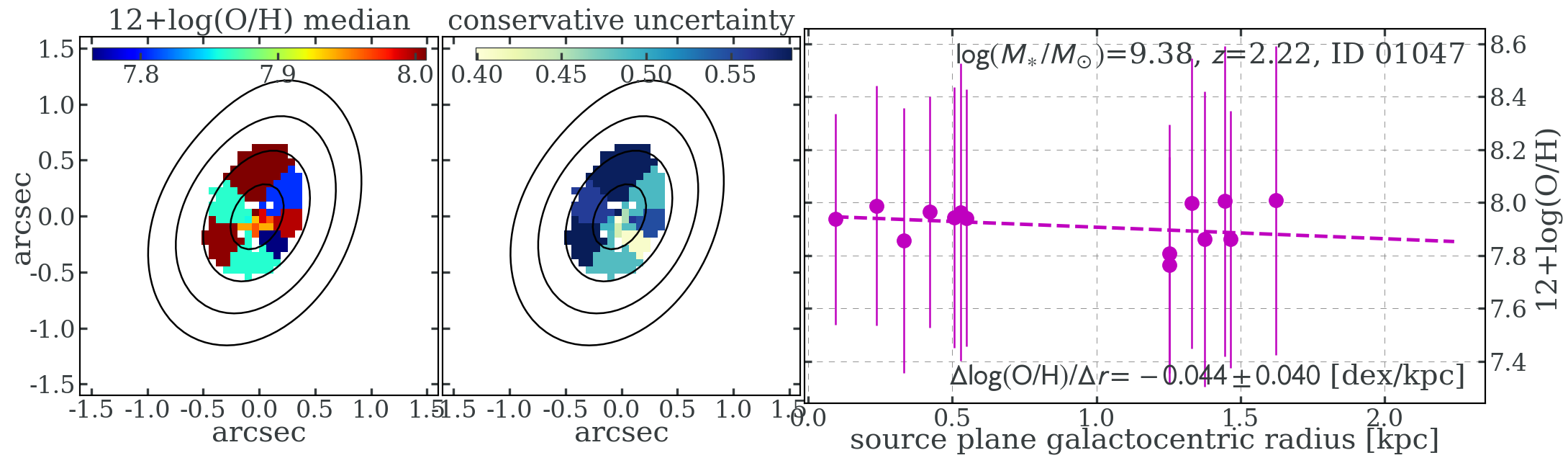}
    \caption{The source ID01047 in the field of \clsi is shown.}
    \label{fig:clM0416_ID01047_figs}
\end{figure*}

\begin{figure*}
    \centering
    \includegraphics[width=\textwidth]{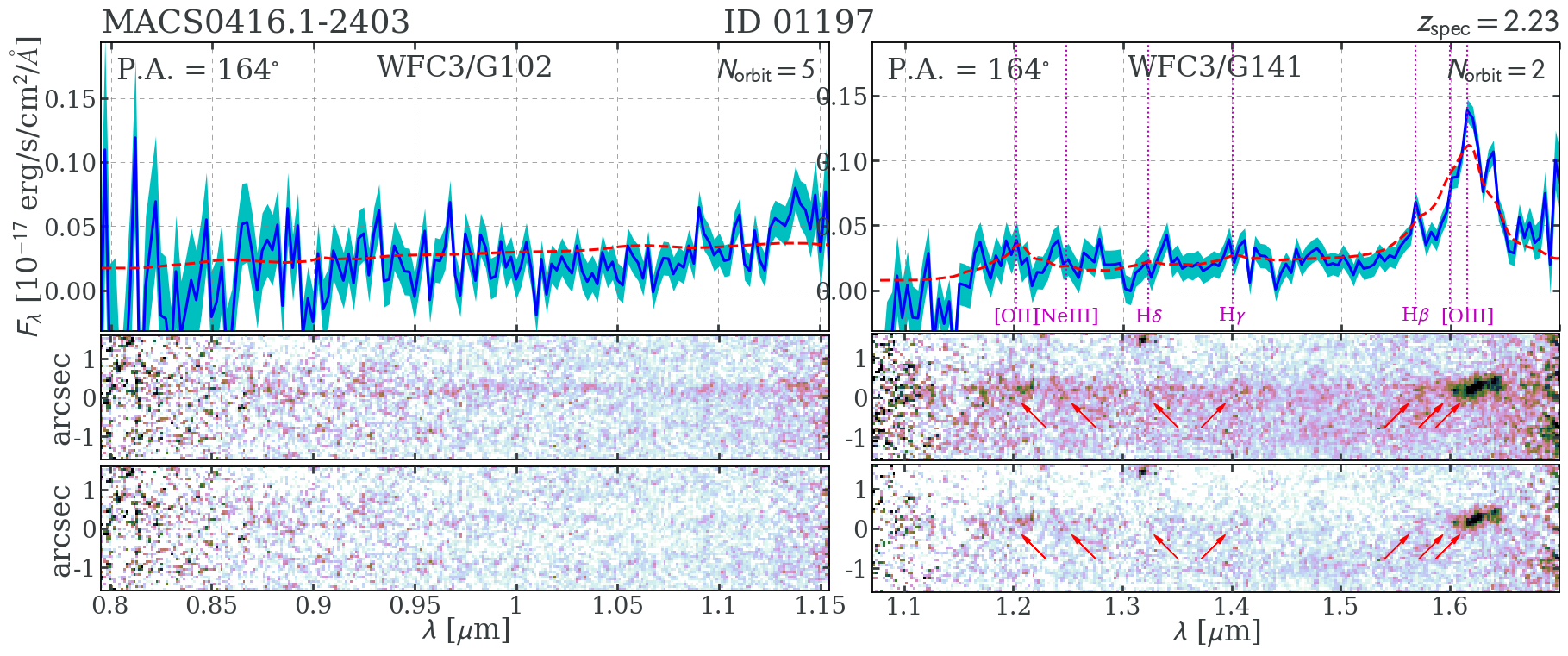}\\
    \includegraphics[width=\textwidth]{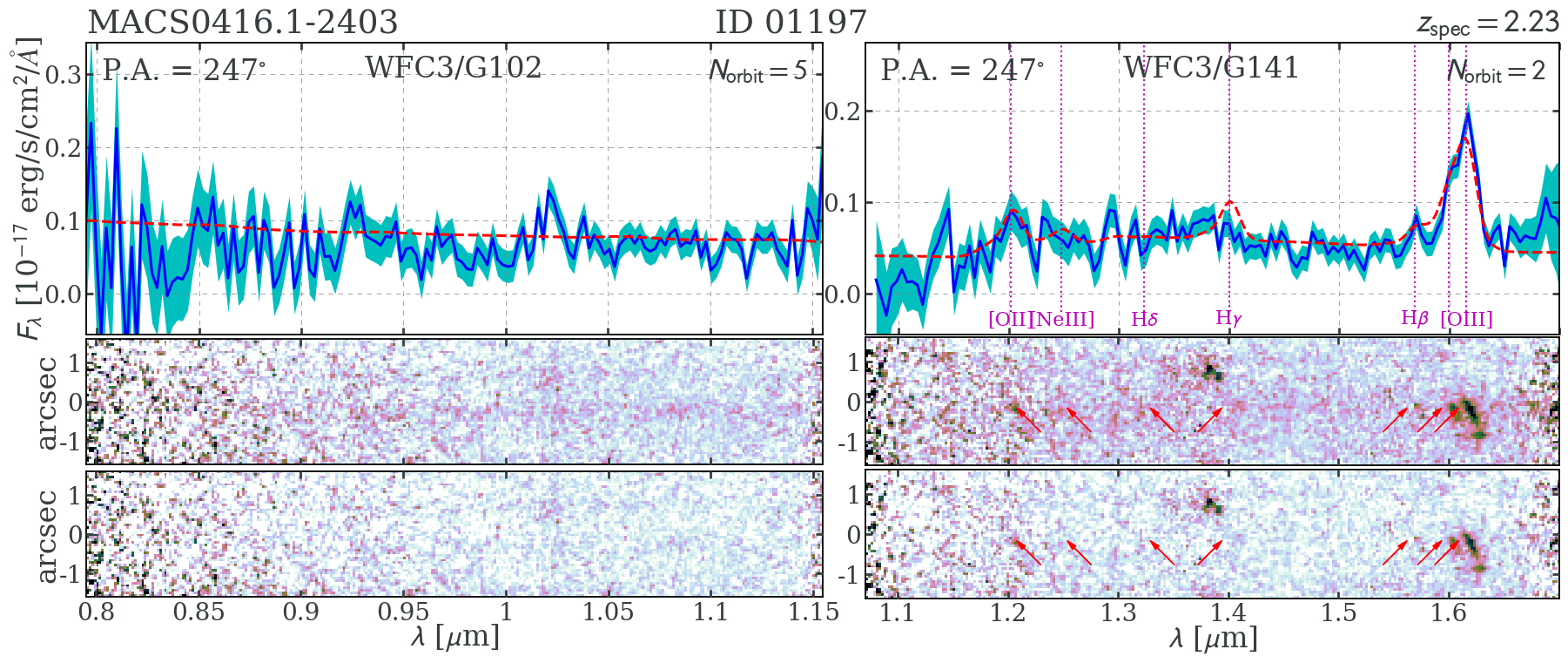}\\
    \includegraphics[width=.16\textwidth]{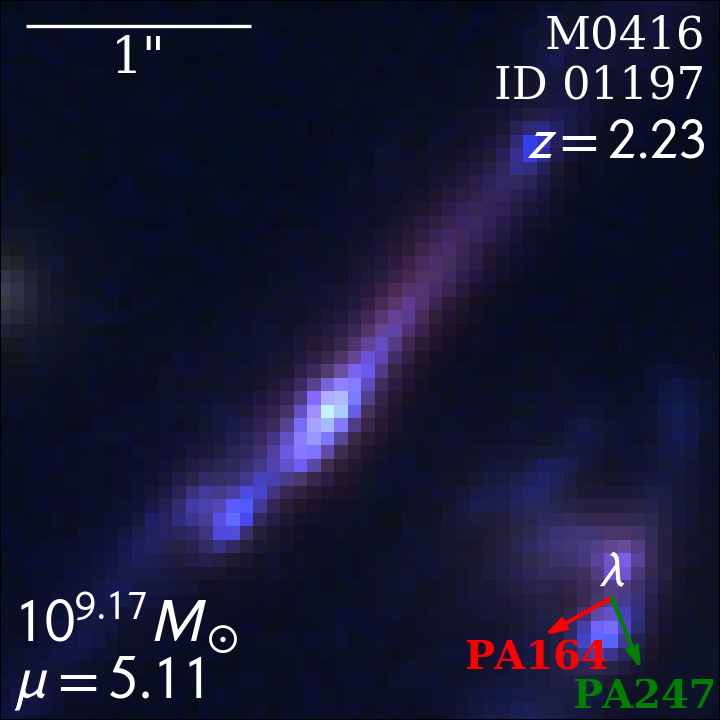}
    \includegraphics[width=.16\textwidth]{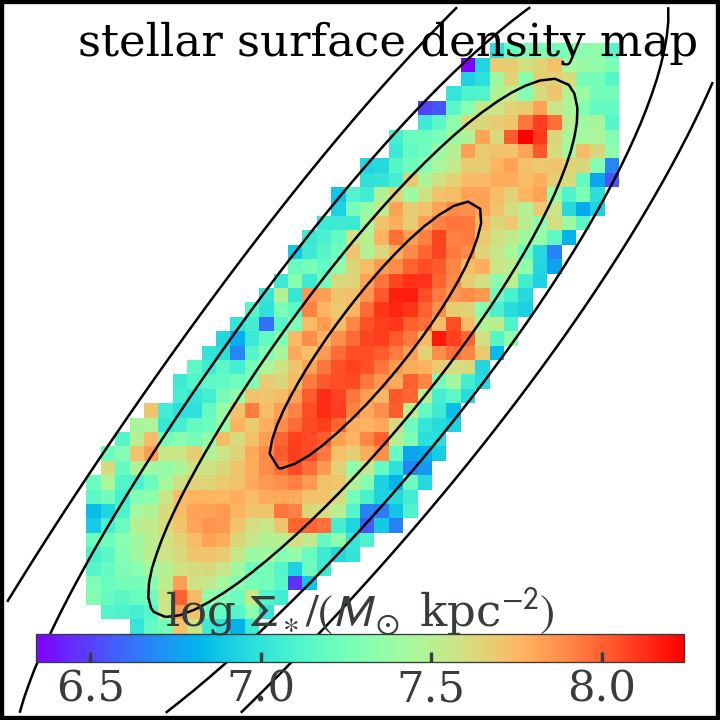}
    \includegraphics[width=.16\textwidth]{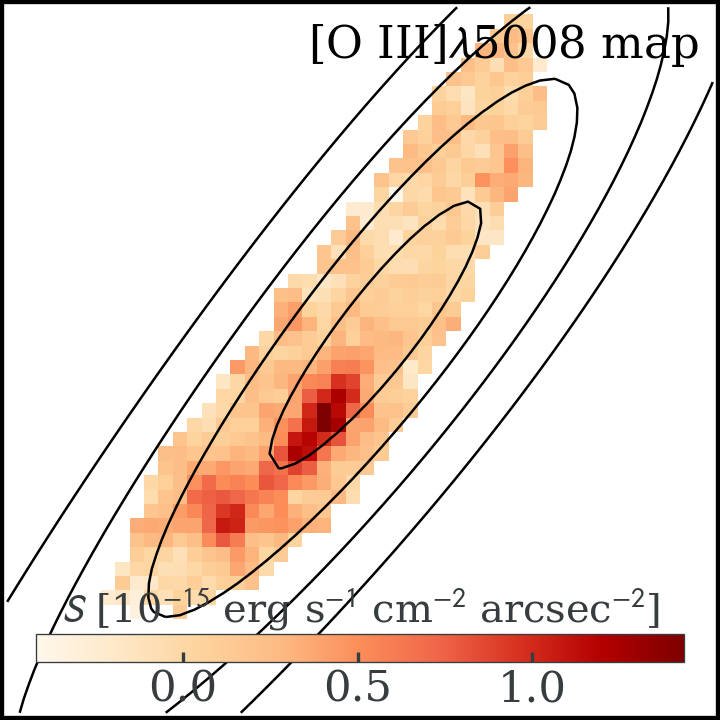}
    \includegraphics[width=.16\textwidth]{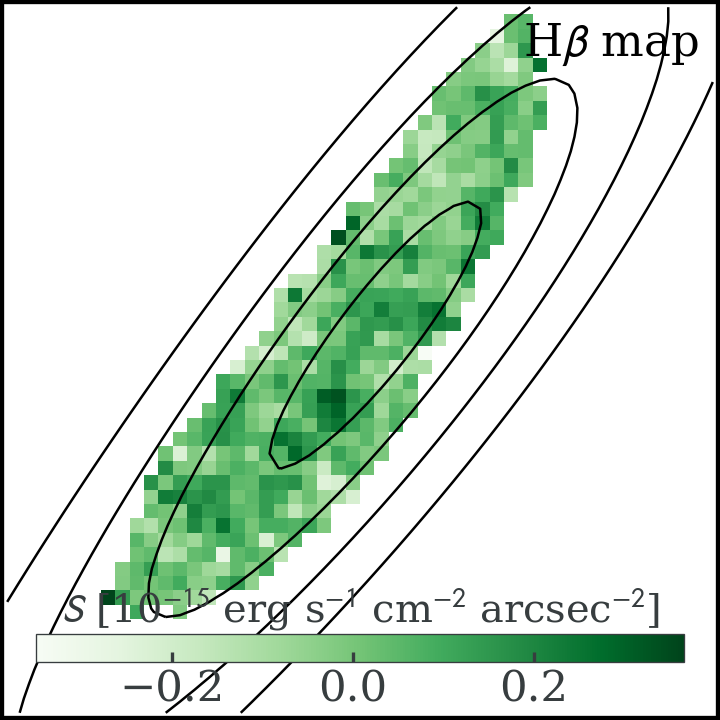}
    \includegraphics[width=.16\textwidth]{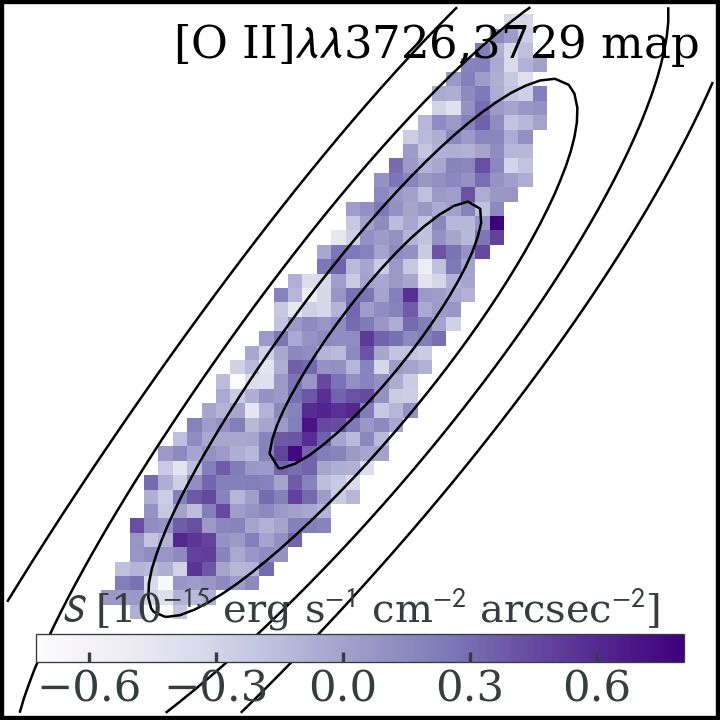}
    \includegraphics[width=.16\textwidth]{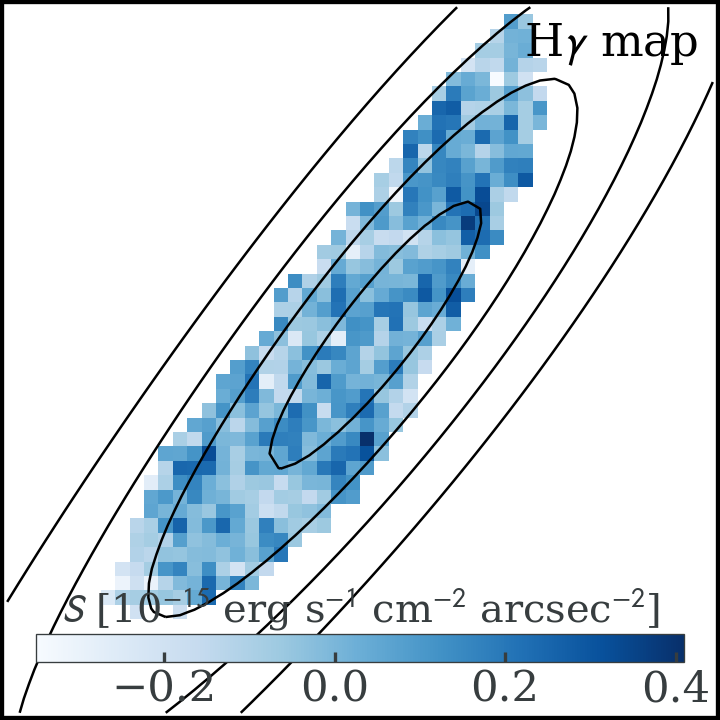}\\
    \includegraphics[width=\textwidth]{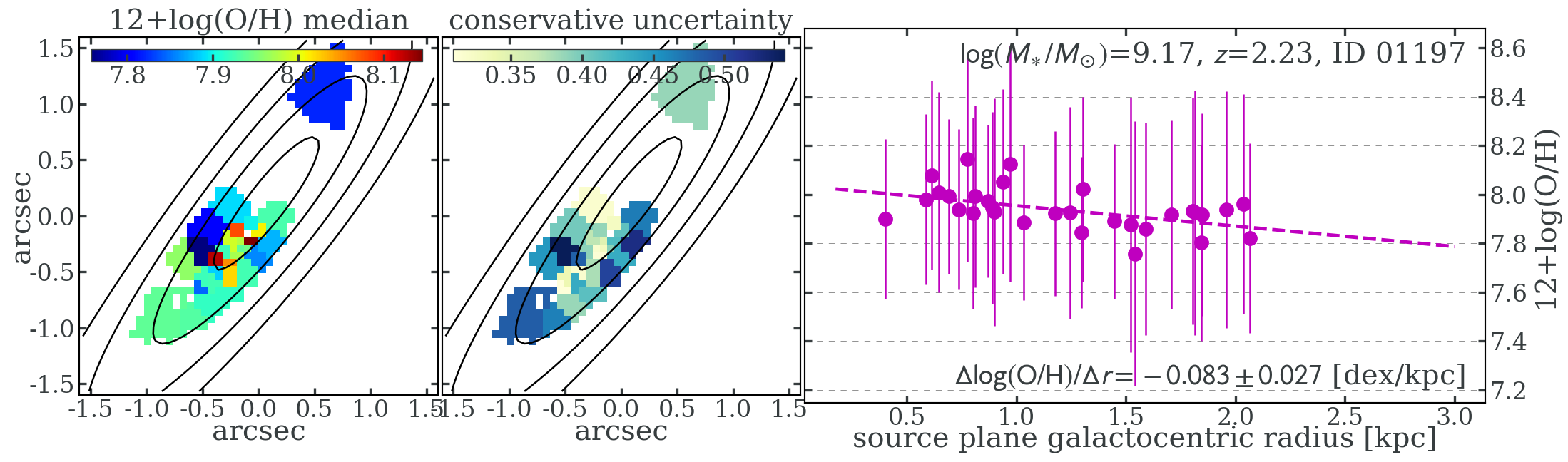}
    \caption{The source ID01197 in the field of \clsi is shown.}
    \label{fig:clM0416_ID01197_figs}
\end{figure*}

\begin{figure*}
    \centering
    \includegraphics[width=\textwidth]{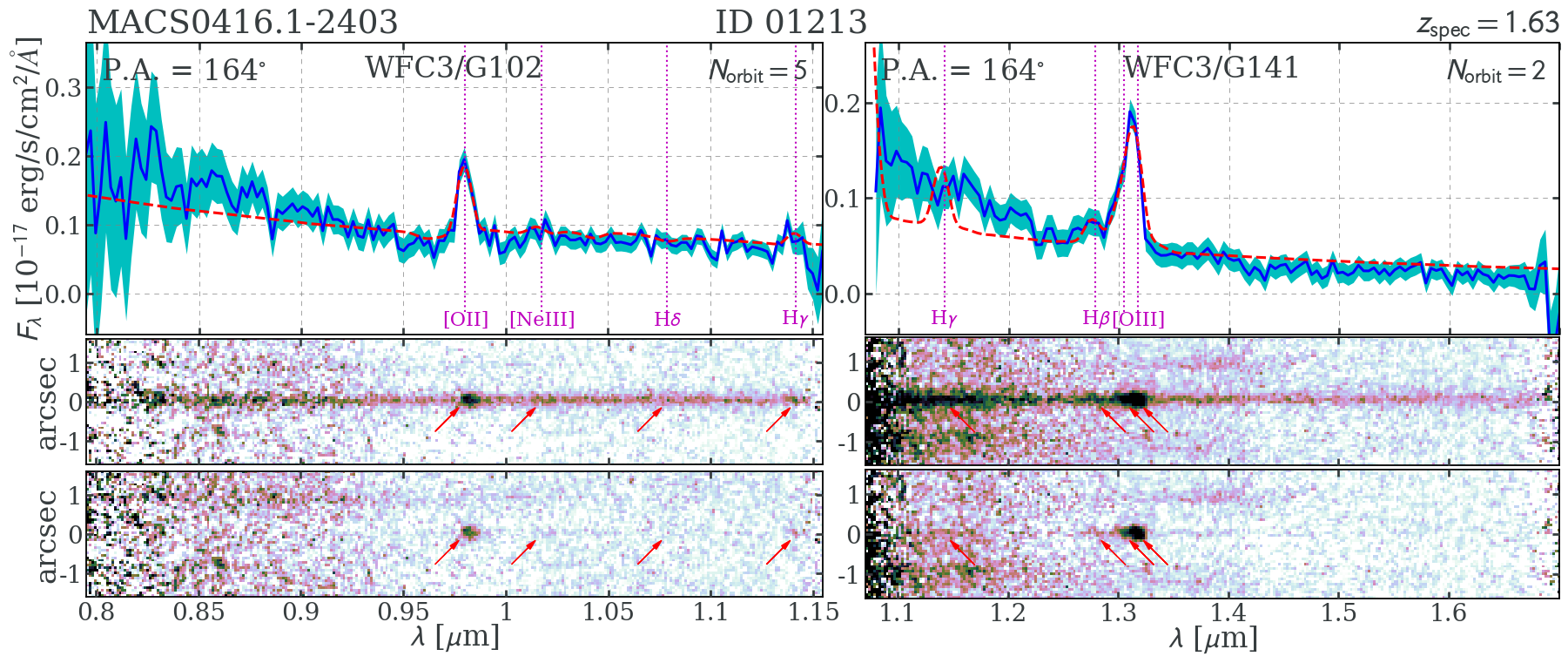}\\
    \includegraphics[width=\textwidth]{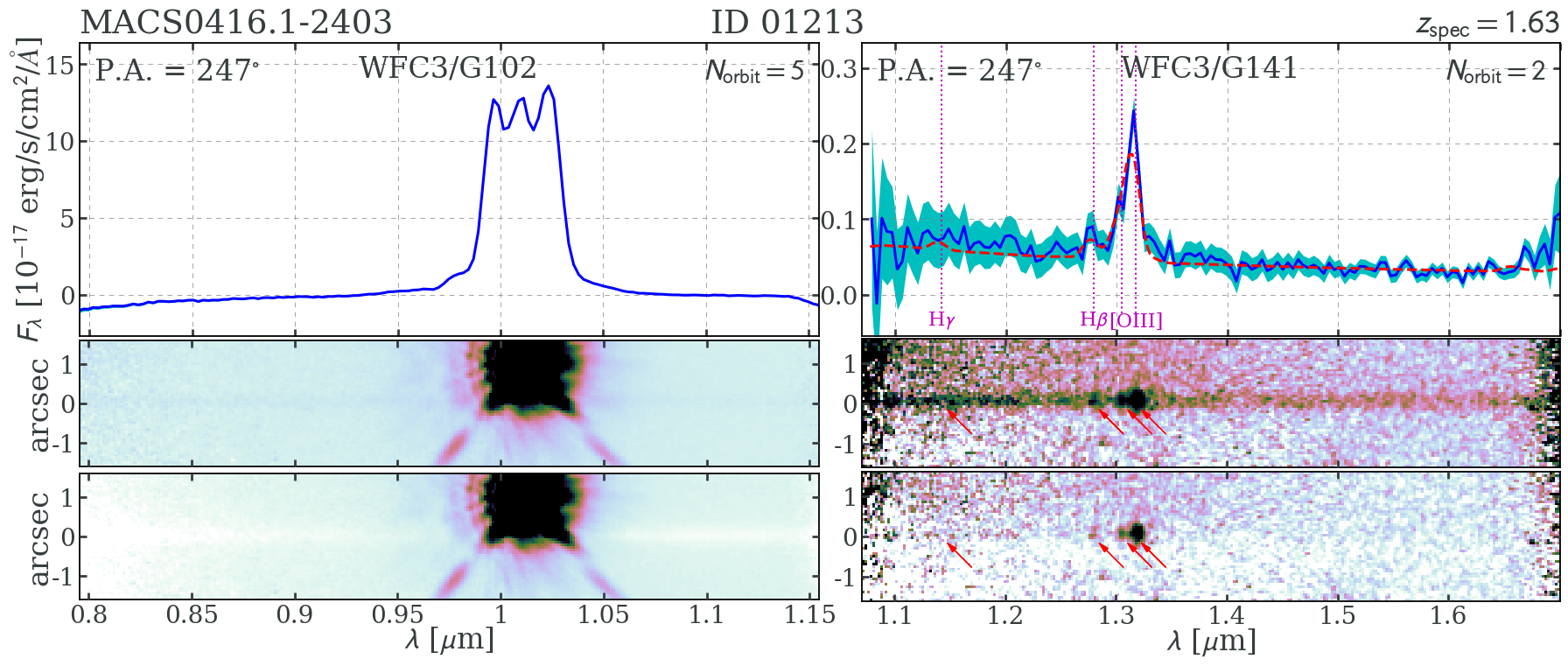}\\
    \includegraphics[width=.16\textwidth]{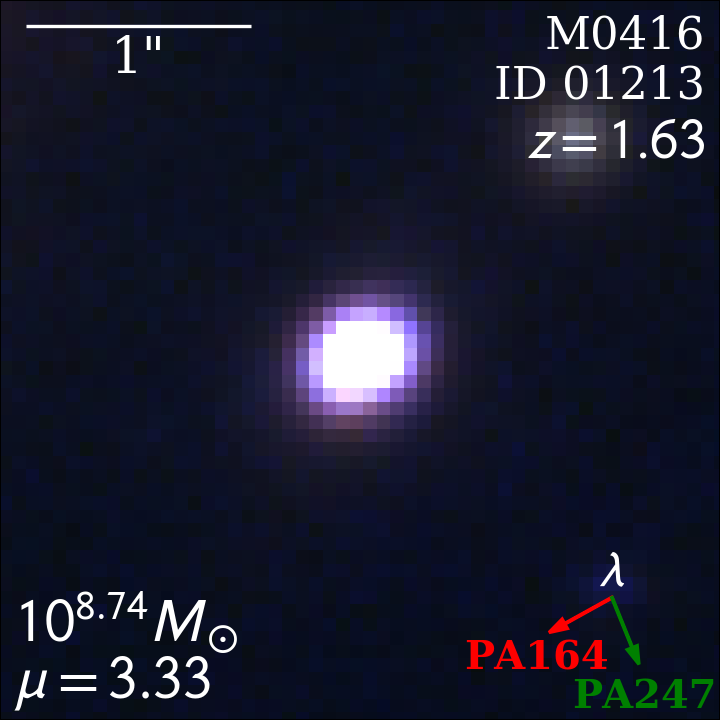}
    \includegraphics[width=.16\textwidth]{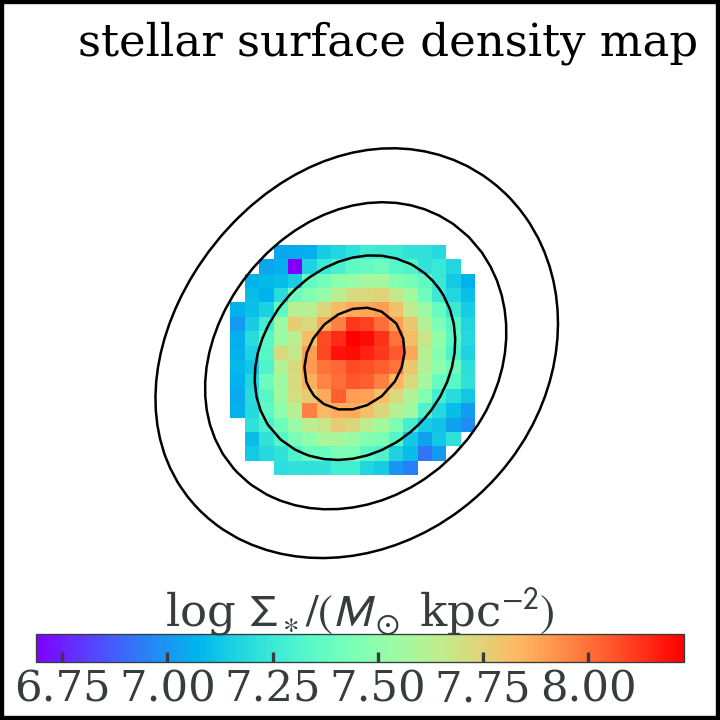}
    \includegraphics[width=.16\textwidth]{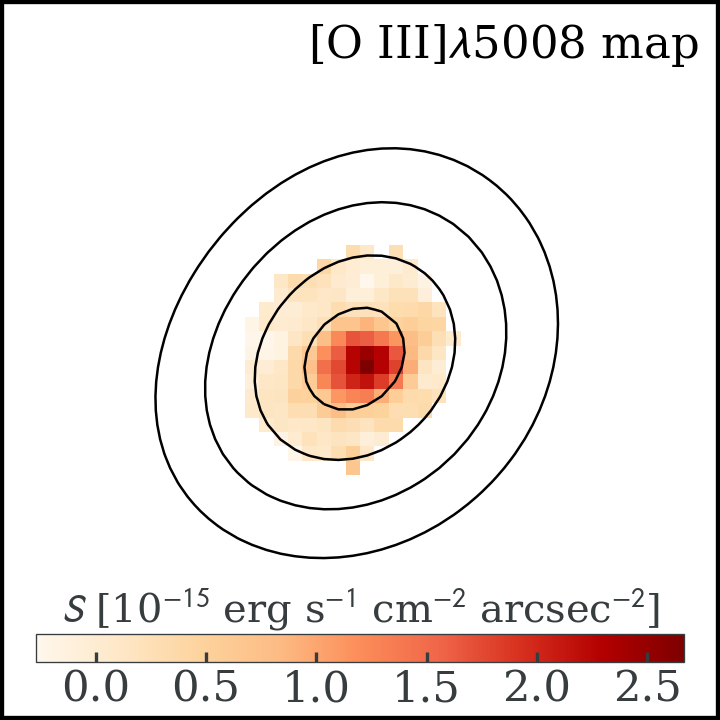}
    \includegraphics[width=.16\textwidth]{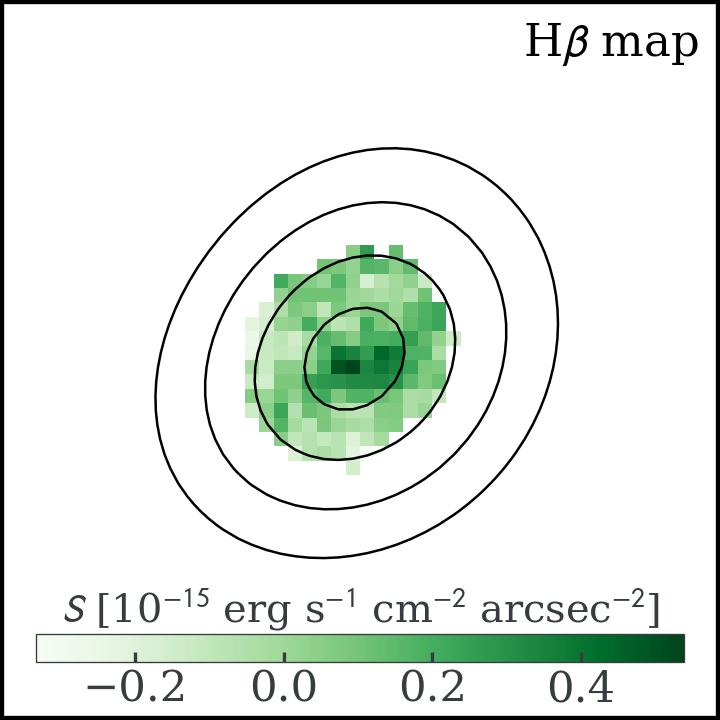}
    \includegraphics[width=.16\textwidth]{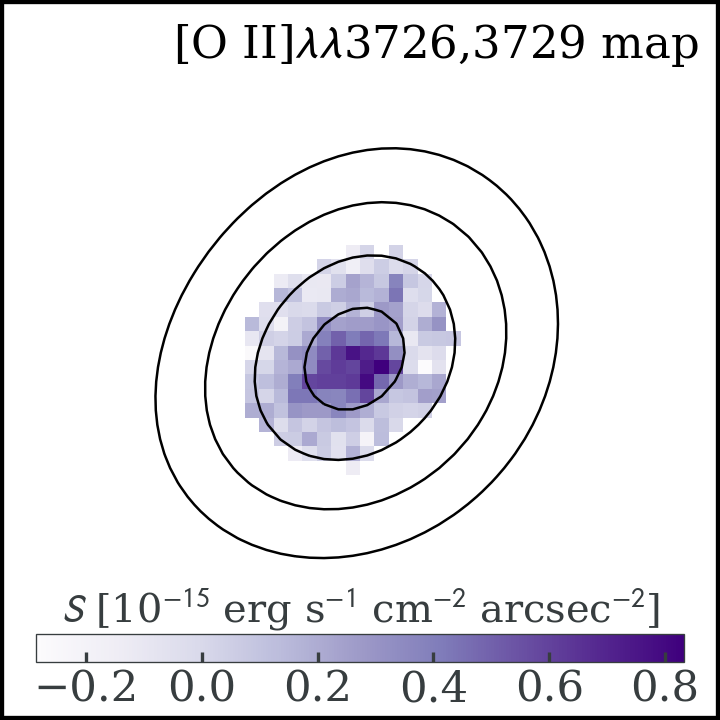}
    \includegraphics[width=.16\textwidth]{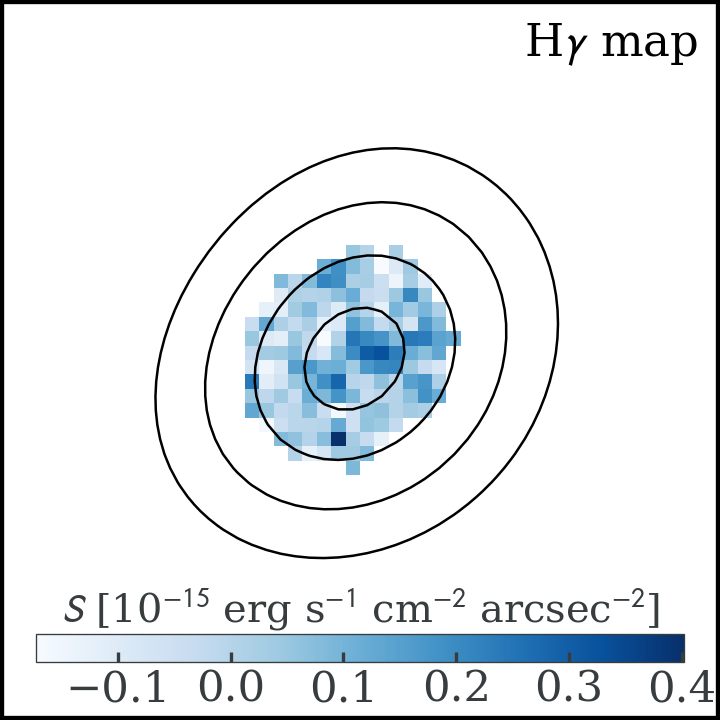}\\
    \includegraphics[width=\textwidth]{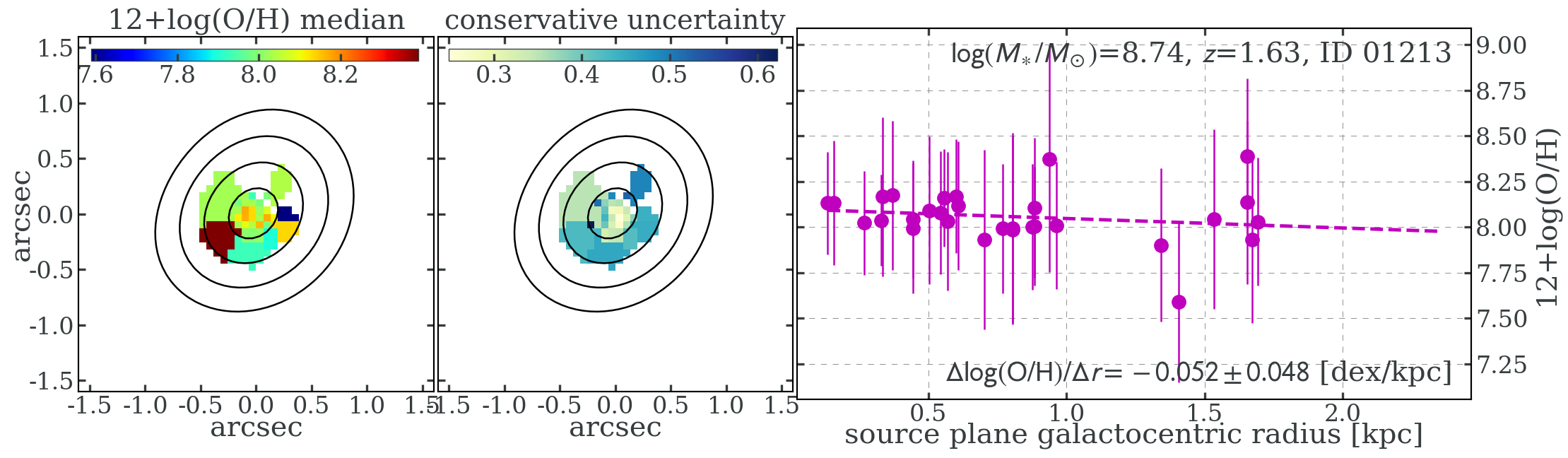}
    \caption{The source ID01213 in the field of \clsi is shown.}
    \label{fig:clM0416_ID01213_figs}
\end{figure*}

\begin{figure*}
    \centering
    \includegraphics[width=\textwidth]{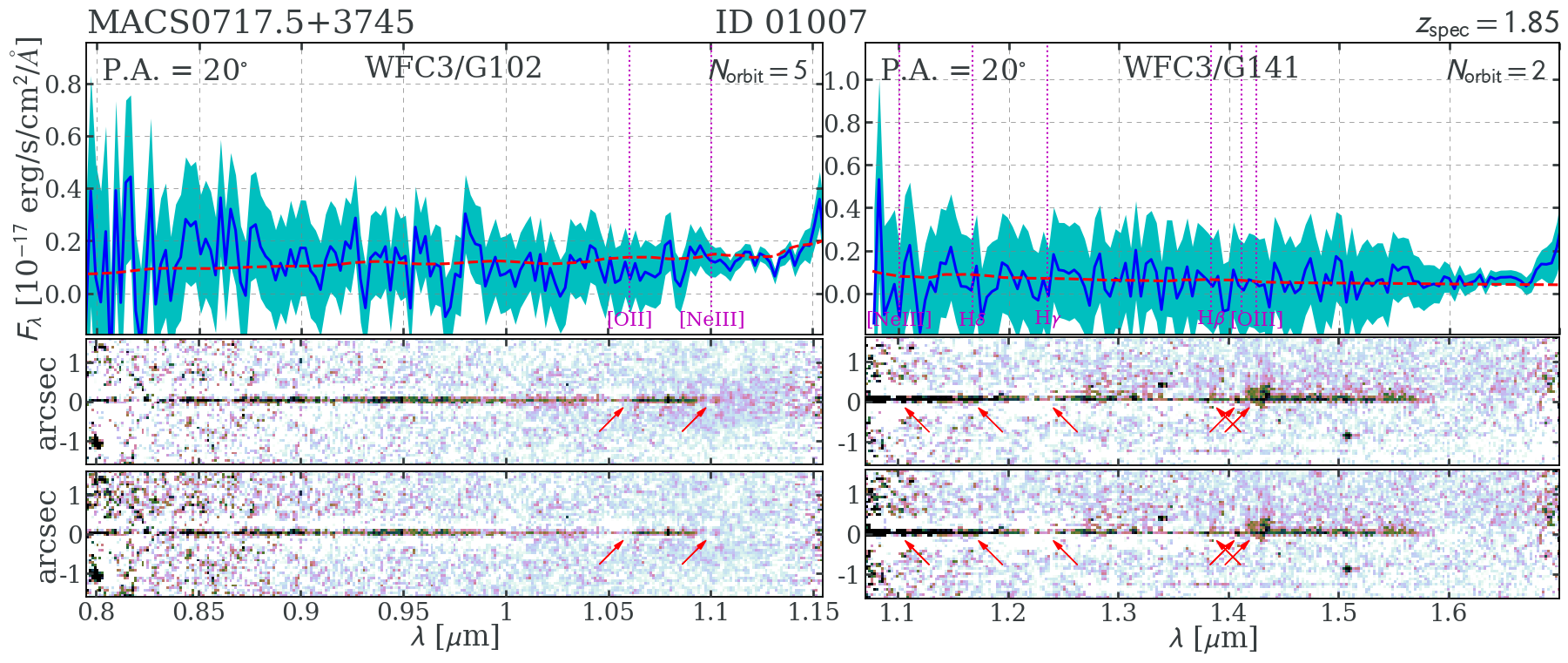}\\
    \includegraphics[width=\textwidth]{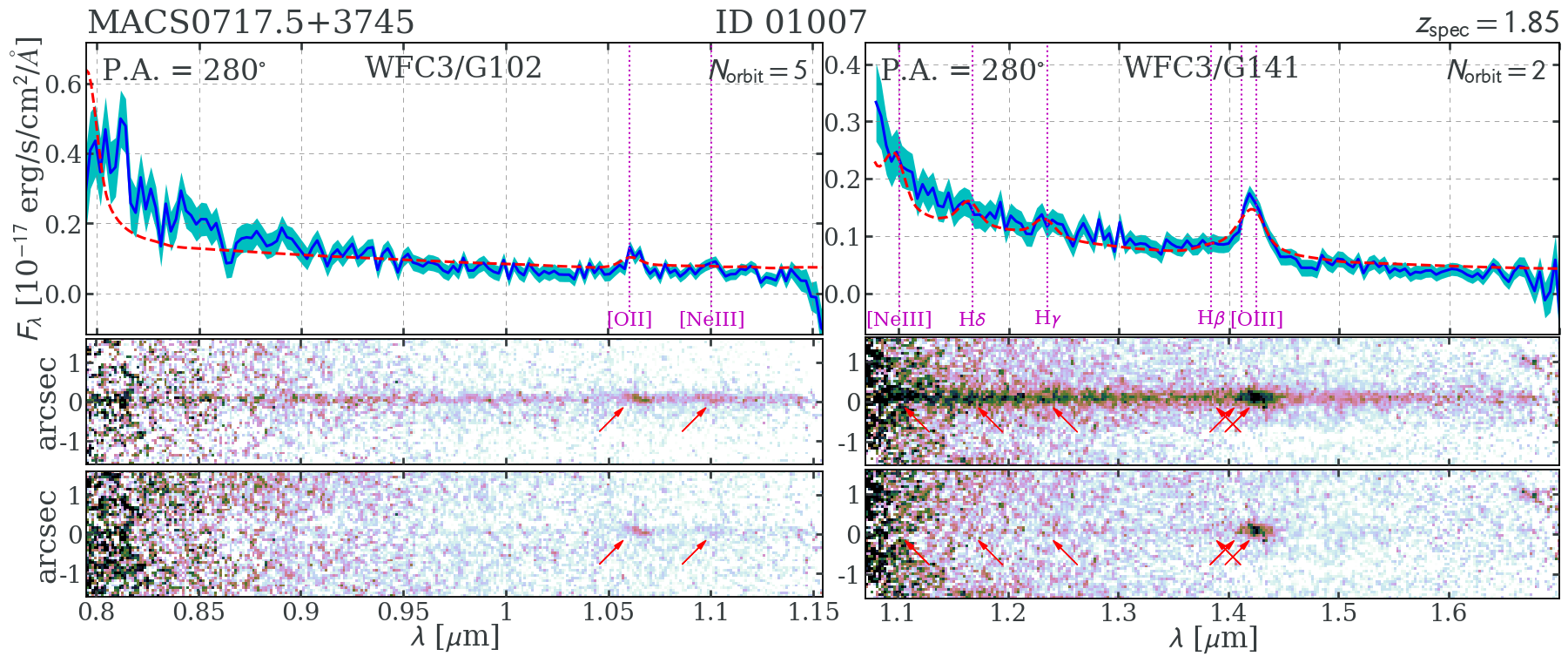}\\
    \includegraphics[width=.16\textwidth]{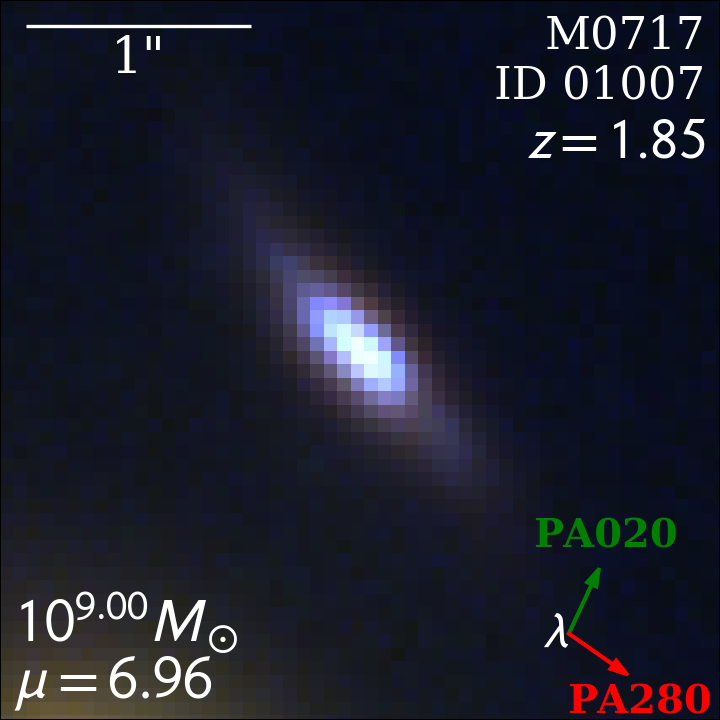}
    \includegraphics[width=.16\textwidth]{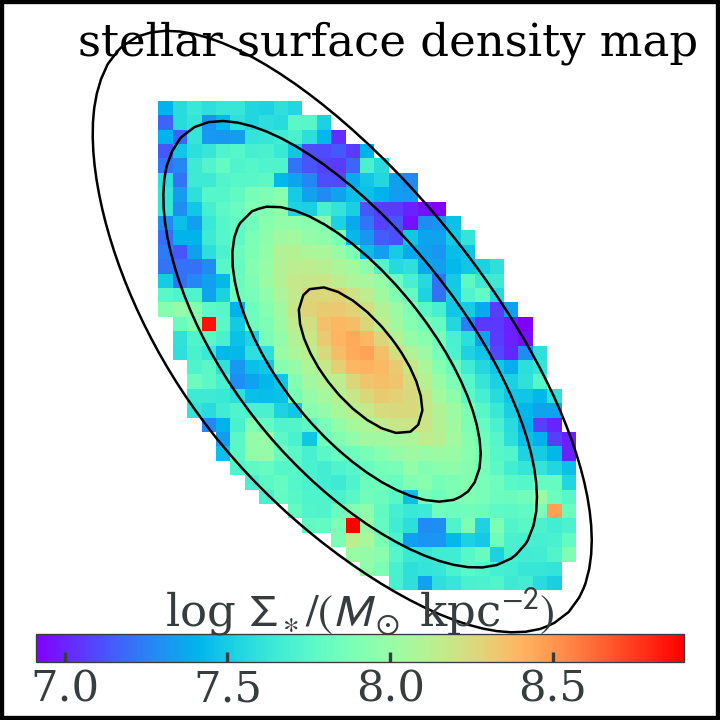}
    \includegraphics[width=.16\textwidth]{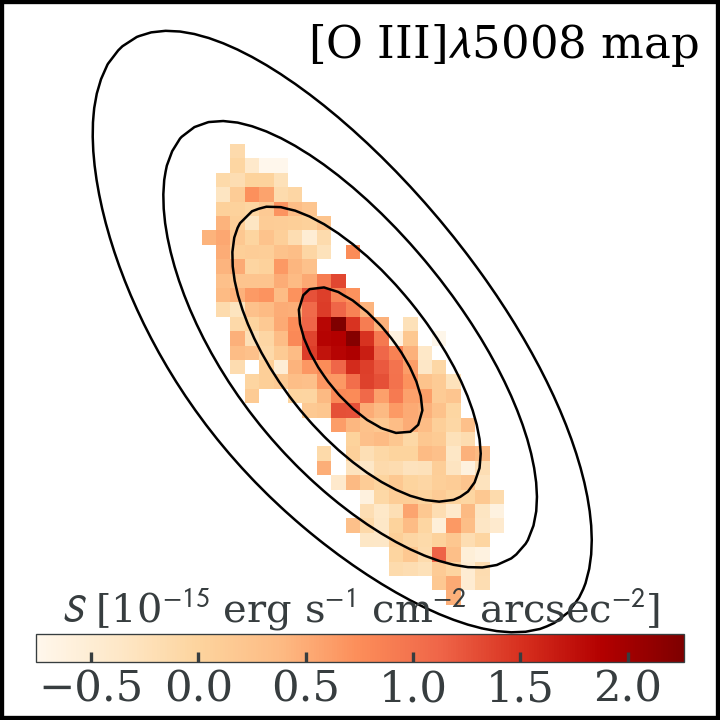}
    \includegraphics[width=.16\textwidth]{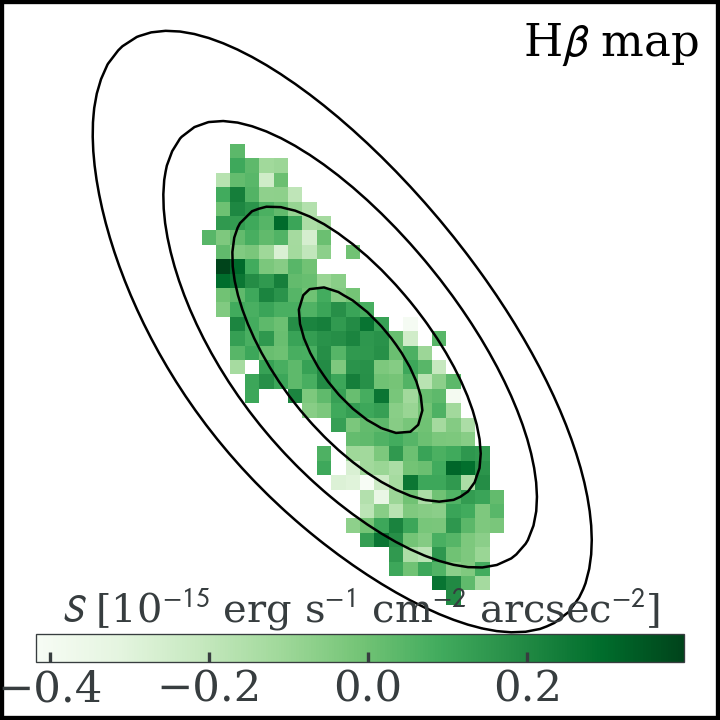}
    \includegraphics[width=.16\textwidth]{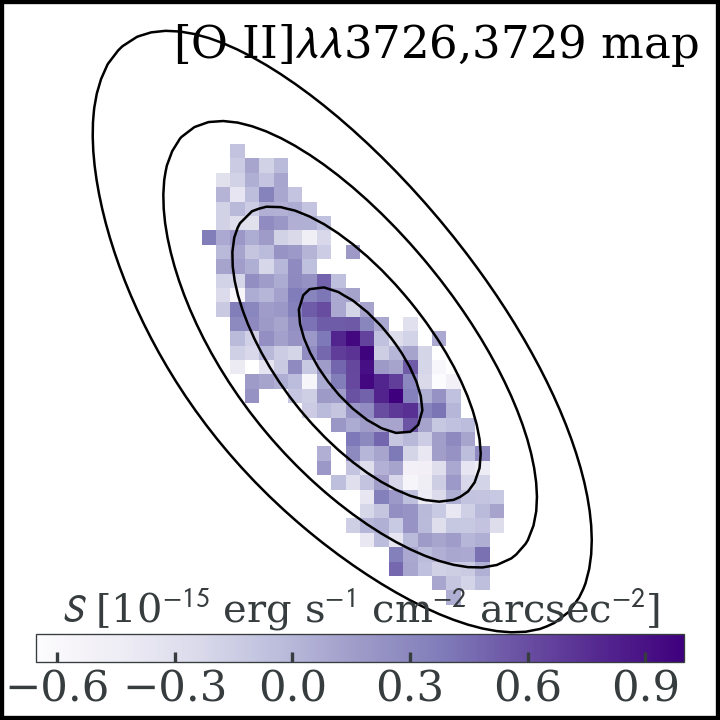}
    \includegraphics[width=.16\textwidth]{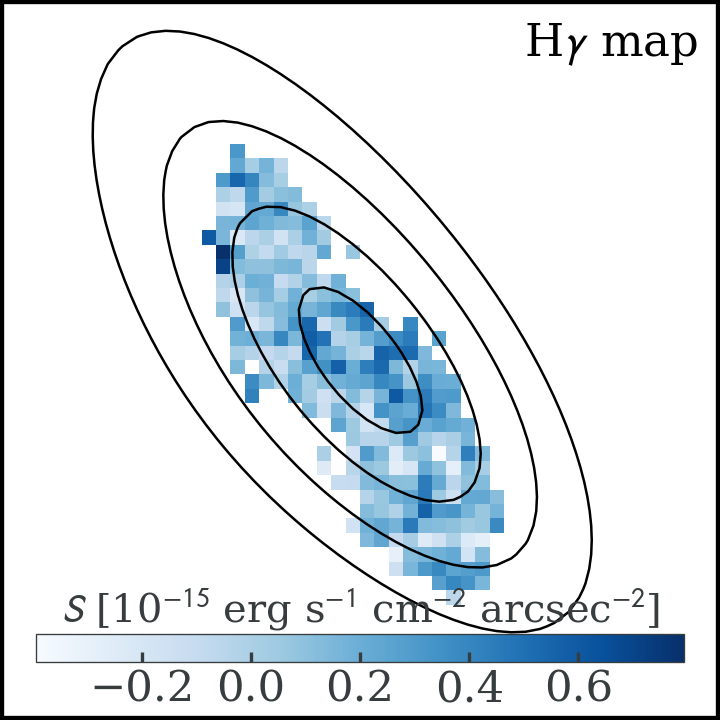}\\
    \includegraphics[width=\textwidth]{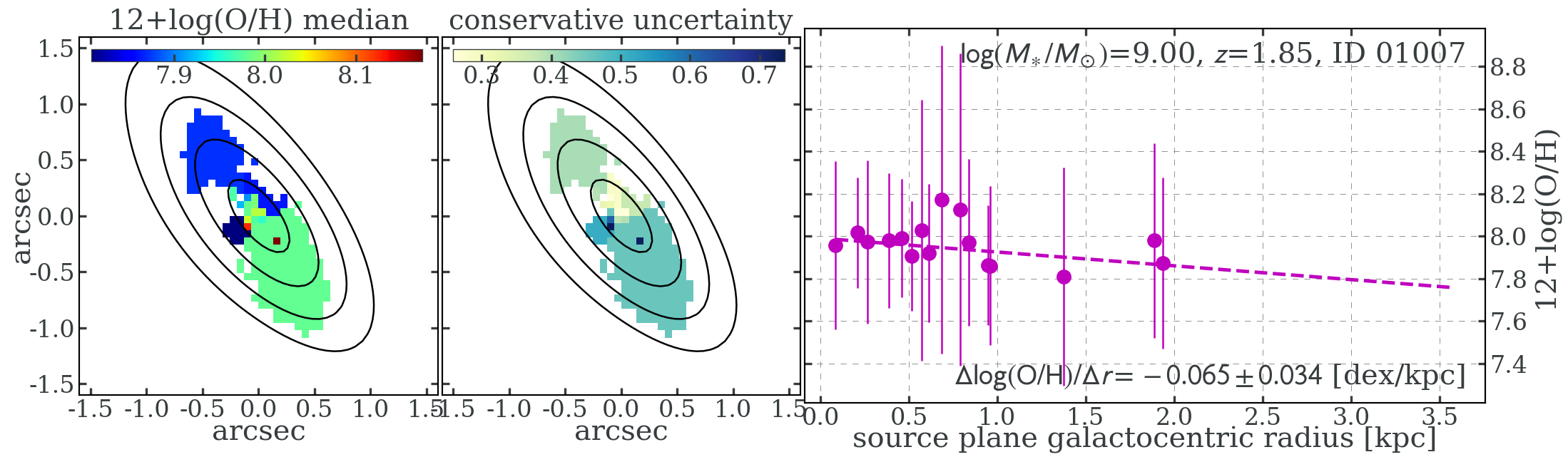}
    \caption{The source ID01007 in the field of \clwu is shown.}
    \label{fig:clM0717_ID01007_figs}
\end{figure*}

\begin{figure*}
    \centering
    \includegraphics[width=\textwidth]{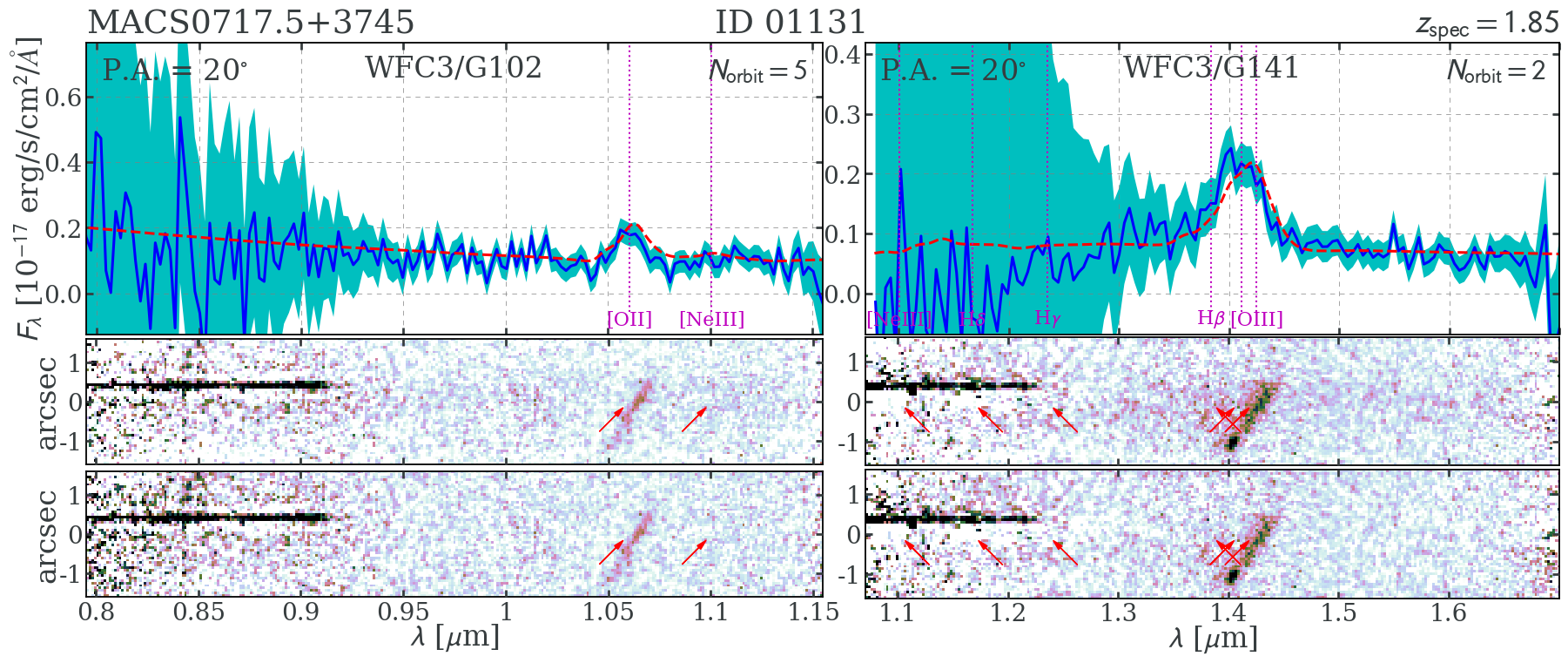}\\
    \includegraphics[width=\textwidth]{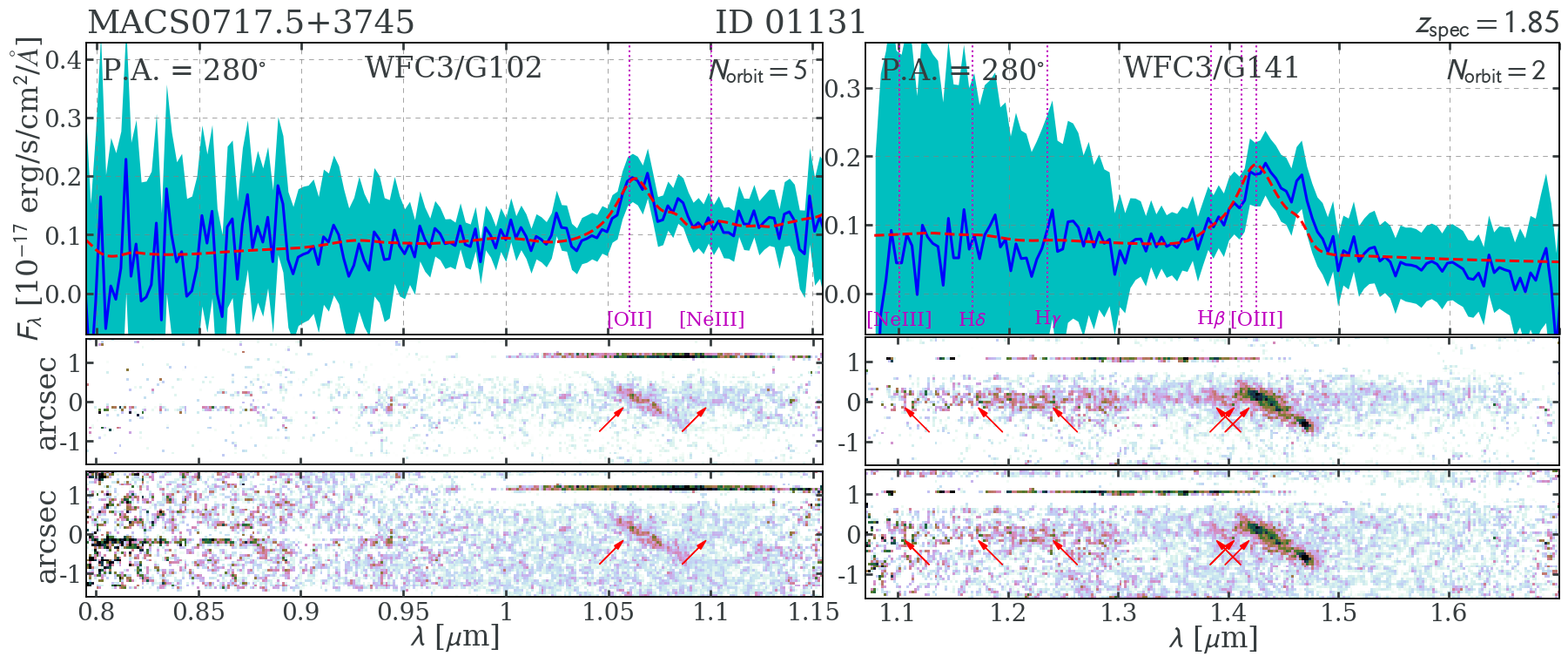}\\
    \includegraphics[width=.16\textwidth]{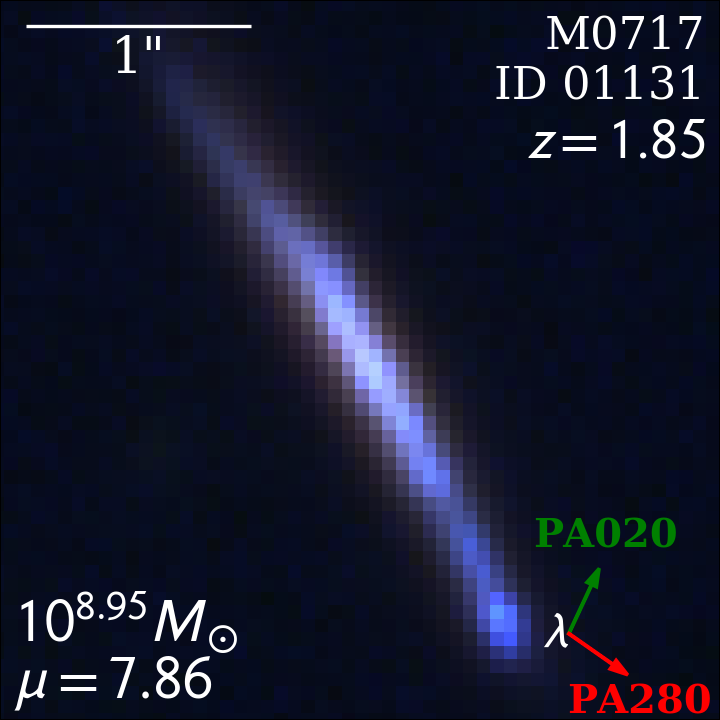}
    \includegraphics[width=.16\textwidth]{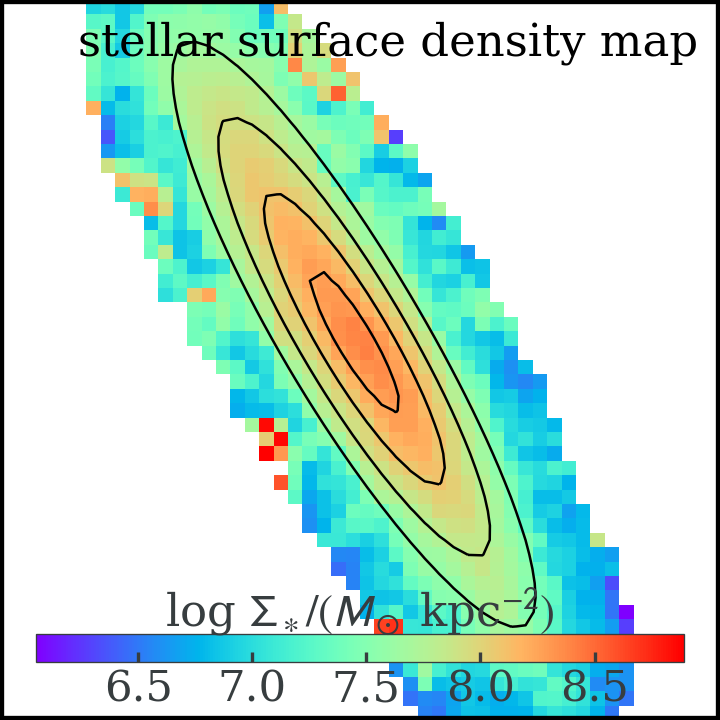}
    \includegraphics[width=.16\textwidth]{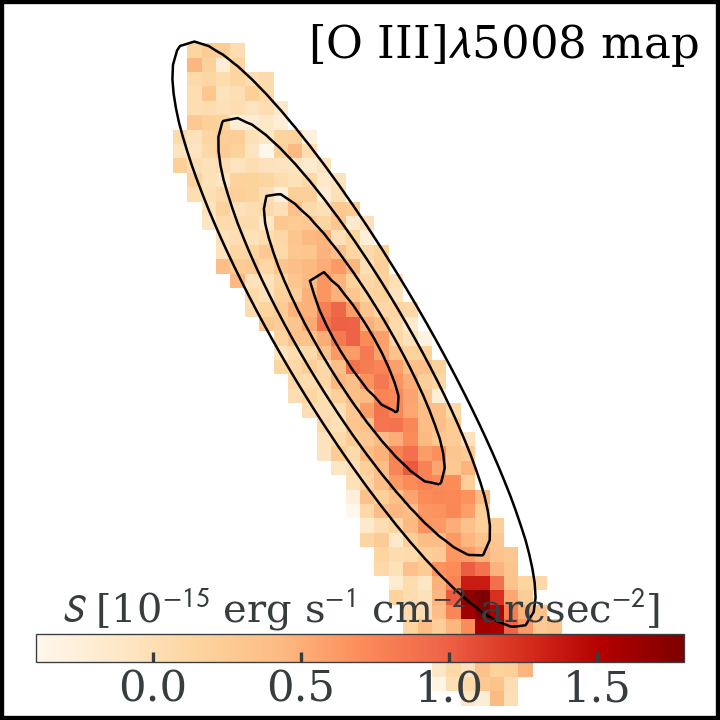}
    \includegraphics[width=.16\textwidth]{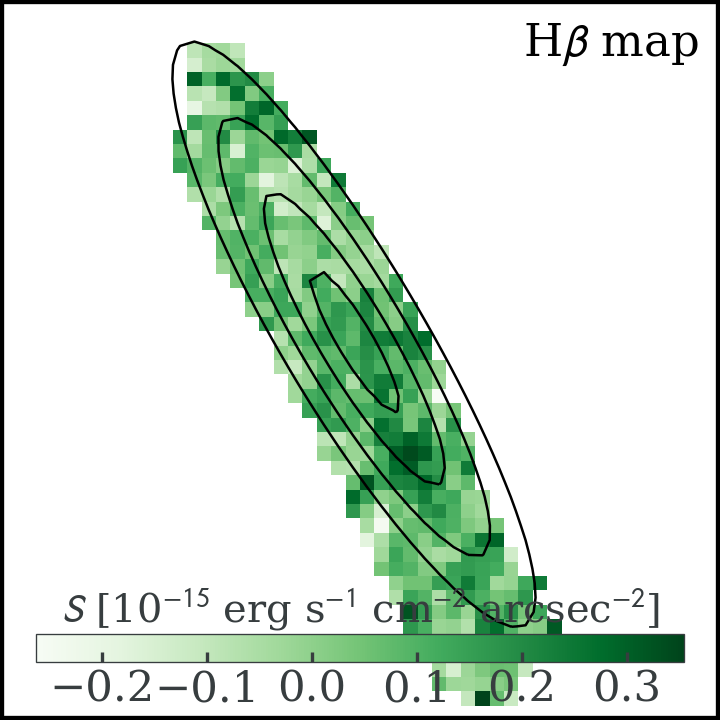}
    \includegraphics[width=.16\textwidth]{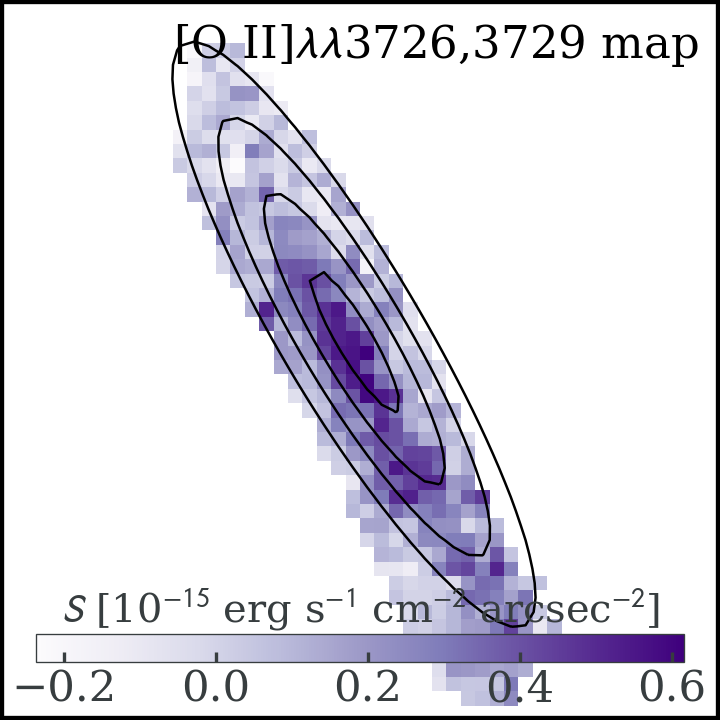}
    \includegraphics[width=.16\textwidth]{fig_ELmaps/baiban.png}\\
    \includegraphics[width=\textwidth]{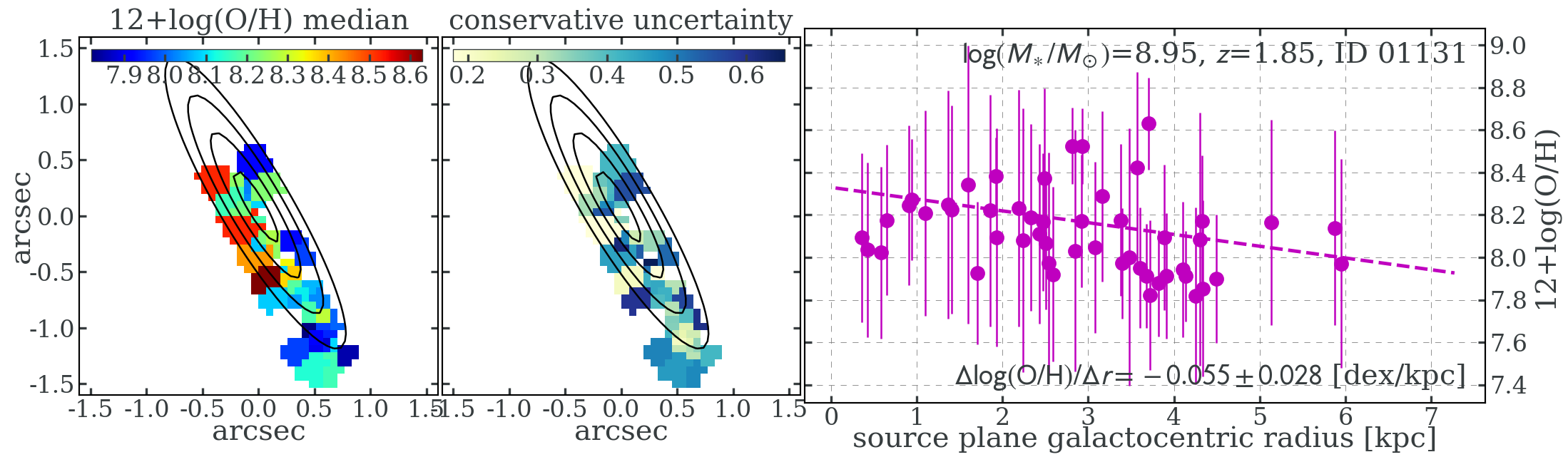}
    \caption{The source ID01131 in the field of \clwu is shown.}
    \label{fig:clM0717_ID01131_figs}
\end{figure*}

\begin{figure*}
    \centering
    \includegraphics[width=\textwidth]{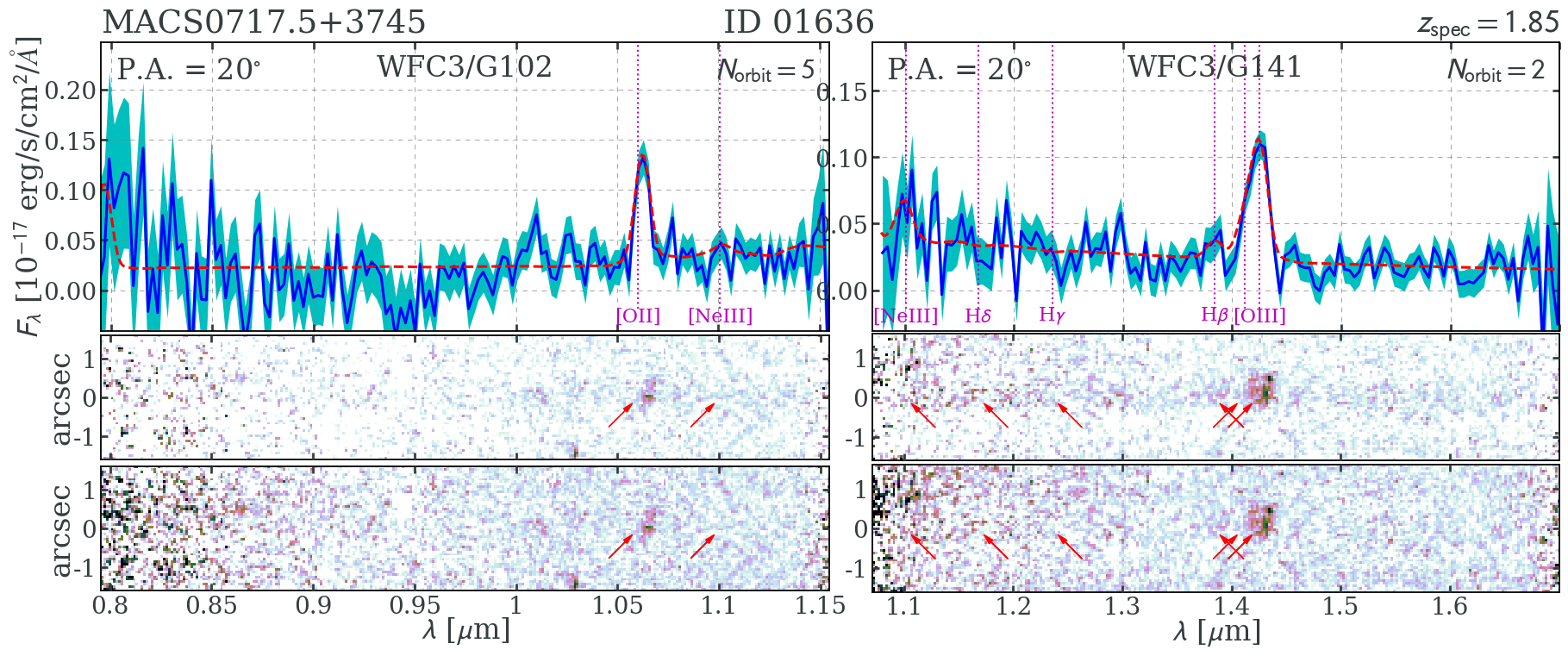}\\
    \includegraphics[width=\textwidth]{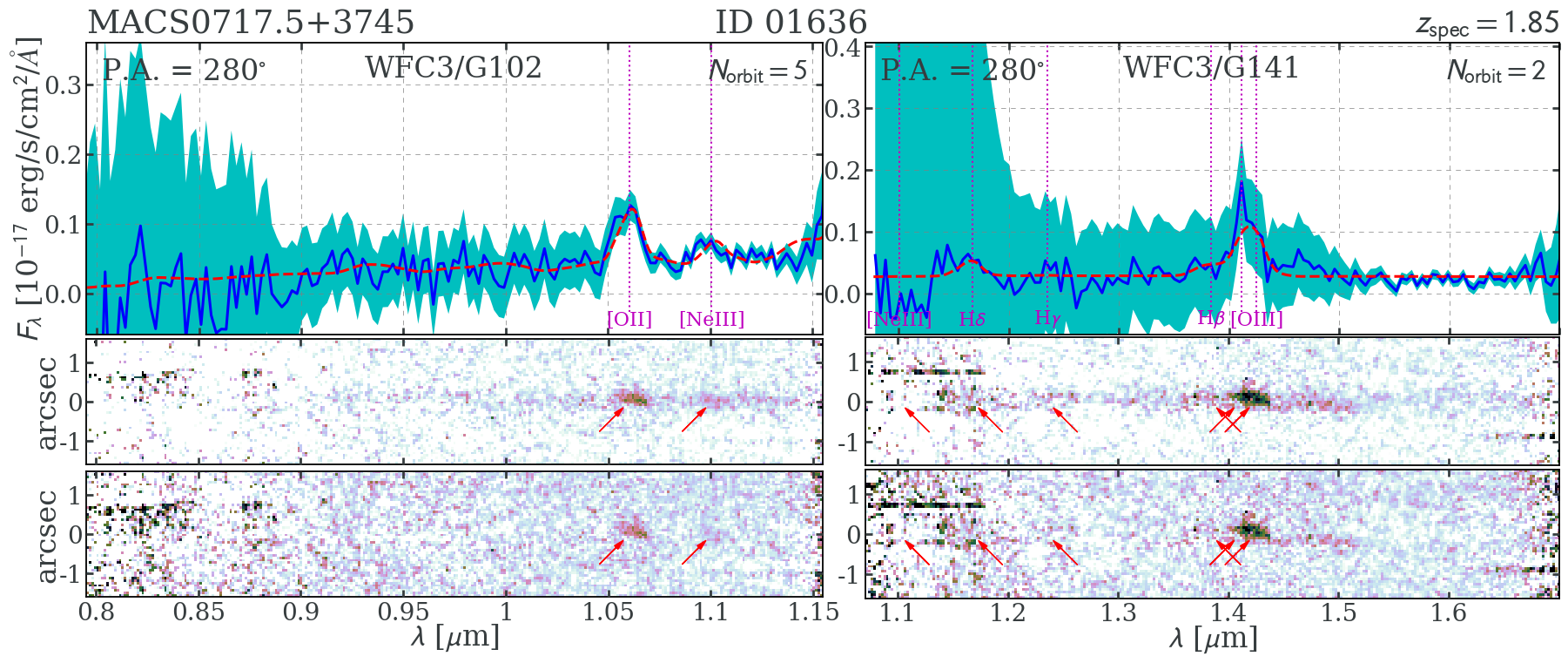}\\
    \includegraphics[width=.16\textwidth]{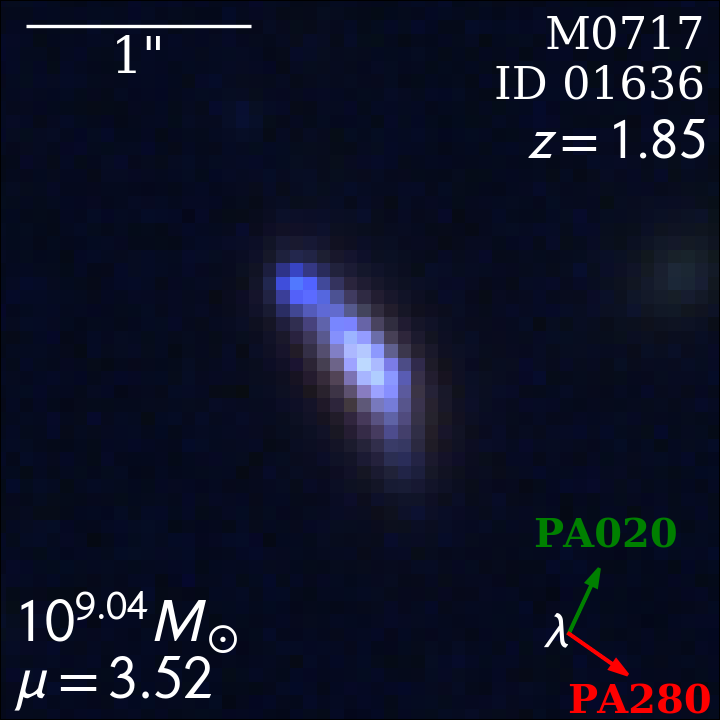}
    \includegraphics[width=.16\textwidth]{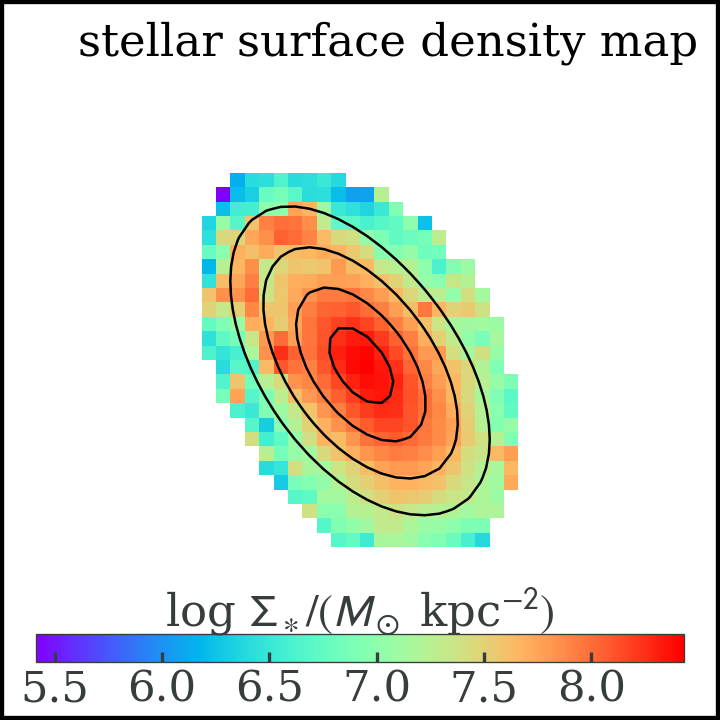}
    \includegraphics[width=.16\textwidth]{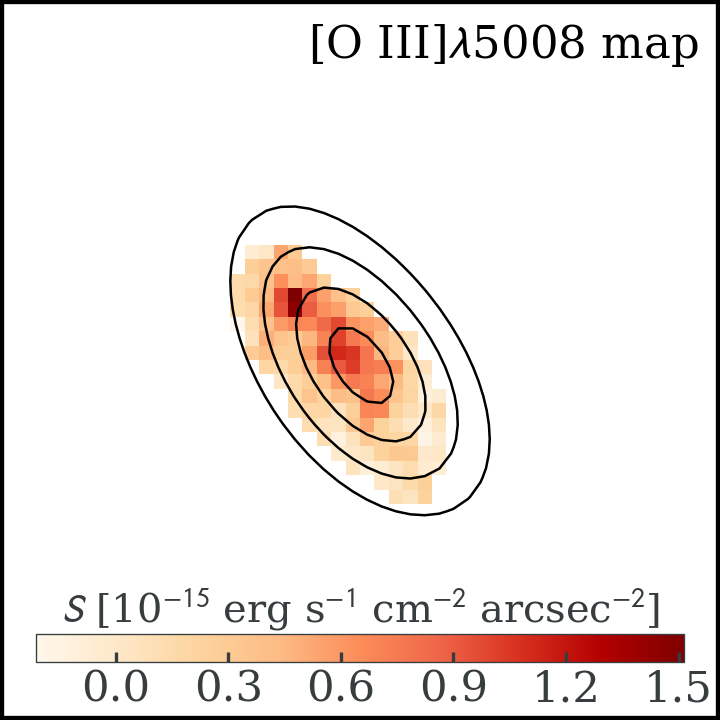}
    \includegraphics[width=.16\textwidth]{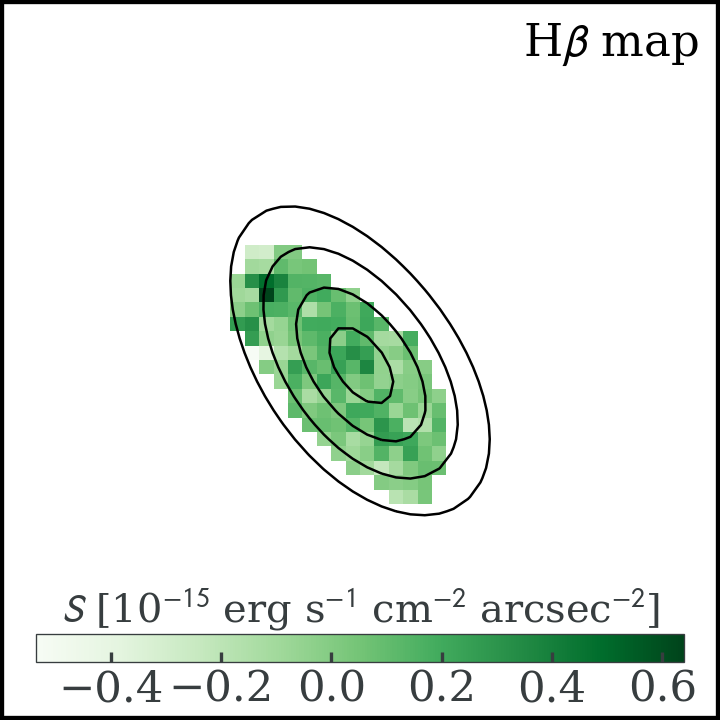}
    \includegraphics[width=.16\textwidth]{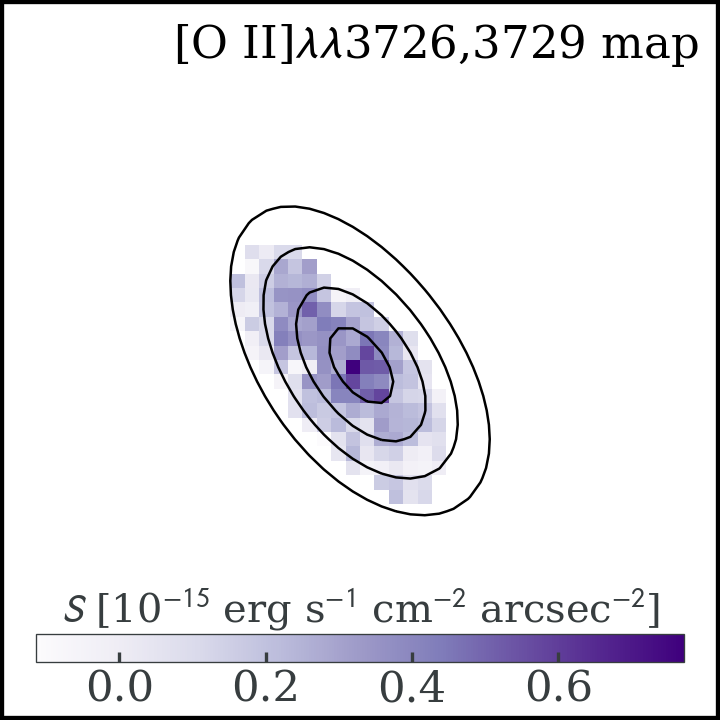}
    \includegraphics[width=.16\textwidth]{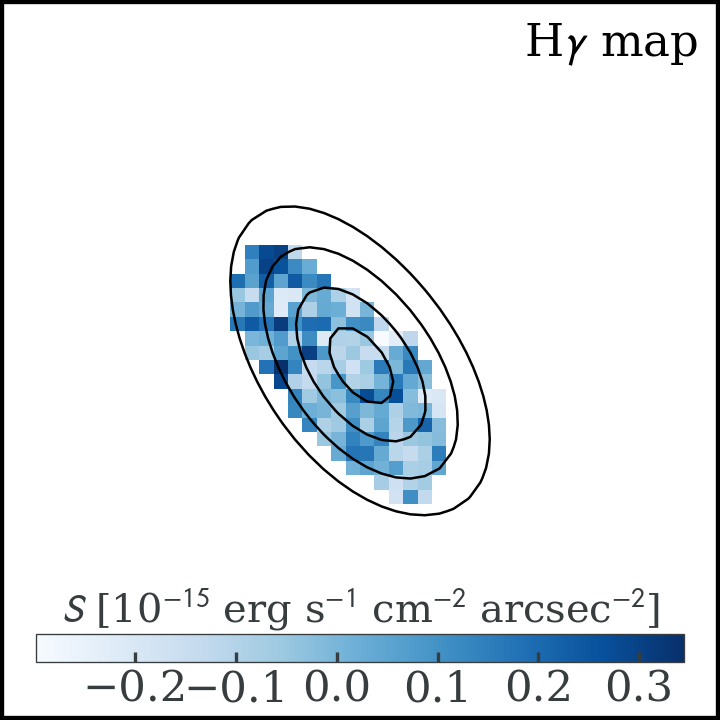}\\
    \includegraphics[width=\textwidth]{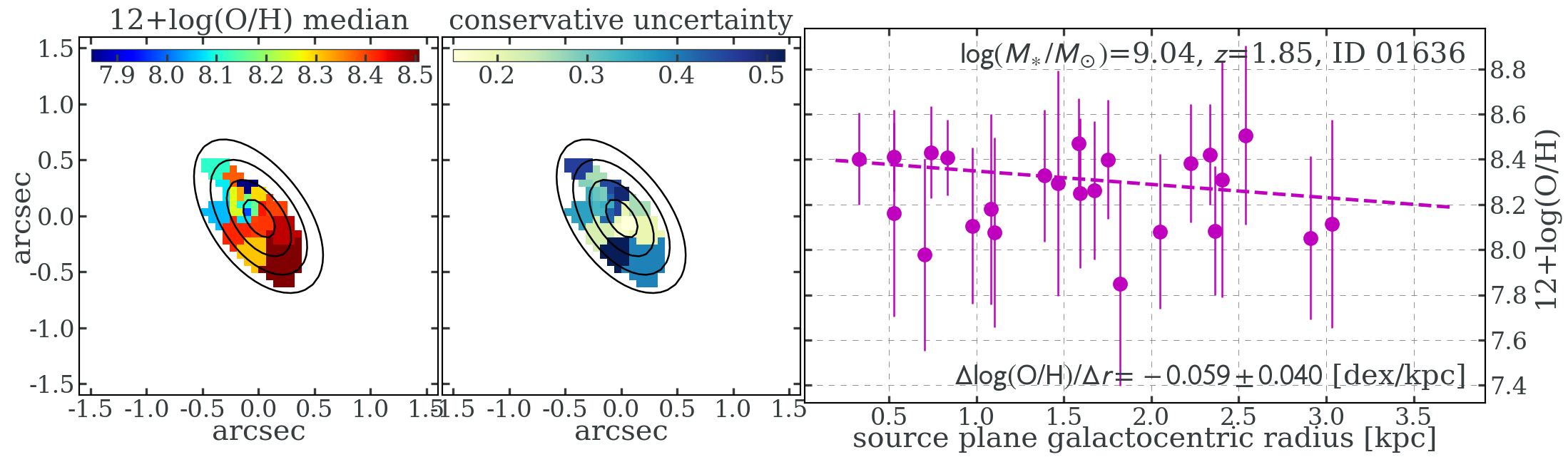}
    \caption{The source ID01636 in the field of \clwu is shown.}
    \label{fig:clM0717_ID01636_figs}
\end{figure*}

\begin{figure*}
    \centering
    \includegraphics[width=\textwidth]{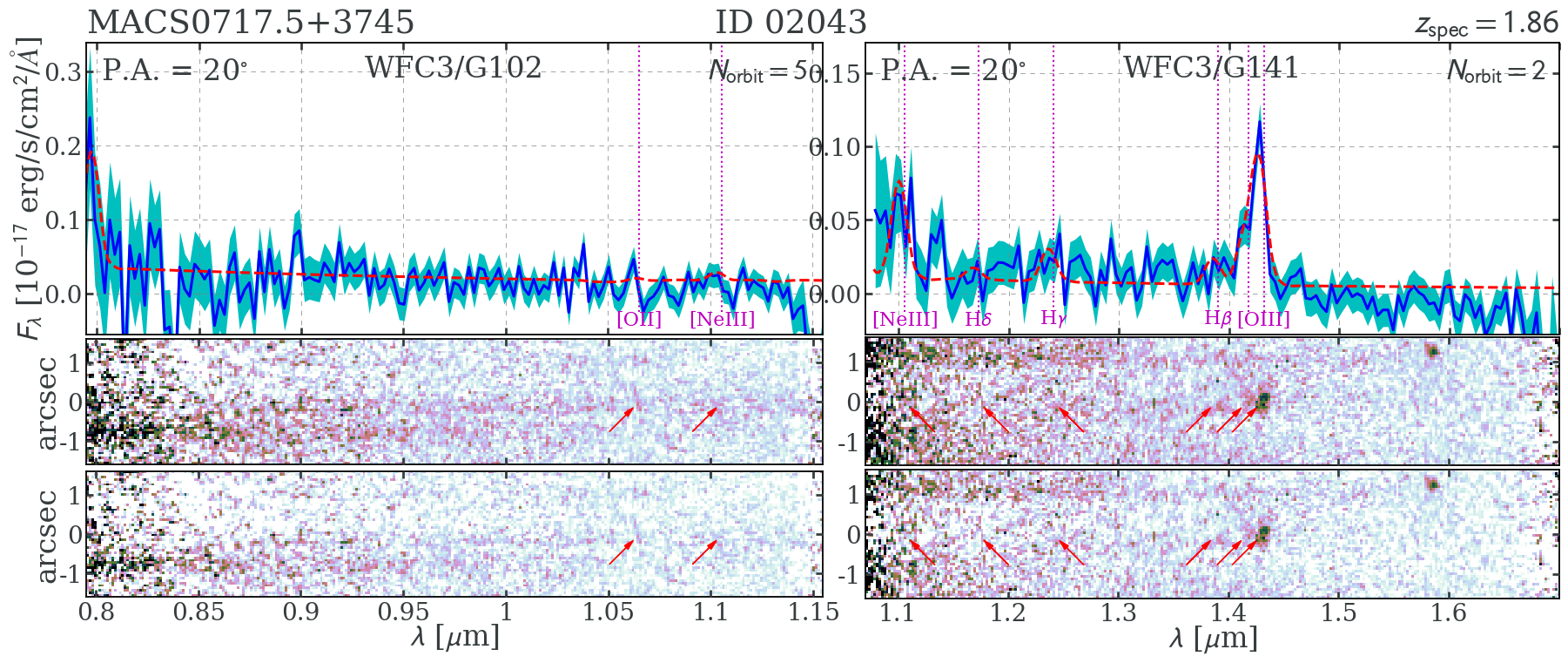}\\
    \includegraphics[width=\textwidth]{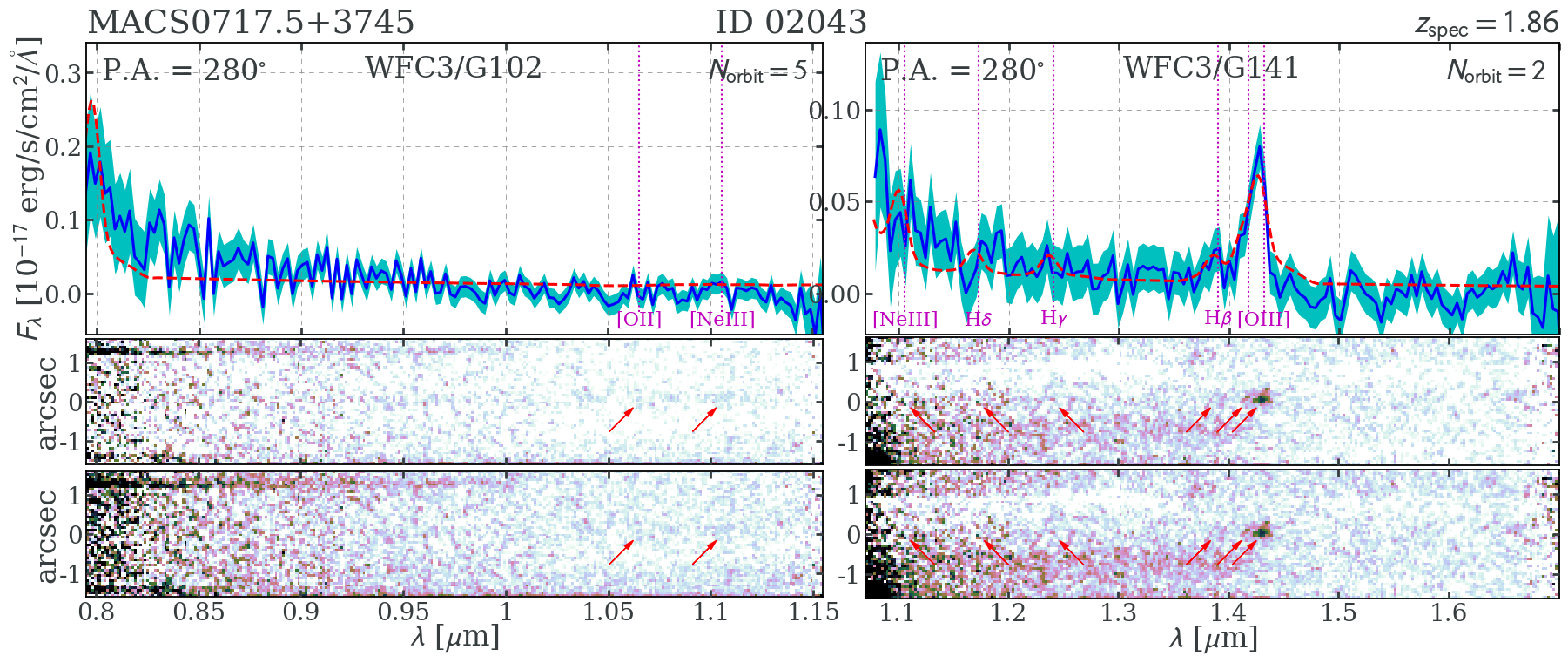}\\
    \includegraphics[width=.16\textwidth]{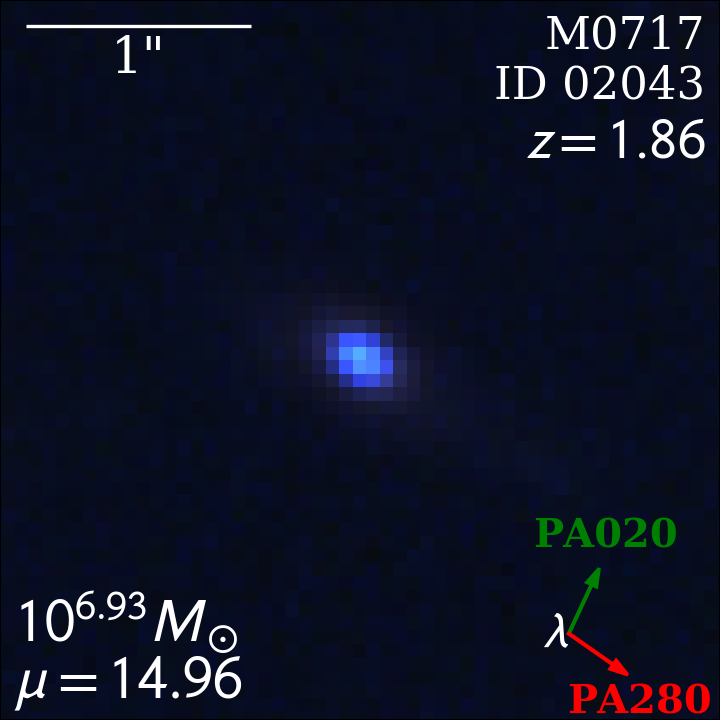}
    \includegraphics[width=.16\textwidth]{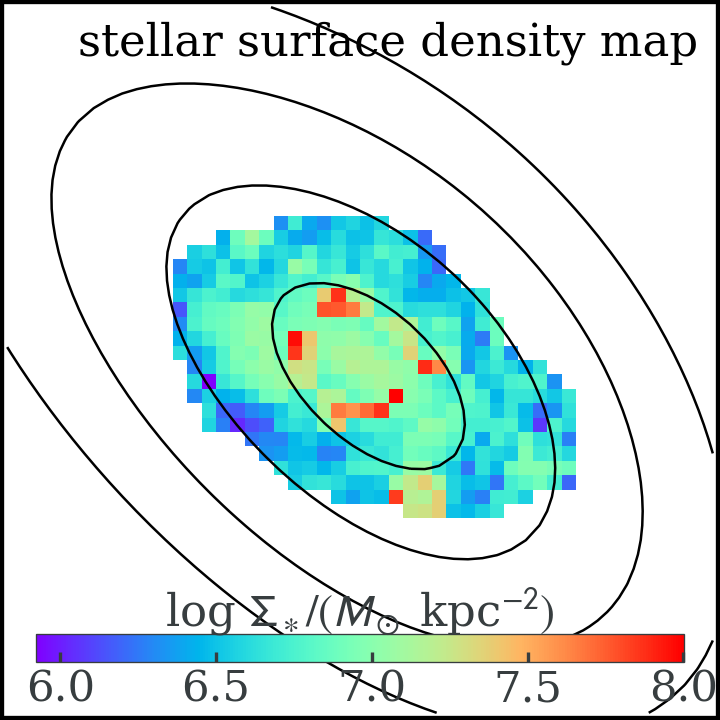}
    \includegraphics[width=.16\textwidth]{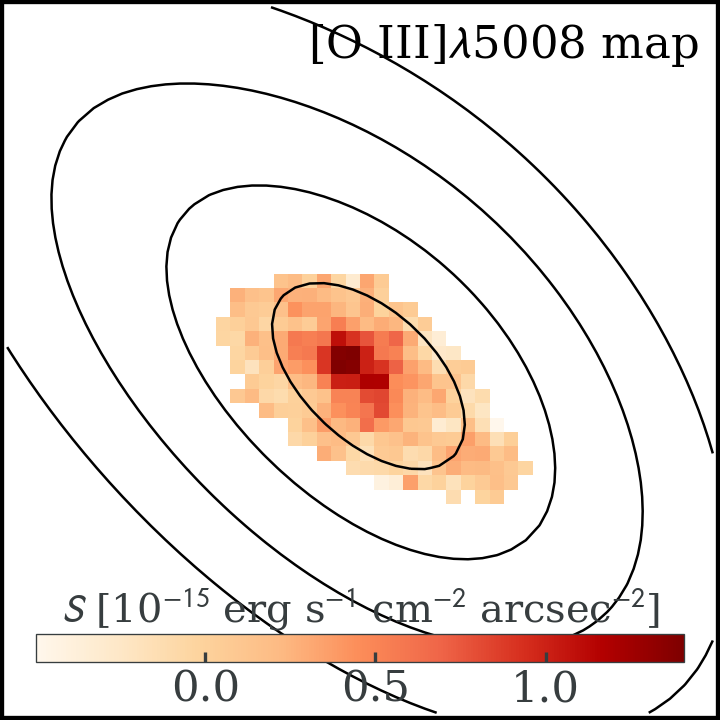}
    \includegraphics[width=.16\textwidth]{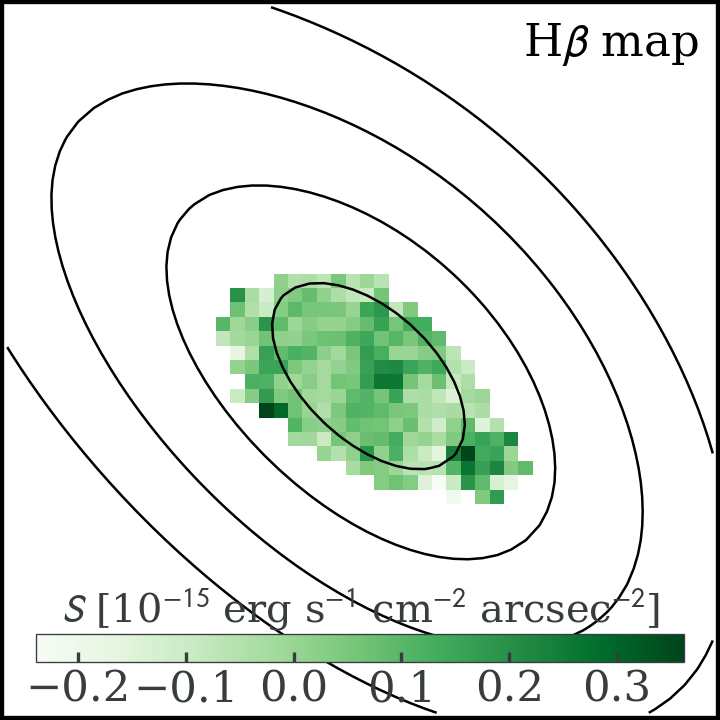}
    \includegraphics[width=.16\textwidth]{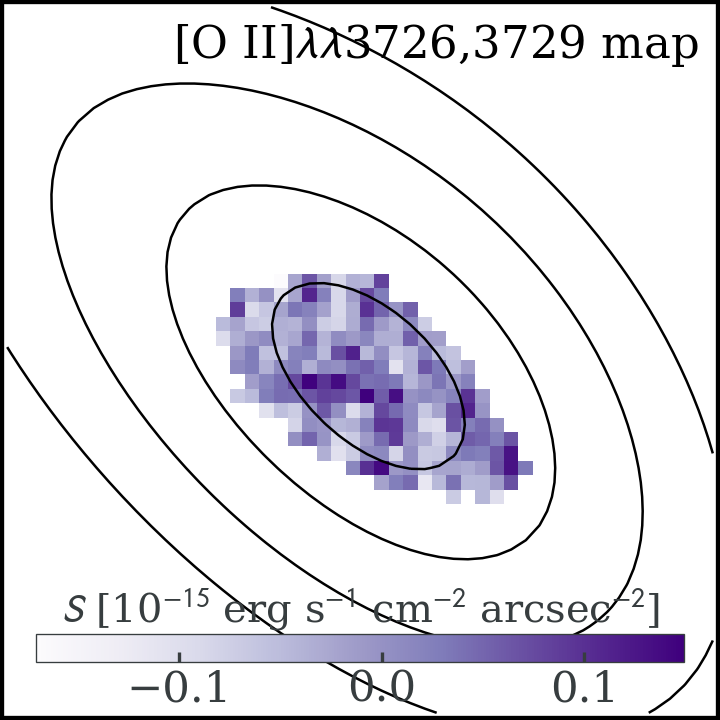}
    \includegraphics[width=.16\textwidth]{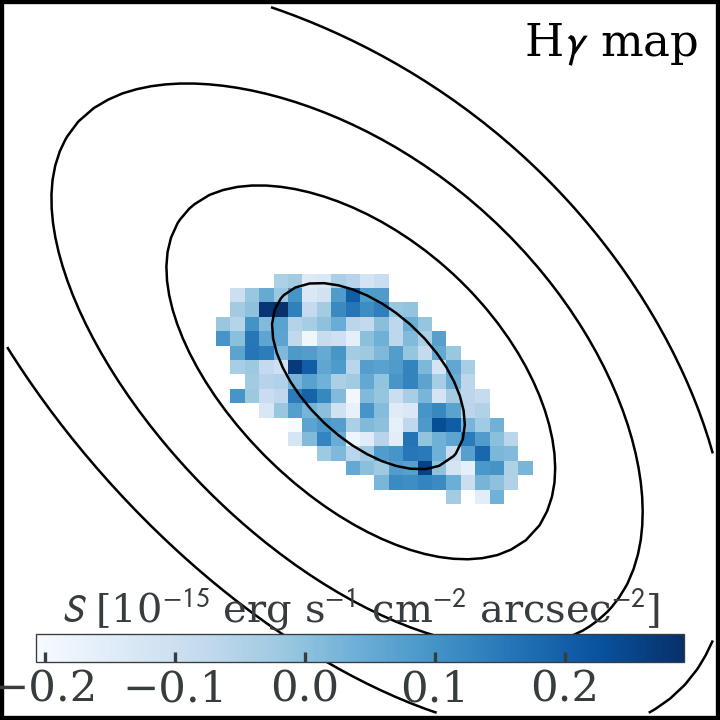}\\
    \includegraphics[width=\textwidth]{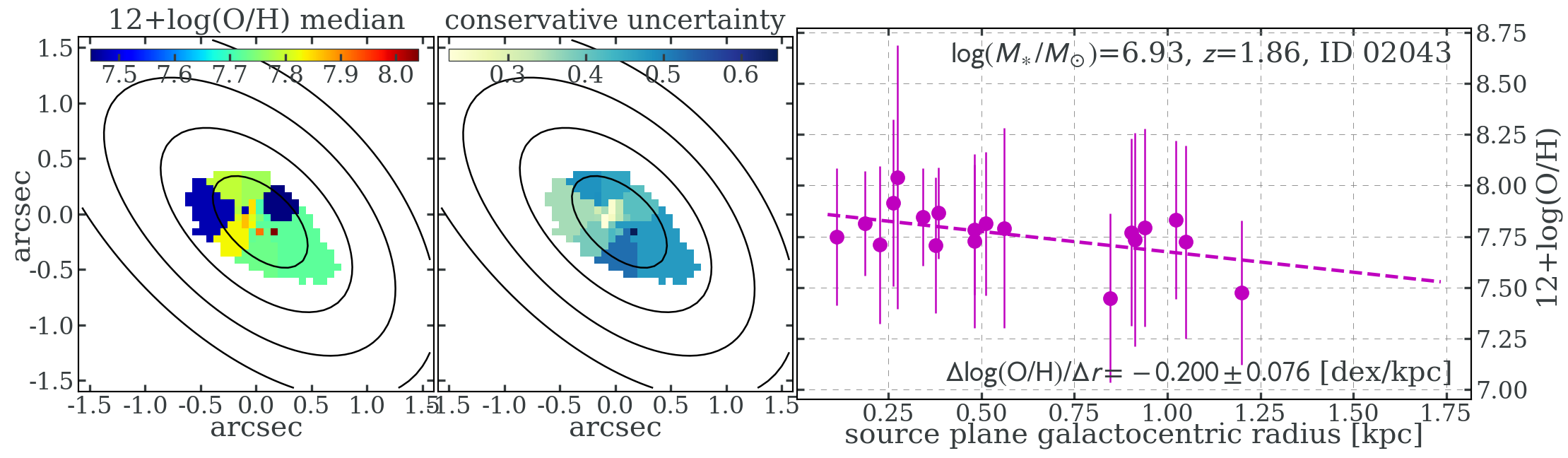}
    \caption{The source ID02043 in the field of \clwu is shown.}
    \label{fig:clM0717_ID02043_figs}
\end{figure*}

\begin{figure*}
    \centering
    \includegraphics[width=\textwidth]{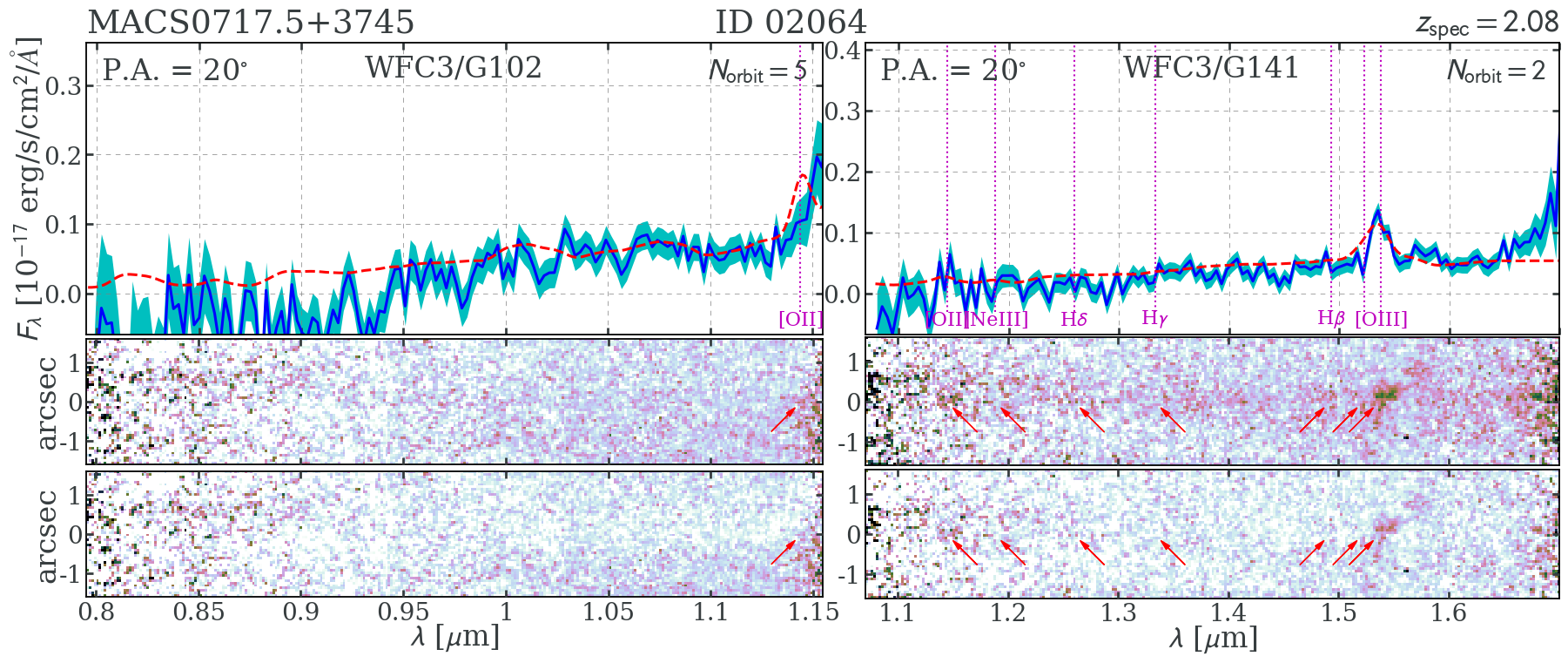}\\
    \includegraphics[width=\textwidth]{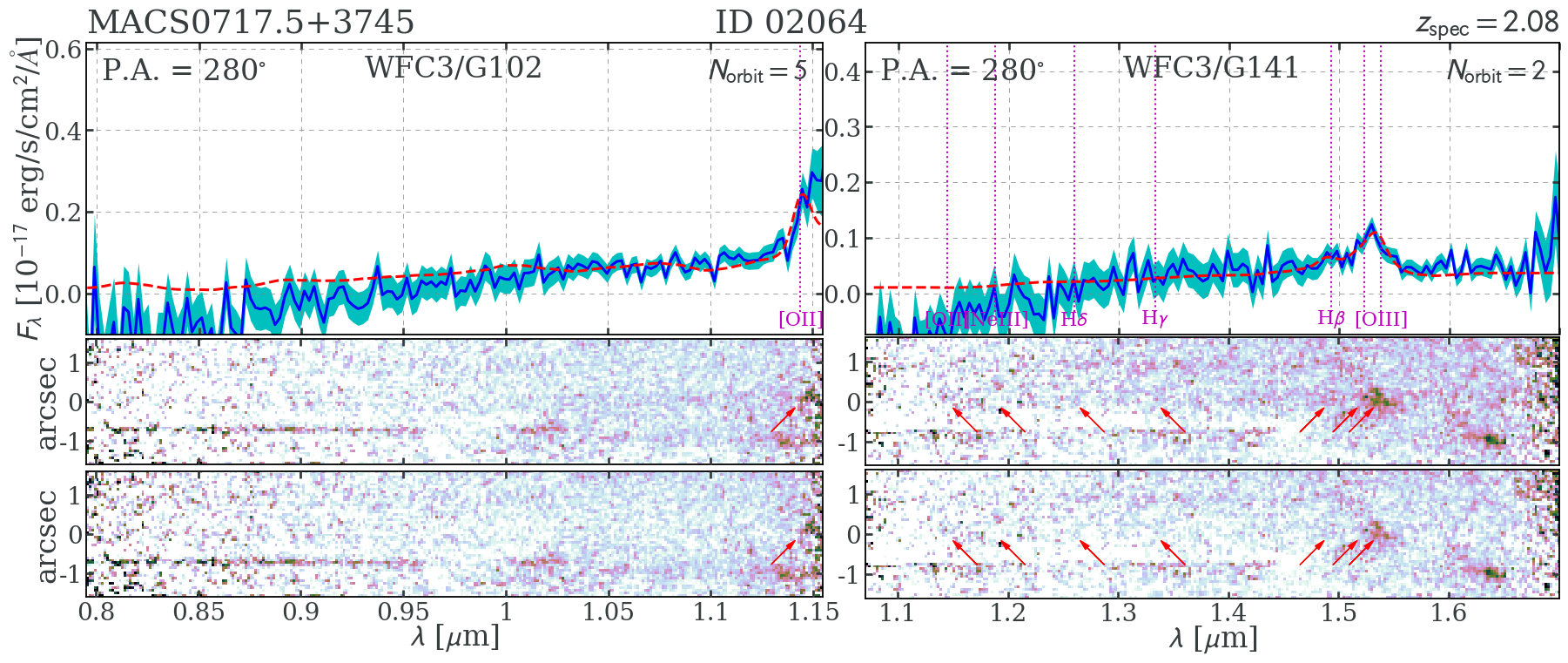}\\
    \includegraphics[width=.16\textwidth]{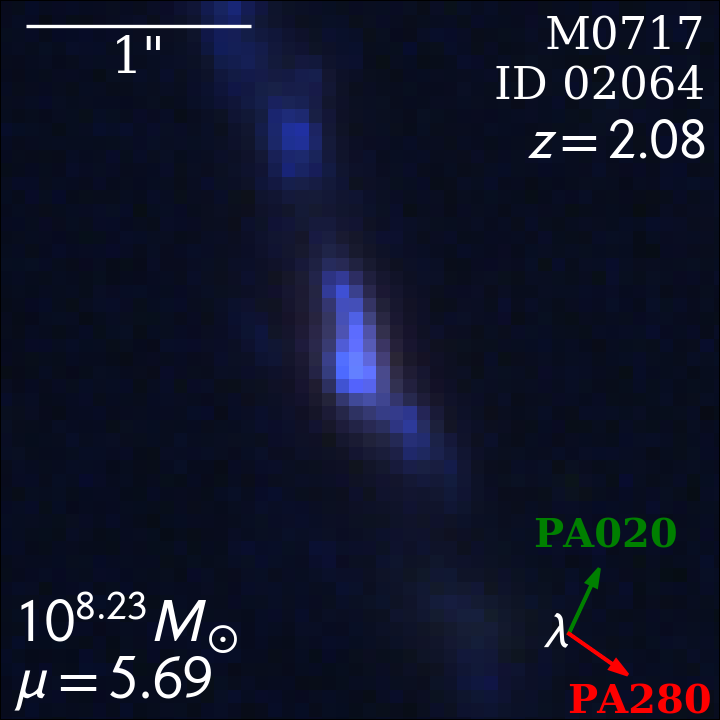}
    \includegraphics[width=.16\textwidth]{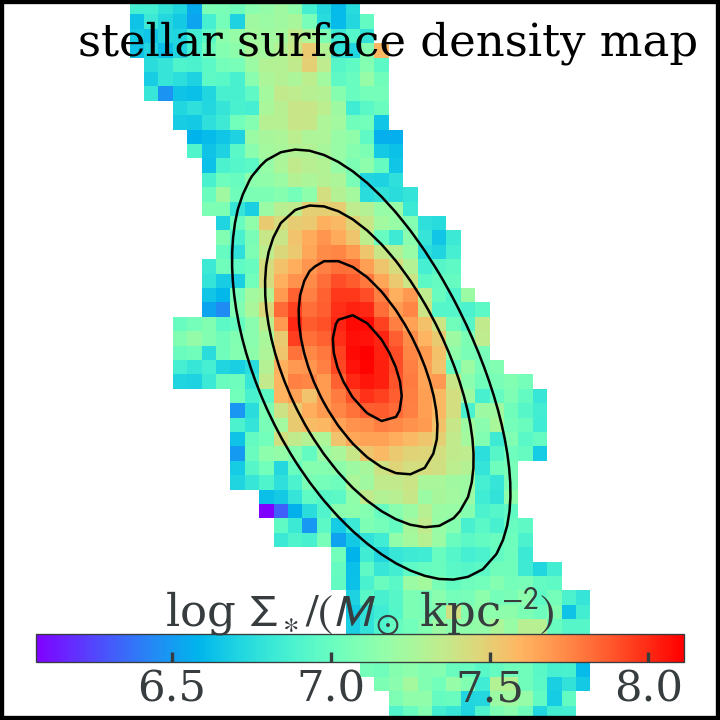}
    \includegraphics[width=.16\textwidth]{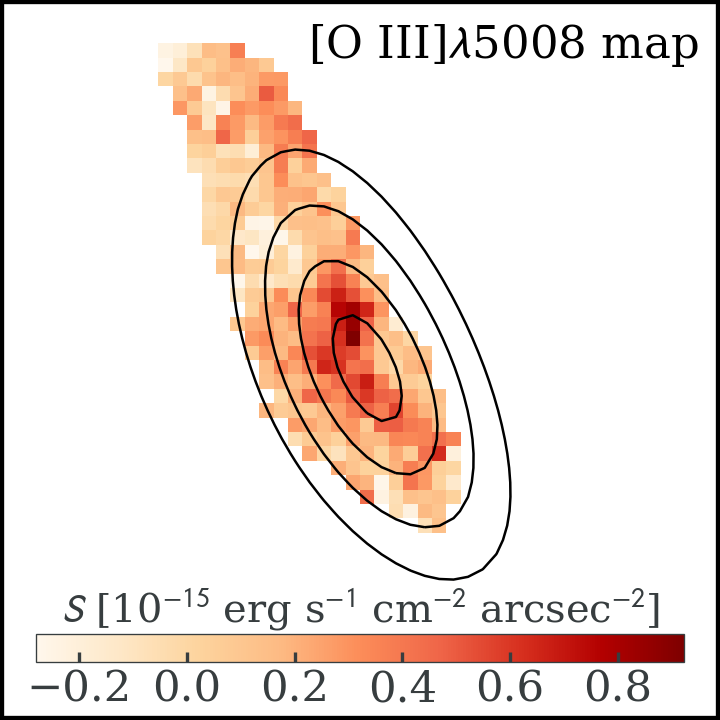}
    \includegraphics[width=.16\textwidth]{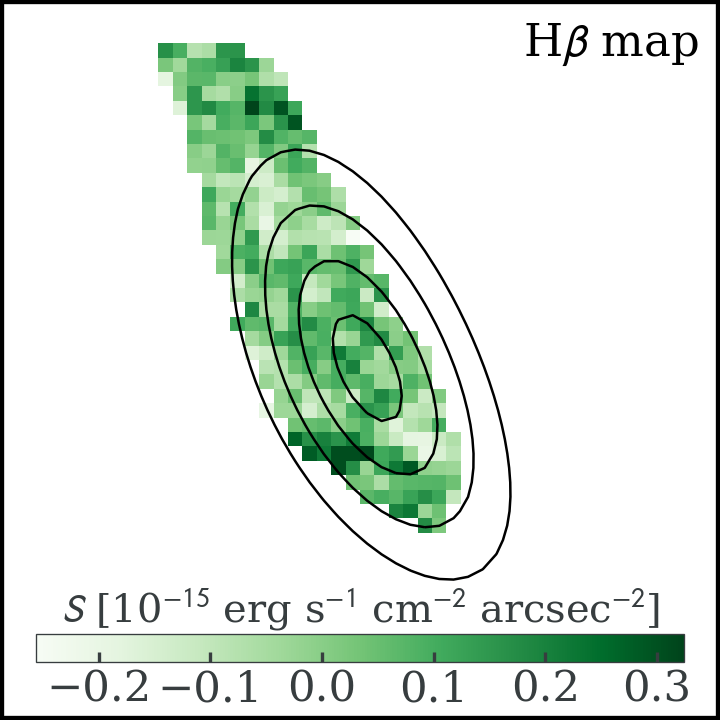}
    \includegraphics[width=.16\textwidth]{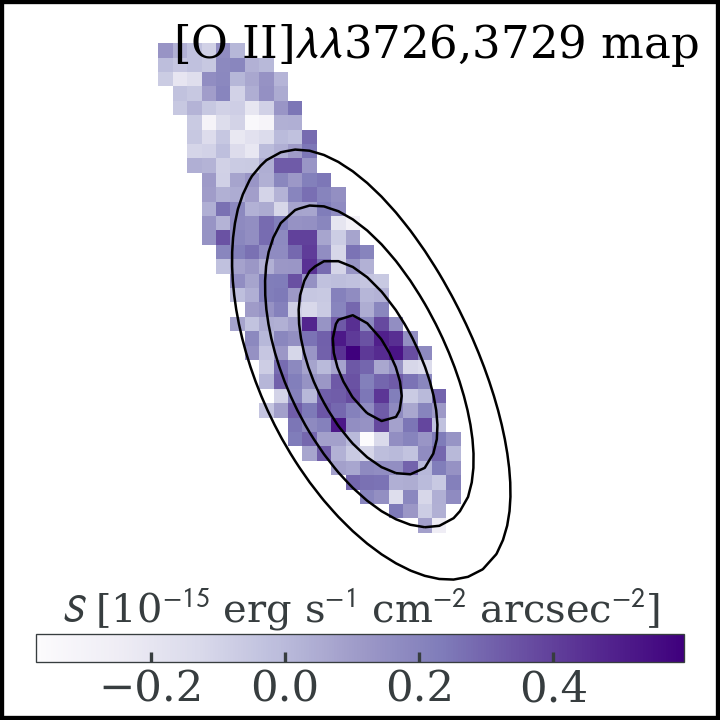}
    \includegraphics[width=.16\textwidth]{fig_ELmaps/baiban.png}\\
    \includegraphics[width=\textwidth]{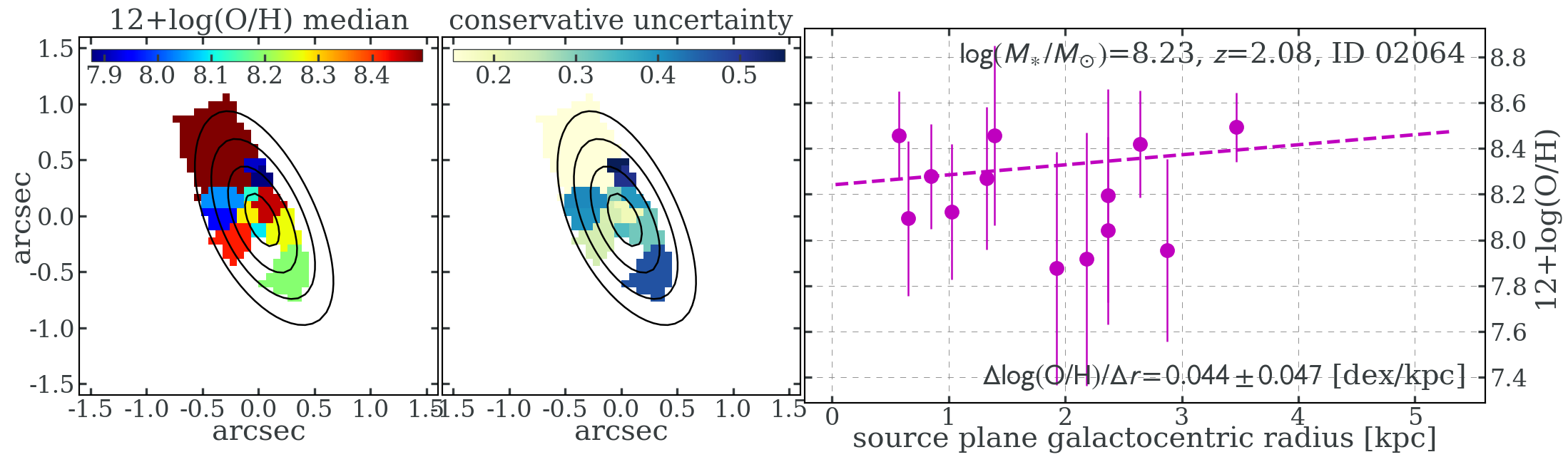}
    \caption{The source ID02064 in the field of \clwu is shown.}
    \label{fig:clM0717_ID02064_figs}
\end{figure*}

\begin{figure*}
    \centering
    \includegraphics[width=\textwidth]{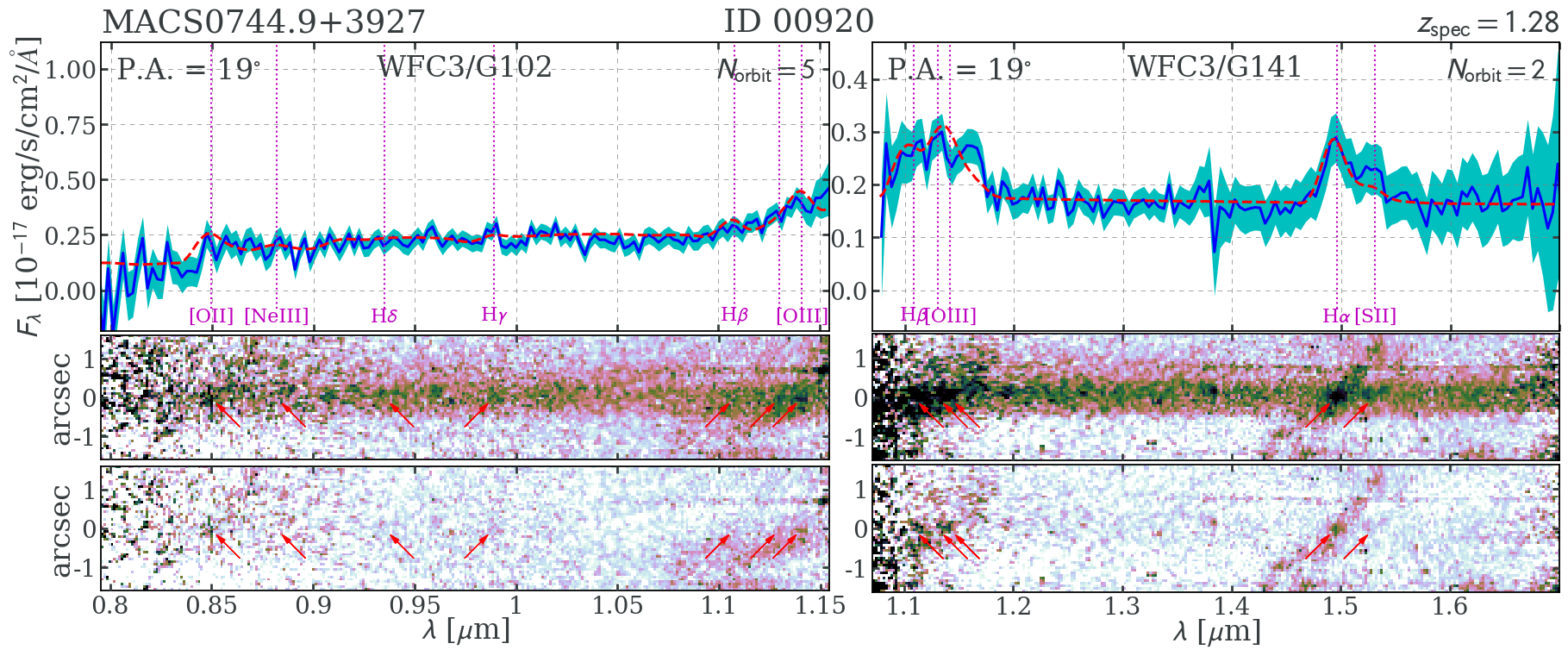}\\
    \includegraphics[width=\textwidth]{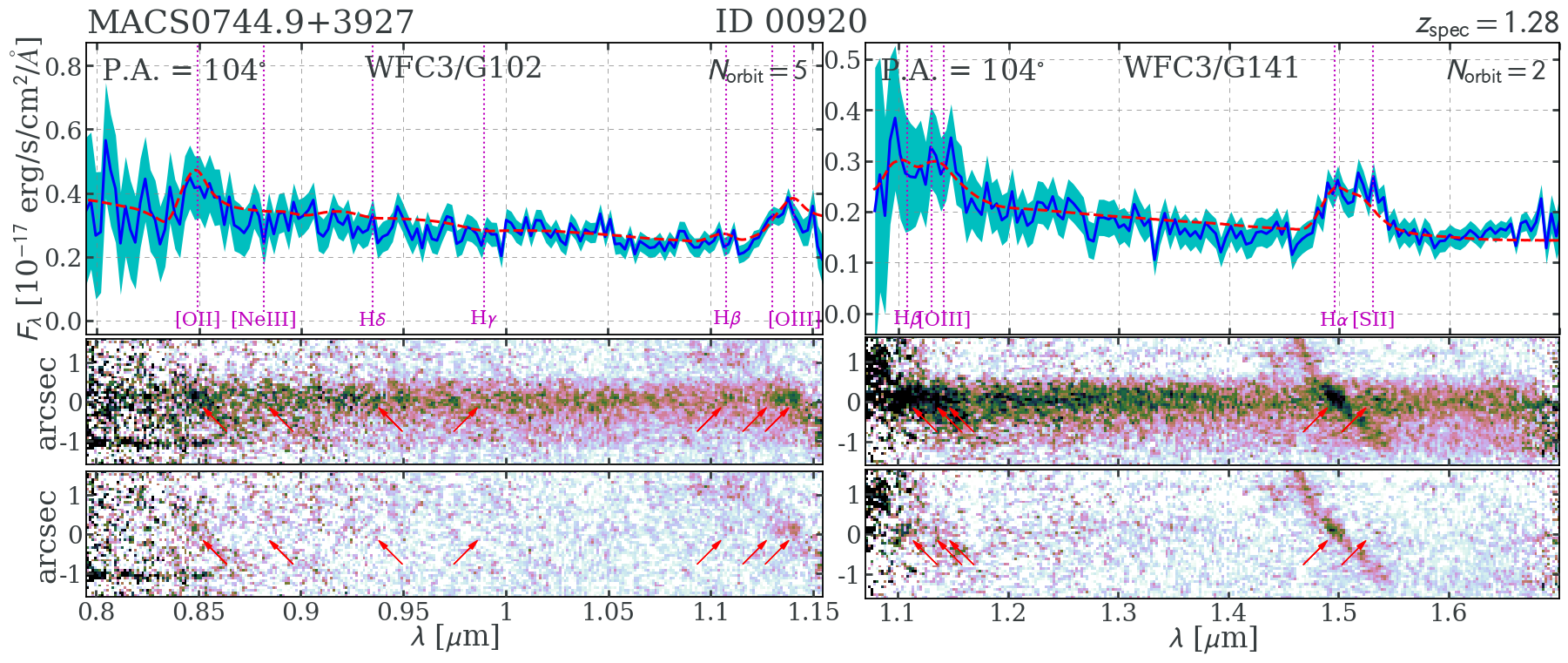}\\
    \includegraphics[width=.16\textwidth]{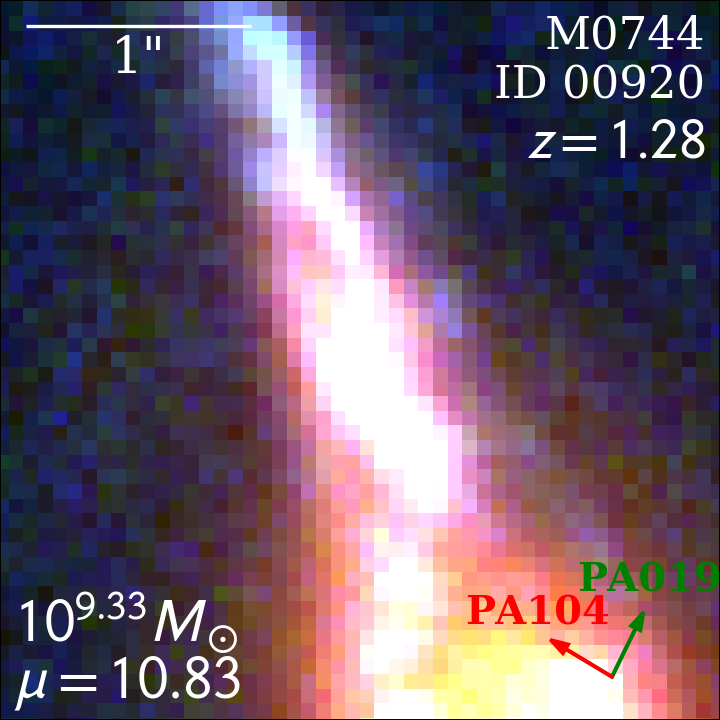}
    \includegraphics[width=.16\textwidth]{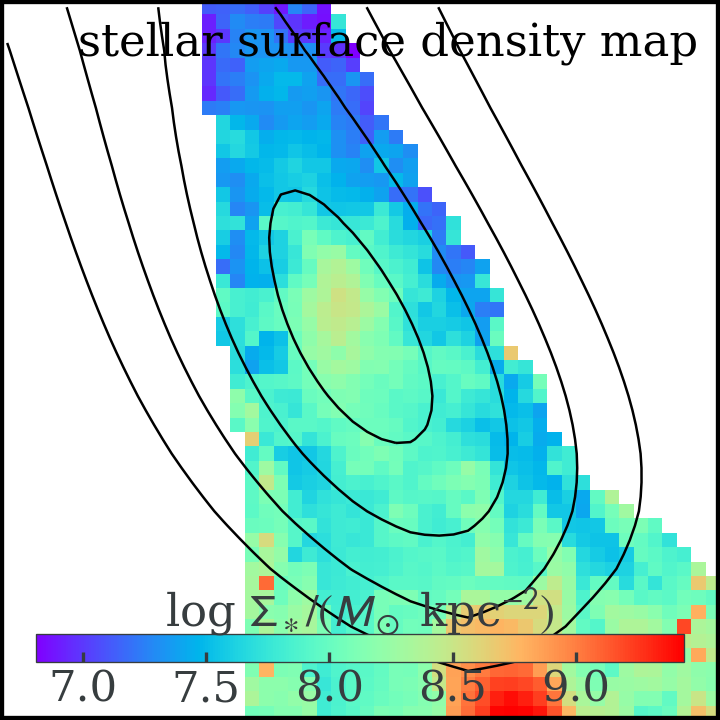}
    \includegraphics[width=.16\textwidth]{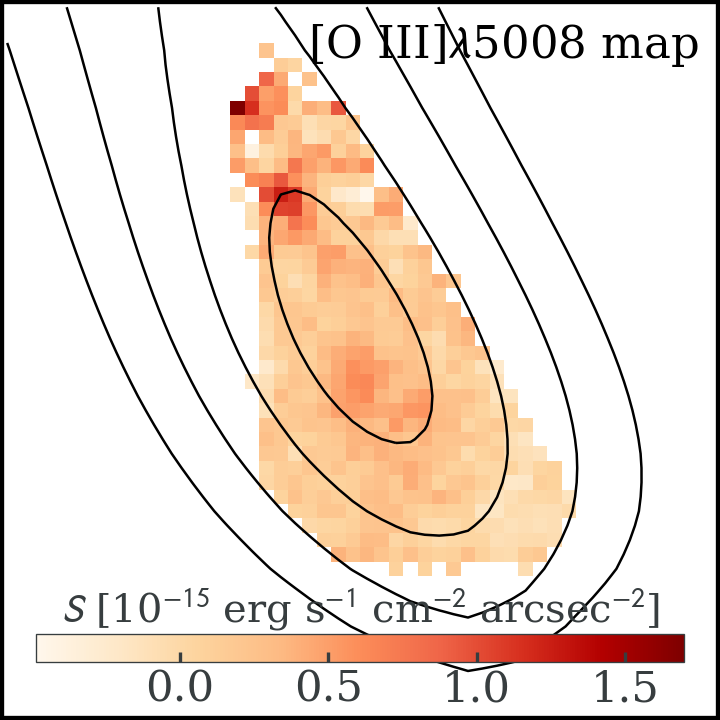}
    \includegraphics[width=.16\textwidth]{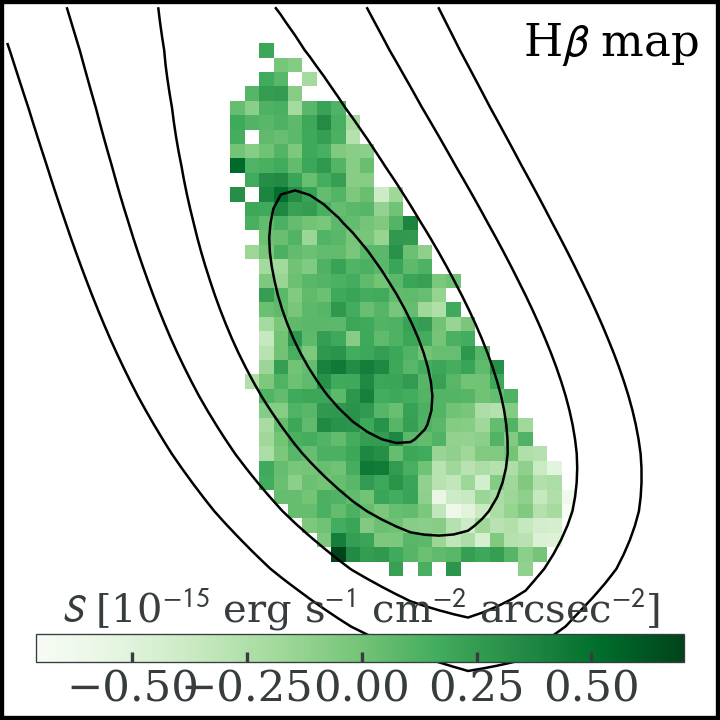}
    \includegraphics[width=.16\textwidth]{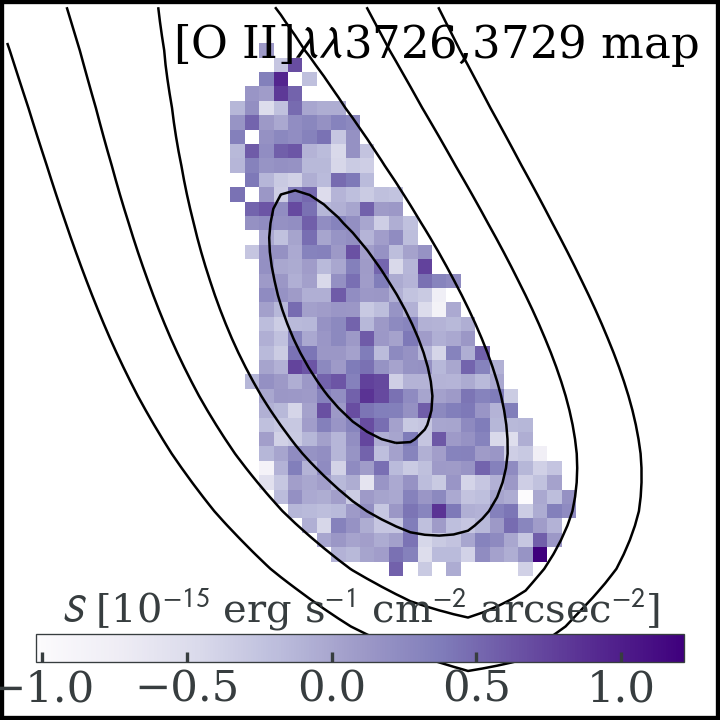}
    \includegraphics[width=.16\textwidth]{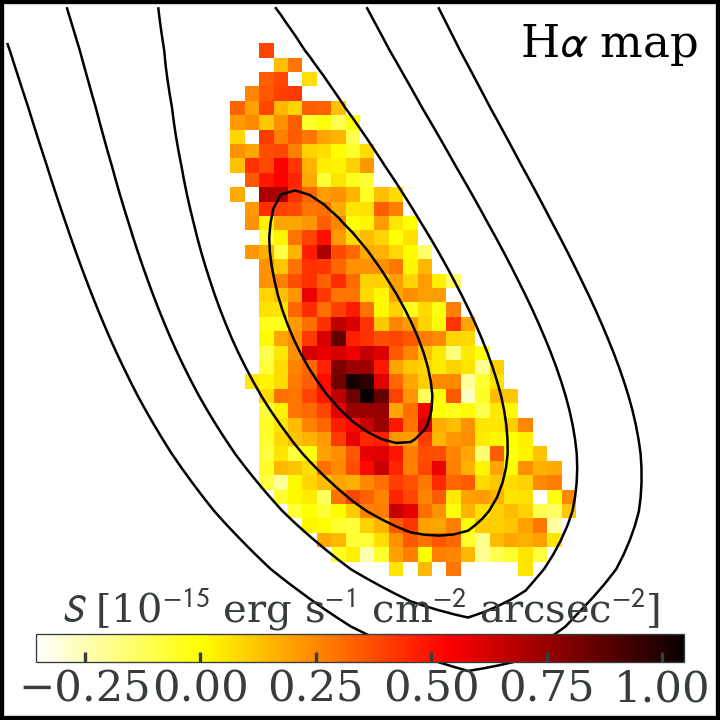}\\
    \includegraphics[width=\textwidth]{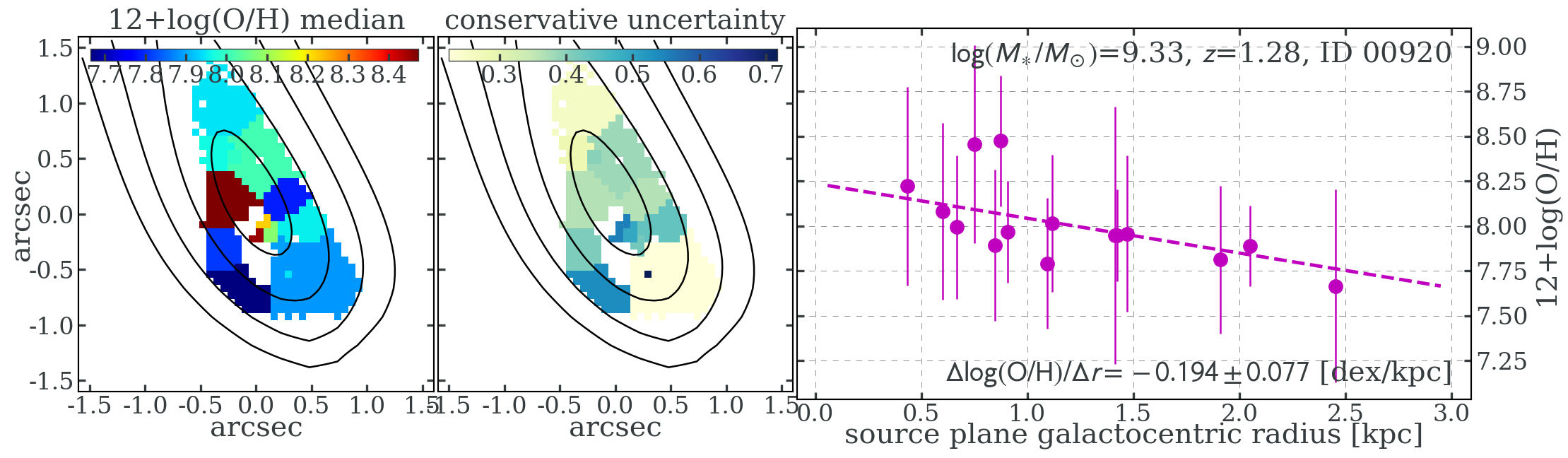}
    \caption{The source ID00920 in the field of \clba is shown.}
    \label{fig:clM0744_ID00920_figs}
\end{figure*}

\begin{figure*}
    \centering
    \includegraphics[width=\textwidth]{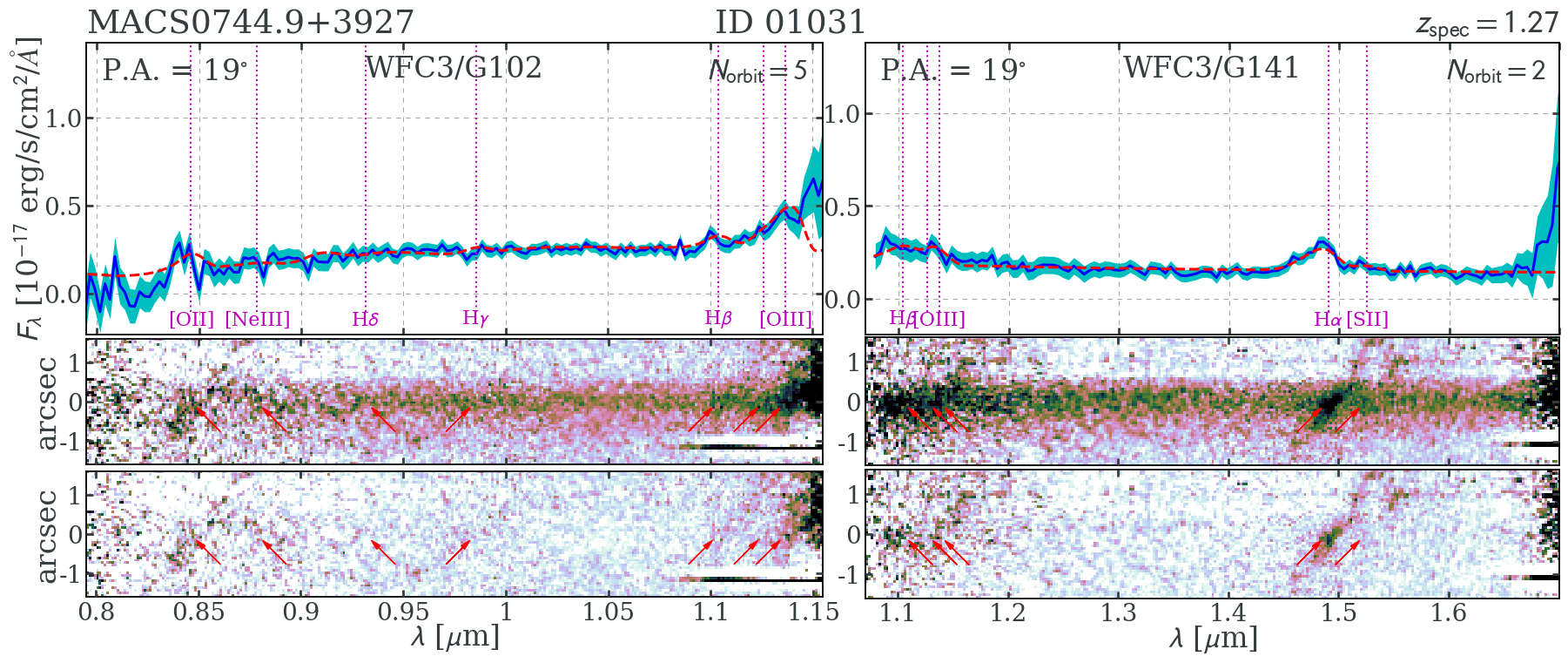}\\
    \includegraphics[width=\textwidth]{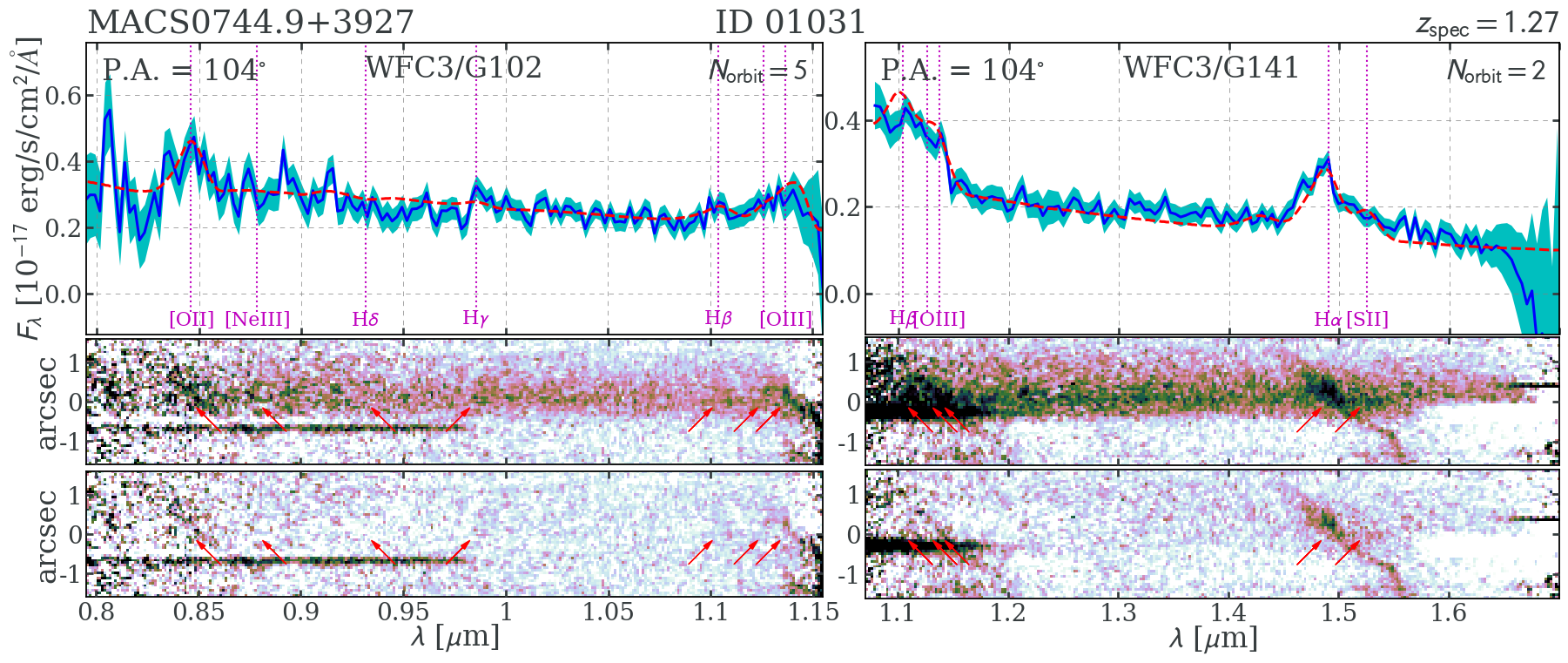}\\
    \includegraphics[width=.16\textwidth]{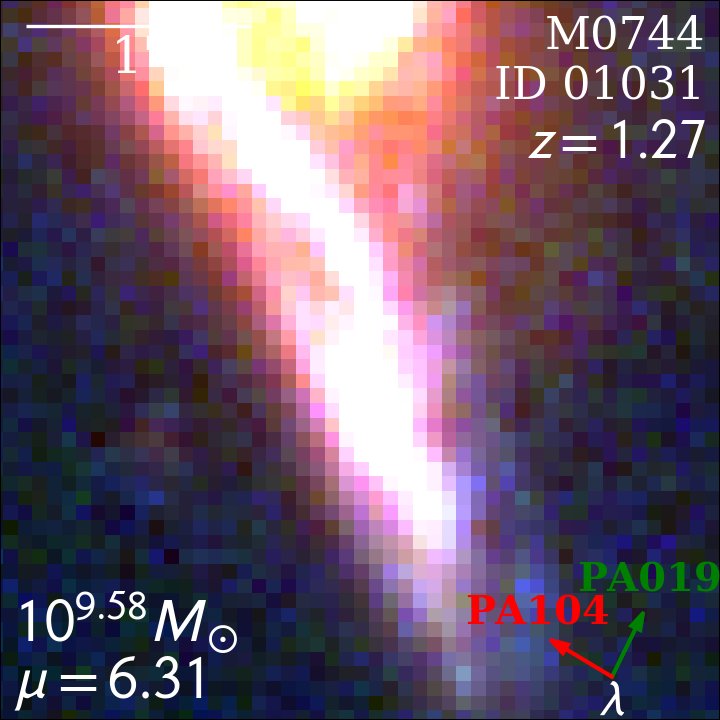}
    \includegraphics[width=.16\textwidth]{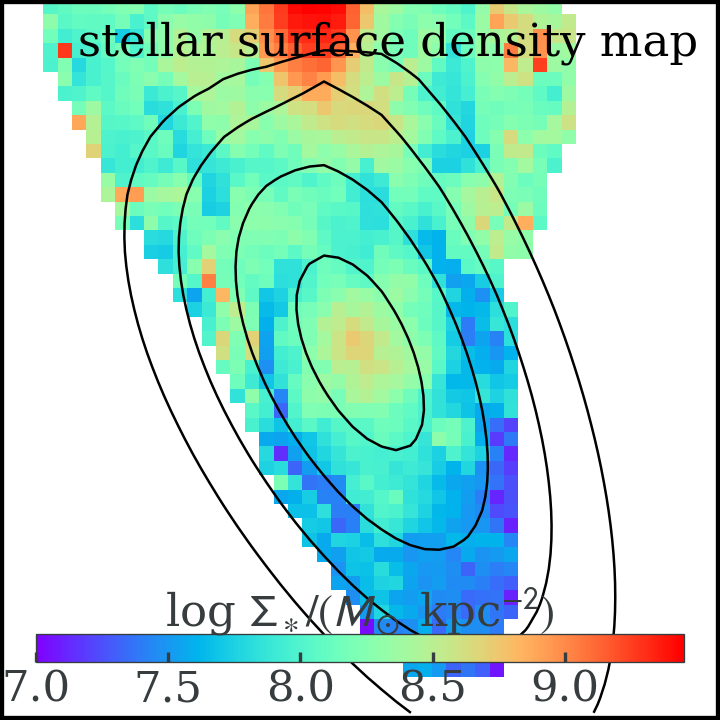}
    \includegraphics[width=.16\textwidth]{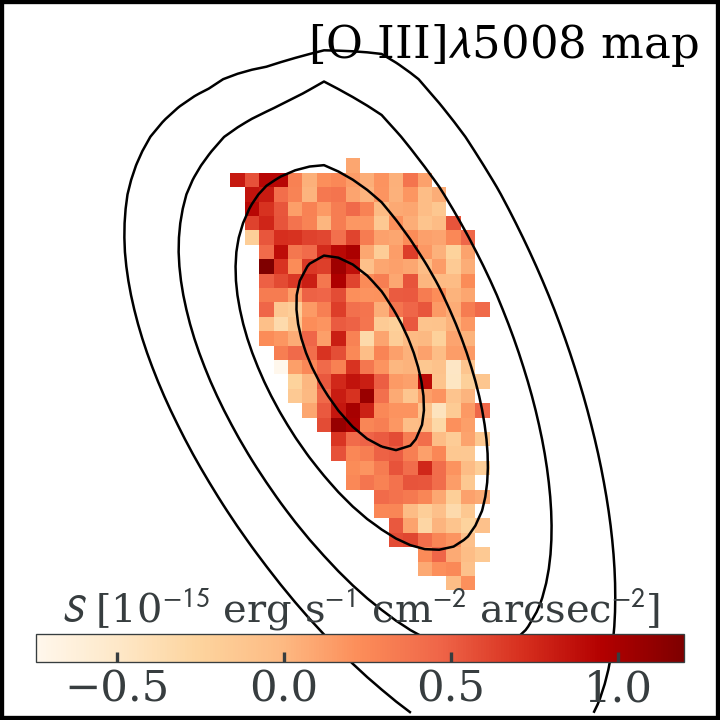}
    \includegraphics[width=.16\textwidth]{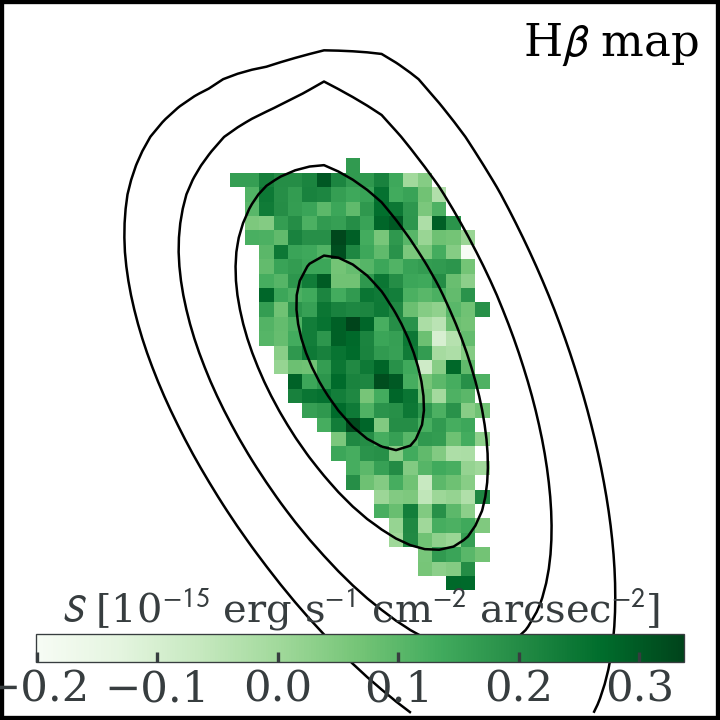}
    \includegraphics[width=.16\textwidth]{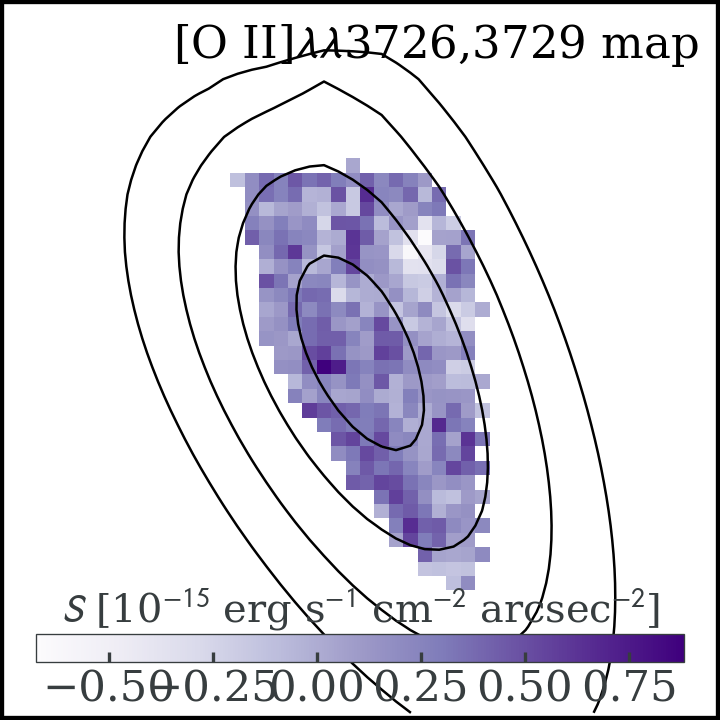}
    \includegraphics[width=.16\textwidth]{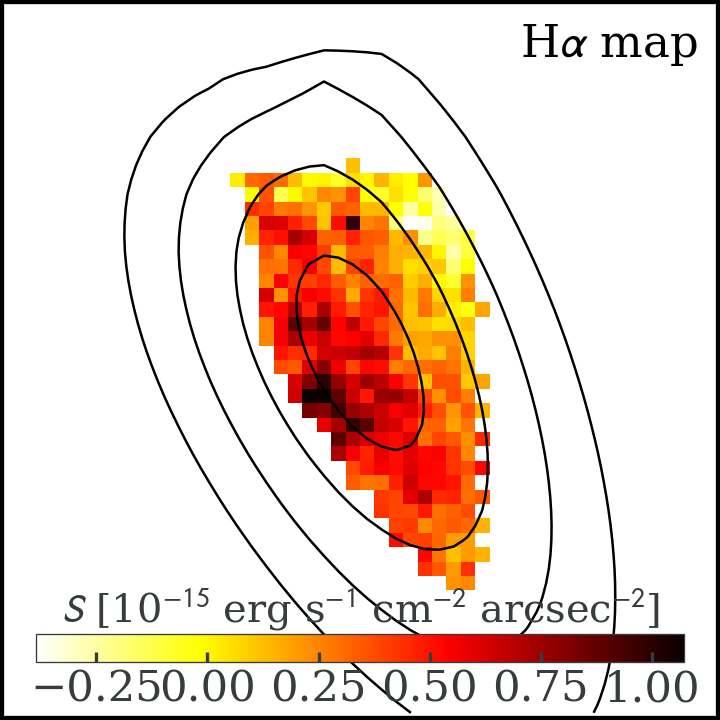}\\
    \includegraphics[width=\textwidth]{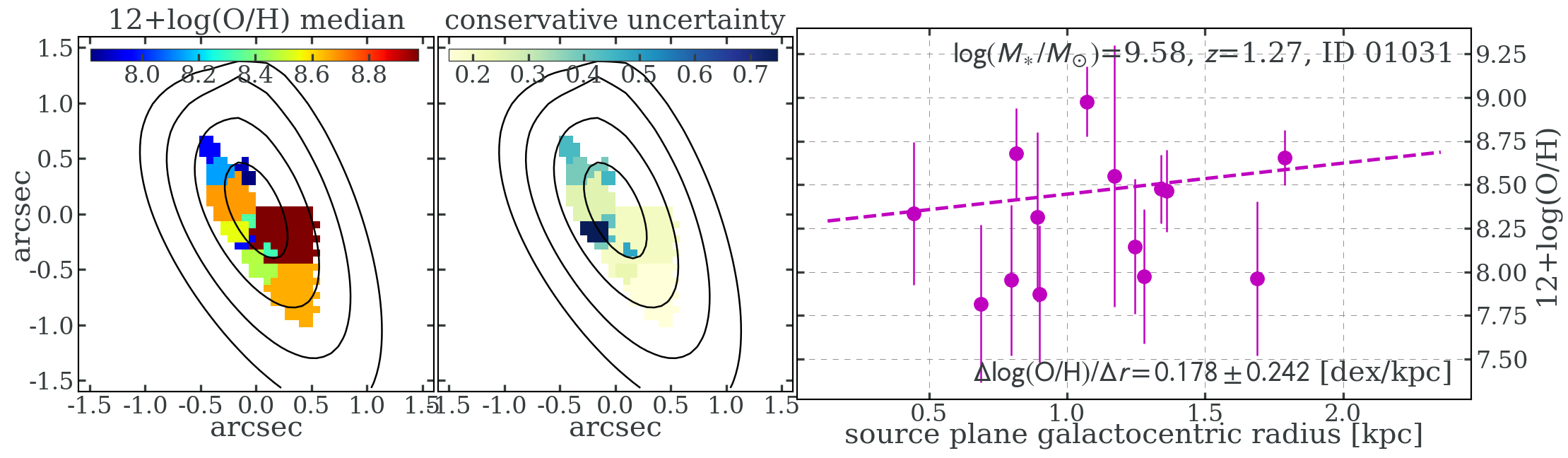}
    \caption{The source ID01031 in the field of \clba is shown.}
    \label{fig:clM0744_ID01031_figs}
\end{figure*}

\begin{figure*}
    \centering
    \includegraphics[width=\textwidth]{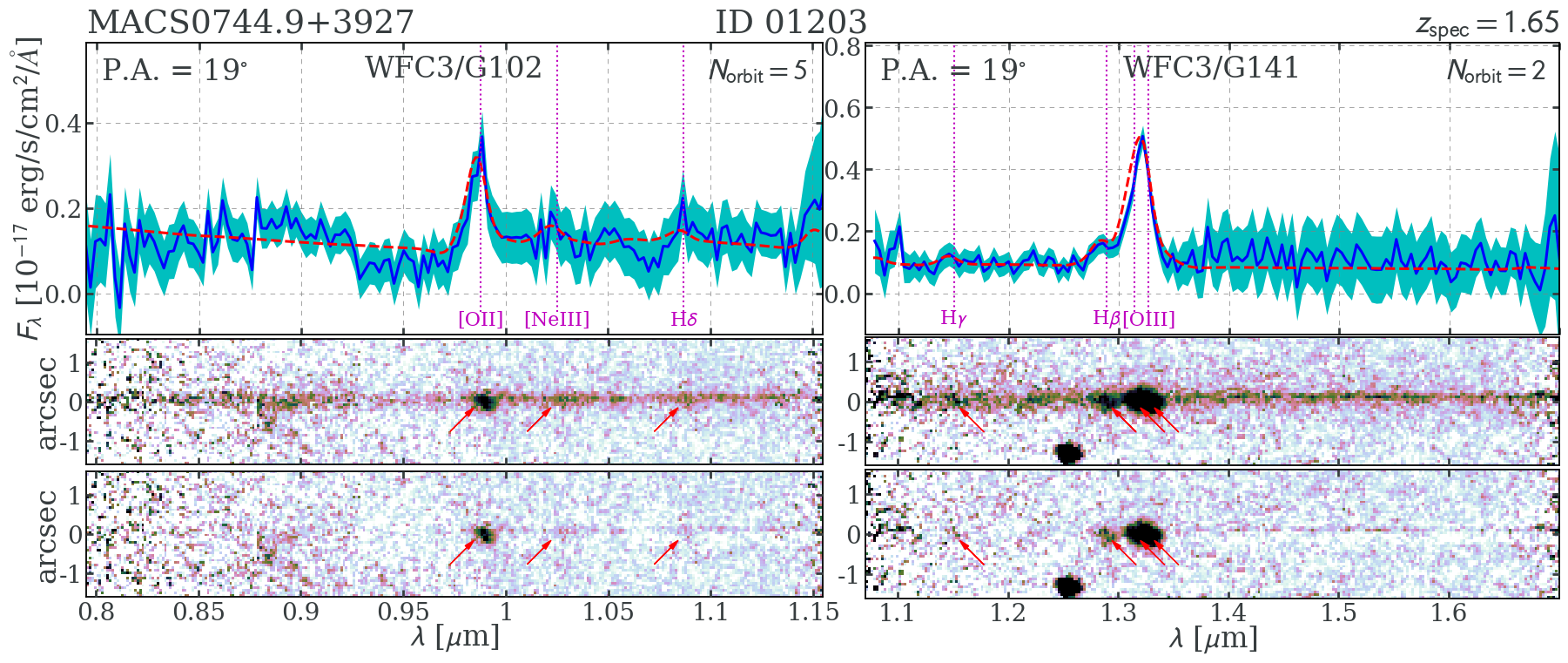}\\
    \includegraphics[width=\textwidth]{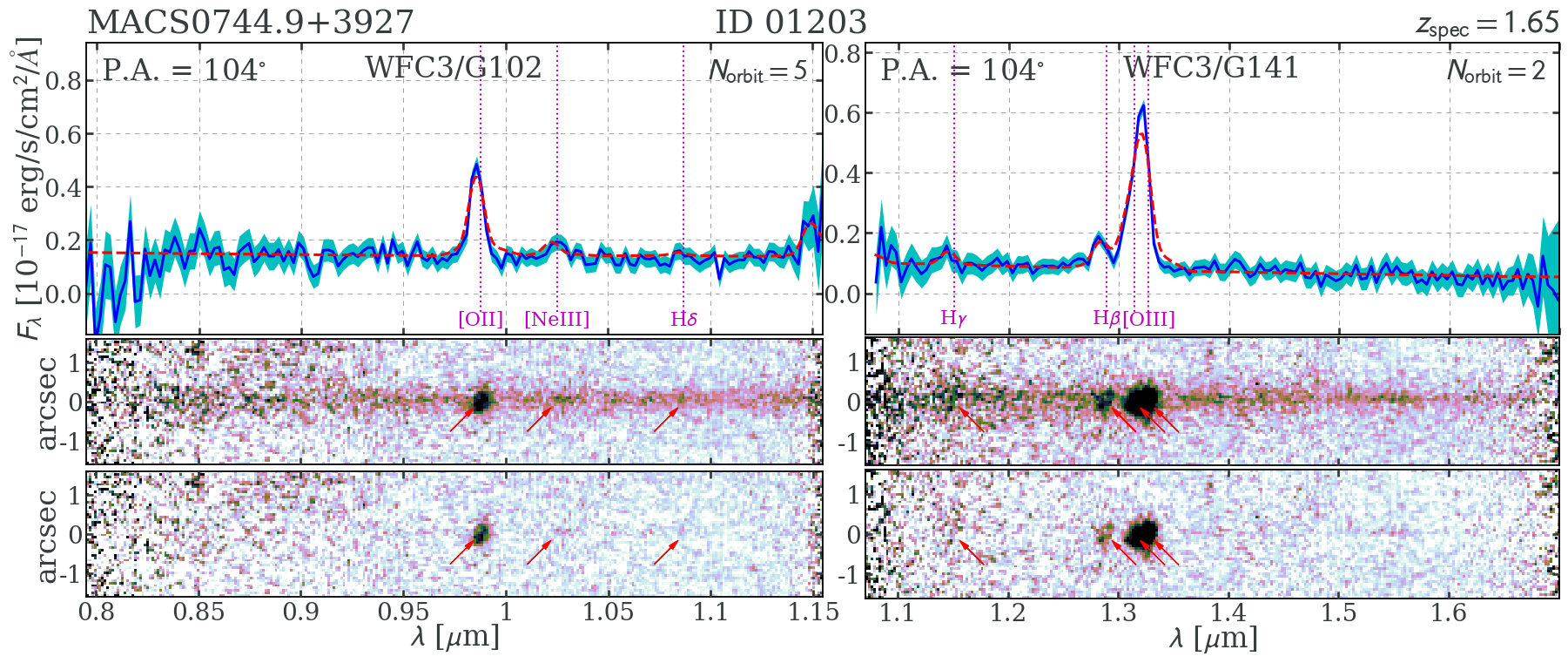}\\
    \includegraphics[width=.16\textwidth]{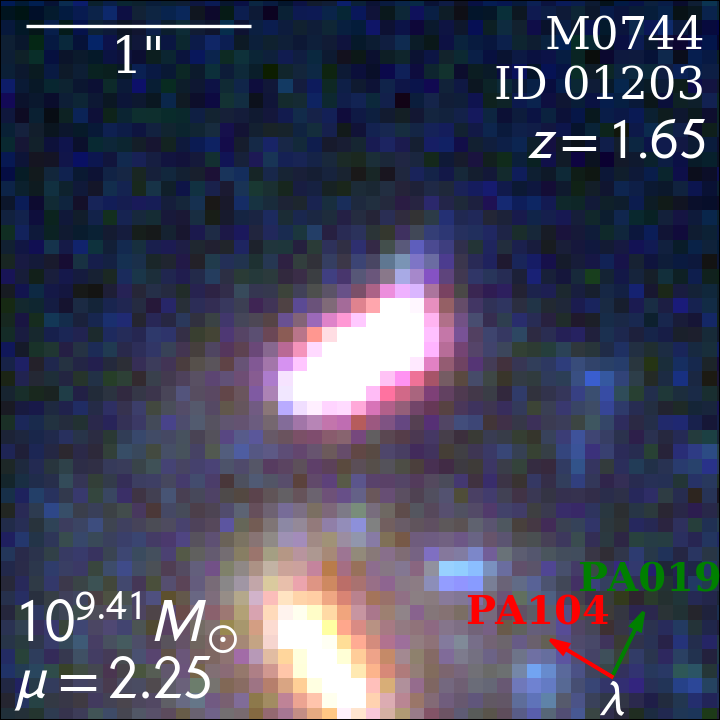}
    \includegraphics[width=.16\textwidth]{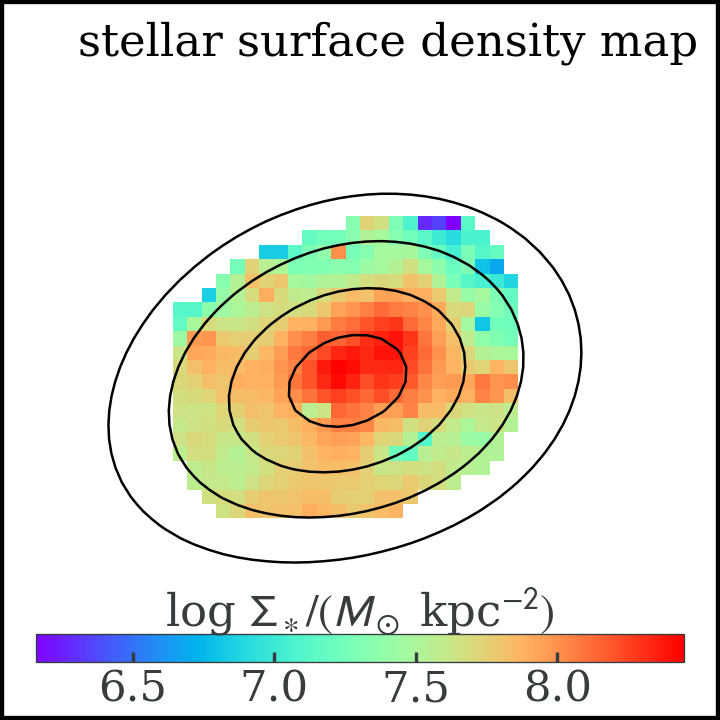}
    \includegraphics[width=.16\textwidth]{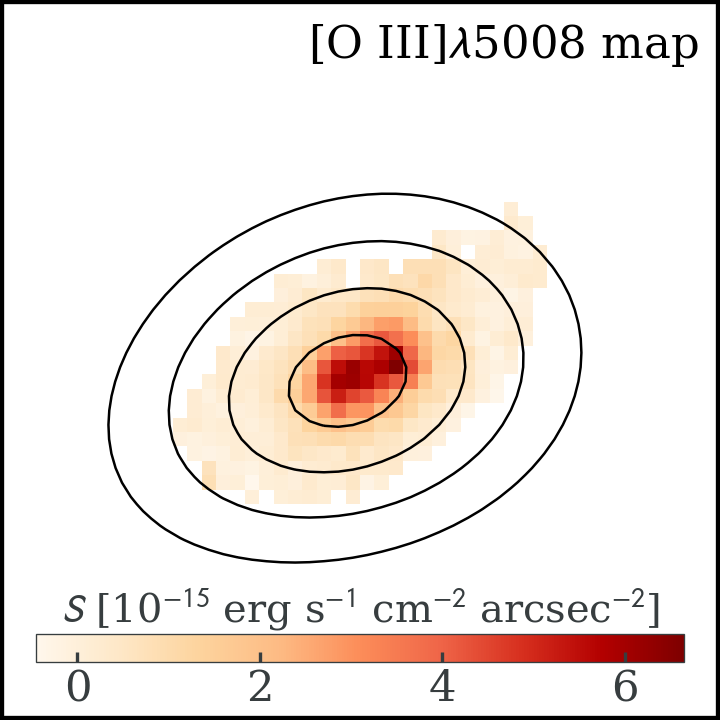}
    \includegraphics[width=.16\textwidth]{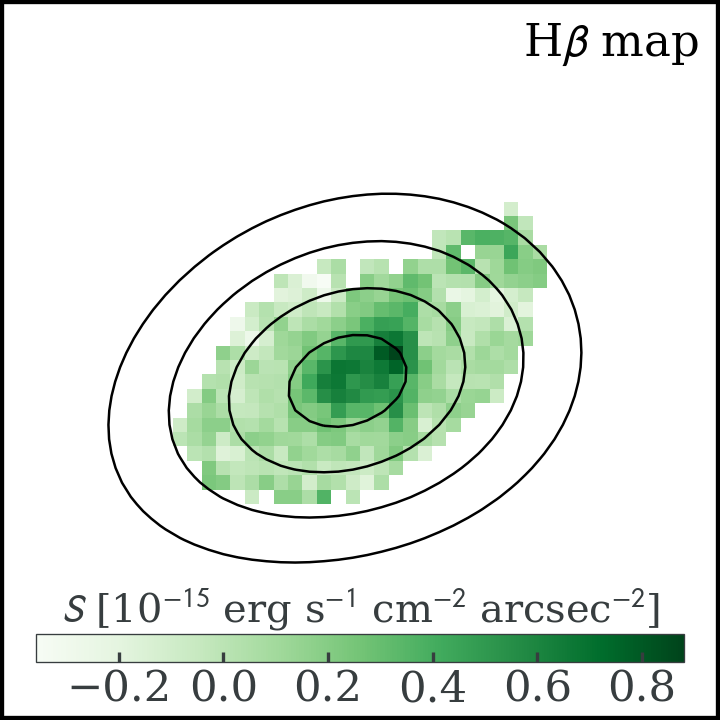}
    \includegraphics[width=.16\textwidth]{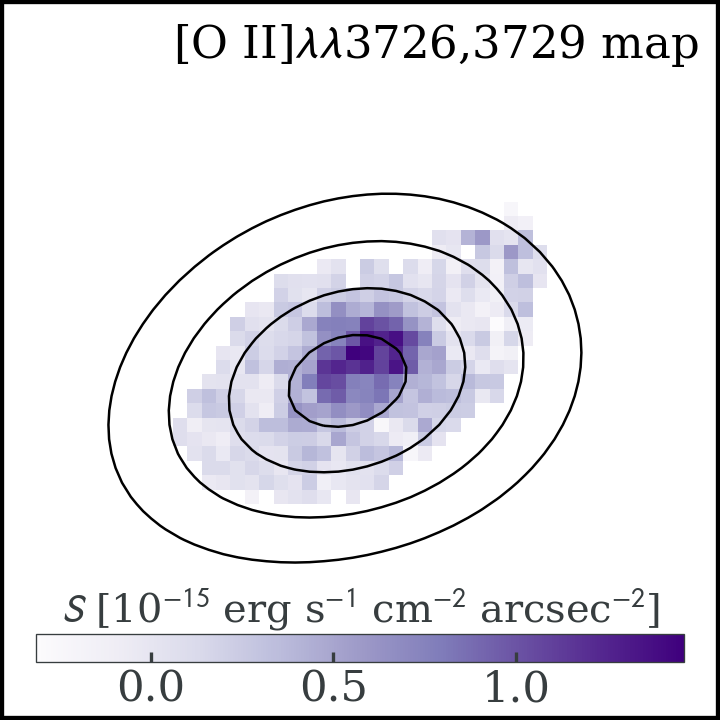}
    \includegraphics[width=.16\textwidth]{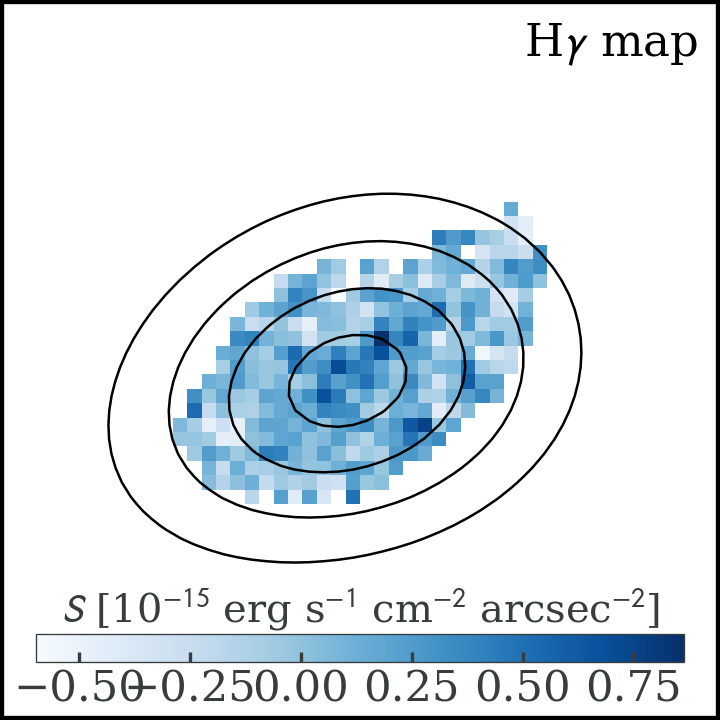}\\
    \includegraphics[width=\textwidth]{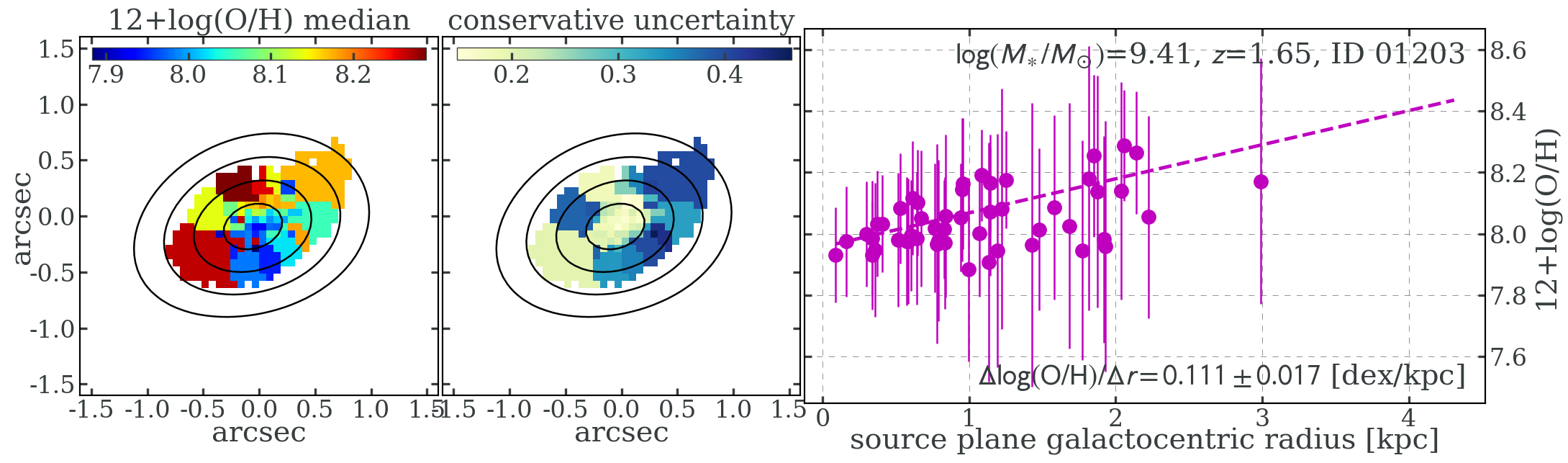}
    \caption{The source ID01203 in the field of \clba is shown.}
    \label{fig:clM0744_ID01203_figs}
\end{figure*}

\begin{figure*}
    \centering
    \includegraphics[width=\textwidth]{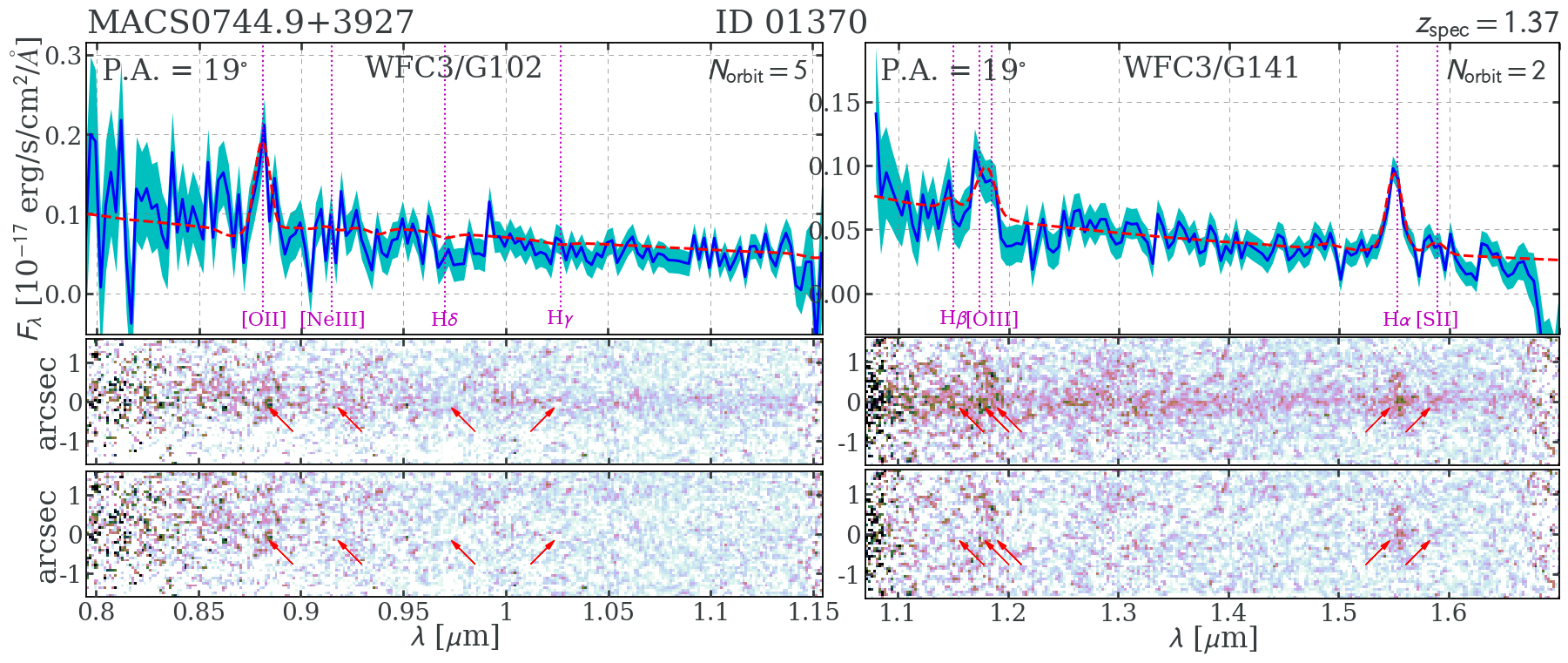}\\
    \includegraphics[width=\textwidth]{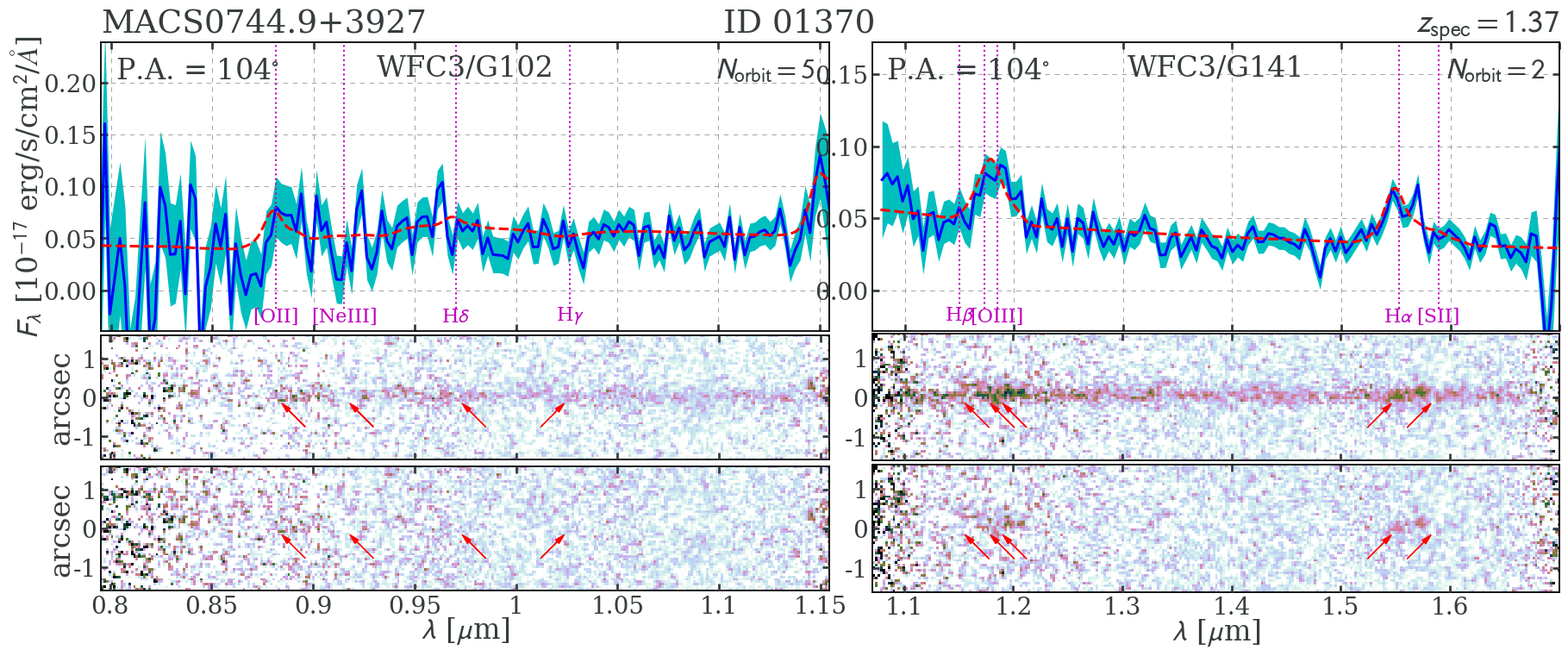}\\
    \includegraphics[width=.16\textwidth]{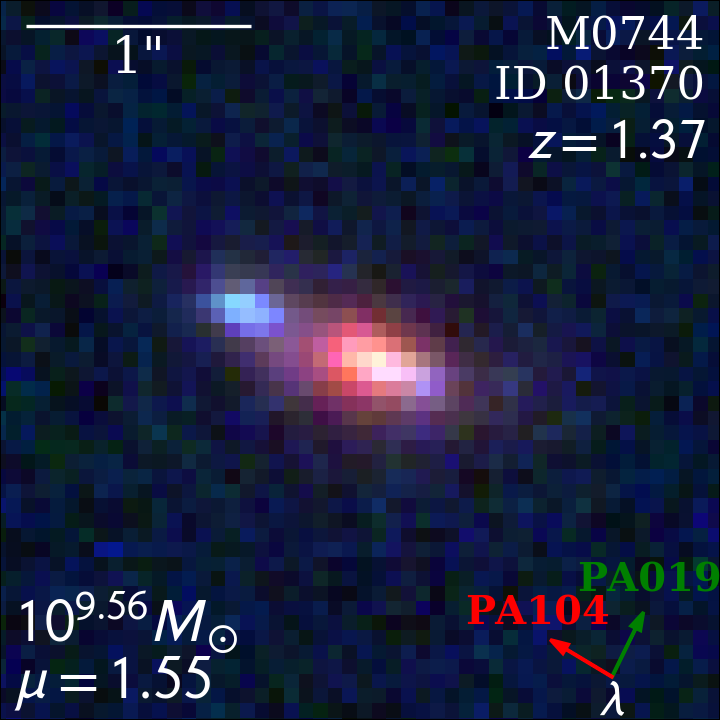}
    \includegraphics[width=.16\textwidth]{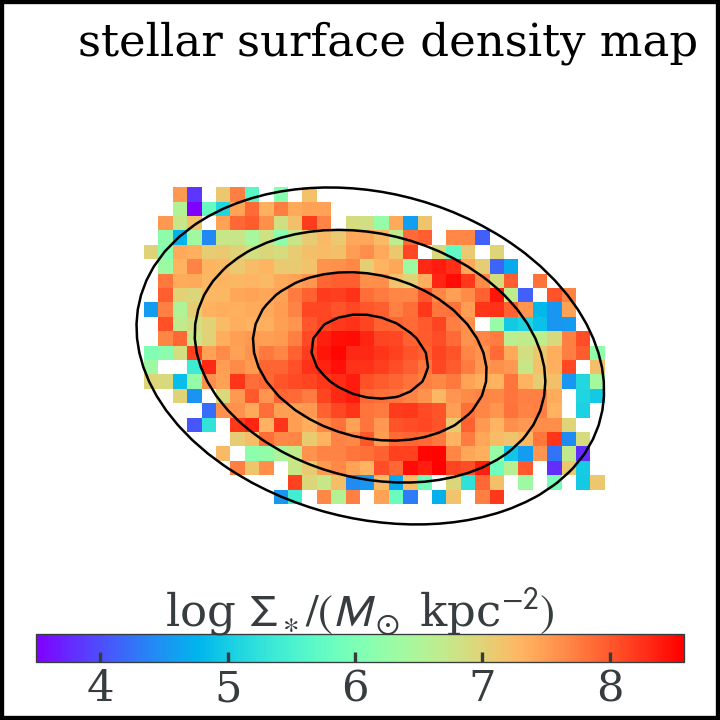}
    \includegraphics[width=.16\textwidth]{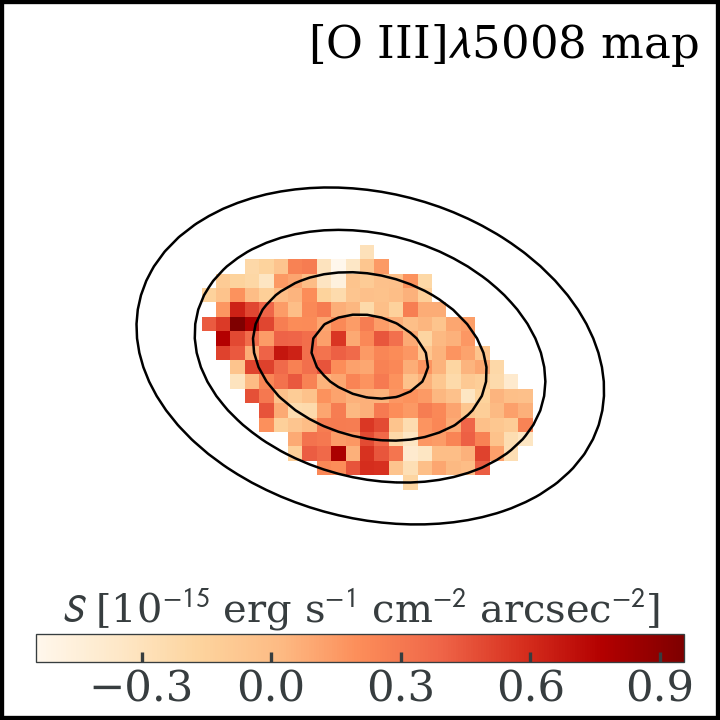}
    \includegraphics[width=.16\textwidth]{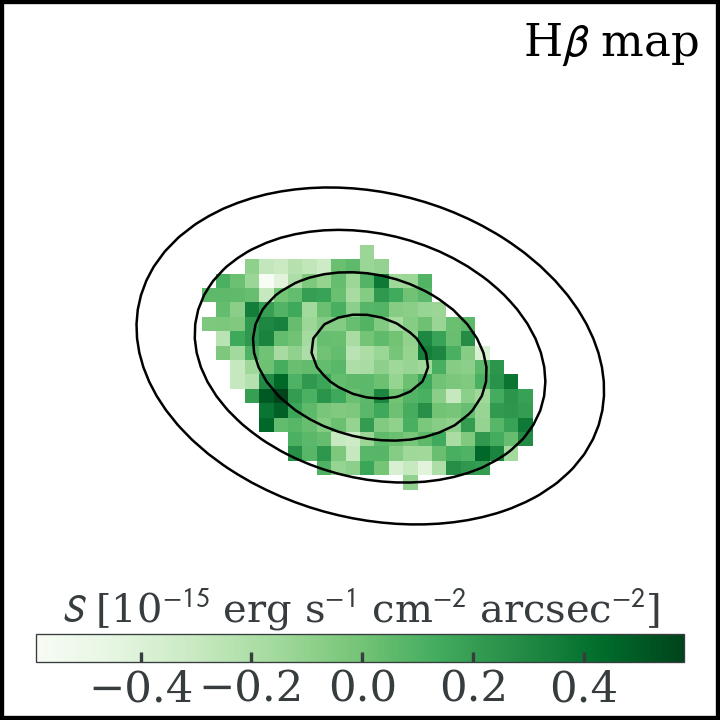}
    \includegraphics[width=.16\textwidth]{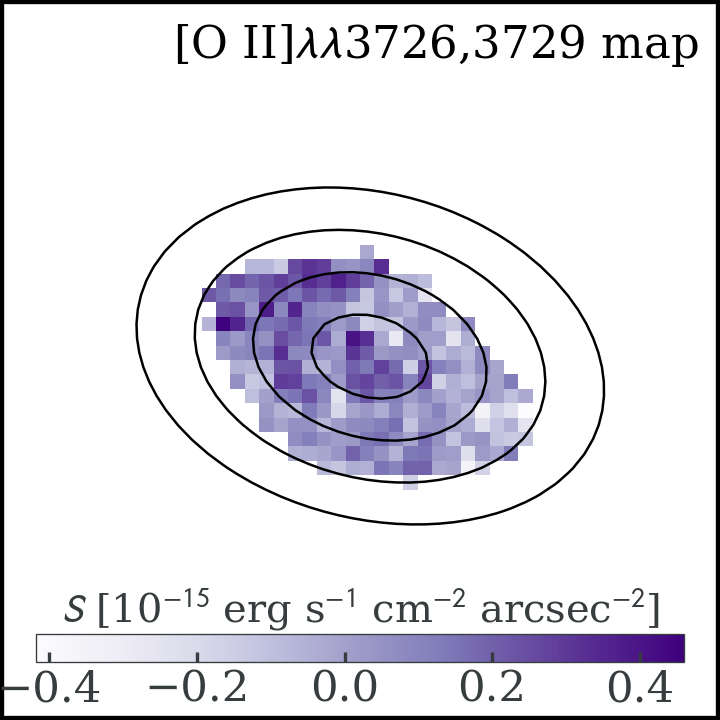}
    \includegraphics[width=.16\textwidth]{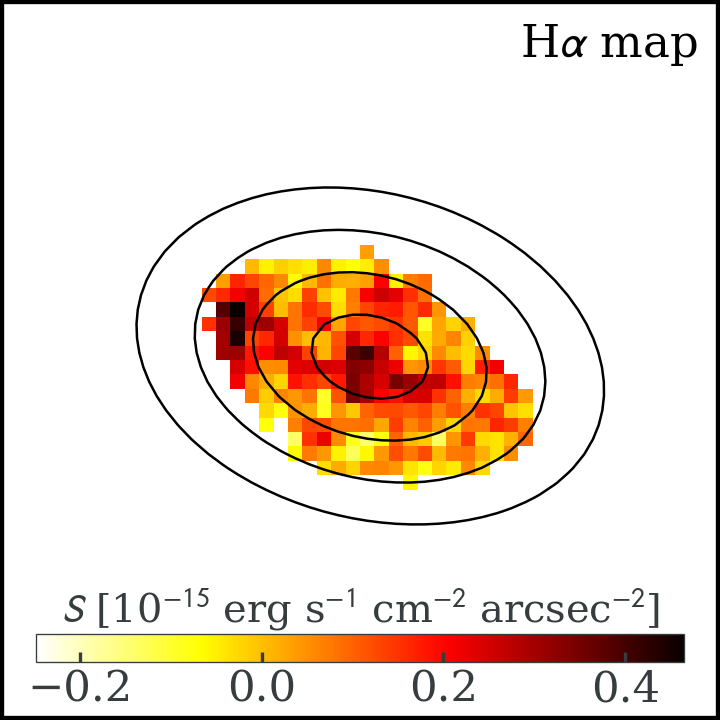}\\
    \includegraphics[width=\textwidth]{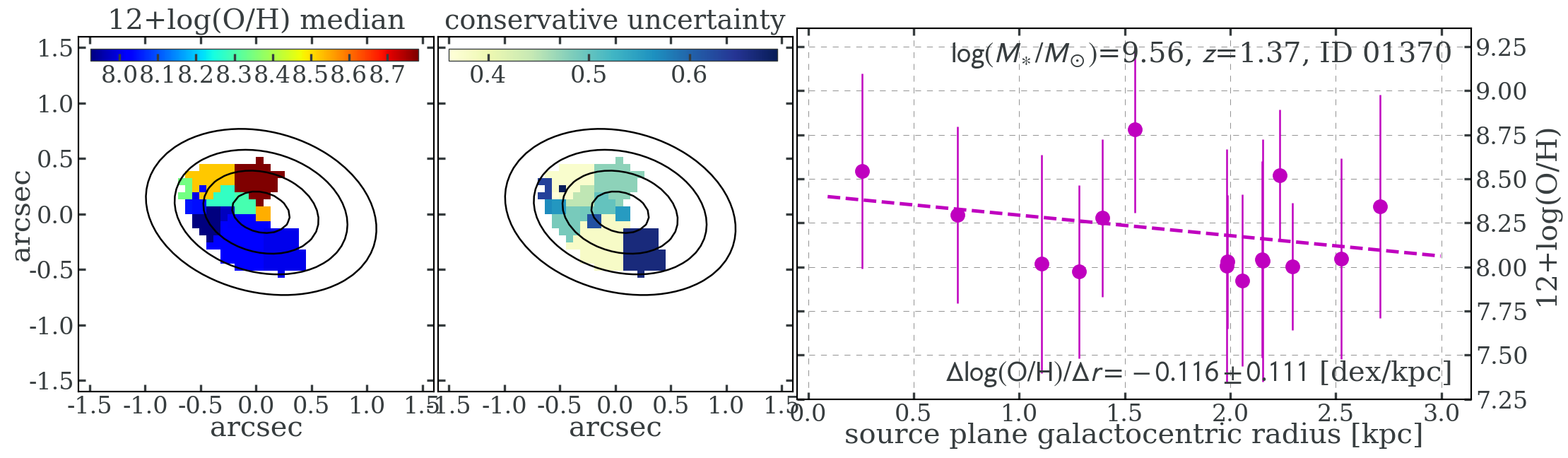}
    \caption{The source ID01370 in the field of \clba is shown.}
    \label{fig:clM0744_ID01370_figs}
\end{figure*}

\begin{figure*}
    \centering
    \includegraphics[width=\textwidth]{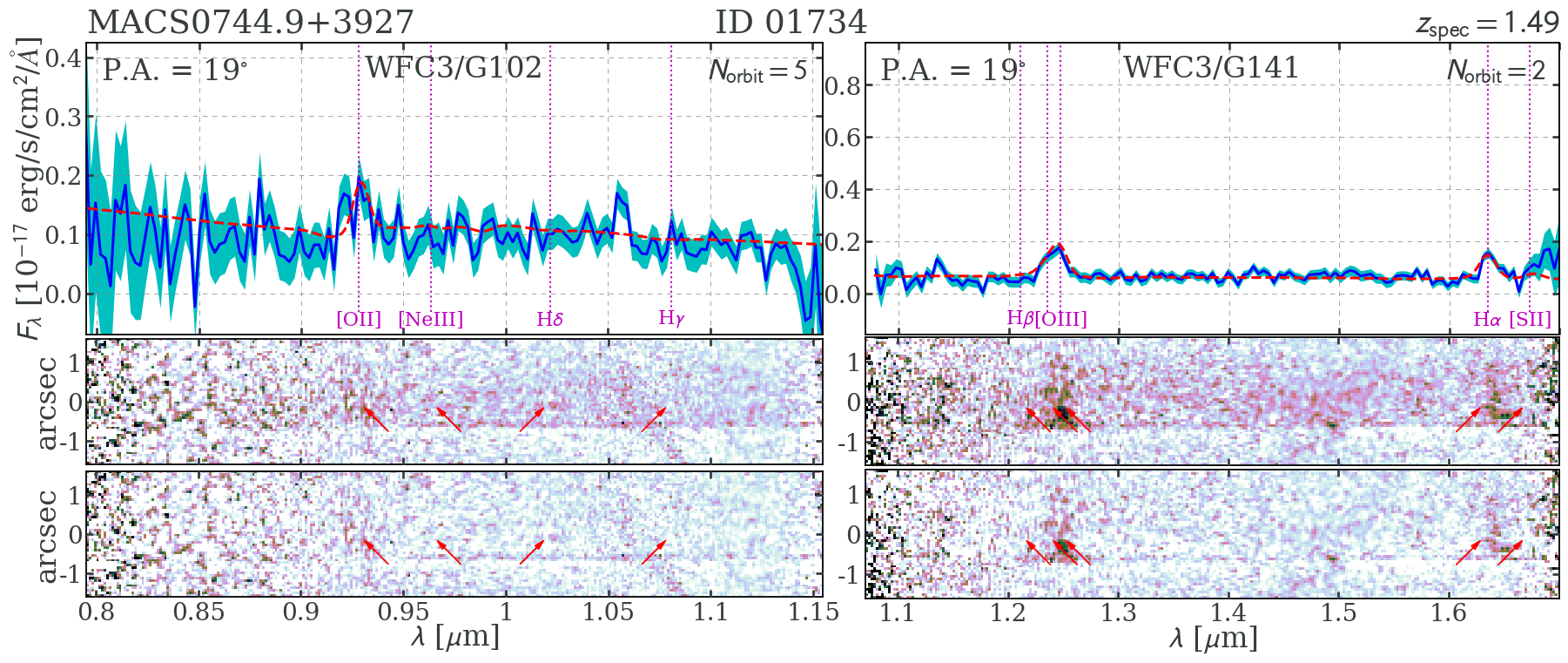}\\
    \includegraphics[width=\textwidth]{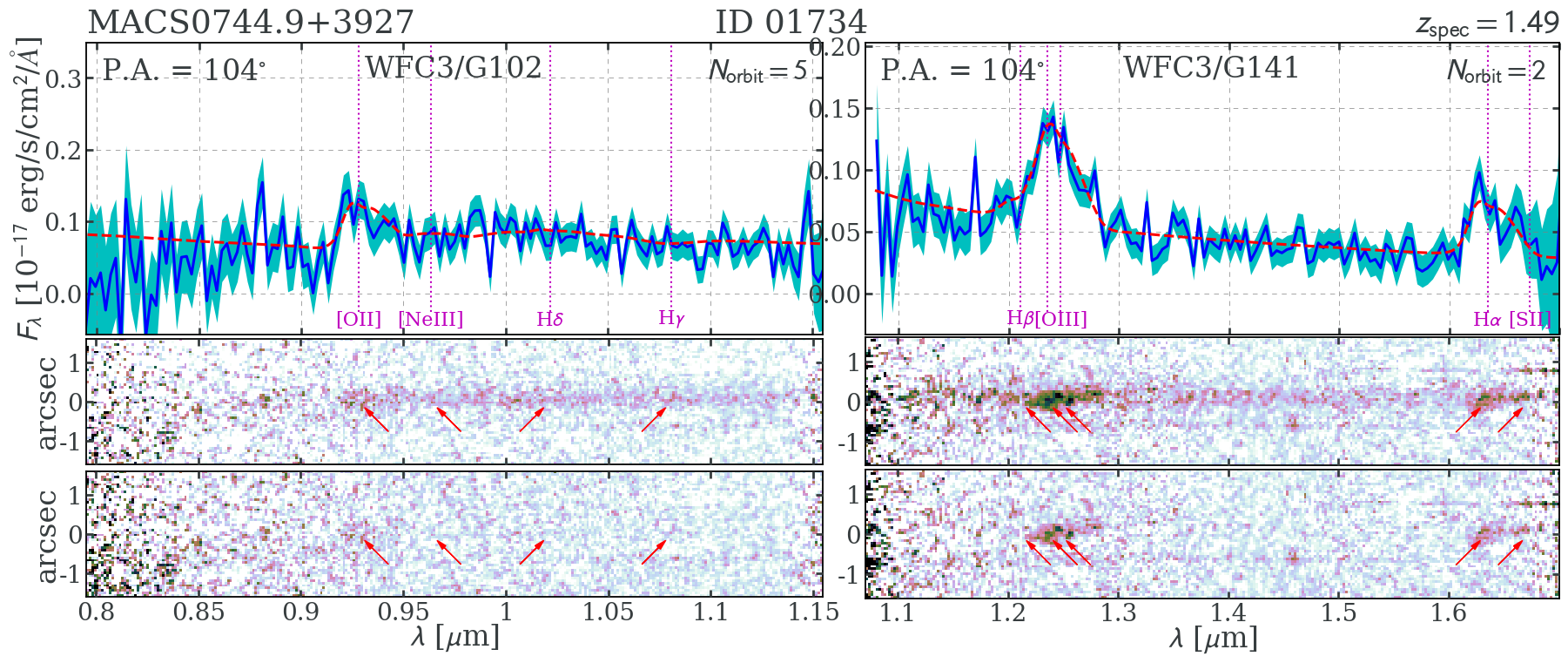}\\
    \includegraphics[width=.16\textwidth]{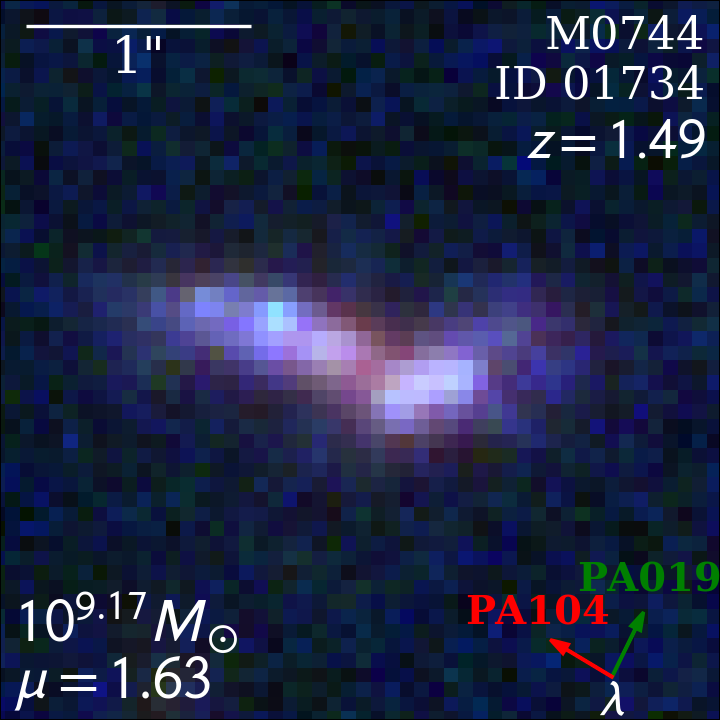}
    \includegraphics[width=.16\textwidth]{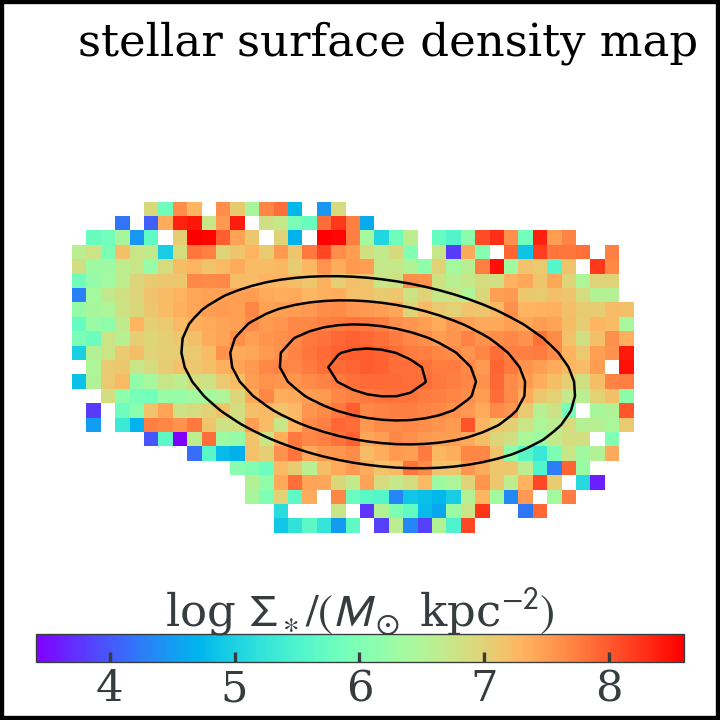}
    \includegraphics[width=.16\textwidth]{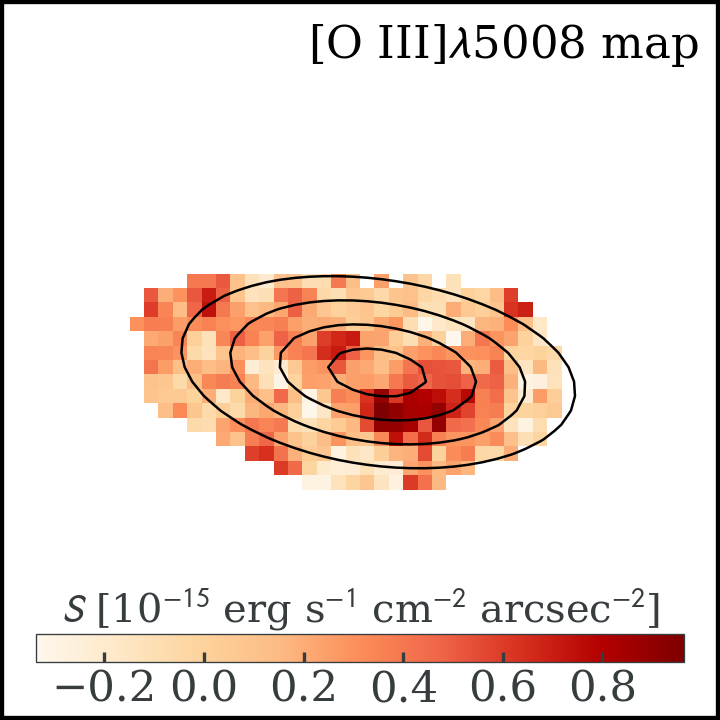}
    \includegraphics[width=.16\textwidth]{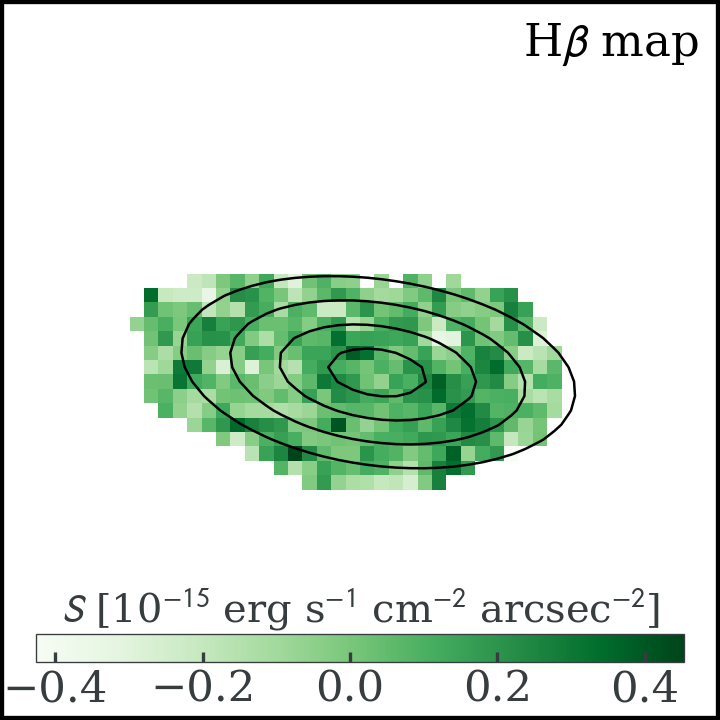}
    \includegraphics[width=.16\textwidth]{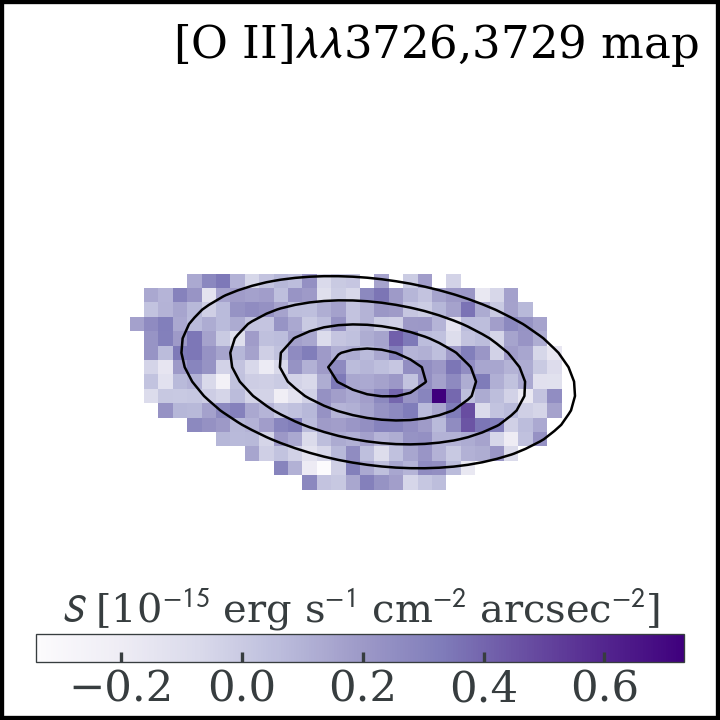}
    \includegraphics[width=.16\textwidth]{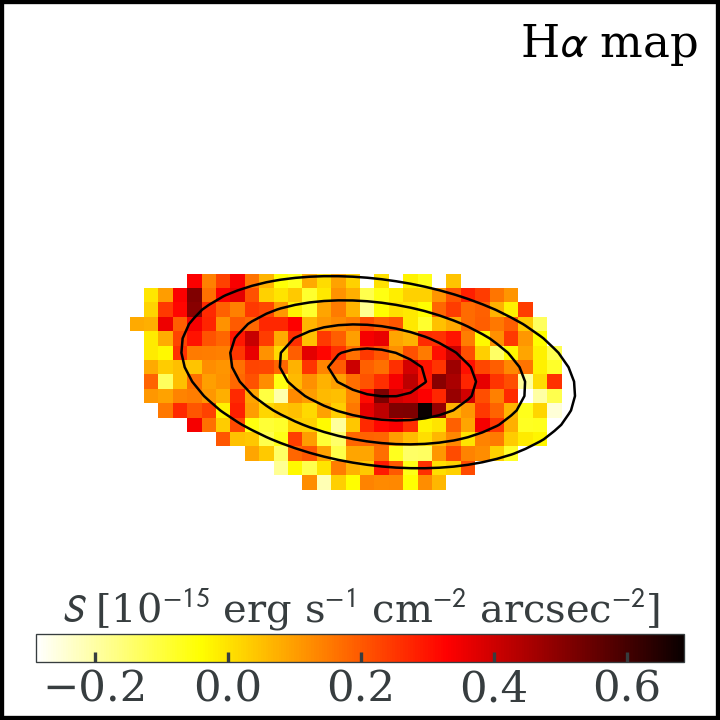}\\
    \includegraphics[width=\textwidth]{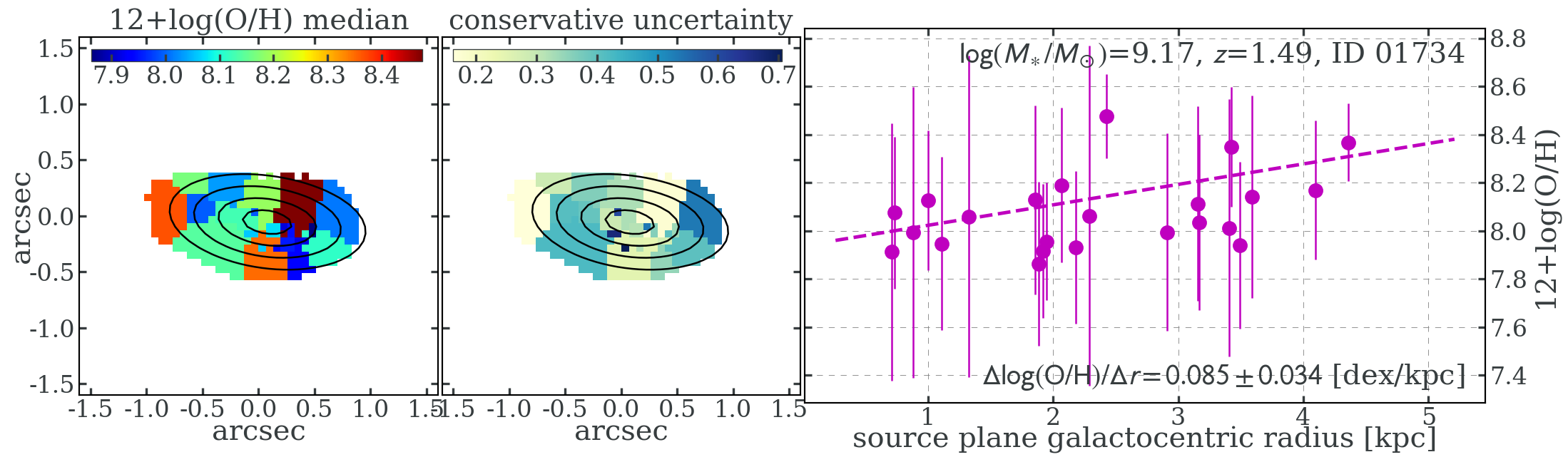}
    \caption{The source ID01734 in the field of \clba is shown.}
    \label{fig:clM0744_ID01734_figs}
\end{figure*}

\begin{figure*}
    \centering
    \includegraphics[width=\textwidth]{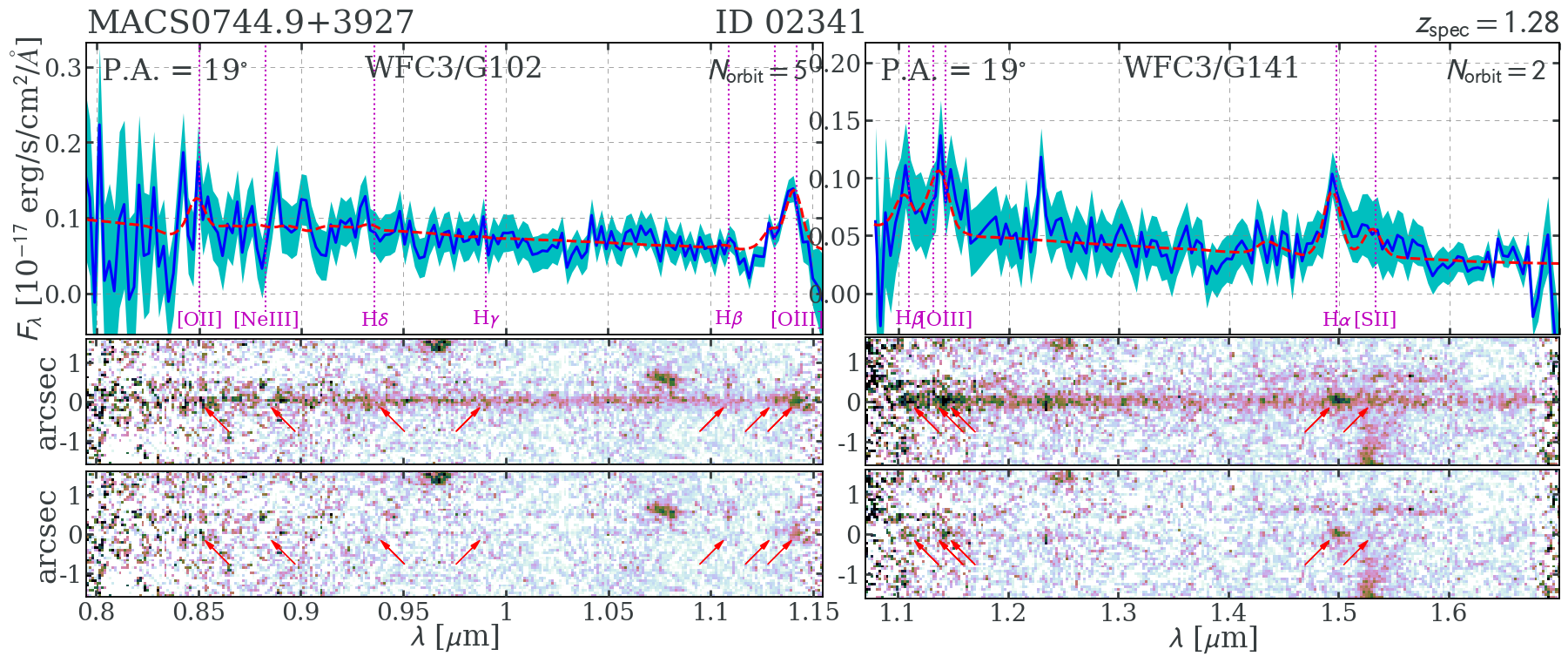}\\
    \includegraphics[width=\textwidth]{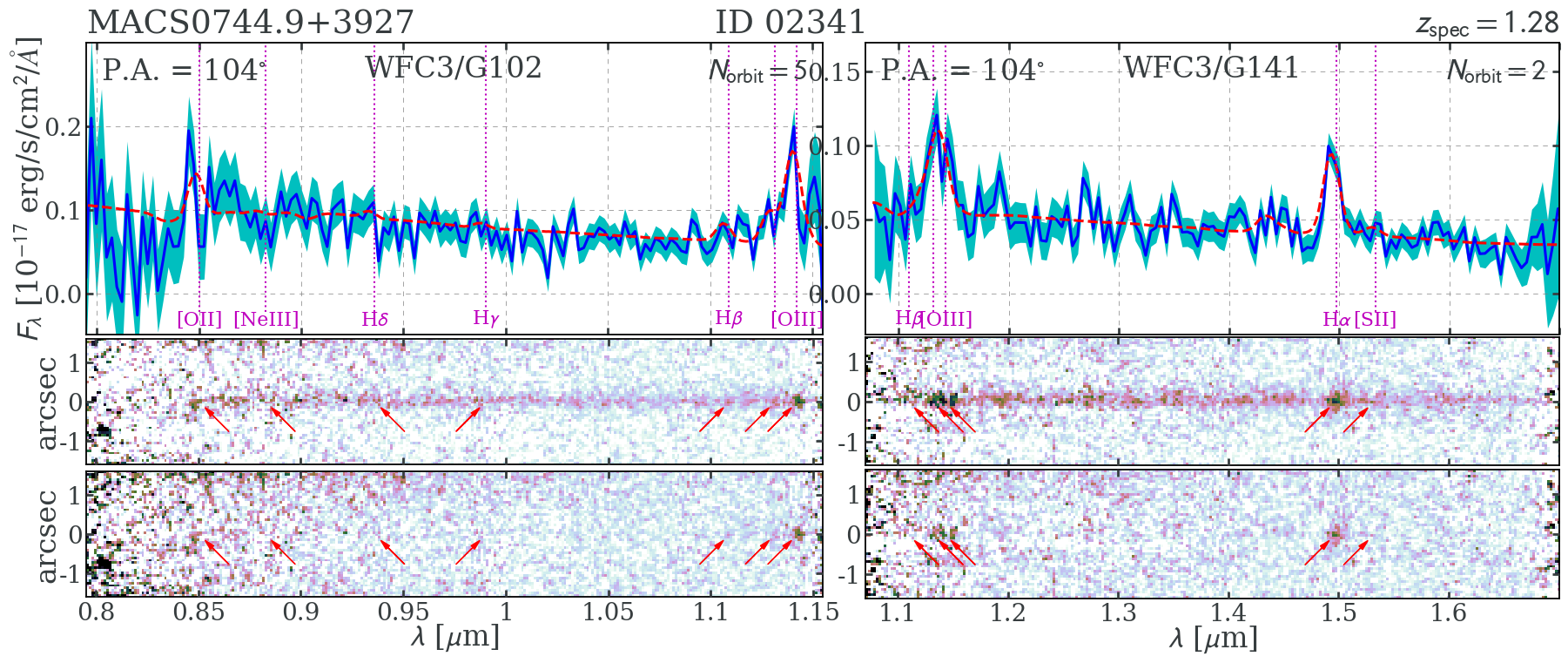}\\
    \includegraphics[width=.16\textwidth]{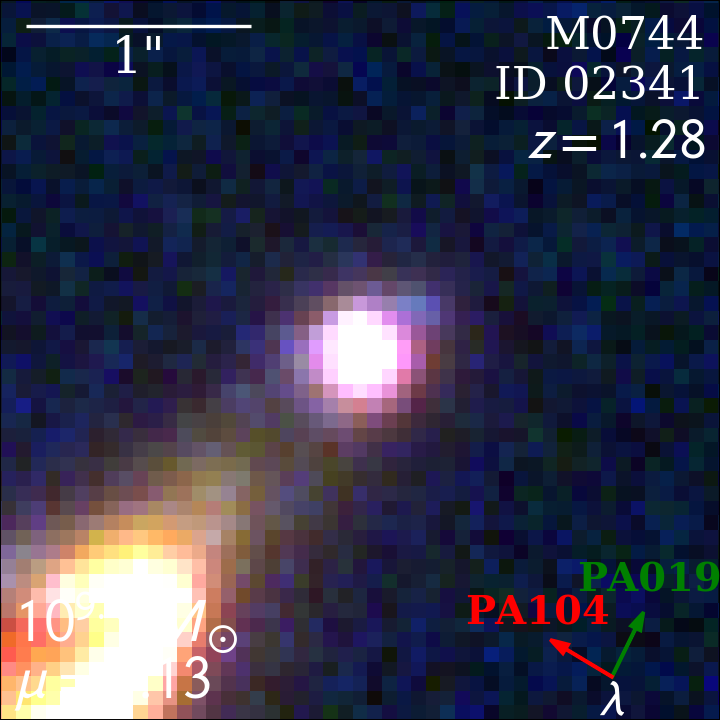}
    \includegraphics[width=.16\textwidth]{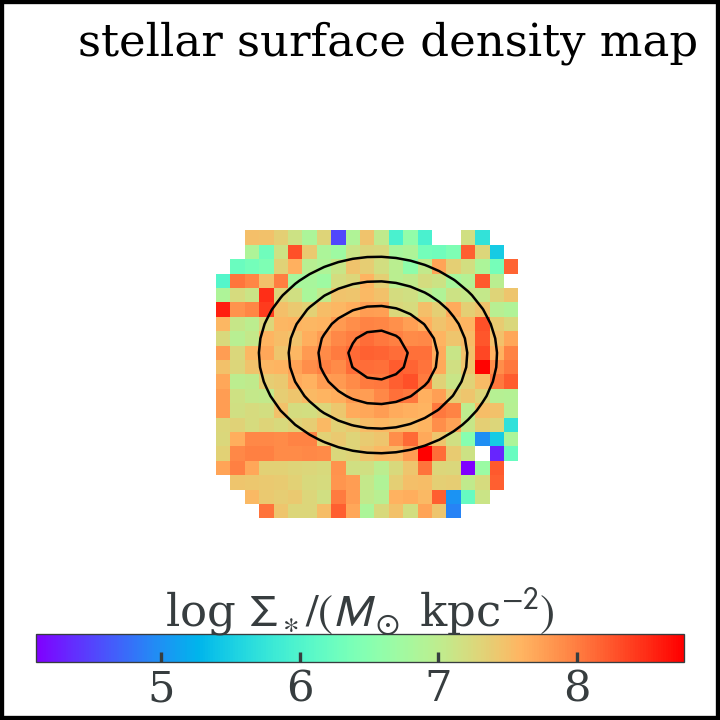}
    \includegraphics[width=.16\textwidth]{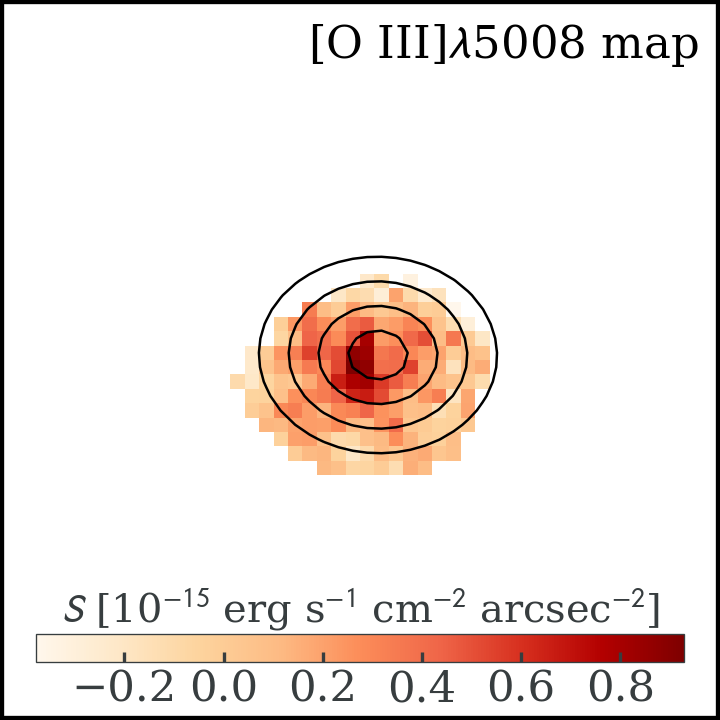}
    \includegraphics[width=.16\textwidth]{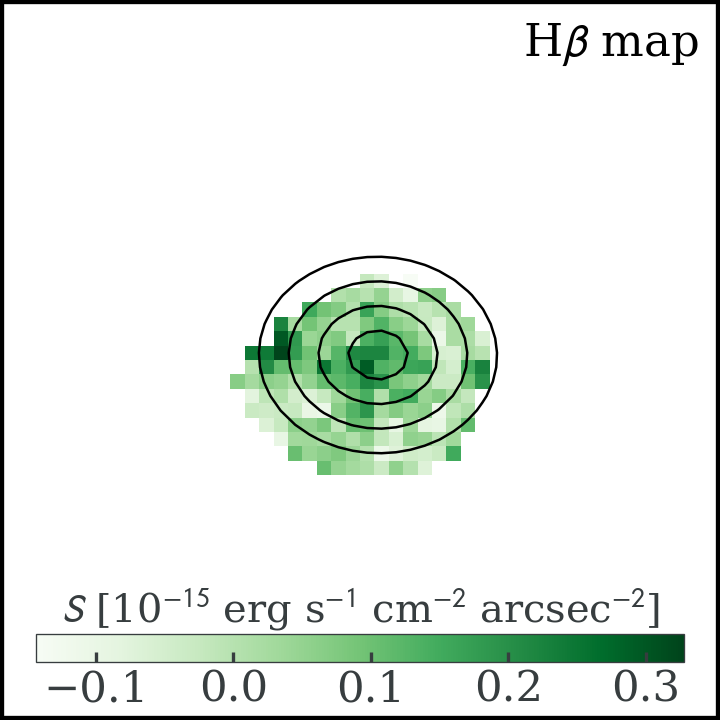}
    \includegraphics[width=.16\textwidth]{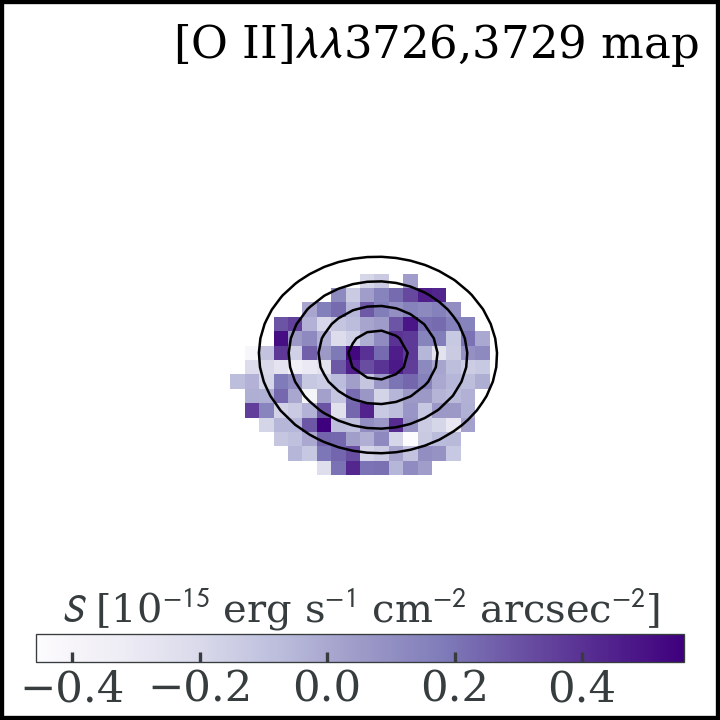}
    \includegraphics[width=.16\textwidth]{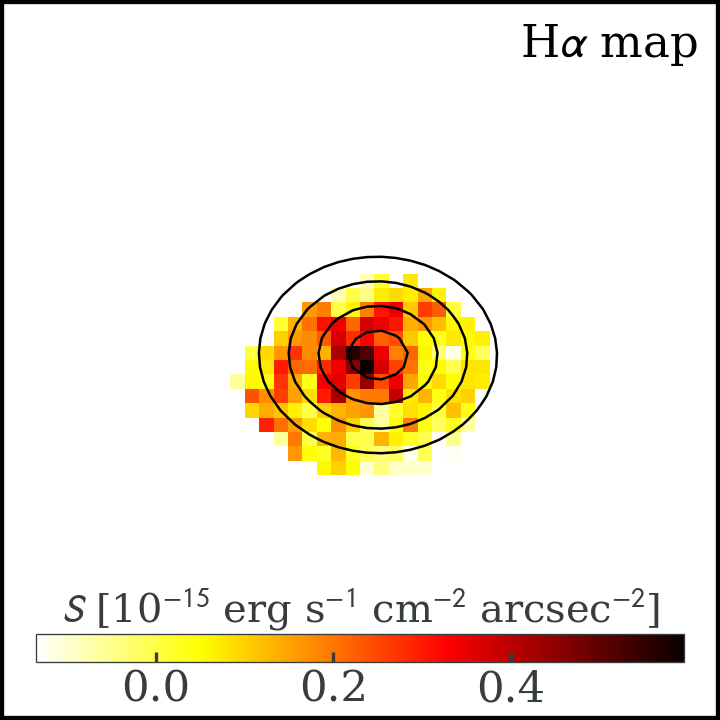}\\
    \includegraphics[width=\textwidth]{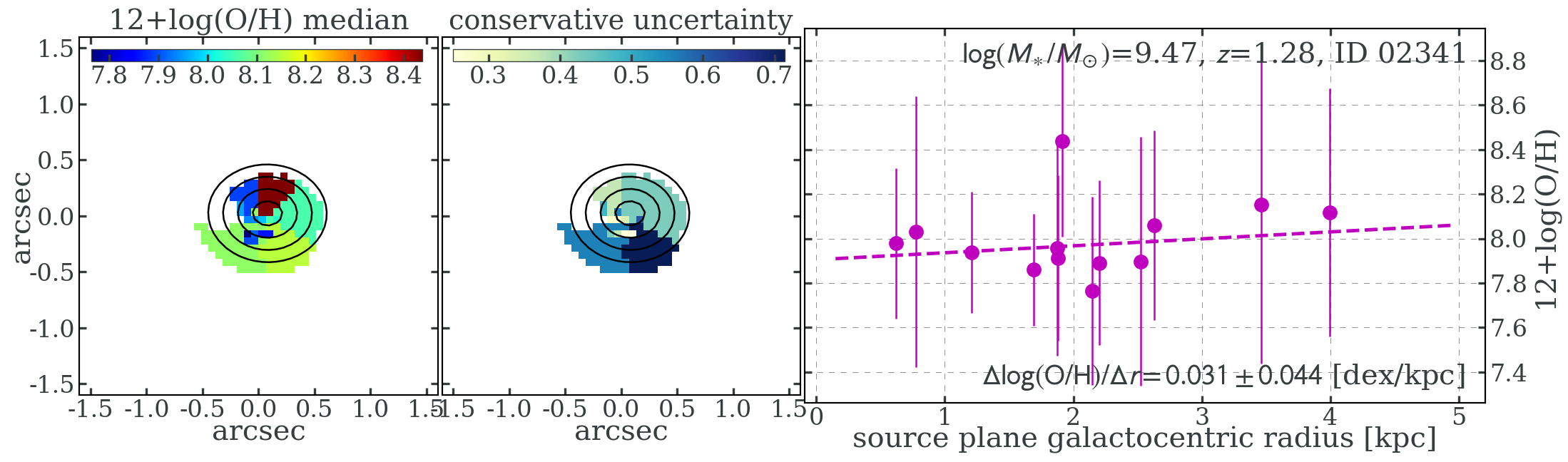}
    \caption{The source ID02341 in the field of \clba is shown.}
    \label{fig:clM0744_ID02341_figs}
\end{figure*}

\begin{figure*}
    \centering
    \includegraphics[width=\textwidth]{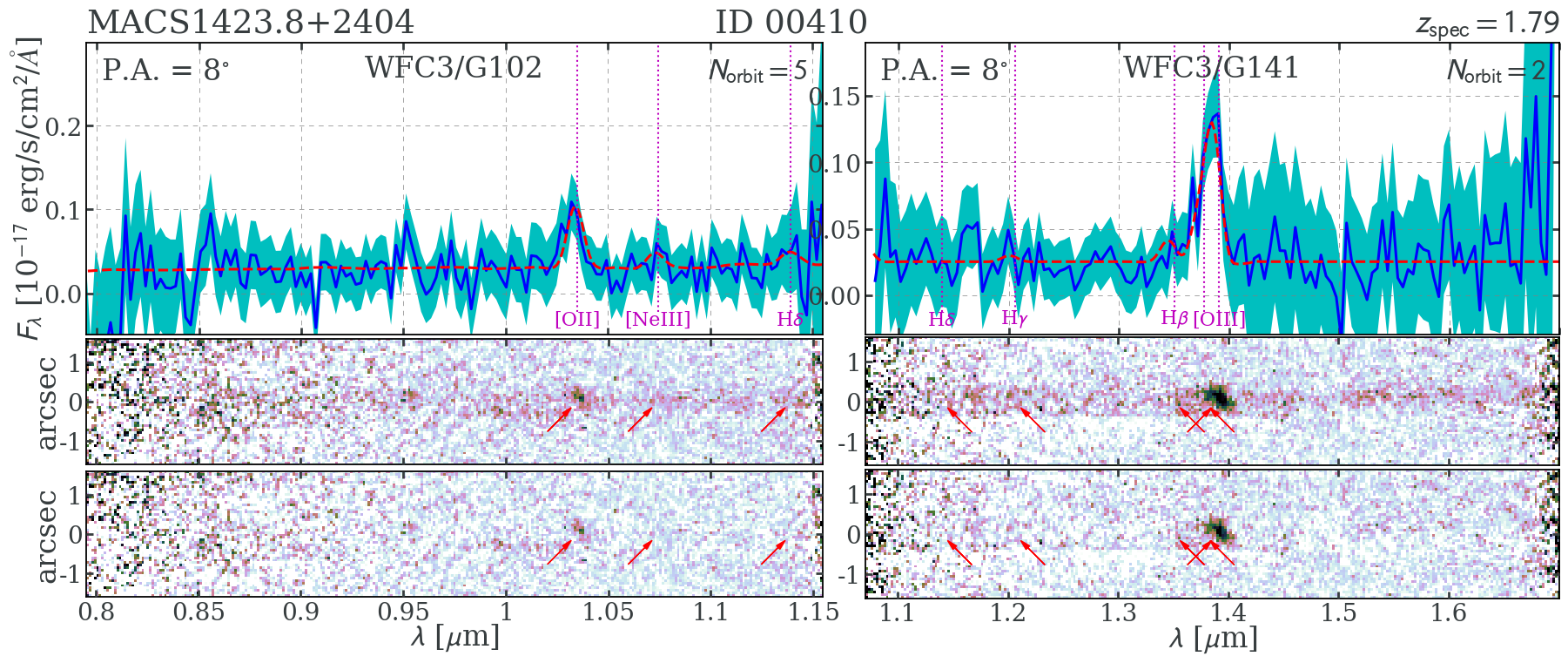}\\
    \includegraphics[width=\textwidth]{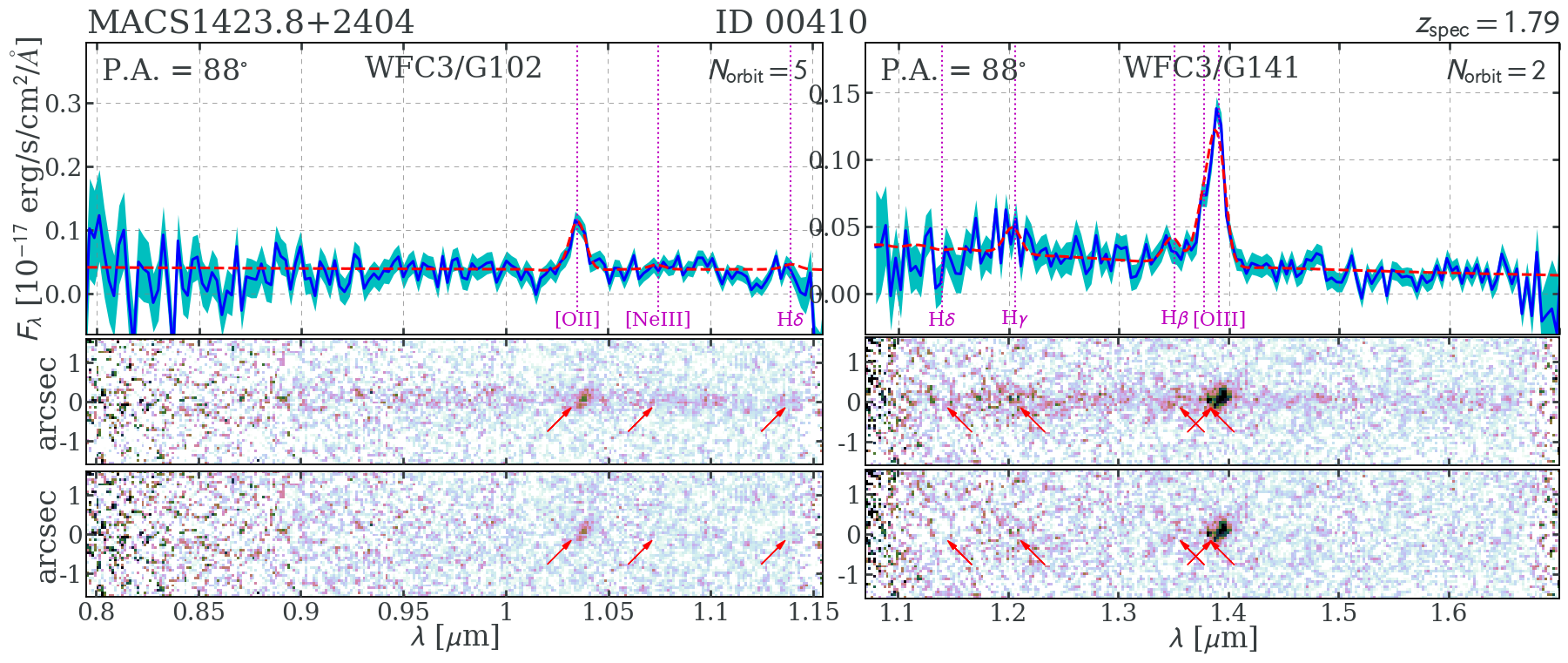}\\
    \includegraphics[width=.16\textwidth]{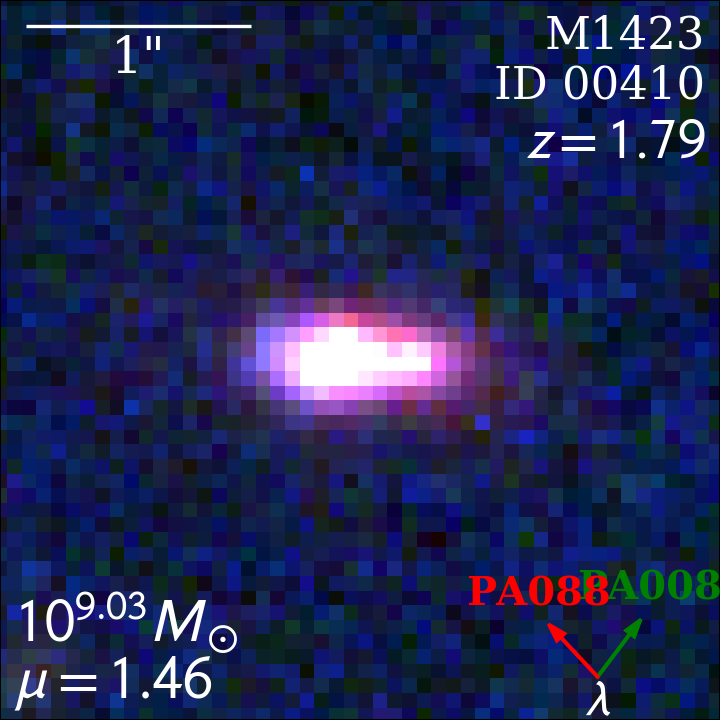}
    \includegraphics[width=.16\textwidth]{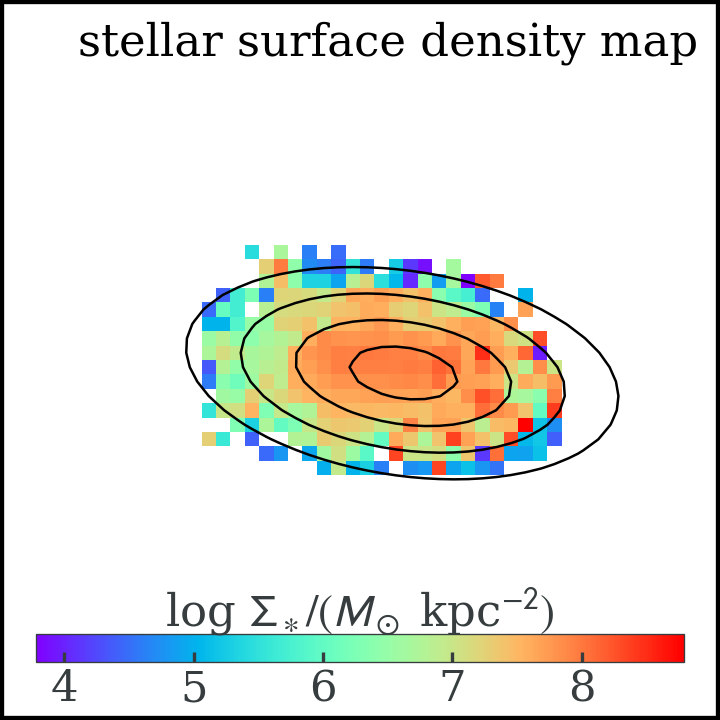}
    \includegraphics[width=.16\textwidth]{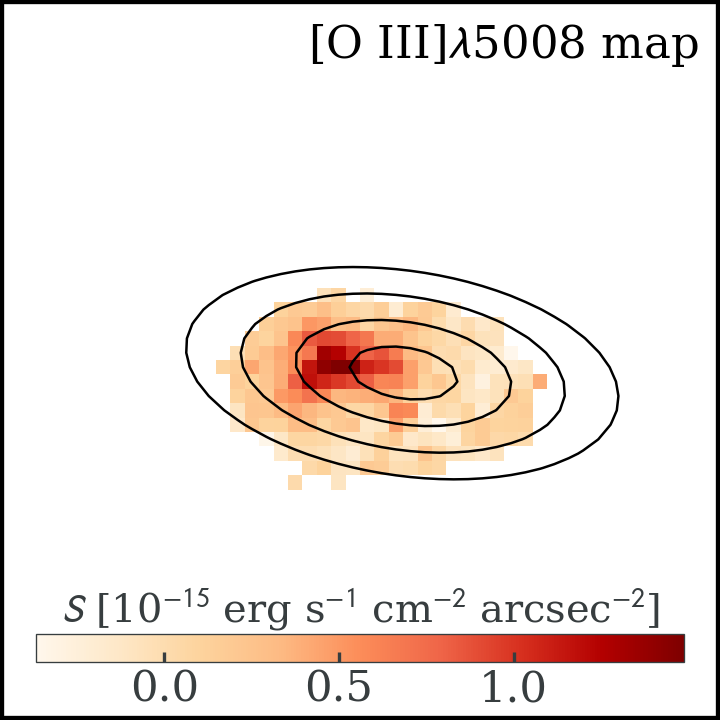}
    \includegraphics[width=.16\textwidth]{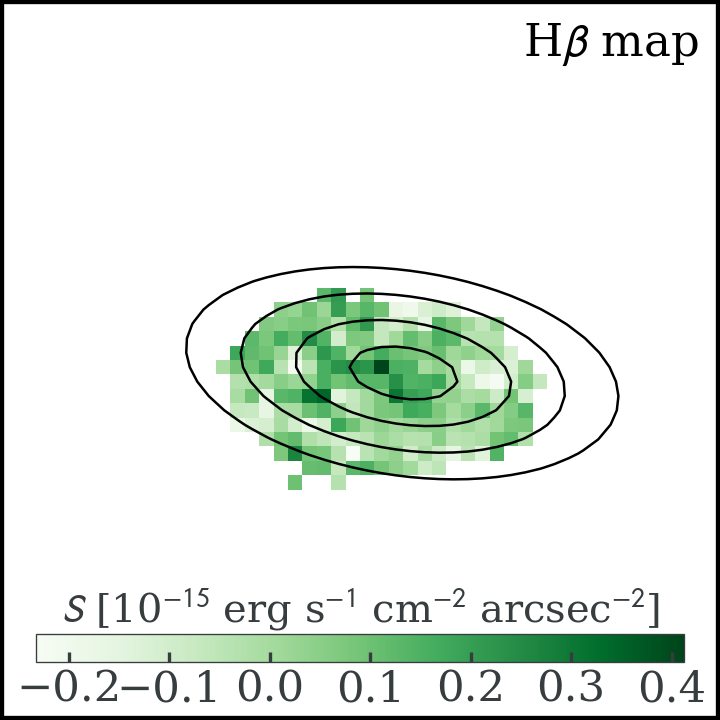}
    \includegraphics[width=.16\textwidth]{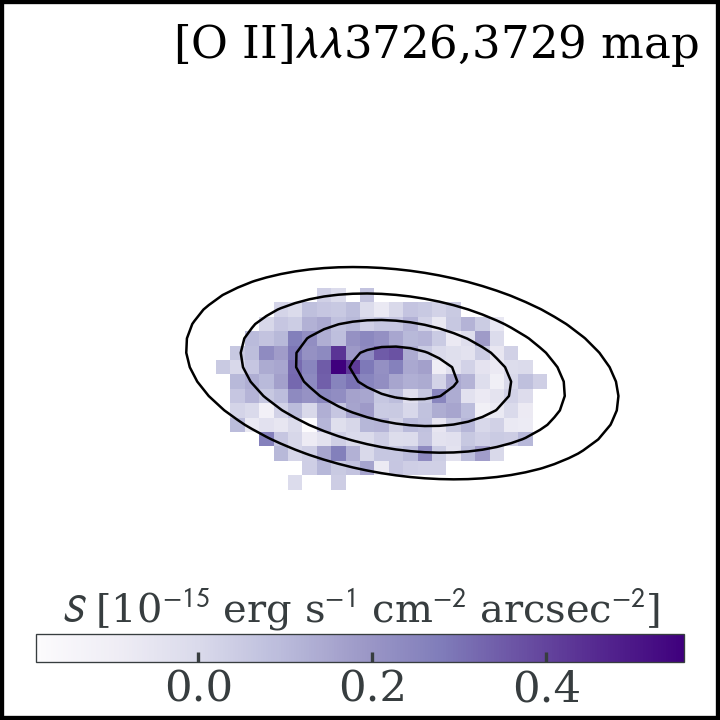}
    \includegraphics[width=.16\textwidth]{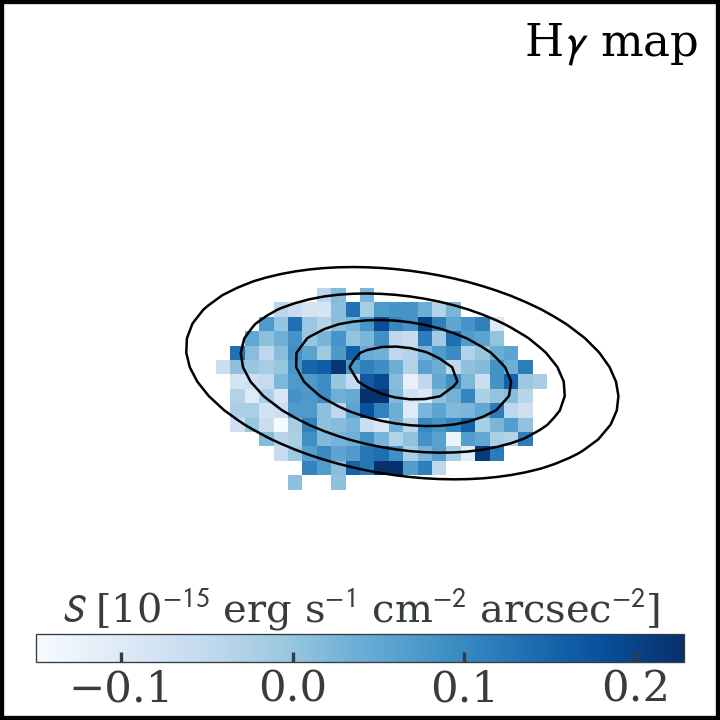}\\
    \includegraphics[width=\textwidth]{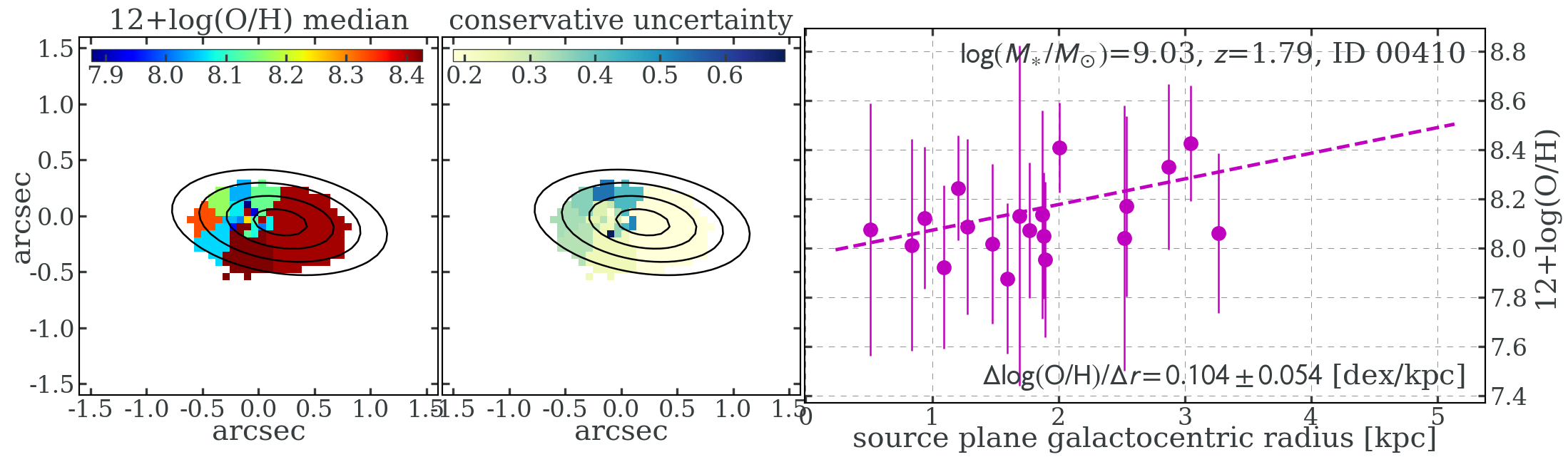}
    \caption{The source ID00410 in the field of \clshi is shown.}
    \label{fig:clM1423_ID00410_figs}
\end{figure*}

\begin{figure*}
    \centering
    \includegraphics[width=\textwidth]{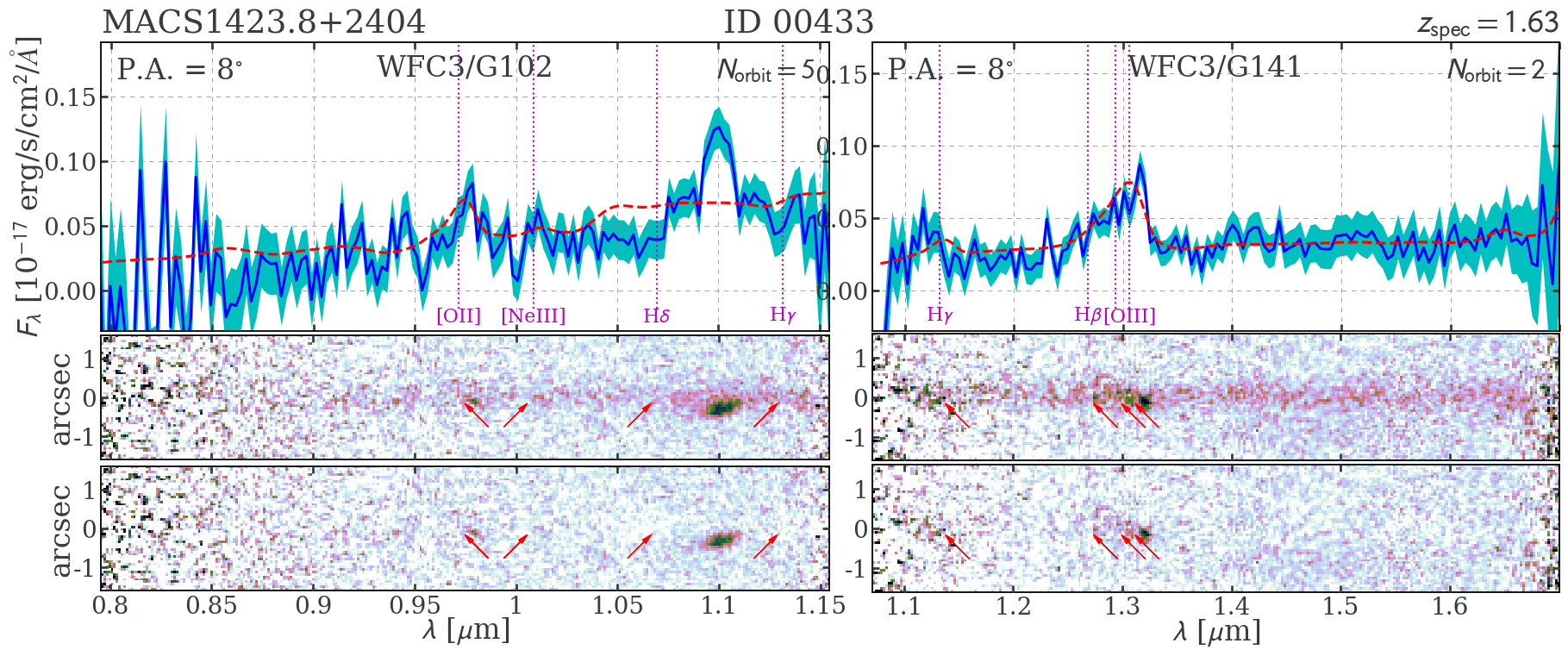}\\
    \includegraphics[width=\textwidth]{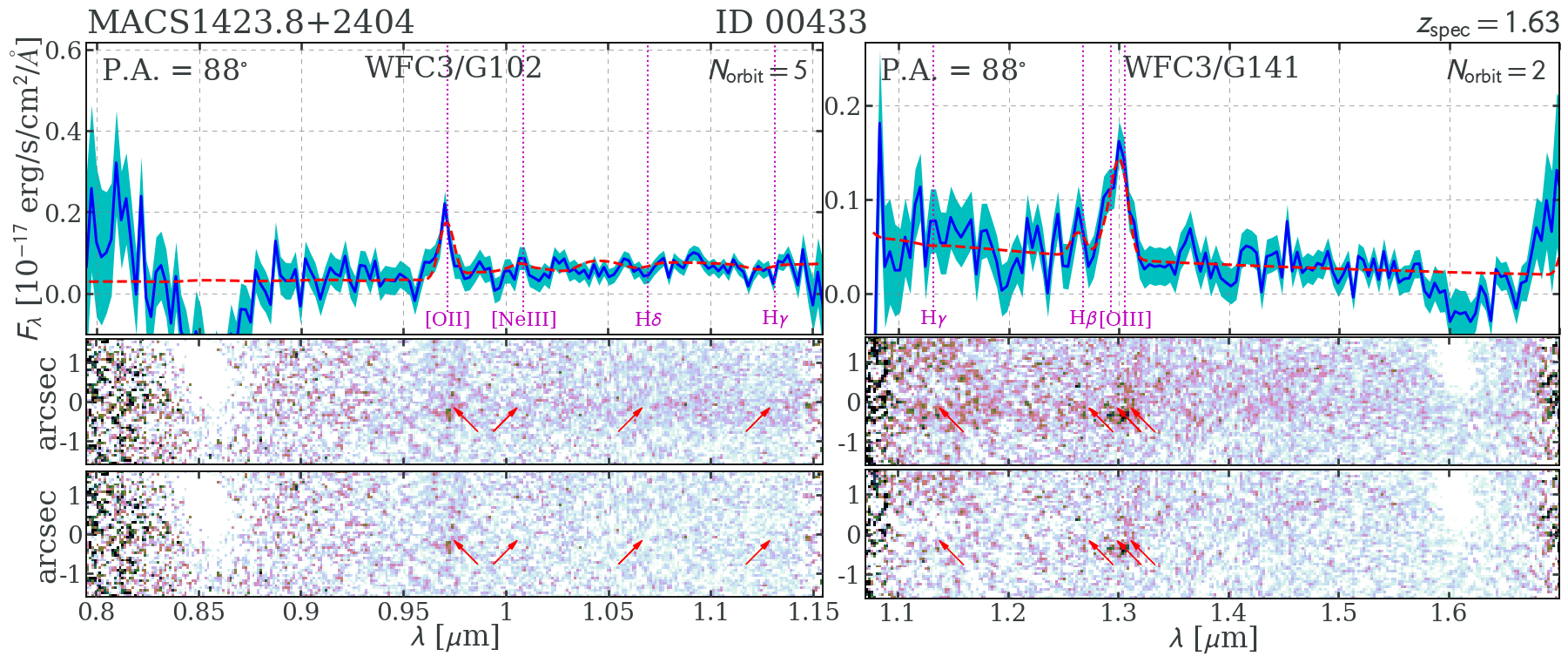}\\
    \includegraphics[width=.16\textwidth]{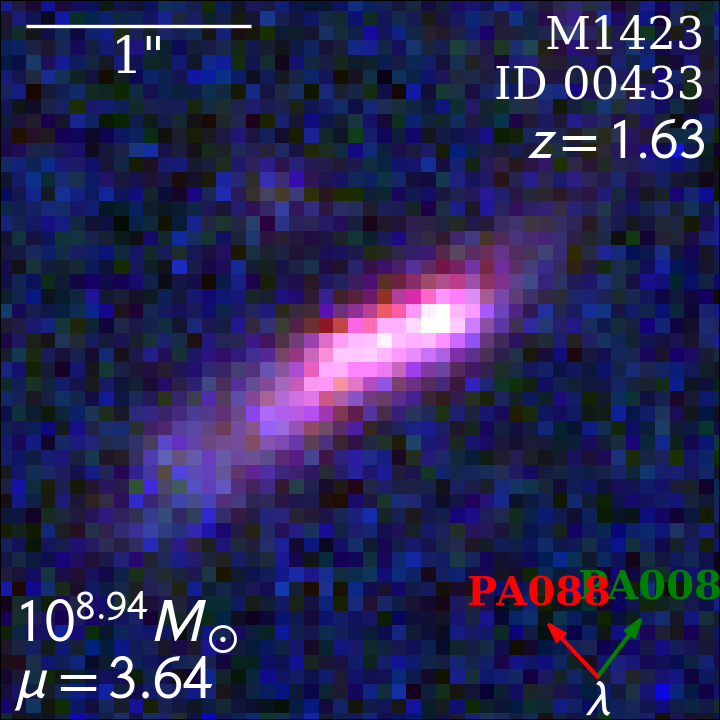}
    \includegraphics[width=.16\textwidth]{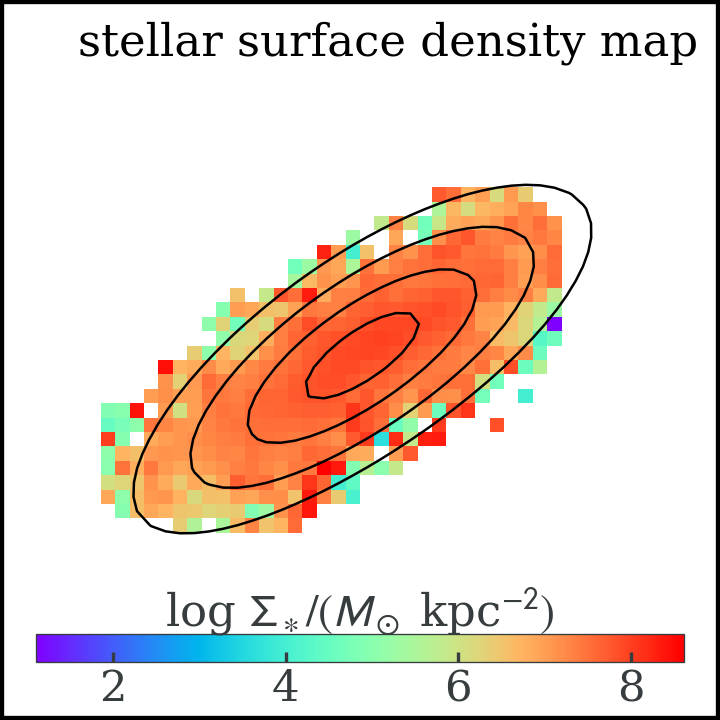}
    \includegraphics[width=.16\textwidth]{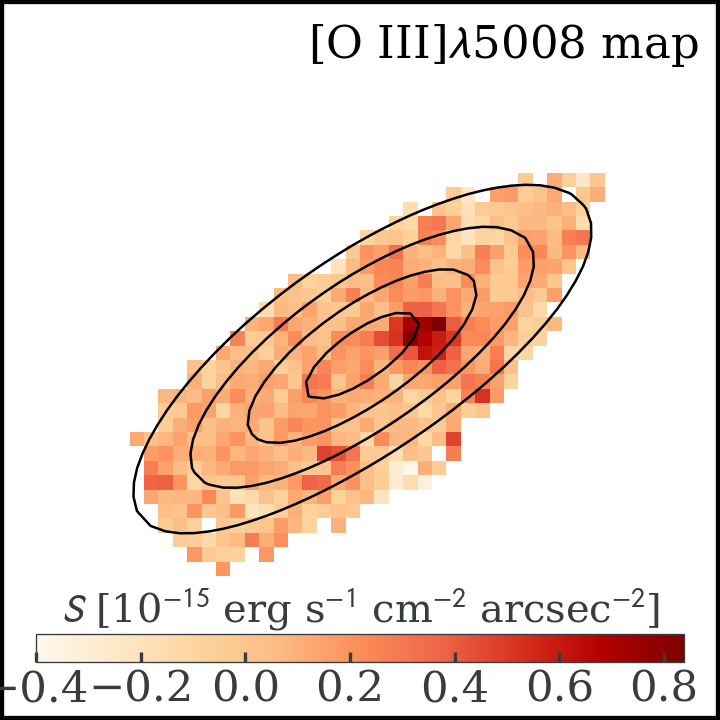}
    \includegraphics[width=.16\textwidth]{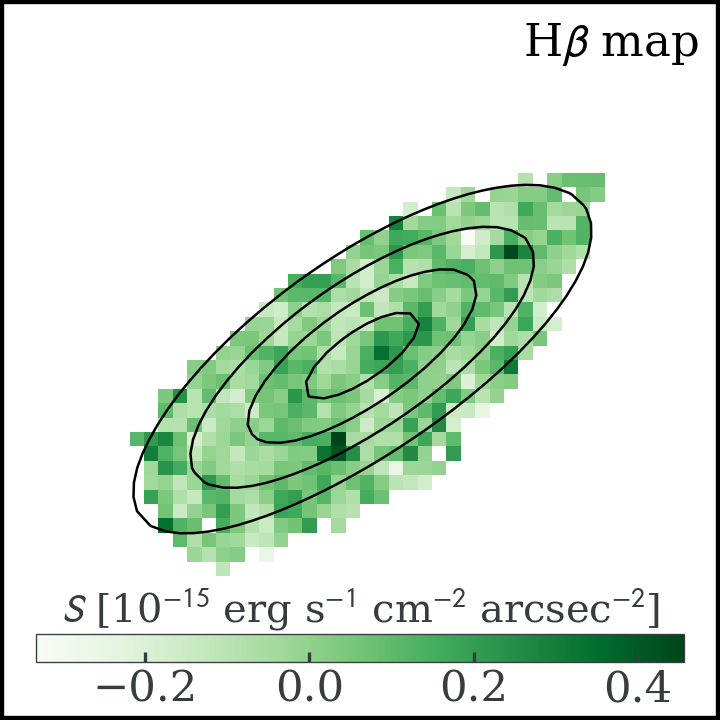}
    \includegraphics[width=.16\textwidth]{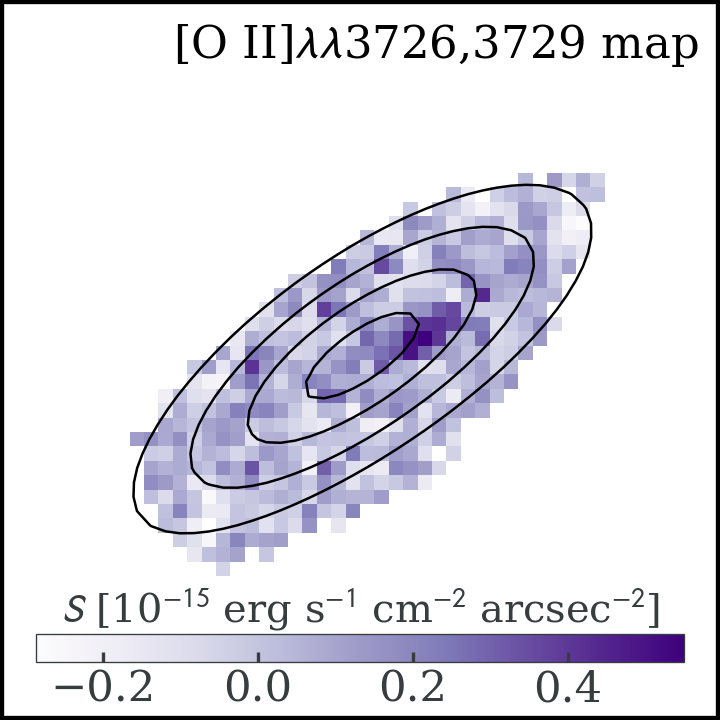}
    \includegraphics[width=.16\textwidth]{fig_ELmaps/baiban.png}\\
    \includegraphics[width=\textwidth]{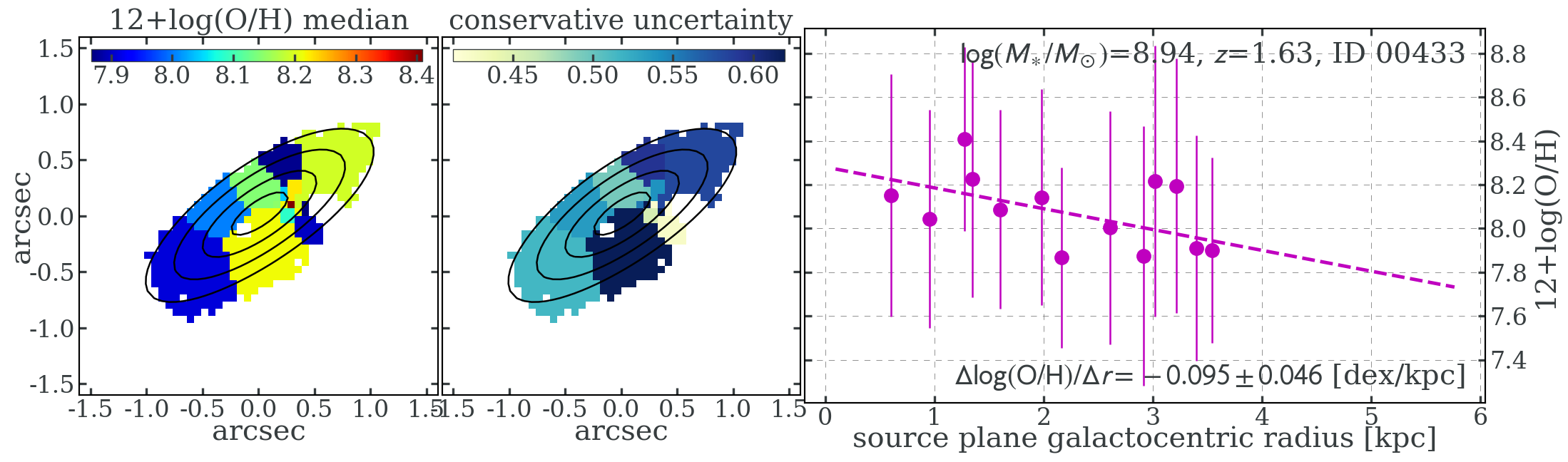}
    \caption{The source ID00433 in the field of \clshi is shown.}
    \label{fig:clM1423_ID00433_figs}
\end{figure*}

\begin{figure*}
    \centering
    \includegraphics[width=\textwidth]{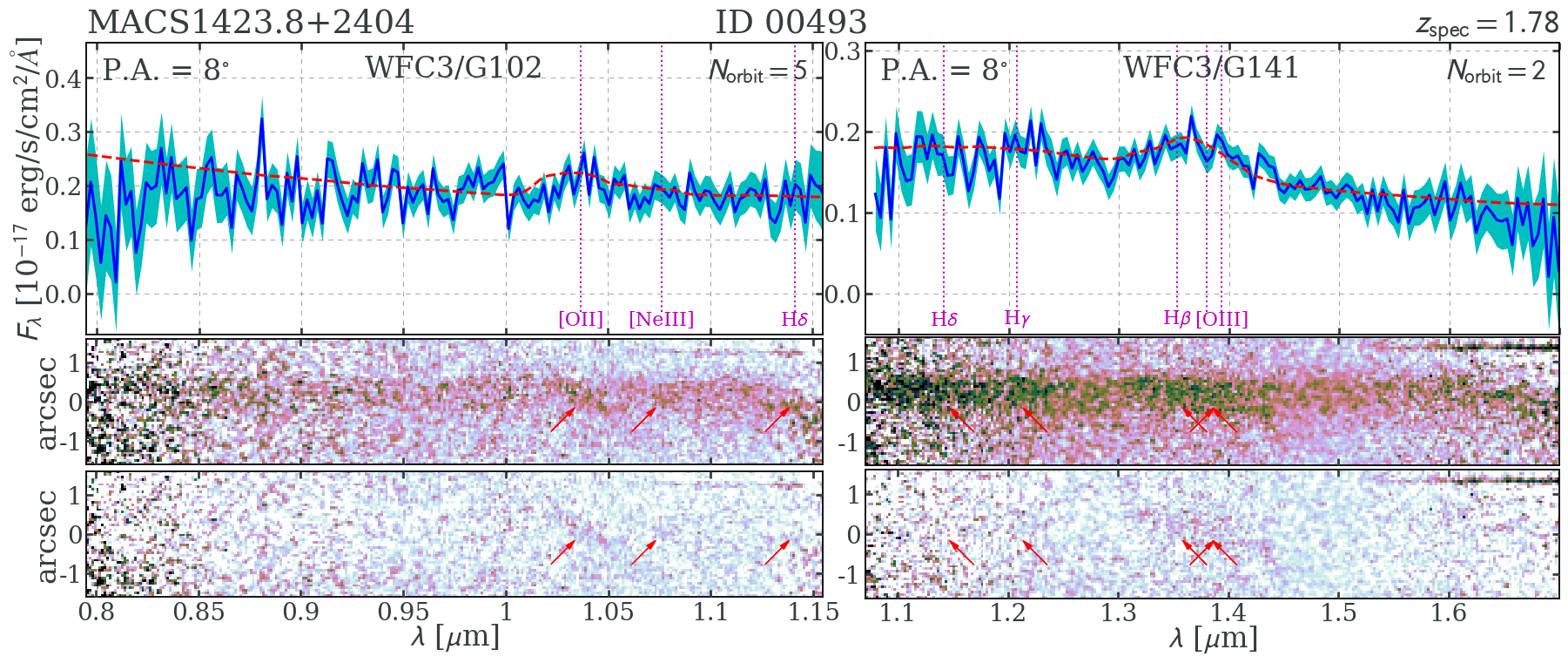}\\
    \includegraphics[width=\textwidth]{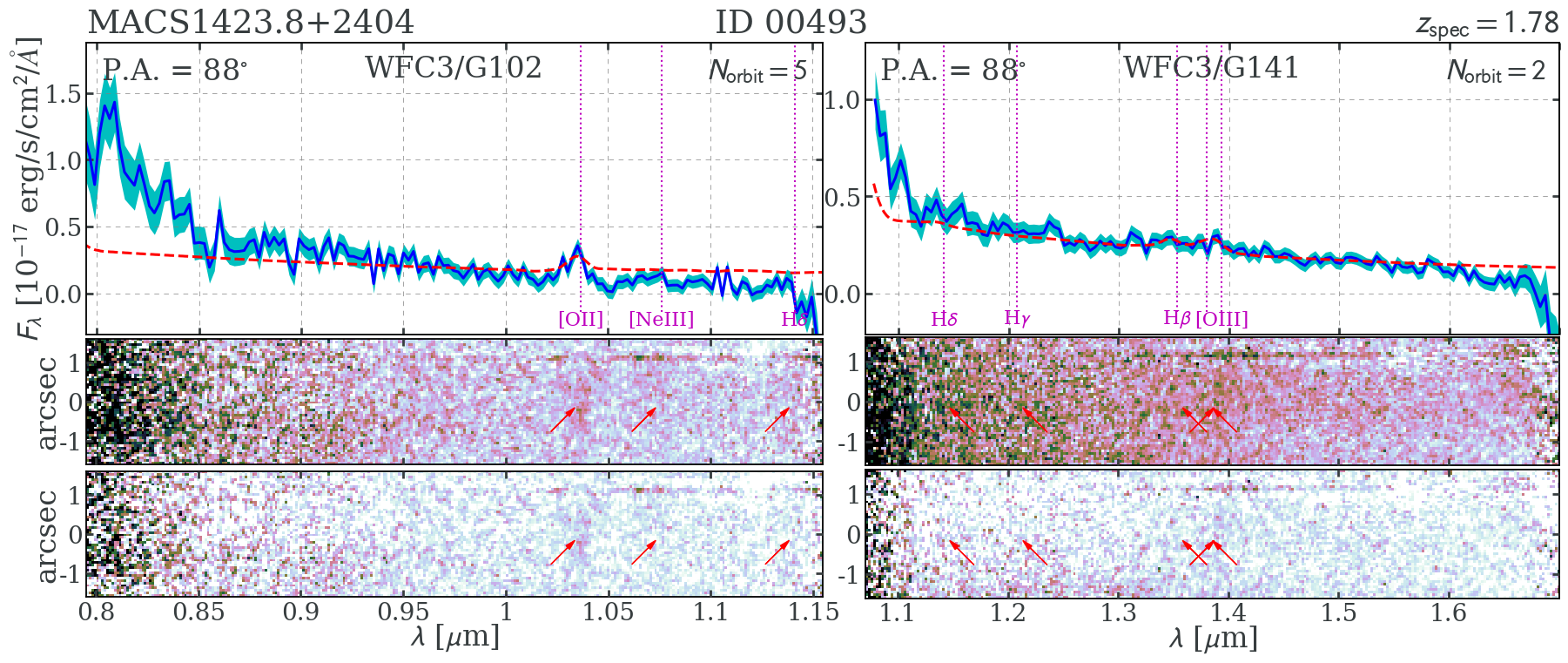}\\
    \includegraphics[width=.16\textwidth]{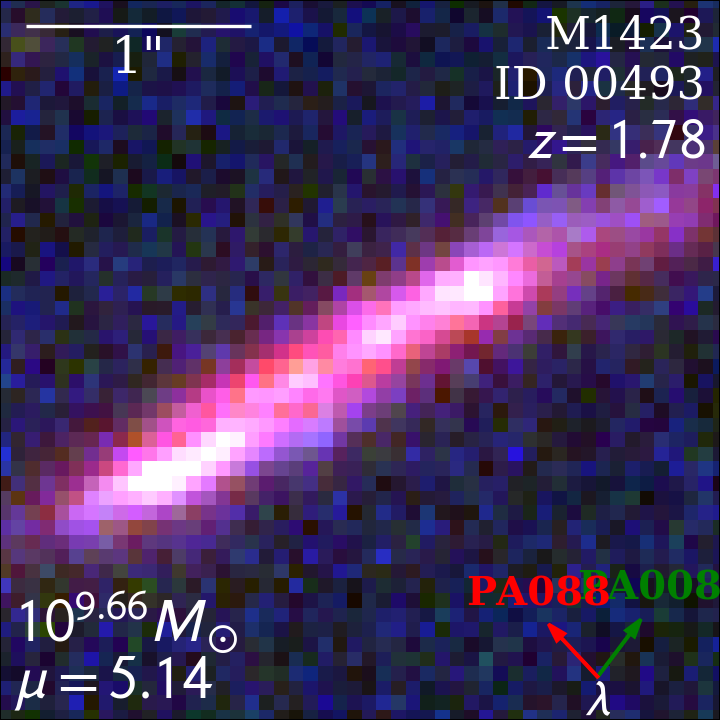}
    \includegraphics[width=.16\textwidth]{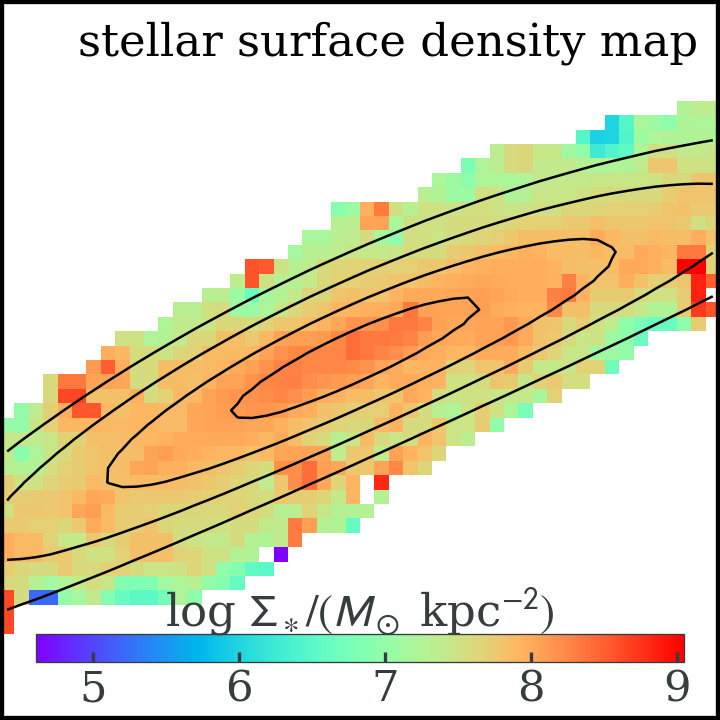}
    \includegraphics[width=.16\textwidth]{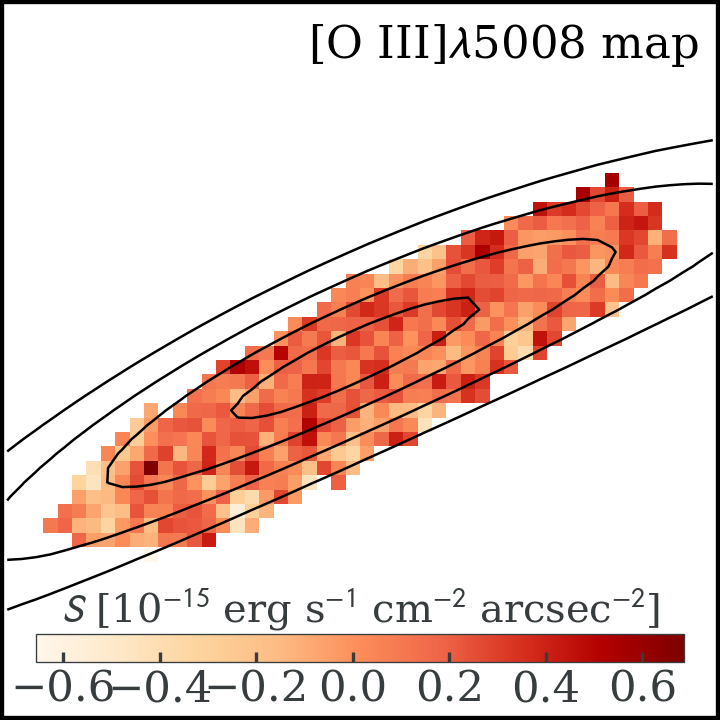}
    \includegraphics[width=.16\textwidth]{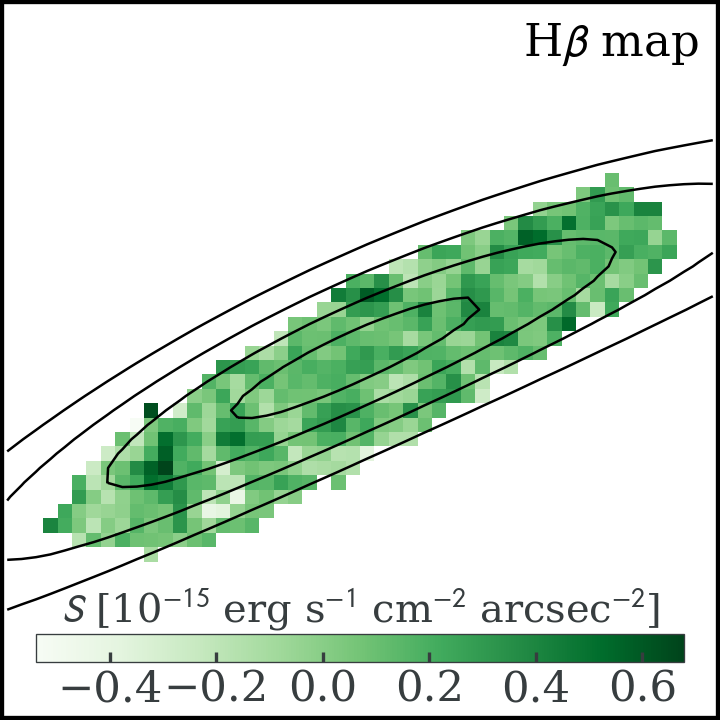}
    \includegraphics[width=.16\textwidth]{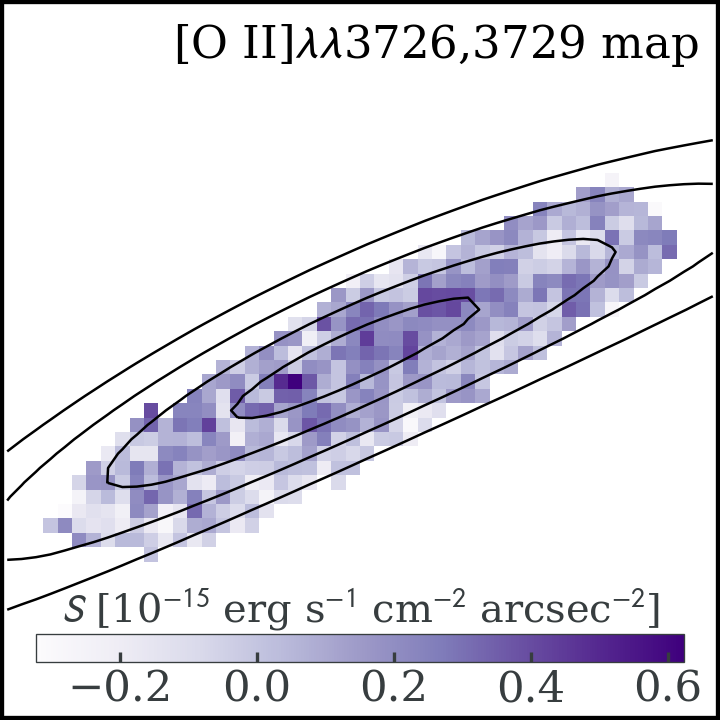}
    \includegraphics[width=.16\textwidth]{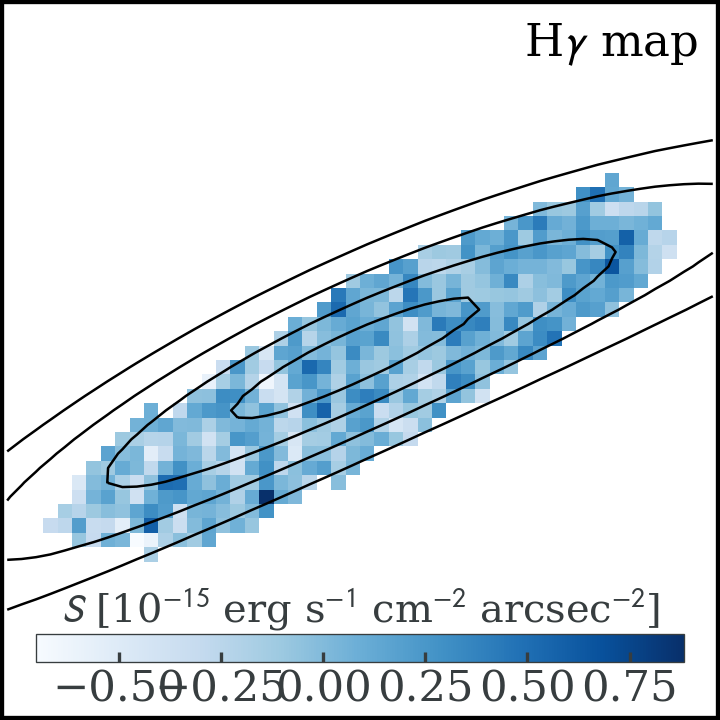}\\
    \includegraphics[width=\textwidth]{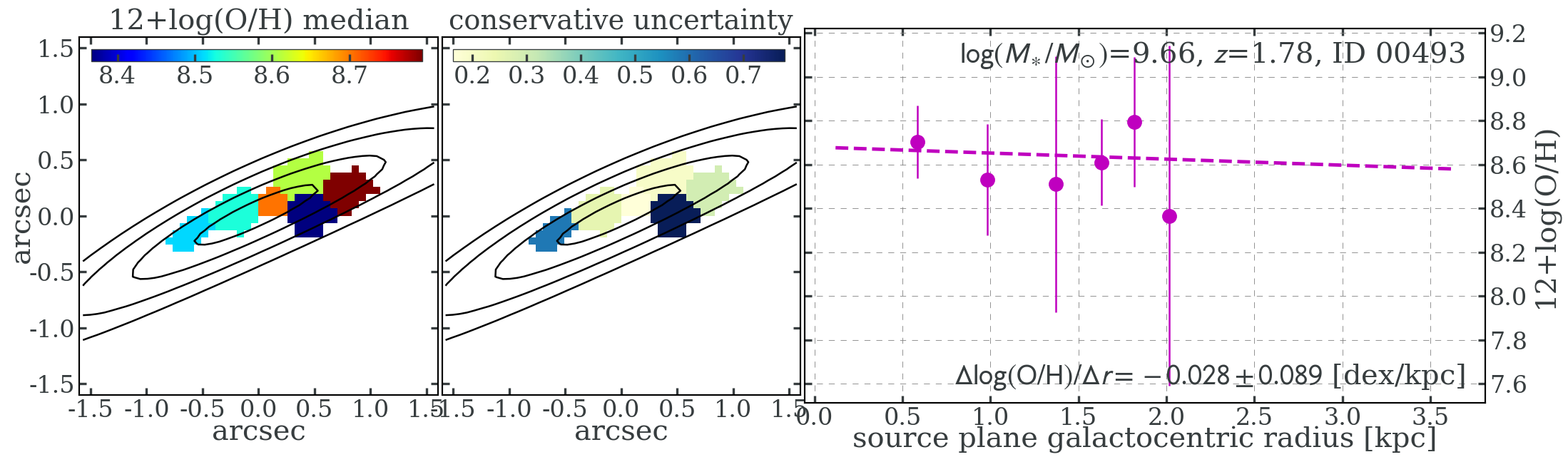}
    \caption{The source ID00493 in the field of \clshi is shown.}
    \label{fig:clM1423_ID00493_figs}
\end{figure*}

\begin{figure*}
    \centering
    \includegraphics[width=\textwidth]{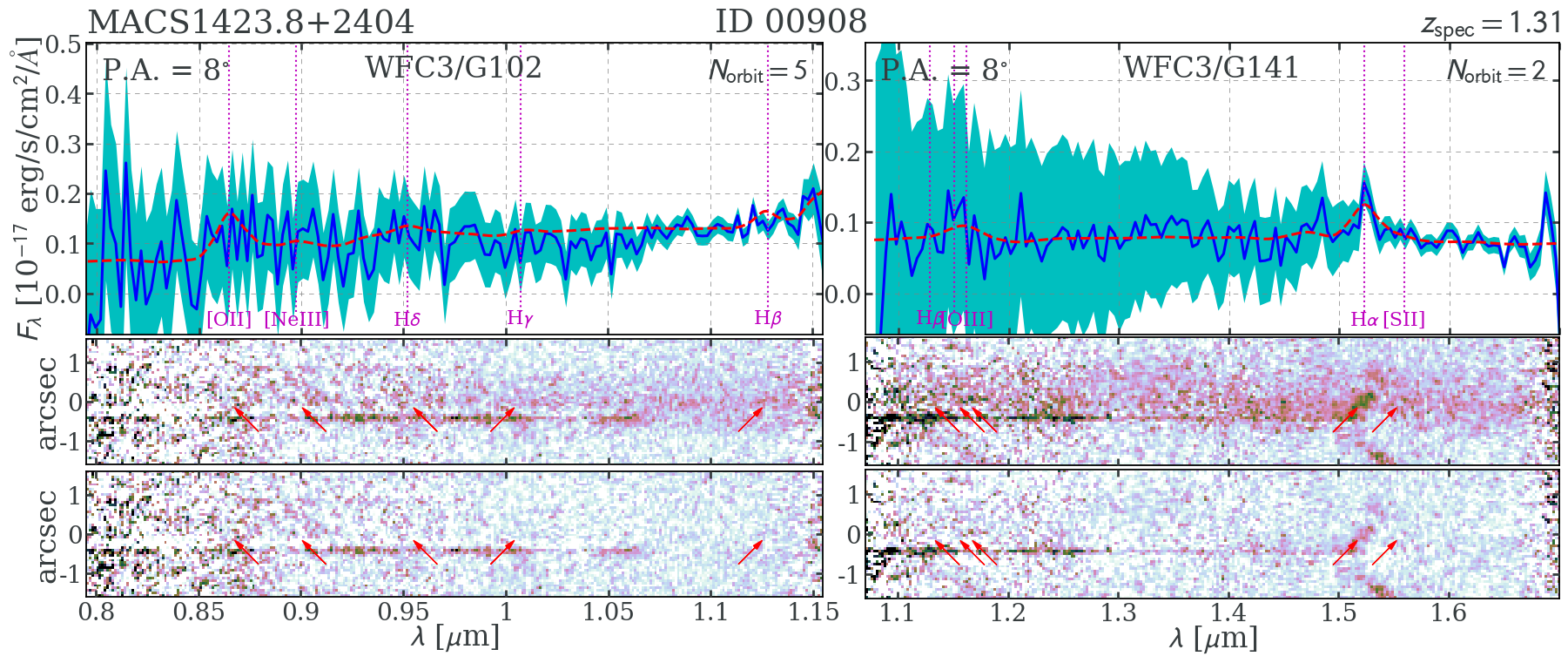}\\
    \includegraphics[width=\textwidth]{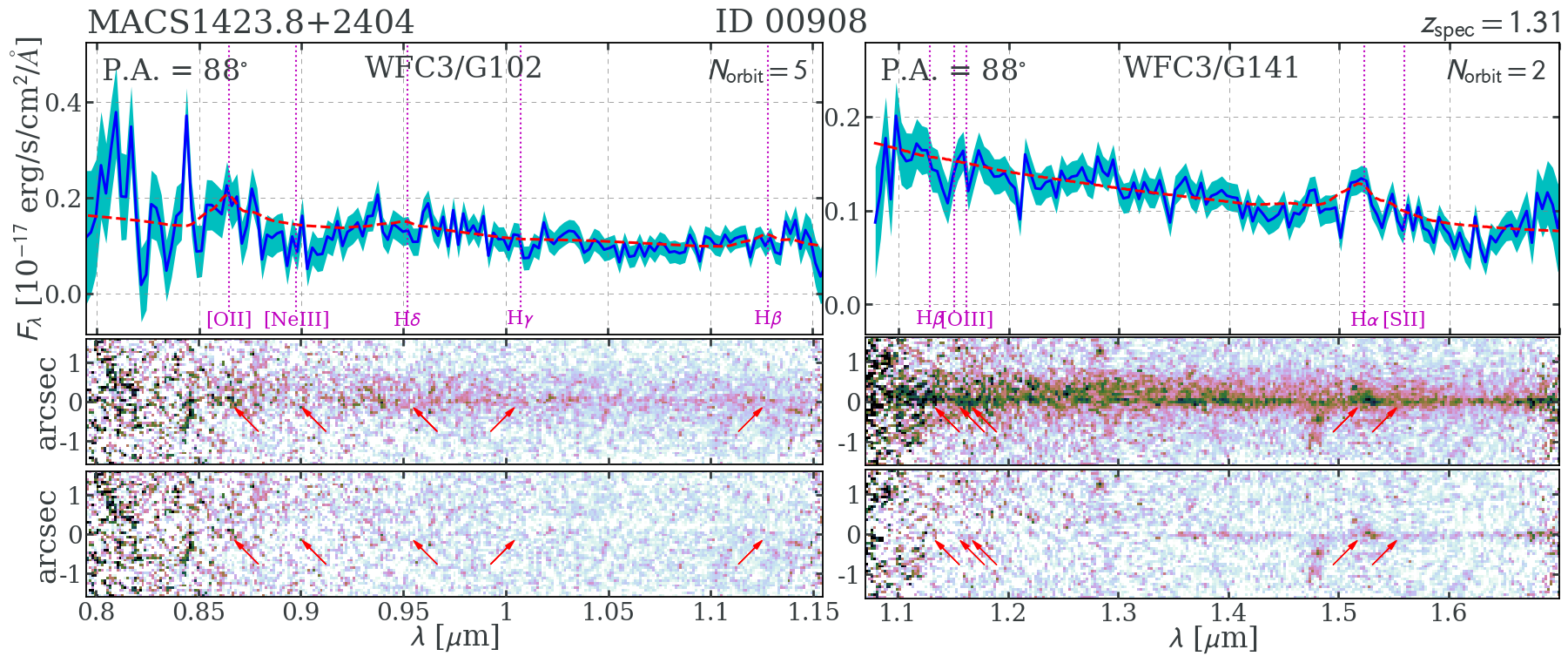}\\
    \includegraphics[width=.16\textwidth]{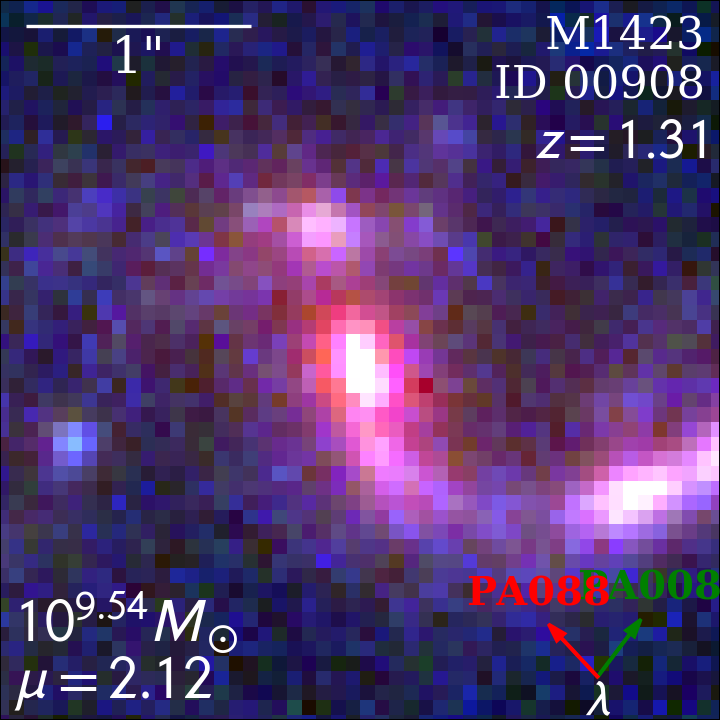}
    \includegraphics[width=.16\textwidth]{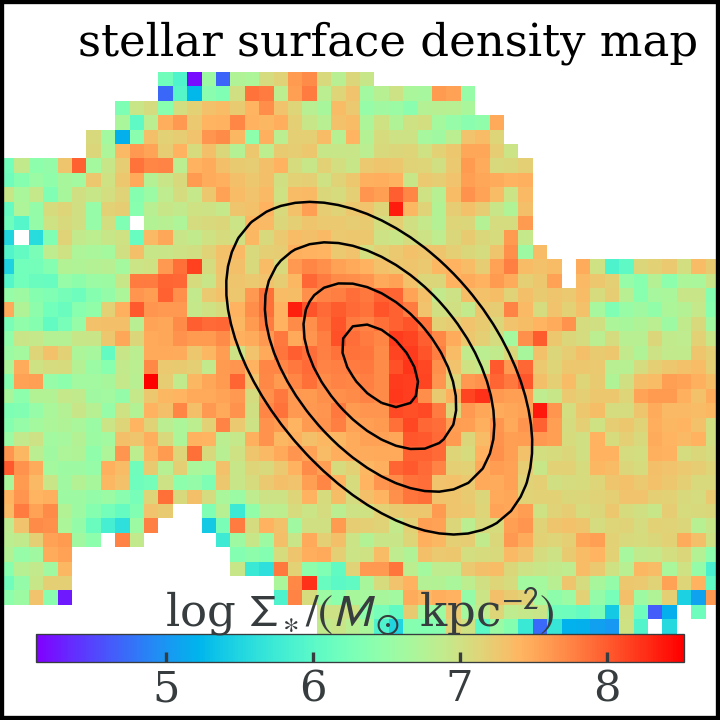}
    \includegraphics[width=.16\textwidth]{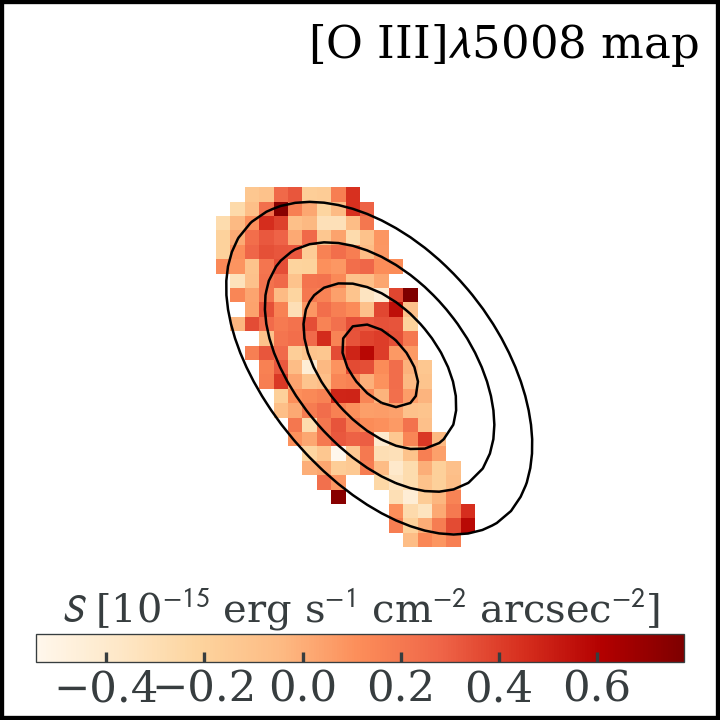}
    \includegraphics[width=.16\textwidth]{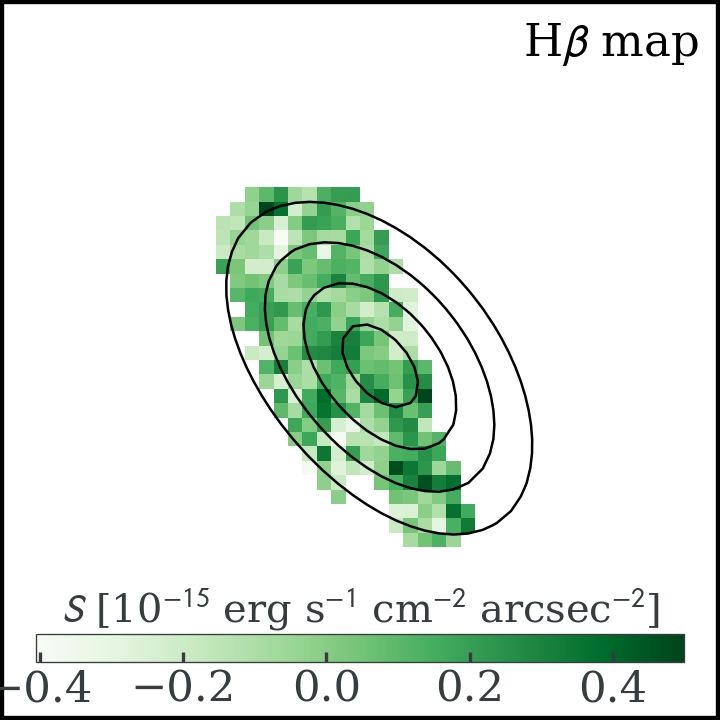}
    \includegraphics[width=.16\textwidth]{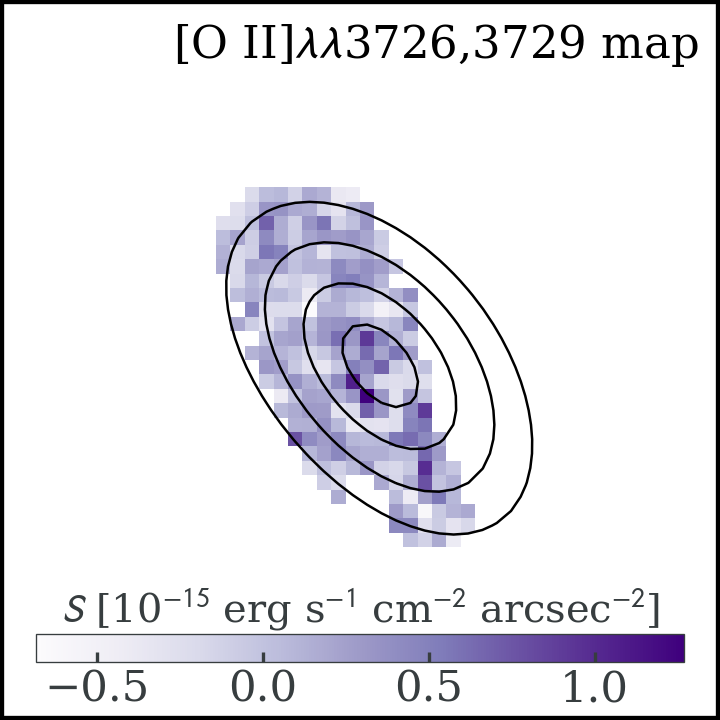}
    \includegraphics[width=.16\textwidth]{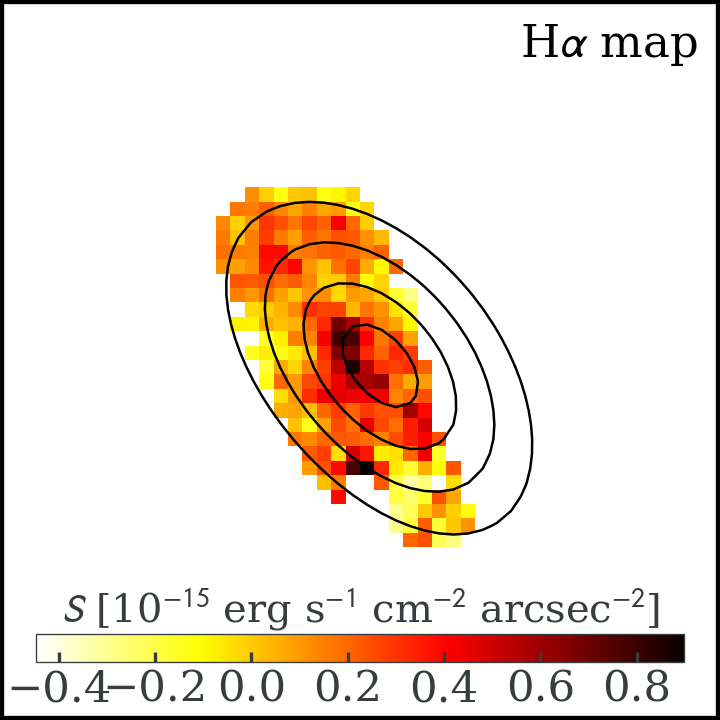}\\
    \includegraphics[width=\textwidth]{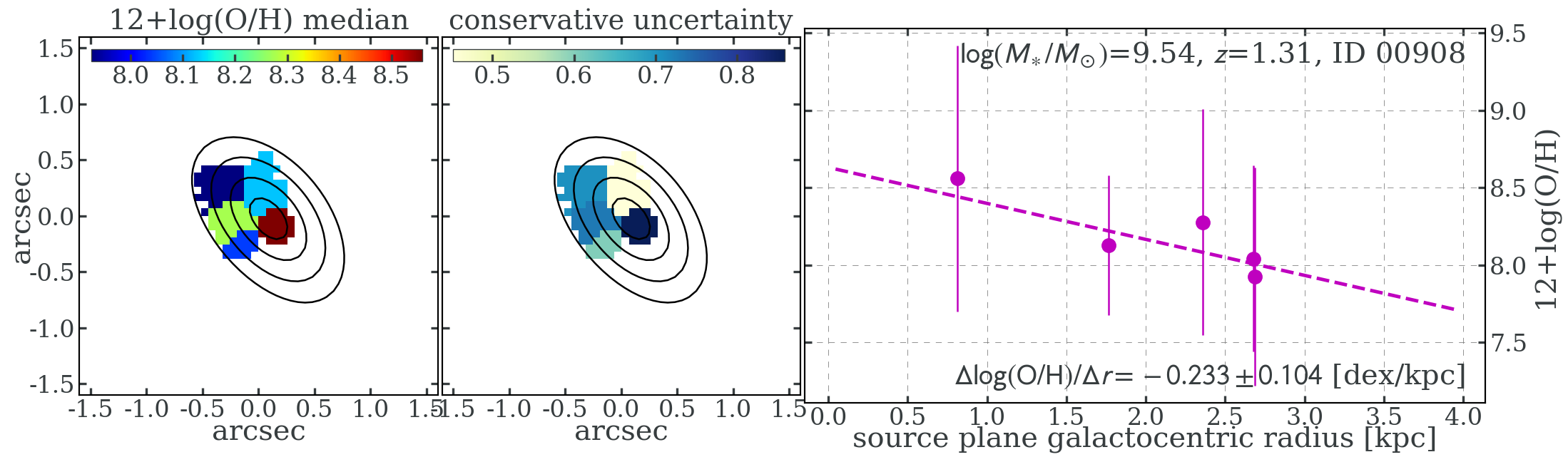}
    \caption{The source ID00908 in the field of \clshi is shown.}
    \label{fig:clM1423_ID00908_figs}
\end{figure*}

\begin{figure*}
    \centering
    \includegraphics[width=\textwidth]{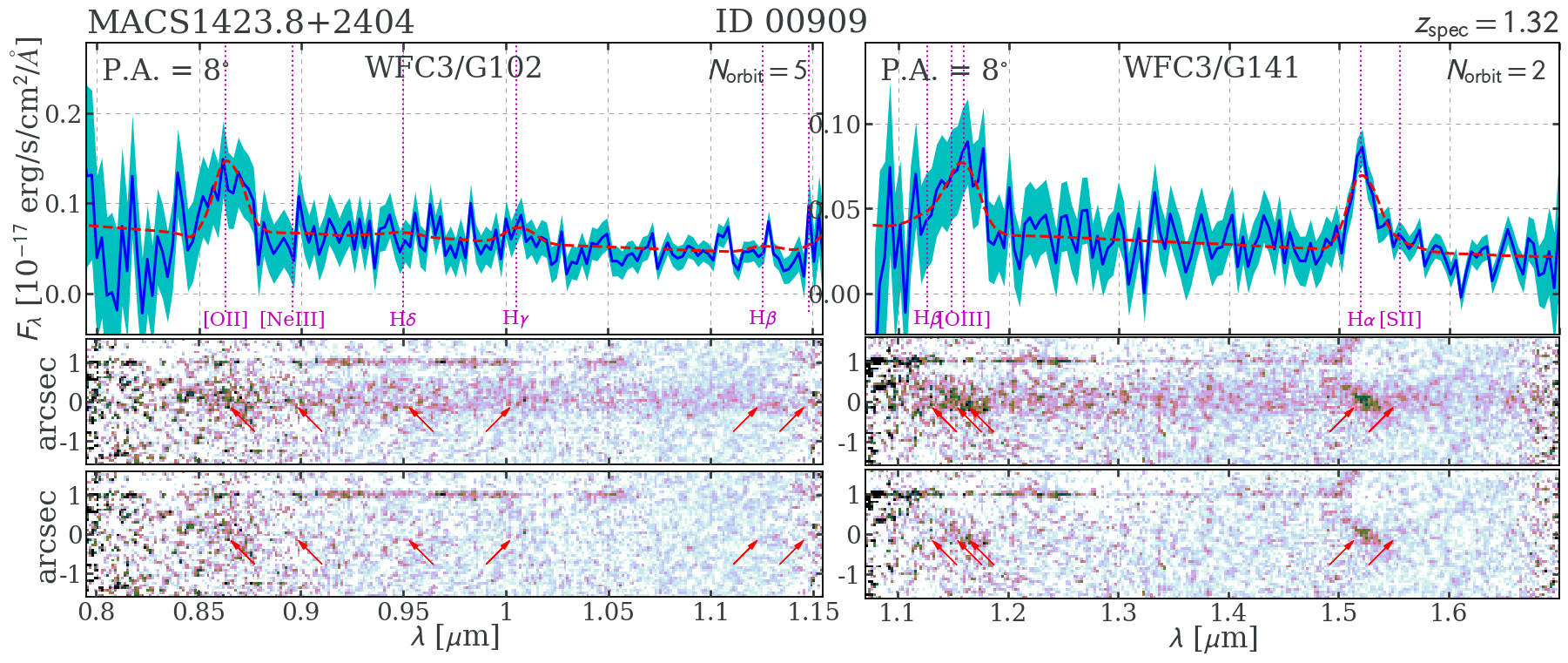}\\
    \includegraphics[width=\textwidth]{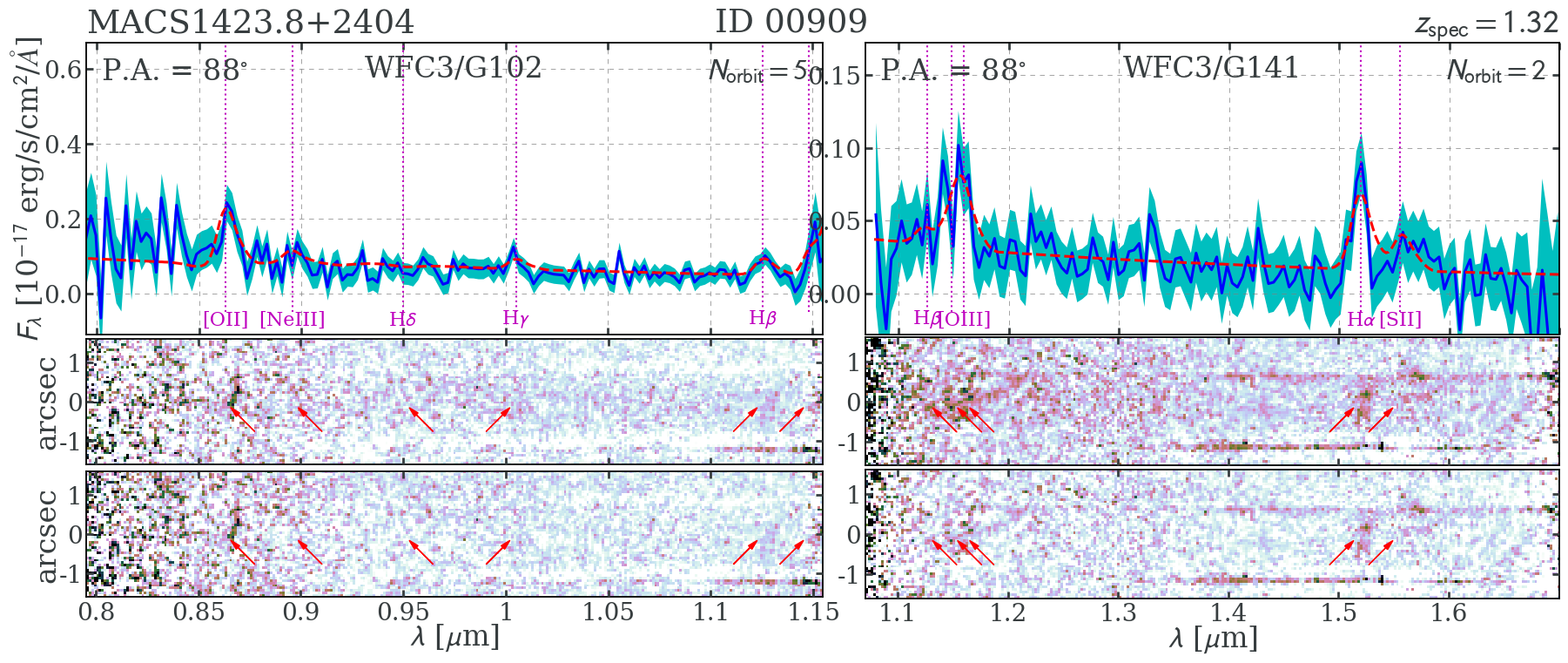}\\
    \includegraphics[width=.16\textwidth]{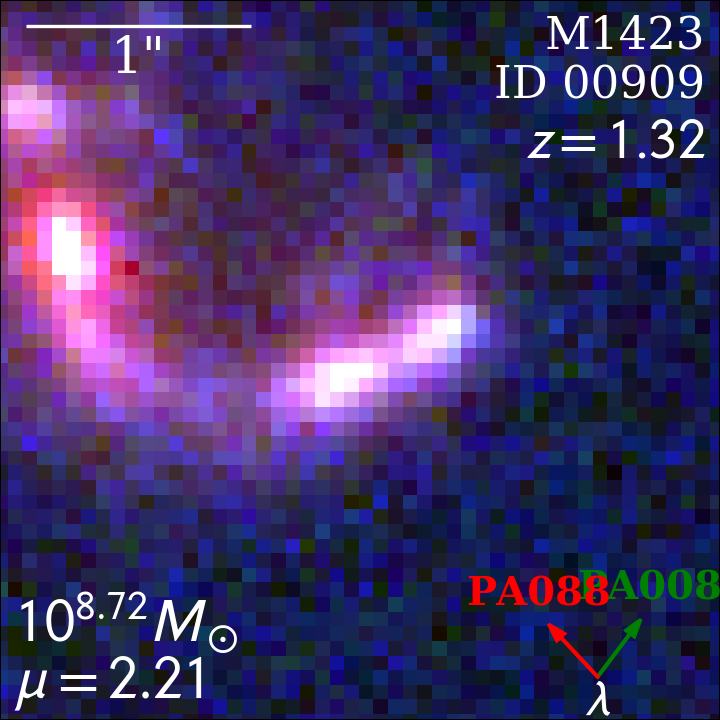}
    \includegraphics[width=.16\textwidth]{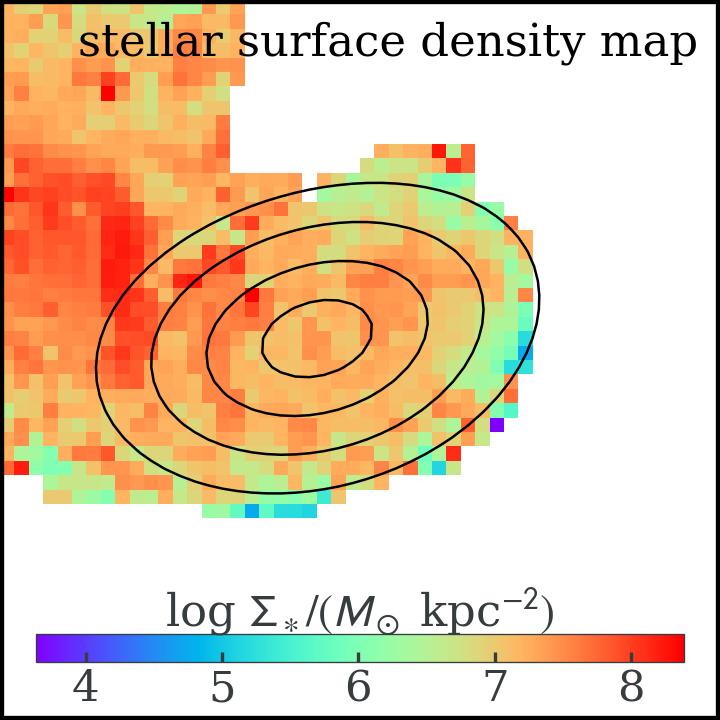}
    \includegraphics[width=.16\textwidth]{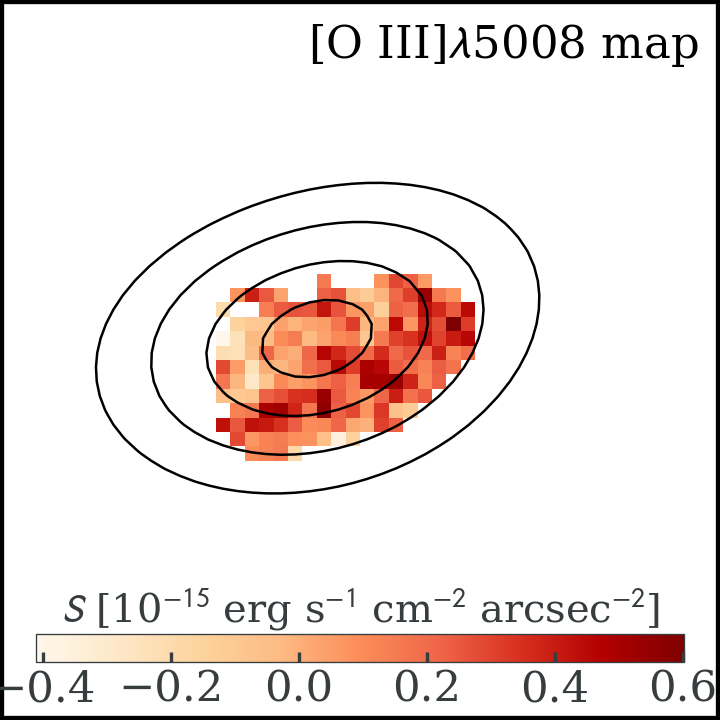}
    \includegraphics[width=.16\textwidth]{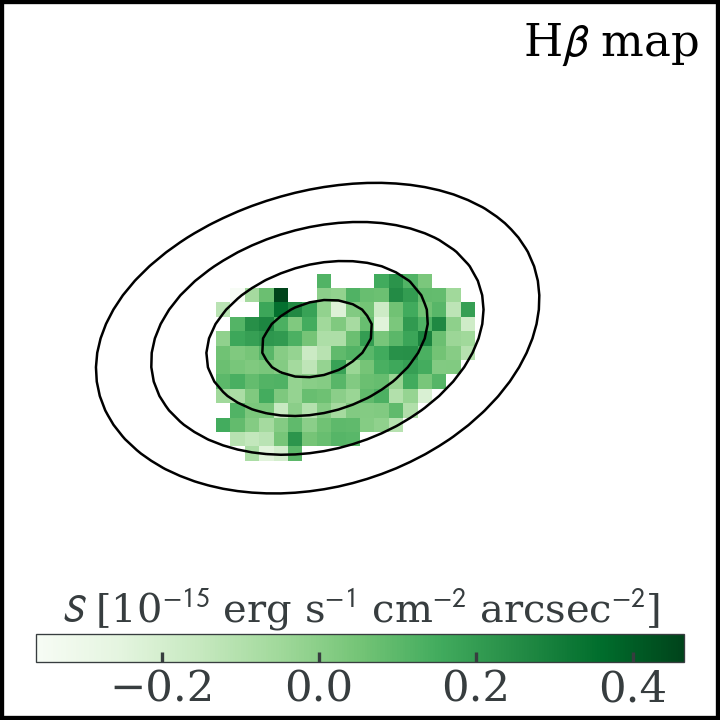}
    \includegraphics[width=.16\textwidth]{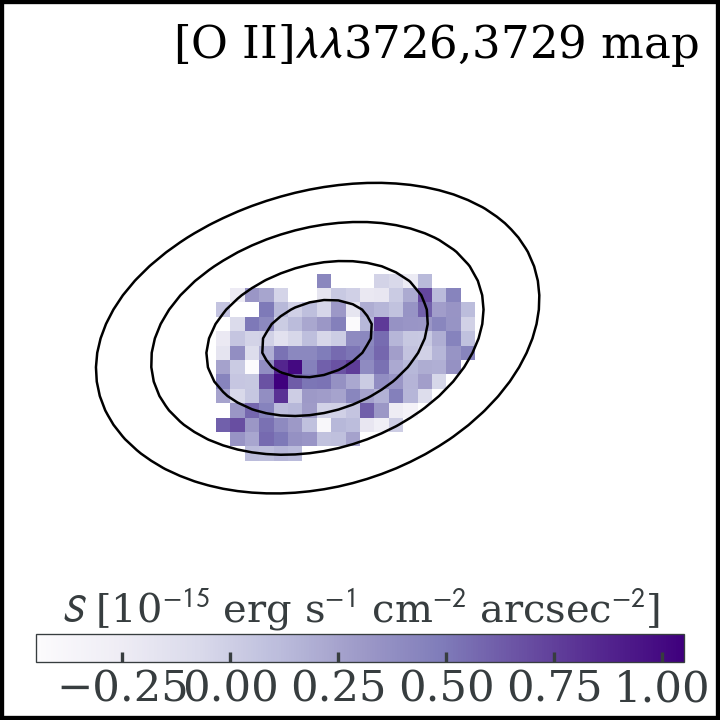}
    \includegraphics[width=.16\textwidth]{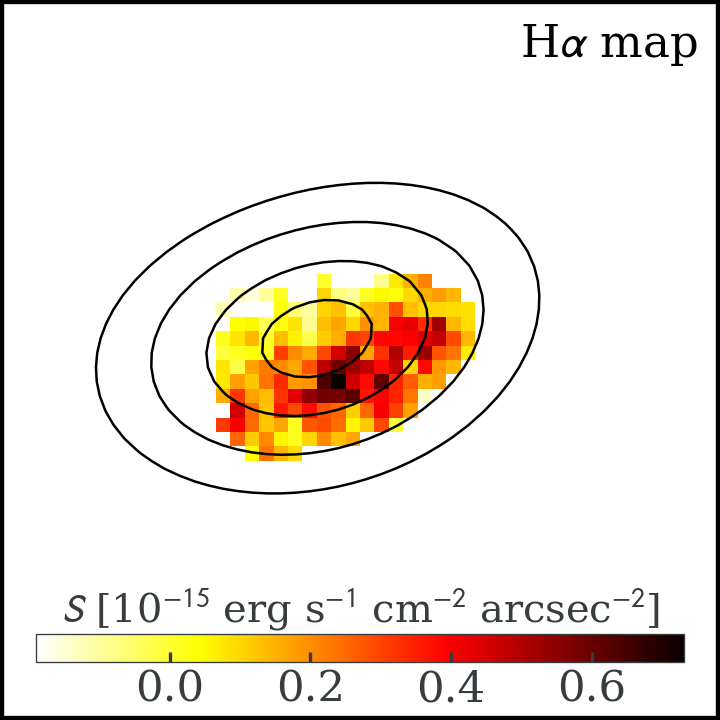}\\
    \includegraphics[width=\textwidth]{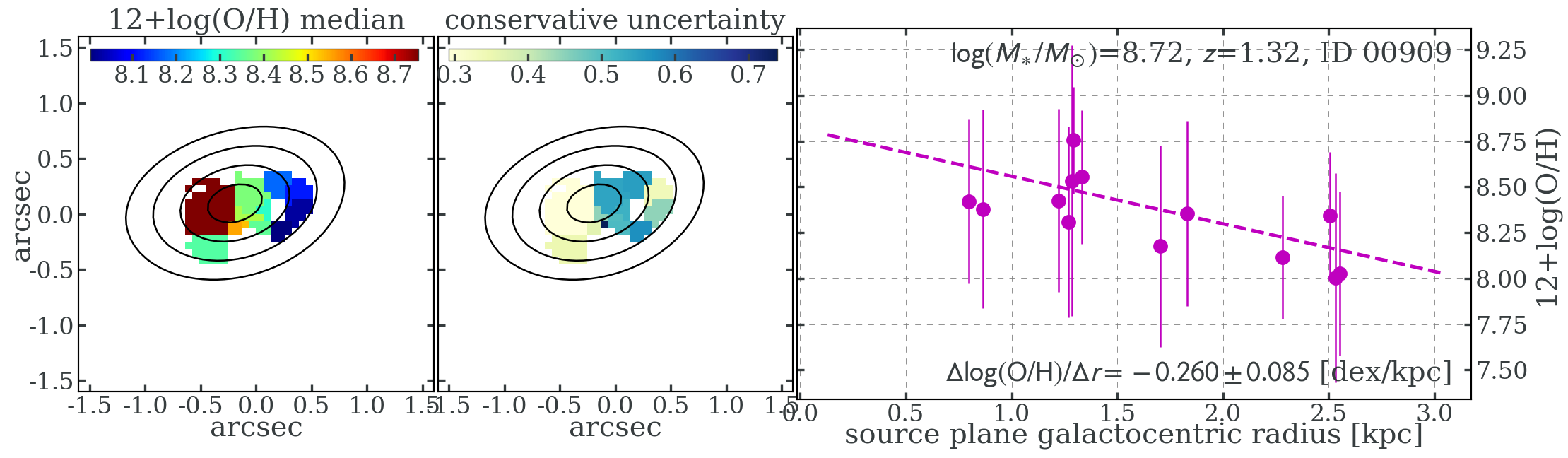}
    \caption{The source ID00909 in the field of \clshi is shown.}
    \label{fig:clM1423_ID00909_figs}
\end{figure*}

\begin{figure*}
    \centering
    \includegraphics[width=\textwidth]{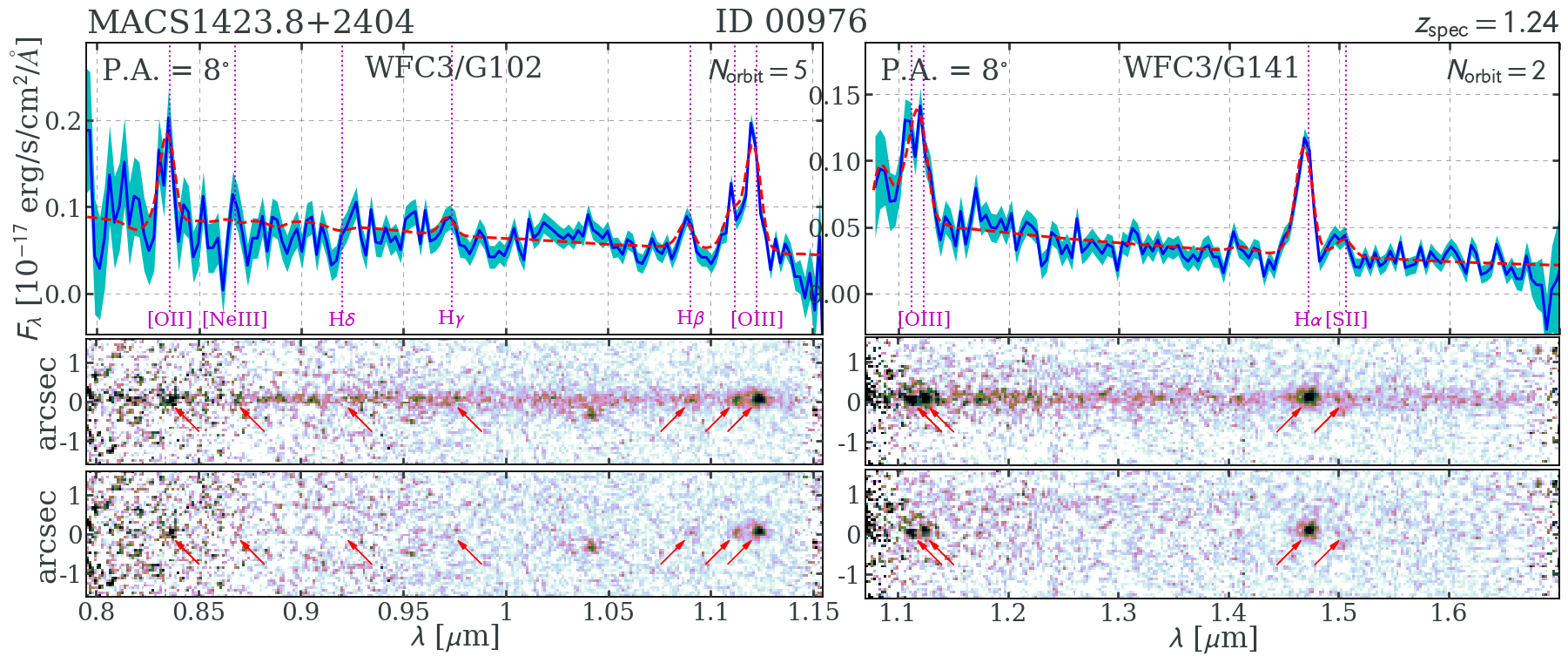}\\
    \includegraphics[width=\textwidth]{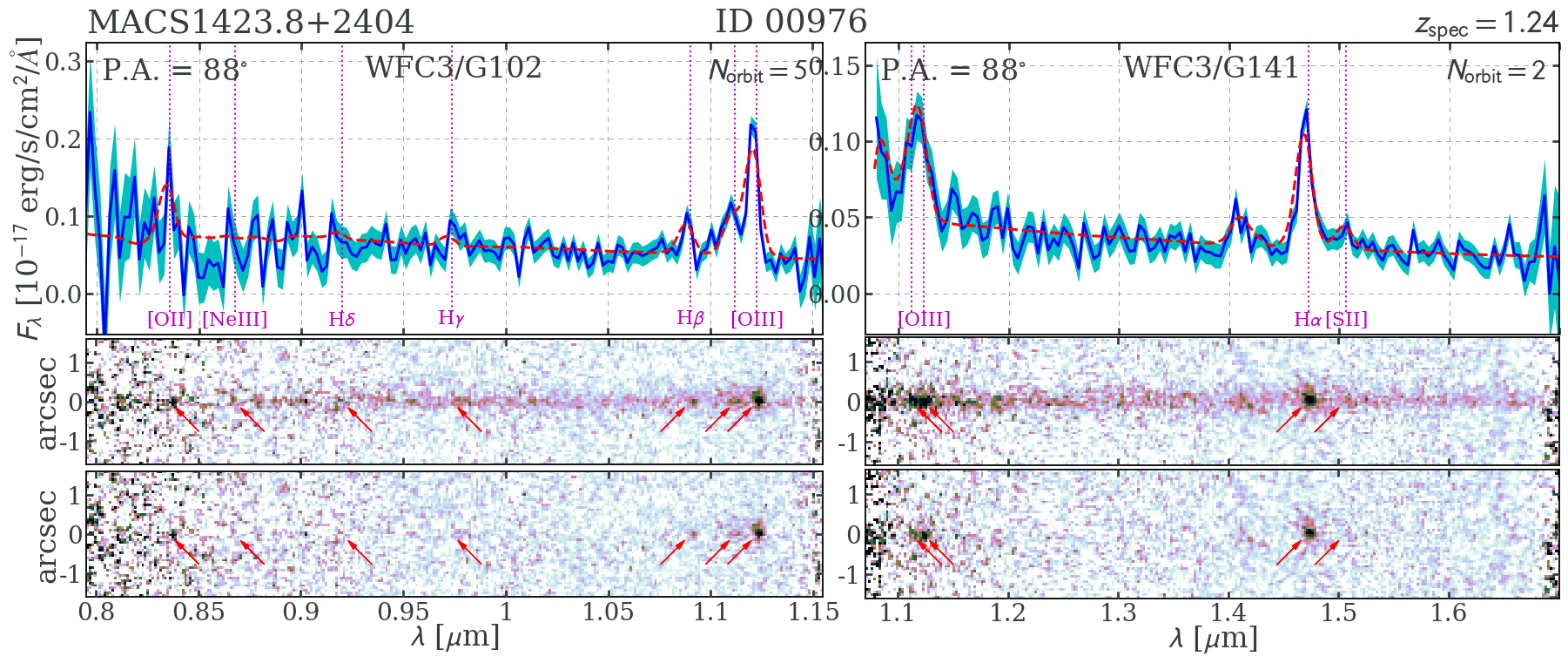}\\
    \includegraphics[width=.16\textwidth]{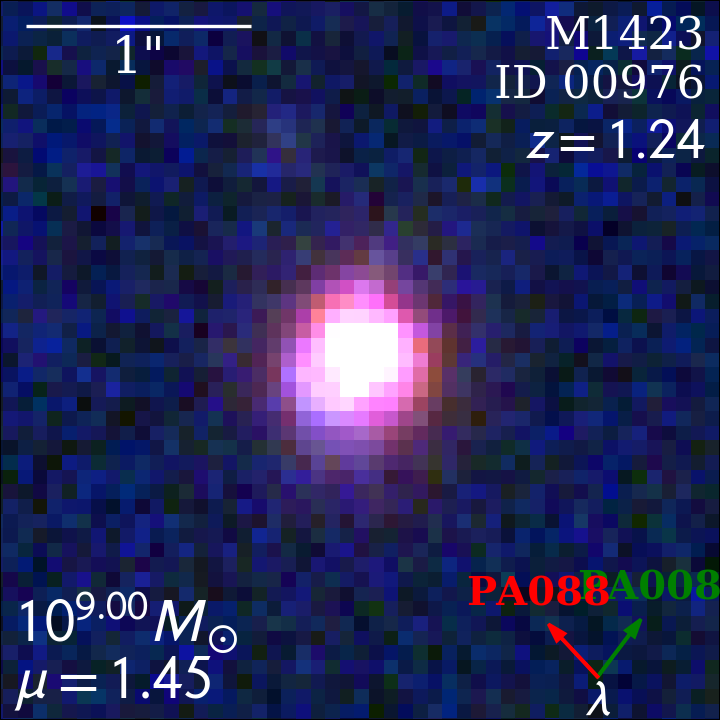}
    \includegraphics[width=.16\textwidth]{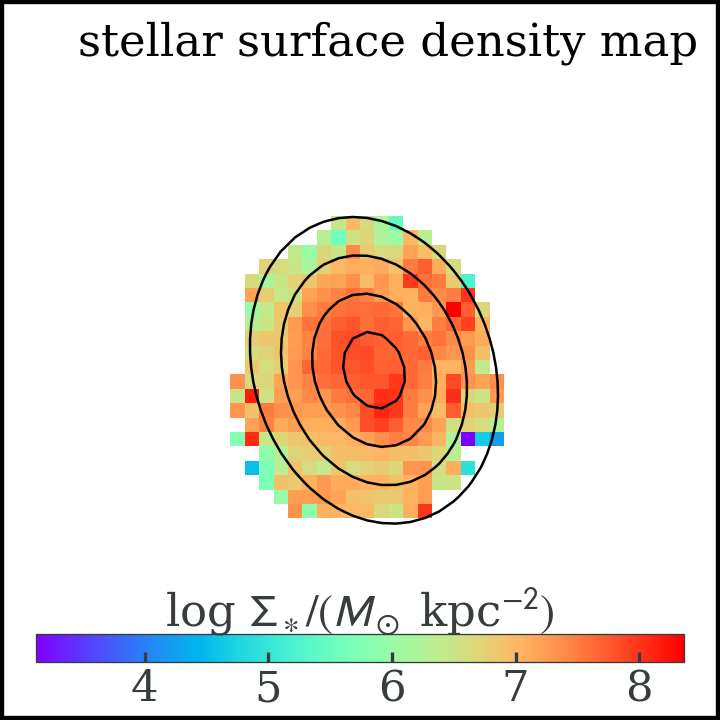}
    \includegraphics[width=.16\textwidth]{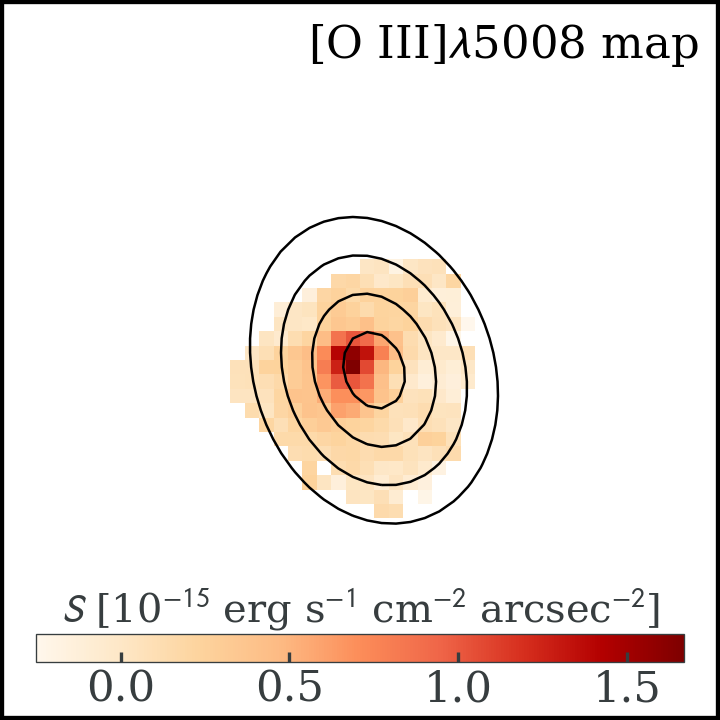}
    \includegraphics[width=.16\textwidth]{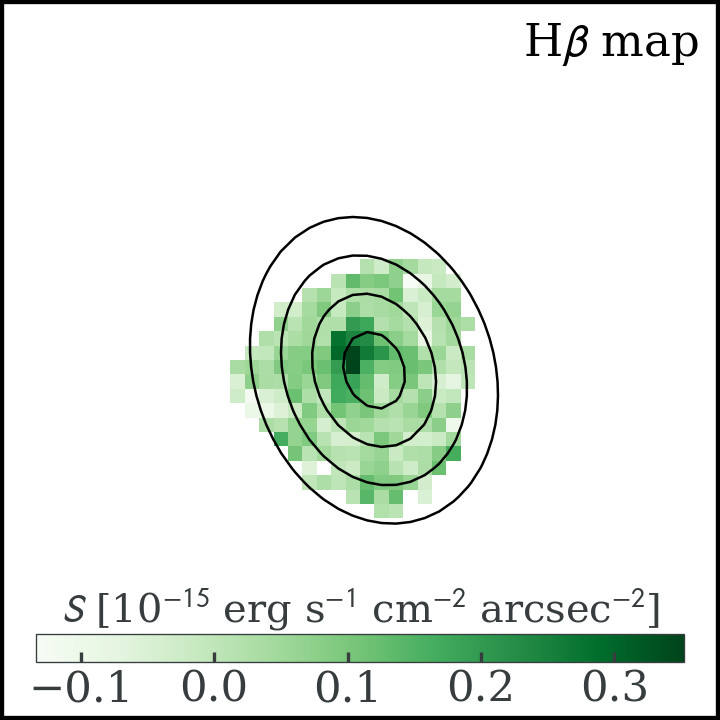}
    \includegraphics[width=.16\textwidth]{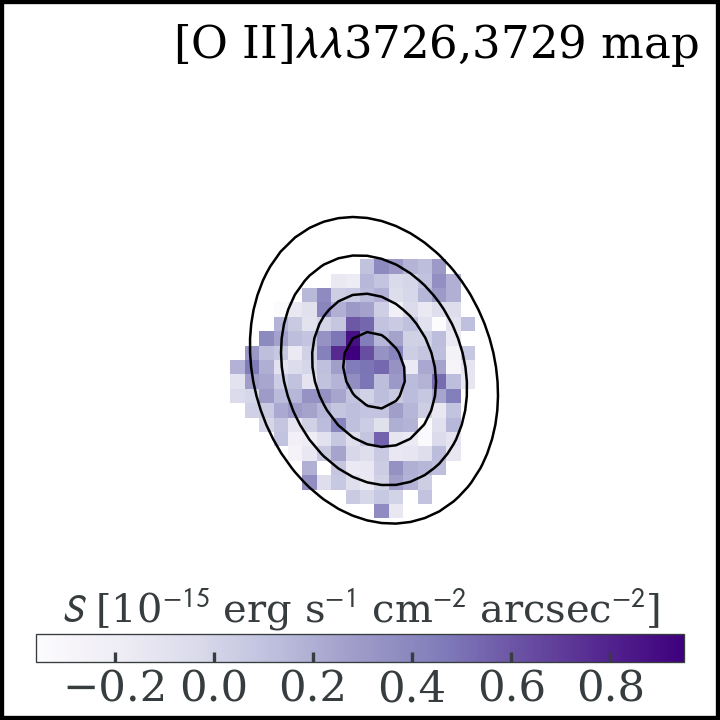}
    \includegraphics[width=.16\textwidth]{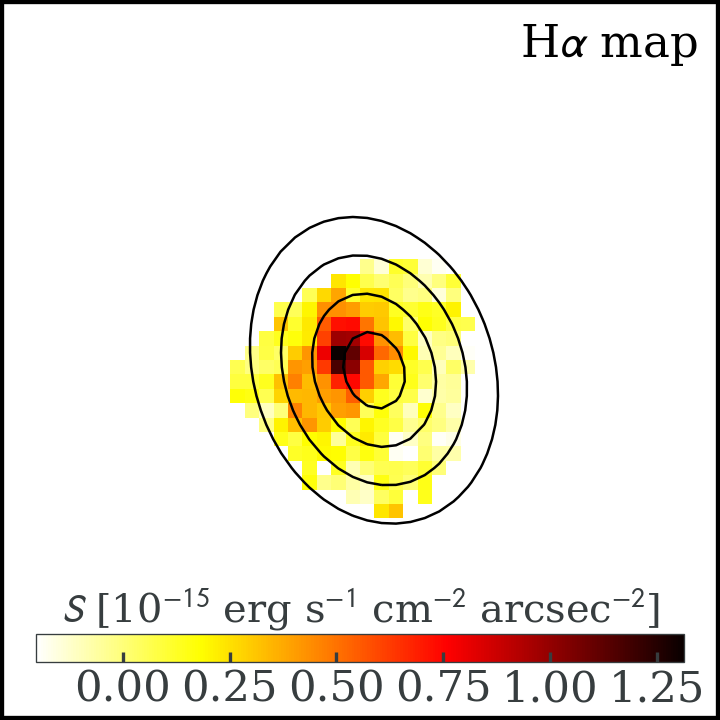}\\
    \includegraphics[width=\textwidth]{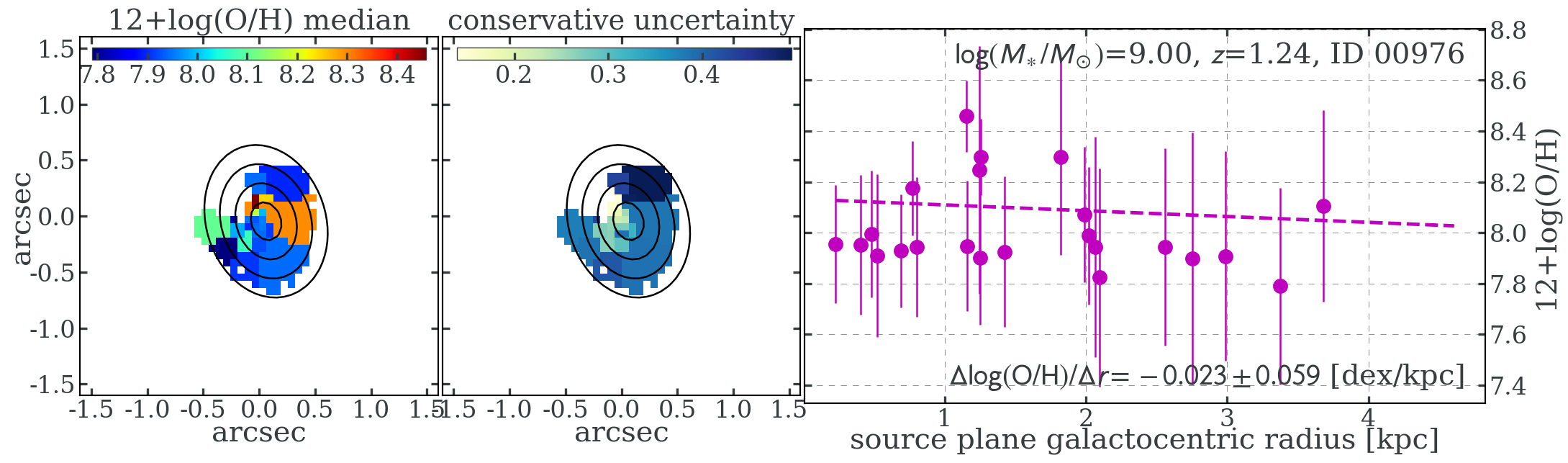}
    \caption{The source ID00976 in the field of \clshi is shown.}
    \label{fig:clM1423_ID00976_figs}
\end{figure*}

\begin{figure*}
    \centering
    \includegraphics[width=\textwidth]{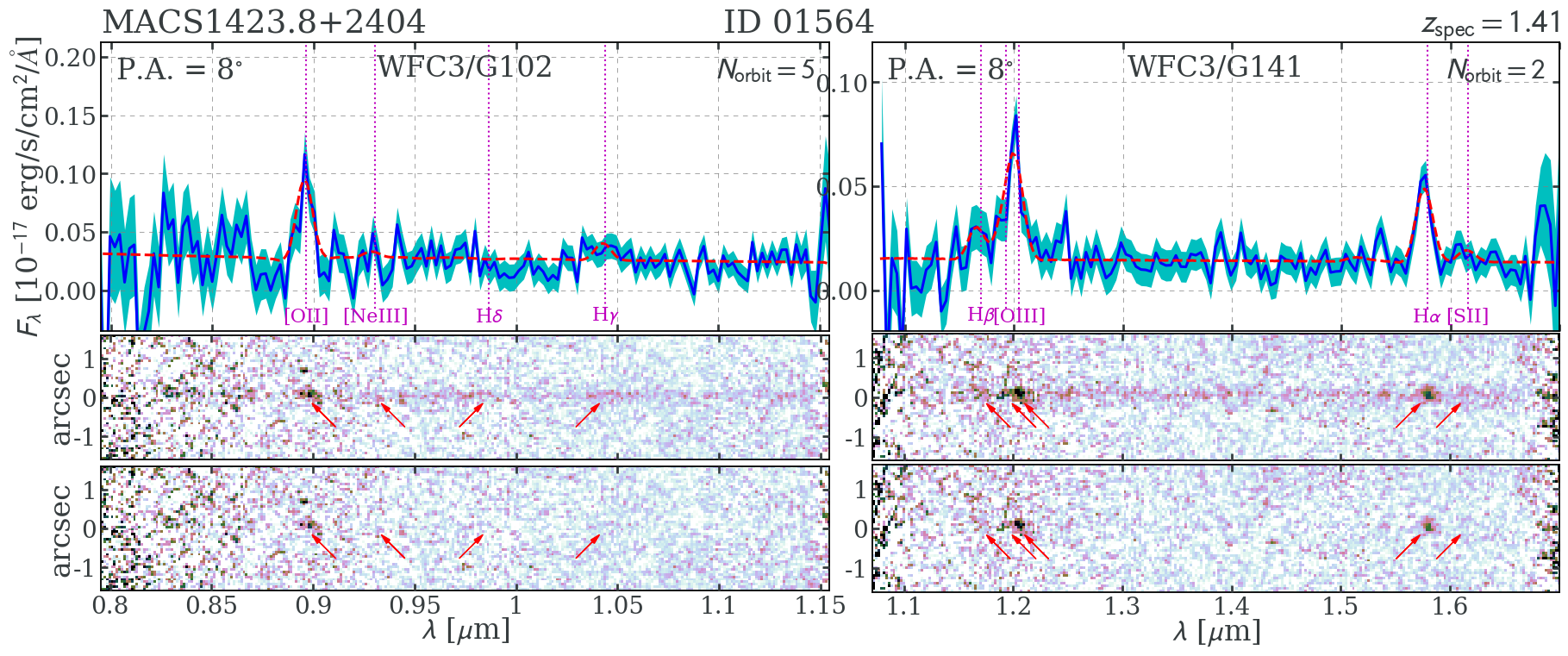}\\
    \includegraphics[width=\textwidth]{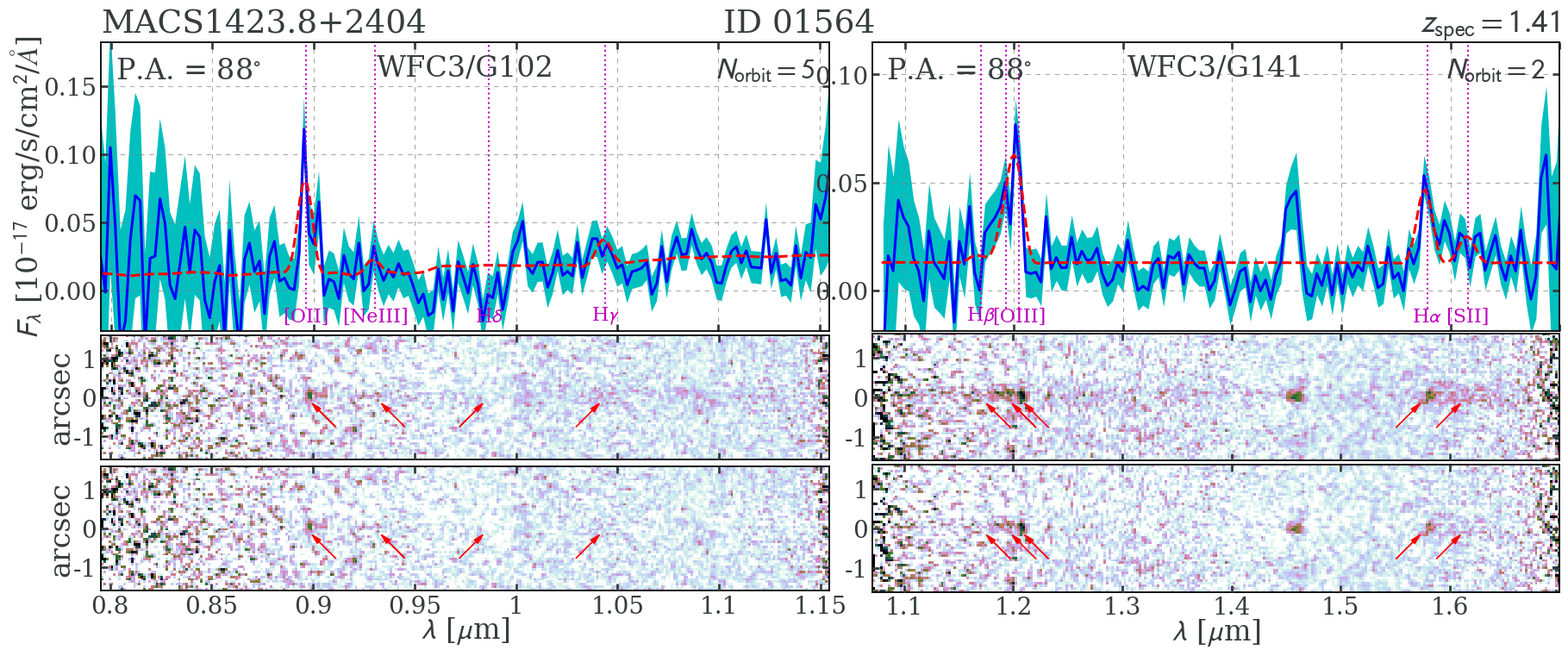}\\
    \includegraphics[width=.16\textwidth]{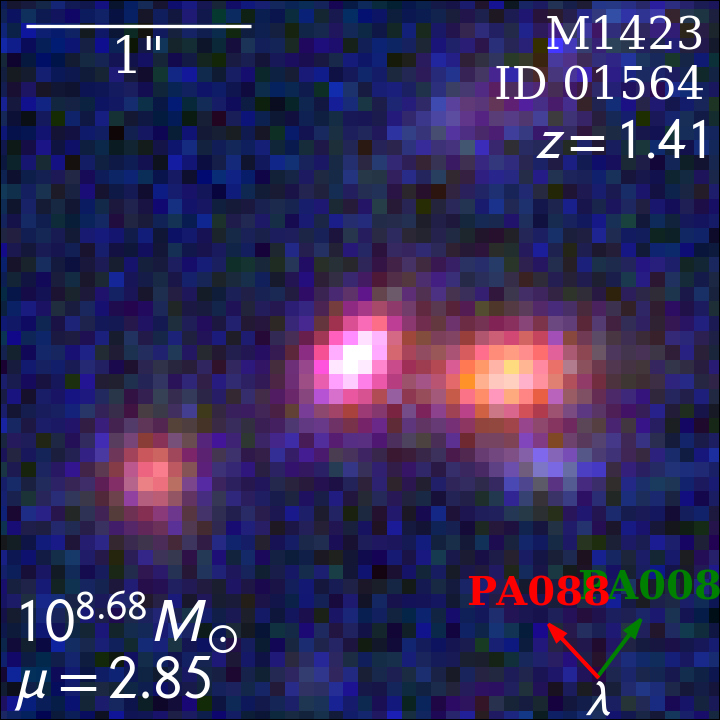}
    \includegraphics[width=.16\textwidth]{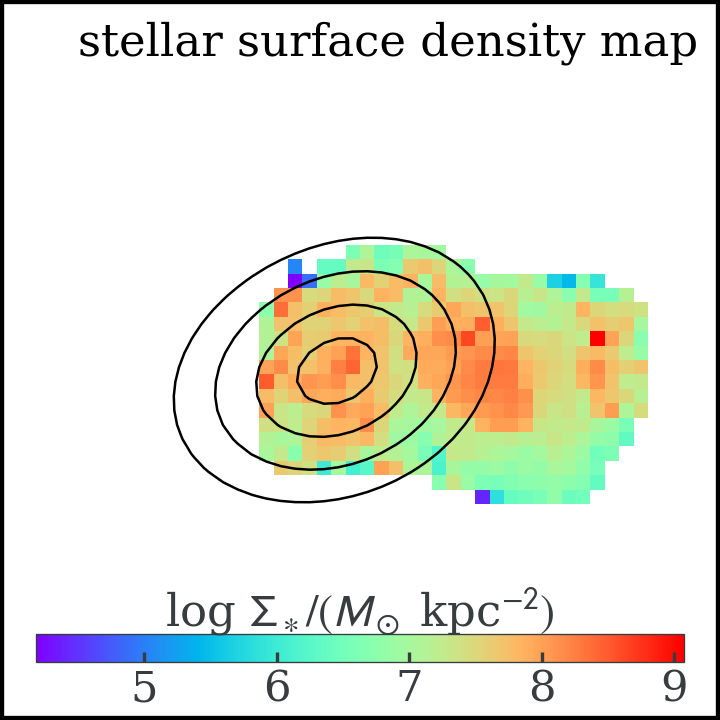}
    \includegraphics[width=.16\textwidth]{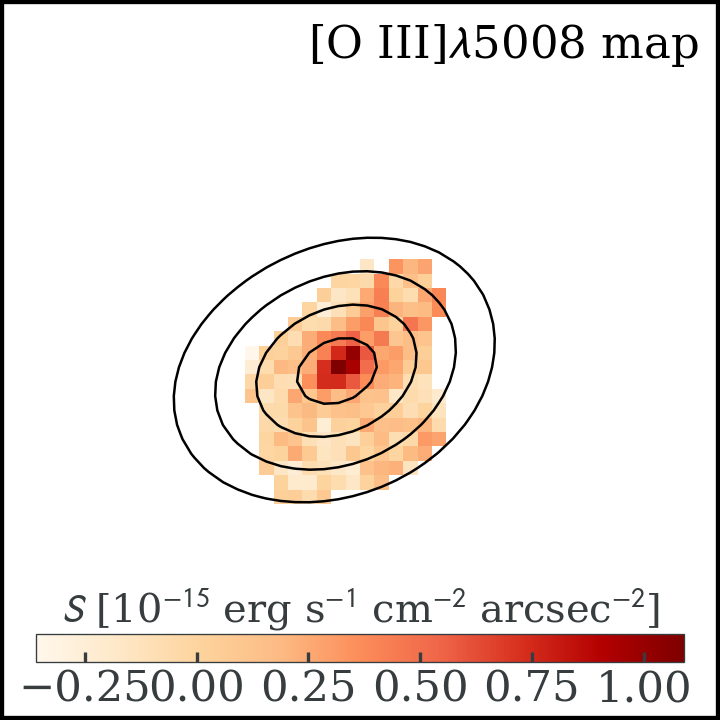}
    \includegraphics[width=.16\textwidth]{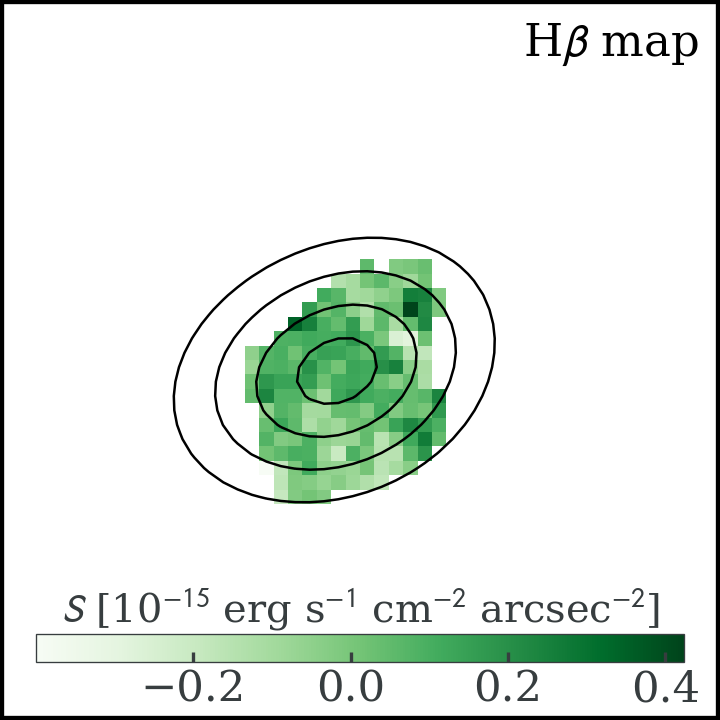}
    \includegraphics[width=.16\textwidth]{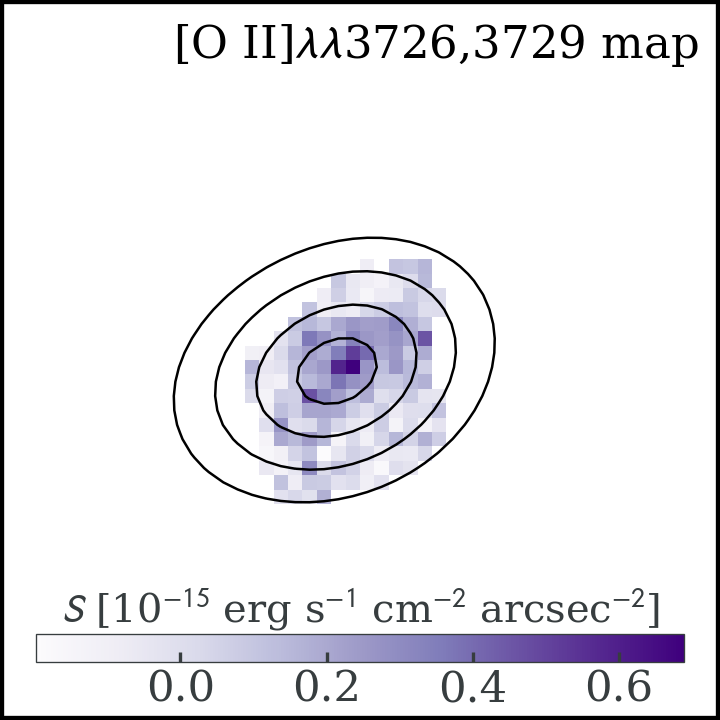}
    \includegraphics[width=.16\textwidth]{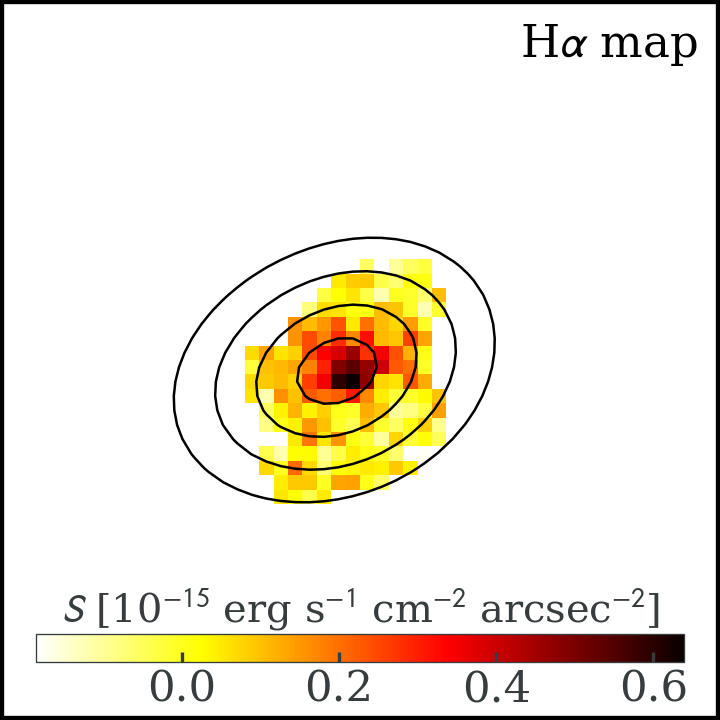}\\
    \includegraphics[width=\textwidth]{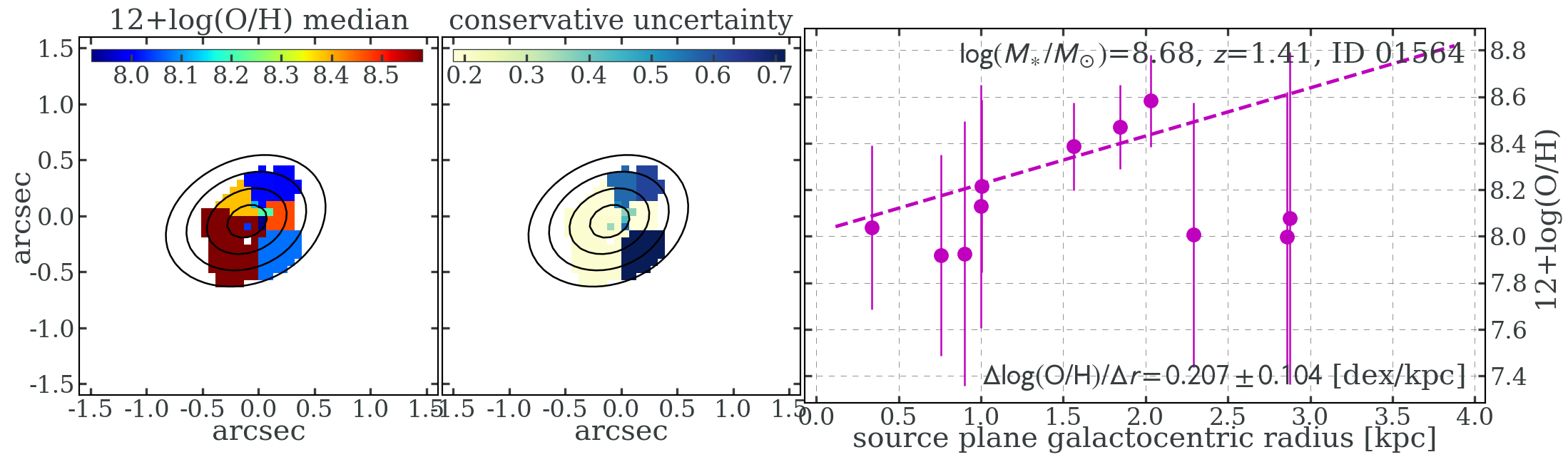}
    \caption{The source ID01564 in the field of \clshi is shown.}
    \label{fig:clM1423_ID01564_figs}
\end{figure*}

\begin{figure*}
    \centering
    \includegraphics[width=\textwidth]{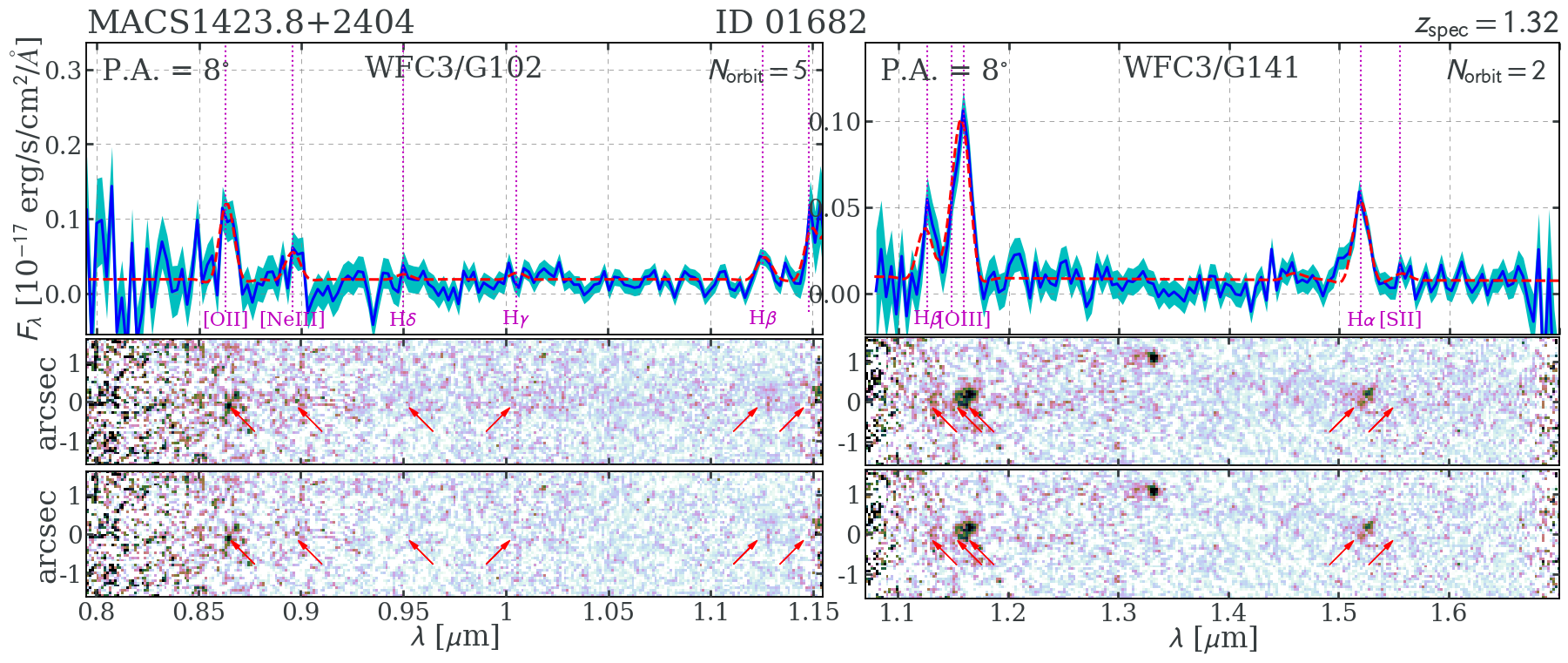}\\
    \includegraphics[width=\textwidth]{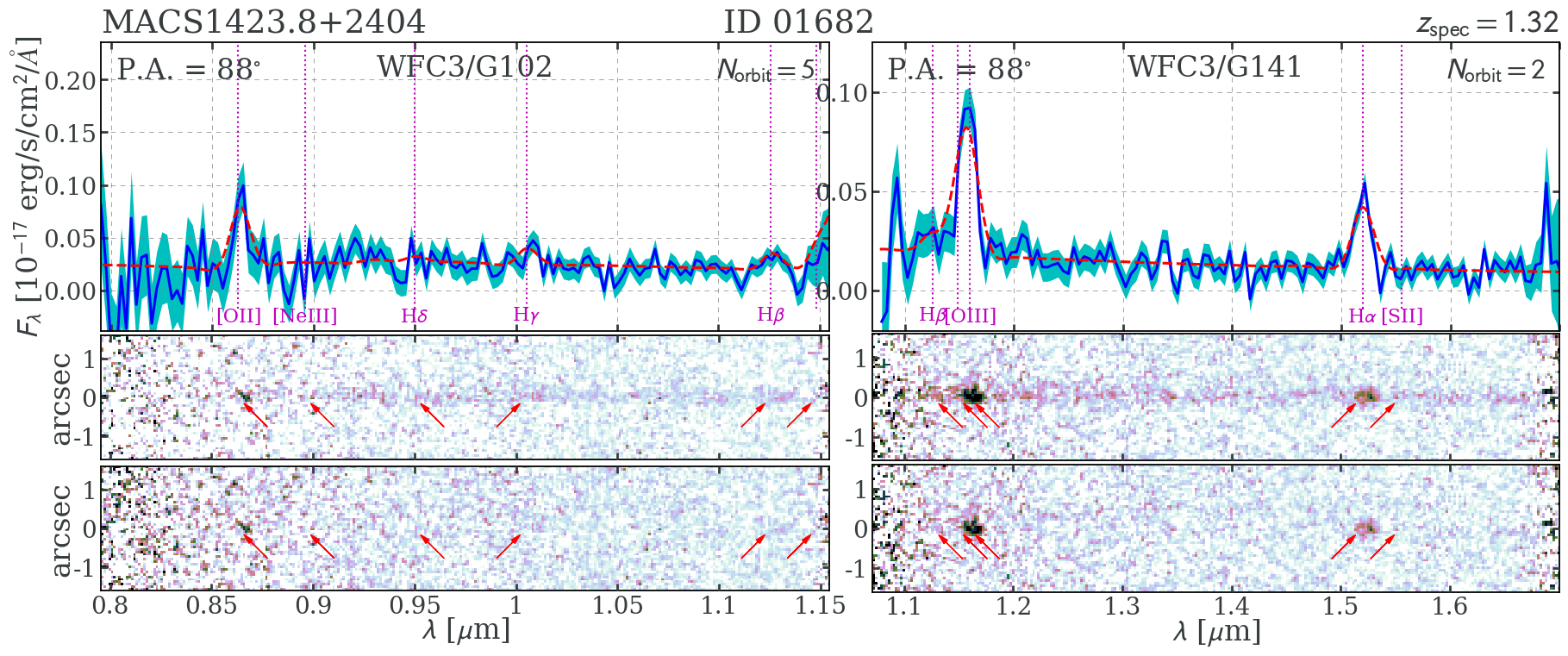}\\
    \includegraphics[width=.16\textwidth]{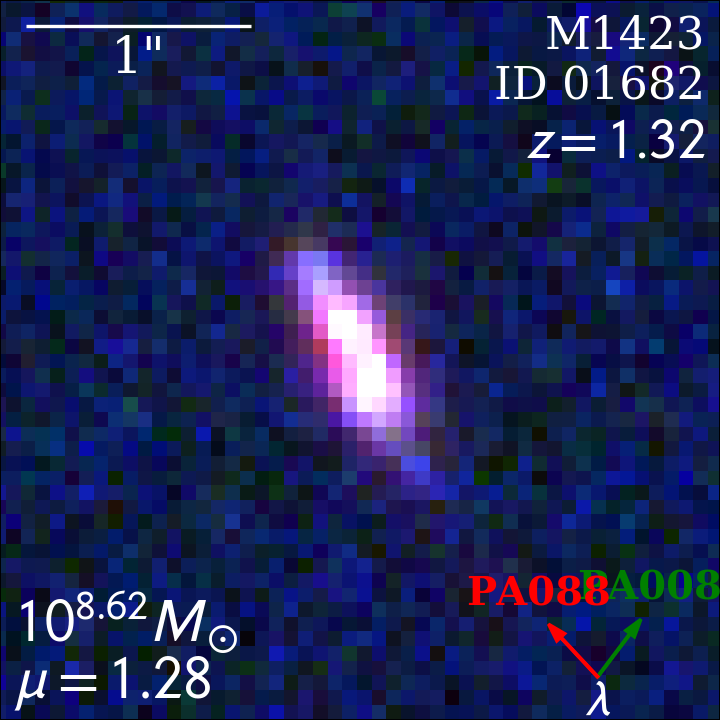}
    \includegraphics[width=.16\textwidth]{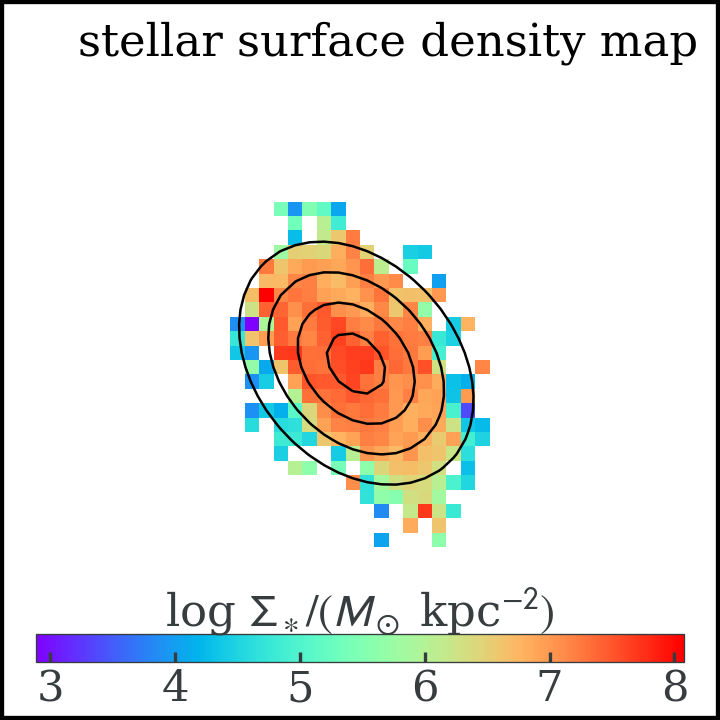}
    \includegraphics[width=.16\textwidth]{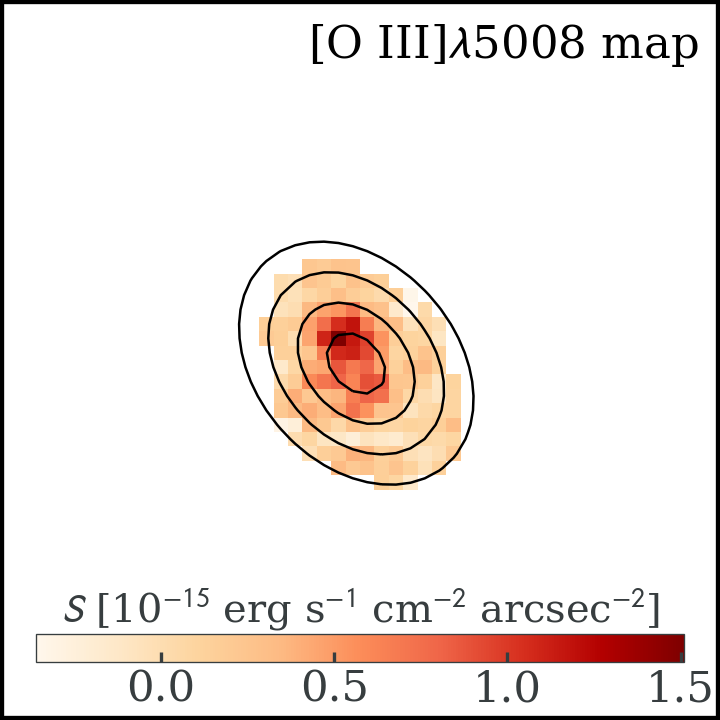}
    \includegraphics[width=.16\textwidth]{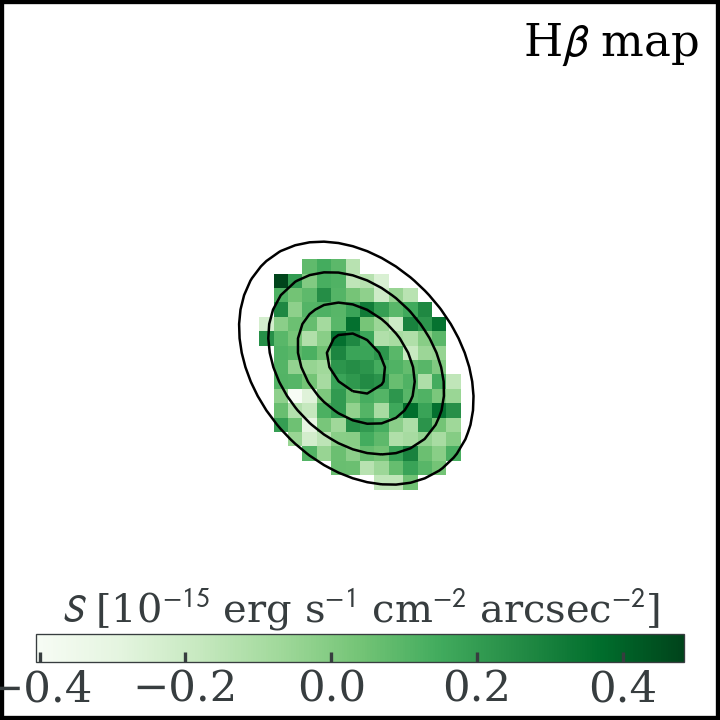}
    \includegraphics[width=.16\textwidth]{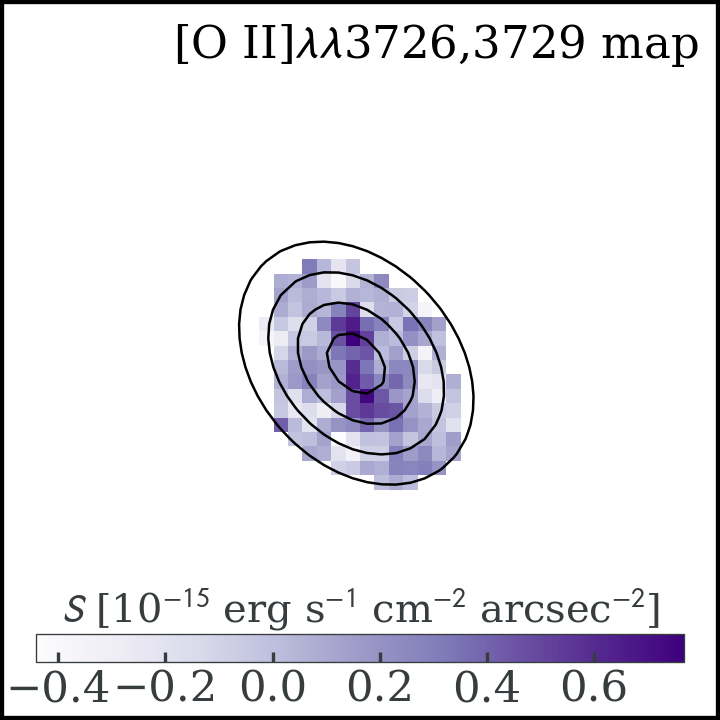}
    \includegraphics[width=.16\textwidth]{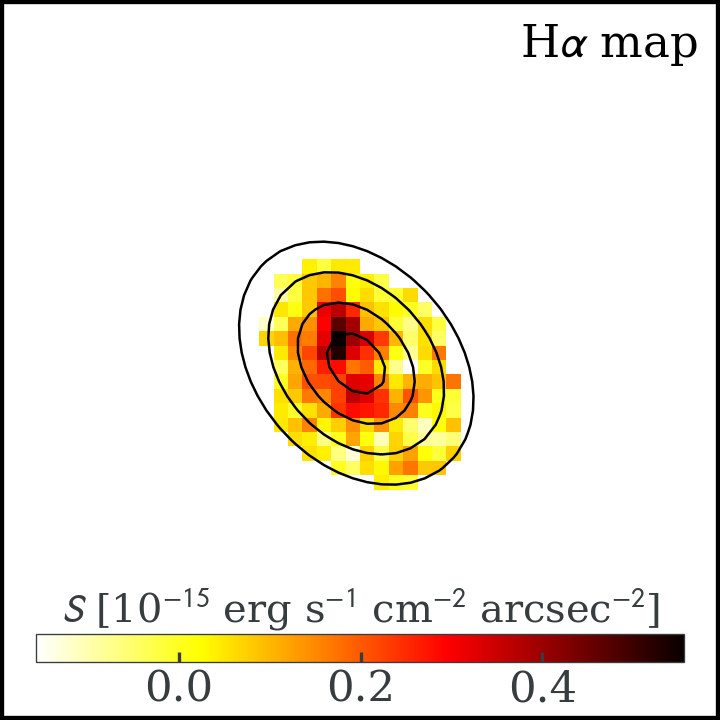}\\
    \includegraphics[width=\textwidth]{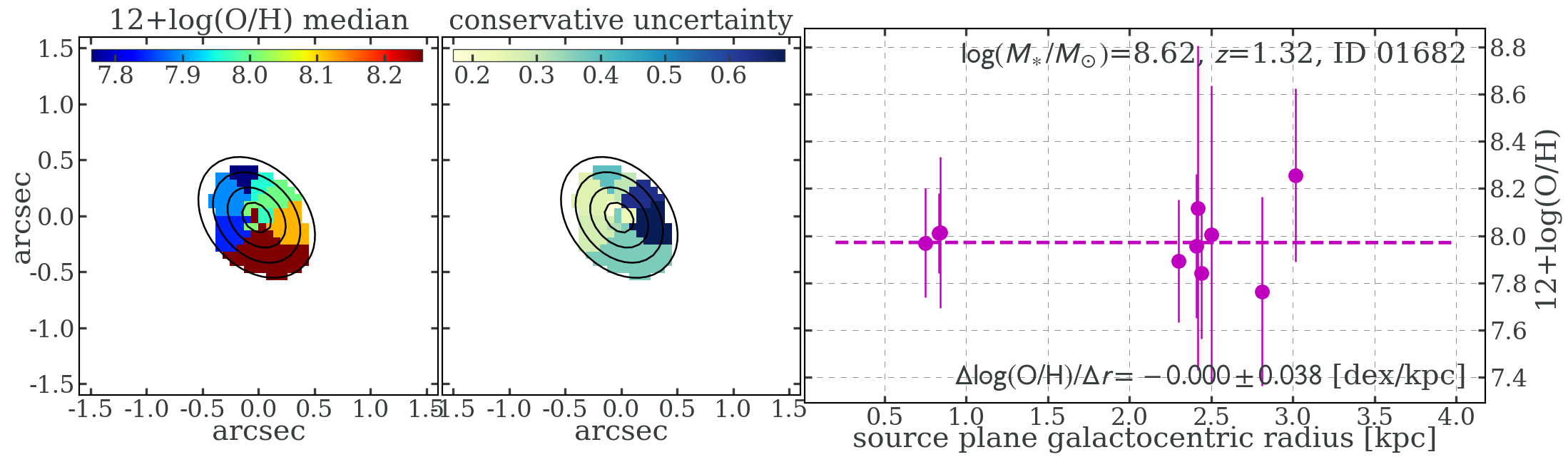}
    \caption{The source ID01682 in the field of \clshi is shown.}
    \label{fig:clM1423_ID01682_figs}
\end{figure*}

\begin{figure*}
    \centering
    \includegraphics[width=\textwidth]{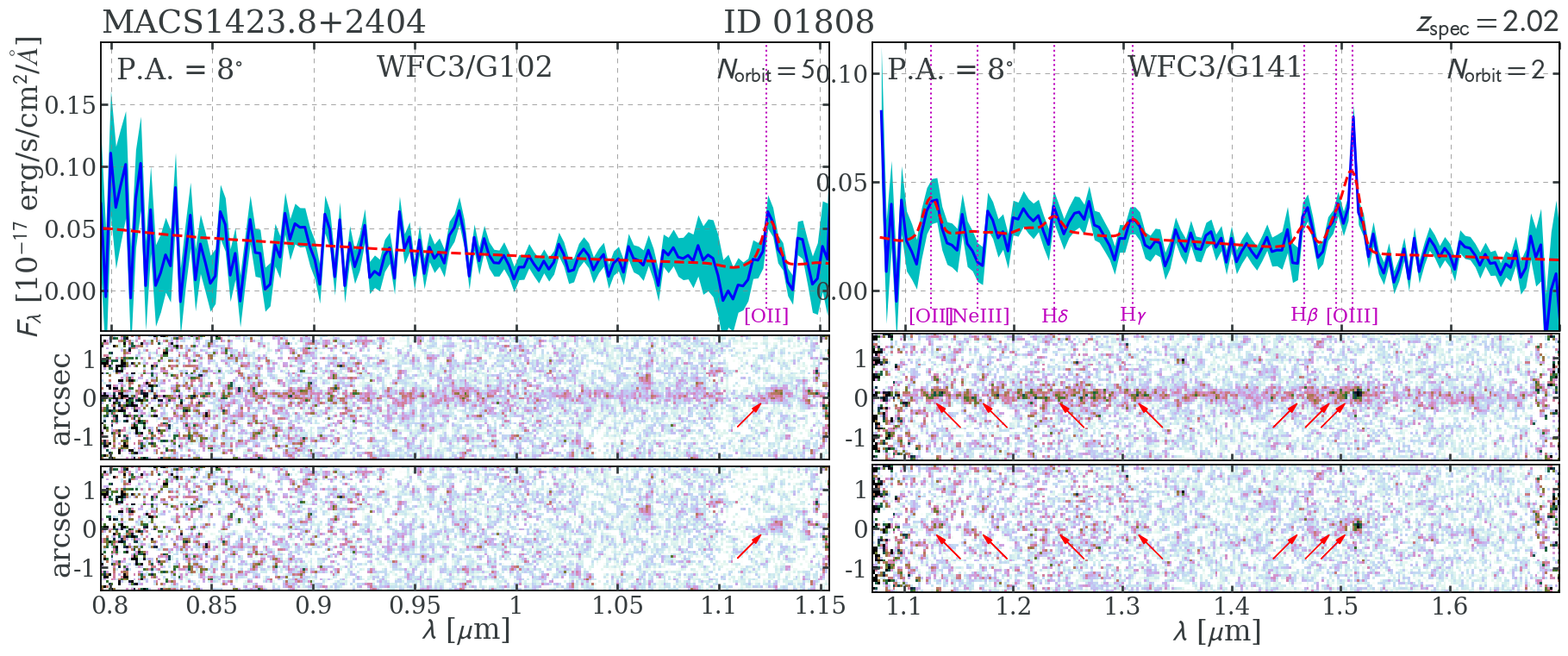}\\
    \includegraphics[width=\textwidth]{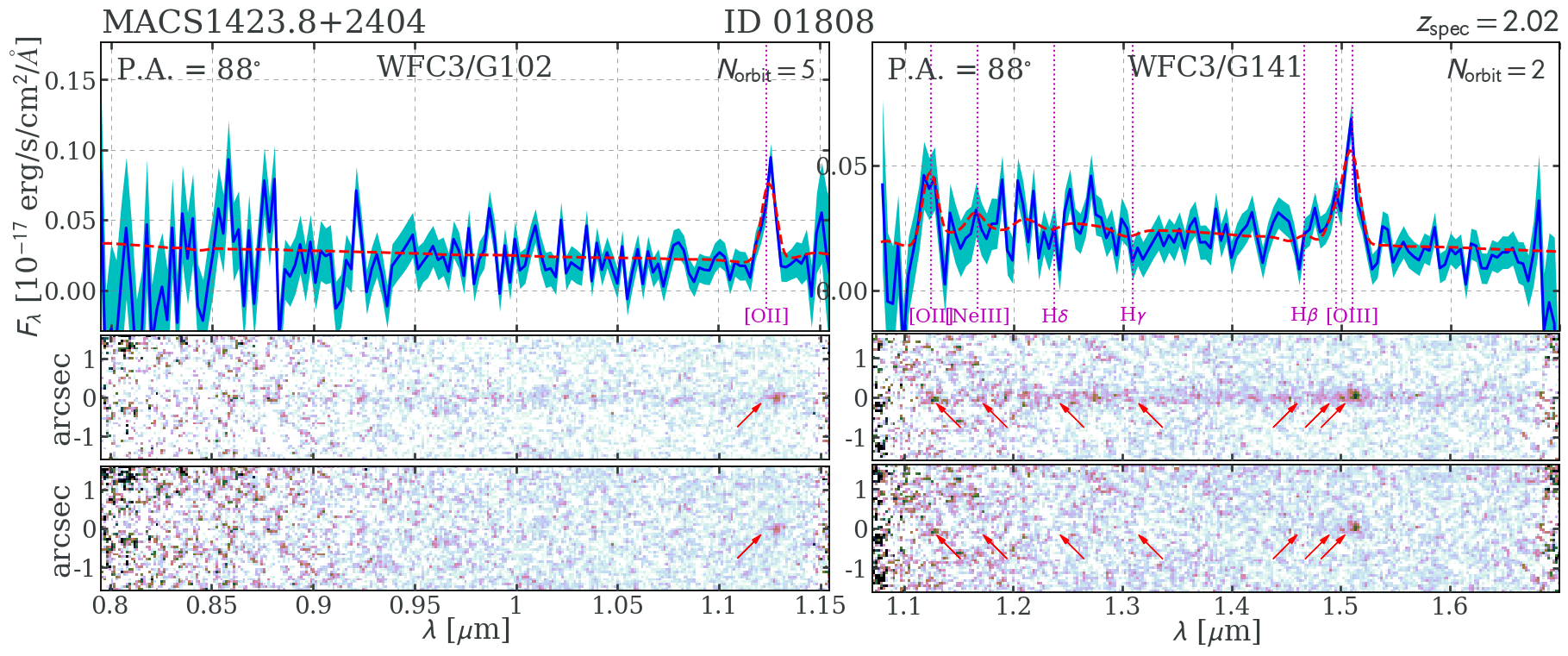}\\
    \includegraphics[width=.16\textwidth]{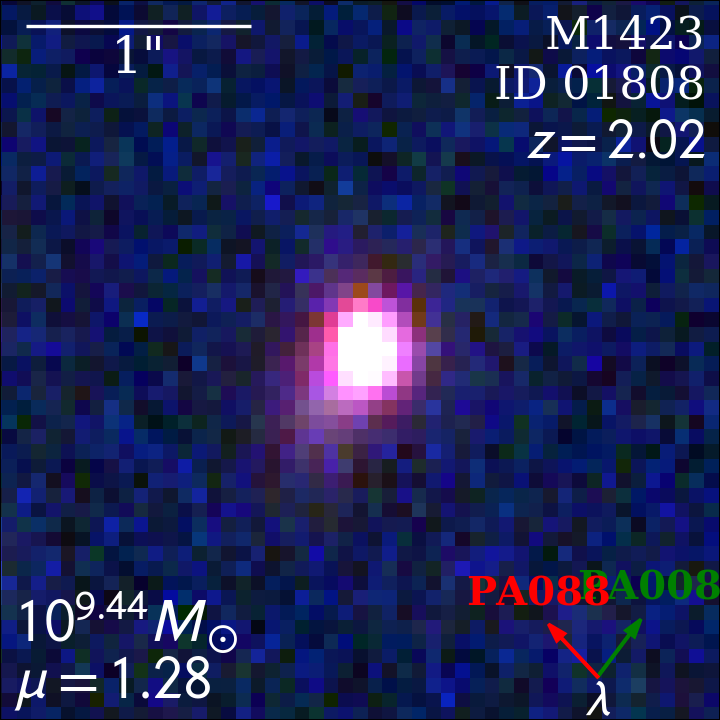}
    \includegraphics[width=.16\textwidth]{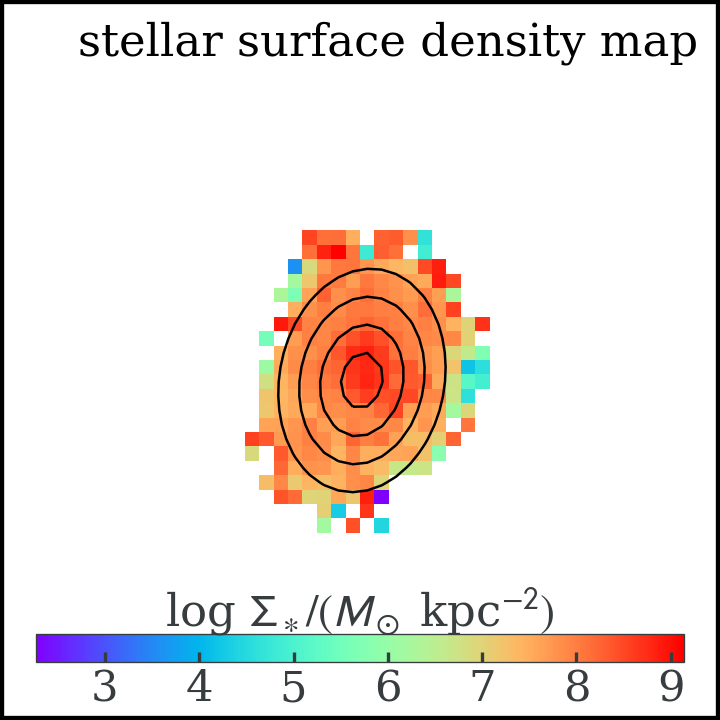}
    \includegraphics[width=.16\textwidth]{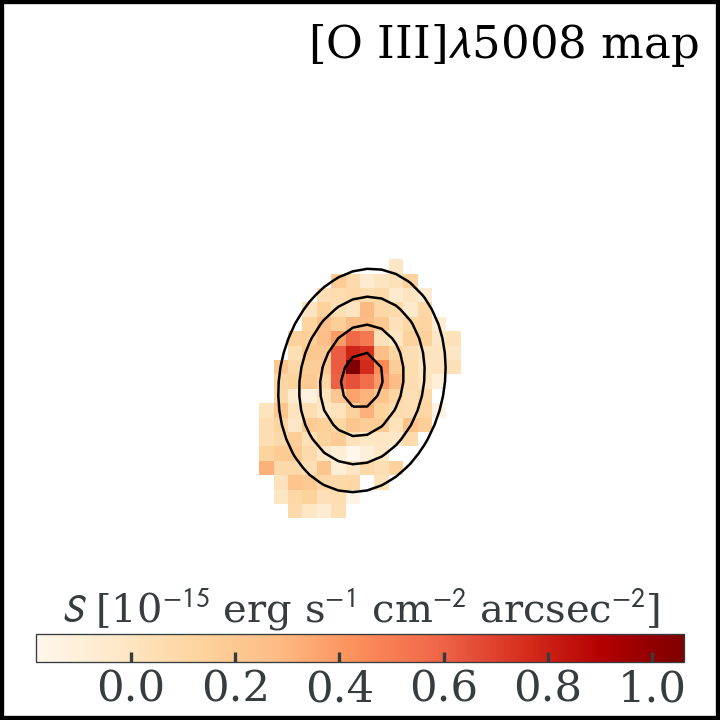}
    \includegraphics[width=.16\textwidth]{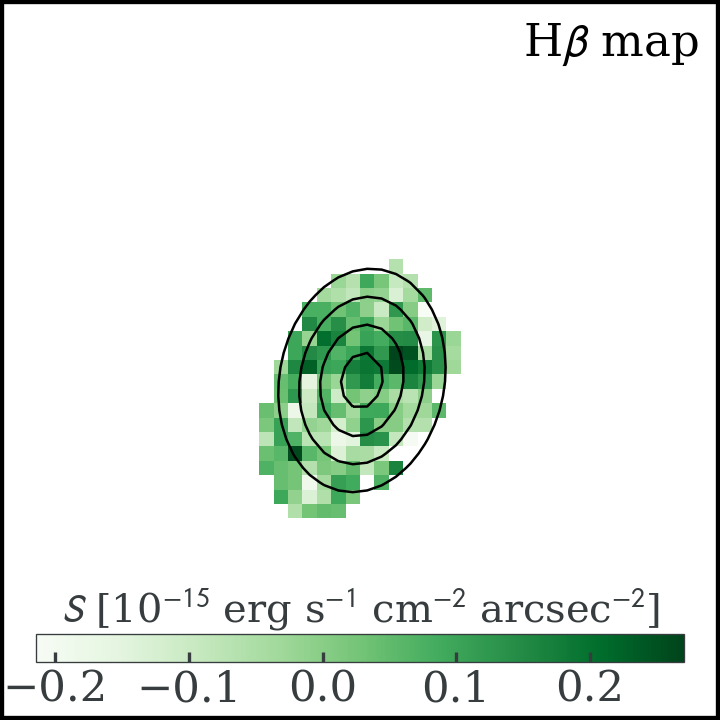}
    \includegraphics[width=.16\textwidth]{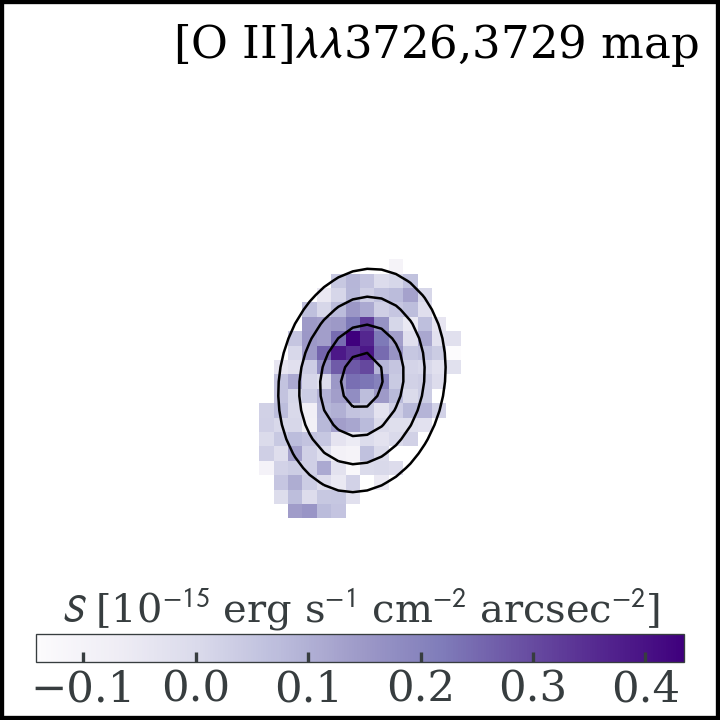}
    \includegraphics[width=.16\textwidth]{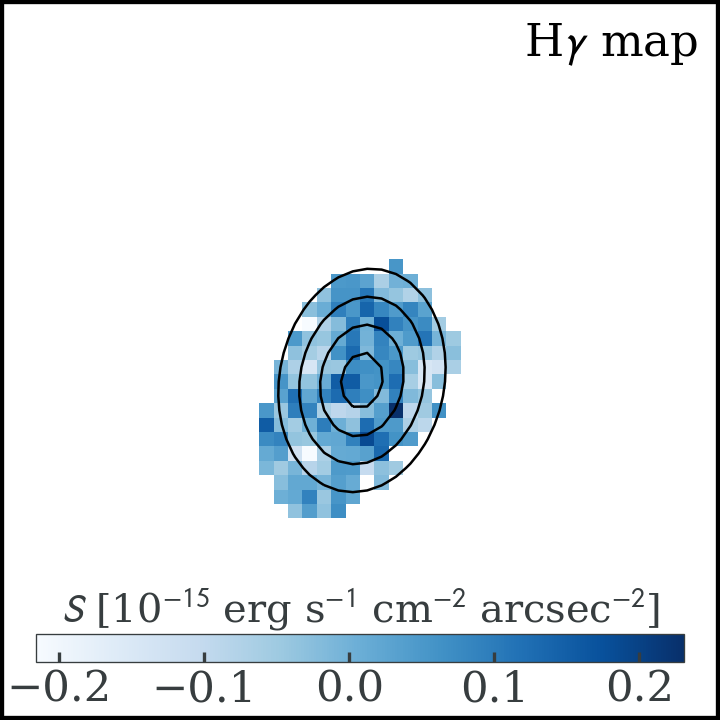}\\
    \includegraphics[width=\textwidth]{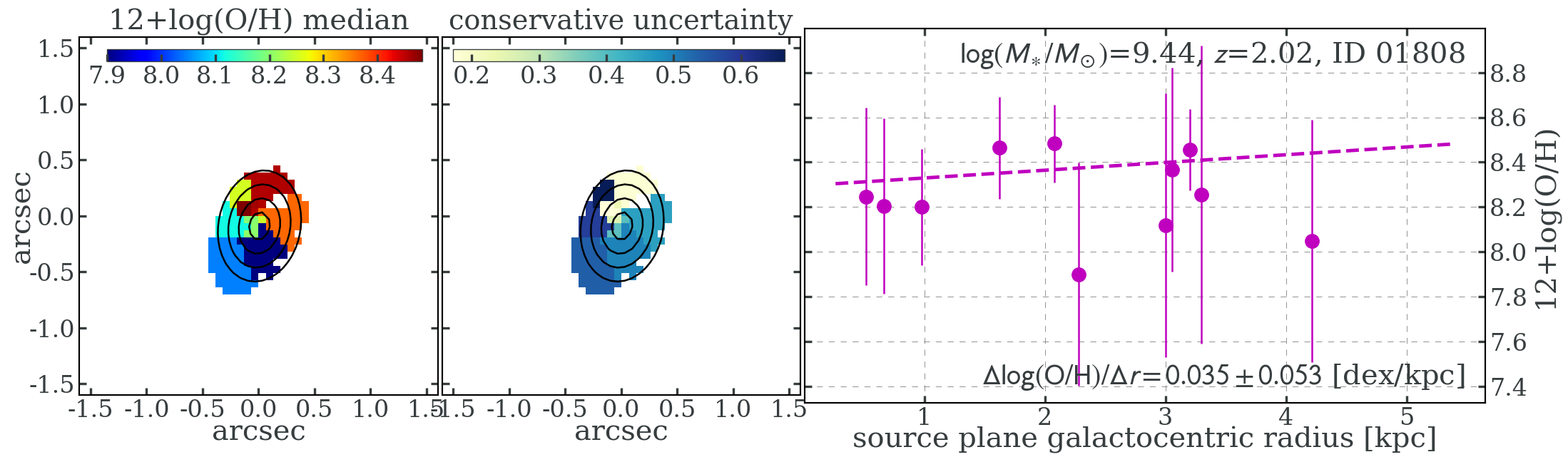}
    \caption{The source ID01808 in the field of \clshi is shown.}
    \label{fig:clM1423_ID01808_figs}
\end{figure*}

\begin{figure*}
    \centering
    \includegraphics[width=\textwidth]{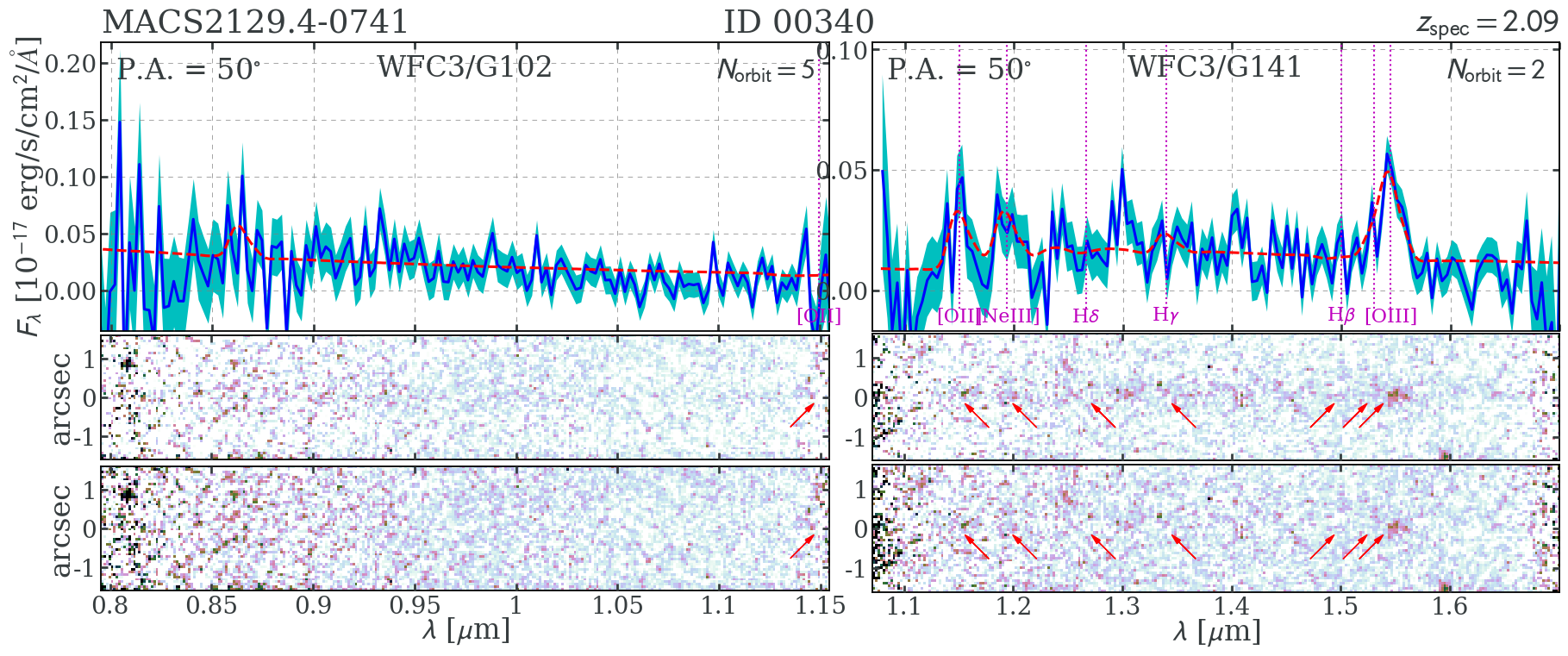}\\
    \includegraphics[width=\textwidth]{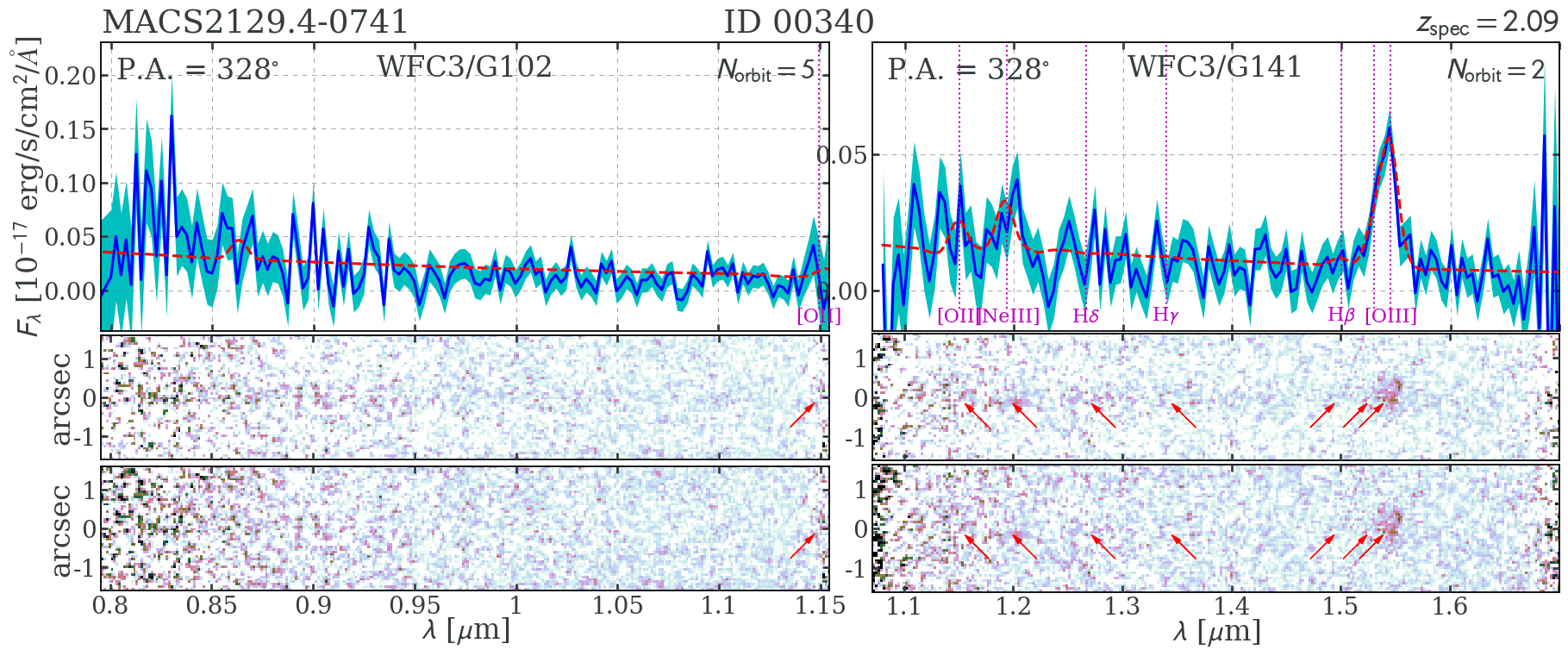}\\
    \includegraphics[width=.16\textwidth]{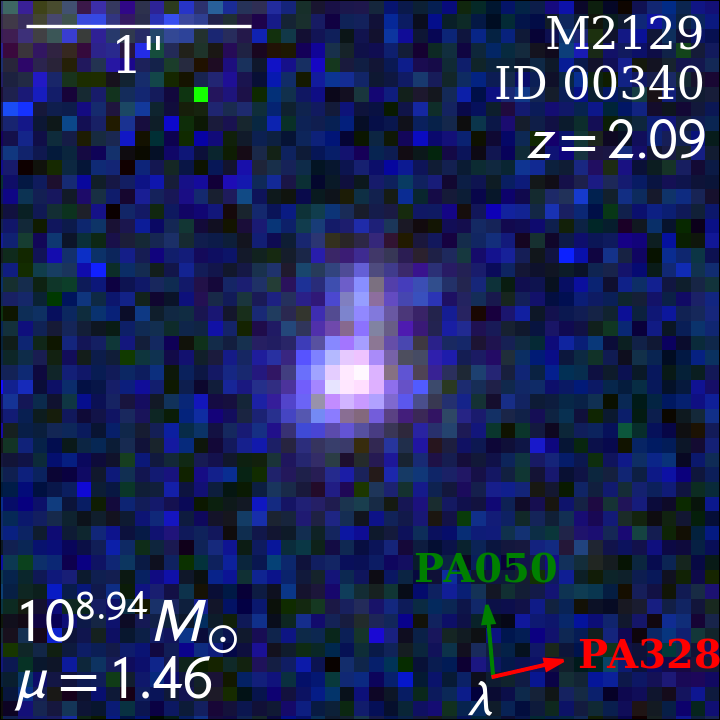}
    \includegraphics[width=.16\textwidth]{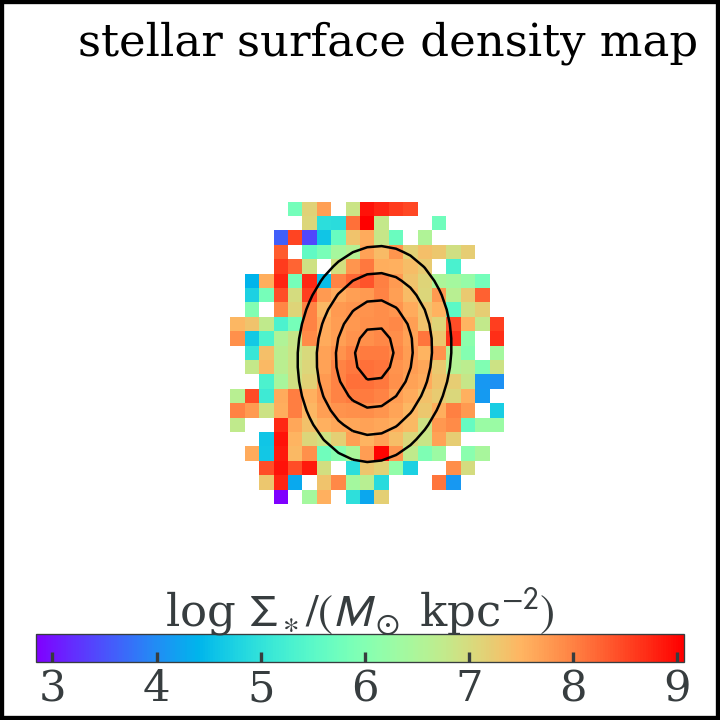}
    \includegraphics[width=.16\textwidth]{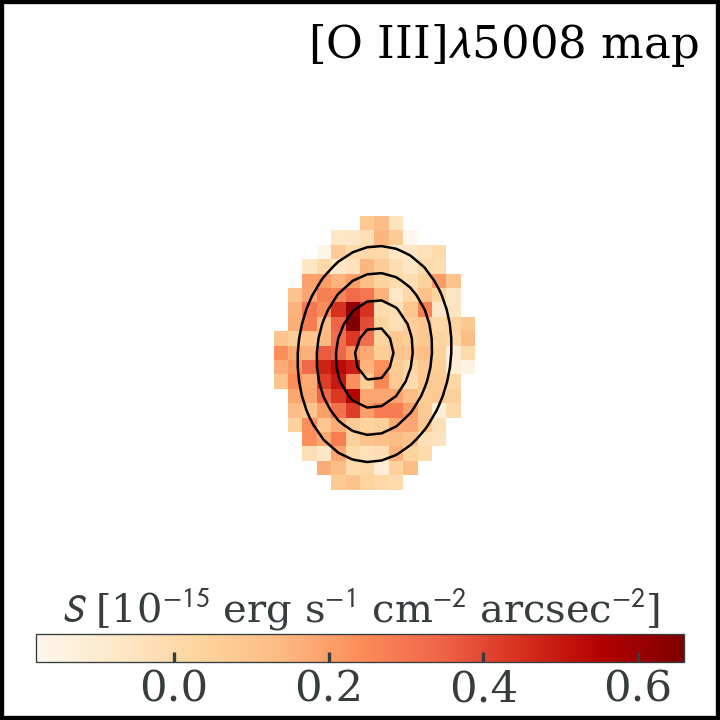}
    \includegraphics[width=.16\textwidth]{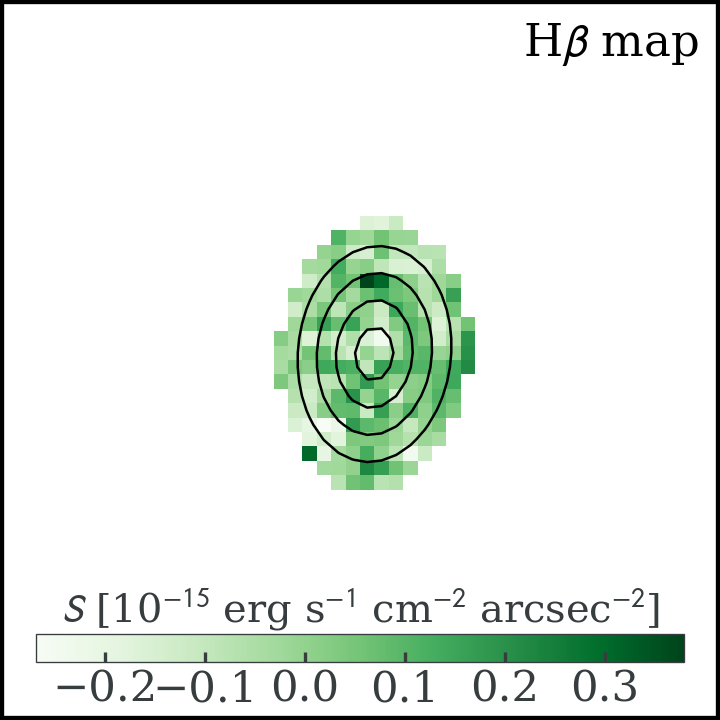}
    \includegraphics[width=.16\textwidth]{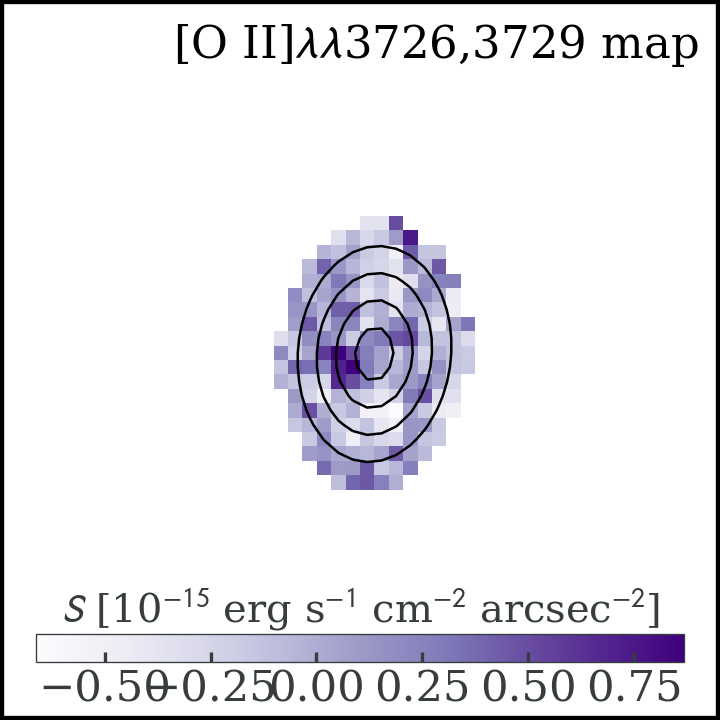}
    \includegraphics[width=.16\textwidth]{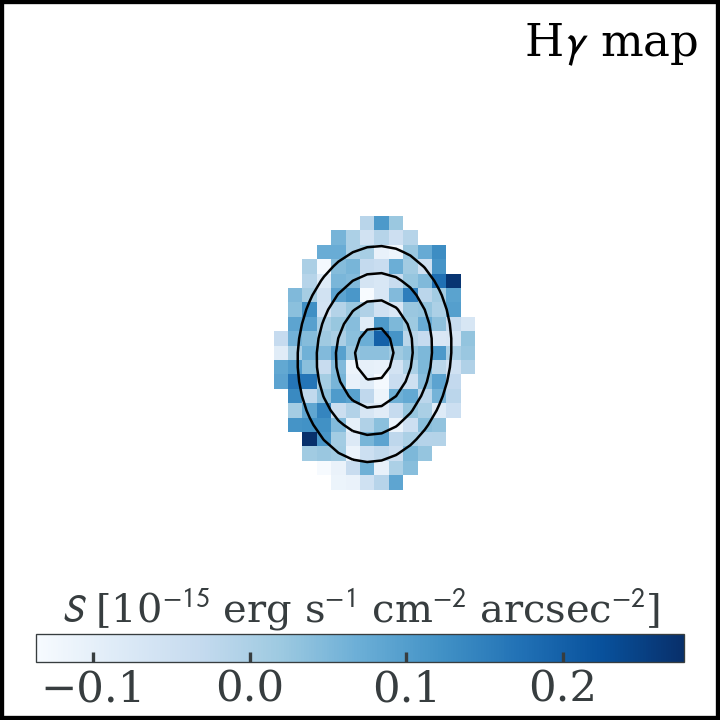}\\
    \includegraphics[width=\textwidth]{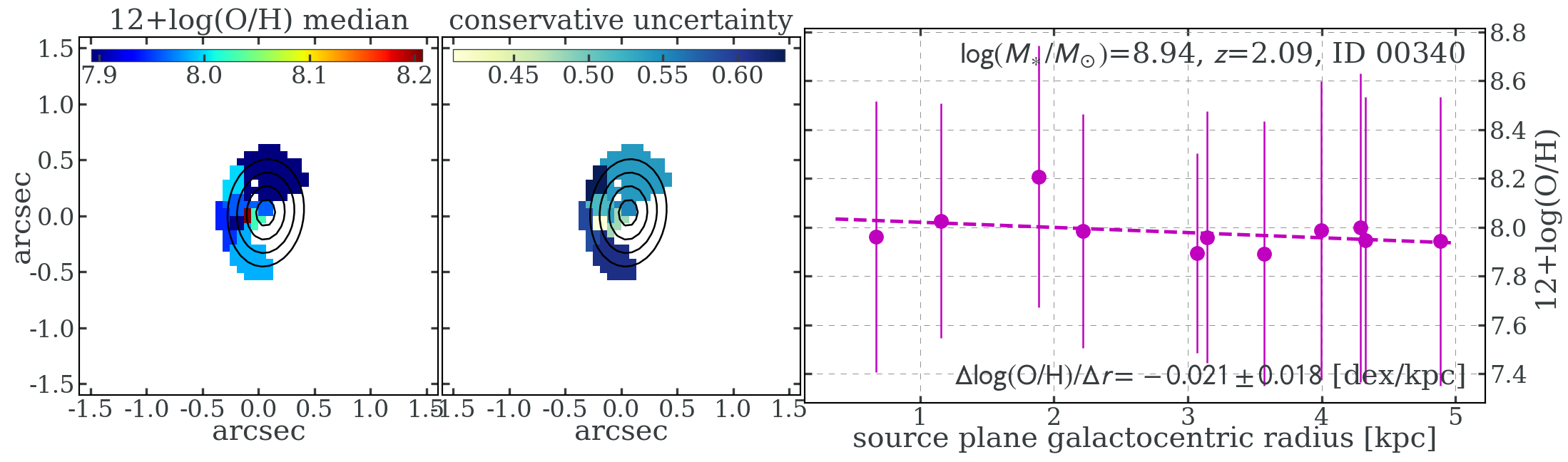}
    \caption{The source ID00340 in the field of \cljiu is shown.}
    \label{fig:clM2129_ID00340_figs}
\end{figure*}

\begin{figure*}
    \centering
    \includegraphics[width=\textwidth]{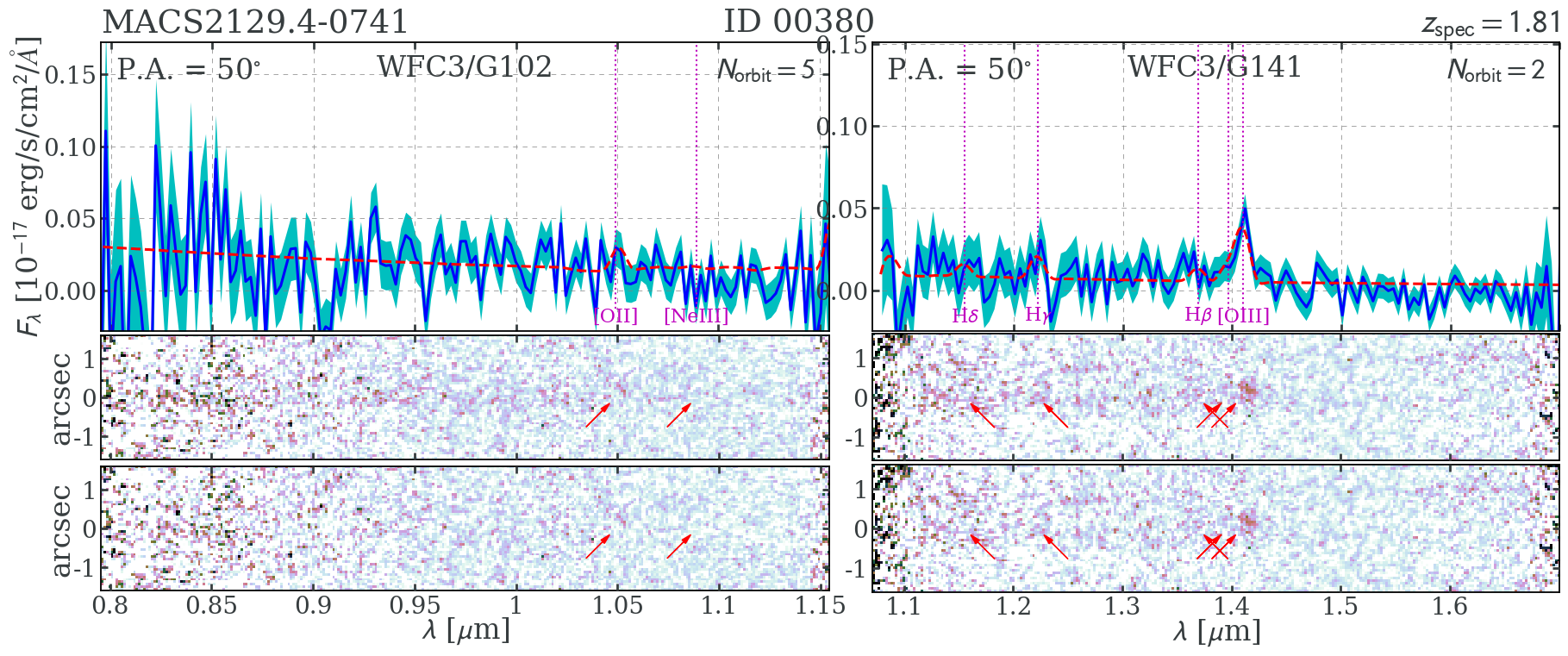}\\
    \includegraphics[width=\textwidth]{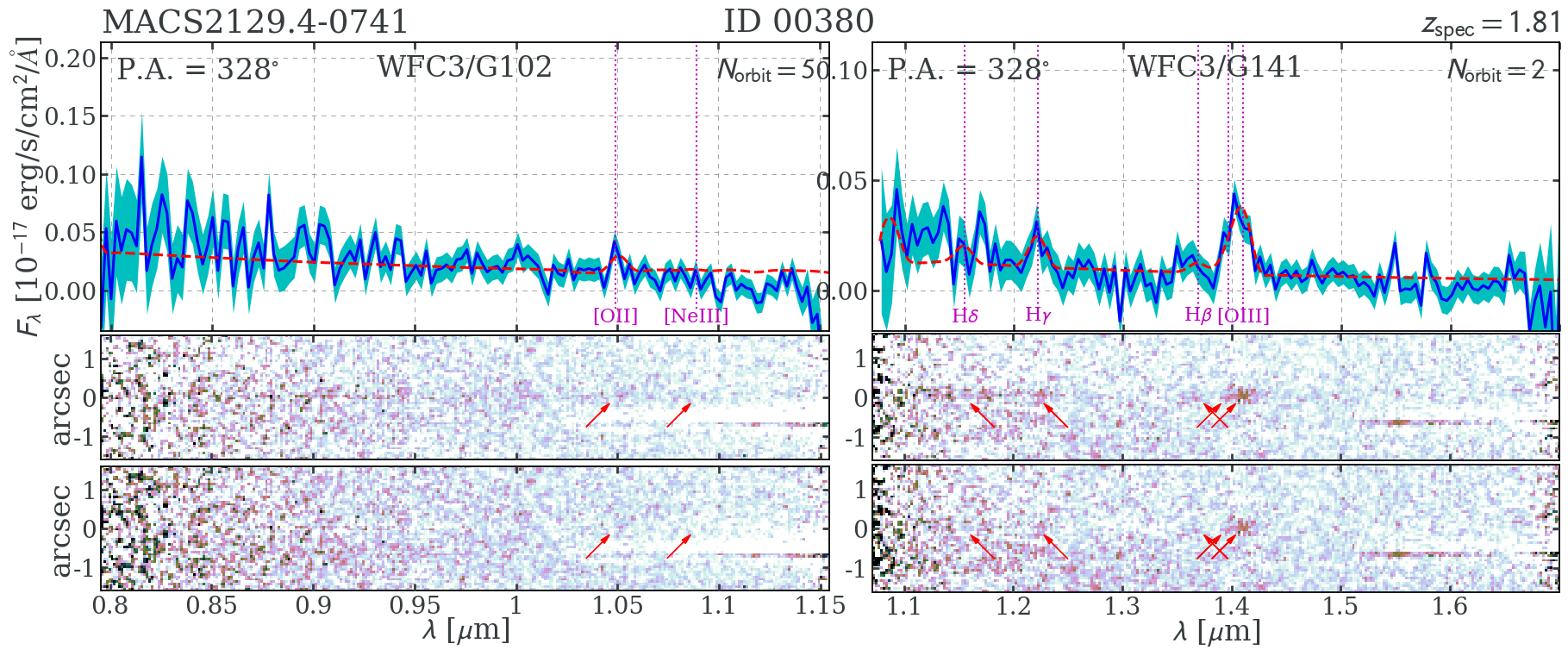}\\
    \includegraphics[width=.16\textwidth]{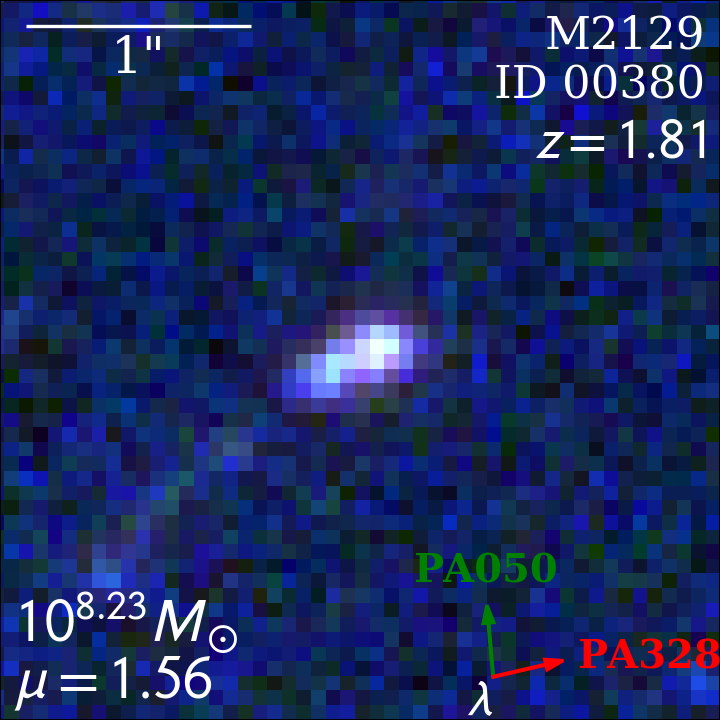}
    \includegraphics[width=.16\textwidth]{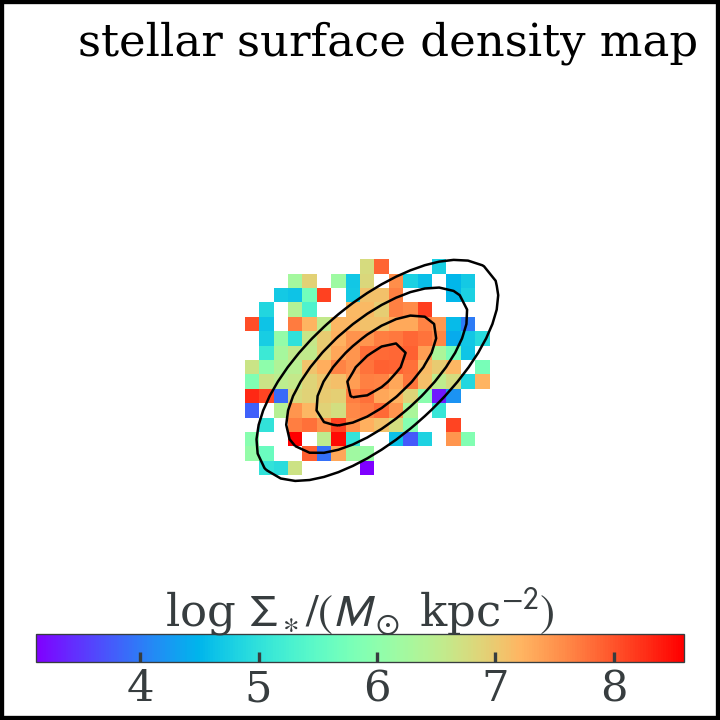}
    \includegraphics[width=.16\textwidth]{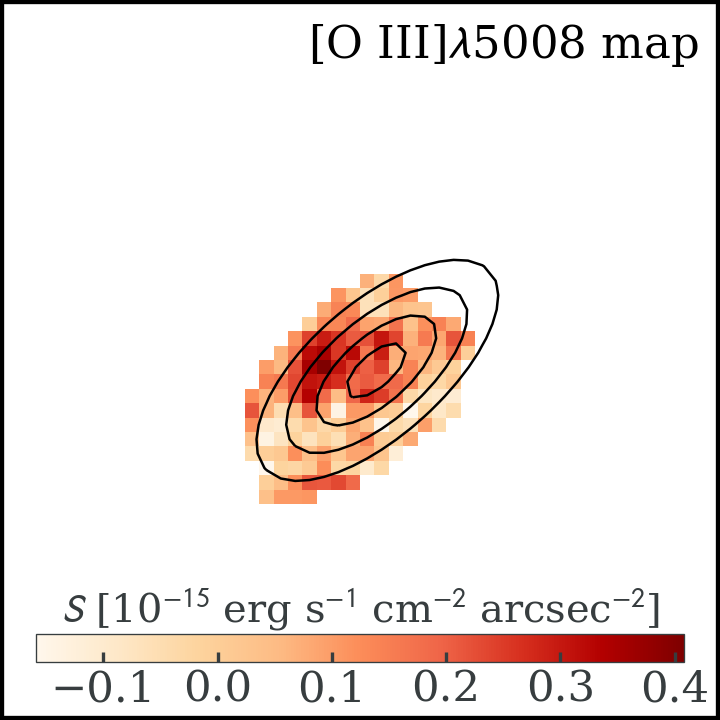}
    \includegraphics[width=.16\textwidth]{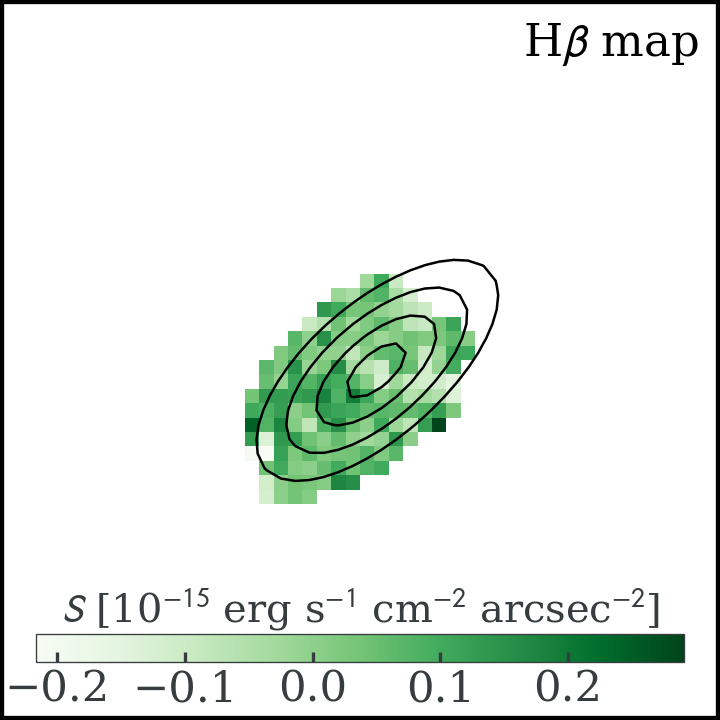}
    \includegraphics[width=.16\textwidth]{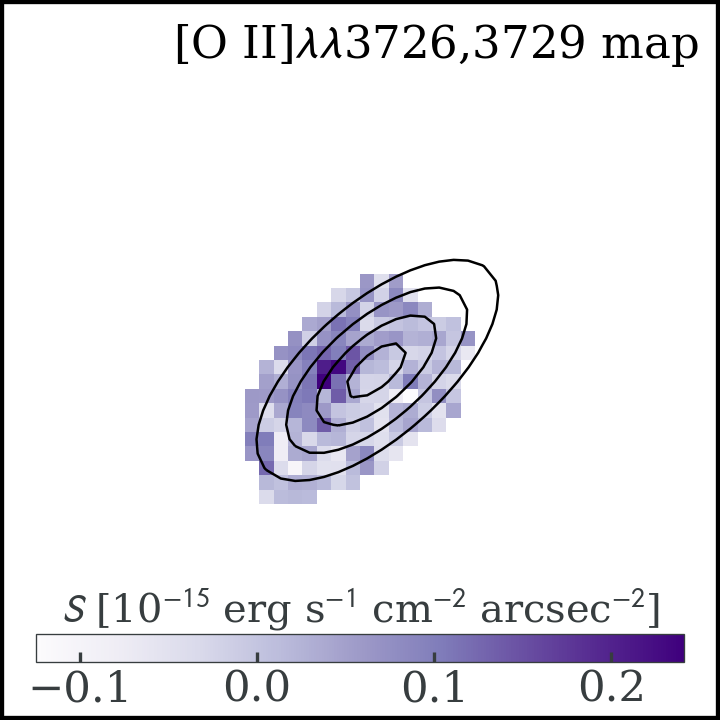}
    \includegraphics[width=.16\textwidth]{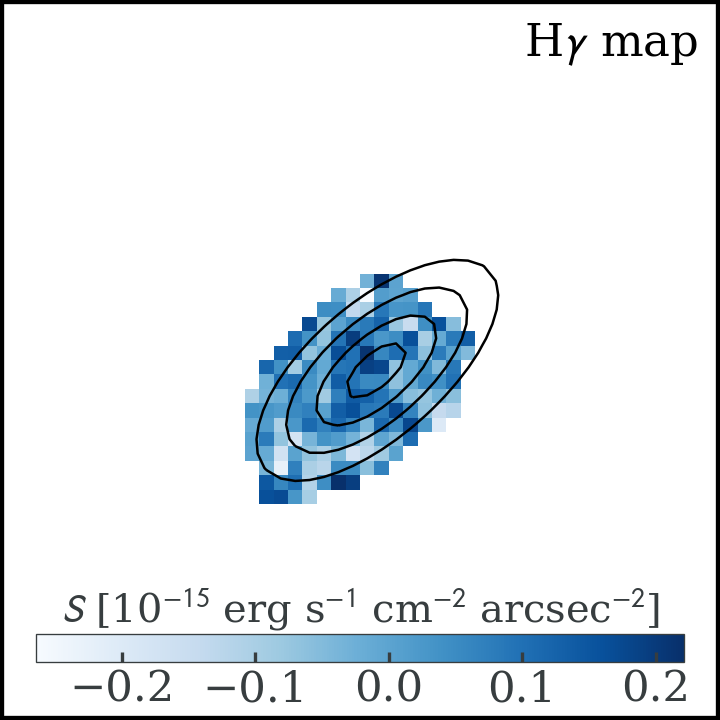}\\
    \includegraphics[width=\textwidth]{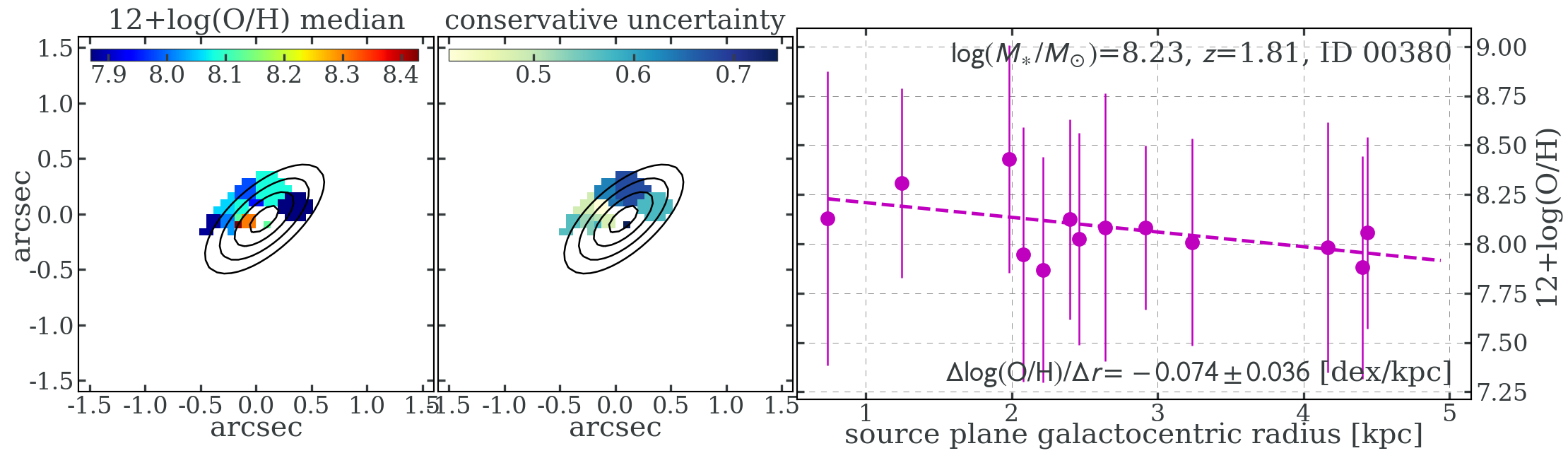}
    \caption{The source ID00380 in the field of \cljiu is shown.}
    \label{fig:clM2129_ID00380_figs}
\end{figure*}

\begin{figure*}
    \centering
    \includegraphics[width=\textwidth]{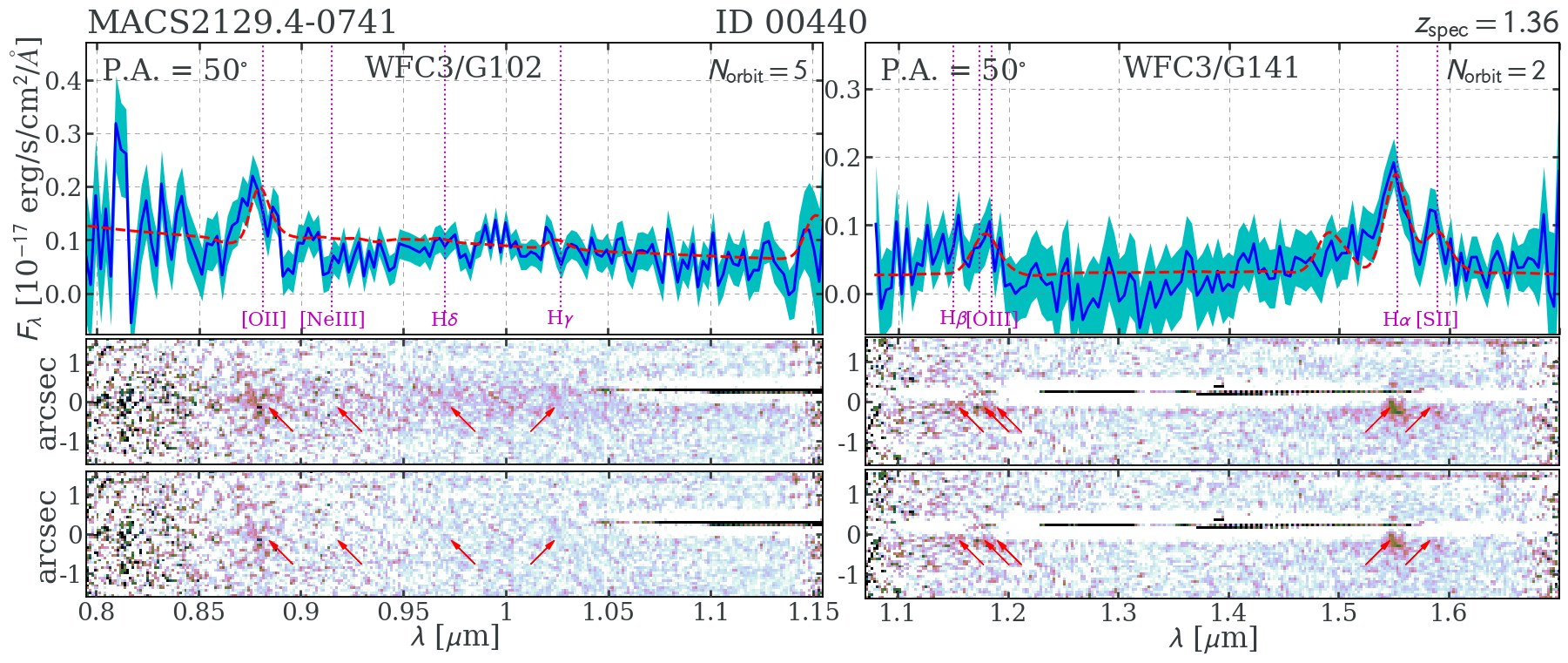}\\
    \includegraphics[width=\textwidth]{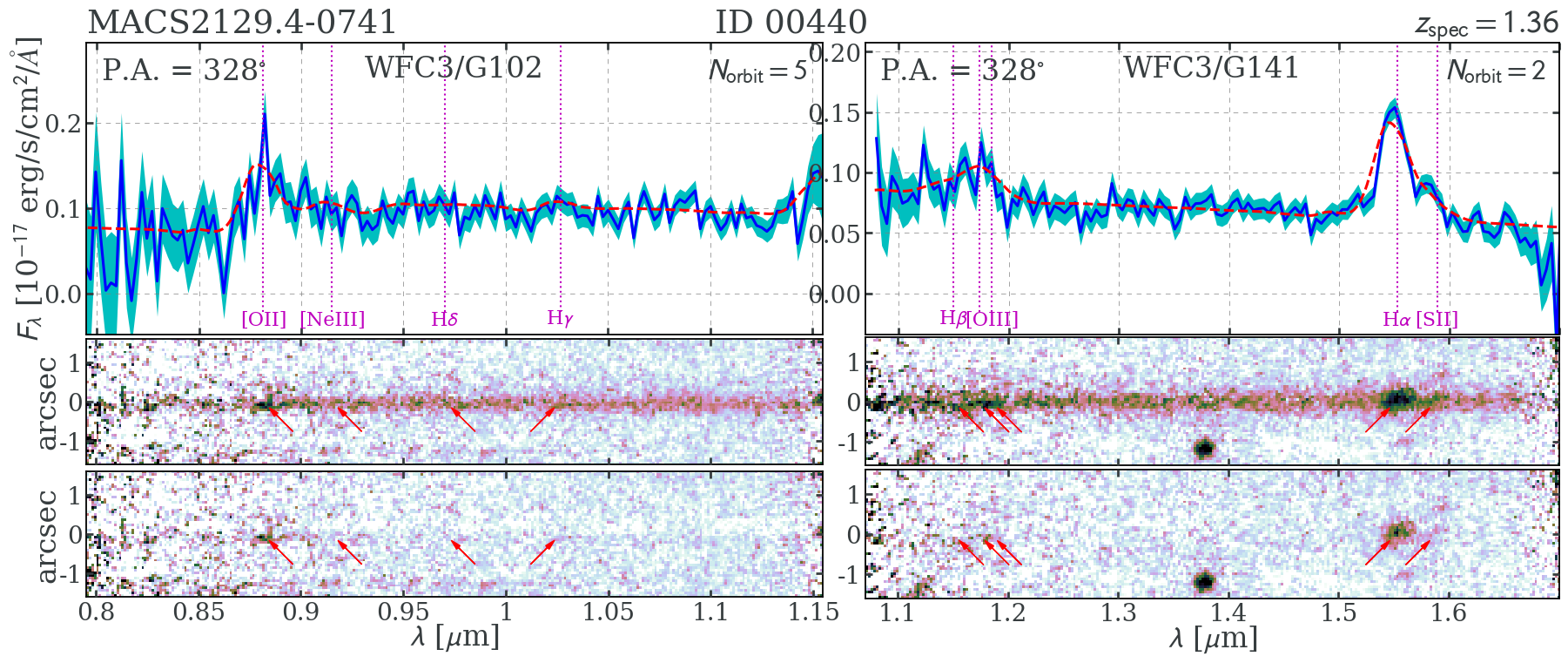}\\
    \includegraphics[width=.16\textwidth]{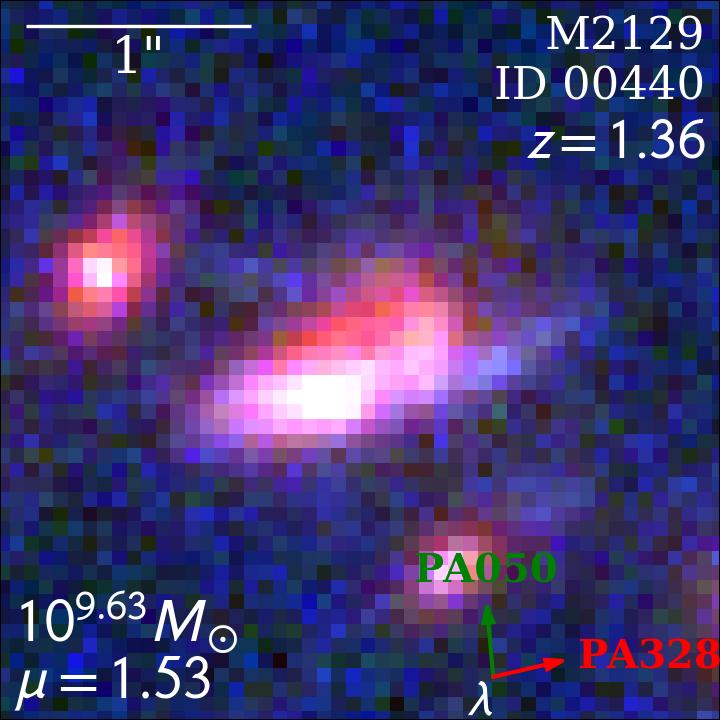}
    \includegraphics[width=.16\textwidth]{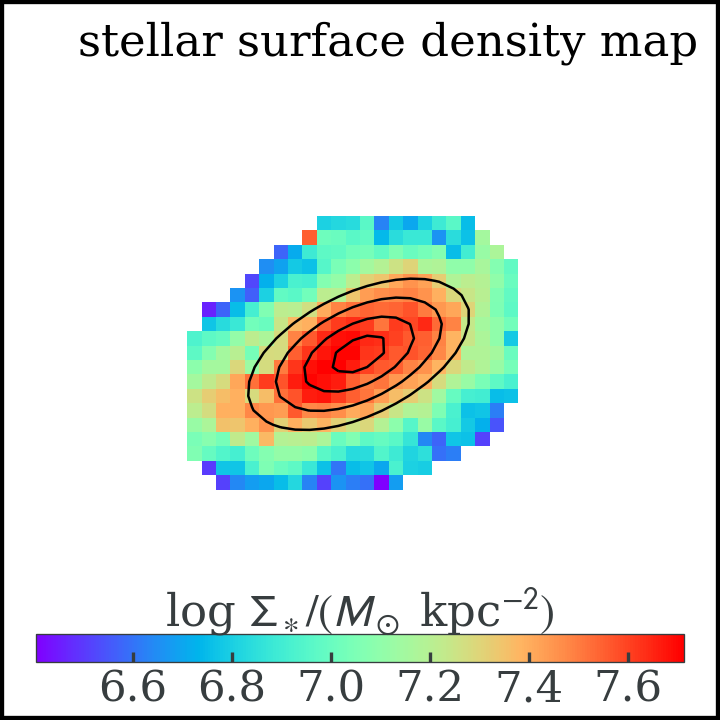}
    \includegraphics[width=.16\textwidth]{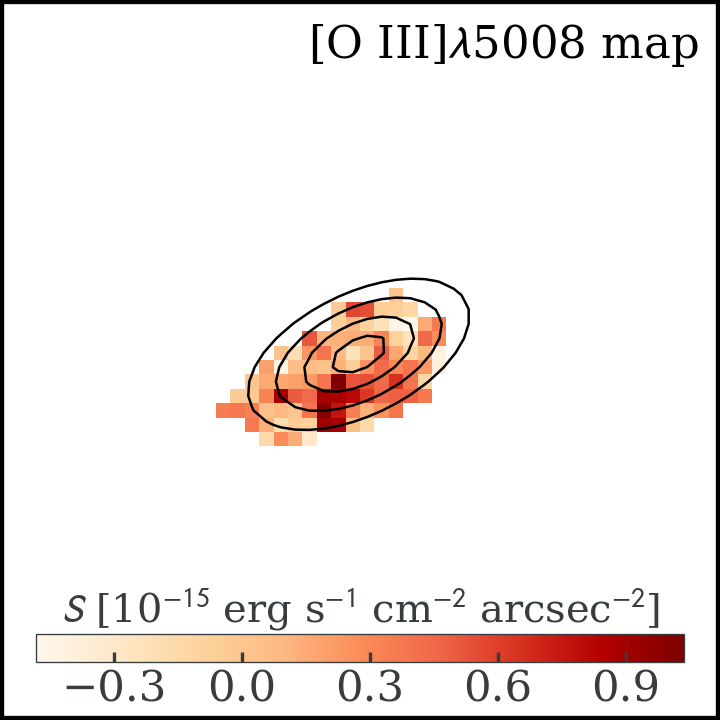}
    \includegraphics[width=.16\textwidth]{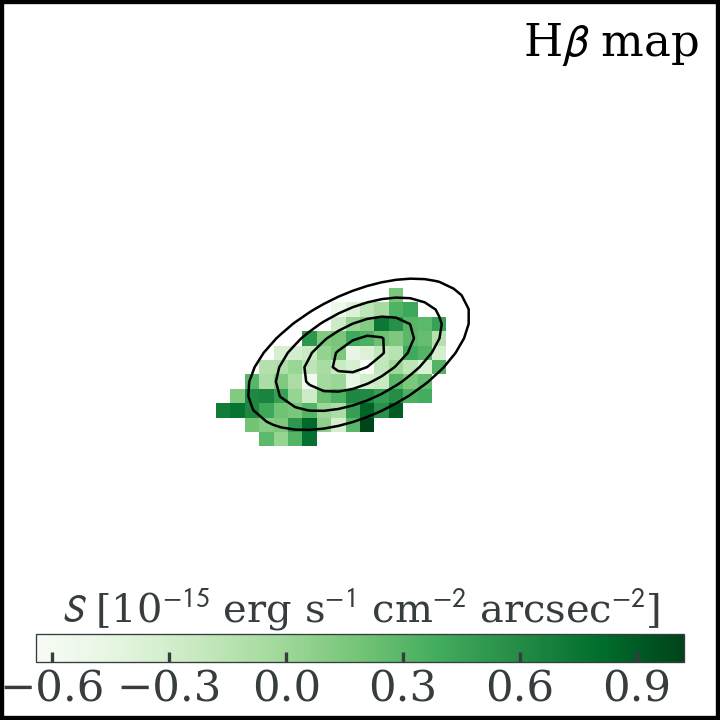}
    \includegraphics[width=.16\textwidth]{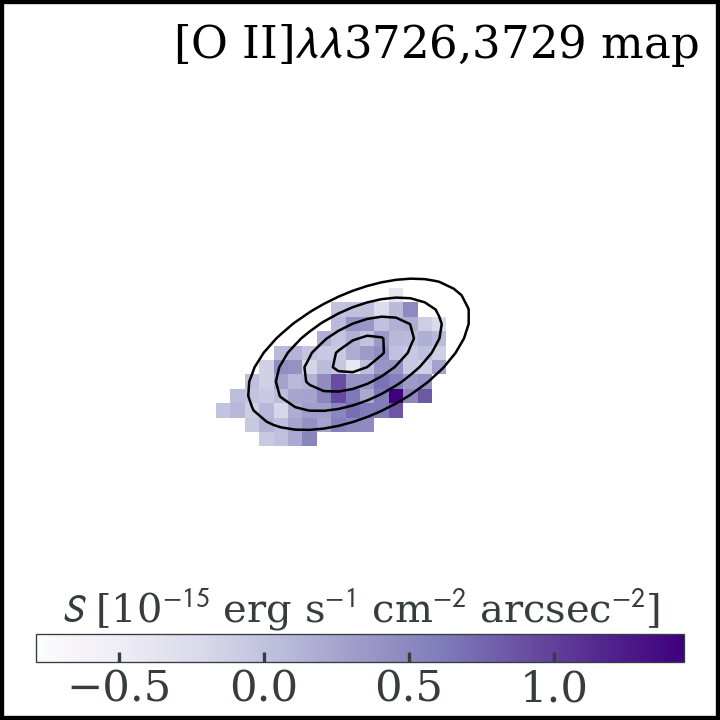}
    \includegraphics[width=.16\textwidth]{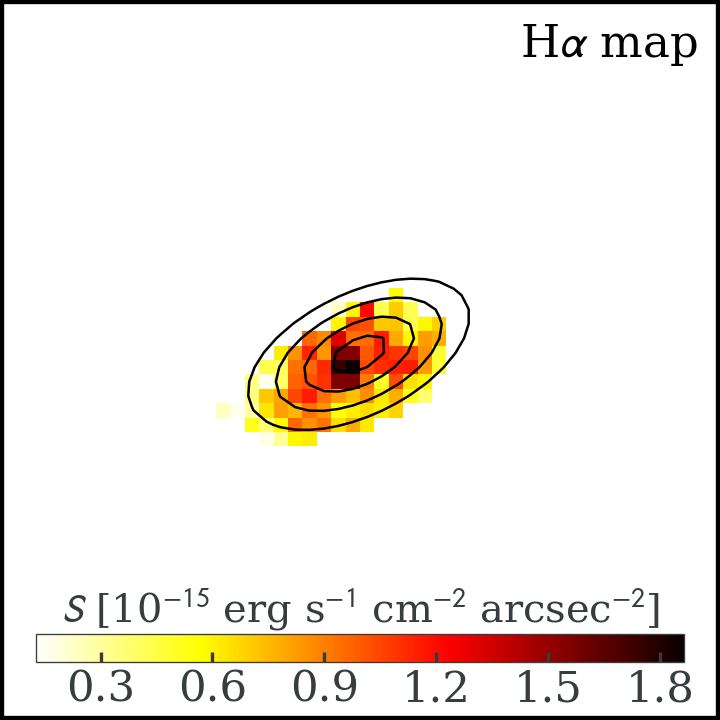}\\
    \includegraphics[width=\textwidth]{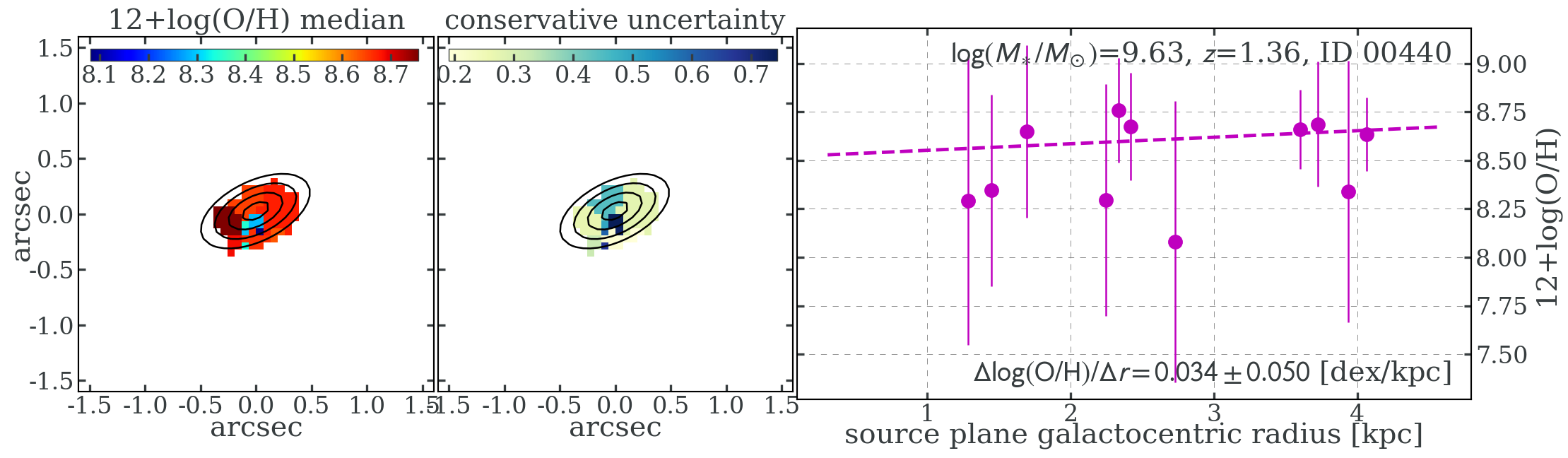}
    \caption{The source ID00440 in the field of \cljiu is shown.}
    \label{fig:clM2129_ID00440_figs}
\end{figure*}

\begin{figure*}
    \centering
    \includegraphics[width=\textwidth]{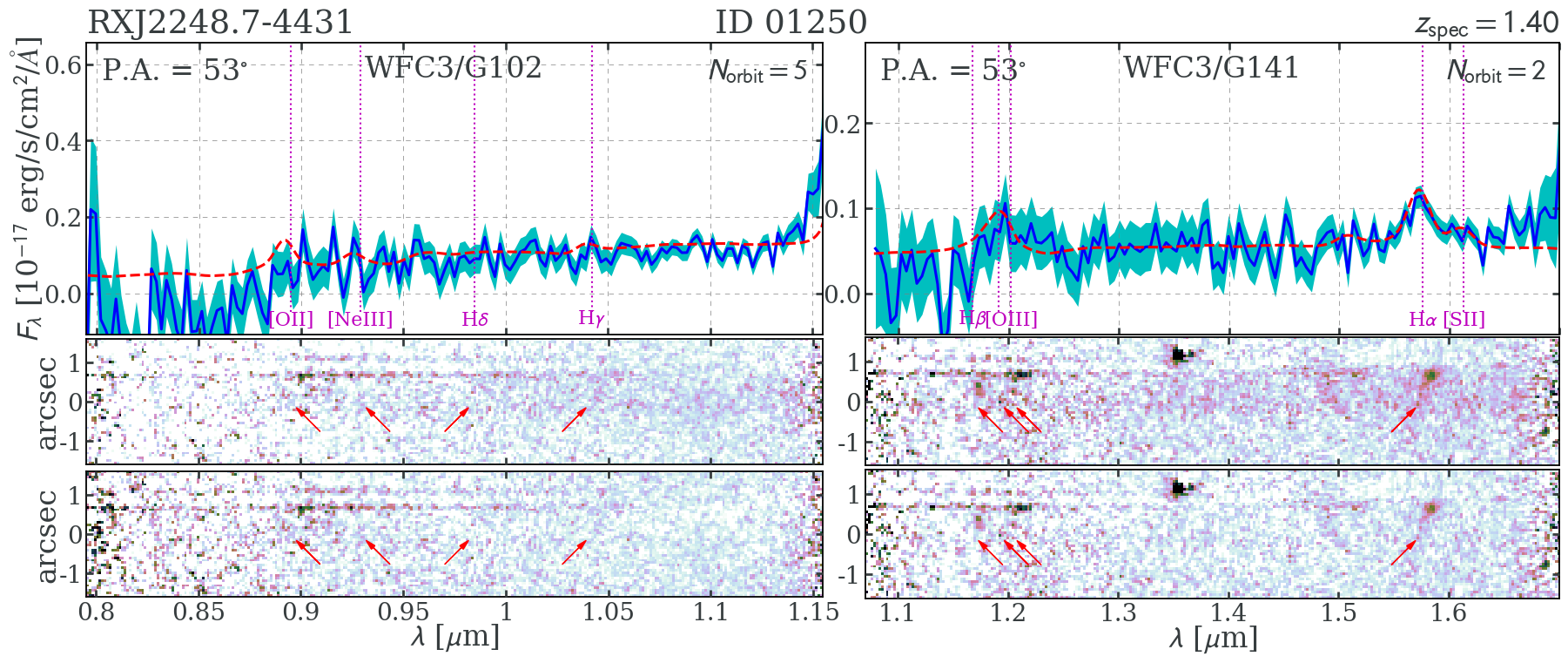}\\
    \includegraphics[width=\textwidth]{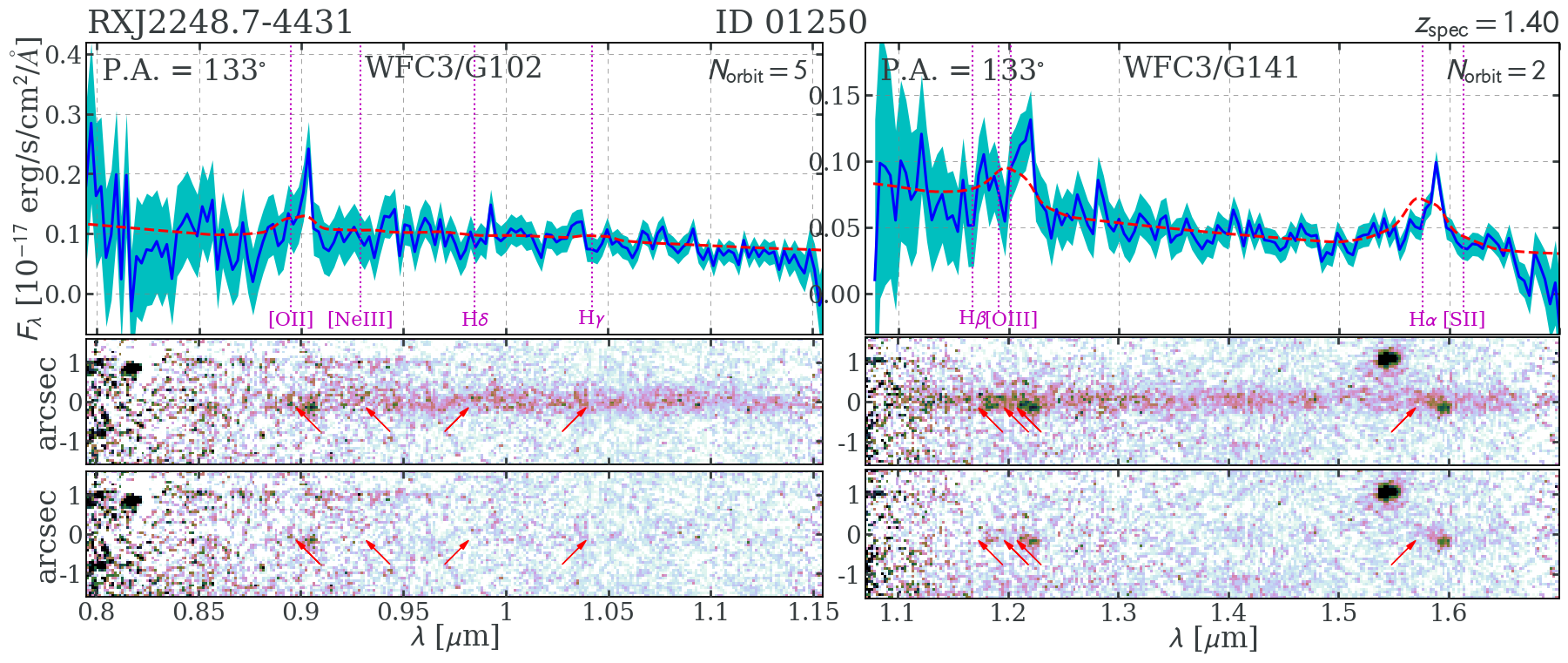}\\
    \includegraphics[width=.16\textwidth]{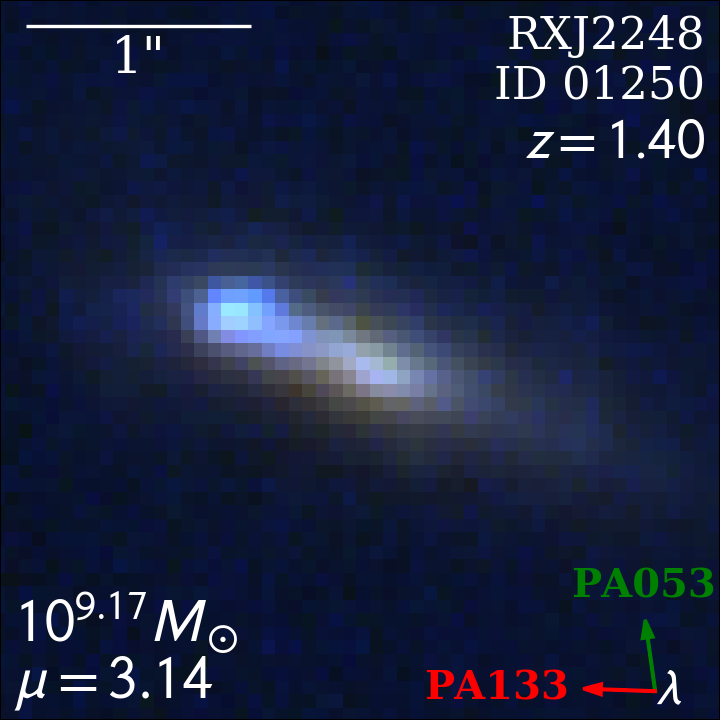}
    \includegraphics[width=.16\textwidth]{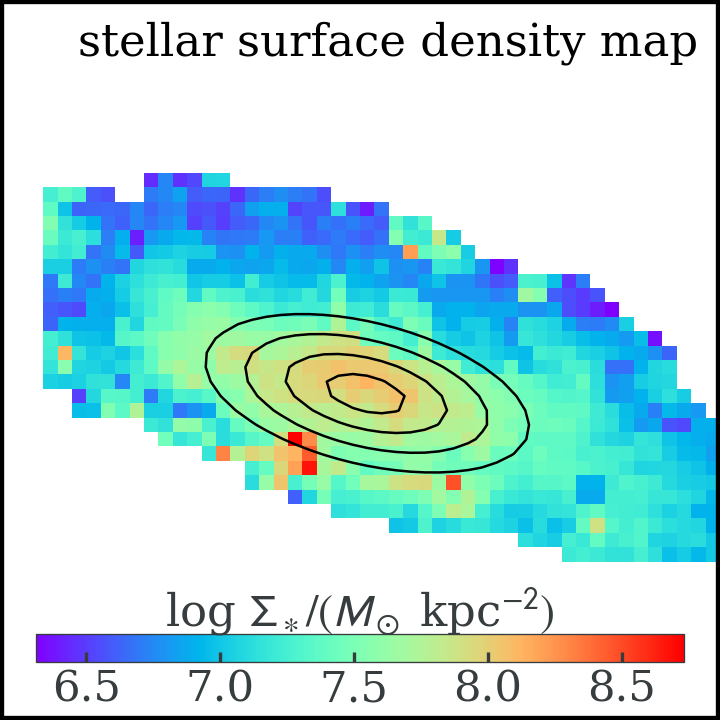}
    \includegraphics[width=.16\textwidth]{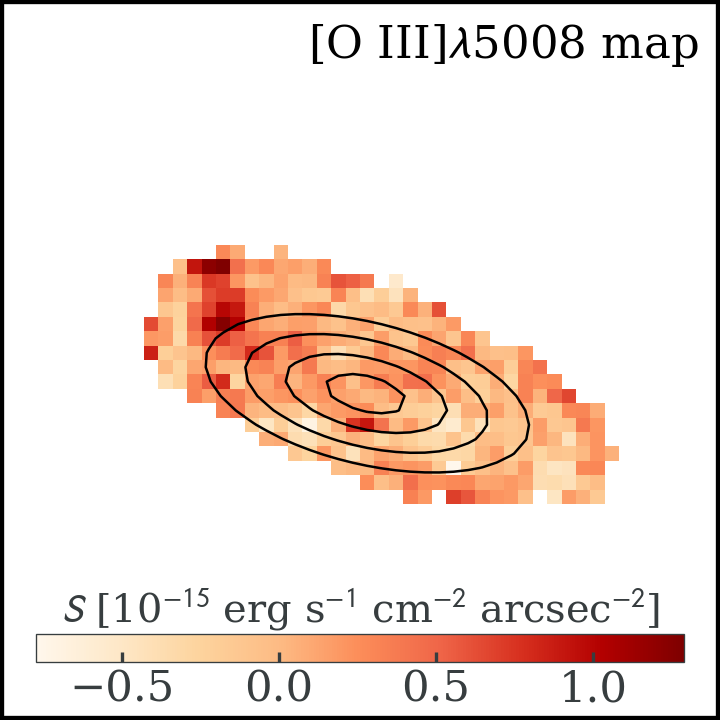}
    \includegraphics[width=.16\textwidth]{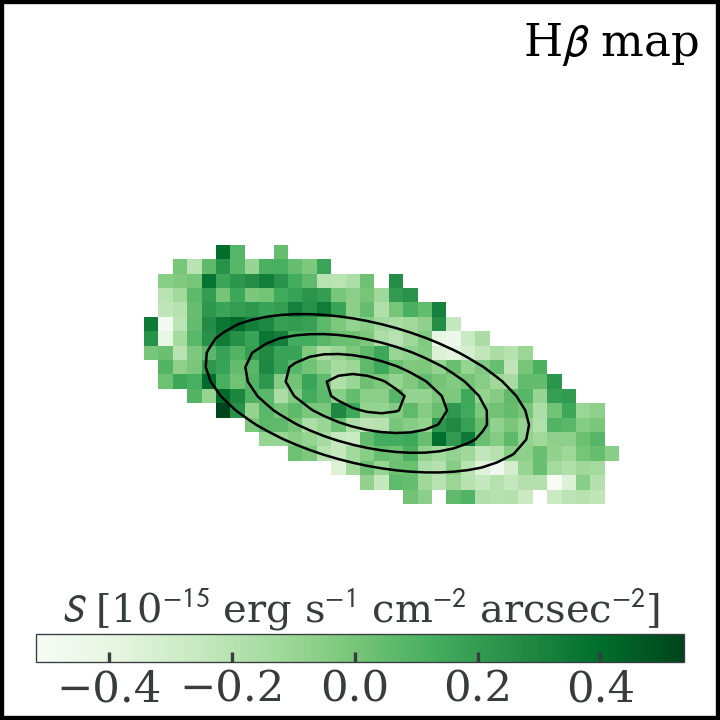}
    \includegraphics[width=.16\textwidth]{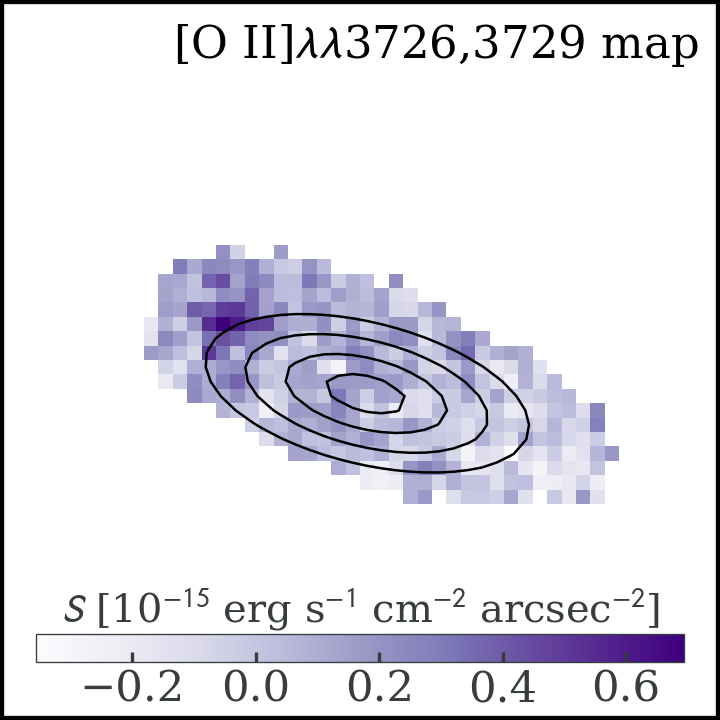}
    \includegraphics[width=.16\textwidth]{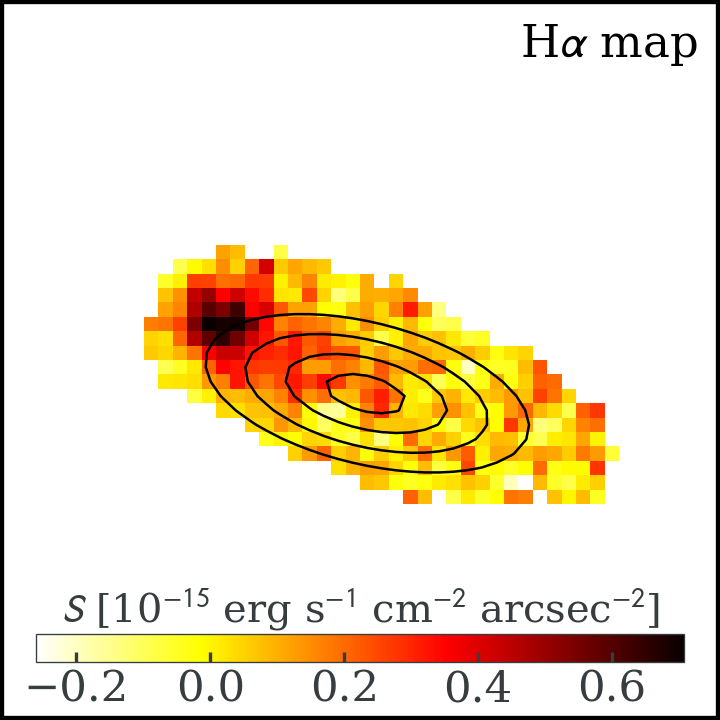}\\
    \includegraphics[width=\textwidth]{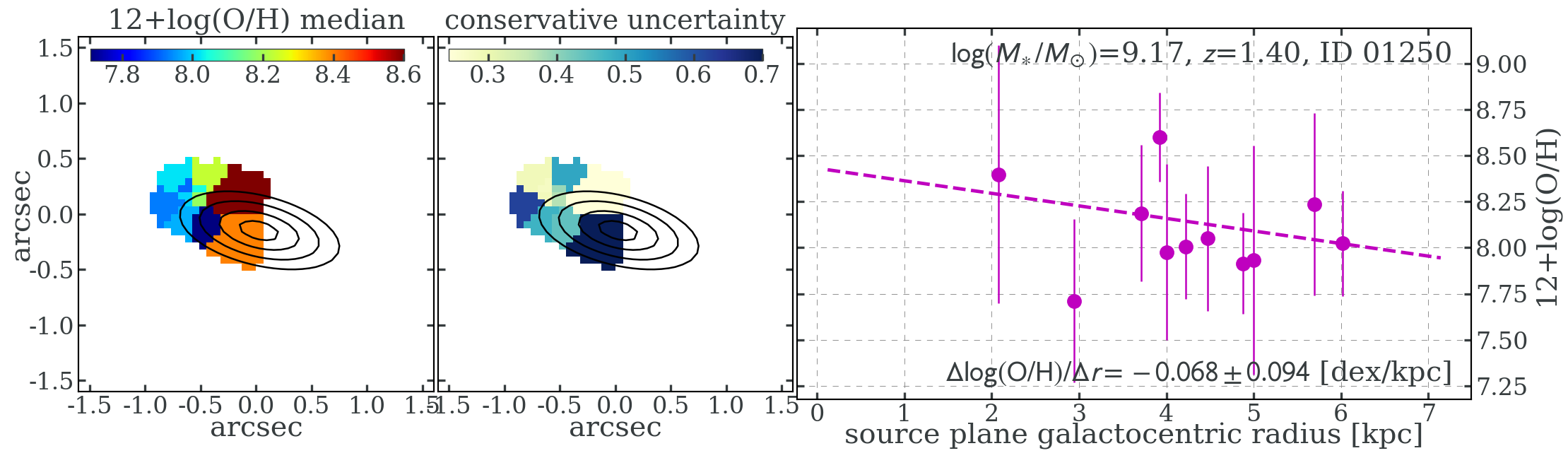}
    \caption{The source ID01250 in the field of \clliu is shown.}
    \label{fig:clRXJ2248_ID01250_figs}
\end{figure*}

In Figures~\ref{fig:clA370_ID01513_figs} through \ref{fig:clRXJ2248_ID01250_figs}, 
we present the source 1D/2D grism spectra, color-composite image, stellar surface density stamp, and EL
maps, as well as metallicity map and radial metallicity gradient measurements for the entire sample.
Following the conventions adopted in Figs.~\ref{fig:ID00955spec} and \ref{fig:combEL_metalgrad}, we first show
the 1D and 2D G102-G141 spectra, at two separate P.A.s.  Note that due to grism defect and/or falling outside
WFC3 FoV, some sources (\ie A2744-ID00144, A2744-ID01897, RXJ1347-ID00664) only have coverage from one of the
orients. Beneath the spectra, we show the source 2D stamps. For sources at $z\lesssim1.6$, we display their
\Ha map, whereas for sources at higher redshifts, \Hg map is shown instead, since \Ha has already redshifted
out of the wavelength coverage of \hst grisms. The bottom panels show the metallicity map and radial gradient
measurements.

\clearpage

\clearpage

\clearpage

\clearpage

\clearpage

\clearpage

\clearpage

\clearpage

\clearpage

\clearpage

\clearpage

\clearpage

\clearpage

\clearpage

\clearpage

\clearpage

\clearpage

\clearpage

\clearpage

\clearpage

\clearpage

\clearpage

\clearpage

\clearpage

\clearpage

\clearpage

\clearpage

\clearpage

\clearpage

\clearpage

\clearpage

\clearpage

\clearpage

\clearpage

\clearpage

\clearpage

\clearpage

\clearpage

\clearpage

\clearpage

\clearpage

\clearpage

\clearpage

\clearpage

\clearpage

\clearpage

\clearpage

\clearpage

\clearpage

\clearpage

\clearpage

\begin{figure*}
    \centering
    \includegraphics[width=\textwidth]{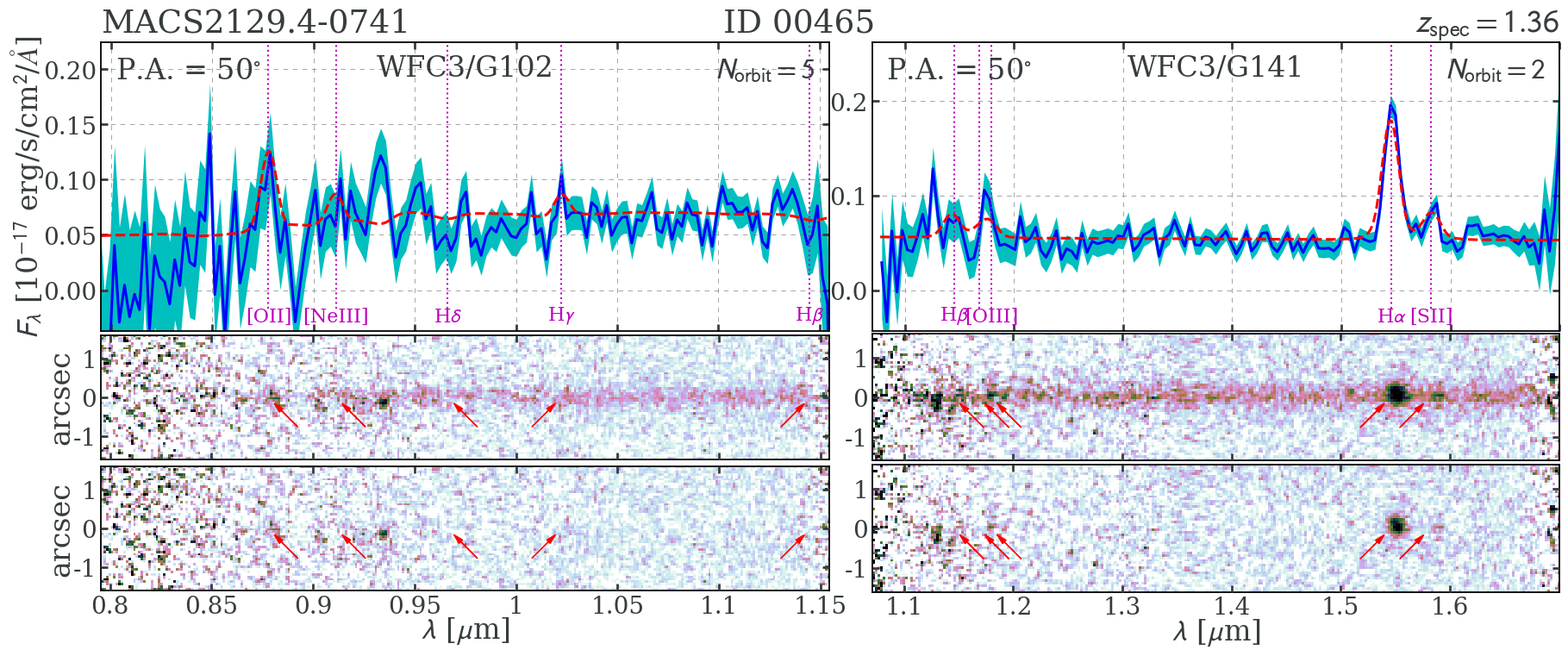}\\
    \includegraphics[width=\textwidth]{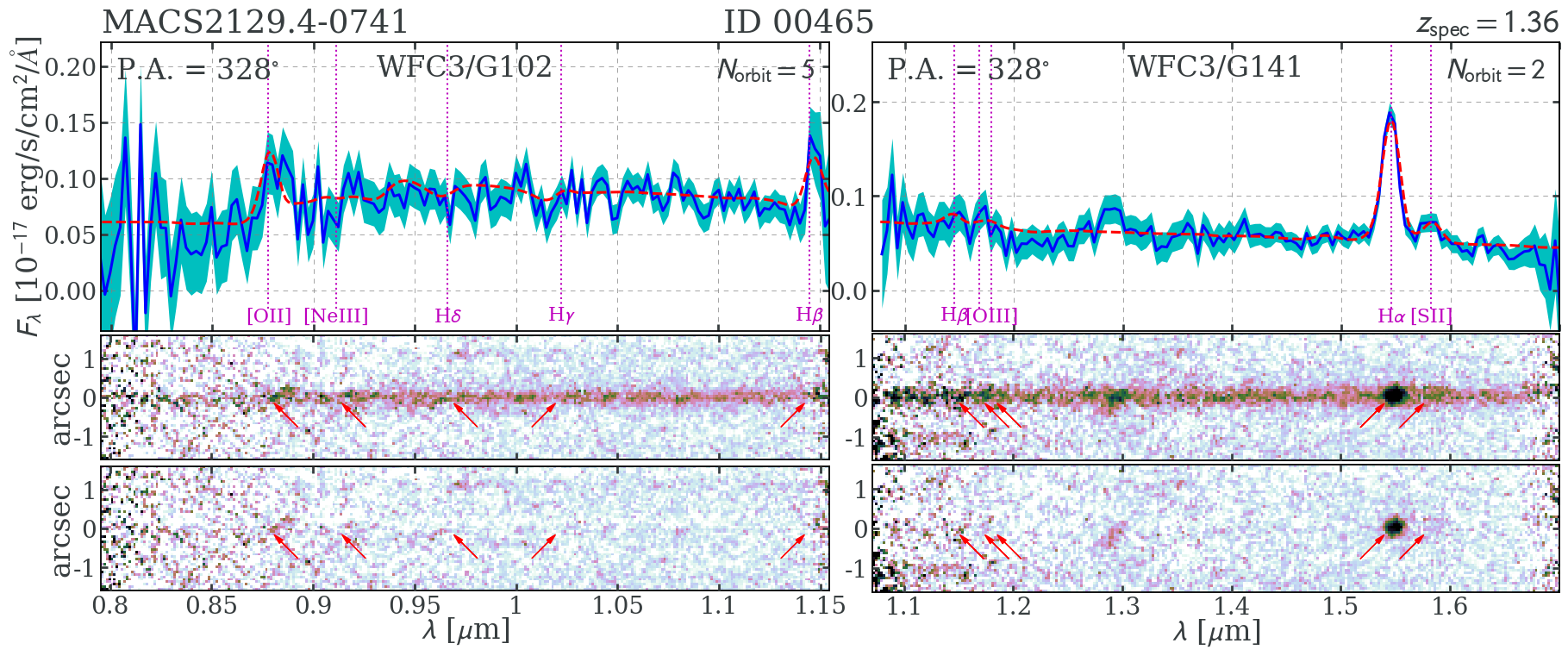}\\
    \includegraphics[width=.16\textwidth]{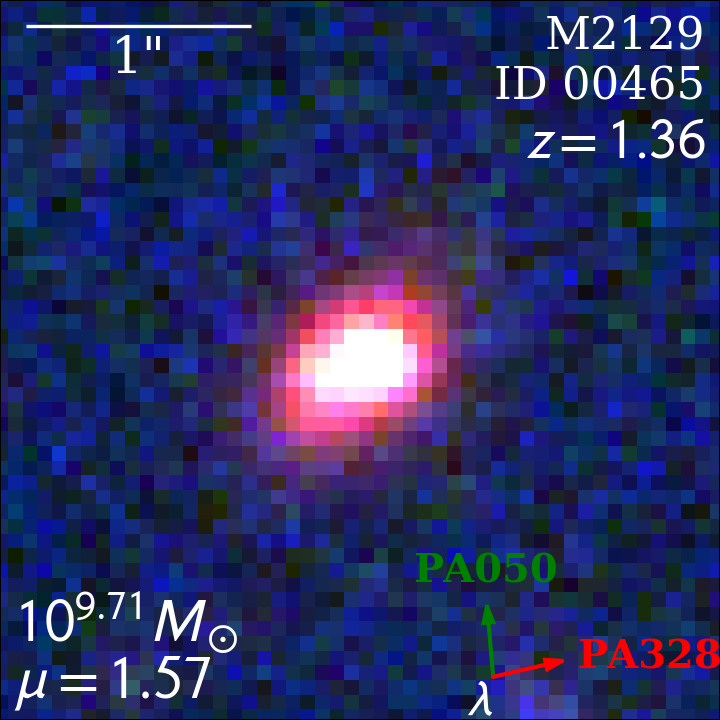}
    \includegraphics[width=.16\textwidth]{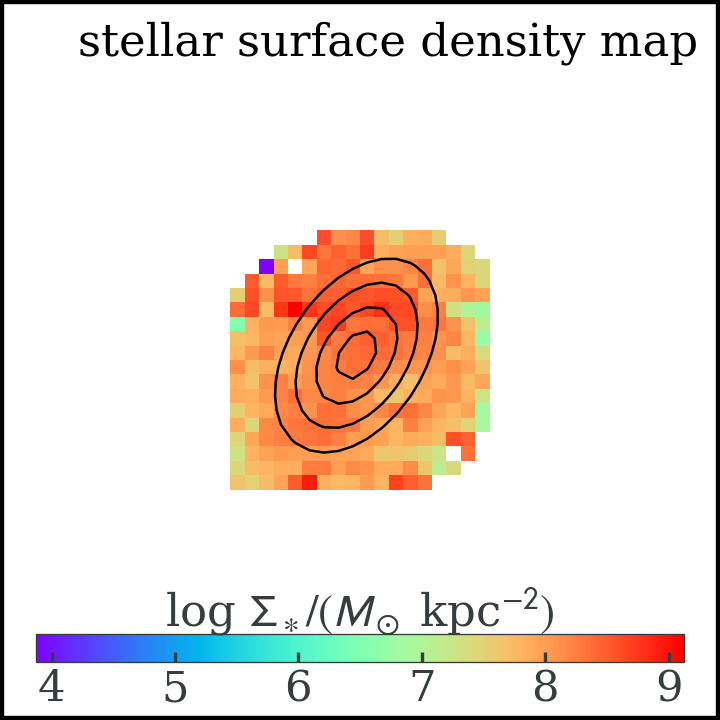}
    \includegraphics[width=.16\textwidth]{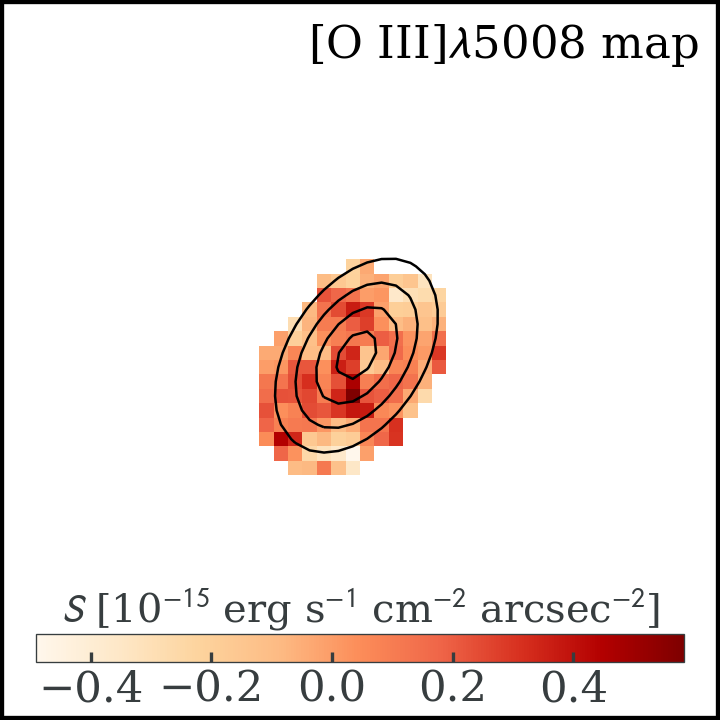}
    \includegraphics[width=.16\textwidth]{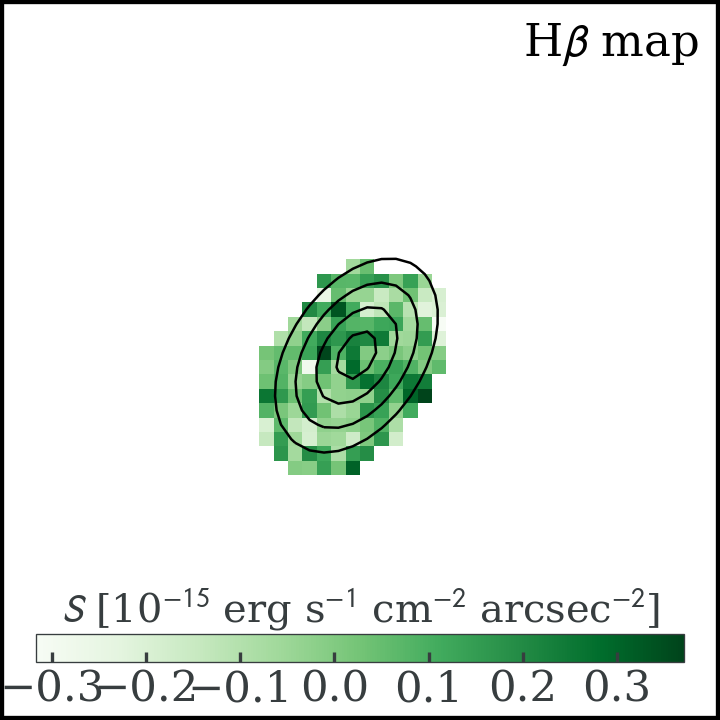}
    \includegraphics[width=.16\textwidth]{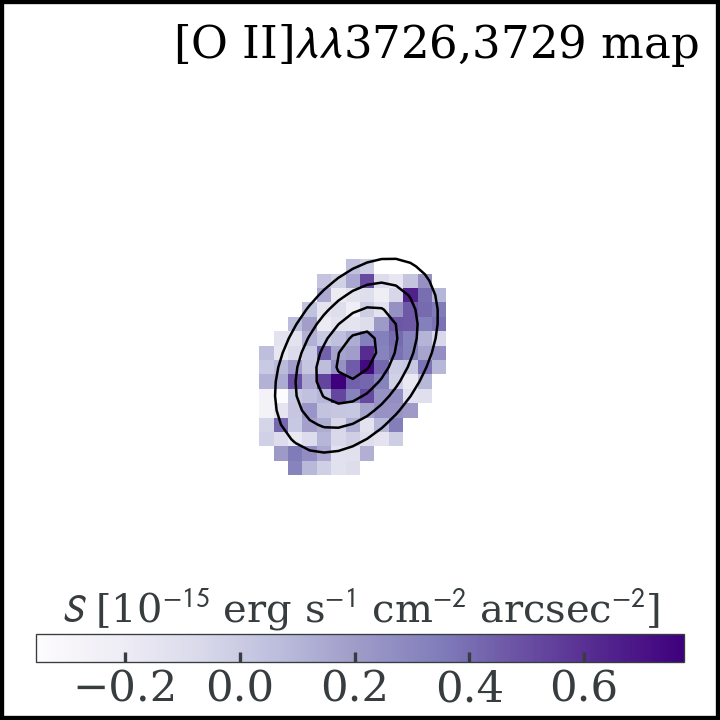}
    \includegraphics[width=.16\textwidth]{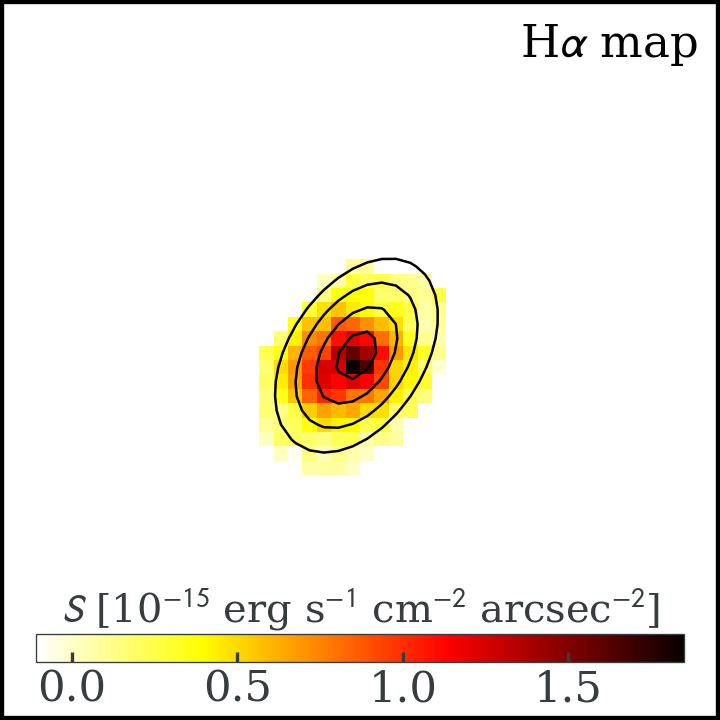}\\
    \includegraphics[width=\textwidth]{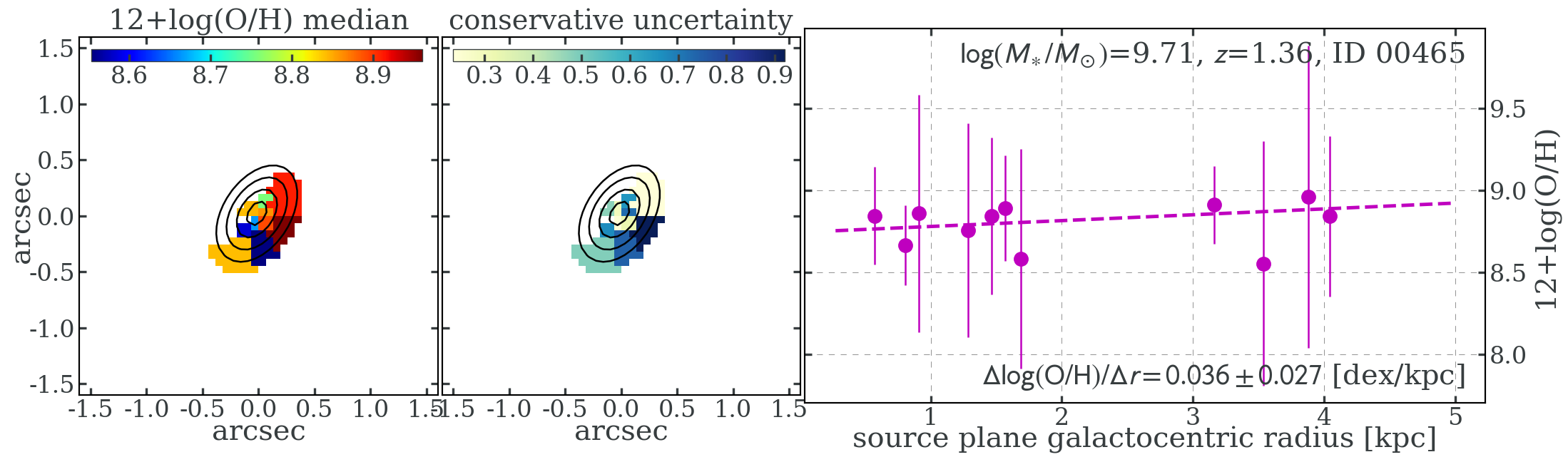}
    \caption{The source ID00465 in the field of \cljiu is shown.}
    \label{fig:clM2129_ID00465_figs}
\end{figure*}
\clearpage

\begin{figure*}
    \centering
    \includegraphics[width=\textwidth]{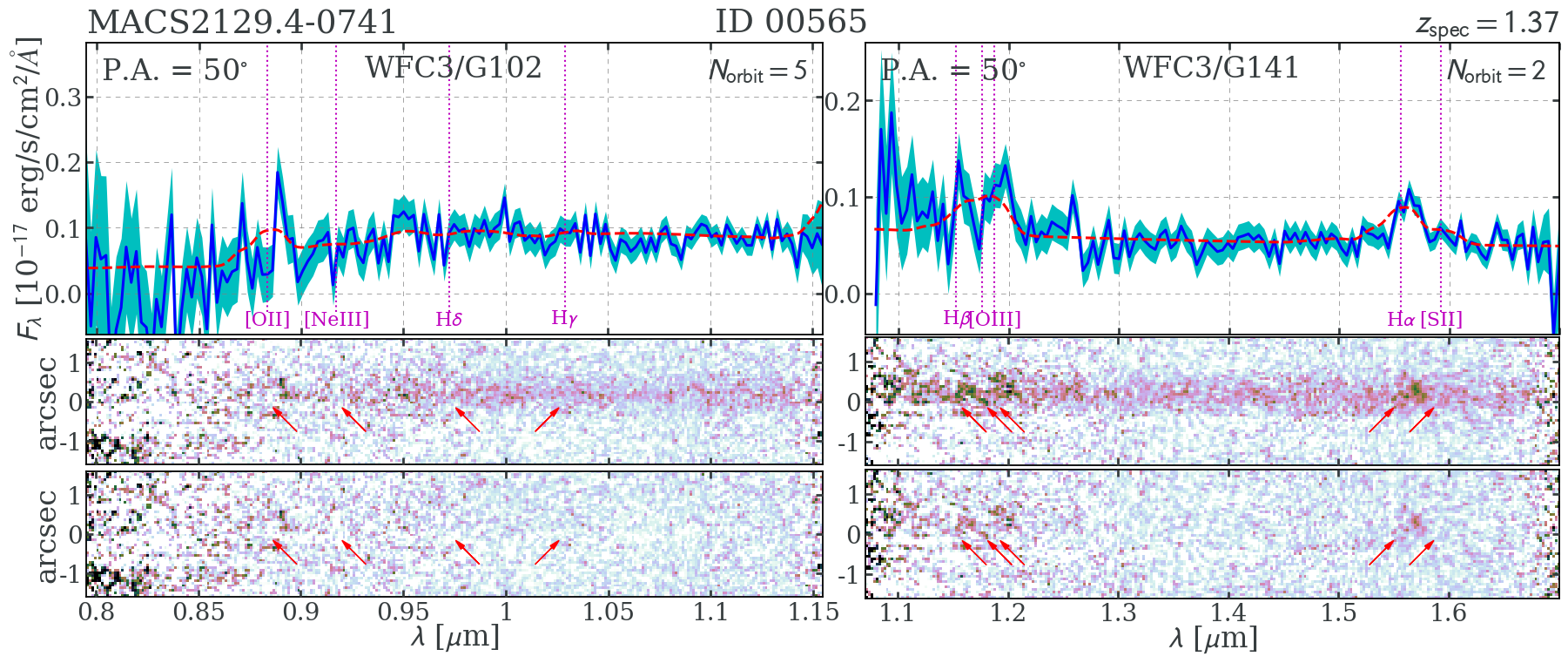}\\
    \includegraphics[width=\textwidth]{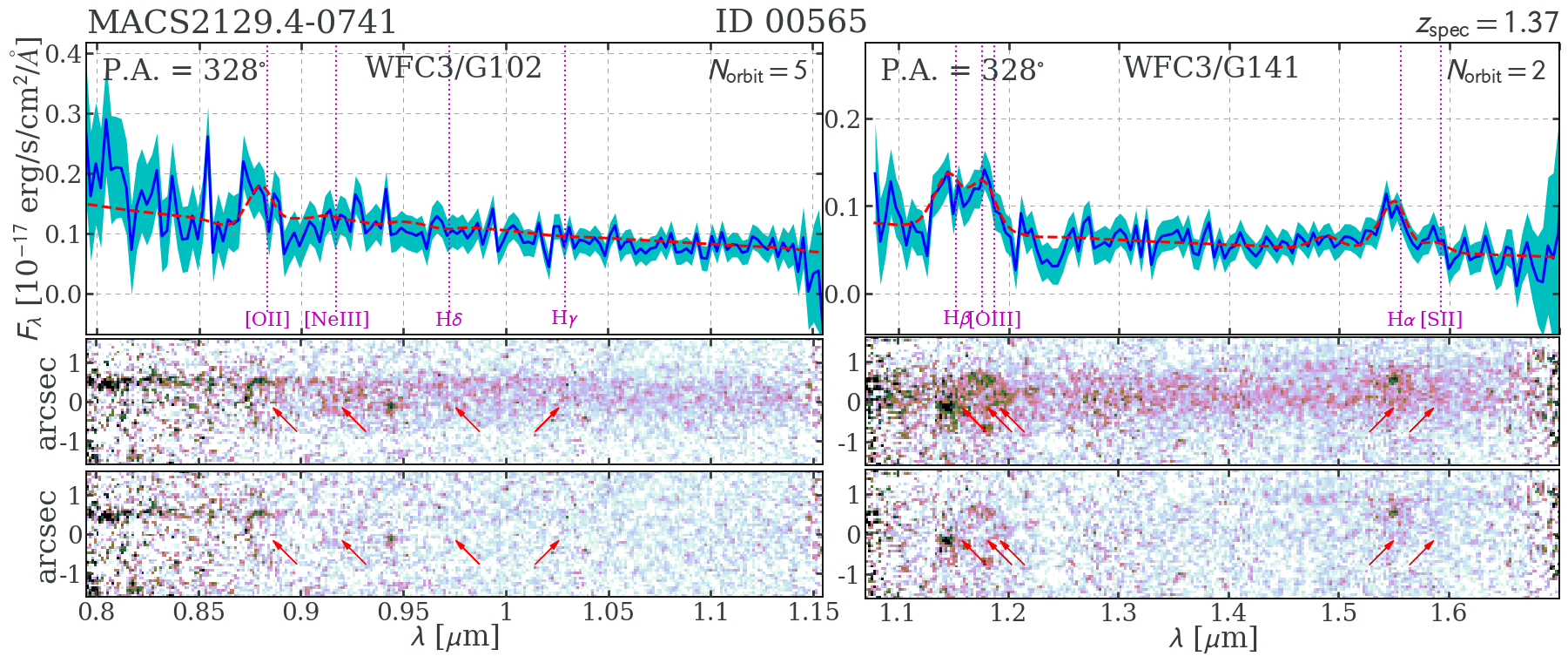}\\
    \includegraphics[width=.16\textwidth]{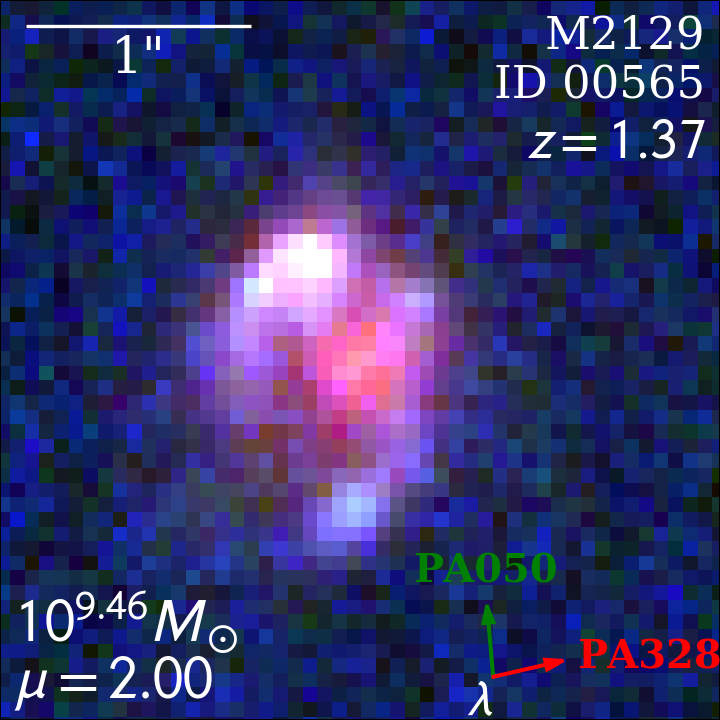}
    \includegraphics[width=.16\textwidth]{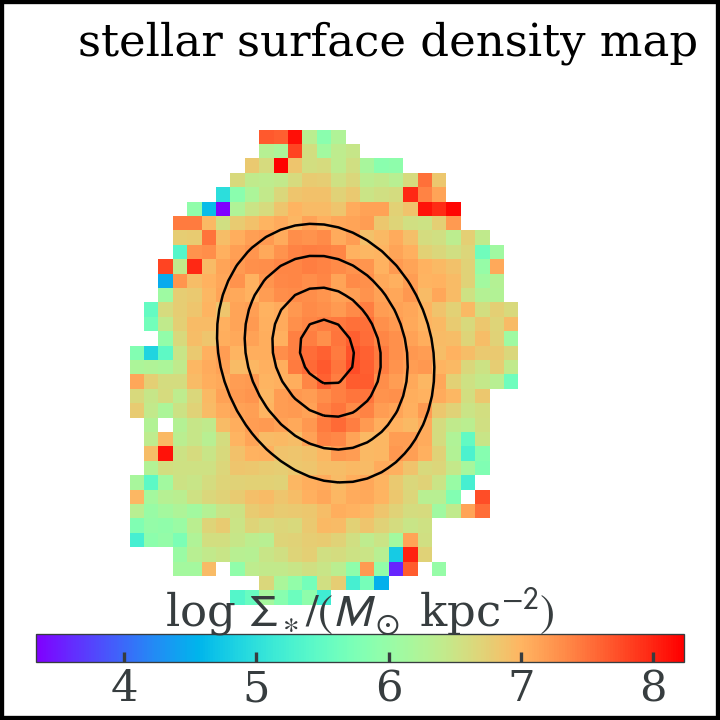}
    \includegraphics[width=.16\textwidth]{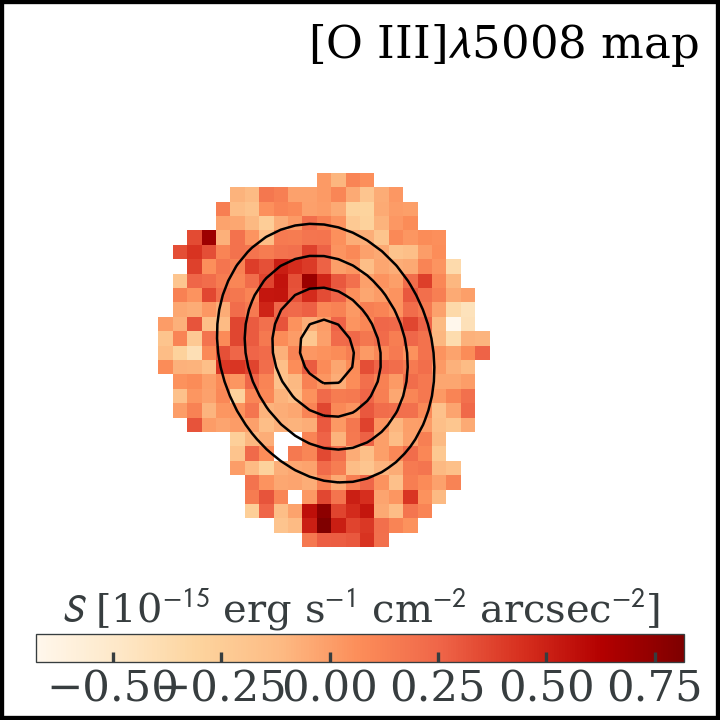}
    \includegraphics[width=.16\textwidth]{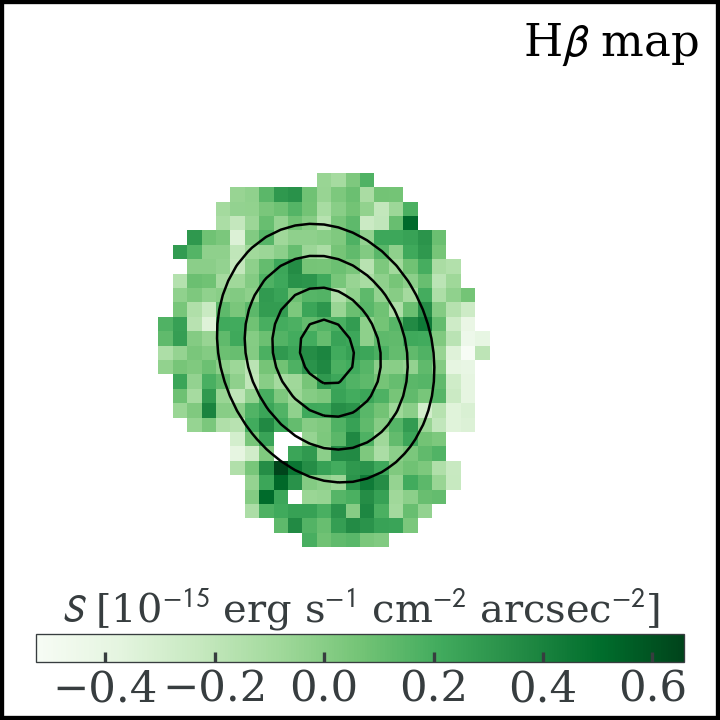}
    \includegraphics[width=.16\textwidth]{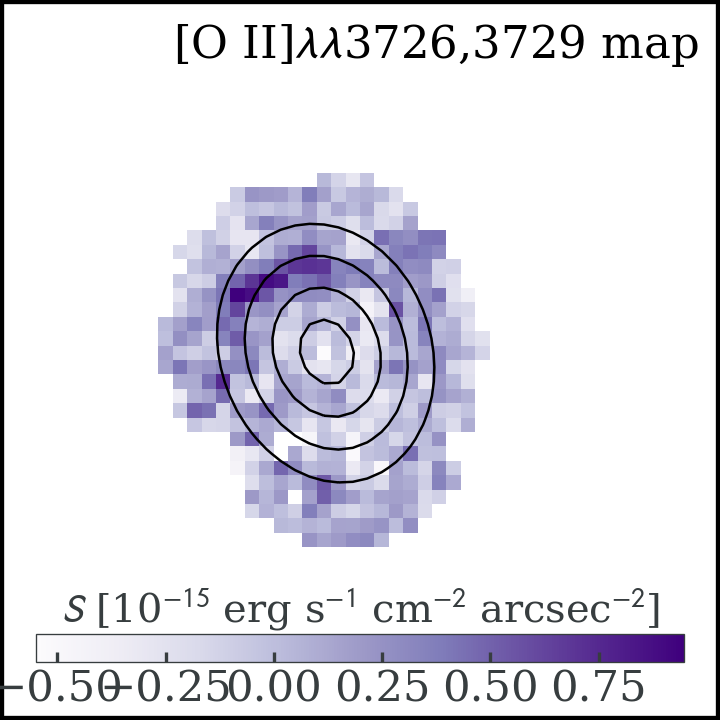}
    \includegraphics[width=.16\textwidth]{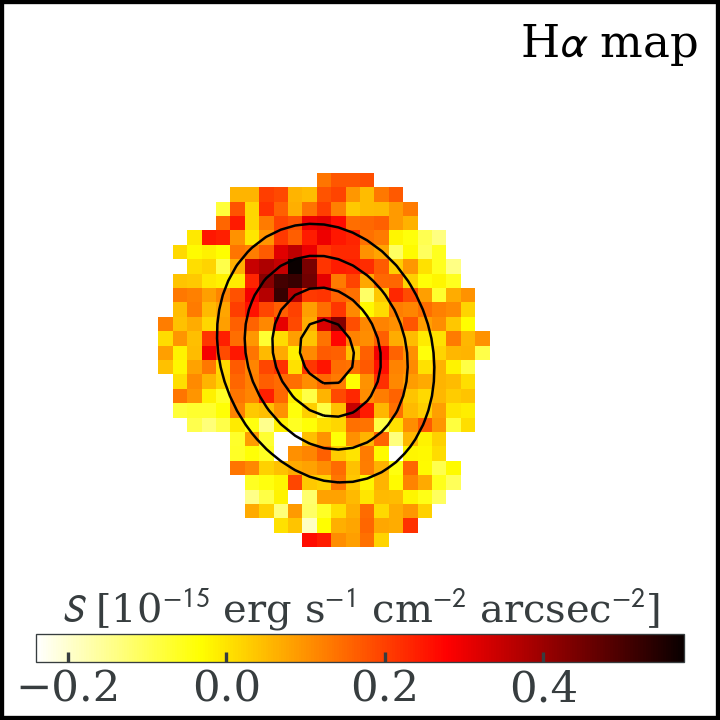}\\
    \includegraphics[width=\textwidth]{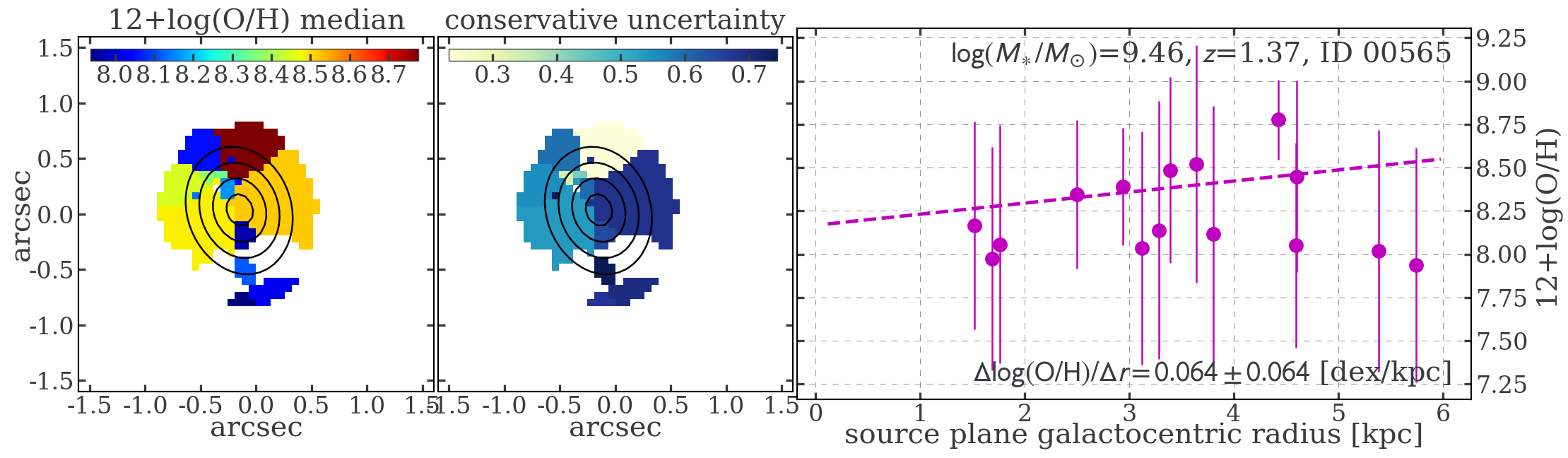}
    \caption{The source ID00565 in the field of \cljiu is shown.}
    \label{fig:clM2129_ID00565_figs}
\end{figure*}
\clearpage

\begin{figure*}
    \centering
    \includegraphics[width=\textwidth]{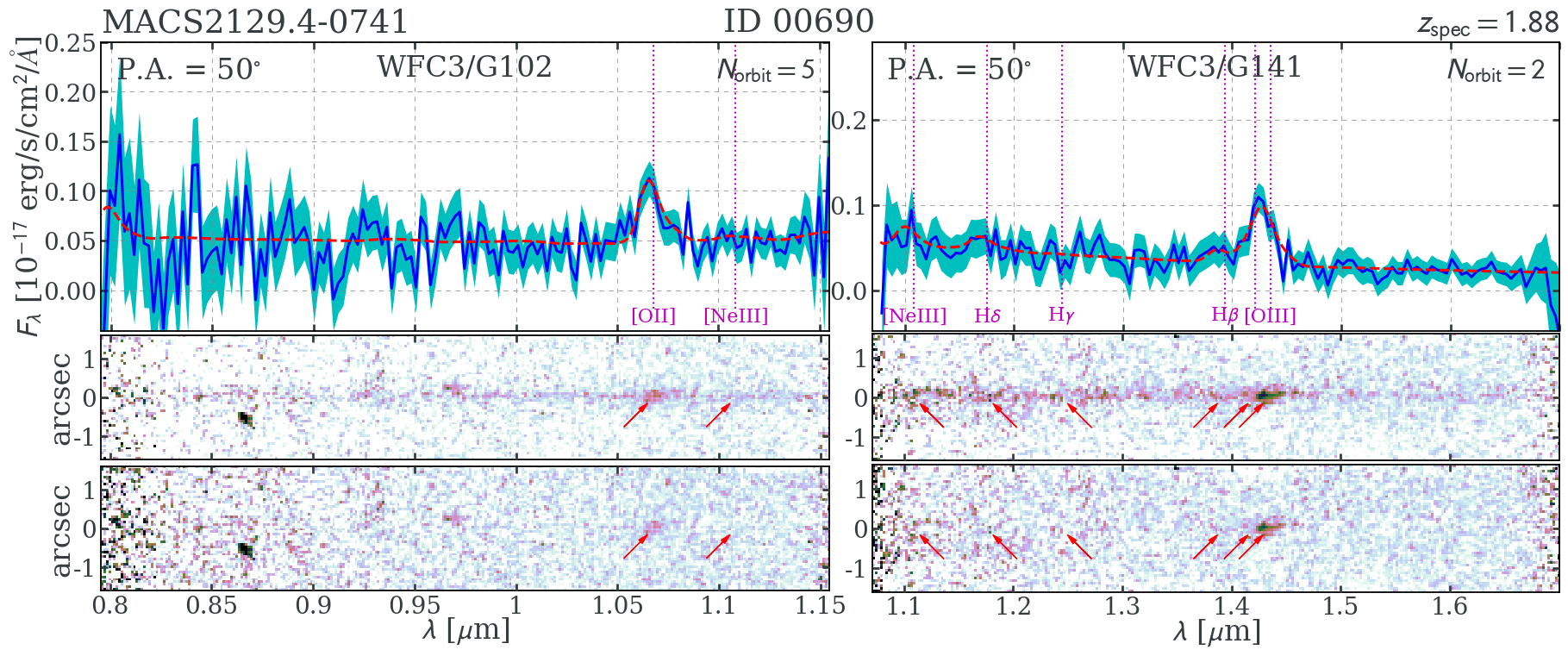}\\
    \includegraphics[width=\textwidth]{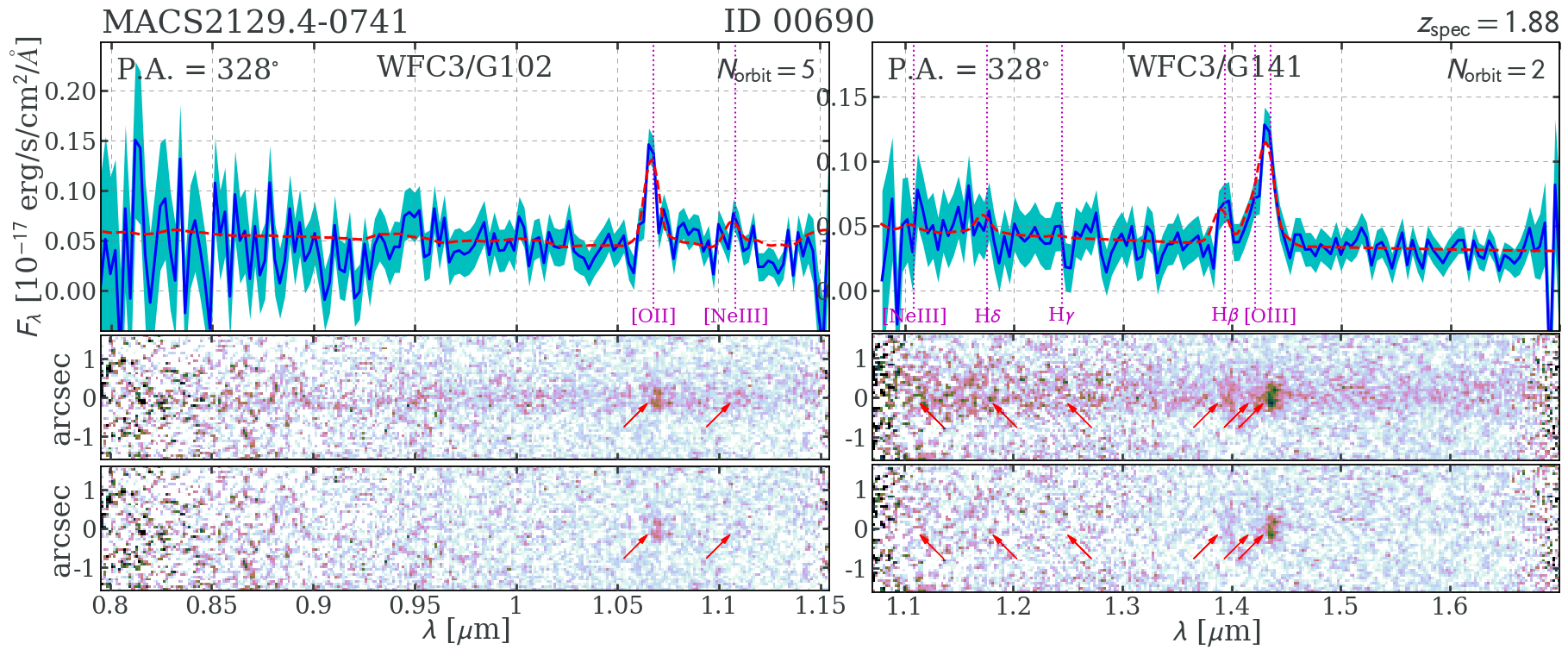}\\
    \includegraphics[width=.16\textwidth]{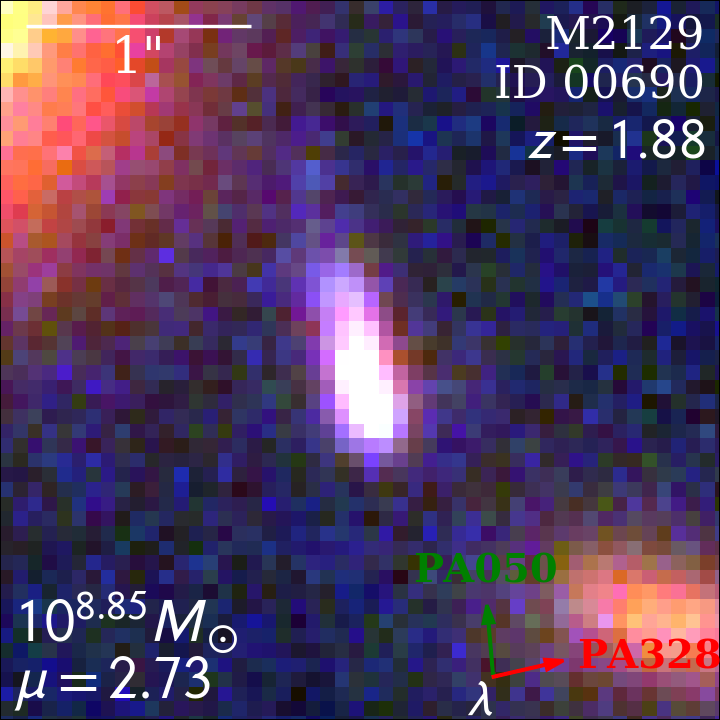}
    \includegraphics[width=.16\textwidth]{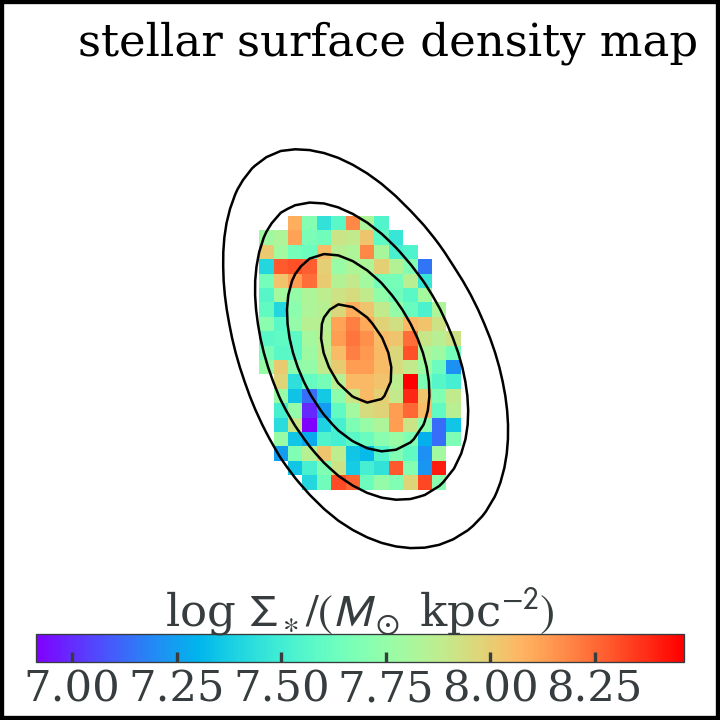}
    \includegraphics[width=.16\textwidth]{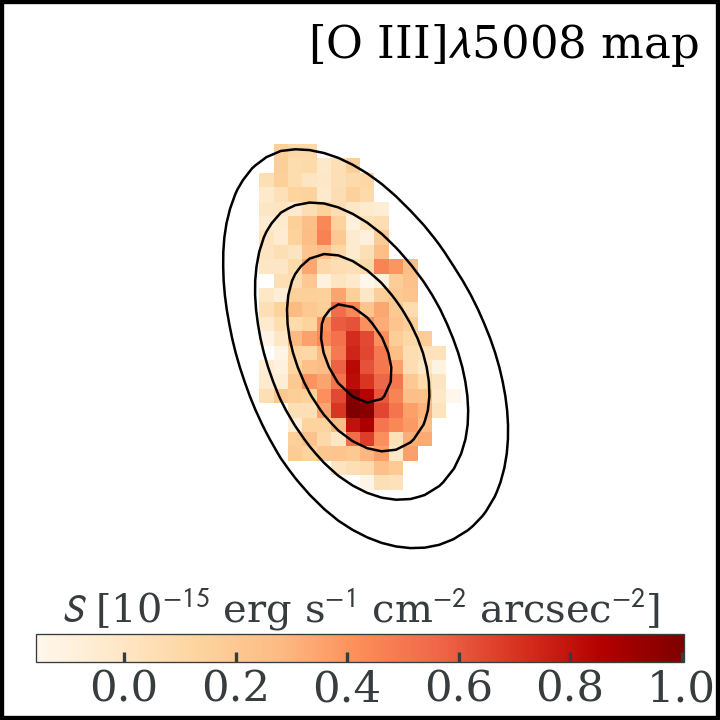}
    \includegraphics[width=.16\textwidth]{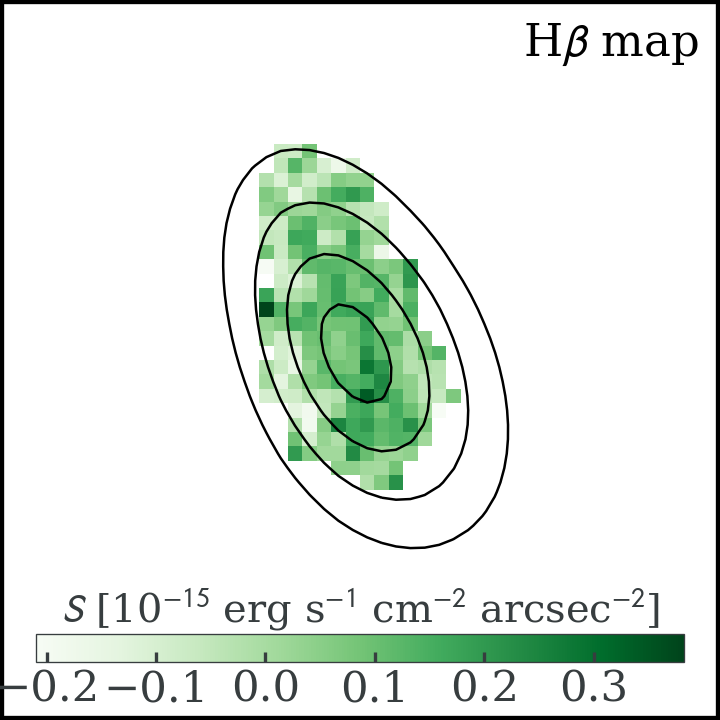}
    \includegraphics[width=.16\textwidth]{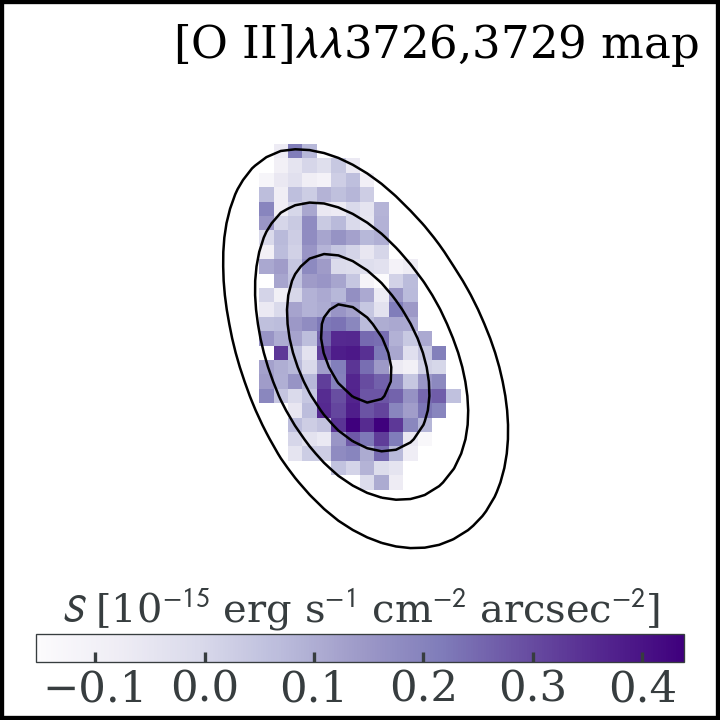}
    \includegraphics[width=.16\textwidth]{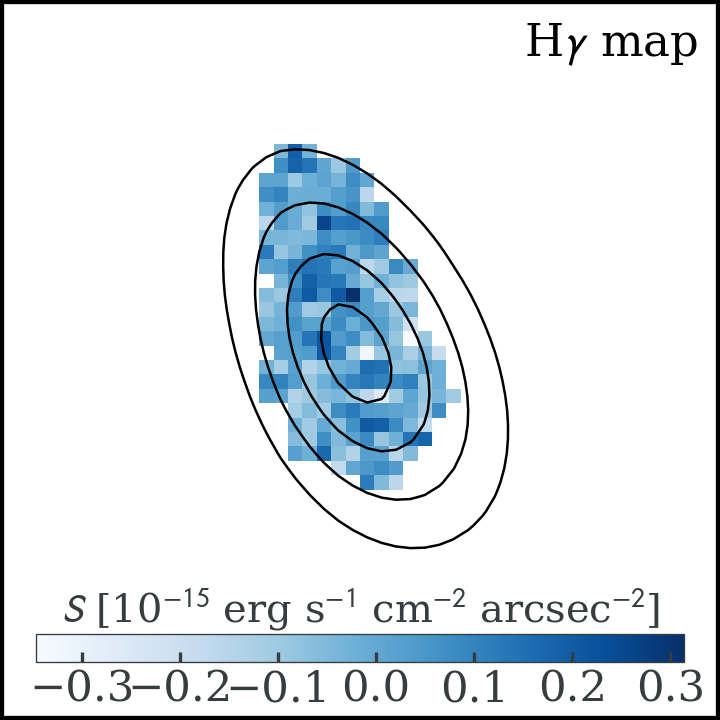}\\
    \includegraphics[width=\textwidth]{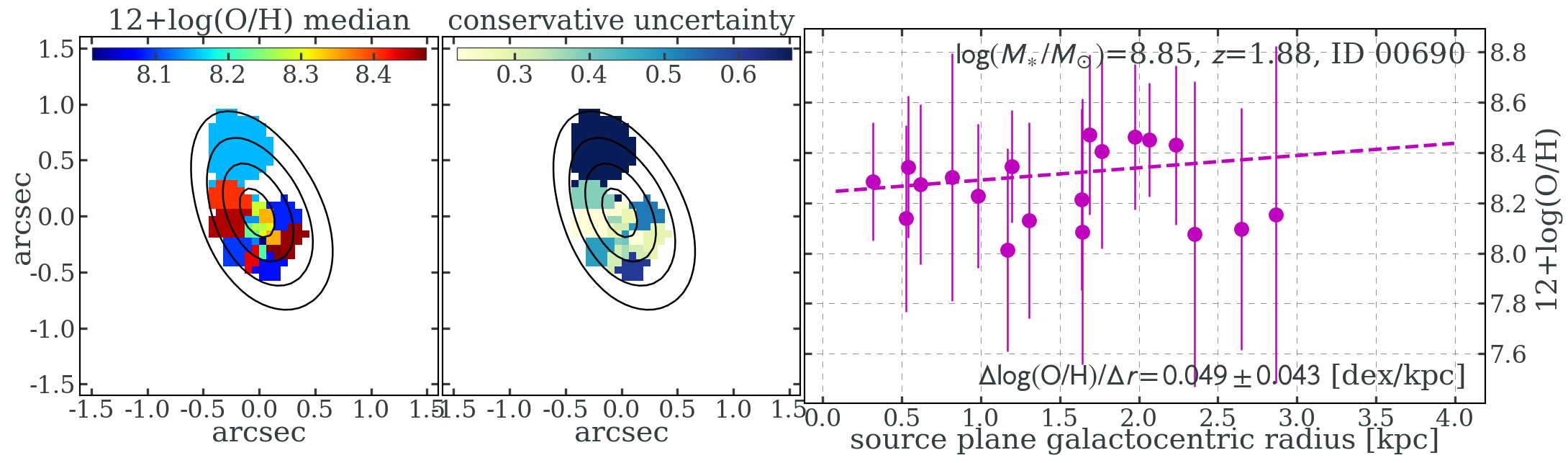}
    \caption{The source ID00690 in the field of \cljiu is shown.}
    \label{fig:clM2129_ID00690_figs}
\end{figure*}
\clearpage

\begin{figure*}
    \centering
    \includegraphics[width=\textwidth]{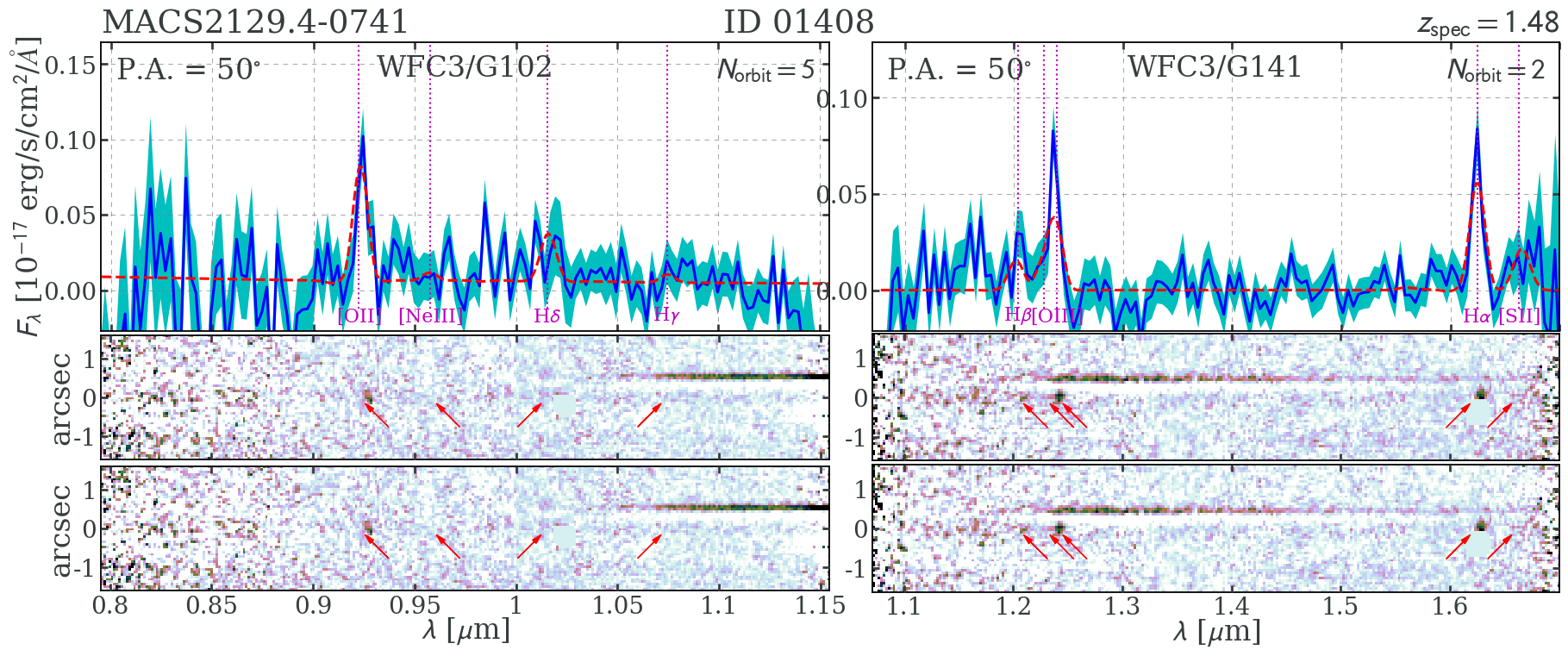}\\
    \includegraphics[width=\textwidth]{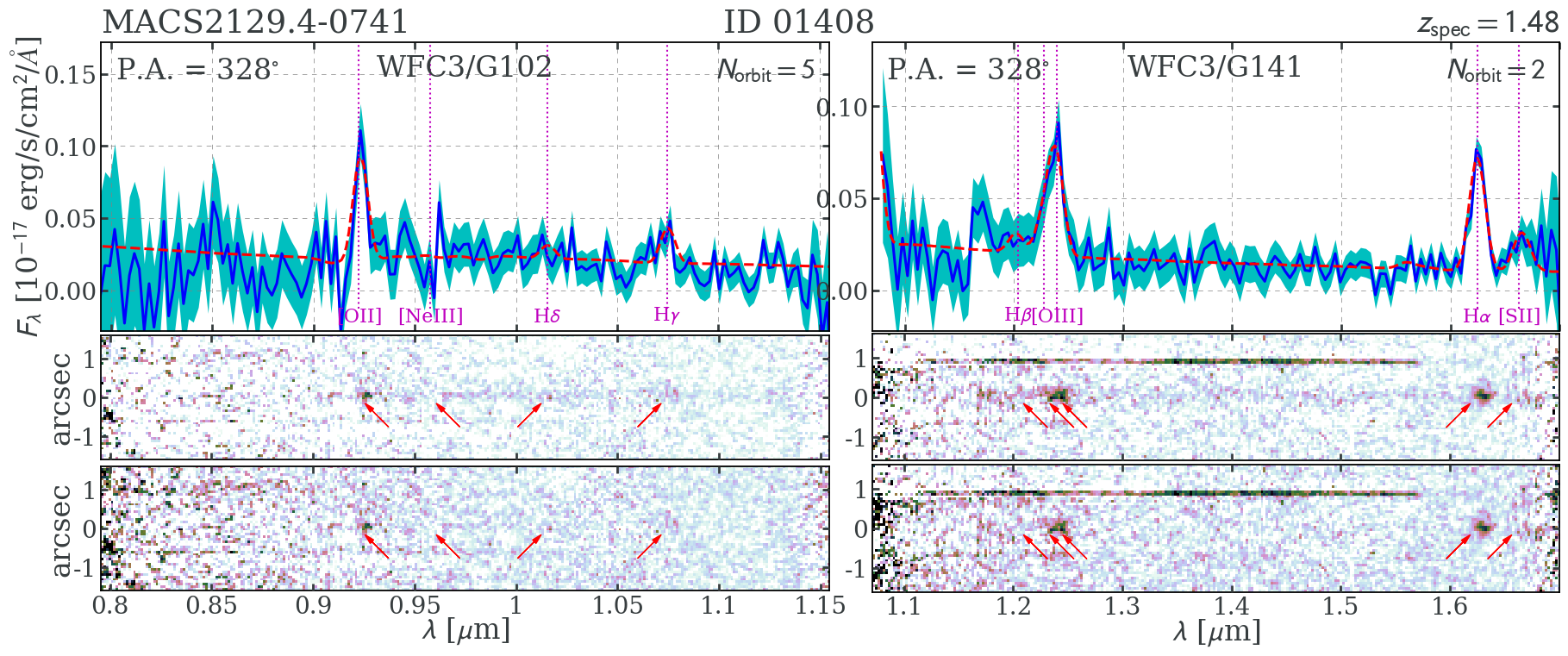}\\
    \includegraphics[width=.16\textwidth]{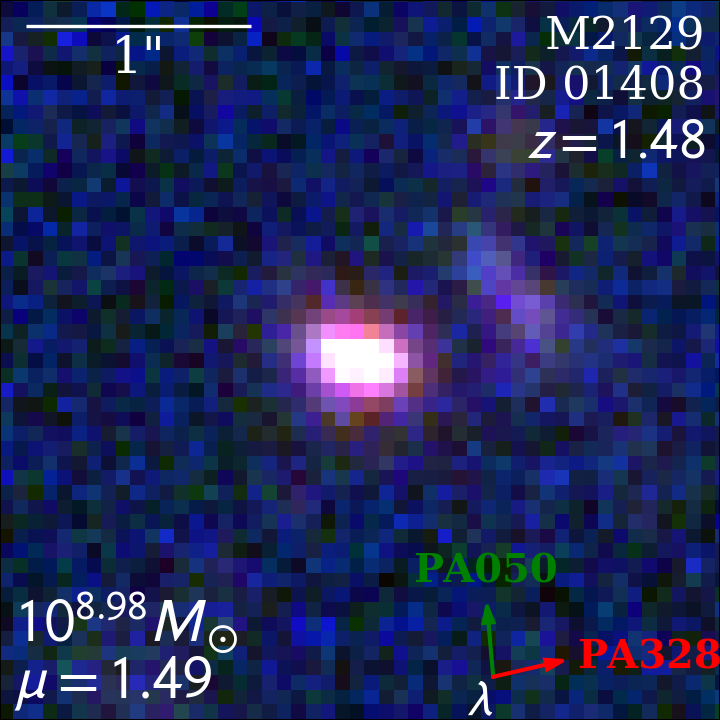}
    \includegraphics[width=.16\textwidth]{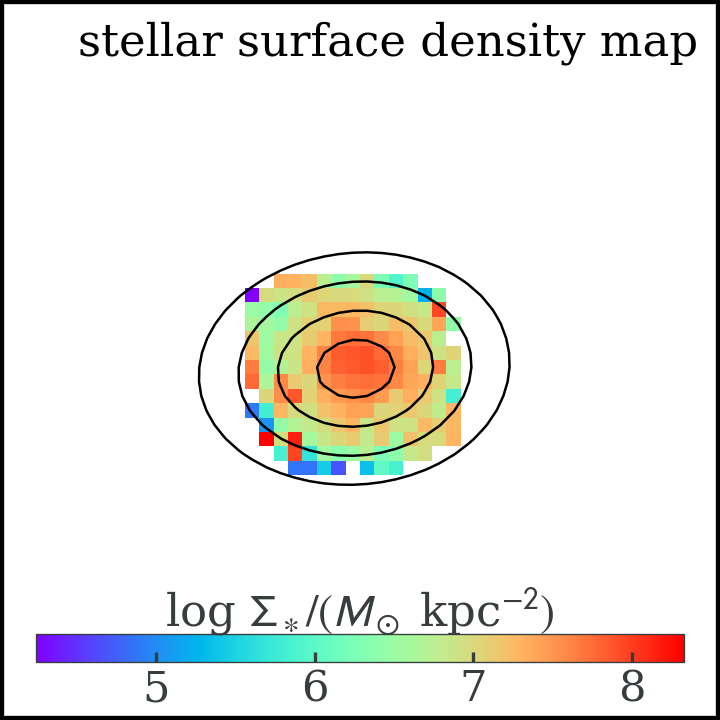}
    \includegraphics[width=.16\textwidth]{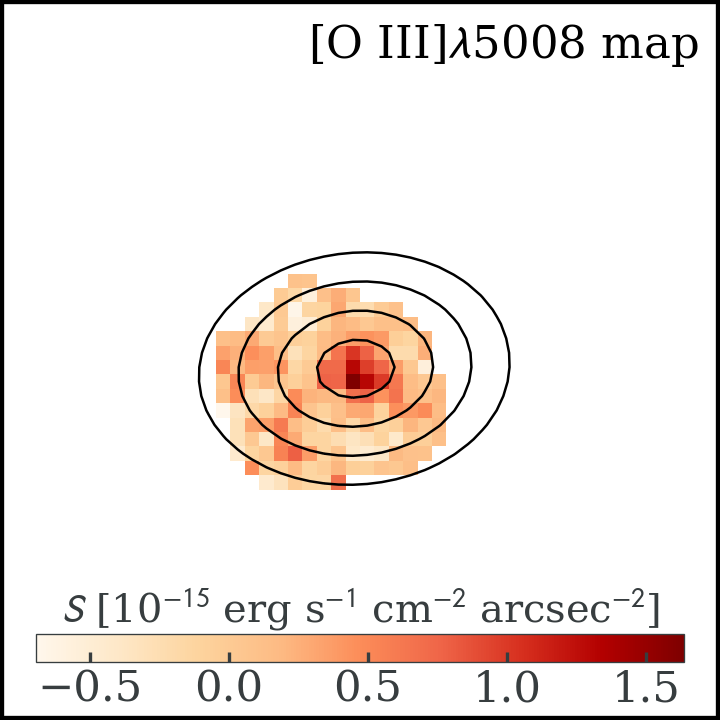}
    \includegraphics[width=.16\textwidth]{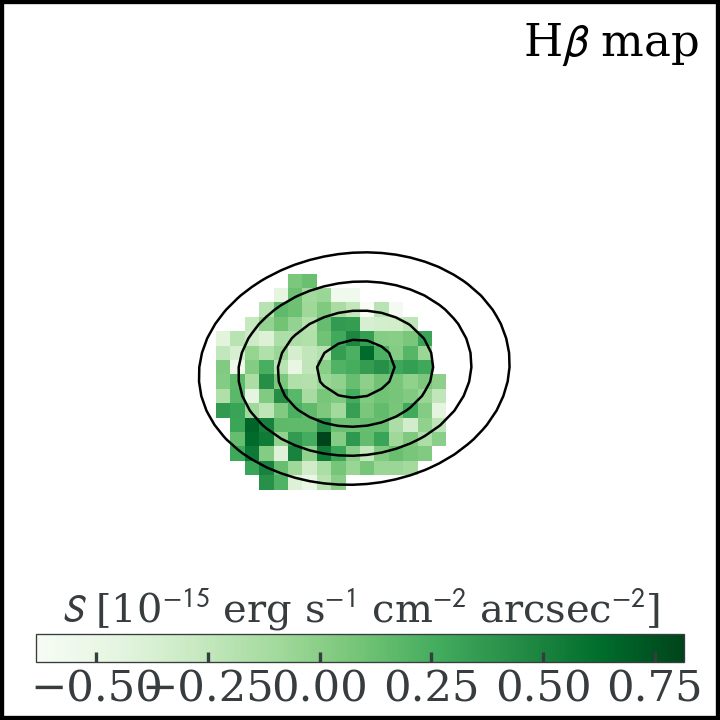}
    \includegraphics[width=.16\textwidth]{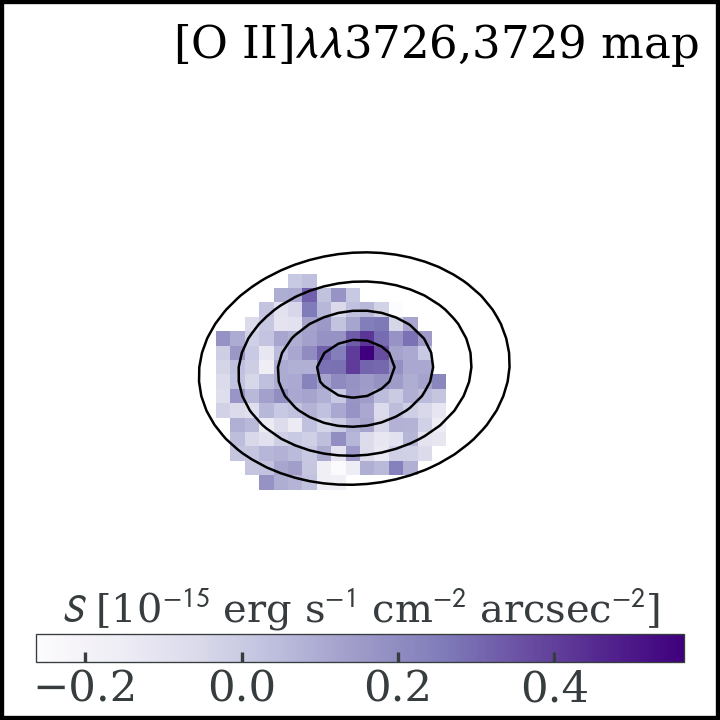}
    \includegraphics[width=.16\textwidth]{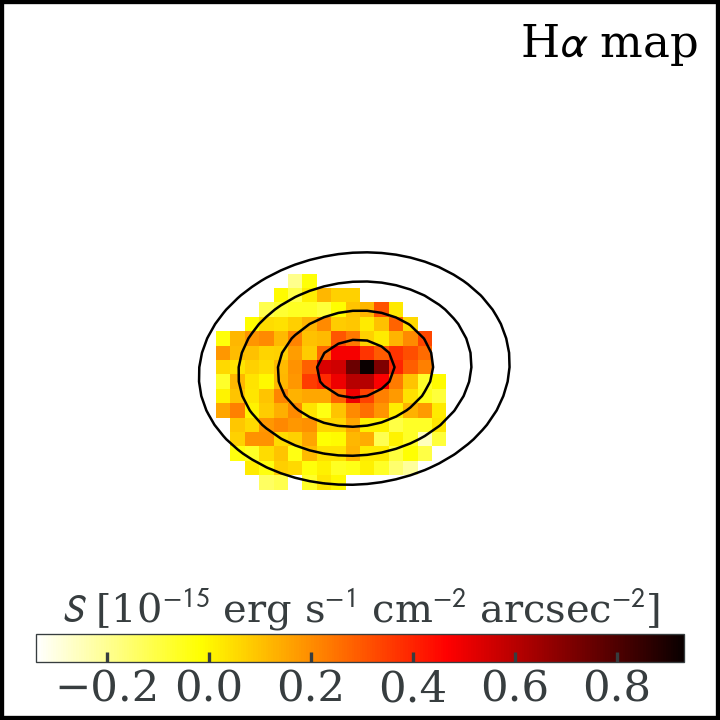}\\
    \includegraphics[width=\textwidth]{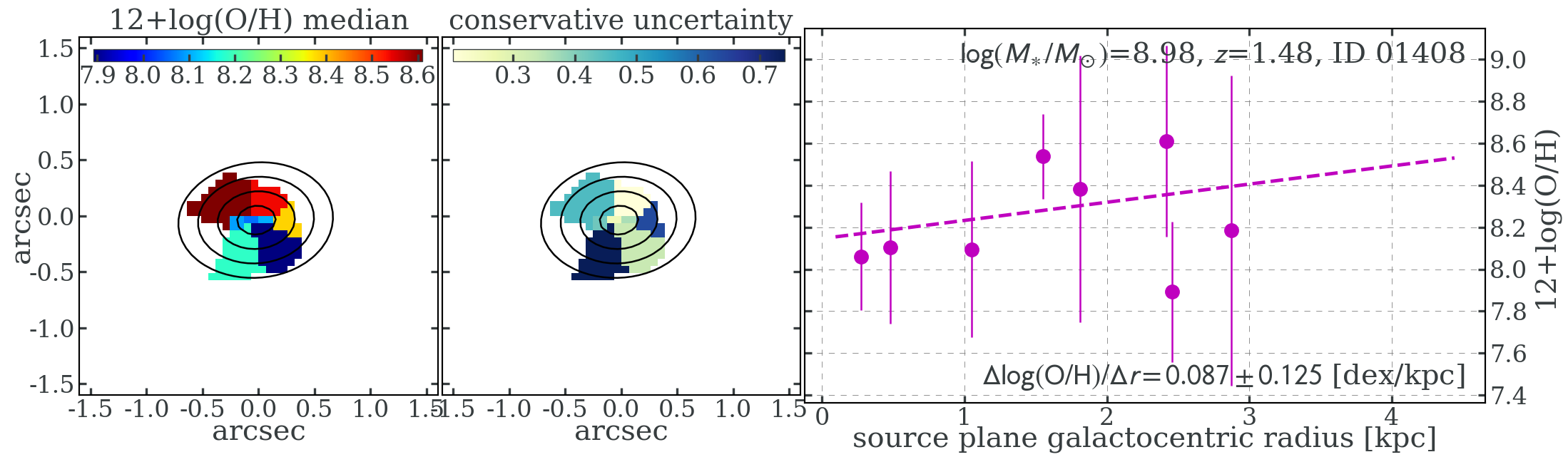}
    \caption{The source ID01408 in the field of \cljiu is shown.}
    \label{fig:clM2129_ID01408_figs}
\end{figure*}
\clearpage

\begin{figure*}
    \centering
    \includegraphics[width=\textwidth]{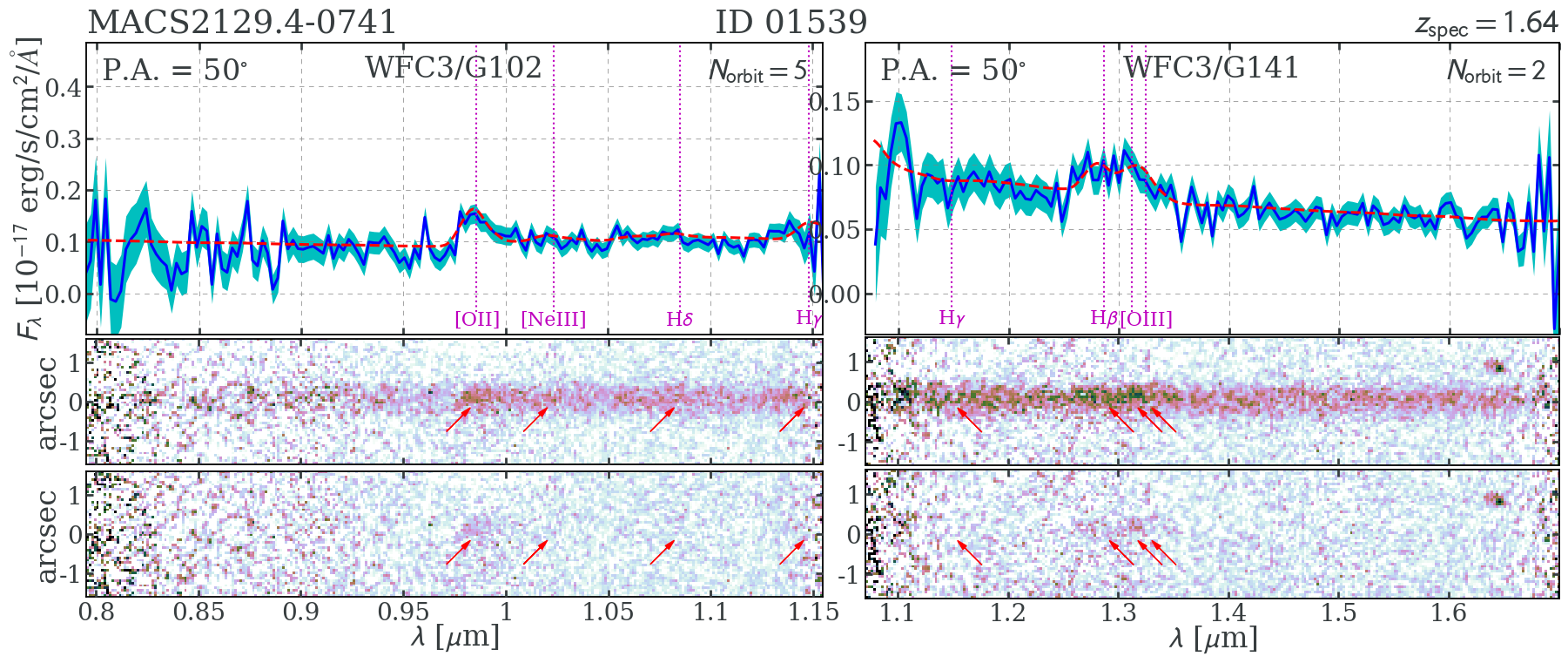}\\
    \includegraphics[width=\textwidth]{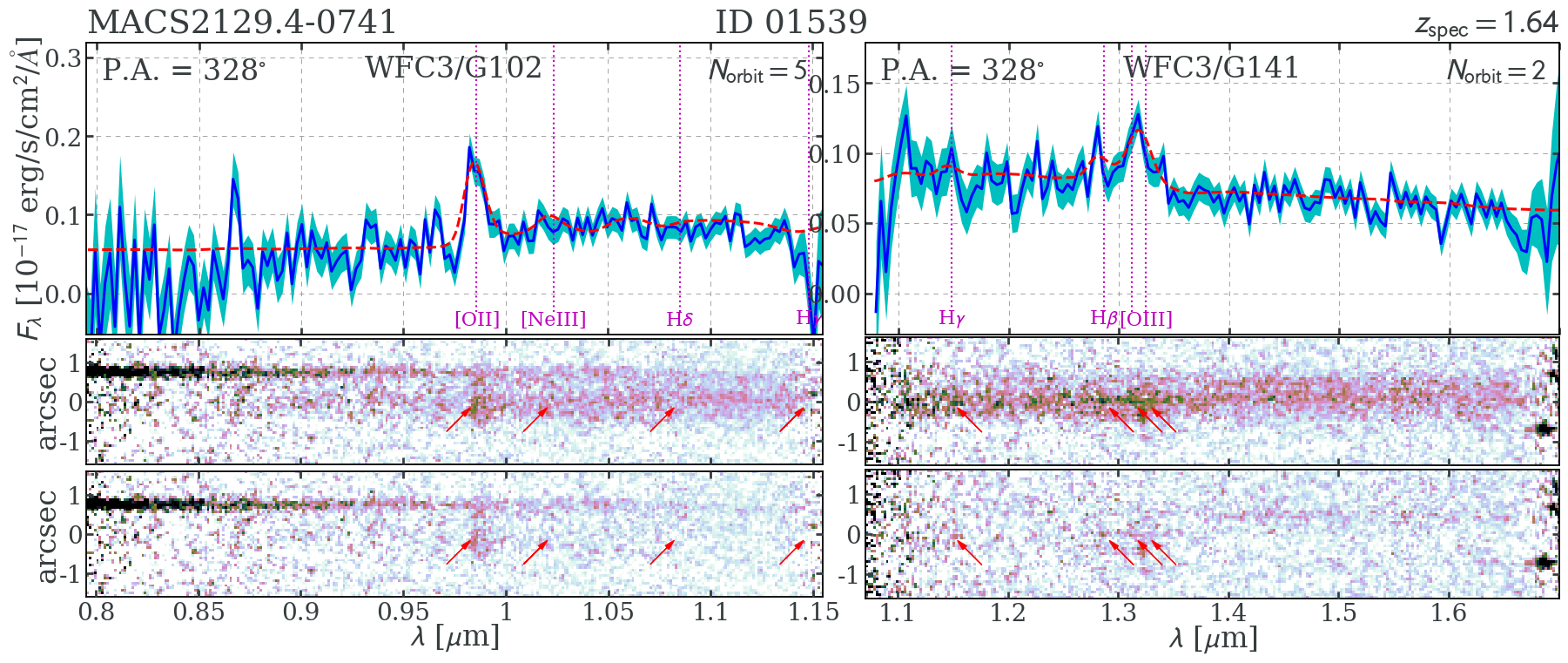}\\
    \includegraphics[width=.16\textwidth]{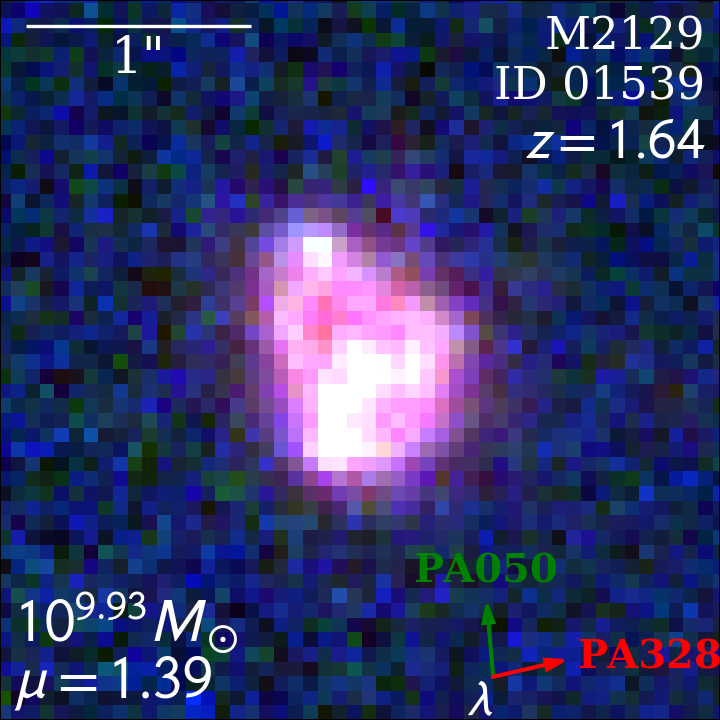}
    \includegraphics[width=.16\textwidth]{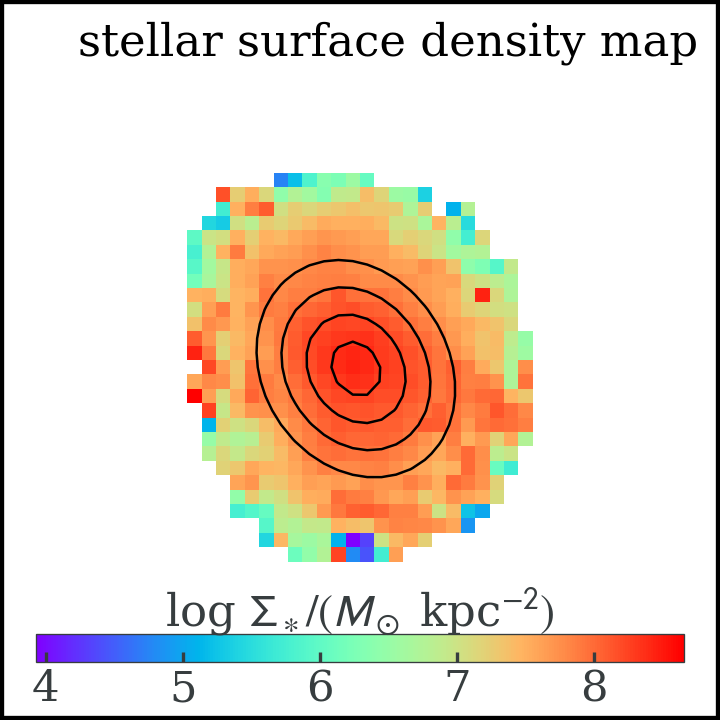}
    \includegraphics[width=.16\textwidth]{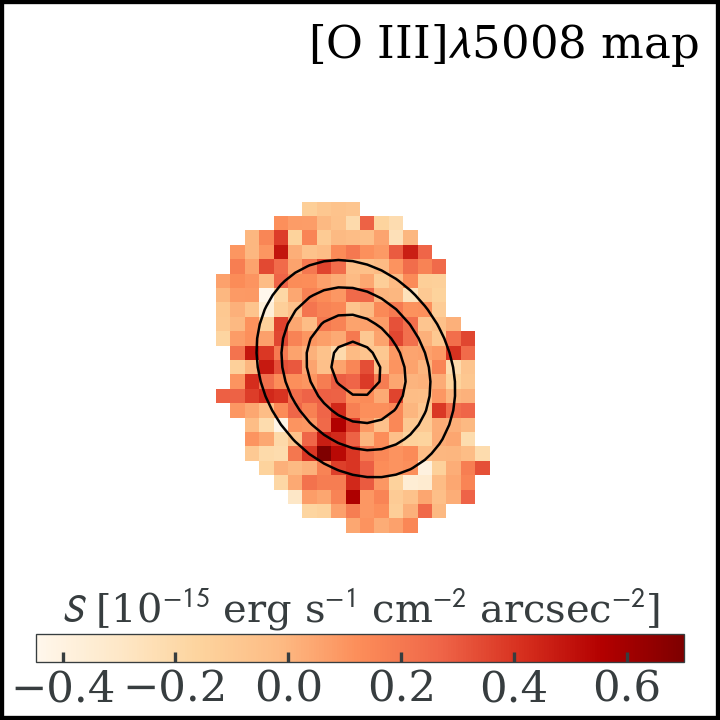}
    \includegraphics[width=.16\textwidth]{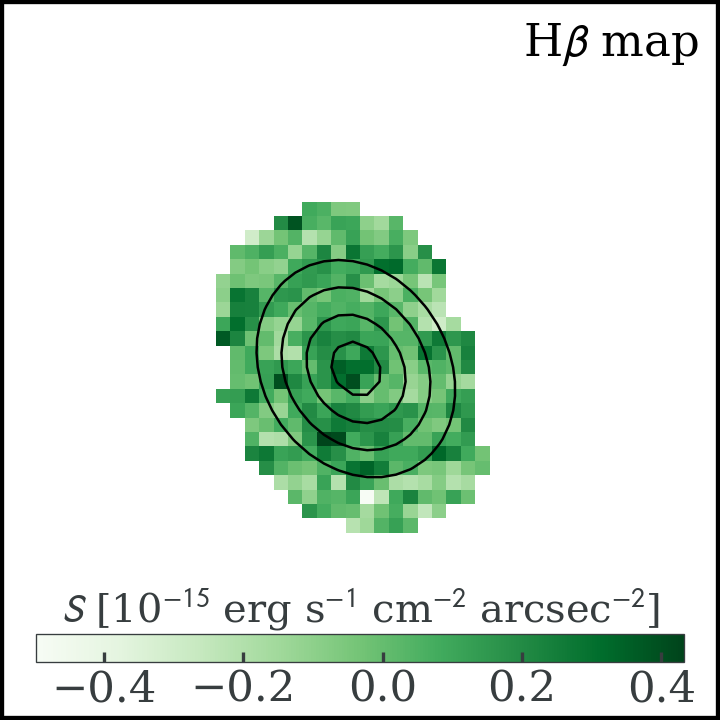}
    \includegraphics[width=.16\textwidth]{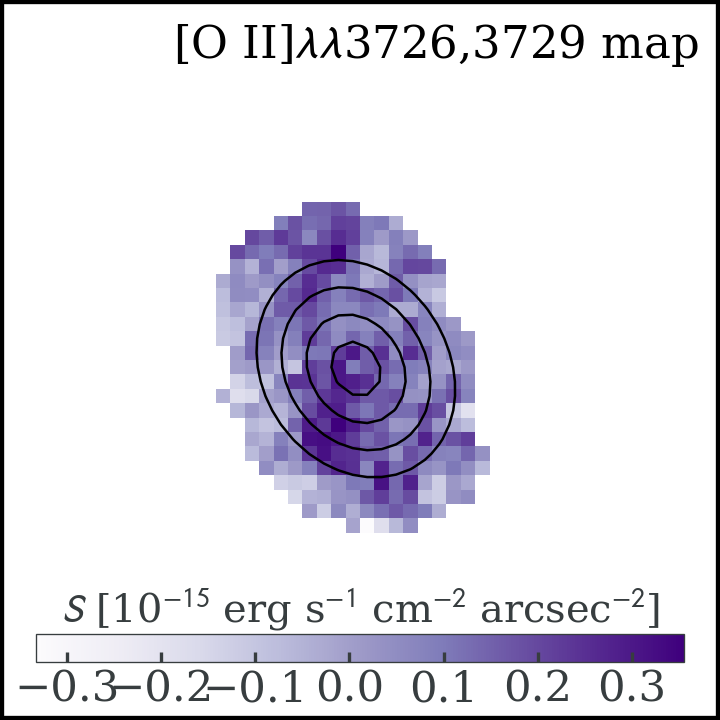}
    \includegraphics[width=.16\textwidth]{fig_ELmaps/baiban.png}\\
    \includegraphics[width=\textwidth]{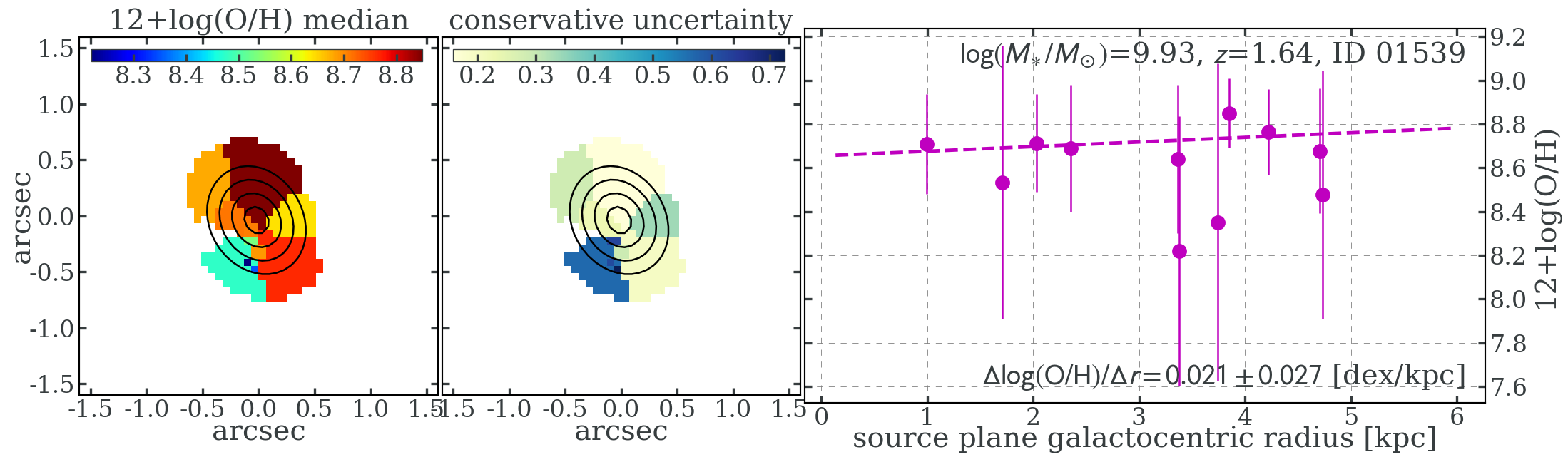}
    \caption{The source ID01539 in the field of \cljiu is shown.}
    \label{fig:clM2129_ID01539_figs}
\end{figure*}
\clearpage

\begin{figure*}
    \centering
    \includegraphics[width=\textwidth]{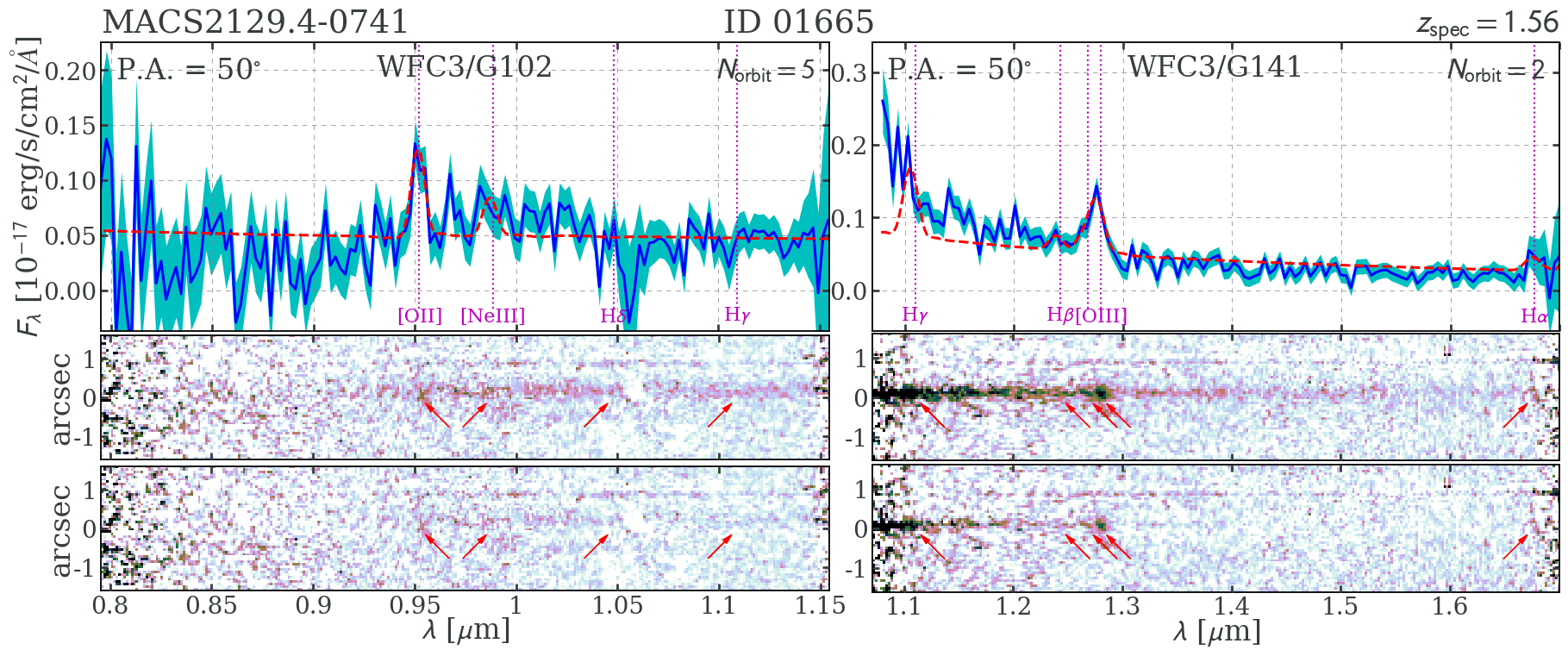}\\
    \includegraphics[width=\textwidth]{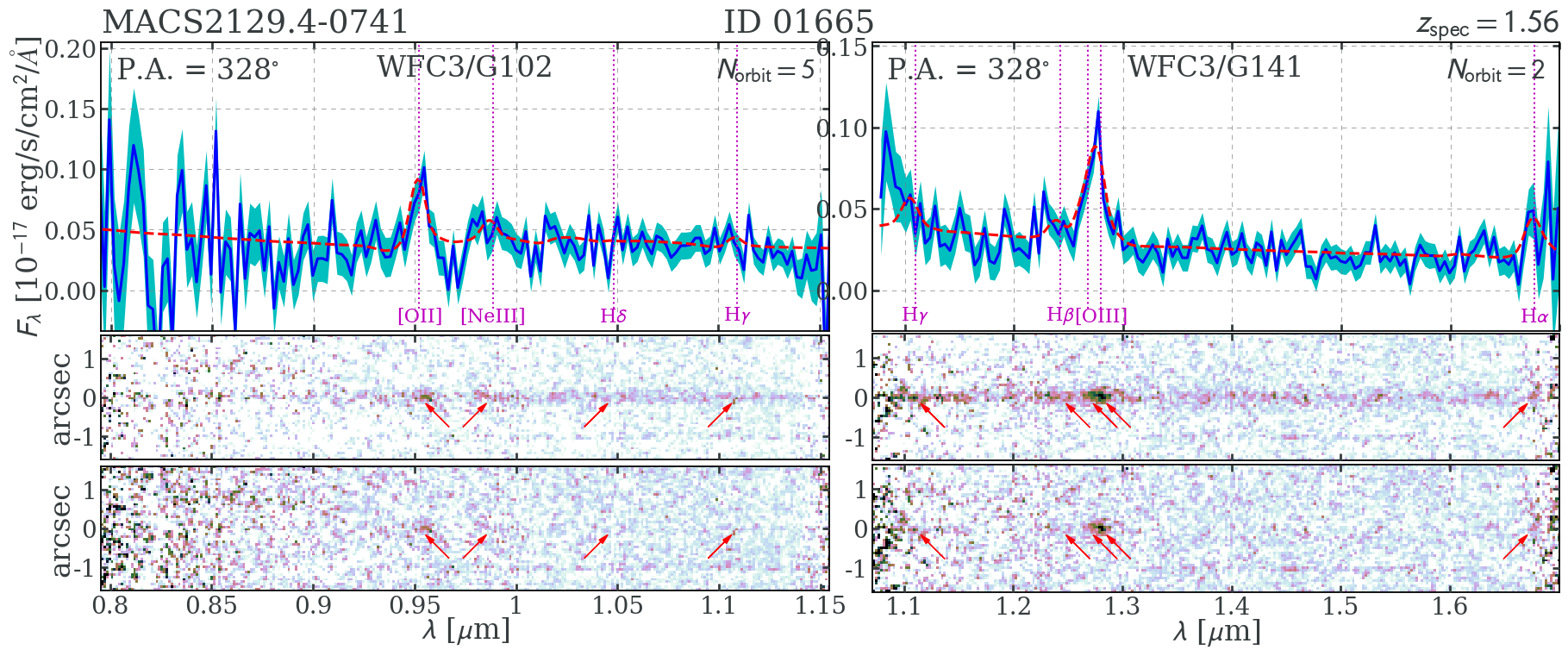}\\
    \includegraphics[width=.16\textwidth]{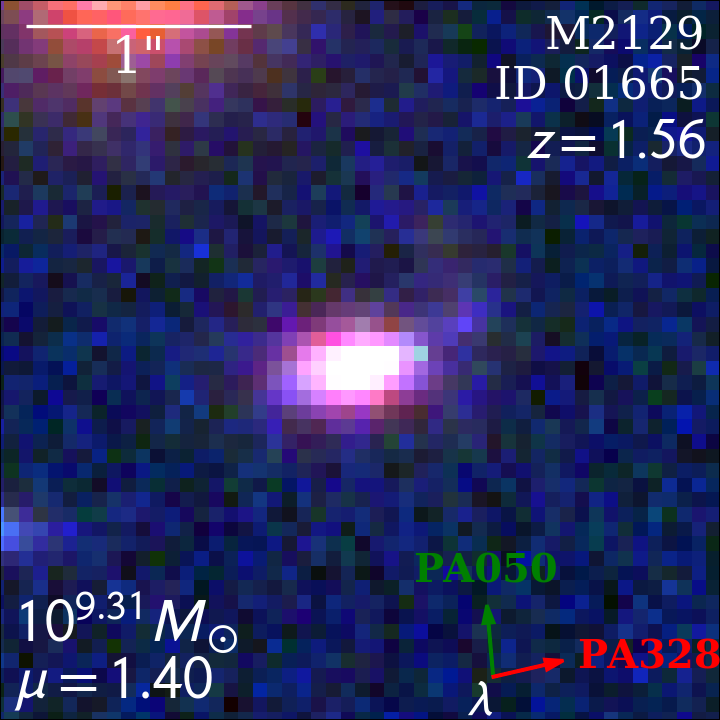}
    \includegraphics[width=.16\textwidth]{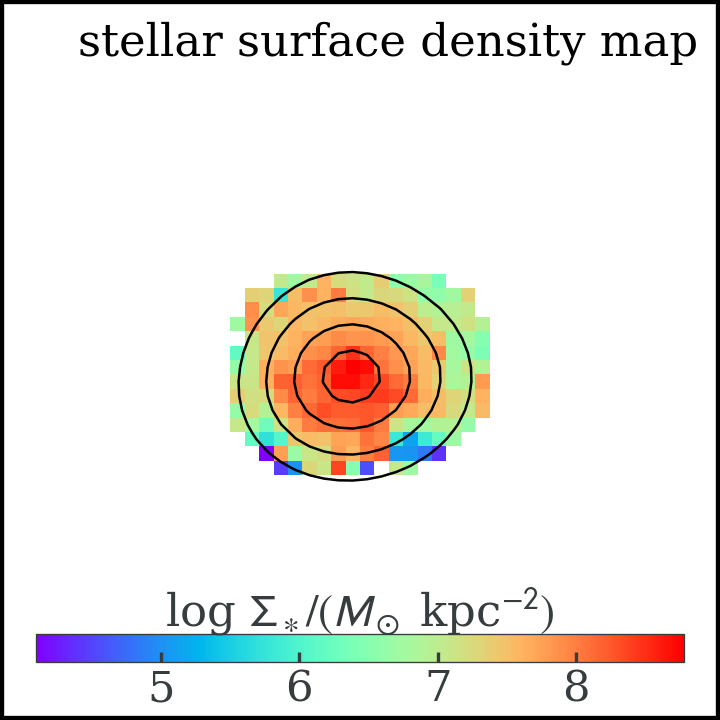}
    \includegraphics[width=.16\textwidth]{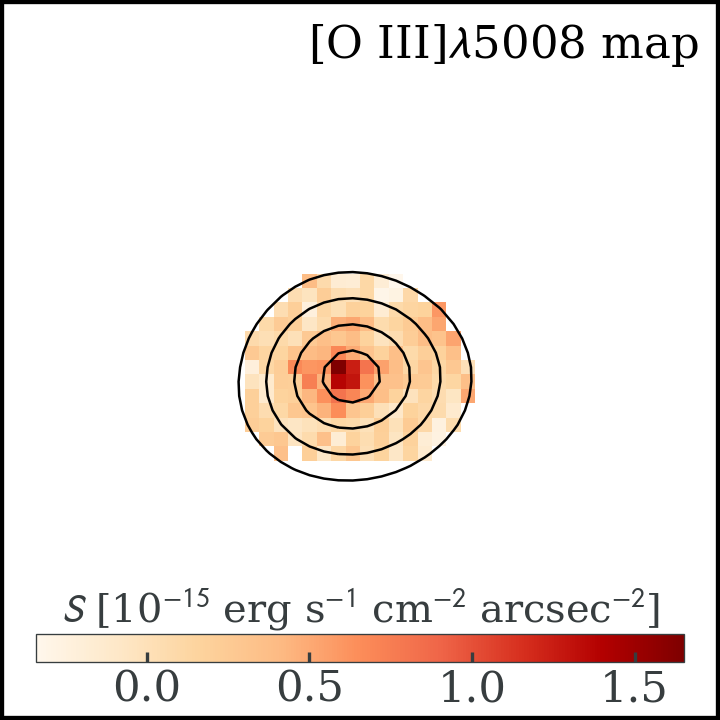}
    \includegraphics[width=.16\textwidth]{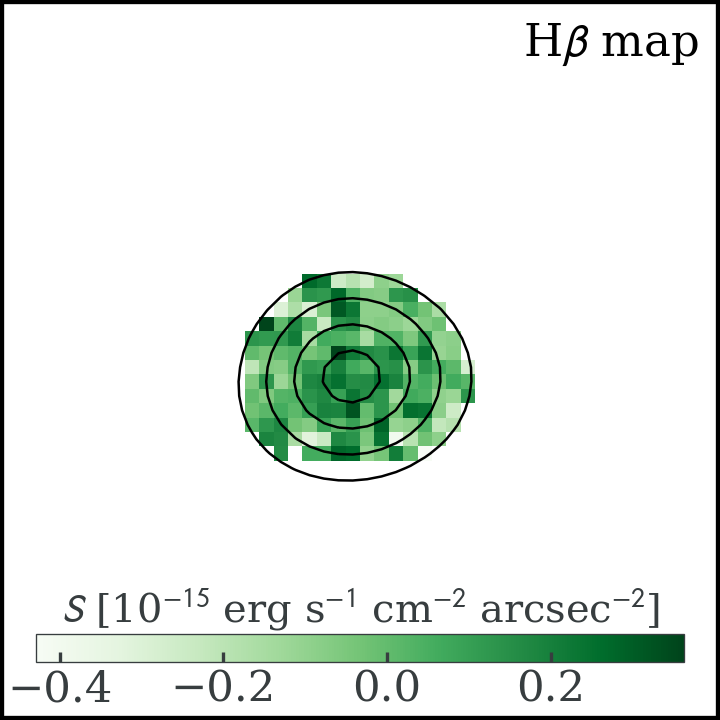}
    \includegraphics[width=.16\textwidth]{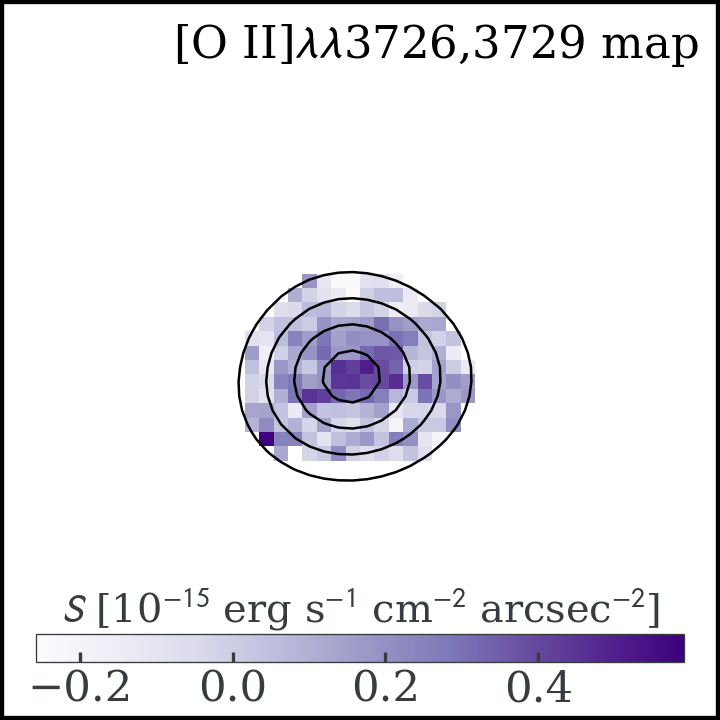}
    \includegraphics[width=.16\textwidth]{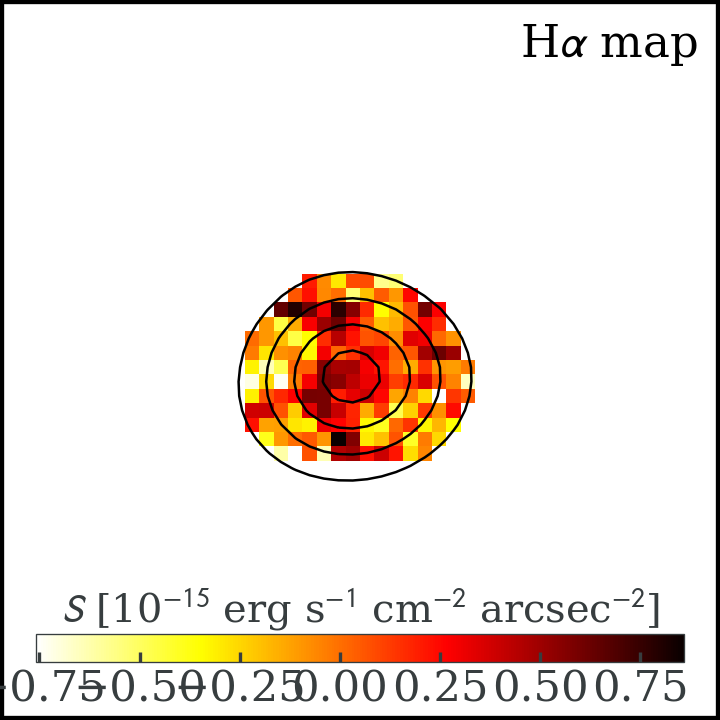}\\
    \includegraphics[width=\textwidth]{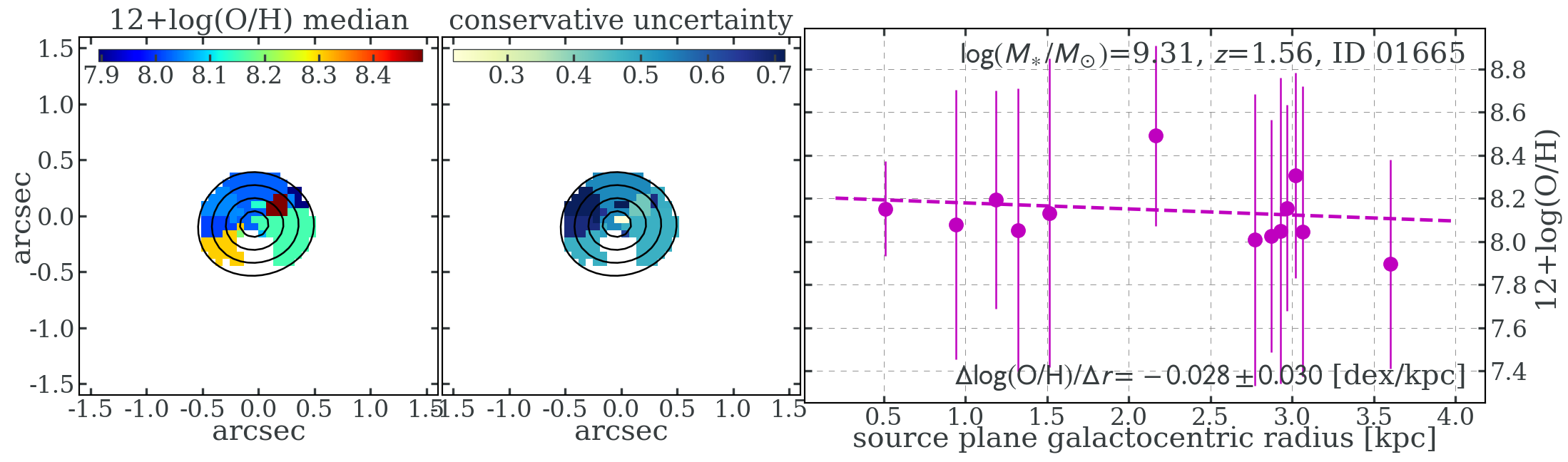}
    \caption{The source ID01665 in the field of \cljiu is shown.}
    \label{fig:clM2129_ID01665_figs}
\end{figure*}
\clearpage

\begin{figure*}
    \centering
    \includegraphics[width=\textwidth]{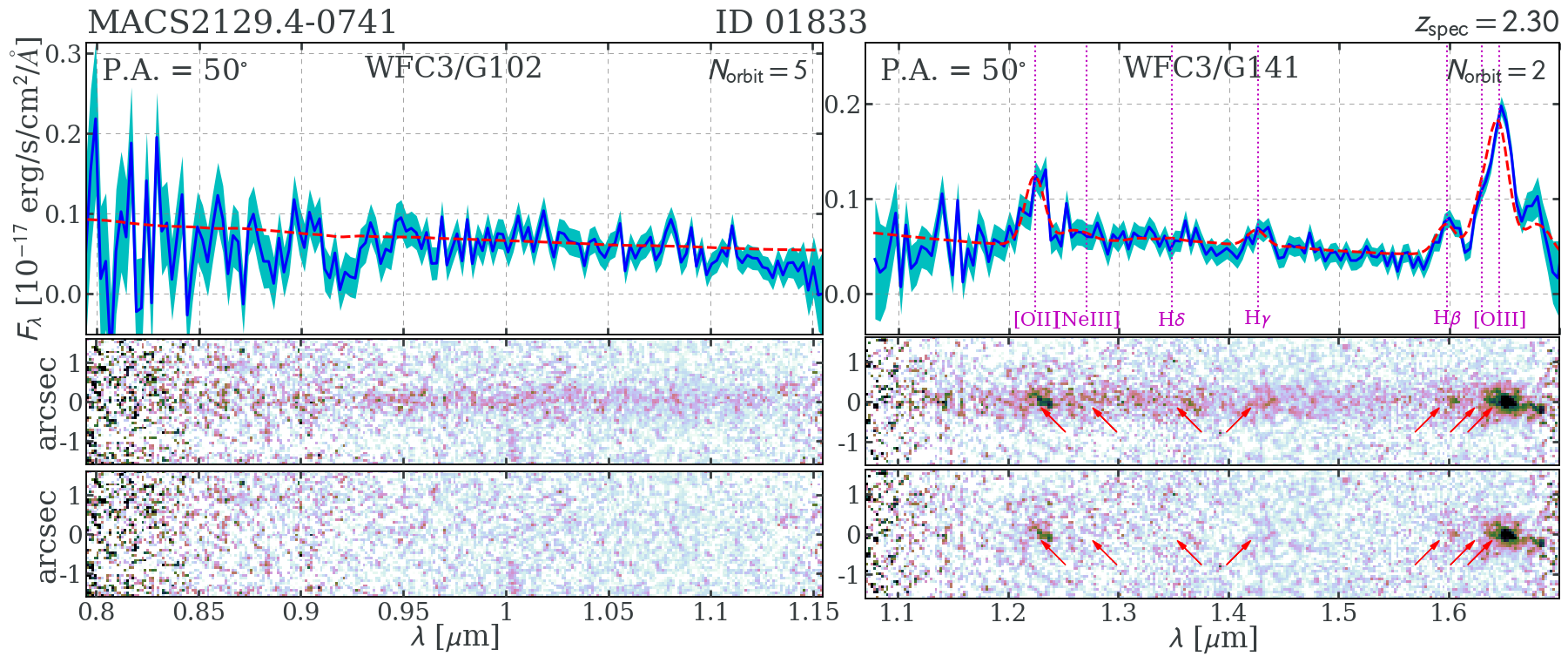}\\
    \includegraphics[width=\textwidth]{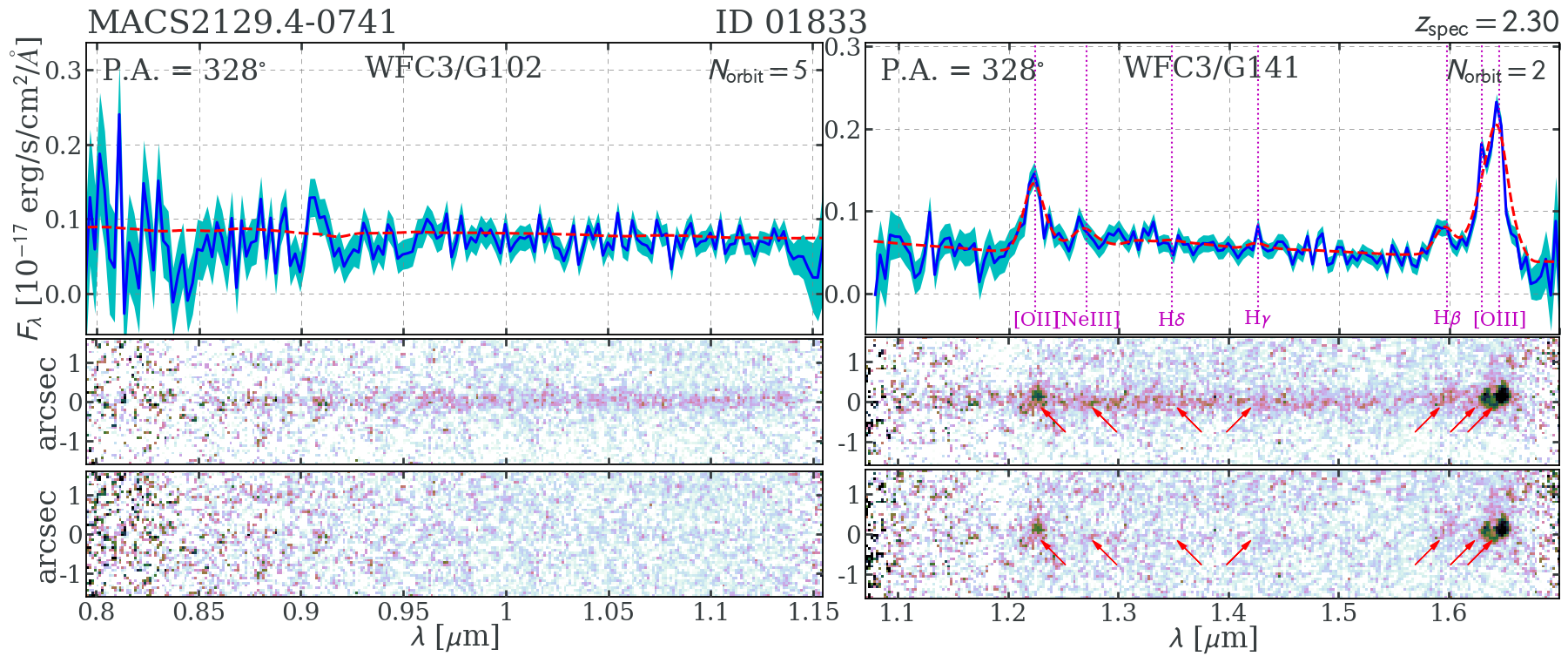}\\
    \includegraphics[width=.16\textwidth]{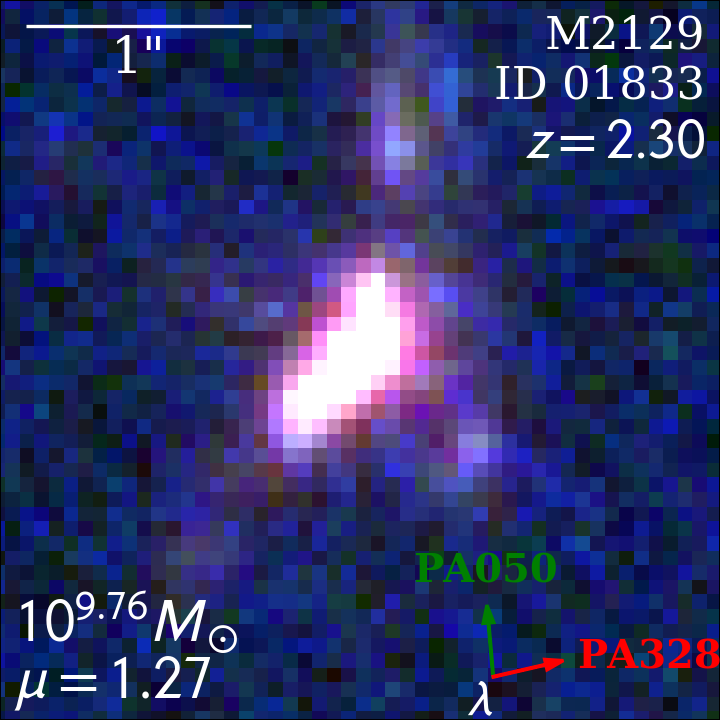}
    \includegraphics[width=.16\textwidth]{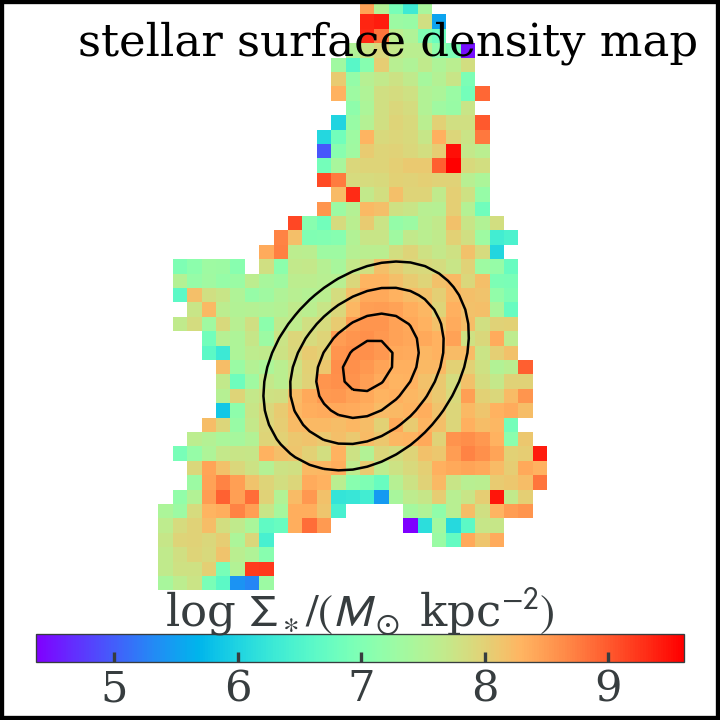}
    \includegraphics[width=.16\textwidth]{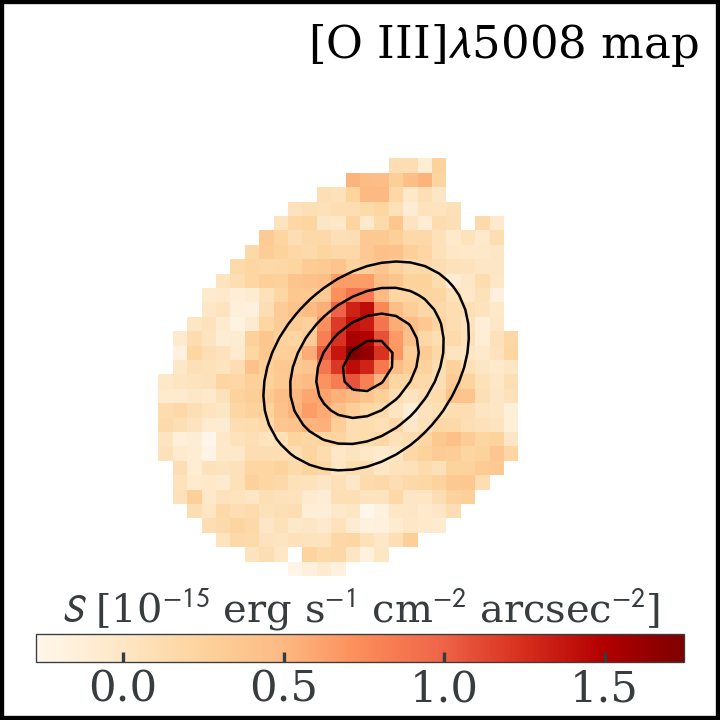}
    \includegraphics[width=.16\textwidth]{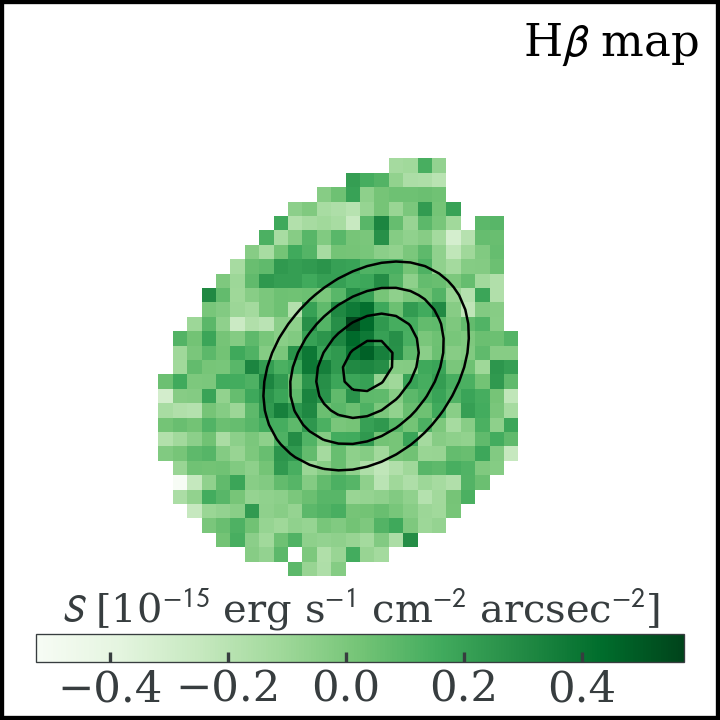}
    \includegraphics[width=.16\textwidth]{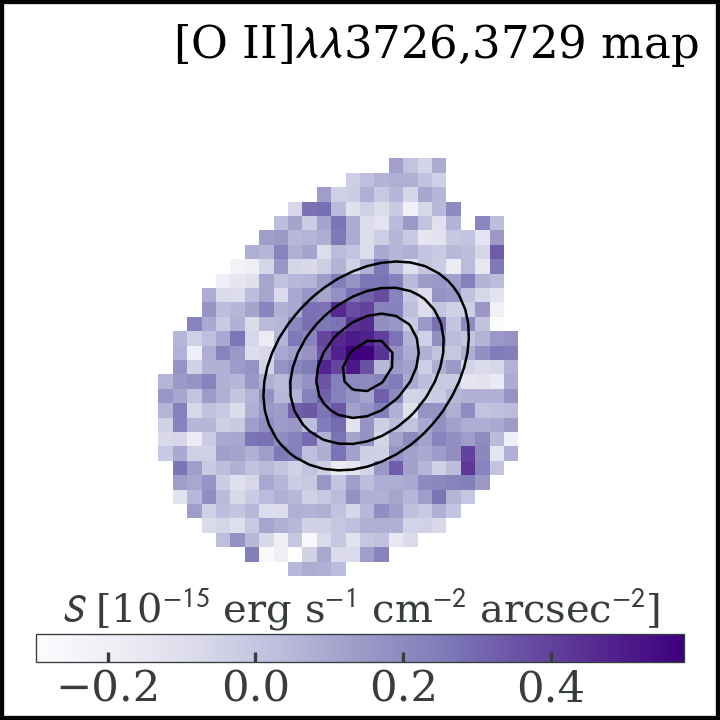}
    \includegraphics[width=.16\textwidth]{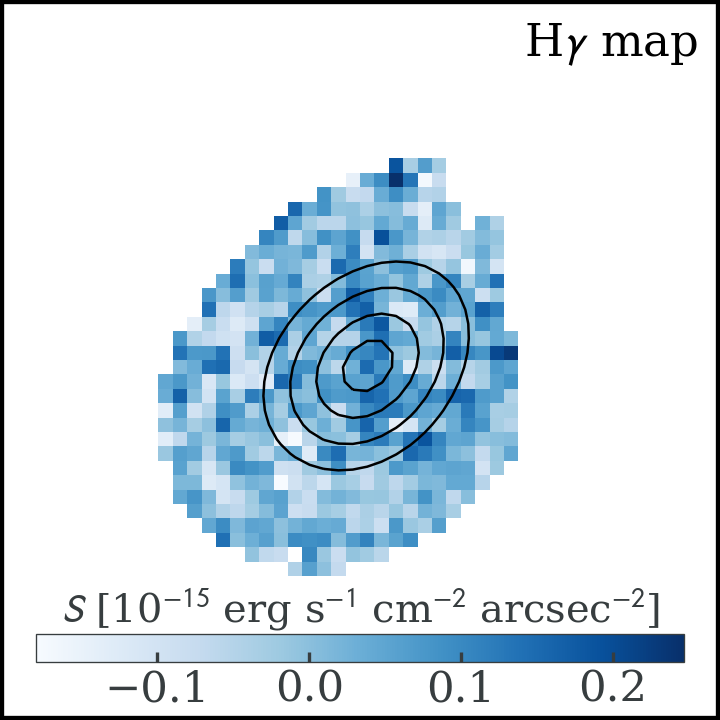}\\
    \includegraphics[width=\textwidth]{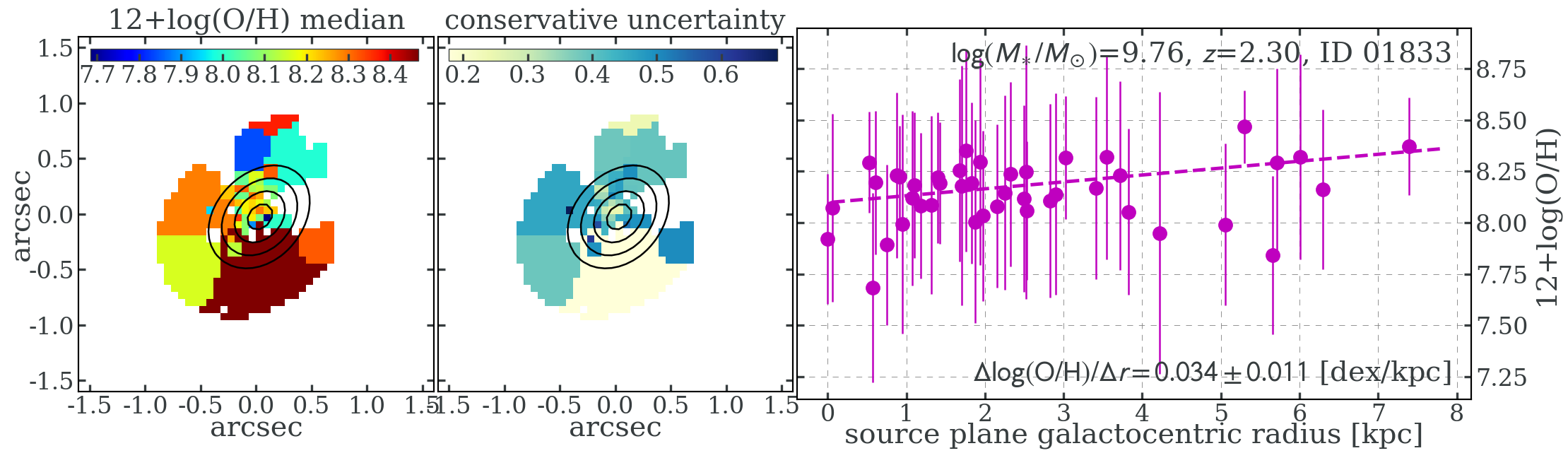}
    \caption{The source ID01833 in the field of \cljiu is shown.}
    \label{fig:clM2129_ID01833_figs}
\end{figure*}
\clearpage

\begin{figure*}
    \centering
    \includegraphics[width=\textwidth]{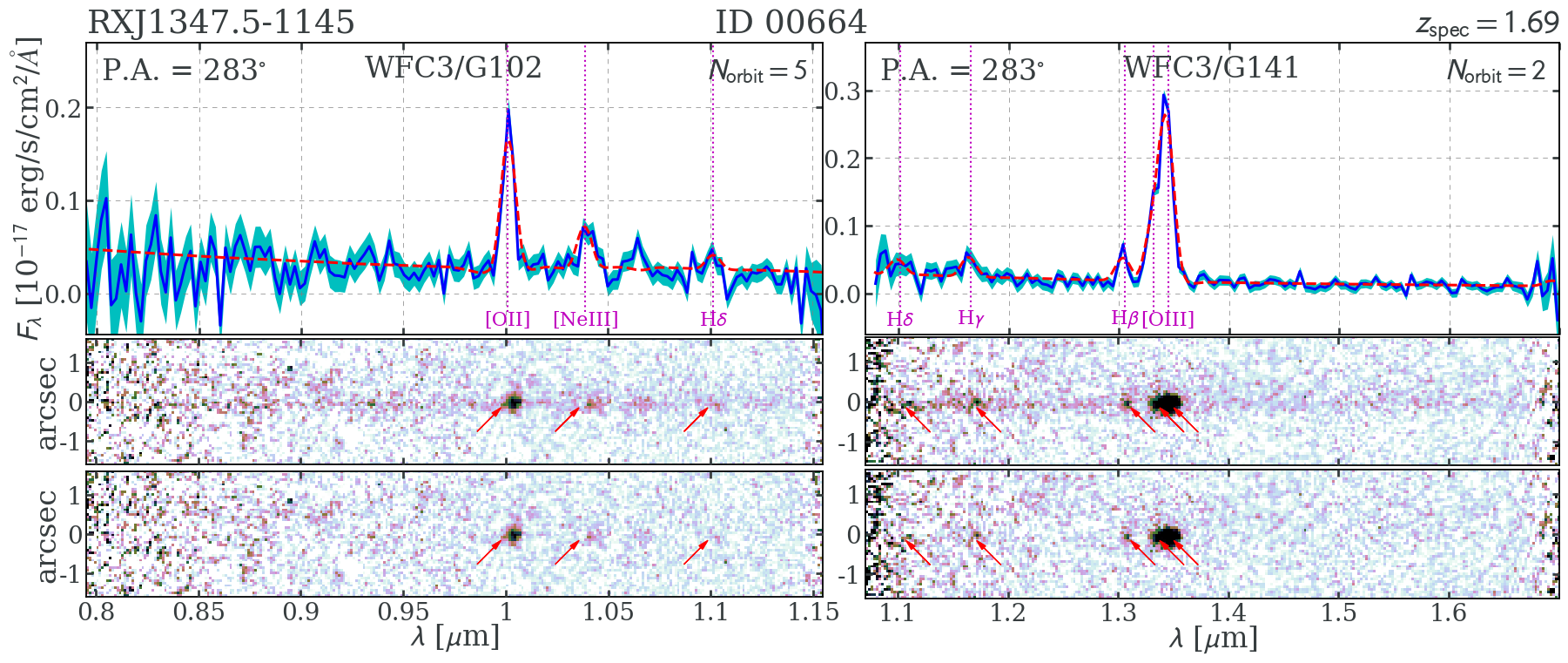}\\
    \includegraphics[width=.16\textwidth]{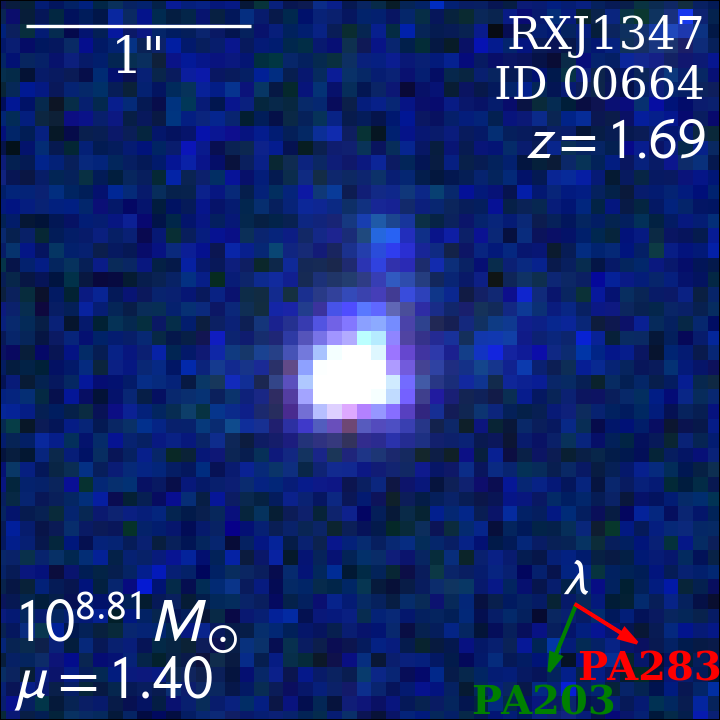}
    \includegraphics[width=.16\textwidth]{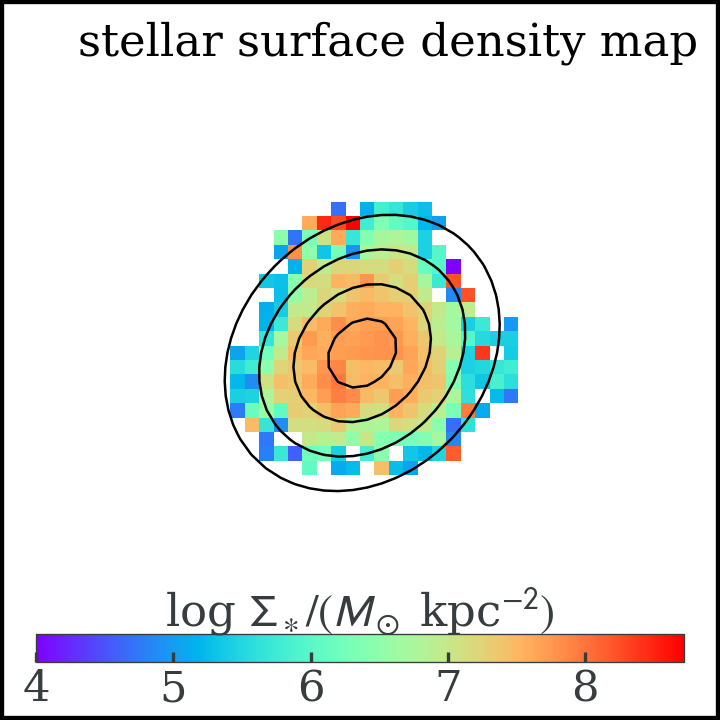}
    \includegraphics[width=.16\textwidth]{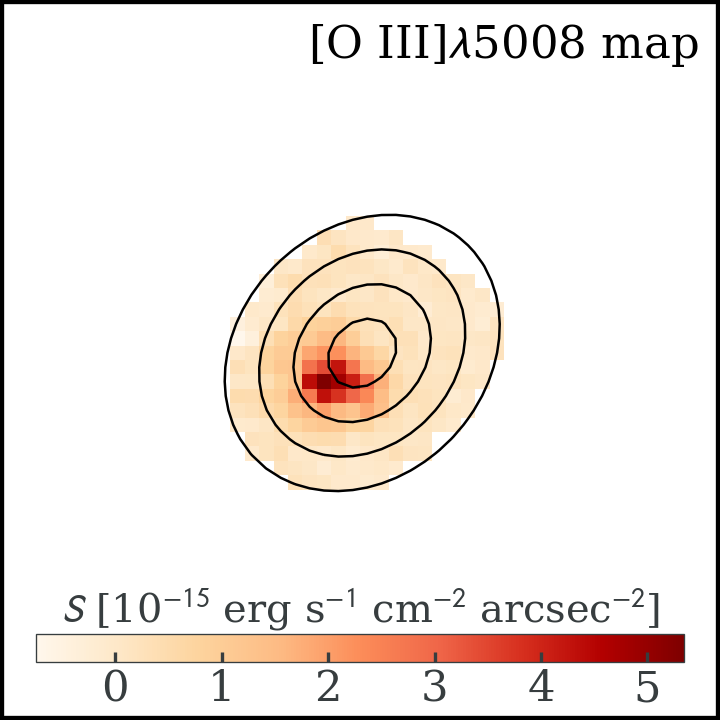}
    \includegraphics[width=.16\textwidth]{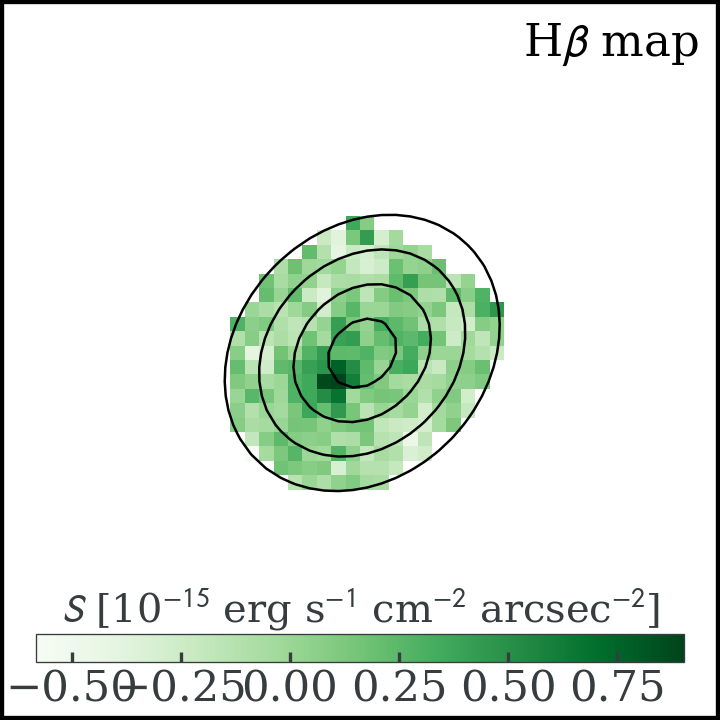}
    \includegraphics[width=.16\textwidth]{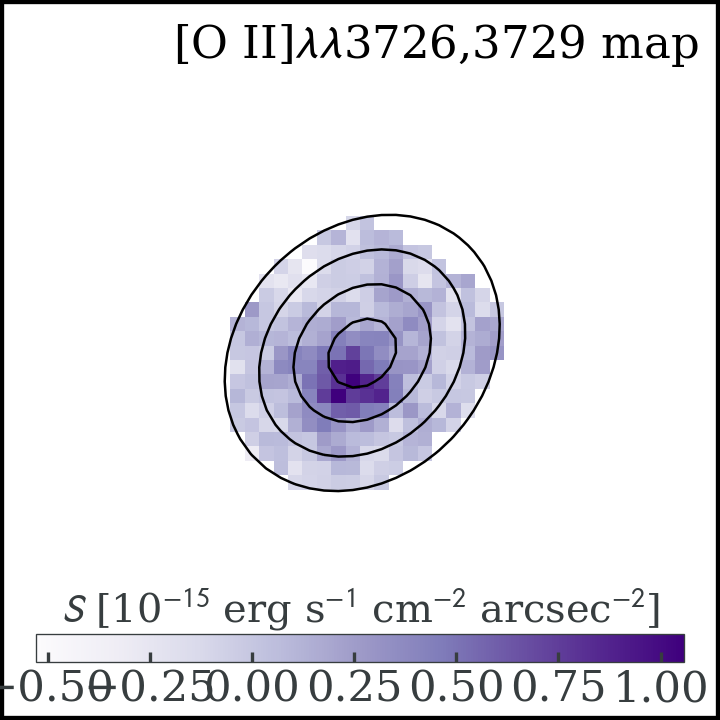}
    \includegraphics[width=.16\textwidth]{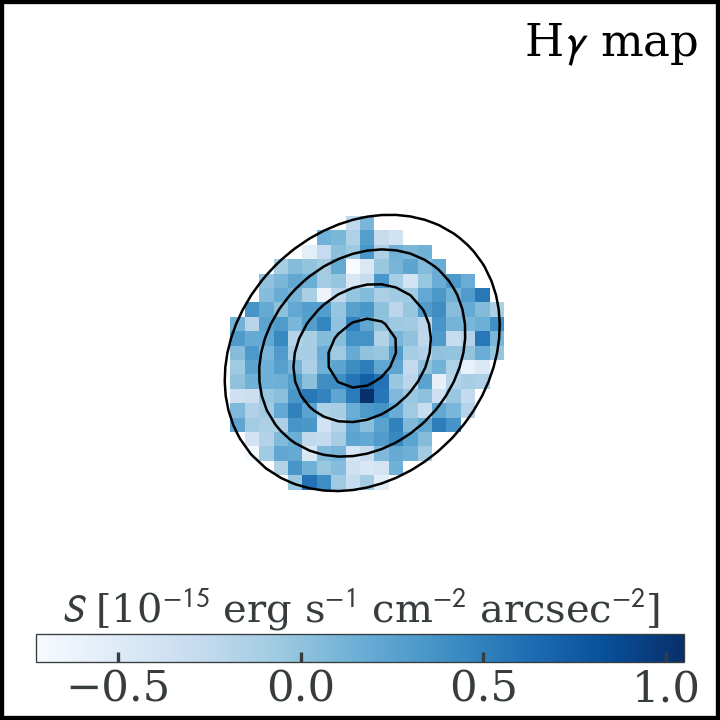}\\
    \includegraphics[width=\textwidth]{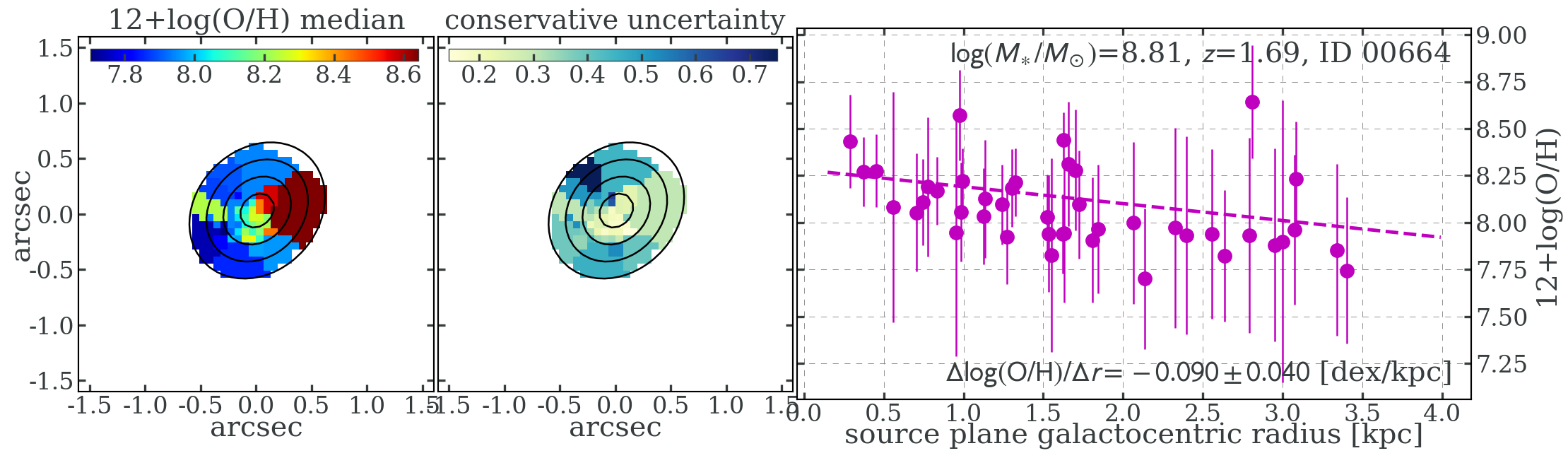}
    \caption{The source ID00664 in the field of \clqi is shown.}
    \label{fig:clRXJ1347_ID00664_figs}
\end{figure*}
\clearpage

\begin{figure*}
    \centering
    \includegraphics[width=\textwidth]{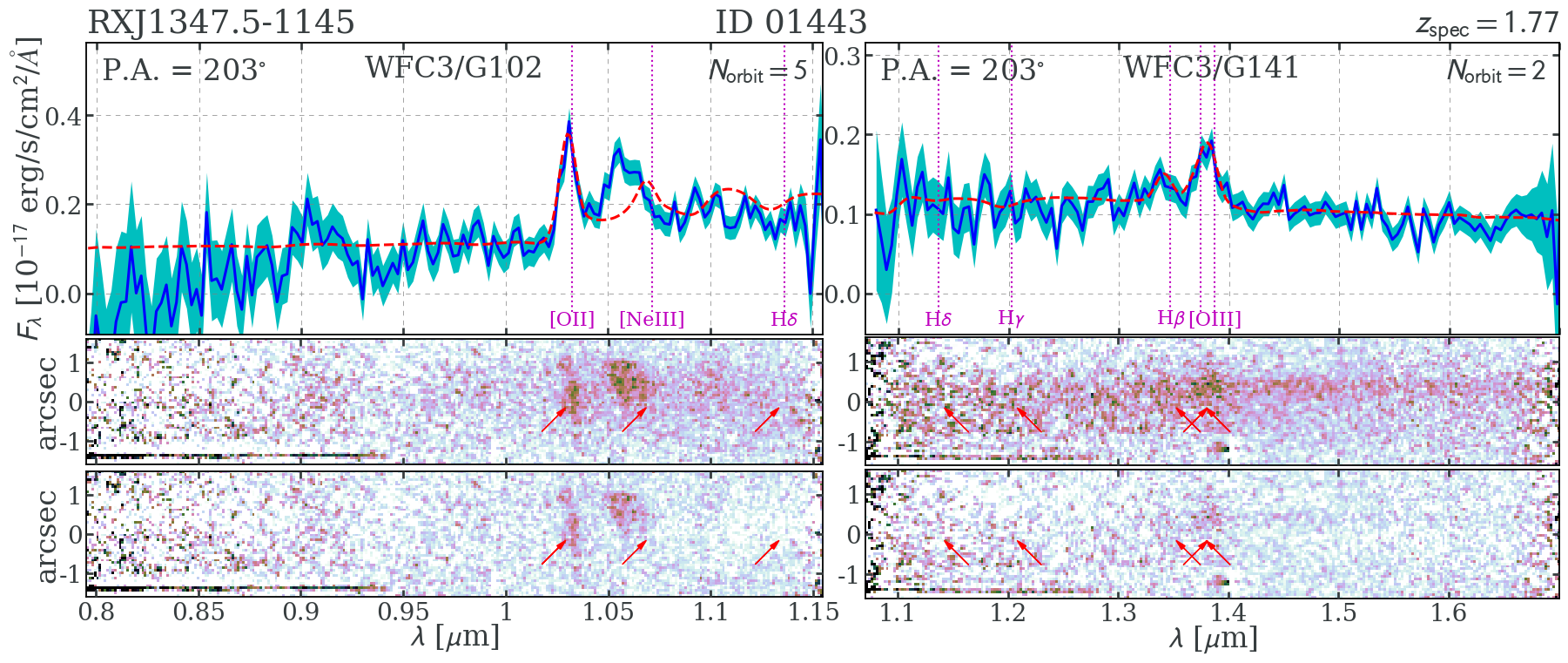}\\
    \includegraphics[width=\textwidth]{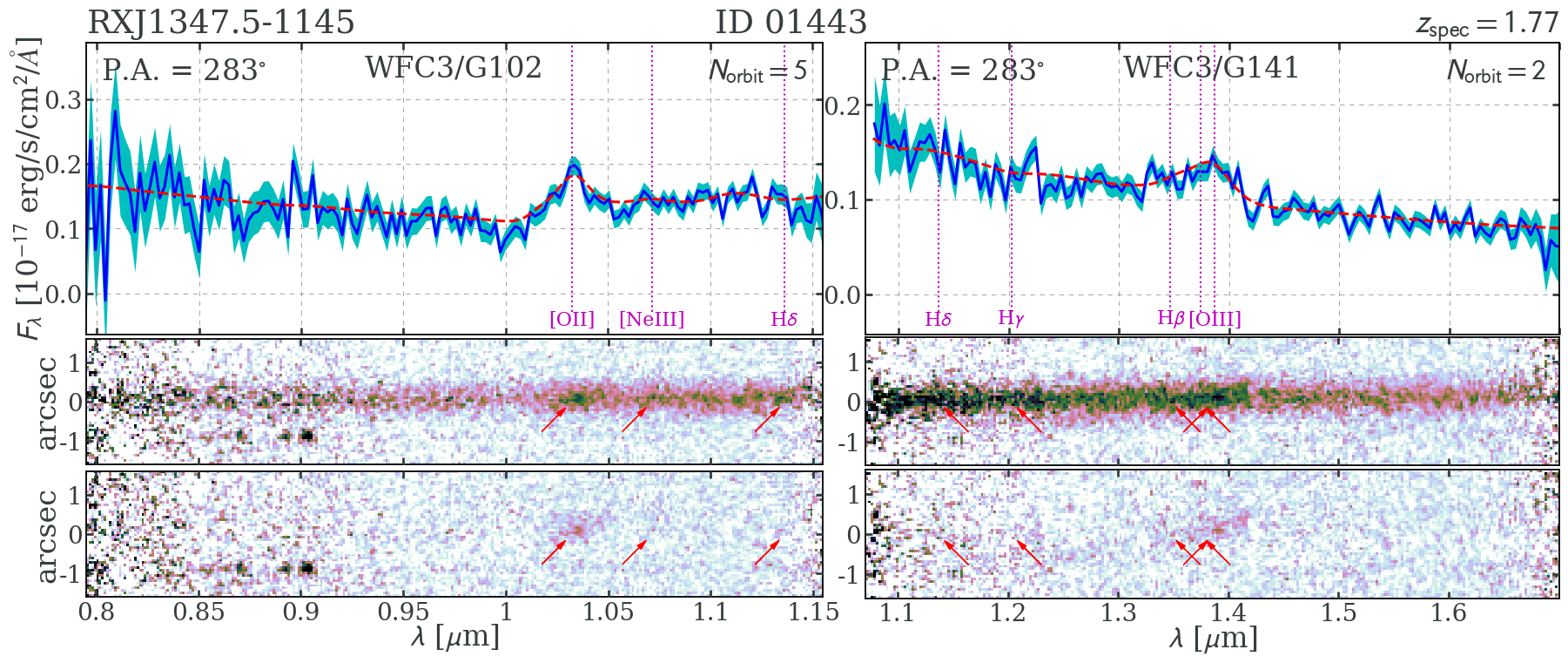}\\
    \includegraphics[width=.16\textwidth]{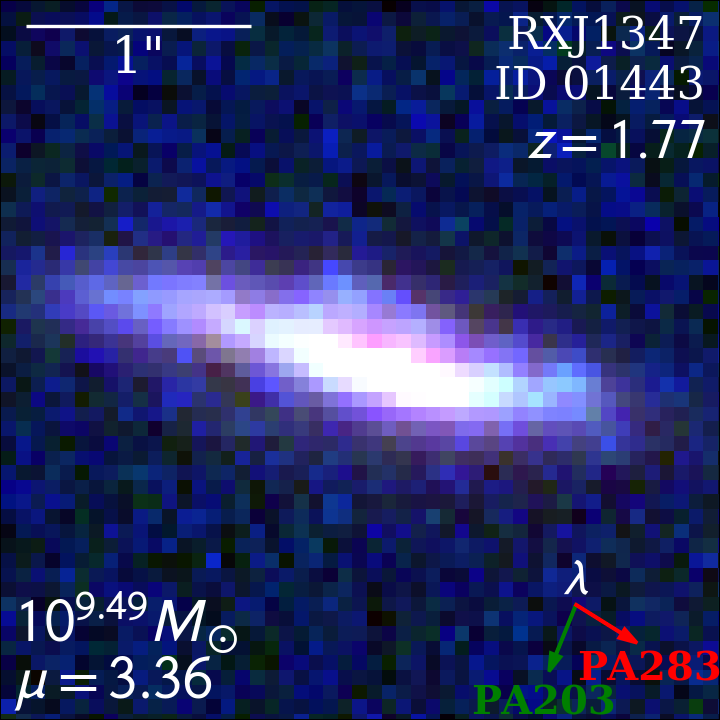}
    \includegraphics[width=.16\textwidth]{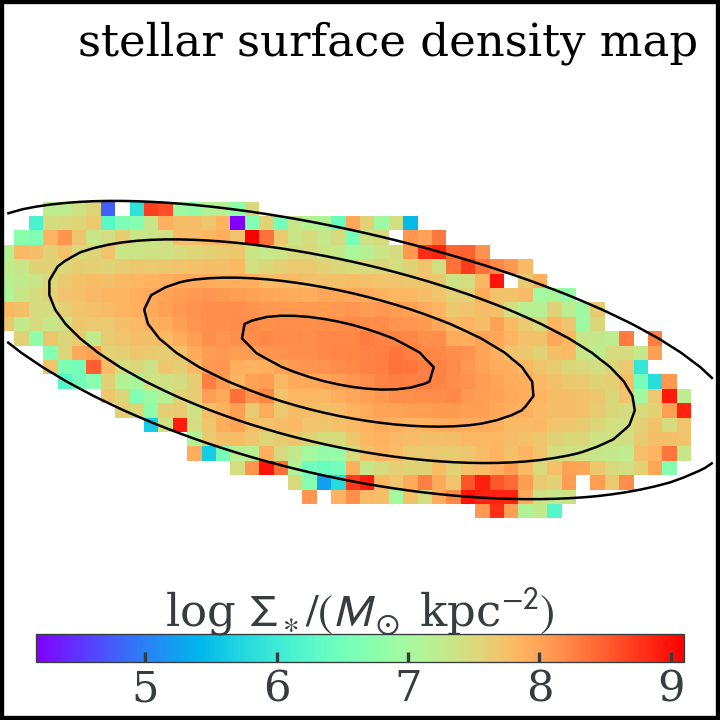}
    \includegraphics[width=.16\textwidth]{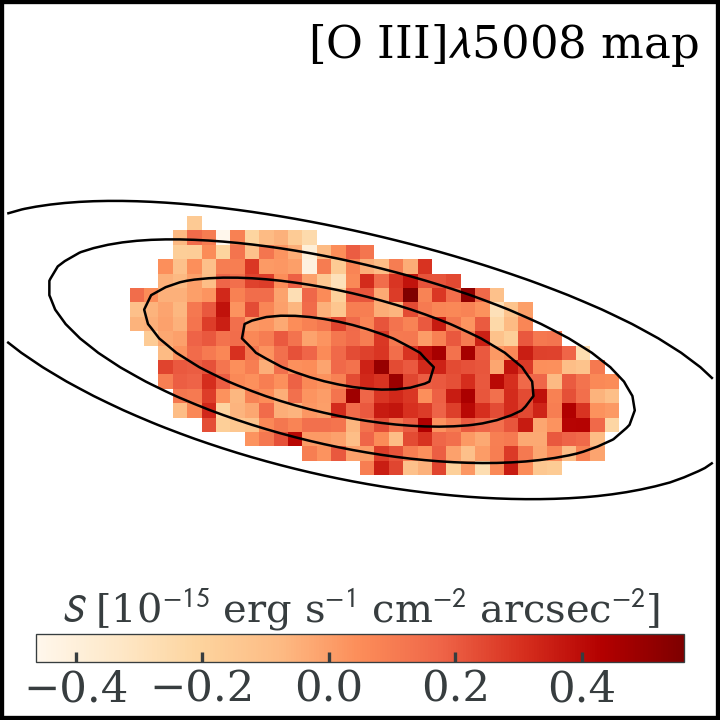}
    \includegraphics[width=.16\textwidth]{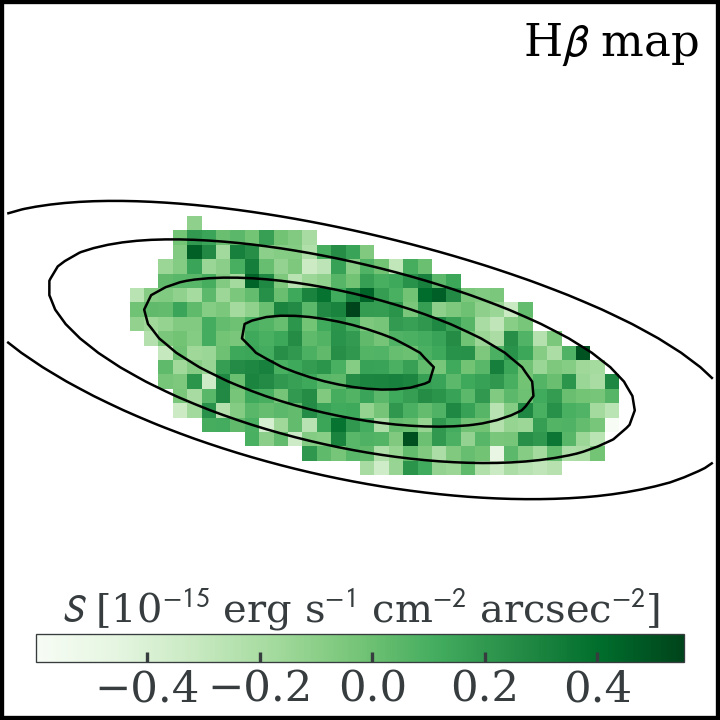}
    \includegraphics[width=.16\textwidth]{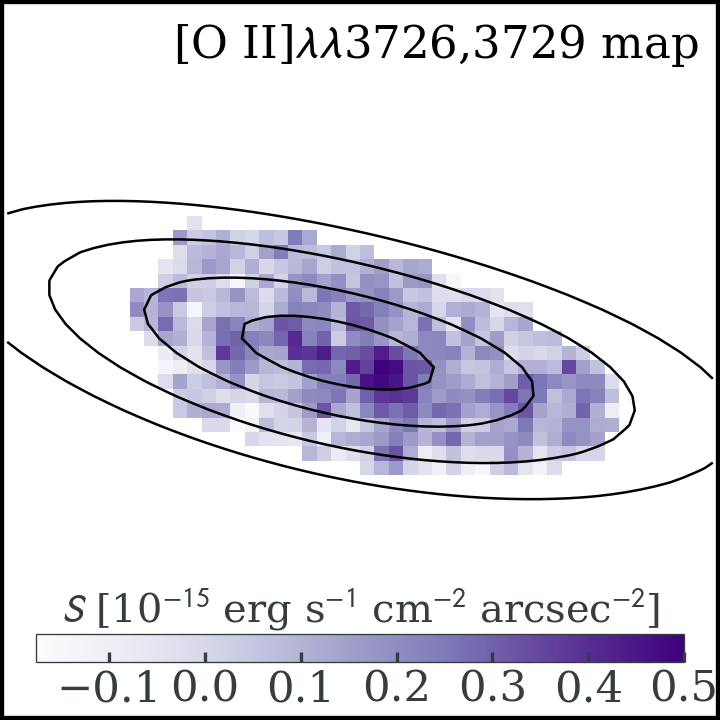}
    \includegraphics[width=.16\textwidth]{fig_ELmaps/baiban.png}\\
    \includegraphics[width=\textwidth]{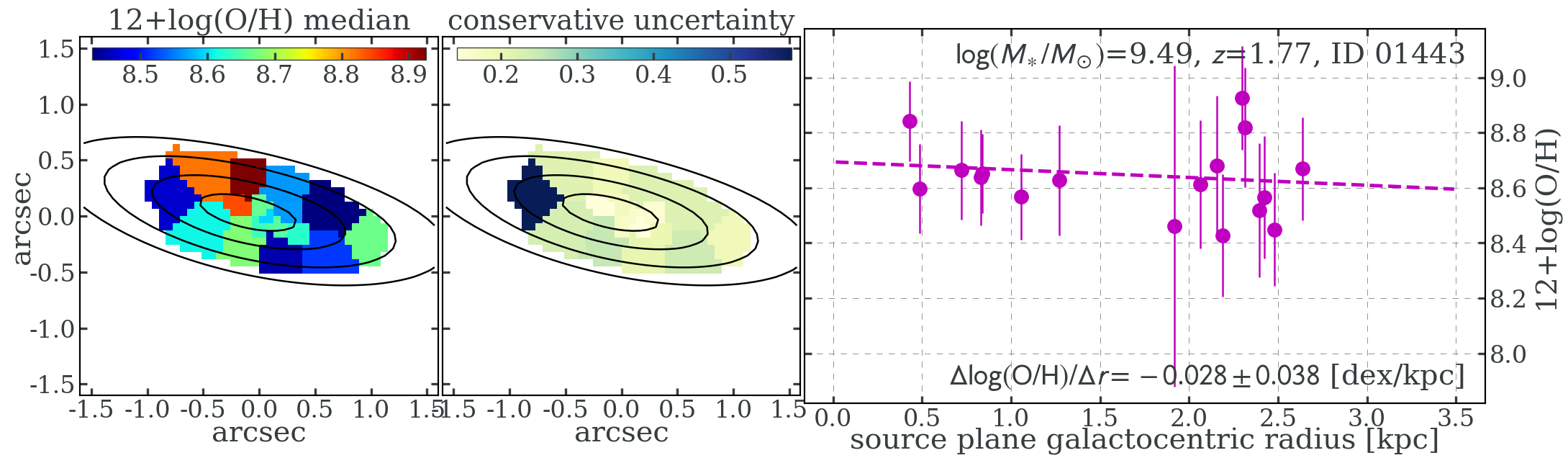}
    \caption{The source ID01443 in the field of \clqi is shown.}
    \label{fig:clRXJ1347_ID01443_figs}
\end{figure*}
\clearpage

\begin{figure*}
    \centering
    \includegraphics[width=\textwidth]{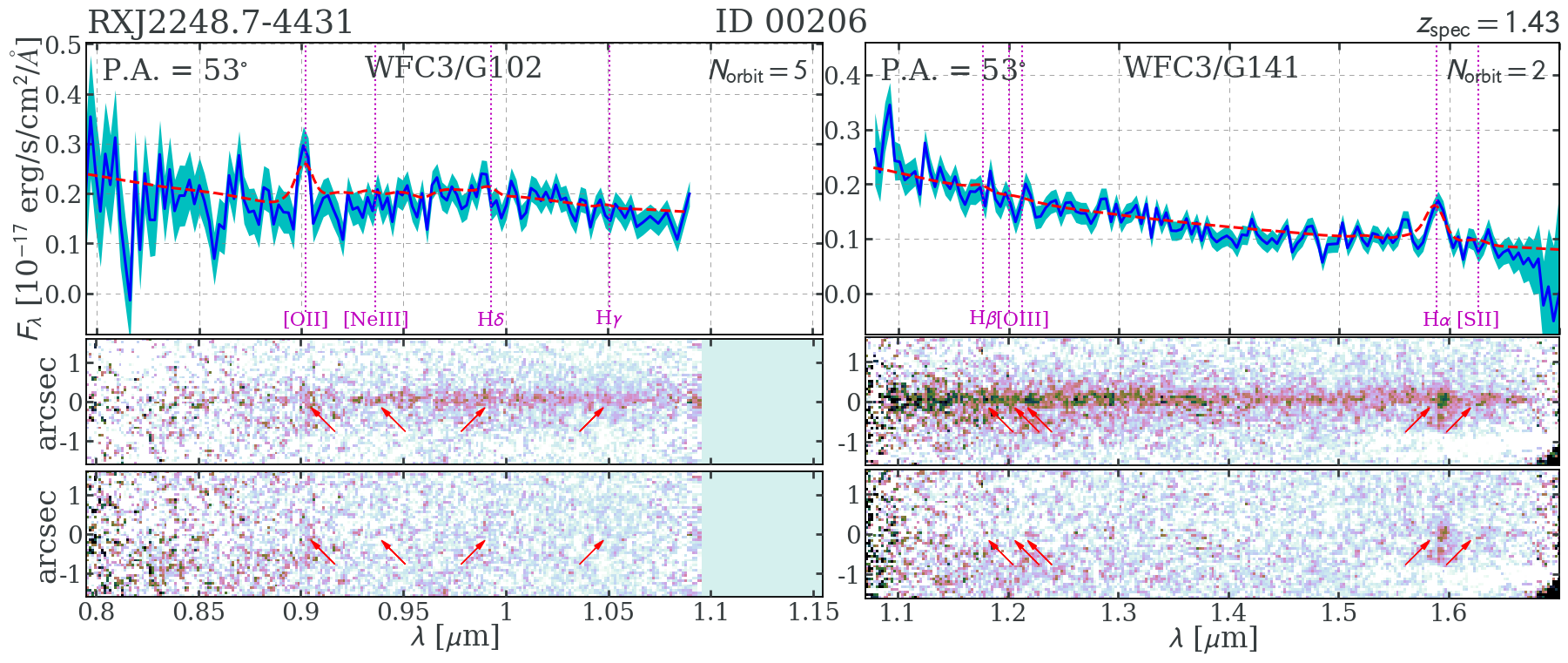}\\
    \includegraphics[width=\textwidth]{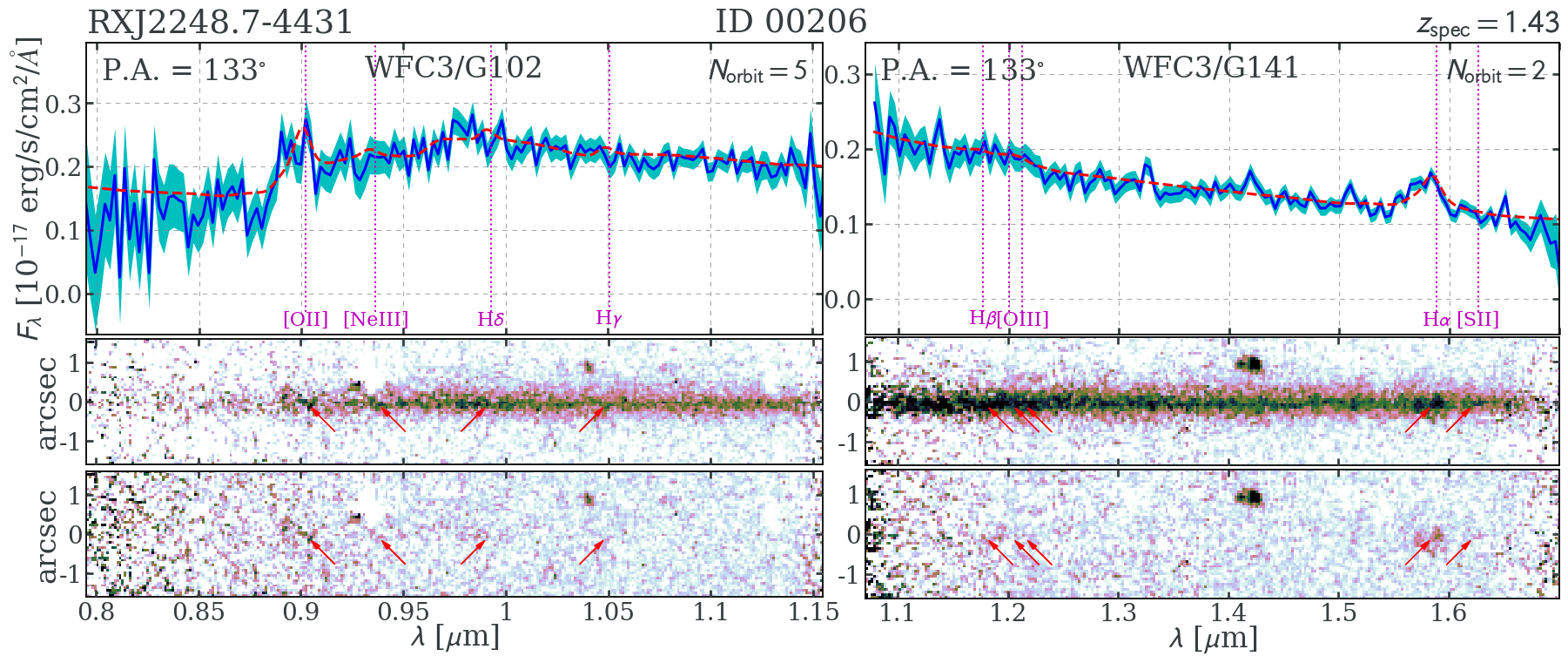}\\
    \includegraphics[width=.16\textwidth]{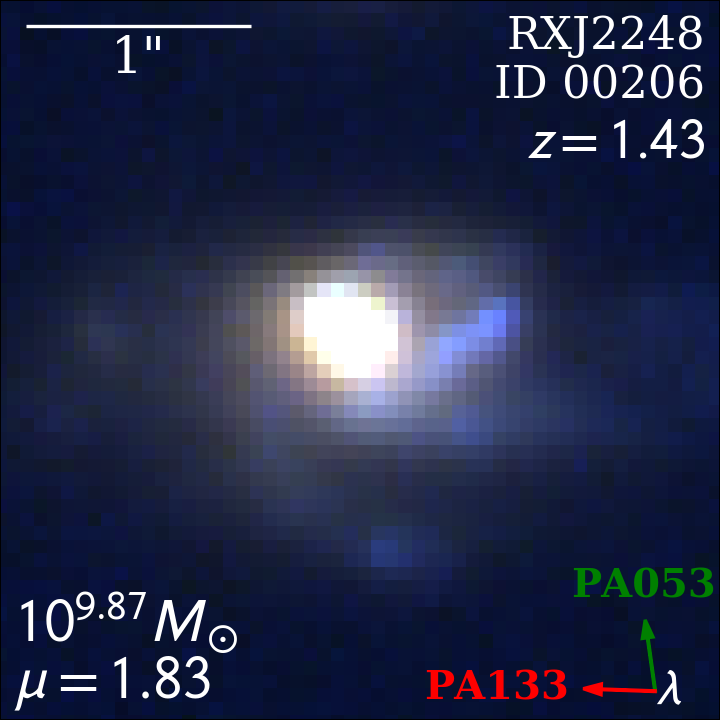}
    \includegraphics[width=.16\textwidth]{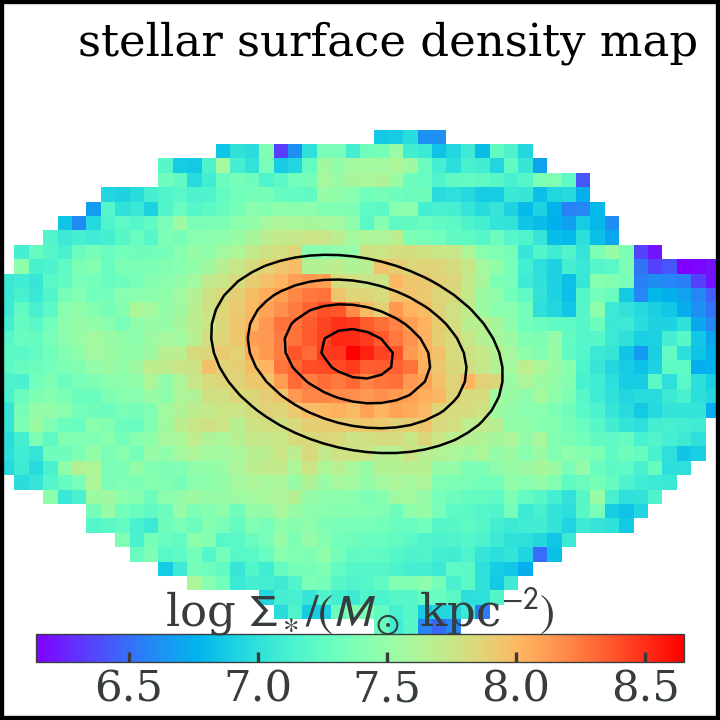}
    \includegraphics[width=.16\textwidth]{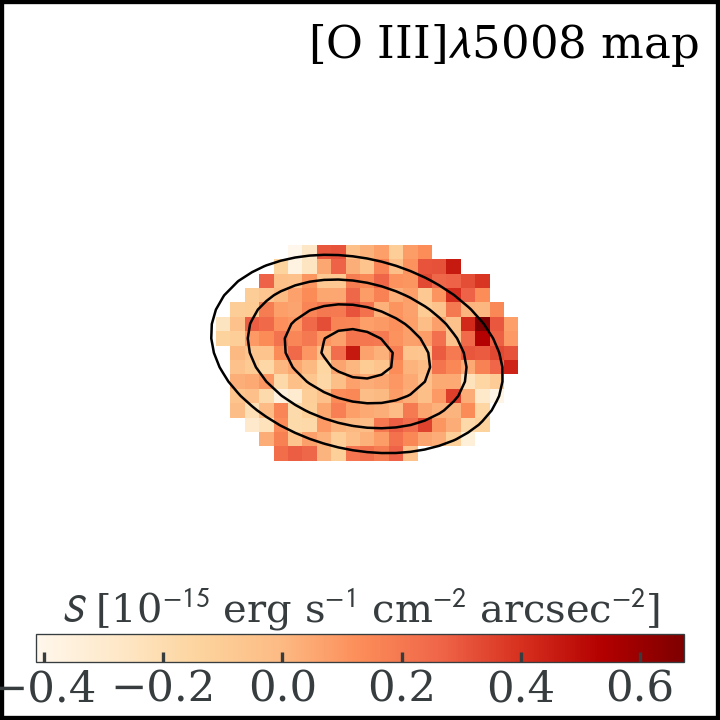}
    \includegraphics[width=.16\textwidth]{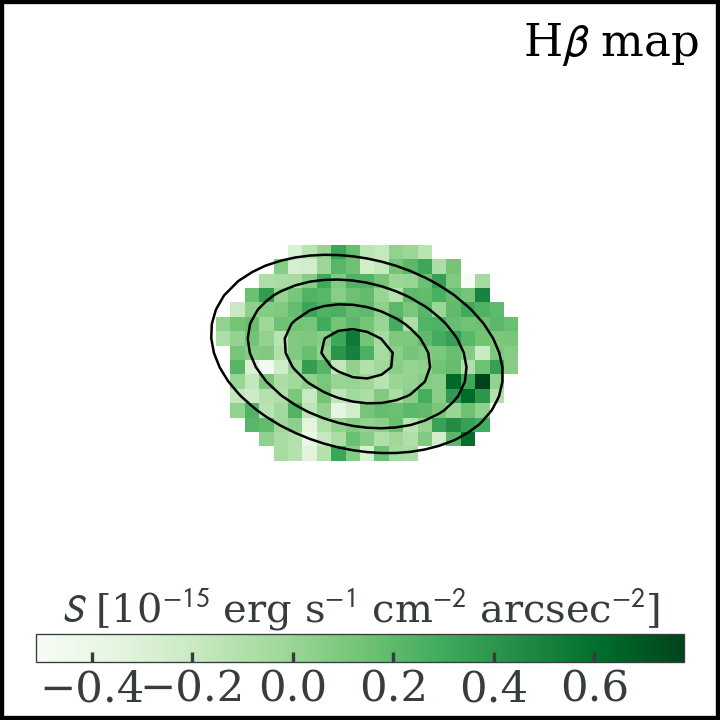}
    \includegraphics[width=.16\textwidth]{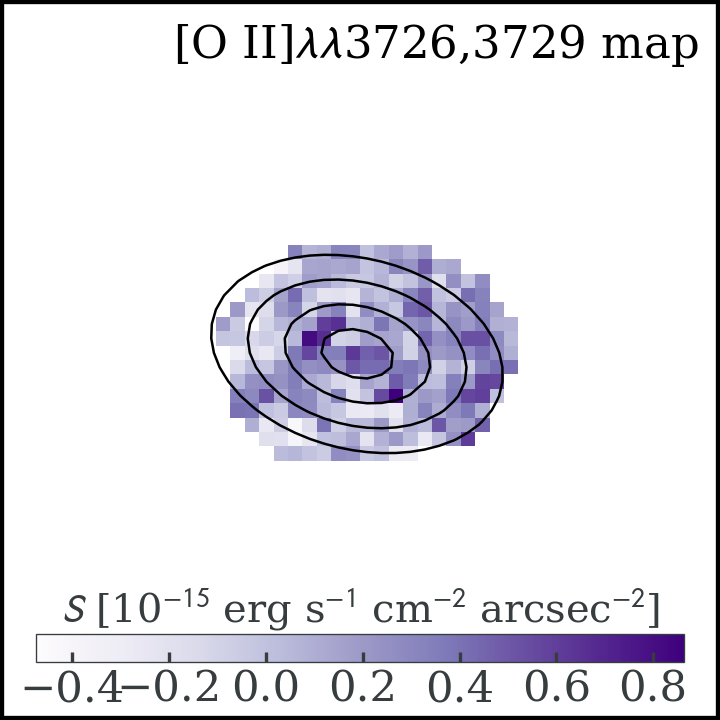}
    \includegraphics[width=.16\textwidth]{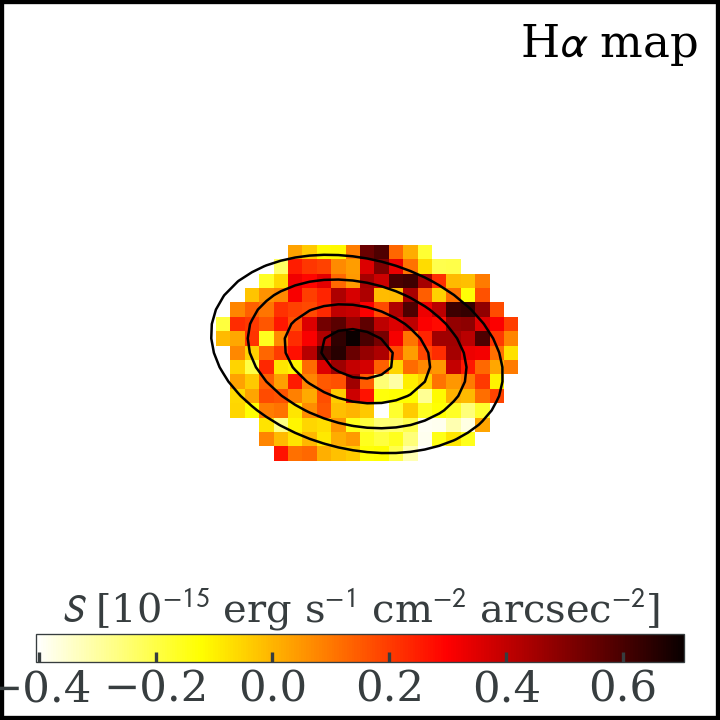}\\
    \includegraphics[width=\textwidth]{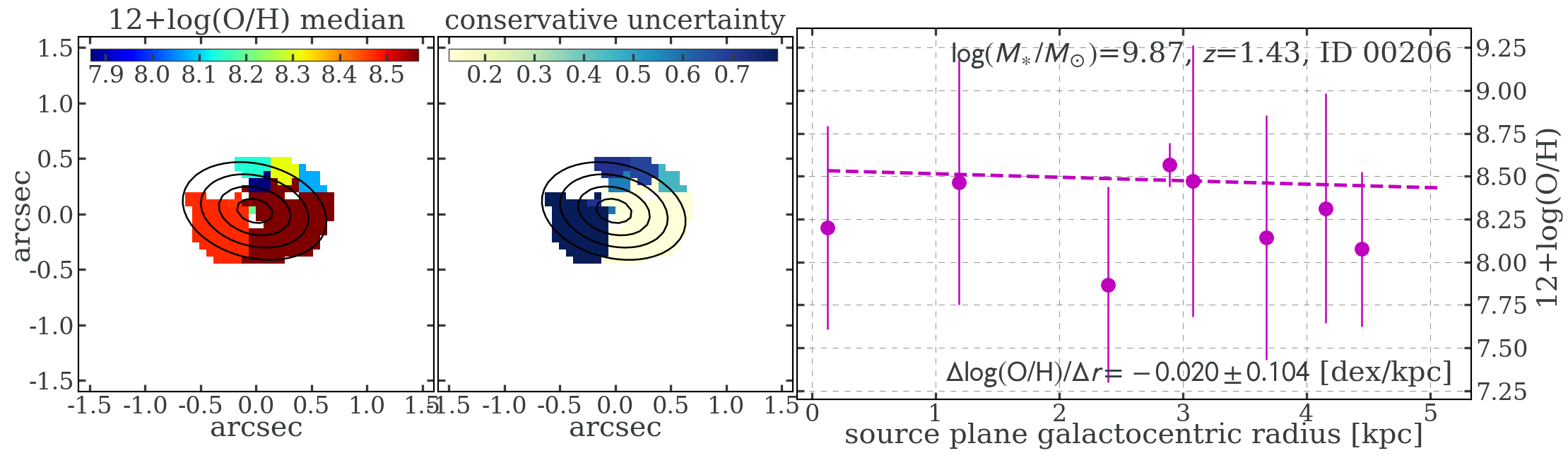}
    \caption{The source ID00206 in the field of \clliu is shown.}
    \label{fig:clRXJ2248_ID00206_figs}
\end{figure*}
\clearpage

\begin{figure*}
    \centering
    \includegraphics[width=\textwidth]{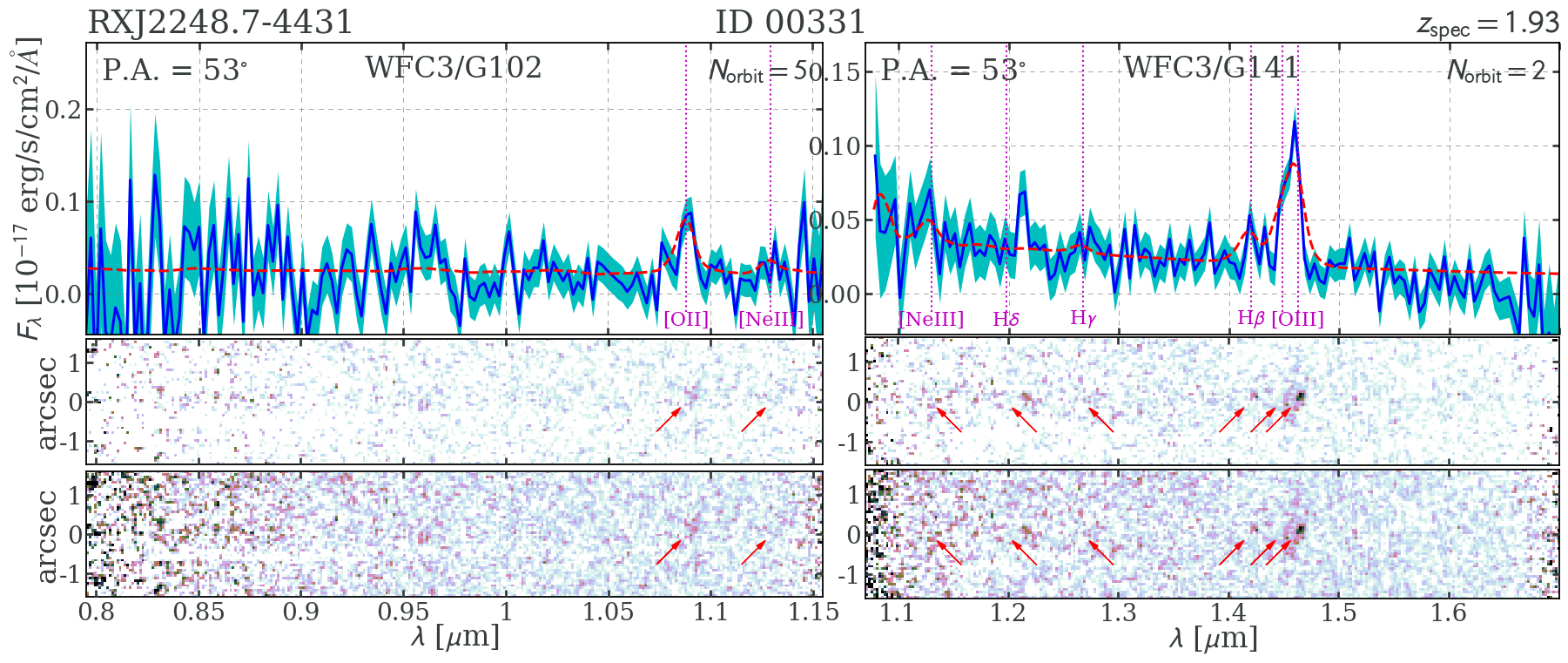}\\
    \includegraphics[width=\textwidth]{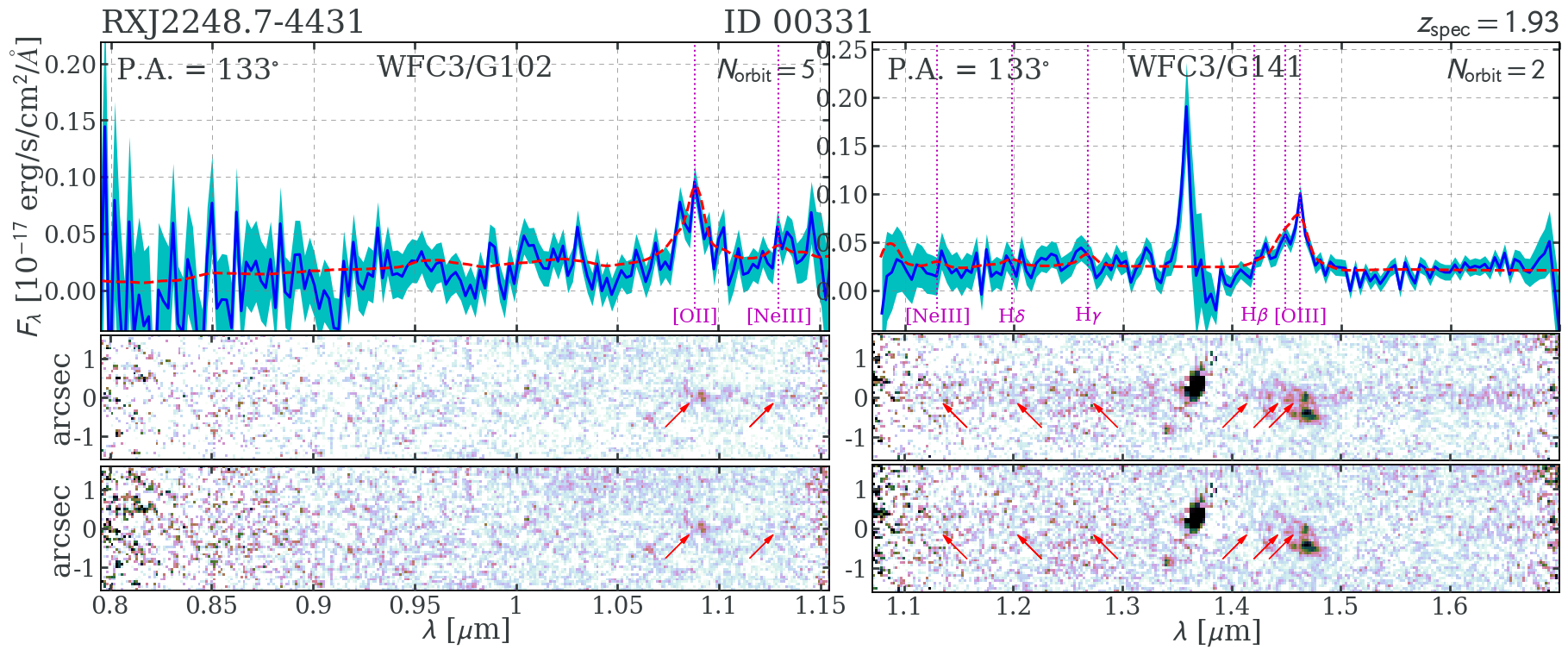}\\
    \includegraphics[width=.16\textwidth]{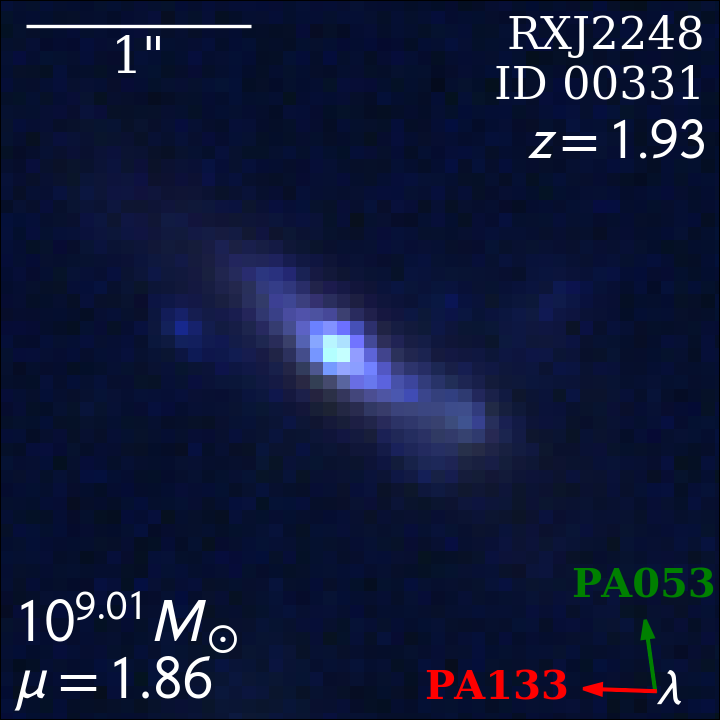}
    \includegraphics[width=.16\textwidth]{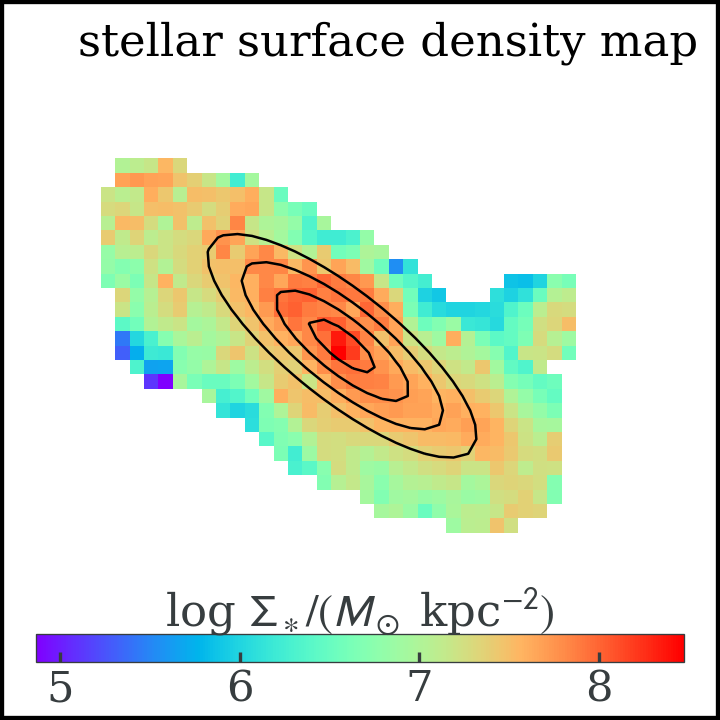}
    \includegraphics[width=.16\textwidth]{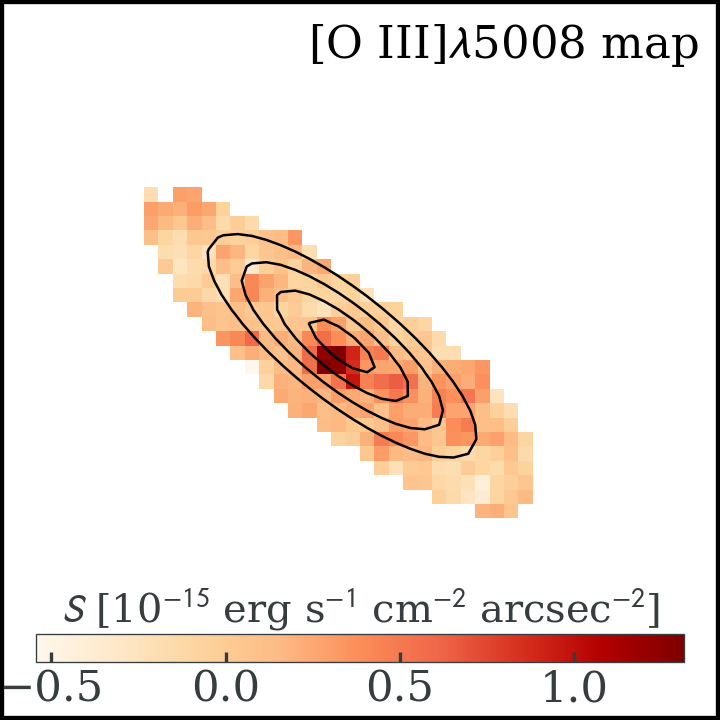}
    \includegraphics[width=.16\textwidth]{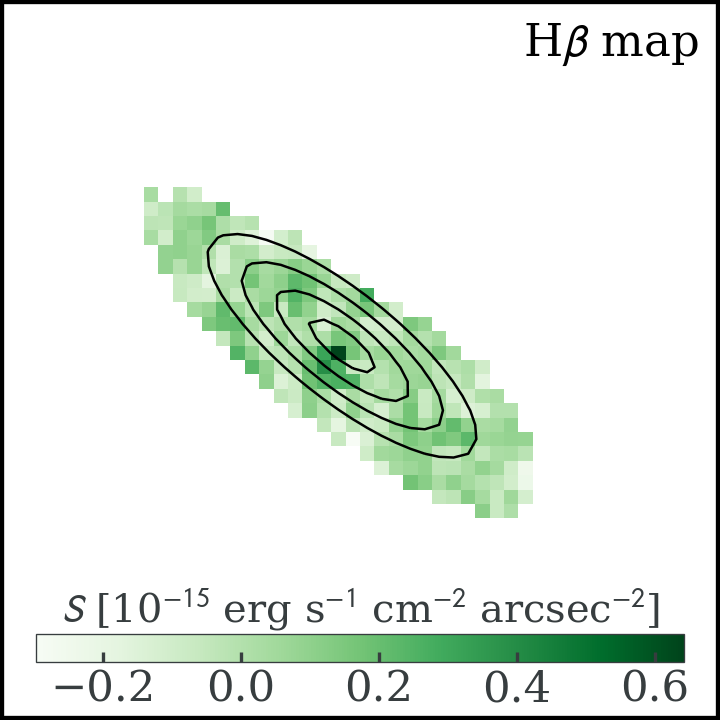}
    \includegraphics[width=.16\textwidth]{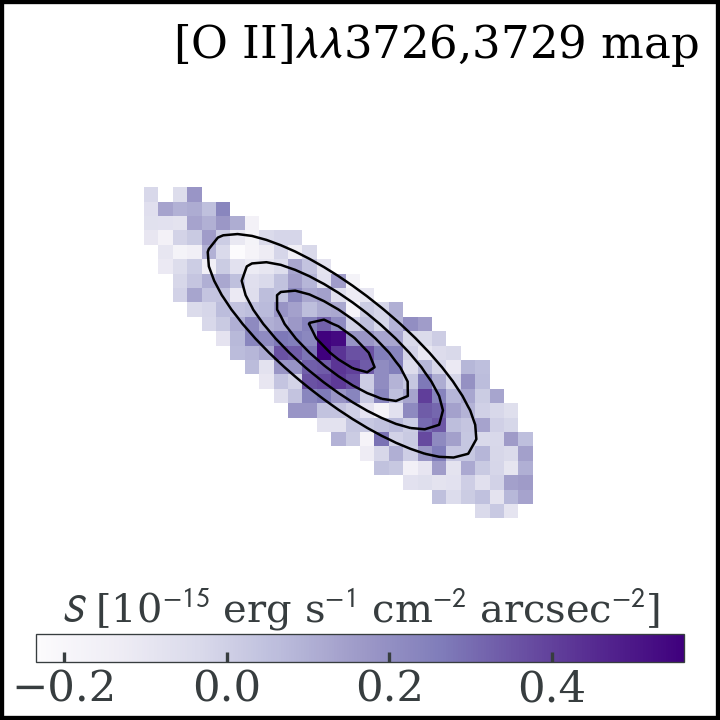}
    \includegraphics[width=.16\textwidth]{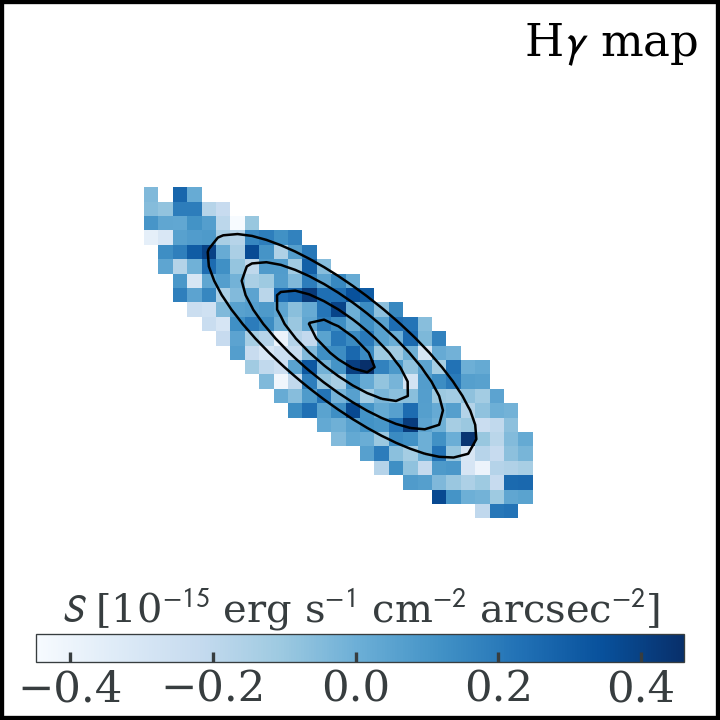}\\
    \includegraphics[width=\textwidth]{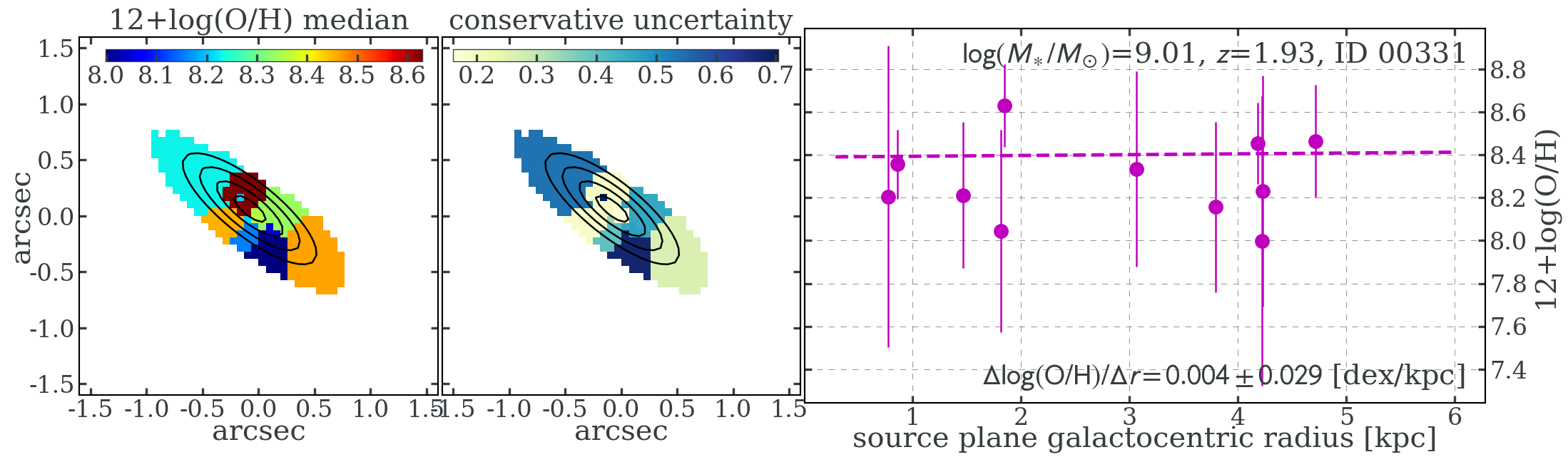}
    \caption{The source ID00331 in the field of \clliu is shown.}
    \label{fig:clRXJ2248_ID00331_figs}
\end{figure*}
\clearpage

\begin{figure*}
    \centering
    \includegraphics[width=\textwidth]{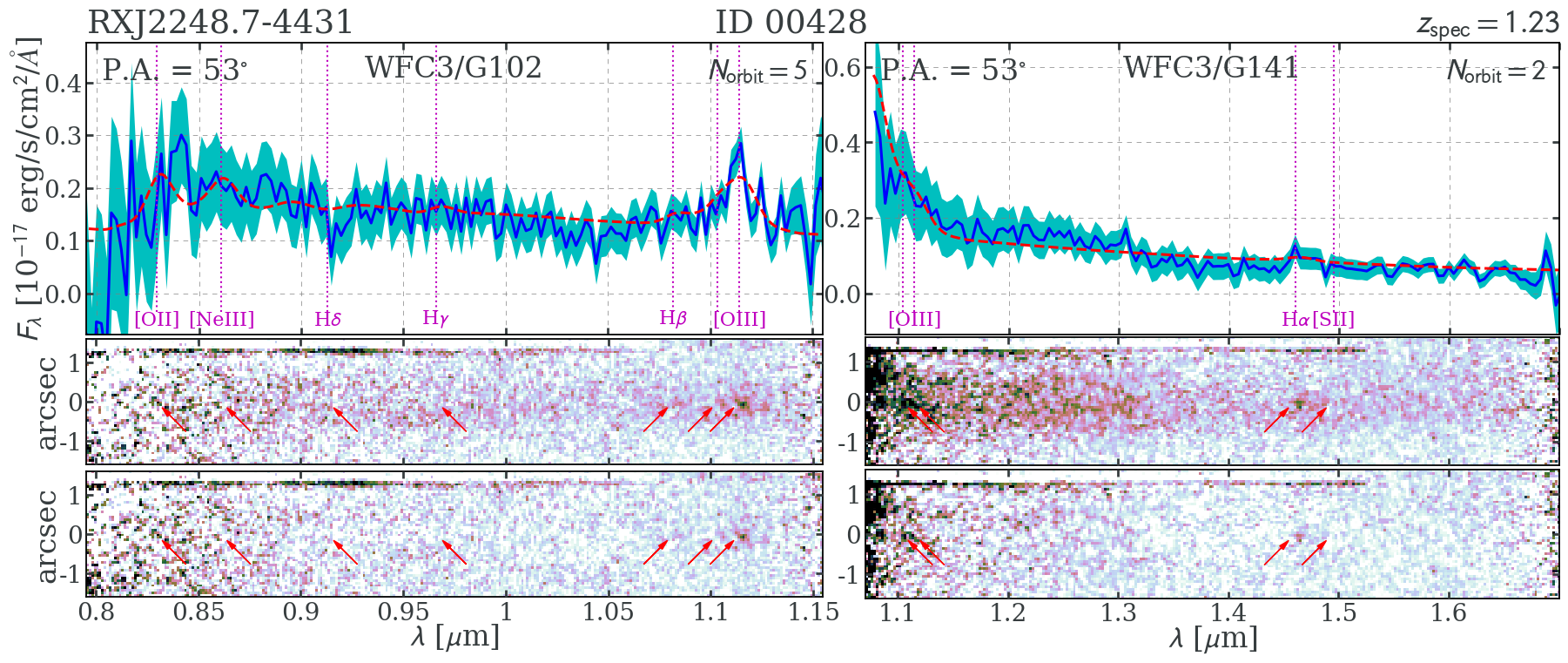}\\
    \includegraphics[width=\textwidth]{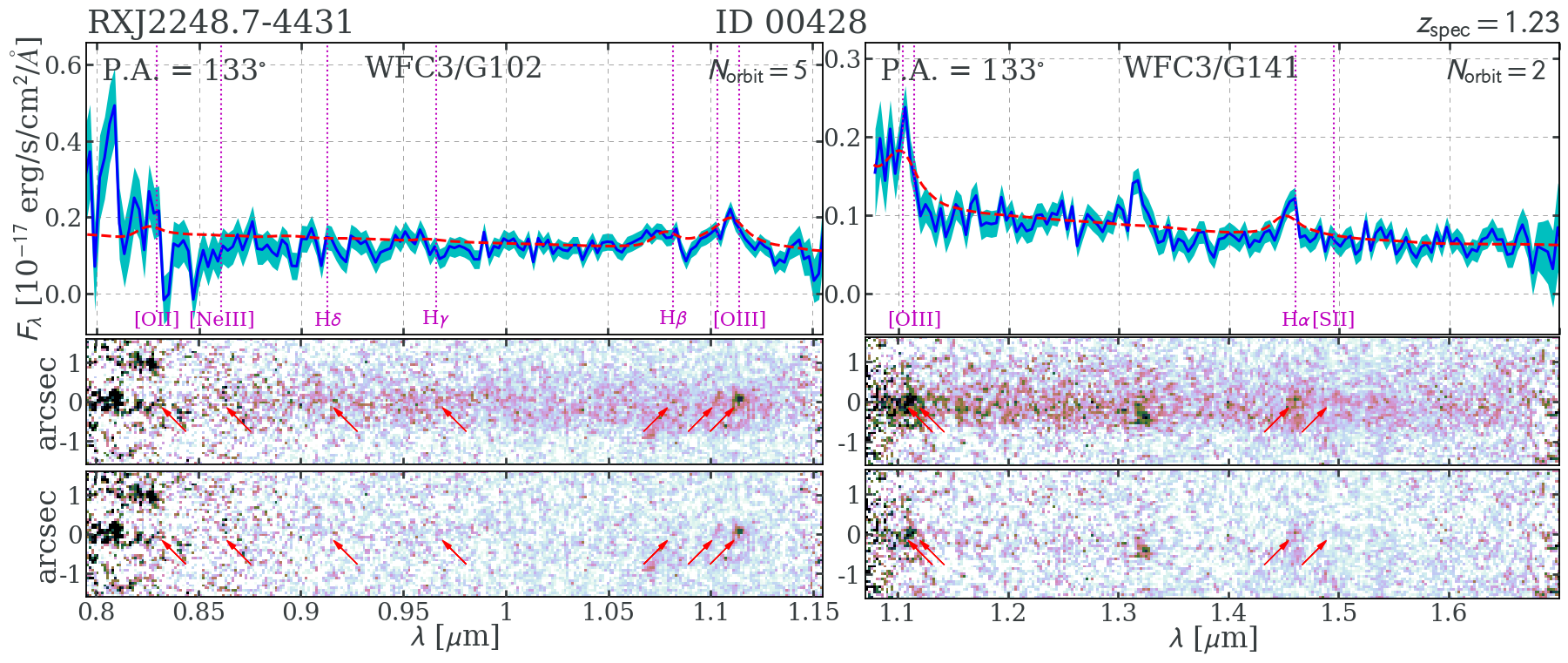}\\
    \includegraphics[width=.16\textwidth]{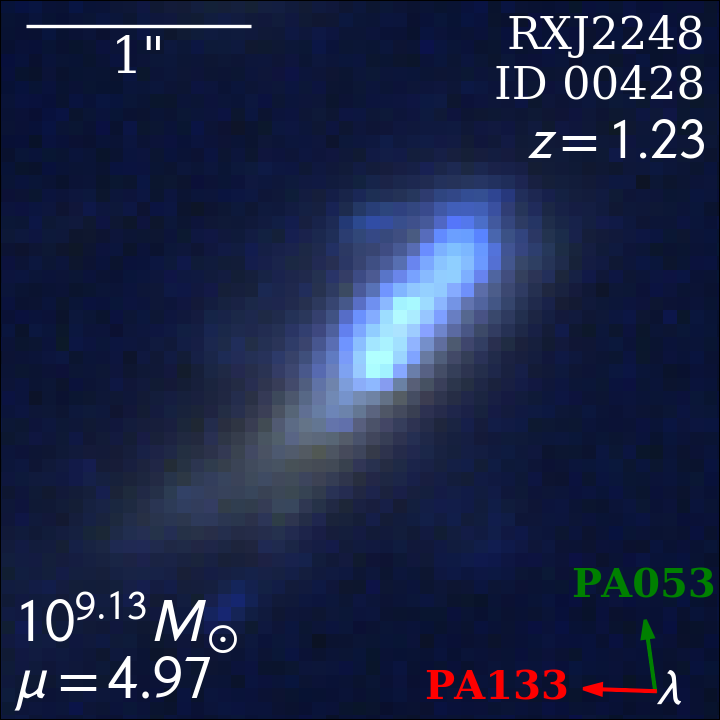}
    \includegraphics[width=.16\textwidth]{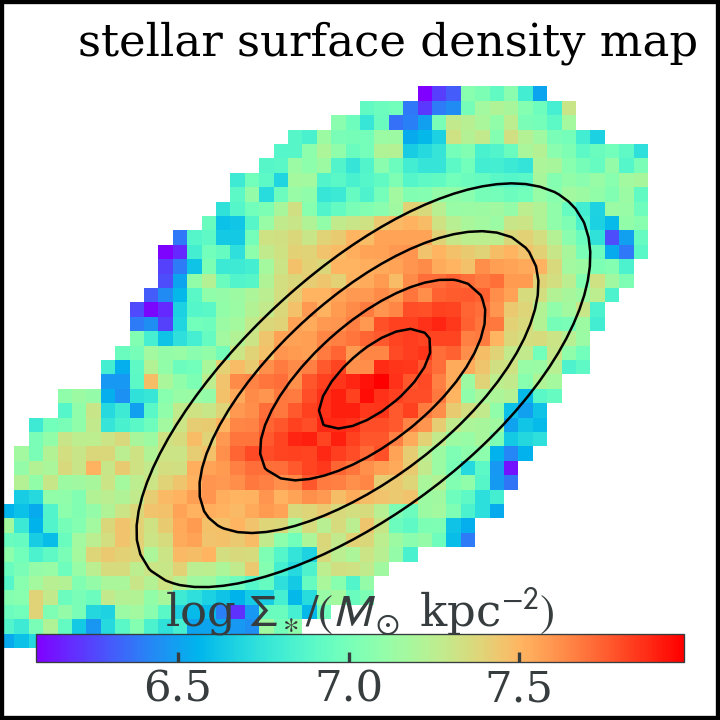}
    \includegraphics[width=.16\textwidth]{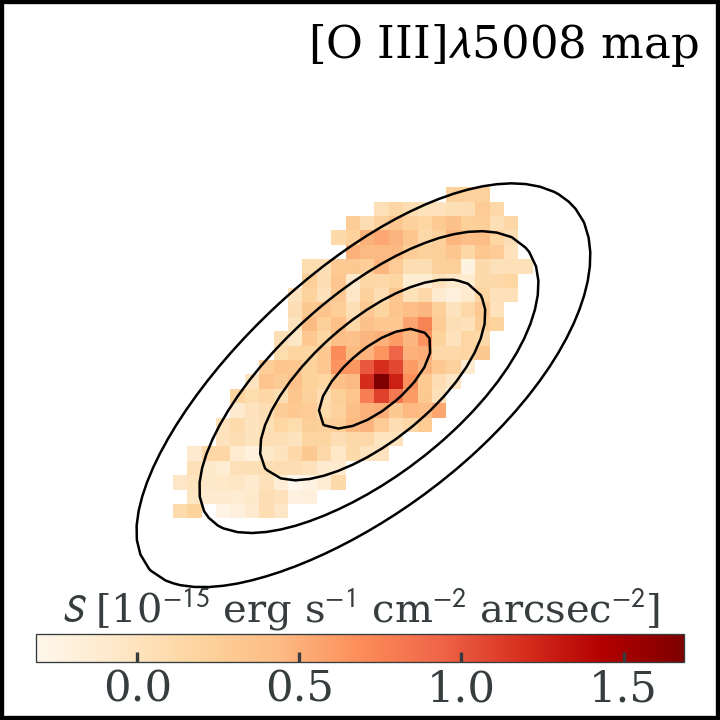}
    \includegraphics[width=.16\textwidth]{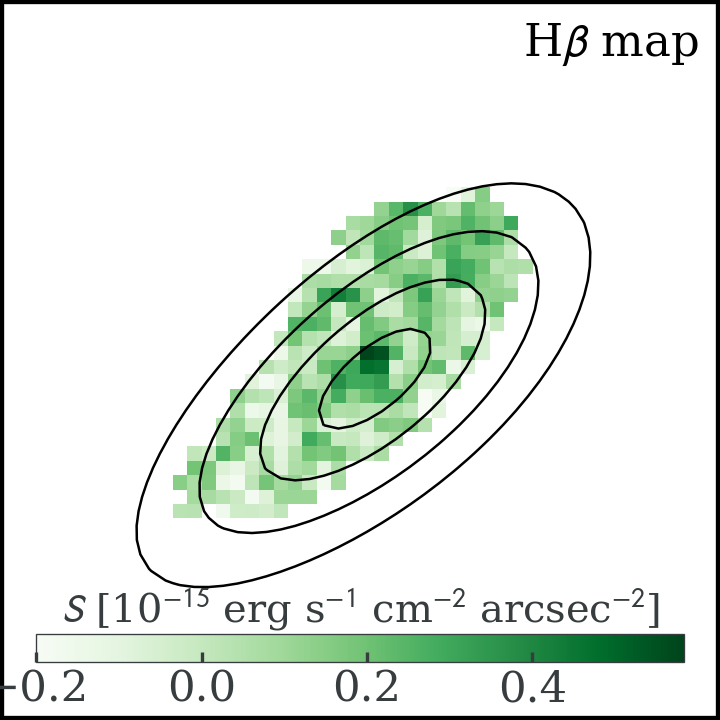}
    \includegraphics[width=.16\textwidth]{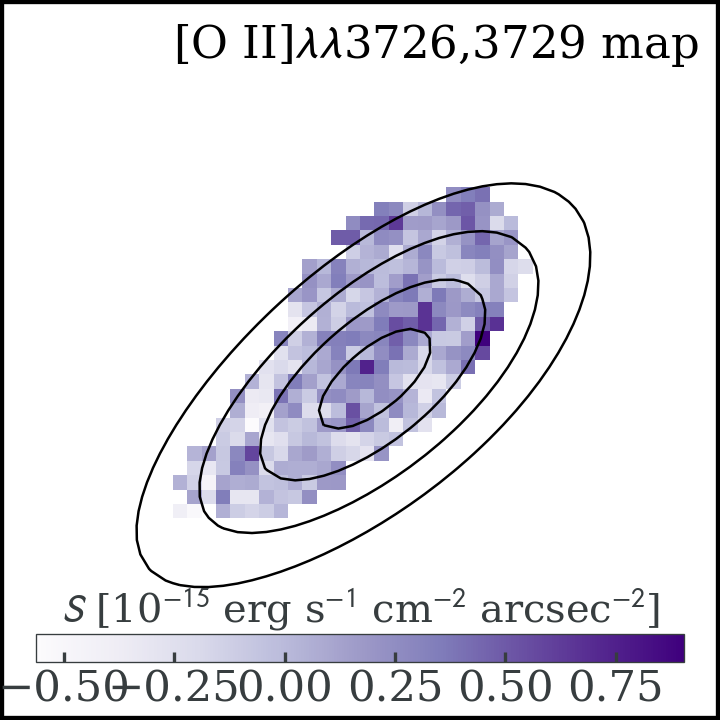}
    \includegraphics[width=.16\textwidth]{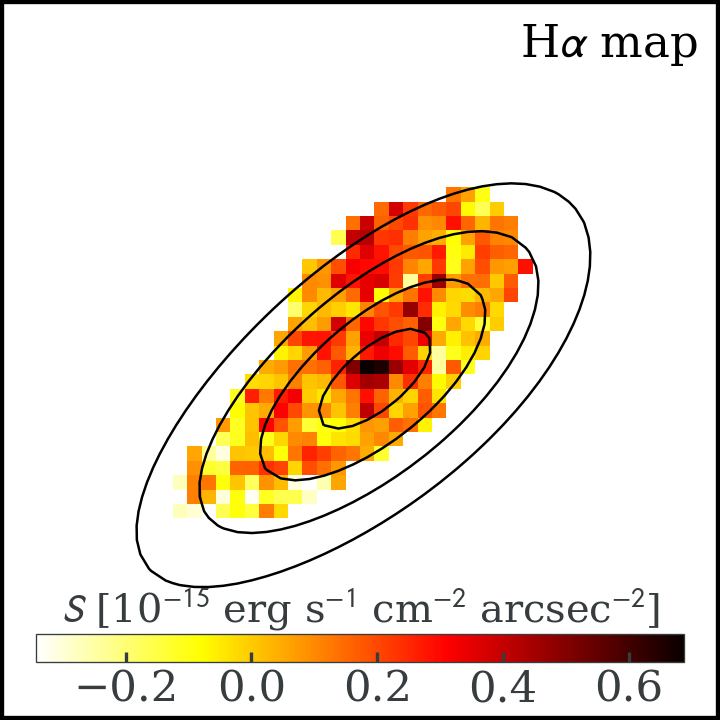}\\
    \includegraphics[width=\textwidth]{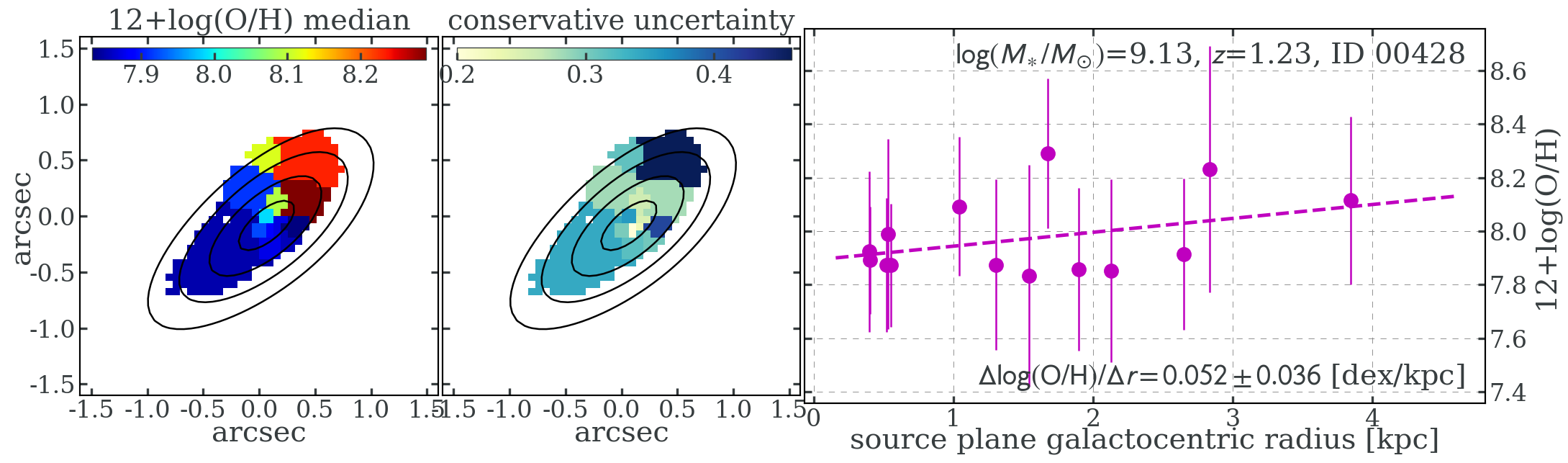}
    \caption{The source ID00428 in the field of \clliu is shown.}
    \label{fig:clRXJ2248_ID00428_figs}
\end{figure*}
\clearpage

\begin{figure*}
    \centering
    \includegraphics[width=\textwidth]{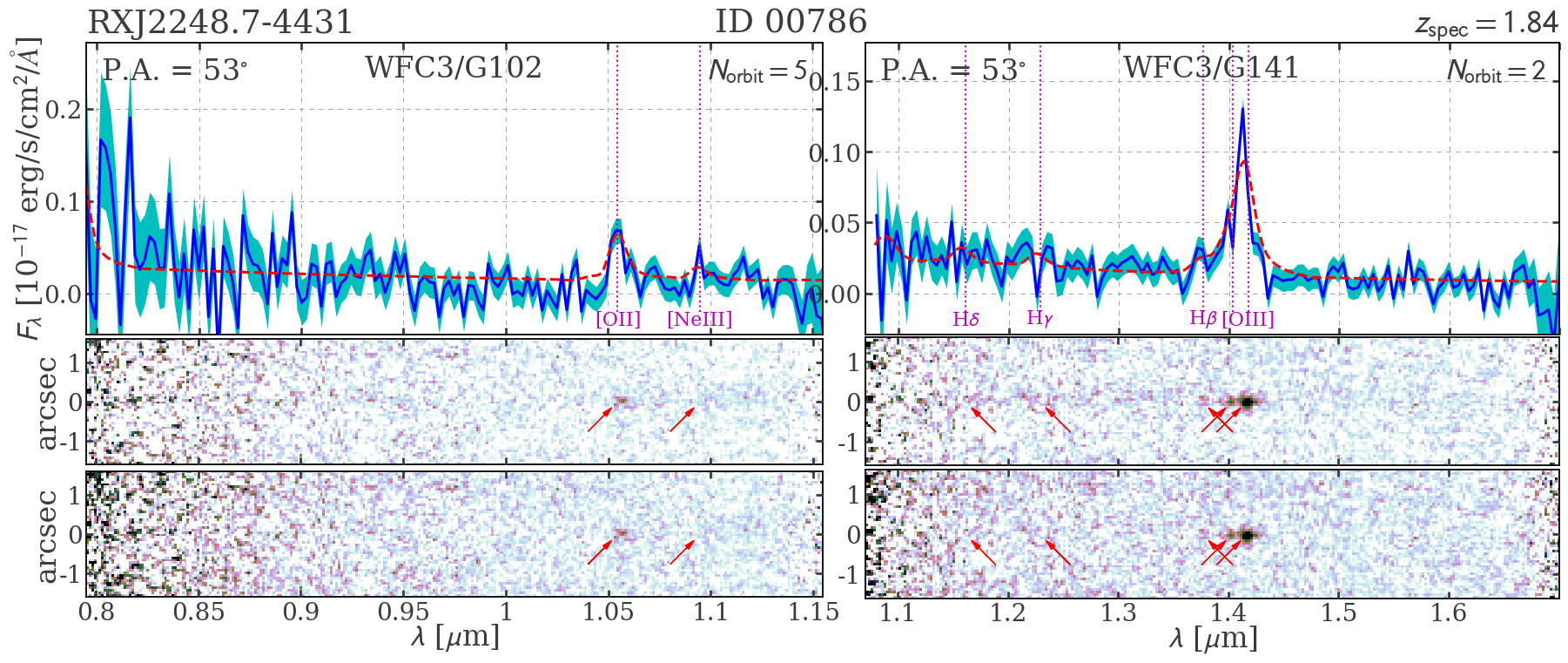}\\
    \includegraphics[width=\textwidth]{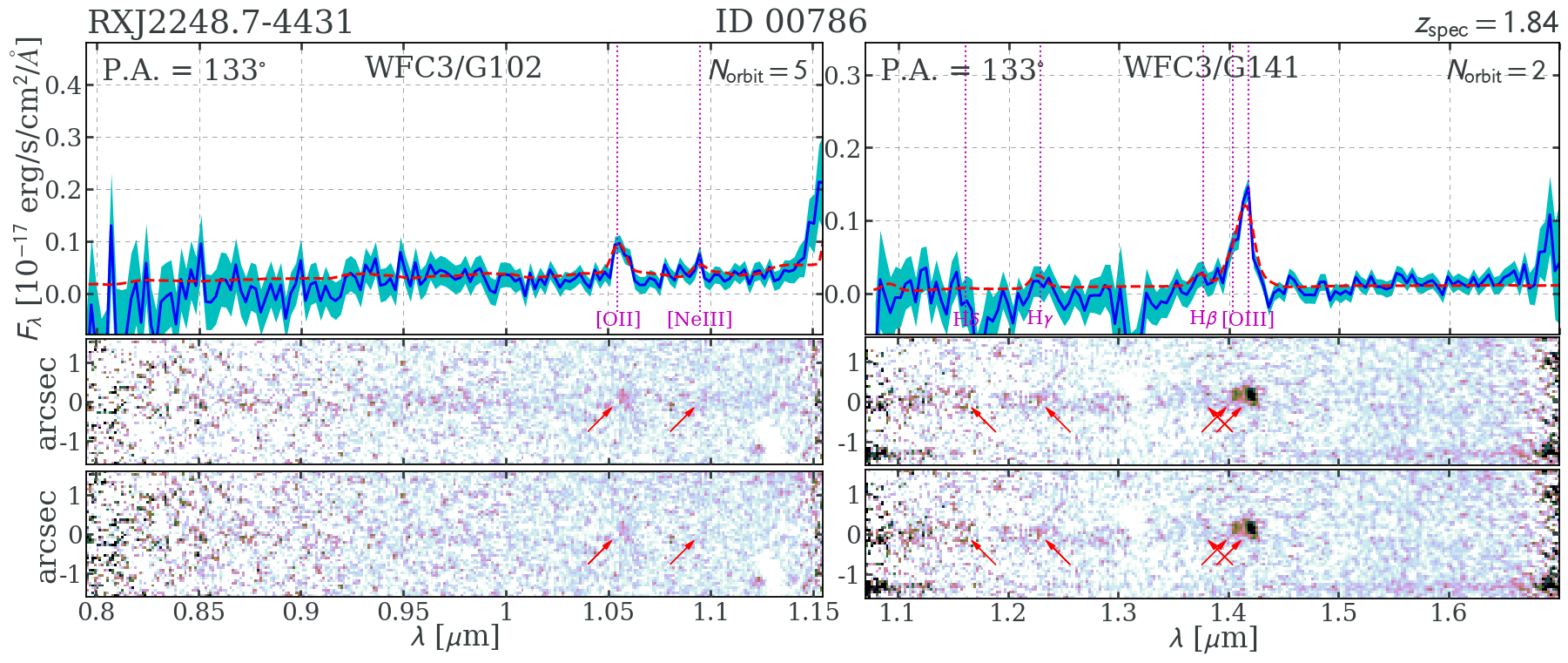}\\
    \includegraphics[width=.16\textwidth]{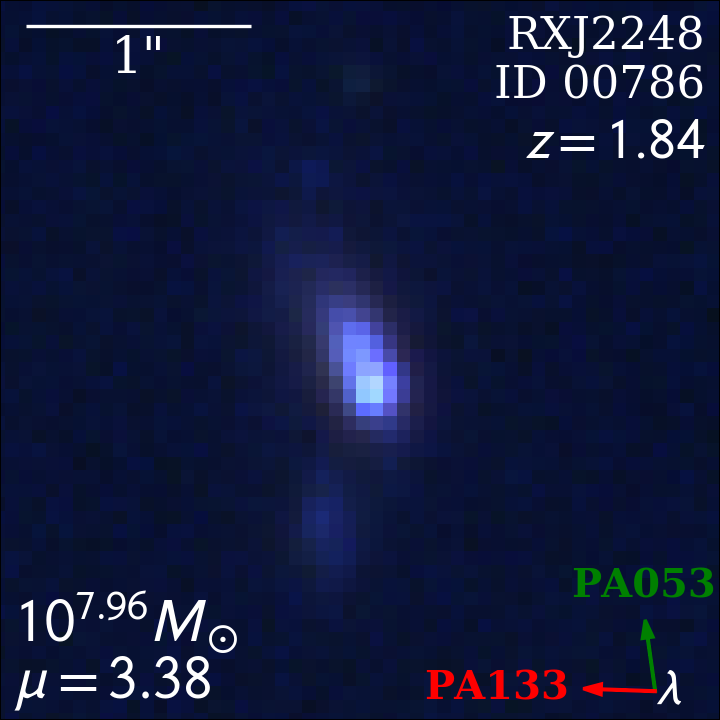}
    \includegraphics[width=.16\textwidth]{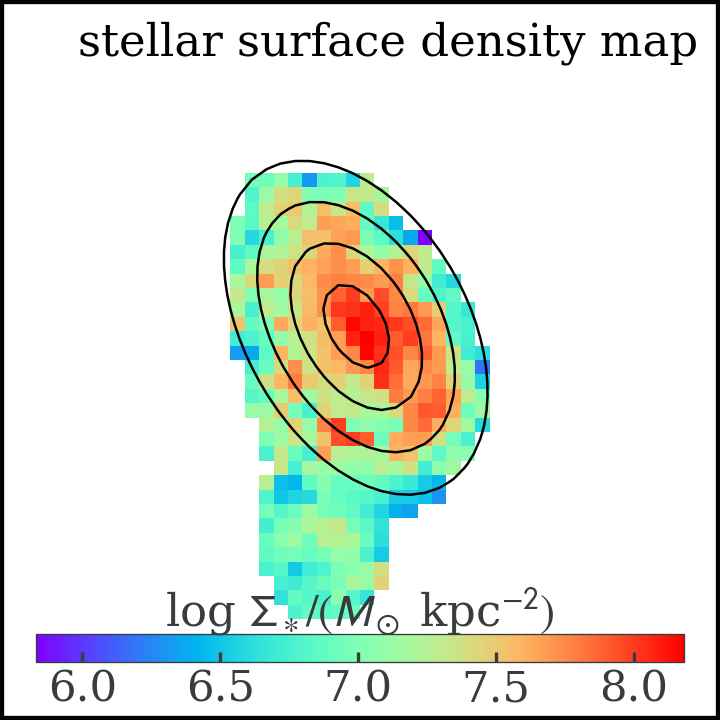}
    \includegraphics[width=.16\textwidth]{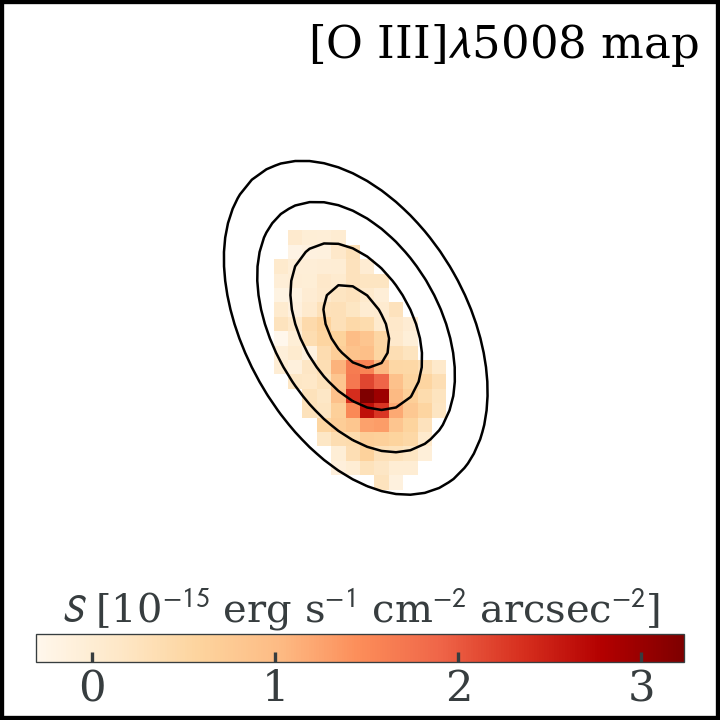}
    \includegraphics[width=.16\textwidth]{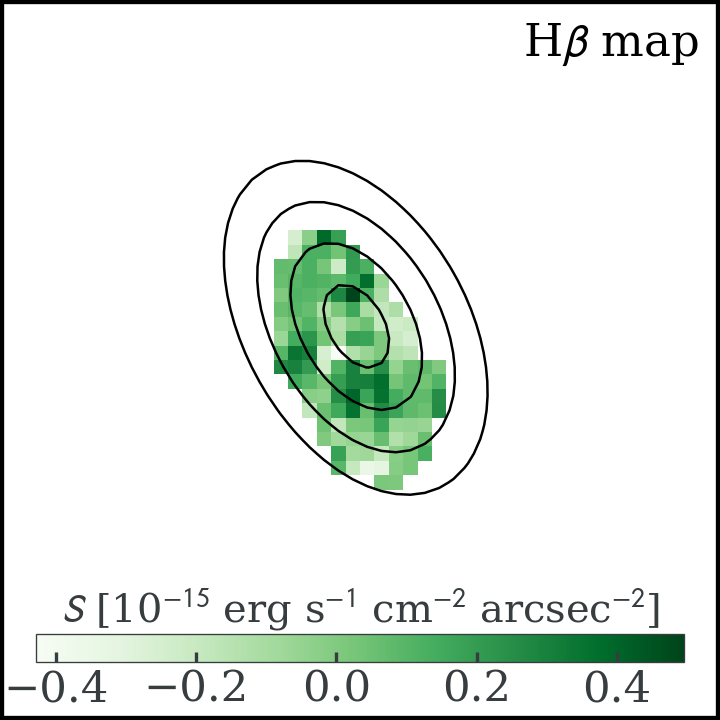}
    \includegraphics[width=.16\textwidth]{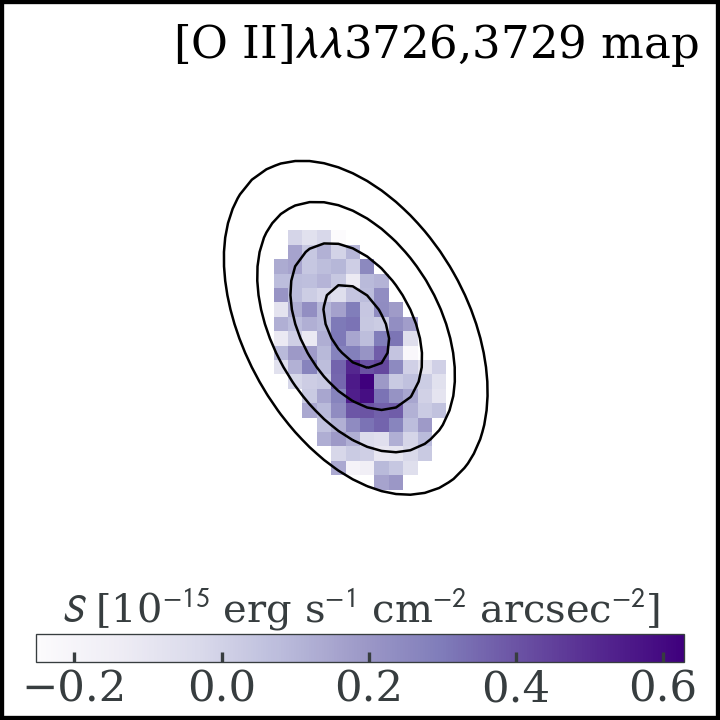}
    \includegraphics[width=.16\textwidth]{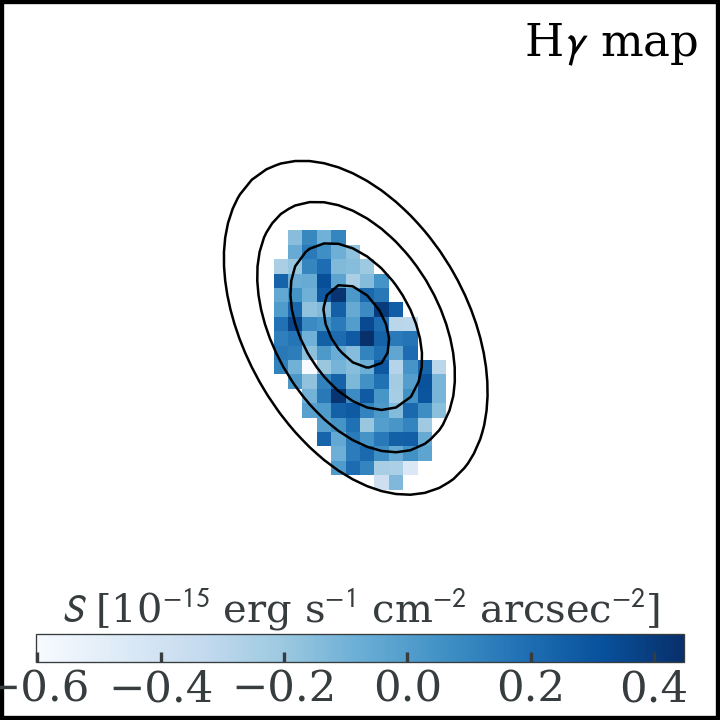}\\
    \includegraphics[width=\textwidth]{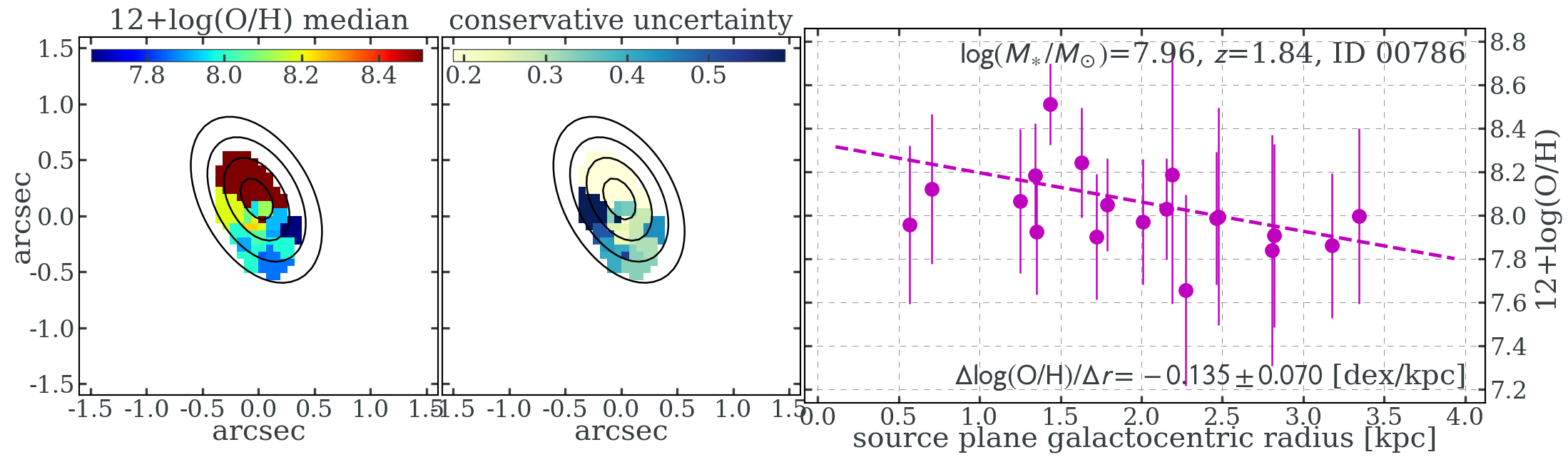}
    \caption{The source ID00786 in the field of \clliu is shown.}
    \label{fig:clRXJ2248_ID00786_figs}
\end{figure*}
\clearpage

\clearpage

\end{appendix}

\begin{longrotatetable}

\tabletypesize{\tiny}
\tabcolsep=2pt
\begin{deluxetable*}{cccccccccccccccccccccc}
    \tablecolumns{22}
    \tablewidth{0pt}
    \tablecaption{Measured quantities of our sample galaxies.}
\tablehead{
    \colhead{Cluster} & 
    \colhead{ID} & 
    \colhead{R.A.} & 
    \colhead{Dec.} & 
    \colhead{$z_\textrm{spec}$} & 
    \colhead{$\Delta\log({\rm O/H})/\Delta r$} & 
    \colhead{F140W\tablenotemark{a}} & 
    \colhead{$r_{\rm F140W}$\tablenotemark{b}} & 
    \multicolumn{6}{c}{Observed integrated EL fluxes [$10^{-17}$\Funit]} &
    \colhead{$\mu$\tablenotemark{c}} & 
    \multicolumn{3}{c}{Stellar continuum SED fitting} &
    \colhead{[N II]/H$\alpha$\tablenotemark{d}} & 
    \multicolumn{3}{c}{Nebular emission diagnostics} \\
    & & [deg.] & [deg.] &  & [dex/kpc] & [ABmag] &  & \multicolumn{6}{c}{\hrulefill} &  & \multicolumn{3}{c}{\hrulefill} &  & \multicolumn{3}{c}{\hrulefill} \\
    & & & & & & & & 
    \colhead{$f_{\rm [OII]}$} & 
    \colhead{$f_{\rm H\gamma}$} & 
    \colhead{$f_{\rm H\beta}$} & 
    \colhead{$f_{\rm [OIII]}$} & 
    \colhead{$f_{\rm H\alpha}$} & 
    \colhead{$f_{\rm [SII]}$} & 
    & 
    \colhead{log(\Mstar\tablenotemark{e}/\Msun)} & 
    \colhead{SFR$^{\rm S}$\tablenotemark{e} [\Msun/yr]} & 
    \colhead{A$_{\rm V}^{\rm S}$} & 
    &
    \colhead{$12+\log({\rm O/H})$} & 
    \colhead{A$_{\rm V}^{\rm N}$} & 
    \colhead{SFR$^{\rm N}$\tablenotemark{e} [\Msun/yr]}
}
\startdata
 A370 & 01513 & 39.958904 & -1.565269 & 1.27 & $-0.230\pm0.026$ & 21.17 & 0.95 & $20.05\pm2.47$ & $1.36\pm1.06$ & $8.53\pm0.80$ & $28.99\pm0.88$ & $37.55\pm1.07$ & $5.20\pm1.10$ & $2.80_{-0.03}^{+0.02}$ & $9.72_{-0.03}^{+0.23}$ & $1.30_{-1.26}^{+58.61}$ & $0.90_{-0.38}^{+0.50}$ & $0.19_{-0.01}^{+0.08}$ & $8.36_{-0.13}^{+0.10}$ & $1.02_{-0.42}^{+0.45}$ & $10.70_{-3.43}^{+5.11}$ \\ 
 A370 & 01590 & 39.965663 & -1.566864 & 1.96 & $0.115\pm0.056$ & 23.85 & 0.57 & $4.61\pm0.79$ & $4.60\pm2.74$ & $2.89\pm2.48$ & $19.48\pm1.05$ & \nodata & \nodata & $17.76_{-7.58}^{+10.14}$ & $7.48_{-0.35}^{+0.32}$ & $0.28_{-0.18}^{+0.55}$ & $0.00_{-0.00}^{+0.21}$ & $0.04_{-0.00}^{+0.00}$ & $8.07_{-0.35}^{+0.21}$ & $<1.06$ & $2.73_{-2.00}^{+13.54}$ \\ 
 A370 & 02056 & 39.959425 & -1.573309 & 1.27 & $0.023\pm0.009$ & 23.30 & 0.91 & $13.75\pm2.00$ & $4.74\pm0.92$ & $3.25\pm0.70$ & $32.83\pm0.83$ & $10.15\pm1.14$ & \nodata & $3.55_{-0.10}^{+0.08}$ & $8.50_{-0.09}^{+0.10}$ & $5.01_{-4.27}^{+2.97}$ & $1.00_{-0.33}^{+0.19}$ & $0.05_{-0.00}^{+0.01}$ & $7.98_{-0.20}^{+0.18}$ & $<0.73$ & $4.03_{-1.74}^{+5.14}$ \\ 
 A370 & 02546 & 39.961942 & -1.577895 & 1.27 & $0.029\pm0.108$ & 24.24 & 0.90 & $6.56\pm1.60$ & $0.58\pm0.72$ & $1.23\pm0.66$ & $6.98\pm0.66$ & $3.81\pm0.68$ & $0.98\pm0.68$ & $3.41_{-0.08}^{+0.05}$ & $8.25_{-0.22}^{+0.13}$ & $0.61_{-0.52}^{+0.58}$ & $0.20_{-0.20}^{+0.32}$ & $0.05_{-0.00}^{+0.00}$ & $8.10_{-0.35}^{+0.25}$ & $<0.64$ & $0.92_{-0.35}^{+1.12}$ \\ 
 A370 & 03097 & 39.959500 & -1.584013 & 1.55 & $-0.016\pm0.054$ & 22.56 & 0.94 & $11.81\pm0.97$ & $5.65\pm0.63$ & $4.03\pm1.31$ & $6.41\pm1.14$ & $34.27\pm1.60$ & \nodata  & $2.55_{-0.08}^{+0.06}$ & $9.28_{-0.06}^{+0.21}$ & $0.73_{-0.73}^{+8.55}$ & $0.30_{-0.30}^{+0.52}$ & $0.11_{-0.01}^{+0.01}$ & $8.78_{-0.12}^{+0.09}$ & $1.55_{-0.30}^{+0.29}$ & $26.80_{-6.63}^{+8.71}$ \\ 
 A370 & 03751 & 39.977361 & -1.591636 & 1.97 & $0.122\pm0.008$ & 21.55 & 0.70 & $29.57\pm0.51$ & $7.21\pm0.67$ & $17.68\pm0.68$ & $111.41\pm0.84$ & \nodata & \nodata & $6.35_{-0.58}^{+0.40}$ & $9.05_{-0.06}^{+0.05}$ & $32.91_{-10.50}^{+3.28}$ & $0.80_{-0.11}^{+0.01}$ & $0.07_{-0.00}^{+0.00}$ & $8.08_{-0.12}^{+0.11}$ & $0.85_{-0.14}^{+0.21}$ & $26.23_{-2.34}^{+3.19}$ \\ 
 A370 & 03834 & 39.979759 & -1.591402 & 1.25 & $0.017\pm0.011$ & 23.56 & 0.91 & $2.51\pm2.71$ & $3.71\pm0.74$ & $2.56\pm0.60$ & $15.80\pm0.79$ & $7.02\pm0.70$ & $0.78\pm0.69$ & $3.18_{-0.12}^{+0.11}$ & $8.58_{-0.08}^{+0.27}$ & $0.33_{-0.24}^{+8.25}$ & $0.90_{-0.26}^{+0.60}$ & $0.06_{-0.01}^{+0.01}$ & $7.88_{-0.22}^{+0.22}$ & $<0.32$ & $1.43_{-0.35}^{+0.77}$ \\ 
 A2744 & 00144 & 3.606322 & -30.380061 & 1.34 & $0.020\pm0.085$ & 22.76 & 0.92 & $7.45\pm2.39$ & $2.28\pm1.12$ & $3.71\pm3.10$ & $6.38\pm1.45$ & $14.13\pm1.17$ & $1.08\pm1.19$ & $1.43_{-0.01}^{+0.01}$ & $9.56_{-0.21}^{+0.17}$ & $15.30_{-14.67}^{+22.41}$ & $1.10_{-0.50}^{+0.42}$ & $0.15_{-0.04}^{+0.04}$ & $8.55_{-0.22}^{+0.13}$ & $0.77_{-0.45}^{+0.57}$ & $7.66_{-2.45}^{+4.52}$ \\ 
 A2744 & 00161 & 3.588863 & -30.380291 & 1.36 & $0.027\pm0.063$ & 23.55 & 0.93 & $4.50\pm1.30$ & $1.95\pm0.65$ & $2.71\pm1.86$ & $12.51\pm1.21$ & $6.31\pm0.70$ & $0.25\pm0.71$ & $2.12_{-0.01}^{+0.02}$ & $8.53_{-0.09}^{+0.18}$ & $2.72_{-2.41}^{+5.94}$ & $0.40_{-0.38}^{+0.32}$ & $0.05_{-0.00}^{+0.01}$ & $8.00_{-0.25}^{+0.21}$ & $<0.27$ & $1.88_{-0.31}^{+0.65}$ \\ 
 A2744 & 00169 & 3.571780 & -30.380437 & 1.68 & $-0.147\pm0.079$ & 23.83 & 0.58 & $4.45\pm0.59$ & $3.94\pm1.00$ & $3.07\pm0.70$ & $18.64\pm0.74$ & \nodata & \nodata & $2.76_{-0.02}^{+0.02}$ & $8.17_{-0.05}^{+0.02}$ & $4.36_{-1.30}^{+0.03}$ & $0.40_{-0.14}^{+0.01}$ & $0.05_{-0.00}^{+0.00}$ & $8.06_{-0.25}^{+0.19}$ & $<1.06$ & $<14.87$ \\ 
 A2744 & 00188 & 3.573408 & -30.380770 & 1.76 & $0.033\pm0.030$ & 21.77 & 0.95 & $16.97\pm0.82$ & $3.00\pm1.43$ & $5.48\pm1.02$ & $13.56\pm1.09$ & \nodata & \nodata & $2.79_{-0.02}^{+0.03}$ & $9.93_{-0.07}^{+0.12}$ & $1.39_{-1.39}^{+20.38}$ & $0.90_{-0.34}^{+0.30}$ & $0.20_{-0.02}^{+0.04}$ & $8.55_{-0.10}^{+0.09}$ & $<0.68$ & $12.86_{-5.74}^{+16.11}$ \\ 
 A2744 & 00263 & 3.591110 & -30.381705 & 1.89 & $0.018\pm0.014$ & 23.02 & 0.65 & $6.68\pm0.55$ & $3.19\pm0.84$ & $4.14\pm0.68$ & $34.01\pm0.75$ & \nodata & \nodata & $2.21_{-0.02}^{+0.02}$ & $8.58_{-0.01}^{+0.12}$ & $10.39_{-4.87}^{+2.19}$ & $0.20_{-0.14}^{+0.10}$ & $0.05_{-0.00}^{+0.00}$ & $8.02_{-0.17}^{+0.14}$ & $0.90_{-0.62}^{+0.88}$ & $21.64_{-10.71}^{+34.81}$ \\ 
 A2744 & 00892 & 3.599948 & -30.393554 & 1.34 & $0.044\pm0.041$ & 23.21 & 0.74 & $14.74\pm0.98$ & $3.14\pm0.48$ & $6.97\pm0.45$ & $37.56\pm1.01$ & $26.46\pm0.54$ & $5.58\pm0.53$ & $1.90_{-0.02}^{+0.02}$ & $8.63_{-0.04}^{+0.03}$ & $9.16_{-3.56}^{+3.59}$ & $0.90_{-0.13}^{+0.10}$ & $0.06_{-0.00}^{+0.00}$ & $8.25_{-0.10}^{+0.09}$ & $0.39_{-0.24}^{+0.31}$ & $8.67_{-1.60}^{+2.53}$ \\ 
 A2744 & 00893 & 3.600091 & -30.393759 & 1.34 & $-0.066\pm0.085$ & 22.58 & 0.90 & $15.59\pm1.46$ & $1.02\pm0.75$ & $3.97\pm0.67$ & $17.64\pm1.43$ & $17.00\pm0.79$ & $4.91\pm0.79$ & $1.90_{-0.02}^{+0.02}$ & $9.24_{-0.12}^{+0.15}$ & $25.81_{-24.47}^{+17.37}$ & $1.20_{-0.52}^{+0.21}$ & $0.10_{-0.01}^{+0.02}$ & $8.39_{-0.12}^{+0.10}$ & $<0.38$ & $5.33_{-1.03}^{+1.83}$ \\ 
 A2744 & 00895 & 3.576755 & -30.393627 & 1.37 & $0.139\pm0.079$ & 22.91 & 0.87 & $14.76\pm1.07$ & $0.05\pm0.64$ & $6.59\pm0.91$ & $22.15\pm1.11$ & $18.12\pm0.64$ & $2.56\pm0.64$ & $3.29_{-0.04}^{+0.05}$ & $8.66_{-0.06}^{+0.15}$ & $3.11_{-2.07}^{+9.15}$ & $0.70_{-0.26}^{+0.30}$ & $0.06_{-0.00}^{+0.01}$ & $8.29_{-0.09}^{+0.09}$ & $1.41_{-0.29}^{+0.31}$ & $7.81_{-1.86}^{+2.54}$ \\ 
 A2744 & 01057 & 3.575089 & -30.399687 & 1.74 & $-0.033\pm0.049$ & 21.88 & 0.90 & $10.73\pm0.79$ & \nodata & $4.28\pm0.93$ & $26.52\pm1.05$ & \nodata & \nodata & $2.64_{-0.03}^{+0.03}$ & $9.52_{-0.15}^{+0.10}$ & $24.44_{-22.43}^{+17.59}$ & $0.50_{-0.37}^{+0.29}$ & $0.13_{-0.02}^{+0.01}$ & $8.21_{-0.13}^{+0.12}$ & $<0.61$ & $10.51_{-4.19}^{+13.74}$ \\ 
 A2744 & 01279 & 3.576955 & -30.400863 & 1.34 & $-0.051\pm0.014$ & 23.67 & 0.83 & $8.63\pm1.34$ & $1.23\pm0.68$ & $5.94\pm0.60$ & $29.19\pm1.27$ & $12.51\pm0.70$ & $1.54\pm0.69$ & $2.71_{-0.04}^{+0.03}$ & $8.12_{-0.03}^{+0.04}$ & $2.80_{-1.10}^{+1.12}$ & $0.40_{-0.17}^{+0.11}$ & $0.05_{-0.00}^{+0.00}$ & $8.00_{-0.15}^{+0.13}$ & $<0.22$ & $2.83_{-0.44}^{+0.79}$ \\ 
 A2744 & 01649 & 3.578893 & -30.409998 & 1.26 & $0.026\pm0.011$ & 22.67 & 0.95 & $8.27\pm4.96$ & $1.07\pm1.39$ & $1.67\pm1.58$ & $23.97\pm1.50$ & $10.04\pm1.54$ & \nodata & $2.46_{-0.03}^{+0.02}$ & $9.10_{-0.17}^{+0.21}$ & $4.35_{-4.30}^{+13.14}$ & $0.60_{-0.42}^{+0.52}$ & $0.10_{-0.02}^{+0.02}$ & $7.95_{-0.24}^{+0.21}$ & $1.33_{-0.86}^{+1.02}$ & $8.35_{-5.01}^{+15.09}$ \\ 
 A2744 & 01773 & 3.576699 & -30.410185 & 1.66 & $0.129\pm0.079$ & 22.95 & 0.88 & $8.68\pm0.72$ & $2.46\pm1.06$ & $2.74\pm0.81$ & $10.98\pm0.85$ & \nodata & \nodata & $2.00_{-0.02}^{+0.02}$ & $9.32_{-0.19}^{+0.10}$ & $8.31_{-8.24}^{+8.41}$ & $0.50_{-0.50}^{+0.33}$ & $0.11_{-0.02}^{+0.01}$ & $8.46_{-0.12}^{+0.12}$ & $<1.01$ & $14.68_{-8.68}^{+32.79}$ \\ 
 A2744 & 01897 & 3.574584 & -30.412278 & 1.37 & $-0.159\pm0.076$ & 22.87 & 0.90 & $15.77\pm1.74$ & $2.18\pm0.89$ & $2.23\pm1.10$ & $7.95\pm1.19$ & $13.62\pm1.02$ & $3.68\pm0.98$ & $1.68_{-0.01}^{+0.01}$ & $9.39_{-0.20}^{+0.12}$ & $10.56_{-10.47}^{+10.17}$ & $0.90_{-0.44}^{+0.38}$ & $0.12_{-0.03}^{+0.02}$ & $8.56_{-0.11}^{+0.09}$ & $<0.51$ & $5.17_{-1.39}^{+2.87}$ \\ 
 M0416 & 00094 & 64.033092 & -24.056326 & 1.36 & $-0.010\pm0.126$ & 22.25 & 0.93 & $8.52\pm1.42$ & $1.83\pm1.04$ & $0.16\pm0.98$ & $7.33\pm1.37$ & $17.85\pm0.85$ & $3.32\pm0.85$ & $1.47_{-0.01}^{+0.02}$ & $9.72_{-0.15}^{+0.08}$ & $10.27_{-10.06}^{+15.13}$ & $0.90_{-0.36}^{+0.27}$ & $0.18_{-0.03}^{+0.02}$ & $8.57_{-0.14}^{+0.15}$ & $1.50_{-0.51}^{+0.59}$ & $14.78_{-5.48}^{+9.31}$ \\ 
 M0416 & 00519 & 64.032451 & -24.068482 & 2.09 & $-0.037\pm0.025$ & 21.52 & 0.89 & $10.61\pm1.08$ & \nodata & $4.81\pm0.94$ & $45.24\pm1.11$ & \nodata & \nodata & $4.74_{-2.63}^{+11.66}$ & $9.85_{-0.54}^{+0.35}$ & $210.95_{-149.96}^{+263.50}$ & $1.80_{-0.00}^{+0.00}$ & $0.16_{-0.07}^{+0.08}$ & $8.03_{-0.12}^{+0.10}$ & $<0.65$ & $11.13_{-9.16}^{+44.59}$ \\ 
 M0416 & 00572 & 64.047481 & -24.068851 & 1.90 & $0.045\pm0.047$ & 22.94 & 0.72 & $9.32\pm0.78$ & $5.94\pm4.44$ & $7.47\pm1.05$ & $26.96\pm1.13$ & \nodata & \nodata & $3.31_{-0.08}^{+0.09}$ & $8.72_{-0.03}^{+0.06}$ & $14.45_{-2.19}^{+3.35}$ & $0.80_{-0.03}^{+0.10}$ & $0.05_{-0.00}^{+0.00}$ & $8.25_{-0.22}^{+0.15}$ & $<1.22$ & $21.50_{-12.22}^{+36.41}$ \\ 
 M0416 & 00889 & 64.026197 & -24.074377 & 1.63 & $0.054\pm0.038$ & 22.91 & 0.83 & $8.79\pm0.42$ & $2.39\pm0.37$ & $2.71\pm0.48$ & $18.55\pm0.55$ & \nodata & \nodata & $3.22_{-0.04}^{+0.03}$ & $8.72_{-0.10}^{+0.09}$ & $7.45_{-4.23}^{+2.96}$ & $0.40_{-0.20}^{+0.12}$ & $0.06_{-0.01}^{+0.00}$ & $8.27_{-0.12}^{+0.11}$ & $<0.55$ & $5.50_{-2.06}^{+6.52}$ \\ 
 M0416 & 00932 & 64.018102 & -24.075620 & 2.18 & $-0.082\pm0.073$ & 22.16 & 0.92 & $13.67\pm1.33$ & \nodata & $4.87\pm0.95$ & $17.30\pm1.08$ & \nodata & \nodata & $1.86_{-0.02}^{+0.02}$ & $10.03_{-0.15}^{+0.06}$ & $74.38_{-72.49}^{+31.79}$ & $1.10_{-0.52}^{+0.16}$ & $0.19_{-0.03}^{+0.02}$ & $8.45_{-0.13}^{+0.11}$ & $<0.94$ & $43.12_{-24.74}^{+88.65}$ \\ 
 M0416 & 00955 & 64.041852 & -24.075813 & 1.99 & $-0.079\pm0.014$ & 23.12 & 0.36 & $7.32\pm0.44$ & $3.52\pm1.09$ & $6.88\pm0.60$ & $57.38\pm0.71$ & \nodata & \nodata & $2.81_{-0.05}^{+0.05}$ & $8.15_{-0.08}^{+0.01}$ & $4.18_{-0.84}^{+0.08}$ & $0.10_{-0.10}^{+0.01}$ & $0.04_{-0.00}^{+0.00}$ & $7.95_{-0.13}^{+0.12}$ & $1.02_{-0.66}^{+0.68}$ & $32.07_{-16.46}^{+36.99}$ \\ 
 M0416 & 01047 & 64.021434 & -24.078456 & 2.22 & $-0.044\pm0.040$ & 22.88 & 0.96 & $4.64\pm0.89$ & $0.05\pm0.65$ & $1.80\pm0.60$ & $14.75\pm0.72$ & \nodata & \nodata & $2.85_{-0.03}^{+0.05}$ & $9.38_{-0.11}^{+0.08}$ & $8.04_{-1.90}^{+2.43}$ & $0.20_{-0.19}^{+0.14}$ & $0.09_{-0.01}^{+0.01}$ & $7.97_{-0.31}^{+0.28}$ & $2.29_{-1.47}^{+1.59}$ & $55.56_{-44.42}^{+242.09}$ \\ 
 M0416 & 01197 & 64.023302 & -24.081517 & 2.23 & $-0.083\pm0.027$ & 22.45 & 0.94 & $9.61\pm1.45$ & $3.83\pm1.08$ & $5.02\pm1.02$ & $40.99\pm1.17$ & \nodata & \nodata & $5.11_{-0.12}^{+0.10}$ & $9.17_{-0.09}^{+0.04}$ & $0.07_{-0.00}^{+38.09}$ & $0.00_{-0.00}^{+1.00}$ & $0.08_{-0.00}^{+0.00}$ & $7.97_{-0.20}^{+0.18}$ & $<1.00$ & $<27.87$ \\ 
 M0416 & 01213 & 64.035828 & -24.081321 & 1.63 & $-0.052\pm0.048$ & 22.65 & 0.77 & $11.39\pm0.83$ & $2.91\pm0.91$ & $4.24\pm0.70$ & $30.81\pm0.77$ & \nodata & \nodata & $3.33_{-0.06}^{+0.07}$ & $8.74_{-0.01}^{+0.08}$ & $1.28_{-0.75}^{+0.02}$ & $0.10_{-0.10}^{+0.07}$ & $0.06_{-0.00}^{+0.00}$ & $8.17_{-0.13}^{+0.12}$ & $<0.55$ & $7.08_{-2.67}^{+8.81}$ \\ 
 M0717 & 01007 & 109.388805 & 37.752159 & 1.85 & $-0.065\pm0.034$ & 22.23 & 0.84 & $8.01\pm1.20$ & $16.05\pm2.03$ & $3.81\pm1.68$ & $32.95\pm1.70$ & \nodata & \nodata & $6.96_{-0.55}^{+0.62}$ & $9.00_{-0.11}^{+0.11}$ & $29.32_{-12.33}^{+2.49}$ & $1.30_{-0.18}^{+0.02}$ & $0.07_{-0.01}^{+0.01}$ & $7.96_{-0.24}^{+0.21}$ & $<1.21$ & $7.27_{-4.10}^{+13.98}$ \\ 
 M0717 & 01131 & 109.381097 & 37.750440 & 1.85 & $-0.055\pm0.028$ & 22.07 & 0.75 & $27.83\pm0.95$ & \nodata & $12.00\pm1.43$ & $53.78\pm1.30$ & \nodata & \nodata & $7.86_{-0.96}^{+1.04}$ & $8.95_{-0.15}^{+0.19}$ & $6.23_{-5.45}^{+17.82}$ & $0.90_{-0.30}^{+0.24}$ & $0.07_{-0.01}^{+0.02}$ & $8.30_{-0.11}^{+0.10}$ & $<0.53$ & $9.07_{-3.87}^{+11.67}$ \\ 
 M0717 & 01636 & 109.376338 & 37.744601 & 1.85 & $-0.059\pm0.040$ & 22.97 & 0.78 & $12.42\pm0.57$ & $1.12\pm0.90$ & $3.41\pm0.71$ & $22.03\pm0.78$ & \nodata & \nodata & $3.52_{-0.09}^{+0.13}$ & $9.04_{-0.18}^{+0.11}$ & $13.91_{-13.08}^{+10.92}$ & $1.00_{-0.42}^{+0.21}$ & $0.08_{-0.02}^{+0.01}$ & $8.30_{-0.11}^{+0.10}$ & $<0.47$ & $7.29_{-2.58}^{+8.48}$ \\ 
 M0717 & 02043 & 109.394456 & 37.739171 & 1.86 & $-0.200\pm0.076$ & 24.49 & 0.21 & $0.07\pm0.59$ & $1.16\pm1.26$ & $3.57\pm0.93$ & $19.27\pm0.81$ & \nodata & \nodata & $14.96_{-1.33}^{+1.25}$ & $6.93_{-0.03}^{+0.04}$ & $0.25_{-0.02}^{+0.02}$ & $0.00_{-0.00}^{+0.00}$ & $0.05_{-0.00}^{+0.00}$ & $7.70_{-0.24}^{+0.29}$ & $2.94_{-1.79}^{+1.63}$ & $17.16_{-14.80}^{+88.69}$ \\ 
 M0717 & 02064 & 109.413018 & 37.738458 & 2.08 & $0.044\pm0.047$ & 23.42 & 0.64 & $15.24\pm1.29$ & \nodata & $3.92\pm1.18$ & $24.27\pm1.20$ & \nodata & \nodata & $5.69_{-0.39}^{+0.42}$ & $8.23_{-0.05}^{+0.15}$ & $4.73_{-3.37}^{+0.46}$ & $0.60_{-0.20}^{+0.07}$ & $0.04_{-0.00}^{+0.00}$ & $8.31_{-0.11}^{+0.10}$ & $<0.49$ & $6.92_{-2.65}^{+8.50}$ \\ 
 M0744 & 00920 & 116.212401 & 39.460342 & 1.28 & $-0.194\pm0.077$ & 20.81 & 0.95 & $13.04\pm6.32$ & $0.01\pm1.30$ & $6.74\pm1.76$ & $41.30\pm1.25$ & $42.86\pm1.22$ & $9.02\pm1.21$ & $10.83_{-0.76}^{+1.01}$ & $9.33_{-0.06}^{+0.15}$ & $3.20_{-1.83}^{+6.28}$ & $0.80_{-0.21}^{+0.27}$ & $0.11_{-0.01}^{+0.03}$ & $7.99_{-0.28}^{+0.22}$ & $1.75_{-0.86}^{+0.79}$ & $5.76_{-3.50}^{+7.20}$ \\ 
 M0744 & 01031 & 116.212064 & 39.459430 & 1.27 & $0.178\pm0.242$ & 20.94 & 0.93 & $24.57\pm2.77$ & $6.57\pm1.25$ & $20.09\pm0.93$ & $36.40\pm1.09$ & $48.45\pm1.15$ & $17.99\pm1.13$ & $6.31_{-0.31}^{+0.36}$ & $9.58_{-0.10}^{+0.15}$ & $4.79_{-1.80}^{+9.09}$ & $0.70_{-0.11}^{+0.32}$ & $0.16_{-0.02}^{+0.03}$ & $8.52_{-0.13}^{+0.11}$ & $1.63_{-0.97}^{+0.78}$ & $22.33_{-14.78}^{+32.02}$ \\ 
 M0744 & 01203 & 116.197585 & 39.456699 & 1.65 & $0.111\pm0.017$ & 21.68 & 0.64 & $34.00\pm0.96$ & $7.06\pm1.00$ & $17.46\pm1.06$ & $117.66\pm1.17$ & \nodata & \nodata & $2.25_{-0.03}^{+0.03}$ & $9.41_{-0.01}^{+0.01}$ & $75.31_{-1.11}^{+1.04}$ & $1.00_{-0.00}^{+0.00}$ & $0.12_{-0.00}^{+0.00}$ & $8.10_{-0.12}^{+0.10}$ & $0.89_{-0.20}^{+0.28}$ & $47.74_{-3.06}^{+5.28}$ \\ 
 M0744 & 01370 & 116.236950 & 39.454058 & 1.37 & $-0.116\pm0.111$ & 22.71 & 0.94 & $5.59\pm1.52$ & \nodata & $2.22\pm1.18$ & $12.37\pm1.39$ & $10.19\pm0.81$ & $2.38\pm0.80$ & $1.55_{-0.02}^{+0.01}$ & $9.56_{-0.06}^{+0.16}$ & $3.71_{-3.69}^{+17.90}$ & $1.10_{-0.32}^{+0.34}$ & $0.15_{-0.01}^{+0.03}$ & $8.29_{-0.17}^{+0.13}$ & $0.54_{-0.37}^{+0.54}$ & $4.57_{-1.24}^{+2.63}$ \\ 
 M0744 & 01734 & 116.219106 & 39.449158 & 1.49 & $0.085\pm0.034$ & 22.62 & 0.77 & $16.67\pm1.45$ & $3.16\pm0.92$ & $7.48\pm1.86$ & $30.72\pm1.53$ & $19.46\pm1.08$ & $10.71\pm2.31$ & $1.63_{-0.01}^{+0.02}$ & $9.17_{-0.01}^{+0.14}$ & $43.51_{-43.43}^{+0.30}$ & $1.10_{-1.00}^{+0.05}$ & $0.09_{-0.00}^{+0.02}$ & $8.25_{-0.11}^{+0.09}$ & $<0.17$ & $8.97_{-1.31}^{+2.11}$ \\ 
 M0744 & 02341 & 116.212719 & 39.441062 & 1.28 & $0.031\pm0.044$ & 22.60 & 0.93 & $4.92\pm1.55$ & $0.69\pm0.68$ & $3.38\pm0.78$ & $11.49\pm0.71$ & $10.74\pm0.65$ & $3.20\pm0.63$ & $1.13_{-0.00}^{+0.00}$ & $9.47_{-0.07}^{+0.11}$ & $3.01_{-2.88}^{+19.34}$ & $0.50_{-0.30}^{+0.42}$ & $0.14_{-0.01}^{+0.02}$ & $8.27_{-0.25}^{+0.14}$ & $0.63_{-0.41}^{+0.55}$ & $6.17_{-1.86}^{+3.65}$ \\ 
 M1423 & 00410 & 215.941735 & 24.088841 & 1.79 & $0.104\pm0.054$ & 23.19 & 0.77 & $7.50\pm0.47$ & $3.10\pm0.71$ & $3.08\pm0.54$ & $18.78\pm0.64$ & \nodata & \nodata & $1.46_{-0.01}^{+0.02}$ & $9.03_{-0.07}^{+0.11}$ & $20.72_{-18.14}^{+6.12}$ & $0.60_{-0.33}^{+0.11}$ & $0.08_{-0.01}^{+0.01}$ & $8.24_{-0.14}^{+0.12}$ & $<0.59$ & $15.47_{-6.11}^{+25.95}$ \\ 
 M1423 & 00433 & 215.954478 & 24.088107 & 1.63 & $-0.095\pm0.046$ & 23.00 & 0.80 & $7.83\pm0.85$ & $0.47\pm0.61$ & $4.28\pm1.70$ & $18.45\pm1.03$ & \nodata & \nodata & $3.64_{-0.13}^{+0.06}$ & $8.94_{-0.16}^{+0.15}$ & $10.22_{-10.03}^{+8.63}$ & $1.00_{-0.63}^{+0.28}$ & $0.08_{-0.01}^{+0.01}$ & $8.27_{-0.21}^{+0.15}$ & $<1.03$ & $8.97_{-5.32}^{+20.72}$ \\ 
 M1423 & 00493 & 215.952044 & 24.083337 & 1.78 & $-0.028\pm0.089$ & 21.75 & 0.93 & $13.40\pm2.17$ & $3.05\pm3.47$ & $19.79\pm2.00$ & $18.92\pm2.61$ & \nodata & \nodata & $5.14_{-0.20}^{+0.42}$ & $9.66_{-0.24}^{+0.24}$ & $6.60_{-6.60}^{+9.50}$ & $0.60_{-0.44}^{+0.40}$ & $0.14_{-0.03}^{+0.05}$ & $8.52_{-0.15}^{+0.13}$ & $1.67_{-0.80}^{+0.39}$ & $>26.26$ \\ 
 M1423 & 00908 & 215.957460 & 24.079767 & 1.31 & $-0.233\pm0.104$ & 21.91 & 0.96 & $12.88\pm4.10$ & \nodata & $4.07\pm2.50$ & $8.01\pm3.07$ & $17.25\pm1.47$ & \nodata & $2.12_{-0.01}^{+0.03}$ & $9.54_{-0.12}^{+0.34}$ & $12.11_{-11.30}^{+78.23}$ & $1.20_{-0.70}^{+0.51}$ & $0.15_{-0.02}^{+0.09}$ & $8.60_{-0.20}^{+0.17}$ & $0.75_{-0.49}^{+0.75}$ & $5.81_{-2.14}^{+4.91}$ \\ 
 M1423 & 00909 & 215.957063 & 24.079622 & 1.32 & $-0.260\pm0.085$ & 23.08 & 0.86 & $16.21\pm1.76$ & $3.73\pm0.82$ & $3.47\pm1.24$ & $12.18\pm1.48$ & $14.55\pm0.78$ & $4.38\pm0.78$ & $2.21_{-0.02}^{+0.04}$ & $8.72_{-0.12}^{+0.15}$ & $6.09_{-5.65}^{+6.44}$ & $0.60_{-0.43}^{+0.26}$ & $0.06_{-0.01}^{+0.01}$ & $8.51_{-0.09}^{+0.08}$ & $<0.28$ & $3.66_{-0.64}^{+1.09}$ \\ 
 M1423 & 00976 & 215.938230 & 24.078815 & 1.24 & $-0.023\pm0.059$ & 22.85 & 0.89 & $9.56\pm1.47$ & $1.96\pm0.60$ & $4.02\pm0.46$ & $17.47\pm0.53$ & $15.17\pm0.57$ & $2.78\pm0.55$ & $1.45_{-0.01}^{+0.01}$ & $9.00_{-0.11}^{+0.17}$ & $3.54_{-2.72}^{+22.83}$ & $0.50_{-0.25}^{+0.50}$ & $0.09_{-0.01}^{+0.01}$ & $8.30_{-0.11}^{+0.09}$ & $0.44_{-0.26}^{+0.29}$ & $5.49_{-1.14}^{+1.61}$ \\ 
 M1423 & 01564 & 215.943001 & 24.069814 & 1.41 & $0.207\pm0.104$ & 23.84 & 0.79 & $5.95\pm0.77$ & $1.94\pm0.43$ & $2.27\pm1.21$ & $9.01\pm0.76$ & $5.97\pm0.48$ & $1.67\pm0.48$ & $2.85_{-0.07}^{+0.04}$ & $8.68_{-0.07}^{+0.01}$ & $0.02_{-0.02}^{+0.24}$ & $0.60_{-0.25}^{+0.12}$ & $0.06_{-0.00}^{+0.00}$ & $8.33_{-0.12}^{+0.10}$ & $<0.22$ & $1.53_{-0.28}^{+0.48}$ \\ 
 M1423 & 01682 & 215.959334 & 24.067604 & 1.32 & $-0.000\pm0.038$ & 23.80 & 0.90 & $8.07\pm1.04$ & $1.59\pm0.47$ & $4.30\pm1.63$ & $17.49\pm0.89$ & $6.85\pm0.47$ & $0.48\pm0.47$ & $1.28_{-0.00}^{+0.01}$ & $8.62_{-0.08}^{+0.10}$ & $2.91_{-1.95}^{+8.45}$ & $0.50_{-0.22}^{+0.31}$ & $0.06_{-0.00}^{+0.01}$ & $8.09_{-0.17}^{+0.13}$ & $<0.16$ & $3.35_{-0.47}^{+0.72}$ \\ 
 M1423 & 01808 & 215.960238 & 24.065007 & 2.02 & $0.035\pm0.053$ & 23.33 & 0.87 & $4.28\pm0.36$ & $0.90\pm0.52$ & $1.92\pm0.79$ & $8.75\pm0.54$ & \nodata & \nodata & $1.28_{-0.00}^{+0.01}$ & $9.44_{-0.13}^{+0.11}$ & $14.51_{-14.01}^{+12.61}$ & $0.60_{-0.48}^{+0.20}$ & $0.11_{-0.01}^{+0.01}$ & $8.33_{-0.15}^{+0.13}$ & $1.04_{-0.73}^{+1.08}$ & $21.46_{-12.51}^{+56.60}$ \\ 
 M2129 & 00340 & 322.352173 & -7.679022 & 2.09 & $-0.021\pm0.018$ & 23.73 & 0.81 & $1.95\pm0.89$ & $1.08\pm0.64$ & $0.64\pm0.57$ & $10.54\pm0.65$ & \nodata & \nodata & $1.46_{-0.01}^{+0.02}$ & $8.94_{-0.04}^{+0.17}$ & $20.70_{-13.92}^{+6.83}$ & $0.80_{-0.28}^{+0.10}$ & $0.07_{-0.00}^{+0.01}$ & $7.84_{-0.38}^{+0.40}$ & $1.75_{-1.16}^{+1.18}$ & $43.84_{-31.85}^{+112.18}$ \\ 
 M2129 & 00380 & 322.364762 & -7.679957 & 1.81 & $-0.074\pm0.036$ & 24.55 & 0.67 & $1.50\pm0.39$ & $2.06\pm0.58$ & $2.22\pm0.77$ & $6.94\pm0.54$ & \nodata & \nodata & $1.56_{-0.01}^{+0.02}$ & $8.23_{-0.07}^{+0.17}$ & $3.94_{-3.46}^{+1.05}$ & $0.40_{-0.31}^{+0.11}$ & $0.05_{-0.00}^{+0.00}$ & $7.89_{-0.52}^{+0.41}$ & $1.47_{-0.94}^{+1.36}$ & $16.98_{-10.93}^{+56.16}$ \\ 
 M2129 & 00440 & 322.366118 & -7.681231 & 1.36 & $0.034\pm0.050$ & 22.16 & 0.88 & $8.43\pm1.81$ & \nodata & $6.34\pm2.65$ & $8.86\pm1.97$ & $33.17\pm1.40$ & $6.90\pm1.34$ & $1.53_{-0.01}^{+0.02}$ & $9.63_{-0.11}^{+0.13}$ & $59.57_{-56.40}^{+42.05}$ & $1.50_{-0.53}^{+0.22}$ & $0.16_{-0.02}^{+0.03}$ & $8.63_{-0.12}^{+0.14}$ & $2.05_{-0.42}^{+0.41}$ & $44.89_{-13.01}^{+17.40}$ \\ 
 M2129 & 00465 & 322.367391 & -7.681865 & 1.36 & $0.036\pm0.027$ & 22.34 & 0.88 & $7.06\pm1.23$ & $1.81\pm0.62$ & $3.08\pm0.85$ & $1.27\pm1.15$ & $27.21\pm0.70$ & $5.59\pm0.68$ & $1.57_{-0.01}^{+0.02}$ & $9.71_{-0.08}^{+0.08}$ & $2.92_{-2.70}^{+99.02}$ & $1.20_{-0.25}^{+0.80}$ & $0.18_{-0.02}^{+0.02}$ & $8.90_{-0.15}^{+0.13}$ & $2.26_{-0.55}^{+0.58}$ & $42.01_{-14.75}^{+24.05}$ \\ 
 M2129 & 00565 & 322.364752 & -7.683770 & 1.37 & $0.064\pm0.064$ & 22.34 & 0.92 & $10.16\pm2.04$ & $1.77\pm1.00$ & $12.87\pm2.50$ & $15.41\pm1.84$ & $17.05\pm1.14$ & $5.64\pm1.14$ & $2.00_{-0.02}^{+0.02}$ & $9.46_{-0.16}^{+0.26}$ & $1.65_{-1.64}^{+41.19}$ & $0.70_{-0.36}^{+0.77}$ & $0.13_{-0.02}^{+0.05}$ & $8.33_{-0.39}^{+0.15}$ & $<0.42$ & $5.60_{-1.34}^{+2.83}$ \\ 
 M2129 & 00690 & 322.346464 & -7.685883 & 1.88 & $0.049\pm0.043$ & 23.08 & 0.78 & $9.45\pm0.59$ & $0.70\pm0.95$ & $4.85\pm0.76$ & $18.93\pm0.82$ & \nodata & \nodata & $2.73_{-0.07}^{+0.02}$ & $8.85_{-0.05}^{+0.10}$ & $15.65_{-10.14}^{+6.52}$ & $0.80_{-0.20}^{+0.10}$ & $0.06_{-0.00}^{+0.01}$ & $8.33_{-0.14}^{+0.11}$ & $<1.01$ & $15.12_{-7.66}^{+27.89}$ \\ 
 M2129 & 01408 & 322.369038 & -7.700541 & 1.48 & $0.087\pm0.125$ & 23.65 & 0.83 & $4.86\pm0.70$ & $1.39\pm0.45$ & $2.27\pm0.73$ & $8.91\pm0.79$ & $8.43\pm0.57$ & $4.23\pm0.76$ & $1.49_{-0.02}^{+0.02}$ & $8.98_{-0.09}^{+0.16}$ & $16.52_{-16.29}^{+9.96}$ & $1.20_{-0.73}^{+0.21}$ & $0.08_{-0.00}^{+0.01}$ & $8.32_{-0.17}^{+0.15}$ & $<0.48$ & $4.81_{-1.27}^{+2.41}$ \\ 
 M2129 & 01539 & 322.363350 & -7.703212 & 1.64 & $0.021\pm0.027$ & 22.08 & 0.94 & $13.52\pm0.95$ & $0.45\pm1.21$ & $6.47\pm1.12$ & $9.61\pm1.12$ & \nodata & \nodata & $1.39_{-0.01}^{+0.01}$ & $9.93_{-0.12}^{+0.08}$ & $45.29_{-44.93}^{+59.20}$ & $1.00_{-0.50}^{+0.30}$ & $0.21_{-0.03}^{+0.03}$ & $8.63_{-0.12}^{+0.13}$ & $1.09_{-0.72}^{+0.75}$ & $31.57_{-19.34}^{+55.80}$ \\ 
 M2129 & 01665 & 322.358149 & -7.706707 & 1.56 & $-0.028\pm0.030$ & 23.26 & 0.79 & $6.84\pm0.68$ & $1.91\pm0.47$ & $3.74\pm0.78$ & $15.71\pm0.85$ & $4.35\pm1.29$ & \nodata & $1.40_{-0.01}^{+0.01}$ & $9.31_{-0.21}^{+0.11}$ & $6.99_{-6.93}^{+7.03}$ & $0.50_{-0.43}^{+0.36}$ & $0.11_{-0.03}^{+0.01}$ & $8.18_{-0.16}^{+0.13}$ & $<0.28$ & $6.23_{-1.41}^{+2.77}$ \\ 
 M2129 & 01833 & 322.362724 & -7.709921 & 2.30 & $0.034\pm0.011$ & 22.45 & 0.91 & $20.08\pm1.32$ & $3.42\pm0.92$ & $8.64\pm0.96$ & $47.71\pm1.15$ & \nodata & \nodata & $1.27_{-0.01}^{+0.01}$ & $9.76_{-0.03}^{+0.05}$ & $150.11_{-16.90}^{+39.63}$ & $1.00_{-0.01}^{+0.10}$ & $0.14_{-0.00}^{+0.01}$ & $8.24_{-0.13}^{+0.11}$ & $<0.65$ & $87.00_{-36.79}^{+109.28}$ \\ 
 RXJ1347 & 00664 & 206.904786 & -11.750602 & 1.69 & $-0.090\pm0.040$ & 23.01 & 0.48 & $12.64\pm0.65$ & $4.88\pm1.09$ & $5.93\pm0.79$ & $51.74\pm0.95$ & \nodata & \nodata & $1.40_{-0.07}^{+0.11}$ & $8.81_{-0.03}^{+0.02}$ & $1.52_{-0.11}^{+0.09}$ & $0.00_{-0.00}^{+0.00}$ & $0.06_{-0.00}^{+0.00}$ & $8.08_{-0.15}^{+0.14}$ & $<0.85$ & $34.43_{-16.28}^{+54.74}$ \\ 
 RXJ1347 & 01443 & 206.882540 & -11.764358 & 1.77 & $-0.028\pm0.038$ & 21.75 & 0.93 & $21.20\pm0.96$ & \nodata & $6.34\pm1.21$ & $18.50\pm1.30$ & \nodata & \nodata & $3.36_{-0.39}^{+0.70}$ & $9.49_{-0.09}^{+0.19}$ & $0.77_{-0.64}^{+46.86}$ & $0.30_{-0.24}^{+0.62}$ & $0.12_{-0.01}^{+0.03}$ & $8.53_{-0.11}^{+0.10}$ & $<0.63$ & $12.80_{-6.71}^{+20.97}$ \\ 
 RXJ2248 & 00206 & 342.175906 & -44.516503 & 1.43 & $-0.020\pm0.104$ & 21.53 & 0.97 & $14.31\pm1.82$ & $0.36\pm1.05$ & $7.84\pm1.80$ & $7.13\pm1.72$ & $12.87\pm1.05$ & \nodata & $1.83_{-0.02}^{+0.01}$ & $9.87_{-0.03}^{+0.04}$ & $0.05_{-0.05}^{+3.05}$ & $0.50_{-0.20}^{+0.25}$ & $0.22_{-0.01}^{+0.01}$ & $8.60_{-0.11}^{+0.10}$ & $<0.23$ & $4.09_{-0.73}^{+1.27}$ \\ 
 RXJ2248 & 00331 & 342.169955 & -44.518922 & 1.93 & $0.004\pm0.029$ & 23.29 & 0.79 & $9.03\pm0.65$ & $2.54\pm0.99$ & $2.96\pm1.29$ & $15.17\pm0.90$ & \nodata & \nodata & $1.86_{-0.02}^{+0.02}$ & $9.01_{-0.04}^{+0.20}$ & $0.58_{-0.35}^{+9.53}$ & $0.00_{-0.00}^{+0.58}$ & $0.07_{-0.00}^{+0.02}$ & $8.36_{-0.13}^{+0.11}$ & $<0.72$ & $16.56_{-7.66}^{+27.33}$ \\ 
 RXJ2248 & 00428 & 342.186462 & -44.521187 & 1.23 & $0.052\pm0.036$ & 22.19 & 0.96 & $8.61\pm5.13$ & $0.95\pm1.51$ & $10.12\pm1.11$ & $24.10\pm1.15$ & $11.30\pm1.33$ & \nodata & $4.97_{-0.09}^{+0.08}$ & $9.13_{-0.10}^{+0.05}$ & $1.92_{-0.42}^{+0.54}$ & $0.20_{-0.15}^{+0.11}$ & $0.10_{-0.01}^{+0.00}$ & $8.05_{-0.24}^{+0.17}$ & $<0.19$ & $1.31_{-0.22}^{+0.36}$ \\ 
 RXJ2248 & 00786 & 342.169402 & -44.527222 & 1.84 & $-0.135\pm0.070$ & 23.81 & 0.46 & $7.24\pm0.67$ & $2.04\pm0.97$ & $2.77\pm0.76$ & $25.73\pm0.86$ & \nodata & \nodata & $3.38_{-0.06}^{+0.07}$ & $7.96_{-0.01}^{+0.03}$ & $2.70_{-0.17}^{+0.05}$ & $0.30_{-0.01}^{+0.00}$ & $0.05_{-0.00}^{+0.00}$ & $8.12_{-0.15}^{+0.13}$ & $<0.73$ & $9.32_{-4.22}^{+15.70}$ \\ 
 RXJ2248 & 01250 & 342.192946 & -44.536578 & 1.40 & $-0.068\pm0.094$ & 22.52 & 0.88 & $8.96\pm2.06$ & \nodata & $0.25\pm2.06$ & $15.49\pm1.92$ & $17.06\pm1.11$ & $4.18\pm1.13$ & $3.14_{-0.05}^{+0.04}$ & $9.17_{-0.06}^{+0.08}$ & $0.16_{-0.16}^{+1.02}$ & $0.20_{-0.20}^{+0.33}$ & $0.09_{-0.01}^{+0.01}$ & $8.41_{-0.12}^{+0.12}$ & $1.14_{-0.54}^{+0.62}$ & $6.71_{-2.47}^{+4.51}$ \\ 
 M1149 & 00270 & 177.385990 & 22.414074 & 1.27 & $-0.160\pm0.030$ & 22.55 & 0.71 & $10.28\pm5.23$ & $4.84\pm8.65$ & $24.87\pm0.94$ & $29.58\pm0.72$ & $13.33\pm0.36$ & $0.36\pm0.35$ & $1.96_{-0.03}^{+0.03}$ & $9.00_{-0.04}^{+0.12}$ & $0.24_{-0.20}^{+13.40}$ & $0.20_{-0.20}^{+0.60}$ & $0.09_{-0.00}^{+0.01}$ & $7.91_{-0.15}^{+0.14}$ & $<0.03$ & $3.04_{-0.13}^{+0.15}$ \\ 
 M1149 & 00593 & 177.406922 & 22.407499 & 1.48 & $-0.010\pm0.020$ & 22.40 & 0.81 & $23.44\pm1.40$ & $6.00\pm0.85$ & $5.72\pm0.44$ & $31.48\pm0.45$ & $29.90\pm0.34$ & $5.31\pm0.41$ & $2.02_{-0.03}^{+0.04}$ & $9.13_{-0.01}^{+0.09}$ & $39.40_{-39.32}^{+0.65}$ & $1.00_{-1.00}^{+0.03}$ & $0.09_{-0.00}^{+0.01}$ & $8.38_{-0.06}^{+0.06}$ & $0.93_{-0.14}^{+0.14}$ & $17.50_{-2.10}^{+2.32}$ \\ 
 M1149 & 00600 & 177.389220 & 22.407583 & 2.31 & $-0.180\pm0.080$ & 24.12 & 0.76 & $1.01\pm0.38$ & $1.33\pm0.29$ & $1.40\pm0.27$ & $8.31\pm0.31$ & \nodata & \nodata & $6.87_{-0.29}^{+0.33}$ & $8.32_{-0.03}^{+0.09}$ & $0.90_{-0.06}^{+0.04}$ & $0.00_{-0.00}^{+0.01}$ & $0.04_{-0.00}^{+0.00}$ & $7.83_{-0.23}^{+0.28}$ & $<0.85$ & $3.10_{-1.55}^{+6.36}$ \\ 
 M1149 & 00676 & 177.415126 & 22.406195 & 1.68 & $0.060\pm0.050$ & 23.31 & 0.78 & $7.48\pm0.77$ & $2.58\pm0.29$ & $2.57\pm0.19$ & $16.16\pm0.22$ & \nodata & \nodata & $1.77_{-0.04}^{+0.04}$ & $8.93_{-0.07}^{+0.16}$ & $18.72_{-17.88}^{+5.91}$ & $0.90_{-0.50}^{+0.10}$ & $0.07_{-0.01}^{+0.01}$ & $8.19_{-0.10}^{+0.09}$ & $<0.18$ & $4.91_{-0.69}^{+1.66}$ \\ 
 M1149 & 00683 & 177.397234 & 22.406181 & 1.68 & $-0.220\pm0.050$ & 24.41 & 0.10 & $3.80\pm1.02$ & $6.97\pm0.48$ & $6.05\pm0.33$ & $28.23\pm0.33$ & \nodata & \nodata & $5.83_{-0.26}^{+0.23}$ & $6.90_{-0.02}^{+0.02}$ & $0.24_{-0.01}^{+0.01}$ & $0.00_{-0.00}^{+0.00}$ & $0.05_{-0.00}^{+0.00}$ & $7.58_{-0.14}^{+0.34}$ & $<0.10$ & $3.20_{-0.37}^{+0.67}$ \\ 
 M1149 & 00862 & 177.403416 & 22.402433 & 1.49 & $-0.040\pm0.020$ & 21.41 & 0.97 & $19.22\pm3.17$ & \nodata & $11.35\pm1.08$ & $5.96\pm1.03$ & $45.12\pm0.70$ & \nodata & $3.63_{-0.06}^{+0.08}$ & $9.79_{-0.15}^{+0.10}$ & $24.57_{-20.09}^{+8.35}$ & $1.00_{-0.31}^{+0.16}$ & $0.19_{-0.04}^{+0.03}$ & $8.96_{-0.08}^{+0.06}$ & $0.72_{-0.29}^{+0.29}$ & $11.96_{-2.56}^{+3.23}$ \\ 
 M1149 & 01058 & 177.391848 & 22.400105 & 1.25 & $-0.030\pm0.030$ & 22.61 & 0.33 & $21.25\pm2.59$ & $16.58\pm1.09$ & $13.49\pm0.69$ & $94.78\pm0.56$ & $40.66\pm0.29$ & $2.06\pm0.28$ & $3.83_{-0.06}^{+0.07}$ & $8.10_{-0.08}^{+0.01}$ & $3.69_{-0.60}^{+0.06}$ & $0.10_{-0.10}^{+0.02}$ & $0.05_{-0.00}^{+0.00}$ & $7.96_{-0.10}^{+0.10}$ & $<0.02$ & $4.31_{-0.13}^{+0.19}$ \\ 
 M1149 & 01322 & 177.392541 & 22.394921 & 1.49 & $-0.010\pm0.020$ & 23.33 & 0.78 & $0.20\pm1.68$ & $5.35\pm0.96$ & $0.82\pm0.61$ & $16.73\pm0.62$ & $7.14\pm0.41$ & $8.37\pm0.80$ & $2.88_{-0.04}^{+0.06}$ & $8.84_{-0.01}^{+0.02}$ & $20.93_{-0.87}^{+0.30}$ & $1.60_{-0.01}^{+0.00}$ & $0.07_{-0.00}^{+0.00}$ & $8.21_{-0.08}^{+0.07}$ & $<0.18$ & $1.78_{-0.21}^{+0.40}$ \\ 
 M1149 & 01468 & 177.406546 & 22.392860 & 1.89 & $-0.070\pm0.040$ & 23.54 & 0.37 & $2.71\pm0.80$ & $3.29\pm0.39$ & $2.58\pm0.30$ & $39.69\pm0.31$ & \nodata & \nodata & $56.29_{-6.27}^{+8.89}$ & $6.22_{-0.06}^{+0.05}$ & $0.05_{-0.01}^{+0.01}$ & $0.00_{-0.00}^{+0.00}$ & $0.05_{-0.00}^{+0.00}$ & $7.92_{-0.08}^{+0.08}$ & $<0.69$ & $0.49_{-0.24}^{+0.72}$ \\ 
 M1149 & 01704 & 177.398643 & 22.387499 & 2.28 & $0.020\pm0.080$ & 24.97 & 0.52 & $3.30\pm0.30$ & $0.83\pm0.22$ & $0.87\pm0.21$ & $7.86\pm0.26$ & \nodata & \nodata & $2.66_{-0.10}^{+0.07}$ & $7.71_{-0.08}^{+0.11}$ & $1.11_{-0.14}^{+0.24}$ & $0.00_{-0.00}^{+0.05}$ & $0.04_{-0.00}^{+0.00}$ & $8.18_{-0.10}^{+0.10}$ & $<0.36$ & $3.86_{-1.04}^{+3.49}$
\enddata
    \tablenotetext{a}{The observed \JH-band magnitude, before accounting for lensing magnification.}
    \tablenotetext{b}{The reduction factor of \JH-band flux, after subtracting the nebular emission that falls within
    the corresponding wavelength window.}
    \tablenotetext{c}{The lensing magnification estimated from various mass models of galaxy clusters.
    For all the \hff clusters, we use the \SJ version 4corr models \citep{Johnson:2014cf}.
    For all the \clash-only clusters except RXJ1347, we use the Zitrin PIEMD+eNFW version 2 models \citep{2015ApJ...801...44Z}.
    For RXJ1347, we use our own model built following closely the approach in \citet{Johnson:2014cf}.}
    \tablenotetext{d}{The flux ratio of \NII and \Ha, estimated from the prescription of \citet{Faisst:2017wv}.
    This value is used to correct for the blended \Ha-\NII fluxes in grism spectra, if necessary.}
    \tablenotetext{e}{Values presented here are corrected for lensing magnification.}
\label{tab:srcprop}
\end{deluxetable*}

\end{longrotatetable}

\bibliographystyle{apj}
\bibliography{bibtexlib}

\end{document}